\documentclass{aa} 
\usepackage{savesym} 
\usepackage{txfonts} 
\usepackage[english]{babel} 
\usepackage{t1enc} 
\usepackage{graphicx} 
\usepackage{pdflscape}
\usepackage{subfig}
\usepackage{fixltx2e} 
\usepackage{natbib} 
\usepackage{footmisc}
\usepackage{url} 
\usepackage{longtable}
\usepackage{multirow}
\usepackage[table]{xcolor}
\usepackage{caption}

\def \CH3CN {CH$_{3}$CN} 
\def \CH3OH {CH$_{3}$OH} 
\def \H2CO {H$_{2}$CO}
\def \C18O {C$^{18}$O}
\def \13CO {$^{13}$CO}
\def \12CO {$^{12}$CO}
\def \13CS {$^{13}$CS}

\begin{document} 
 
\title{Diversity of chemistry and excitation conditions\\ in the high-mass star forming complex W33\thanks{Integrated intensity line and continuum maps of the APEX telescope and the SMA are available as FITS files at the CDS via anonymous ftp to cdsarc.u-strasbg.fr (130.79.128.5)
or via http://cdsweb.u-strasbg.fr/cgi-bin/qcat?J/A+A/}}                                                    
\author{K. Immer\inst{1} \and R. Galv\'an-Madrid\inst{2,3} \and C. K\"onig\inst{1} 
 \and H.~B. Liu\inst{4} \and K.~M. Menten\inst{1}}                                             
 
\institute{ 
Max-Planck-Institut f\"ur Radioastronomie, Auf dem H\"ugel 69, D-53121
Bonn, Germany                                                           
\and 
European Organisation for Astronomical Research in the Southern Hemisphere (ESO), 
Karl-Schwarzschild-Straße 2, D-85748 Garching, Deutschland
\and
Centro de Radioastronom\'ia y Astrof\'isica, Universidad Nacional Aut\'onoma de M\'exico, Morelia 58090, Mexico
\and
Academia Sinica, Institute of Astronomy and Astrophysics, P.O. Box 23-141, Taipei 106, Taiwan, R.O.C.}                                                     
\date{Received xxxx; accepted xxxx} 
 
\abstract{The object W33 is a giant molecular cloud that contains star forming regions at various evolutionary stages from quiescent clumps to developed \ion{H}{ii} regions. Since its star forming regions are located at the same distance and the primary material of the birth clouds is probably similar, we conducted a comparative chemical study to trace the chemical footprint of the different phases of evolution. We observed six clumps in W33 with the Atacama Pathfinder Experiment (APEX) telescope at 280 GHz and the Submillimeter Array (SMA) at 230 GHz. We detected 27 transitions of 10 different molecules in the APEX data and 52 transitions of 16 different molecules in the SMA data. The chemistry on scales larger than $\sim$0.2 pc, which are traced by the APEX data, becomes more complex and diverse the more evolved the star forming region is. On smaller scales traced by the SMA data, the chemical complexity and diversity increase up to the hot core stage. In the \ion{H}{ii} region phase, the SMA spectra resemble the spectra of the protostellar phase. Either these more complex molecules are destroyed or their emission is not compact enough to be detected with the SMA. Synthetic spectra modelling of the H$_{2}$CO transitions, as detected with the APEX telescope, shows that both a warm and a cold component are needed to obtain a good fit to the emission for all sources except for W33\,Main1. The temperatures and column densities of the two components increase during the evolution of the star forming regions. The integrated intensity ratios N$_{2}$H$^{+}$(3$-$2)/CS(6$-$5) and N$_{2}$H$^{+}$(3$-$2)/H$_{2}$CO(4$_{2,2}$$-$3$_{2,1}$) show clear trends as a function of evolutionary stage, luminosity, luminosity-to-mass ratio, and H$_{2}$ peak column density of the clumps and might be usable as chemical clocks.}

 
\authorrunning{K. Immer et al.} 
\titlerunning{W33} 
 
\maketitle 
 
\section{Introduction}  
\label{Intro}
  
Over the past decades, a commonly accepted picture of massive star formation has emerged \citep[for recent reviews see][]{McKee2007, Zinnecker2007, Beuther2007,Tan2014}. The earliest stage is the gravitational collapse of a cold dark cloud, which forms a prestellar core of size 0.03--0.2 pc \citep{Bergin2007}. The prestellar core contracts further and a protostar emerges. With time, the protostar heats the surrounding material and a molecular hot core\footnote{In this publication, a hot core describes a structure of warm and dense gas around the new born star before the excitation of an \ion{H}{ii} region.} can be observed \citep[e.g.][]{Kurtz2000}. The protostar keeps accreting material through an accretion disk. The angular momentum of the infalling material is removed through outflows and likely viscosity in the accretion disk \citep{Lynden-Bell1974, Arce2007}. Furthermore, water, methanol, and hydroxyl masers are observed close to the protostar and in the outflows. When the protostar is energetic enough it starts to ionise its birth cloud \citep[e.g.][]{Wood1989}. Once the expansion of the ionised gas is not quenched anymore, a hypercompact \ion{H}{ii} region is observed, which expands into an ultracompact and then compact \ion{H}{ii} region \citep{Hoare2007}. When all the material around the star is destroyed or blown away by stellar winds, the high-mass star becomes detectable at optical wavelengths.

Studying the chemical evolution of star forming regions from young infrared dark clouds (IRDCs) to evolved \ion{H}{ii} regions yields insights into the chemical and physical processes at work during star formation. Spectral line surveys give information about the chemical composition, the temperature, the density, and the kinematical structures of the star forming material. Many spectral line surveys have been conducted over the last years, focusing on either complete spectral coverage of a certain region like Sgr B2 \citep[e.g.][]{Turner1991,Belloche2013} or Orion-KL \citep[e.g.][]{Turner1991,Beuther2005} or spectroscopically limited surveys of targets selected by their evolutionary stage (IRDCs, e.g. \citet{Vasyunina2014}, \citet{Sanhueza2012} or hot cores, e.g. \citet{Beuther2009}). Recently, \citet{Gerner2014} conducted a 1 mm and 3 mm spectral line 
survey of 59 high-mass star forming regions in four different evolutionary stages from IRDCs to ultra-compact (UC) \ion{H}{ii} regions. They found that the abundances of detected molecules tend to increase during the evolution of the star forming regions. In addition, they show that the overall detection rate of different molecules is much higher for hot cores than for UC \ion{H}{ii} regions.
However, the number of line surveys which cover different evolutionary stages in samples of high-mass star forming regions is still low and often these observed regions are not connected. They are located at different distances and originate from different giant molecular clouds (GMCs), which complicates a quantitative comparison of the detected chemical features due to different spatial scales and different initial star forming material probed. 

\subsection{The W33 complex}
\label{ResultsW33}

\begin{figure*}
	\caption{High-mass star forming complex W33 and its surroundings. The background images show infrared and radio continuum emission at 8 $\mu$m (panel a), 70 $\mu$m (panel b), and 90 cm (panel c), respectively. The contours show dust emission at 870 $\mu$m from the ATLASGAL survey (levels 1, 2, 4, 8, 16, and 32 Jy beam$^{-1}$). Black and red crosses mark 6.7 GHz methanol and 22 GHz water masers, respectively. The positions of star clusters are indicated with black Xs. The pink circles in panel (a) and the pink squares in panel (b) show the sizes of the SMA mosaics and the covered areas of the OTF APEX maps, respectively.}
	\centering
	\subfloat[Infrared emission at 8 $\mu$m (\emph{Spitzer}/GLIMPSE survey).\label{8um}]{\includegraphics[width=8cm]{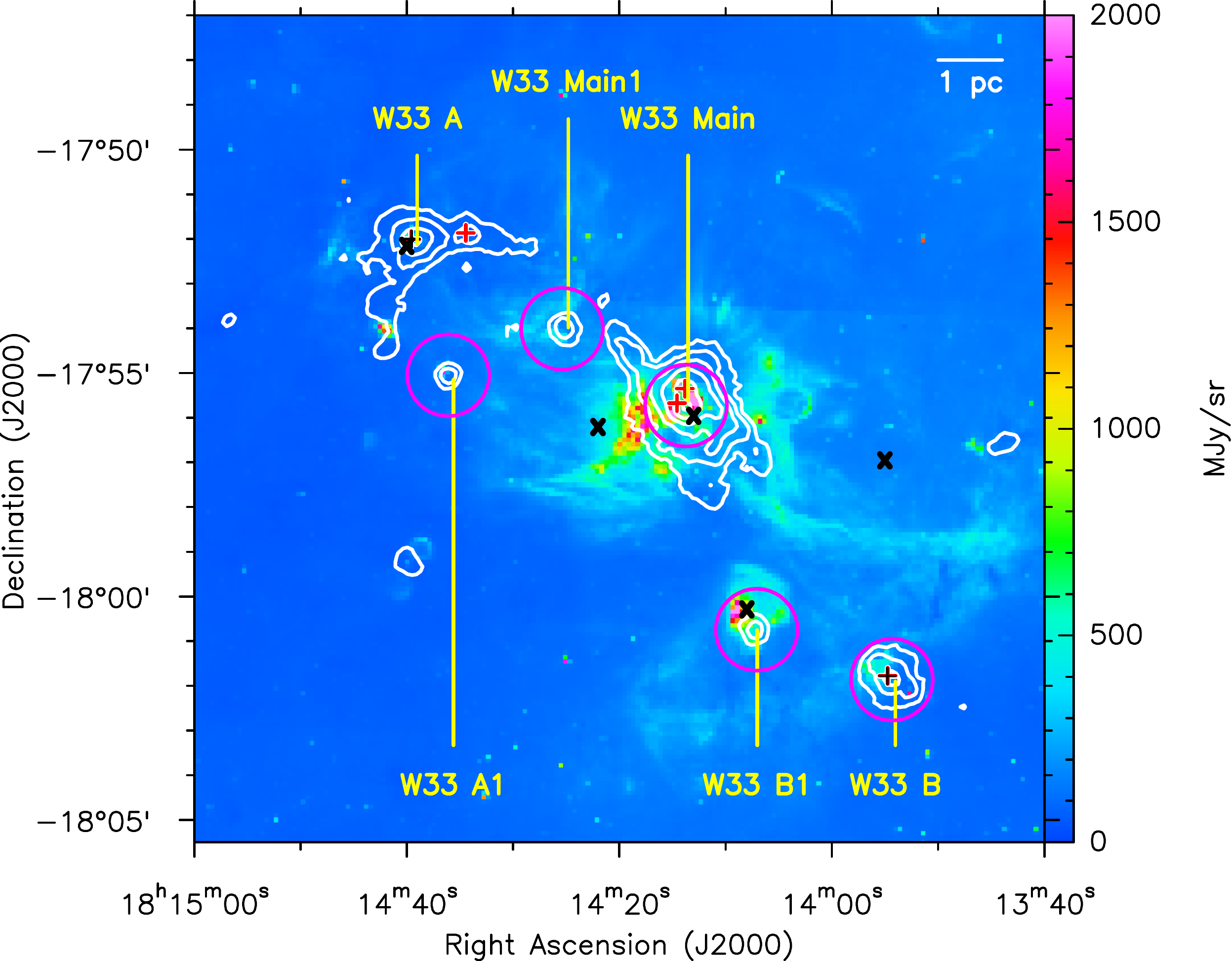}}
	\subfloat[Infrared emission at 70 $\mu$m \emph{(Herschel}/Hi-GAL survey).\label{70um}]{\includegraphics[width=8cm]{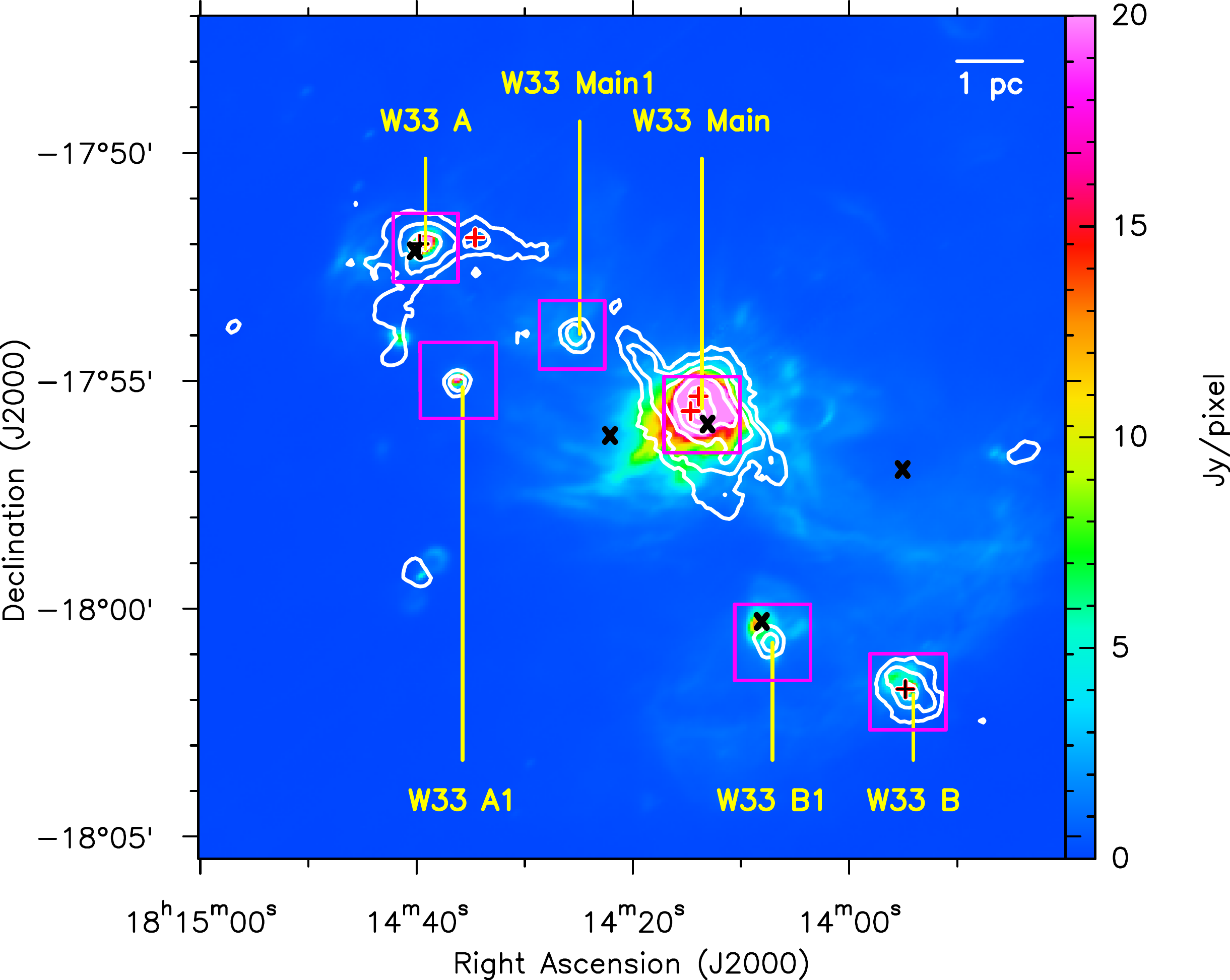}}\\
	\subfloat[Radio emission at 90 cm (VLA survey).\label{330MHz}]{\includegraphics[width=8cm]{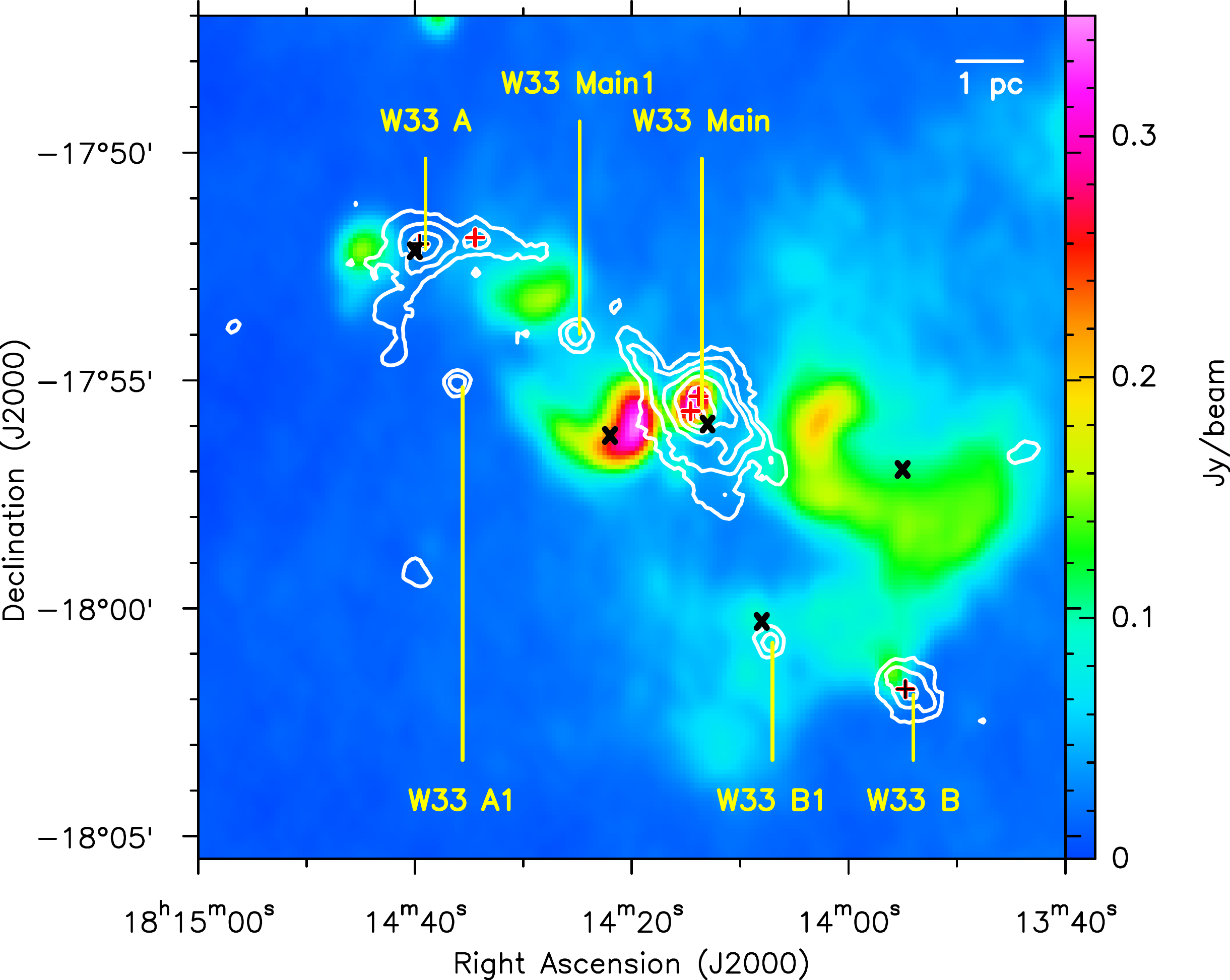}}	
	\label{W33_multi}
\end{figure*}

The W33 complex was first detected as a thermal radio source in the 1.4 GHz survey of \citet{Westerhout1958}. The object W33 is located in the Scutum spiral arm in the first quadrant of the Galaxy \citep[e.g.][]{Immer2013}. Observations of the dust emission in the W33 complex made in the course of the Atacama Pathfinder Experiment (APEX) Telescope Large Area Survey of the GALaxy \citep[ATLASGAL,][]{Schuller2009} at 870 $\mu$m show three large molecular clumps\footnote{In the following publication, the dust condensations on the APEX scales are called ``clumps'' and on the SMA scales ``cores'', roughly following the nomenclature of \citet{Bergin2007}.} (W33\,B, W33\,A, and W33\,Main) and three smaller dust clumps (W33\,Main1, W33\,A1, and W33\,B1; contours in Fig. \ref{W33_multi}). The coordinates of the dust emission peaks in these six clumps, as obtained from the ATLASGAL maps, are given in Table \ref{SourcesW33}. 

From observations of radio recombination lines, H$_{2}$CO absorption, and CO emission, a peculiar velocity field was detected in the W33 complex \citep[e.g.][]{Gardner1972, Gardner1975, Goldsmith1983}. Towards W33\,A and W33\,Main, emission/absorption peaks were observed at a radial velocity of $\sim$36 km s$^{-1}$ while the emission/absorption was maximal at a radial velocity of $\sim$58 km s$^{-1}$ towards W33\,B. However, the CO observations of \citet{Goldsmith1983} show emission of both velocity components throughout the whole complex. Different explanations for this velocity field have been proposed. \citet{Bieging1978} suggest that W33 is a  single connected star forming complex with internal motions of about 20 km s$^{-1}$ at a near-kinematic distance of 3.7 kpc (corresponding to the 36 km s$^{-1}$ velocity component). \citet{Goss1978}, however, favour a superposition of two unrelated star forming regions along the line of sight. Trigonometric parallax observations of the water masers in W33\,B, W33\,Main, and W33\,A yield distances, which are consistent with 2.4 kpc and show that W33 is one connected star forming complex \citep{Immer2013}.

At a distance of 2.4 kpc, the angular size of W33 corresponds to a physical size of $\sim$10 pc.
Revising the results of \citet{Stier1984} for a distance of 2.4 kpc, the W33 complex has a total bolometric luminosity of $\sim$8 $\cdot$ 10$^{5}$ L$_{\sun}$. \citet{Haschick1983} detected the radio continuum emission of a cluster of zero age main sequence stars with spectral types between O7.5 and B1.5 (revised for a distance of 2.4 kpc) in W33\,Main.

Masing transitions of water and methanol have been detected in W33\,A, W33\,Main, and W33\,B \citep[e.g.][]{Genzel1977, Jaffe1981, Menten1986, Haschick1990, Immer2013}, and OH masers have been observed in W33\,A and W33\,B \citep{Wynn-Williams1974, Caswell1998} (positions listed in Table \ref{SourcesW33}). No masing transitions were observed in the clumps W33\,Main1, W33\,A1, and W33\,B1. The methanol masers in W33\,B and W33\,A are radiation pumped (Class II, 6.7 GHz) while the methanol maser in W33\,Main is pumped by collisions (Class I, 25 GHz; see \citet{Menten1991} for the difference between Class I and Class II methanol masers and their different pumping). In W33\,B and W33\,A, the methanol, water, and hydroxyl masers are separated by less than 8$\arcsec$ and 2$\arcsec$, respectively. No OH and 6.7 GHz CH$_{3}$OH masers were detected in W33\,Main.

A cluster of three infrared sources, residing in W33\,Main, was detected by \citet{Dyck1977} at 3.4--33 $\mu$m. Another infrared source was observed in W33\,A with deep absorption features at 3 and 10 $\mu$m \citep{Dyck1977, Capps1978, deWit2010}. \citet{Stier1984} conducted far-infrared observations of W33 at 40--250 $\mu$m and detected four sources in the complex (W33\,B, W33\,B1, W33\,Main, and W33\,A). In the center of W33\,Main, infrared emission peaks at 3.6$-$8 $\mu$m and at 70$-$500 $\mu$m are observed in the \emph{Spitzer}/GLIMPSE and the \emph{Herschel}/Hi-GAL maps (Fig. \ref{8um} and Fig. \ref{70um}), respectively. 
The two water masers are located between these infrared peaks. To the east of 
W33\,Main, two arcs of strong mid-infrared emission can be seen in Fig. \ref{8um}. In the west, 
the mid-infrared emission forms a bubble and a $\sim$5 pc long filament. Both are visible at 70 and 160 $\mu$m in the \emph{Herschel}/Hi-GAL maps (Fig. \ref{70um}) but not at longer wavelengths.
Strong mid- and far-infrared emission is also detected at the center of W33\,A and at the border of W33\,A1 and W33\,B1. Weaker and more diffuse emission is observed at the positions of W33\,B and W33\,Main1. 

Star clusters, signs of advanced star formation in the W33 complex, 
are detected at the border of the W33\,B1 clump, in the centers of 
W33\,Main and W33\,A, and east 
and west of W33\,Main \citep[][and references therein; black Xs in 
Fig. \ref{W33_multi}]{Morales2013}. The coordinates of the star clusters, which are labeled SC-1 to -6, are listed in Table \ref{SourcesW33}. The vicinity of the star clusters outside of W33\,Main 
to the arc- and filament-shaped infrared emission suggests that this emission is influenced by the high-mass stars in these clusters. 

Observations at radio wavelengths show W33\,Main as a compact source, which is embedded in weaker and extended emission. The objects W33\,A and W33\,B are located at the edges of this extended emission \citep[e.g.][]{Wynn-Williams1981, Stier1982, Ho1986, Longmore2007}. 
\citet{Brogan2006} observed the W33 complex and its surroundings at 330 MHz with the VLA (Fig. \ref{330MHz}) to detect supernova remnants (coordinates in Table \ref{SourcesW33}) in the region. Two supernova remnants are found in the north-west of the 
W33 complex at a distance of about 8 pc from the W33\,Main clump (outside of Fig. \ref{W33_multi}).
One strong radio emission peak is detected at the center of W33\,Main, which is located between 
the infrared peaks. The object W33\,Main is also the only source that shows strong emission at 5 GHz in the Co-Ordinated Radio 'N' Infrared Survey for High-mass star formation (CORNISH) VLA survey \citep{Hoare2012}.
East of W33\,Main, strong radio emission at 330 MHz coincides with 
the arc-shaped infrared emission. \citet{Ho1986} suggested that the extended emission is an ionisation front, excited by an earlier generation of high-mass stars that penetrates the molecular material around W33\,Main. They conclude that shocks from high-mass stars can compress molecular clouds on short time scales ($\sim$10$^{4}$ yr) if these stars are not too luminous. The star cluster SC-5 could be a candidate for containing this earlier generation of stars. A comparison of the ages of SC-5 and the star cluster in W33\,Main (SC-4) could provide information if the star formation in W33\,Main was triggered by an earlier generation of stars.
Weaker emission at 330 MHz is detected at the position of 
the infrared bubble and the filament, forming a bow-shaped structure around the star 
cluster SC-2. Diffuse emission is observed at the border of the W33\,B and W33\,B1 clumps.
No radio emission at 330 MHz is found in the clumps W33\,Main1, W33\,A1, and W33\,A.

The object W33\,A was studied by \citet{GalvanMadrid2010} with the Submillimeter Array (SMA) at 219.3--221.3 and 229.3--231.3 GHz. They detected two dust cores, which are surrounded by parsec-scale filaments of cold molecular gas. They suggest that the star formation activity within the cores was triggered by an interaction of the filaments. Evidence for a rotating disk was found  in the brighter core which is orientated perpendicular to a strong outflow. The more massive second core has a less powerful outflow and probably is in an earlier evolutionary stage than the first core.

Combining the continuum and maser information, we can sort our sources in different groups (Table \ref{EvolSequW33}), which mirrors their evolutionary stages. The sources are sorted along an evolutionary sequence which goes from top to bottom. In the last column, we tentatively 
assigned evolutionary stages to the sources that have been previously characterised in other star forming regions. In the rest of this publication, we assume that the six clumps follow similar evolutionary tracks despite their different masses and luminosities, and, thus, a comparison of the chemical composition of the different clumps along the evolutionary sequence is valid.

We observed these six molecular clumps with the APEX\footnote{The APEX project is a collaborative effort among the Max Planck 
Institute for Radioastronomy, the Onsala Space Observatory, and the European Southern 
Observatory.} telescope \citep{Guesten2006} and the SMA\footnote{The Submillimeter Array is a 
joint project between the Smithsonian Astrophysical Observatory and the Academia Sinica 
Institute of Astronomy and Astrophysics and is funded by the Smithsonian Institution and the
Academia Sinica.} \citep{Ho2004}. The goal of this project is to study the chemical composition 
of the star forming regions along a possible evolutionary sequence. The two data sets have 
different resolutions; thus, they provide information about the chemical composition of the star 
forming regions on different scales. Since the high-mass star forming regions in the W33 complex are located at very similar distances and probably originate from the same GMC, we probe the same scales and also likely very similar initial abundance ratios in the different star forming regions, which allows a quantitative comparison of the detected chemical compositions.

The paper has the following structure. In Sect. \ref{Obs}, we describe the observations and complementary data sets that were analysed in this paper. In Sect. \ref{KinematicsW33}, we focus on the kinematics in the W33 complex. Section \ref{TempCloudMass} gives dust temperatures and total masses of the studied molecular clumps. In Sects. \ref{ResultsAPEX} and \ref{ResultsSMA}, we present results from the APEX and SMA observations. In Sect. \ref{TempDens}, we infer column densities and gas temperatures from the APEX and SMA observations for each clump. Section \ref{ChemDiv} discusses the chemical diversity in the W33 clumps. In Sect. \ref{Summary}, we give a short summary of our results. In the online Appendices \ref{ResultsAPEXApp} and \ref{SMAResApp}, we describe the source-specific results of our APEX and SMA observations. The online Appendix \ref{WeedsApp} gives a detailed description of the Weeds modelling and the obtained results.

\begin{table*}
\setlength{\tabcolsep}{2pt}
  \centering 
\begin{footnotesize}
 \caption{Sources in the W33 complex.}   
 \label{SourcesW33} 
  \begin{tabular}{lcccc}
Source & R.A. (hh mm ss) & Dec (dd $\arcmin$$\arcmin$ $\arcsec$$\arcsec$) & Object & Reference\\ \hline
W33\,Main1 & 18 14 25.021 & $-$17 53 58.13 & Molecular clump & \citet{Contreras2013}; this work \\
W33\,A1      & 18 14 36.087 & $-$17 55 04.95 & Molecular clump & \citet{Contreras2013}; this work\\
W33\,B1       & 18 14 07.144 & $-$18 00 45.10 & Molecular clump & \citet{Contreras2013}; this work\\
W33\,B         & 18 13 54.368 & $-$18 01 51.92 & Molecular clump & \citet{Contreras2013}; this work\\
W33\,A        & 18 14 39.059 & $-$17 52 02.72 & Molecular clump & \citet{Contreras2013}; this work\\ 
W33\,Main   & 18 14 13.531 & $-$17 55 47.47 & Molecular clump & \citet{Contreras2013}; this work\\ \hline
W33\,B-OH & 18 13 54.79 & $-$18 01 47.9 & OH maser & \citet{Caswell1998}\\
W33\,A-OH & 18 14 39.53 & $-$17 52 01.1 & OH maser & \citet{Caswell1998}\\ \hline
W33\,B-H$_{2}$O & 18 13 54.7457 & $-$18 01 46.588 & H$_{2}$O maser &\citet{Immer2013}\\
W33\,A-H$_{2}$O-a & 18 14 34.4366 & $-$17 51 51.891 & H$_{2}$O maser &\citet{Immer2013}\\
W33\,A-H$_{2}$O-b & 18 14 39.5714 & $-$17 52 00.382 & H$_{2}$O maser &\citet{Immer2013}\\ 
W33\,Main-H$_{2}$O-a & 18 14 13.8283 & $-$17 55 21.035 & H$_{2}$O maser &\citet{Immer2013}\\
W33\,Main-H$_{2}$O-b & 18 14 14.2315 & $-$17 55 40.573 & H$_{2}$O maser &\citet{Immer2013}\\ \hline
W33\,Main-CH$_{3}$OH & 18 14 11.09 & $-$17 55 57.4 & CH$_{3}$OH 25 GHz maser (Class I) & \citet{Menten1986}\\
W33\,B-CH$_{3}$OH & 18 13 54.75 & $-$18 01 46.6 & CH$_{3}$OH 6.7 GHz maser (Class II) & \citet{Green2010}\\
W33\,A-CH$_{3}$OH & 18 14 39.53 & $-$17 52 00.0 & CH$_{3}$OH 6.7 GHz maser (Class II) & \citet{Green2010}\\ \hline
SNR-1 & 18 13 19 & $-$17 54 & Supernova remnant & \citet{Brogan2006}\\
SNR-2 & 18 13 37 & $-$17 49 & Supernova remnant & \citet{Brogan2006}\\ \hline
SC-1 & 18 14 08.0 & $-$18 00 15 & Star cluster & \citet{Morales2013}\\
SC-2 & 18 13 55.0 & $-$17 56 55 & Star cluster & \citet{Morales2013}\\
SC-3 & 18 13 24.0 & $-$17 53 31 & Star cluster & \citet{Morales2013}\\
SC-4 & 18 14 13.0 & $-$17 55 55 & Star cluster & \citet{Morales2013}\\
SC-5 & 18 14 22.0 & $-$17 56 10 & Star cluster & \citet{Morales2013}\\
SC-6 & 18 14 40.0 & $-$17 52 07 & Star cluster & \citet{Morales2013}\\
\end{tabular}
\end{footnotesize}
\end{table*}

\begin{table*}
\caption{Evolutionary sequence of star forming regions in W33 derived from continuum and maser observations.}
\centering
\begin{tabular}{lccccc}
Source & Submillimeter & Infrared & Maser & Radio & Evol. Stage\\\hline
\rowcolor{gray!20}
W33\,Main1 & $\checkmark$ & weak & -- & --& High-mass protostellar object\\
\rowcolor{gray!20}
W33\,A1 & $\checkmark$ & weak & -- & --& High-mass protostellar object\\
\rowcolor{gray!20}
W33\,B1 & $\checkmark$ & weak & -- & --& High-mass protostellar object\\
\rowcolor{gray!40}
W33\,B & $\checkmark$ & weak & $\checkmark$ & --& Hot Core\\
\rowcolor{gray!40}
W33\,A & $\checkmark$ & $\checkmark$ & $\checkmark$ & -- & Hot Core\\
\rowcolor{gray!60}
W33\,Main & $\checkmark$ & $\checkmark$ & $\checkmark$ & $\checkmark$ & \ion{H}{ii} region\\
\end{tabular}
\label{EvolSequW33}
\end{table*}

\section{Observations and data reduction} 
\label{Obs}

\subsection{Submillimeter Array data set}
We observed five molecular clumps of the W33 complex (W33\,Main1, W33\,A1, W33\,B1, 
W33\,B, and W33\,Main; see contours in Fig. \ref{W33_multi}) at 230 GHz with 
the SMA. The observations were taken in compact 
configuration in 2011 May, covering baselines from 16 to 69 k$\lambda$ 
(corresponding to detected physical scales of $\sim$0.04 to $\sim$0.18 pc). The full width at 
half maximum (FWHM) of the primary beam is 55$\arcsec$. The five sources were observed with 
hexagonal seven-pointing mosaics. The two side bands have a bandwidth of 4 GHz each 
and cover a frequency range of 216.9$-$220.9 GHz (lower sideband, LSB) and 
228.9$-$232.9 GHz (upper sideband, USB). The 
observations have a channel spacing of 0.8125 MHz (corresponding to 1.1 km s$^{-1}$). 
Table \ref{ObsPar} lists the coordinates of the mosaic centers (Columns 2, 3) and the noise levels for the continuum and the line emission per source (Columns 5, 6). 

The data reduction and analysis was conducted with the software packages
MIR\footnote{\url{https://www.cfa.harvard.edu/~cqi/mircook.html}}, 
Miriad\footnote{\url{http://www.cfa.harvard.edu/sma/miriad/}} \citep{Sault1995}, and CASA\footnote{\url{http://casa.nrao.edu/gettingstarted.shtml}} \citep{McMullin2007, Jaeger2008}. 
From observations of Titan, the absolute amplitude was derived. The quasar 3C279 served 
as bandpass calibrator. The observations alternated between the 
target sources and the phase calibrators 1733$-$130 and 1911$-$201.
The continuum data set was constructed from linear baseline fits to 
line-free channels in the upper and lower 
sidebands and subtracted from the whole data set to produce a continuum-free line 
data cube. We imaged the continuum and the line emission of 
all five sources with CASA's interactive cleaning task. 
We made moment 0 (integrated intensity) and moment 1 (velocity field) maps of all the detected lines in each source.

The object W33\,A was previously observed with a similar setup by \citet{GalvanMadrid2010}. Their observations have a slightly coarser spectral resolution than our observations of 3 km s$^{-1}$. The data cube was cleaned with natural weighting with the MIRIAD CLEAN task, resulting in a synthesised beam of 3.9$\arcsec$~x~2.3$\arcsec$, which gives a slightly better spatial resolution than our data sets. However, the observations only have a bandwidth of 2 GHz per sideband. After flagging bad edge channels, the data cube covers the effective frequency ranges 219.38$-$221.21 GHz and 229.38$-$231.31 GHz.

In 2011 March, W33\,Main was observed with the SMA in compact configuration at 345 GHz as part of a filler project. To cover the same area as our 230 GHz observations, W33\,Main was observed with 23 pointings in a hexagonal mosaic. In this publication, we 
focus on the continuum data of these observations. The data reduction was conducted 
in a similar manner as for the 230 GHz data set. As amplitude, bandpass, and phase calibrators, 
the sources Titan, 3C279, 1733$-$130, and 1911$-$201 were observed. The on-source observing time of W33\,Main was $\sim$1~h. The line-free and continuum-free data sets were produced by fitting a linear baseline to line-free channels in both sidebands and subtracting it from the whole data set. Then, we imaged the continuum with CASA's interactive cleaning task.
The image has an rms level of 65 mJy beam$^{-1}$. The size of the synthesised beam is 2.9$\arcsec$~x~1.9$\arcsec$ with a position angle of 151$\degr$.

\begin{table*}
 \caption{Pointing center coordinates and rms noise levels of APEX and SMA observations.}   
 \label{ObsPar} 
  \centering 
  \begin{tabular}{lcccccc}
& & & \multicolumn{1}{c}{APEX} & \multicolumn{3}{c}{SMA (230 GHz)}\\
Source\tablefootmark{a} & R.A. & Dec & rms$_{\textnormal{Line}}$ & rms$_{\textnormal{Cont}}$\tablefootmark{b} & \multicolumn{2}{c}{rms$_{\textnormal{Line}}$}\\ 
 & (hh mm ss.ss) & (dd $\arcmin$$\arcmin$ $\arcsec$$\arcsec$.$\arcsec$) & (mK)  & (mK) & (mK) & (mK)\\
& & & & & LSB & USB\\\hline
W33\,Main1 & 18 14 25.39 & $-$17 54 00.6 & 13   & 11   & 240 & 330\\
W33\,A1 	  & 18 14 36.10 & $-$17 55 03.0 & 16   & 9     & 250 & 340 \\
W33\,B1 	  & 18 14 07.07 & $-$18 00 44.7 & 14   & 9     & 250 & 330 \\
W33\,B 	  & 18 13 54.33 & $-$18 01 51.7 & 26   & 18   & 240 & 310 \\
W33\,A 	  & 18 14 39.00 & $-$17 52 05.0 & 35   & 13   & 150 & 110 \\
W33\,Main   & 18 14 13.72 & $-$17 55 43.6 & 31  & 126  & 260 & 350\\
   \end{tabular}
\tablefoot{
\tablefoottext{a}{In this table and in the rest of this publication, the W33 sources are sorted along the evolutionary sequence that is established in Sects. \ref{ResultsAPEX}, \ref{ResultsSMA}, and \ref{ChemDiv}.}
\tablefoottext{b}{The conversion factor from mJy beam$^{-1}$ to mK is 1.8.} 
}
\end{table*}

\subsection{Atacama Pathfinder Experiment telescope data set}

On-the-fly (OTF) maps of the 
six molecular clumps W33\,Main1, W33\,A1, W33\,B1, W33\,B, W33\,A, and W33\,Main
were observed at 280 GHz at the APEX telescope with the FLASH receiver and the facility Fast Fourier Transform Spectrometer (FFTS). The covered area is 1.8$\arcmin$~x~1.8$\arcmin$. The full width at half maximum (FWHM) of the beam is $\sim$23$\arcsec$. The observations have two sidebands with a bandwidth of 4 GHz each, 
centered at 280.2 GHz and 292.2 GHz, and a smoothed spectral resolution of 1.6 km s$^{-1}$  (to obtain a higher signal-to-noise ratio). 
The calibrated data set was analysed with the software CLASS\footnote{\url{http://iram.fr/IRAMFR/GILDAS/doc/html/class-html/}} from the GILDAS\footnote{\url{http://www.iram.fr/IRAMFR/GILDAS/}} \citep{Pety2005} package. We 
determined the continuum level from the line-free channels and subtracted it as a linear 
baseline from the data cube, generating a continuum-free line data set.
We generated spectra of each source by averaging the spectra of the OTF maps within an area of 24$\arcsec$~x~24$\arcsec$, covering the area of the beam around the (0,0) position. For each source, we produced moment 0 maps of each detected line. Column 4 of Table \ref{ObsPar} gives the noise level of the line data cubes for each source.

\subsection{Institut de Radioastronomie Millim\'etrique 30m telescope data set}

To obtain zero-spacing information for the SMA $^{13}$CO and C$^{18}$O observations of W33\,Main, the molecular clump was observed with the HERA instrument 
at the Institut de Radioastronomie Millim\'etrique (IRAM) 30m telescope using the new FFTS	 backend in 2012 February. 
The observations were centered on the (2$-$1) transitions 
of the CO isotopologues $^{13}$CO and C$^{18}$O at 220.398 and 219.56 GHz, 
respectively. On-the-fly maps were taken in the frequency-switching mode, covering an area of 
6$\arcmin$~x~6$\arcmin$. 
The calibrated data set was further reduced and analysed with the CLASS software to obtain a continuum-free line database. First, baselines of order 7 and 3 were fitted to the line-free channels and subtracted from the data; the spectra were folded, and then again a linear baseline was fitted to the line-free channels and subtracted from the data. To combine the IRAM30m data sets with the SMA data sets, we smoothed the data to a spectral resolution of $\sim$1.2 km s$^{-1}$ and re-imaged the SMA data of $^{13}$CO and C$^{18}$O with the same spectral resolution. The SMA and IRAM30m data cubes were combined in the image plane with the CASA task \textit{feather}. 

\subsection{Additional data sets}

\begin{table*}
 \caption{Resolutions and sensitivities of the infrared to submillimeter observations that were used for the construction of the SEDs in Sect. \ref{TempCloudMass}.}   
 \label{SurveyPar} 
  \centering 
  \begin{tabular}{lccccc}
Survey & Telescope/Instrument & Wavelength [$\mu$m] & Resolution [$\arcsec$] & Sensitivity\\\hline
MSX & MSX & 8.3   & 18.3 & 1 MJy sr$^{-1}$\\
MSX & MSX & 12.1 & 18.3 & 16 MJy sr$^{-1}$\\
MSX & MSX & 14.7 & 18.3 & 11 MJy sr$^{-1}$\\
MSX & MSX & 21.3 & 18.3 & 35 MJy sr$^{-1}$\\
Hi-GAL & \emph{Herschel}/PACS & 70 & 9.0\tablefootmark{a} & 20.2 mJy beam$^{-1}$\tablefootmark{b}\\
Hi-GAL & \emph{Herschel}/PACS & 160 & 13.5\tablefootmark{a} & 45.2 mJy beam$^{-1}$\tablefootmark{b}\\
Hi-GAL & \emph{Herschel}/SPIRE & 250 & 18.1\tablefootmark{c} & 12.1 mJy beam$^{-1}$\tablefootmark{b}\\
Hi-GAL & \emph{Herschel}/SPIRE & 350 & 25.2\tablefootmark{c} & 10.0 mJy beam$^{-1}$\tablefootmark{b}\\
Hi-GAL & \emph{Herschel}/SPIRE & 500 & 36.6\tablefootmark{c} & 14.4 mJy beam$^{-1}$\tablefootmark{b}\\
ATLASGAL & APEX/LABOCA & 870 & 19.2 & 40$-$60 mJy beam$^{-1}$\\
\end{tabular}
\tablefoot{
\tablefoottext{a}{\citet{Poglitsch2010}}
\tablefoottext{b}{Spectral and Photometric Imaging Receiver (SPIRE) Photodetector Array Camera and Spectrometer (PACS) Parallel Mode Observers' Manual (\url{http://herschel.esac.esa.int/Docs/PMODE/html/parallel_om.html})}
\tablefoottext{c}{\citet{Griffin2010}}
}
\end{table*}

To construct spectral energy distributions of the clumps in the W33 complex, we used observations from the Midcourse Space Experiment (MSX) survey \citep{Price2001} at 8.3, 12.1, 14.7, and 21.3 $\mu$m, the \emph{Herschel} infrared Galactic plane survey \citep[Hi-GAL,][]{Molinari2010} at 70, 160, 250, 350, and 500 $\mu$m, and the ATLASGAL survey \citep{Schuller2009} at 870 $\mu$m. The resolutions and sensitivities of the different surveys 
are listed in Table \ref{SurveyPar}. 

Additional infrared observations of the W33 complex were obtained from the 
\emph{Spitzer}/GLIMPSE survey, which were conducted at 3.6, 4.5, 5.8, and 8.0 $\mu$m with 
the InfraRed Array Camera (IRAC) with a resolution of <~2$\arcsec$.

To look for supernova remnants in W33, tracing a former generation of 
high-mass star formation, we used a 330 MHz map taken with the Very Large Array 
(VLA) in B, C, and D configuration by \citet{Brogan2006}. The observations have a spatial resolution of 42$\arcsec$ and an rms of $\sim$5 mJy beam$^{-1}$.

\section{Kinematics in the W33 complex}
\label{KinematicsW33}

From proper motion measurements of water masers in W33\,B, W33\,A, and W33\,Main, \citet{Immer2013} showed that W33\,A is moving tangentially to W33\,Main in the plane of the sky with a total speed of 17 km s$^{-1}$. The radial velocities of W33\,B and W33\,Main differ by 22 km s$^{-1}$. In the plane of the sky, W33\,B moves with a velocity of 22 km s$^{-1}$ relative to W33\,Main, yielding a total speed of 31 km s$^{-1}$ of W33\,B relative to W33\,Main.
 
To determine if W33\,A and W33\,B are gravitationally bound to
W33\,Main, we compare the gravitational and kinetic energies of the clumps.
Using the projected distances between W33\,Main and W33\,A and W33\,Main and W33\,B, respectively, and our mass estimate for W33\,Main of 3965 M$_{\sun}$ (see Sect. \ref{TempCloudMass}), we estimate the escape speed 
needed to leave the gravitational field of W33\,Main. For a projected distance 
of 5.5 pc between W33\,B and W33\,Main and 4.9 pc between W33\,A and 
W33\,Main, we calculate values of 2.5 and 2.6 km s$^{-1}$ 
for the escape speed. The total speed of W33\,B and W33\,A are 
larger than these values. We conclude that both clumps are not gravitationally bound to W33\,Main and thus, the large clumps in the W33 complex drift apart with time. However, these results have to be checked carefully with better determined masses of the clumps.

\section{Dust temperatures and cloud masses}
\label{TempCloudMass}

\begin{figure*}
	\caption{SED fits of the six clumps W33\,Main1, W33\,A1, W33\,B1, W33\,B, W33\,A, and W33\,Main.}
	\centering
	\subfloat{\includegraphics[width=7.8cm]{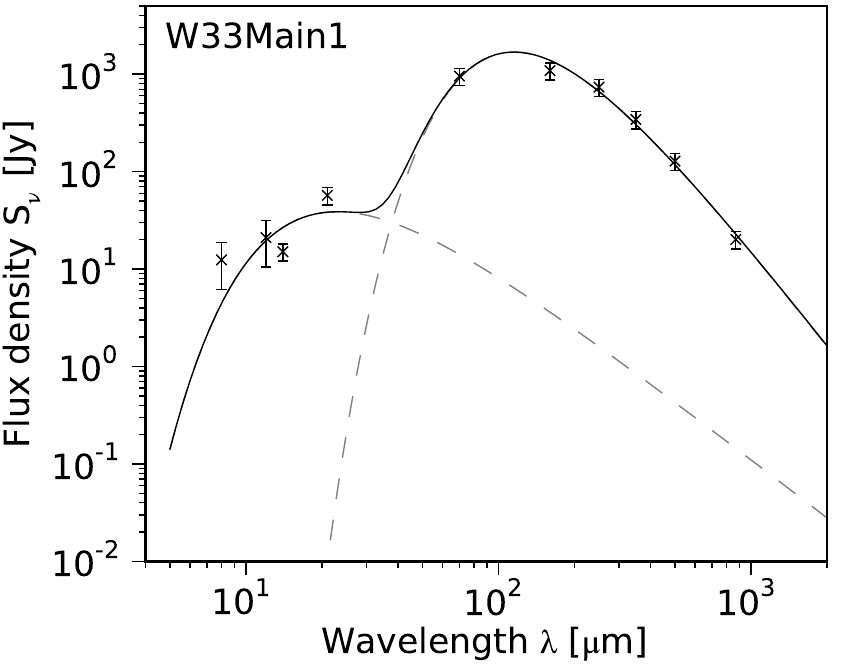}}\hspace{0.2cm} 
	\subfloat{\includegraphics[width=7.8cm]{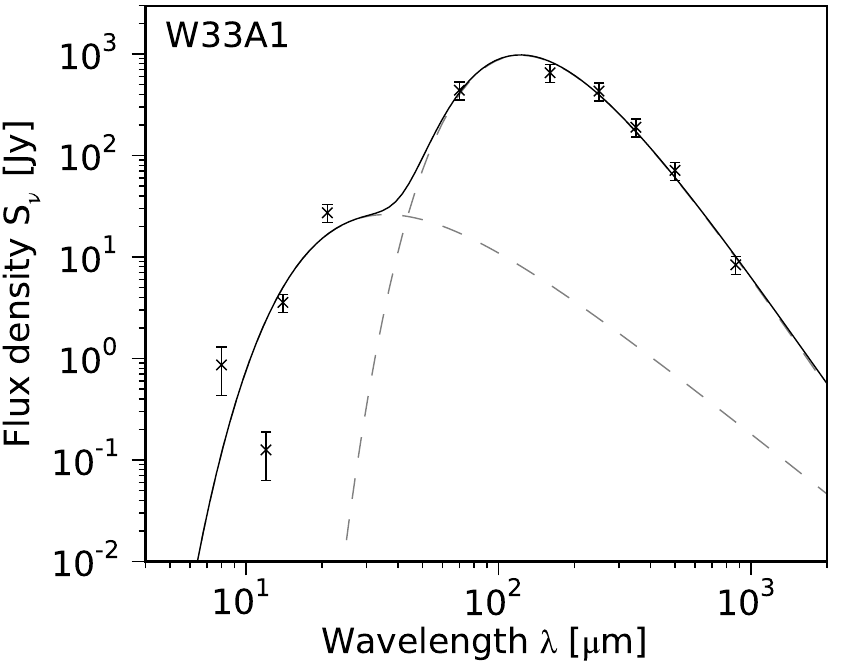}}\\	
	\subfloat{\includegraphics[width=7.8cm]{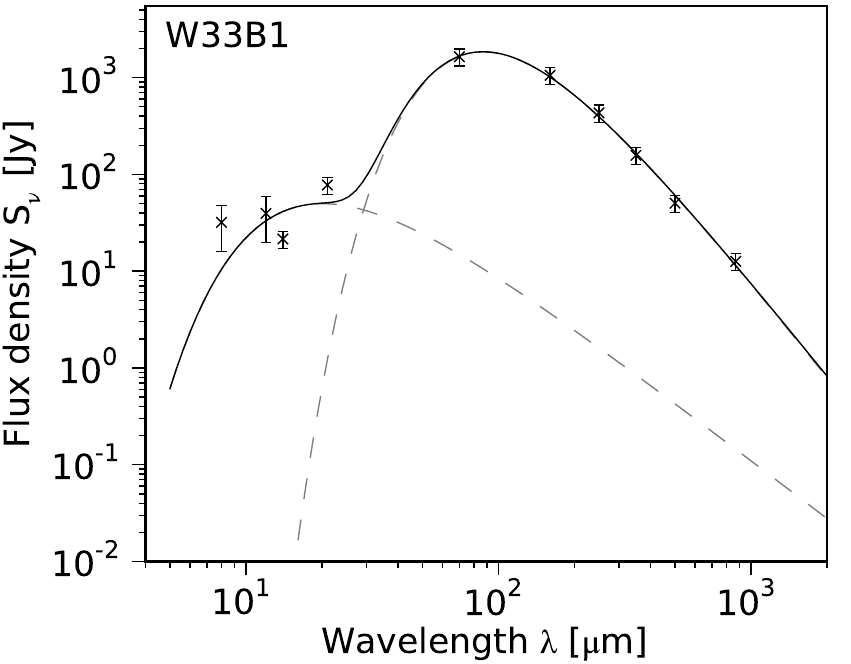}} \hspace{0.2cm}
	\subfloat{\includegraphics[width=7.8cm]{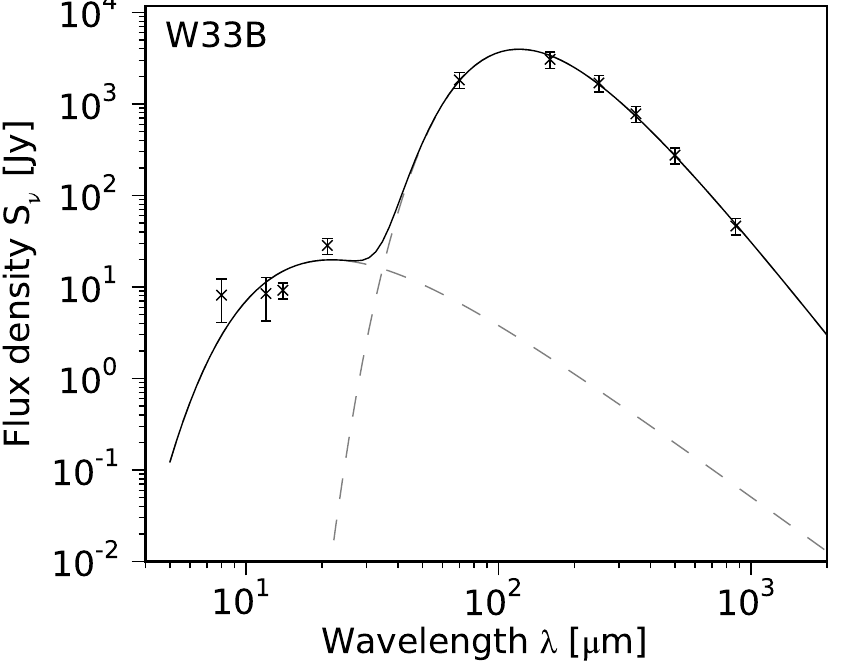}}	\\
	\subfloat{\includegraphics[width=7.8cm]{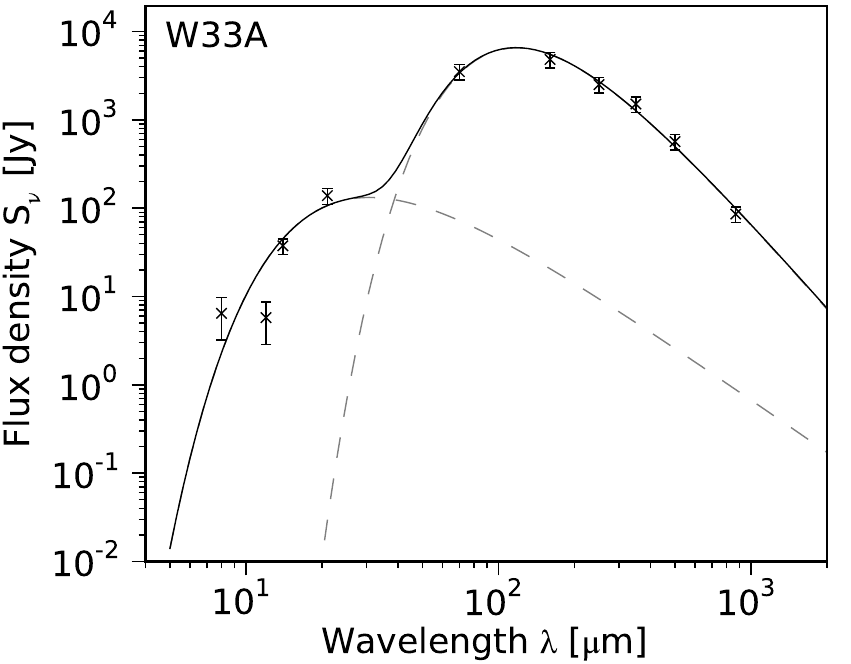}}\hspace{0.2cm}		
	\subfloat{\includegraphics[width=7.8cm]{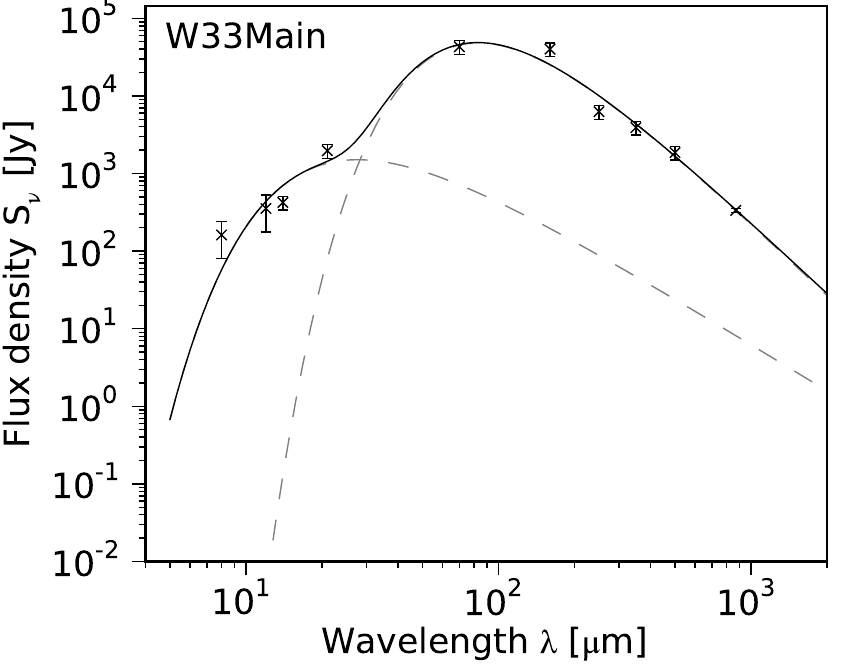}} 
	\label{W33_SEDs}
\end{figure*}

To determine the dust temperatures of the six W33 clumps, we fit the spectral 
energy distributions (SEDs) of the clumps with two-component grey-body and black-body models. The SEDs are constructed from infrared maps of the MSX and the \emph{Herschel}/Hi-GAL surveys and the submillimeter ATLASGAL map, covering a wavelength range of 8$-$870 
$\mu$m. For all wavelengths, we chose the same circular aperture over which the emission 
was integrated. The radius of the aperture was selected to correspond to the deconvolved 
radius of the clumps in the ATLASGAL compact source catalog \citep[][Column 2 in Table \ref{Masses-W33}]{Contreras2013}. To estimate the background 
emission at all wavelengths, we integrated the emission in apertures with a radius of 
40$\arcsec$ at 20 positions around the W33 complex. Then, we averaged these values 
and subtracted the background emission from the total emission of each
clump at each wavelength, following formula 1 of \citet{Anderson2012}. To account for 
the intrinsic instrument error and the source extraction uncertainties, we assumed uncertainties of 20\% for flux densities at wavelengths longer than 14 $\mu$m.
Since the broad-bands of the MSX receivers at 8 and 12 $\mu$m often show  emission/absorption features due to polycyclic aromatic hydrocarbons, very small grains, and silicates \citep[e.g.][]{Churchwell2009}, we increased the flux uncertainties to 50\%. However, 
from the photometry at 14 and 20 $\mu$m, it is clear that a warm component is needed to fit
the short-wavelengths part of the SEDs (Fig. \ref{W33_SEDs}).

Thus, the generated SEDs are fitted with a ``cold'' plus a ``warm'' component. The detection 
of two temperature components suggests that all six clumps have a heating source inside 
or nearby. The cold and the warm components were then fitted with a two-component model, fitting a grey-body and a black body simultaneously. From the fits, we determine the bolometric 
luminosities, the warm and cold temperatures of the clumps, and the spectral emissivity index $\beta$. The uncertainties of the fitting parameters are calculated from the covariance matrix, which is returned from the Levenberg-Marquardt fitting algorithm. The SED fits are shown in Fig. \ref{W33_SEDs}. The cold temperatures and luminosities are listed in Table \ref{Masses-W33} (Columns 6$-$7). The cold temperatures range between 25 and 43~K with the highest values of T$_{cold}$ being estimated for W33\,Main and 
W33\,B1. While the temperatures of both components are similar for all sources, the luminosity covers a wide range of values from 6~$\cdot$~10$^{3}$~L$_{\sun}$ in W33\,A1 to 4.5~$\cdot$~10$^{5}$~L$_{\sun}$ in W33\,Main. 
The total bolometric luminosity of the W33 complex is 5.5~$\cdot$~10$^{5}$~L$_{\sun}$ which is only slightly smaller than the revised value (8~$\cdot$~10$^{5}$~L$_{\sun}$) of \citet{Stier1984}. The spectral emissivity index quantifies the frequency-dependence of the emissivity in the Rayleigh-Jeans part of the spectrum \citep{Dent1998}. This parameter ranges from 1.2 to 1.9 and shows no clear trend with the evolutionary stage of the clump. 

 The masses and column densities estimated for the warm component have large uncertainties and their contribution to the total masses and column densities is only a couple percent, compared to the cold component. We thus neglected its contribution in the following calculations.
Assuming no temperature gradient, the total mass of a clump is given by 
\small
\[
\frac{M_{\textnormal{tot}}}{\textnormal{M}_{\sun}} = \frac{2.0 \cdot 10^{-2}}{J_{\nu}(T_{\textnormal{cold}})} \frac{a}{0.1 \ \mu \textnormal{m}} \frac{\rho}{\textnormal{3 g cm}^{-3}} \frac{R}{100} \frac{F_{\nu, \textnormal{source}}}{\textnormal{Jy}} \left(\frac{d}{\textnormal{kpc}} \right)^{2} \left(\frac{\nu}{\textnormal{1.2 THz}}\right)^{-3-\beta}\,
\]
\normalsize
\citep{Hildebrand1983, Beuther2005Erratum}, where 
\[
J_{\nu}(T_{\textnormal{cold}}) =  \exp\left(\frac{h \nu}{k T_{\textnormal{cold}}}\right) -1.
\]
The column density, averaged over the whole clump, can be calculated with \citep[e.g.][]{Beuther2005Erratum,Kauffmann2008}
\[
\frac{N_{\textnormal{H}_{2}}}{\textnormal{cm}^{-2}} = \frac{1.25 \cdot 10^{12}}{\Omega_{S} J_{\nu}(T_{\textnormal{cold}})}  \frac{a}{0.1 \ \mu \textnormal{m}} \frac{\rho}{\textnormal{3 g cm}^{-3}} \frac{R}{100} \frac{F_{\nu, \textnormal{source}}}{\textnormal{Jy}} \left(\frac{\nu}{\textnormal{1.2 THz}}\right)^{-3-\beta}.
\]
In both formulas, $a$ is the grain size normalisation constant, $\rho$ the grain mass density, $R$ the gas-to-dust ratio, $F_{\nu, \textnormal{source}}$ the flux density at 345 GHz integrated over the whole clump, $d$ the distance,  $\nu$ the frequency of the observations (345 GHz), $\beta$ the spectral emissivity index, and $\Omega_{S}$ the source solid angle at 345 GHz. These calculations correspond to a mass absorption coefficient $\kappa$ of 0.852 ($\beta$~=~1.8) $-$ 2.296~cm$^{2}$~g$^{-1}$ ($\beta$~=~1.2) at 870 $\mu$m, depending on the value of the spectral emissivity index $\beta$.
The peak column density was estimated by substituting F$_{\nu, \textnormal{source}}$ with the peak flux density of each clump and $\Omega_{S}$ with the beam solid angle of the APEX telescope at 870 $\mu$m. To derive the molecular abundances in Sect. \ref{TempDens}, we have to estimate the H$_{2}$ column densities from the same area as the column densities of the APEX data, which are averaged over an area of 24$\arcsec$~x~24$\arcsec$. Thus, we integrated the flux density over this area and replaced $\Omega_{S}$ with the solid angle of this area. 

To determine the flux densities of the sources at 870 $\mu$m, we integrated the continuum emission of each clump down to the 5~$\sigma$ level (=0.35 Jy beam$^{-1}$). The area over which the emission was integrated was then translated into the source solid angle $\Omega_{S}$ 
and an equivalent radius r$_{equiv}$ for each clump. 
For the gas-to-dust ratio, the grain size normalisation constant, and the grain mass density, we assumed typical values of 100, 0.1 $\mu$m, and 3 g cm$^{-3}$, respectively \citep[e.g.][]{Hildebrand1983, Kauffmann2008}. The distance to the W33 complex is 2.4 kpc \citep{Immer2013}. 
Table \ref{Masses-W33} lists the source names (Column 1), the aperture radii (Column 2), equivalent radii (Column 3), the peak flux densities at 870 $\mu$m (Column 4), the integrated flux densities at 870 $\mu$m over an area of 24$\arcsec$~x~24$\arcsec$ (Column 5) and the whole source (Column 6), the dust temperatures of the cold component (Column 7), the luminosities of the clumps (Column 8), the spectral emissivity index $\beta$ (Column 9), the total masses of the clumps (Column 10), the peak column densities (Column 11), and the column densities averaged over an area of 24$\arcsec$~x~24$\arcsec$ around the peak position of each clump (Column 12) and over the whole source (Column 13). For the uncertainties of the flux densities integrated over an area of 24$\arcsec$~x~24$\arcsec$ and the whole source, and the peak flux densities, we assume a value of 15\%.

\begin{table*}
 \centering
\begin{footnotesize}
\setlength{\tabcolsep}{2.5pt}
   \caption{Source parameters of the six W33 clumps derived from continuum observations at 870 $\mu$m.}
  \begin{tabular}{lcccccccccccc}
	Clump 	& r$_{aperture}$ &r$_{equiv}$ & F$_{\textnormal{peak}}$\tablefootmark{a} & F$_{24\arcsec}$\tablefootmark{a} & F$_{\textnormal{source}}$\tablefootmark{a} & T$_{\textnormal{cold}}$ & L$_{\textnormal{bol}}$ & $\beta$ & M$_{\textnormal{source}}$ & N$_{\textnormal{H}_{2},\textnormal{peak}}$ & N$_{\textnormal{H}_{2}, 24\arcsec}$ &N$_{\textnormal{H}_{2},\textnormal{source}}$ \\ 
                         & ($\arcsec$)/(pc) & ($\arcsec$)/(pc) & (Jy) & (Jy) & (Jy) & (K) & (10$^{3}$ L$_{\sun}$) & & (10$^{3}$ M$_{\sun}$) & (10$^{23}$ cm$^{-2}$) & (10$^{23}$ cm$^{-2}$) & (10$^{22}$ cm$^{-2}$) \\\hline
        W33\,Main1 	& 58/0.7 &53/0.6 & 3.4 & 4.0  & 21.3 & 28.6~$\pm$~~~5.6 &  11   & 1.4~$\pm$~0.5 & 0.5~$\pm$~0.3 & 0.9~$\pm$~0.2  & 0.7~$\pm$~0.2   & 2.4~$\pm$~0.7\\
        W33\,A1 		& 40/0.5 &32/0.4 & 3.2 & 3.9  & 8.6 & 25.0~$\pm$~~~7.3 &  6     & 1.8~$\pm$~0.9 & 0.4~$\pm$~0.5   & 1.7~$\pm$~0.6 & 1.2~$\pm$~0.4   & 5.4~$\pm$~2.3\\
        W33\,B1 		& 49/0.6 & 48/0.6 & 3.1 & 3.4  & 14.8 &38.6~$\pm$~11.4  & 16   & 1.3~$\pm$~0.5 & 0.2~$\pm$~0.1   & 0.5~$\pm$~0.2 & 0.4~$\pm$~0.1   & 1.2~$\pm$~0.5\\
        W33\,B 		& 75/0.9 &86/1.0 & 5.4 & 6.6  & 60.7 &26.5~$\pm$~~~3.9  & 22   & 1.6~$\pm$~0.4 & 1.9~$\pm$~1.1 & 2.1~$\pm$~0.5 & 1.6~$\pm$~0.4   & 3.7~$\pm$~0.9\\
        W33\,A 		& 82/1.0 & 128/1.5 & 9.2 & 11.2 & 153.4 & 28.6~$\pm$~~~5.3  &41   & 1.4~$\pm$~0.5 & 3.4~$\pm$~2.3 & 2.5~$\pm$~0.6 & 2.0~$\pm$~0.6 &3.0~$\pm$~0.9\\
        W33\,Main		& 102/1.2 & 126/1.5 & 36.5 & 43.9 & 383.4\tablefootmark{b}  & 42.5~$\pm$~12.6  & 449 & 1.2~$\pm$~0.4 & 4.0~$\pm$~2.5 & 4.6~$\pm$~1.6 & 3.6~$\pm$~1.4 & 3.8~$\pm$~1.5\\\hline
   \end{tabular}
\tablefoot{
\tablefoottext{a}{Assumed uncertainty: 15\%}
\tablefoottext{b}{$\sim$18 Jy from free-free emission}\\
Column 1: clump name; Column 2: radius of the aperture, used for photometry; Column 3: equivalent radius, characterising the clump area at the 5$\sigma$ level; Column 4: peak flux density; Column 5: flux density, integrated over an area of 24$\arcsec$~x~24$\arcsec$; Column 6: flux density of the whole clump; Column 7: temperature of the cold component; Column 8: bolometric luminosity; Column 9: spectral emissivity index $\beta$; Column 10: total mass of the clump; Column 11: H$_{2}$ peak column density; Column 12: H$_{2}$ column density over an area of 24$\arcsec$~x~24$\arcsec$; Column 13: H$_{2}$ column density of the whole clump
}
   \label{Masses-W33}
\end{footnotesize}
\end{table*}

Since W33\,Main contains an \ion{H}{ii} region, part of the emission at 345 GHz possibly 
originates in free-free emission. To approximate this fraction, we first obtained an estimate of 
the radio flux density of W33\,Main. Summing the emission of compact components in 
W33\,Main detected by \citet{Haschick1983} yields a flux density of 23.9 Jy at 14.7 GHz. 
Assuming a typical spectral index of $\alpha$ = $-$0.1 between 14.7 and 345 GHz, we estimate 
a flux density of $\sim$18 Jy at 345 GHz, originating from free-free emission, 
which corresponds to 5\% of the total flux density at 345 GHz.

The total masses of the clumps range from 195 M$_{\sun}$ for W33\,B1 to 3965 M$_{\sun}$ for W33\,Main, yielding a total mass of the W33 complex of $\sim$10200 M$_{\sun}$.
Overall, the sources follow the trend of increasing luminosity with increasing total mass except W33\,B1, which shows a comparably small total mass that can partly be explained by the higher dust temperature. The source-averaged column densities range between 1.2 and 5.4~$\cdot$~10$^{22}$~cm$^{-2}$ but do not show a trend with the evolutionary state of the clump, which indicates that we probe similar material in the sources as was also recently seen in \citet{Hoq2013}. However, the peak column density is correlated with the evolutionary stage of the targets, being largest in the most evolved source W33\,Main. Thus, the substructure with the highest peak column density is formed from the most massive clump. In addition, the most massive clump (W33\,Main) cannot have formed very late compared to the other clumps since it is the most evolved source in the sample. 

\citet{Kauffmann2010} recently proposed an empirical limit for high-mass star formation in 
molecular clumps, comparing the total mass of the clump $m(r)$ with its radius $r$:
\[m(r) > 870 \cdot \textnormal{M}_{\sun} \left(\frac{r}{\textnormal{pc}}\right)^{1.33}.
\]
Inserting the total mass and the radii from Table \ref{Masses-W33} in 
this formula, we conclude that all W33 clumps fulfill this criterion within the mass uncertainties and thus, in principle, have the potential to form high-mass stars.

\section{Results of the APEX observations}
\label{ResultsAPEX}

\begin{figure*}
	\caption{APEX spectra generated over an area of 24$\arcsec$~x~24$\arcsec$ in W33\,Main1, W33\,A1, and W33\,B1. The grey areas mark atmospheric bands in the APEX spectra that were not completely removed by our calibration.}
	\centering
	\subfloat{\includegraphics[width=9cm]{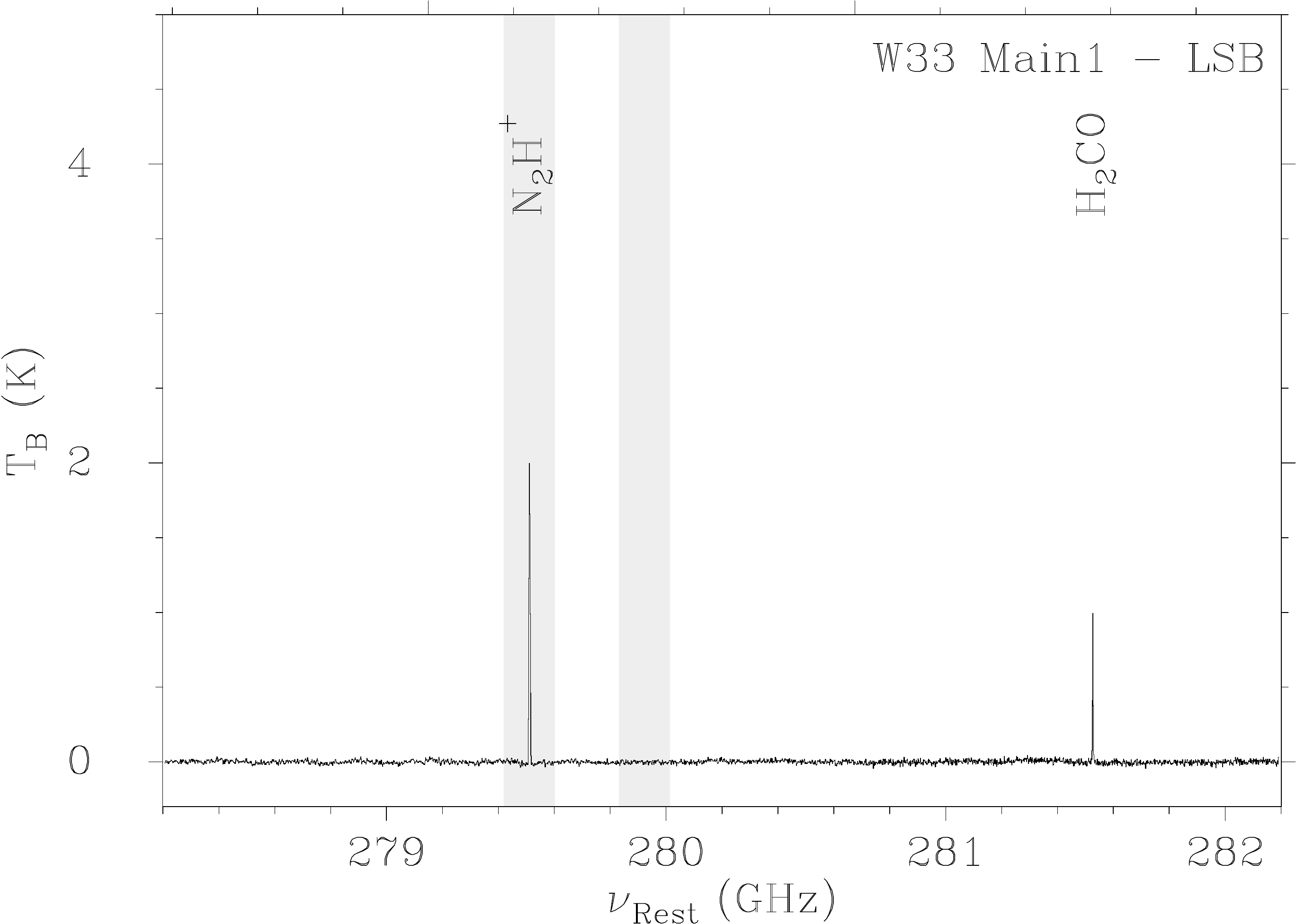}}\hspace{0.2cm} 
	\subfloat{\includegraphics[width=9cm]{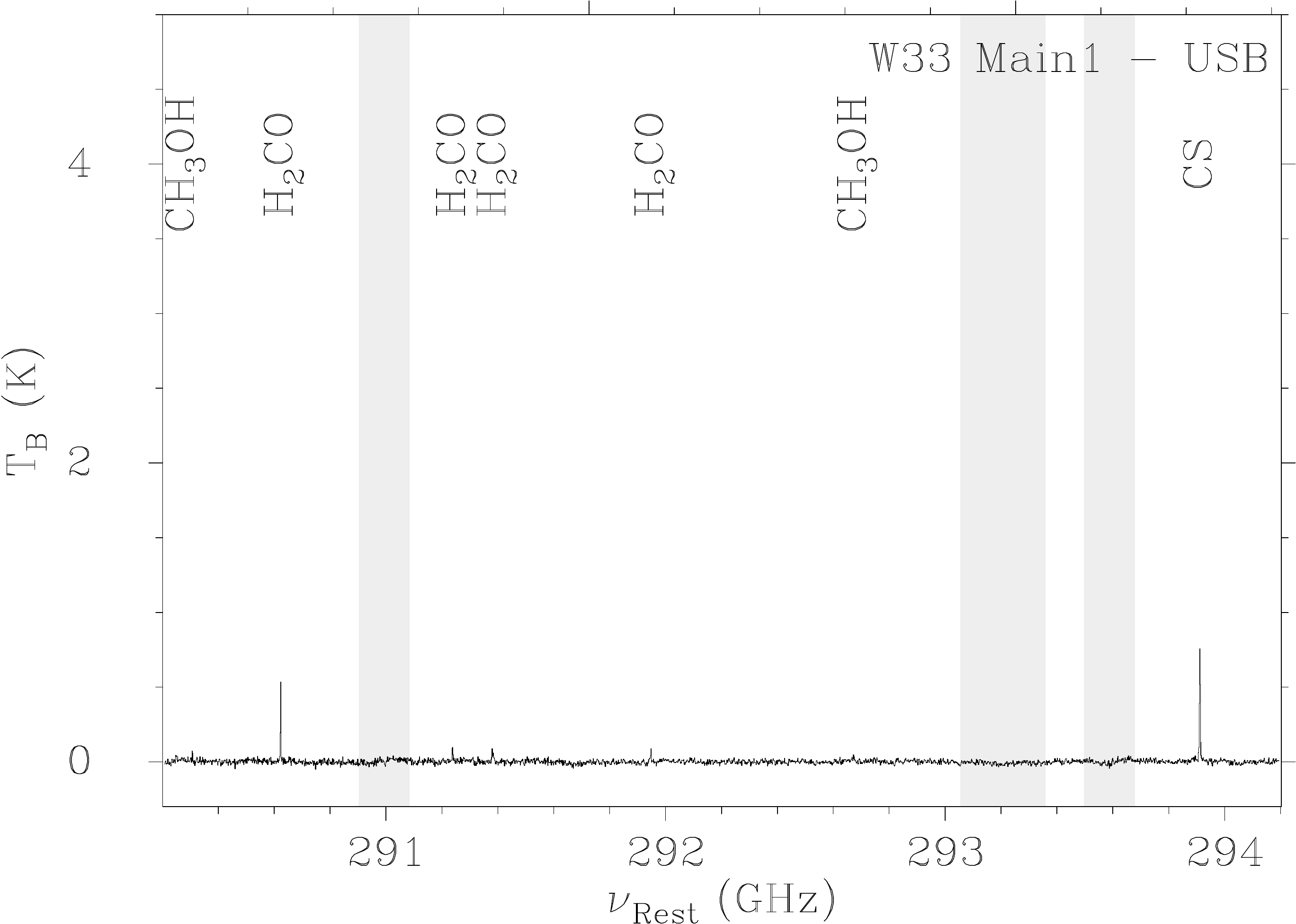}}\\
	\subfloat{\includegraphics[width=9cm]{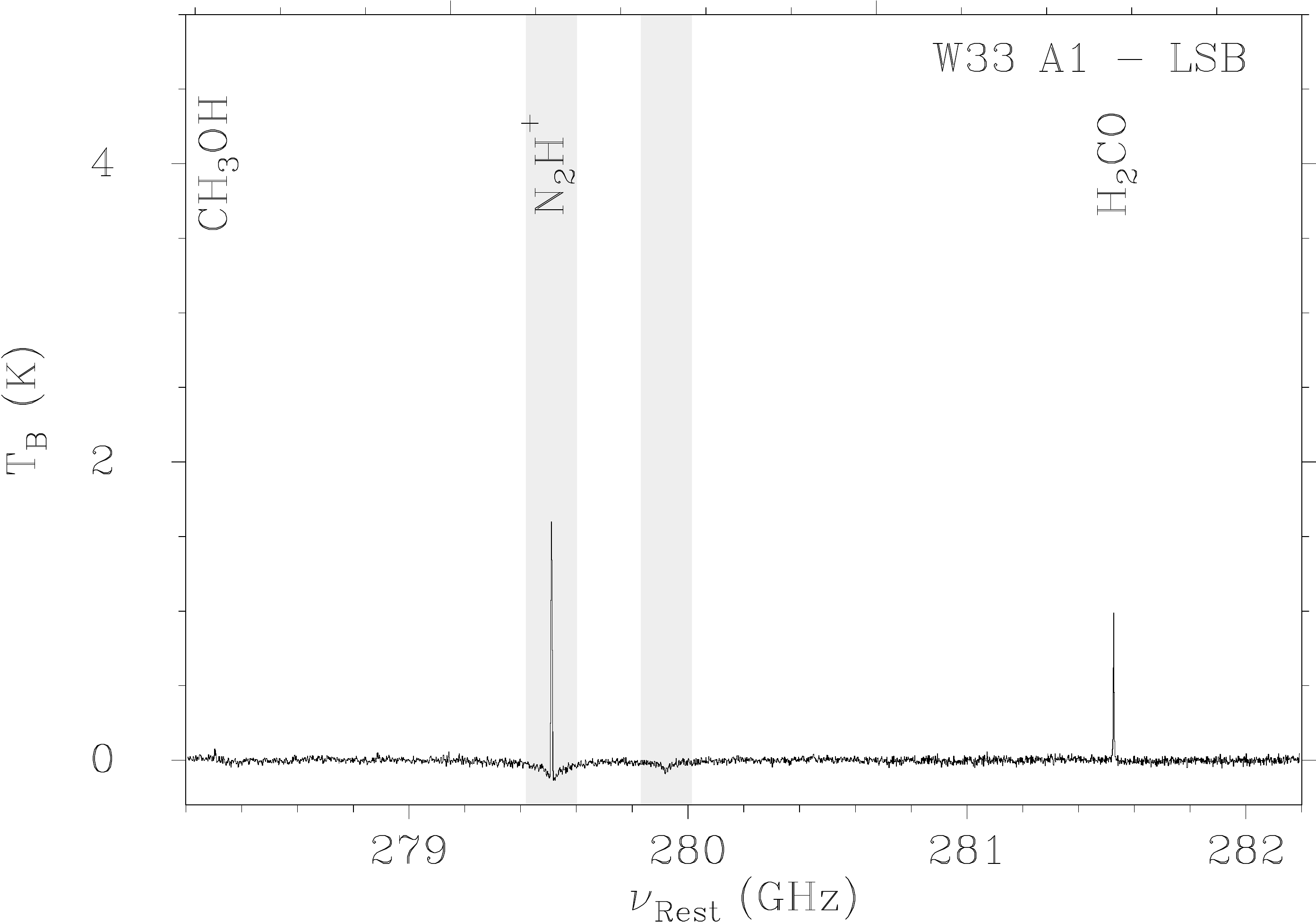}}\hspace{0.2cm} 			
	\subfloat{\includegraphics[width=9cm]{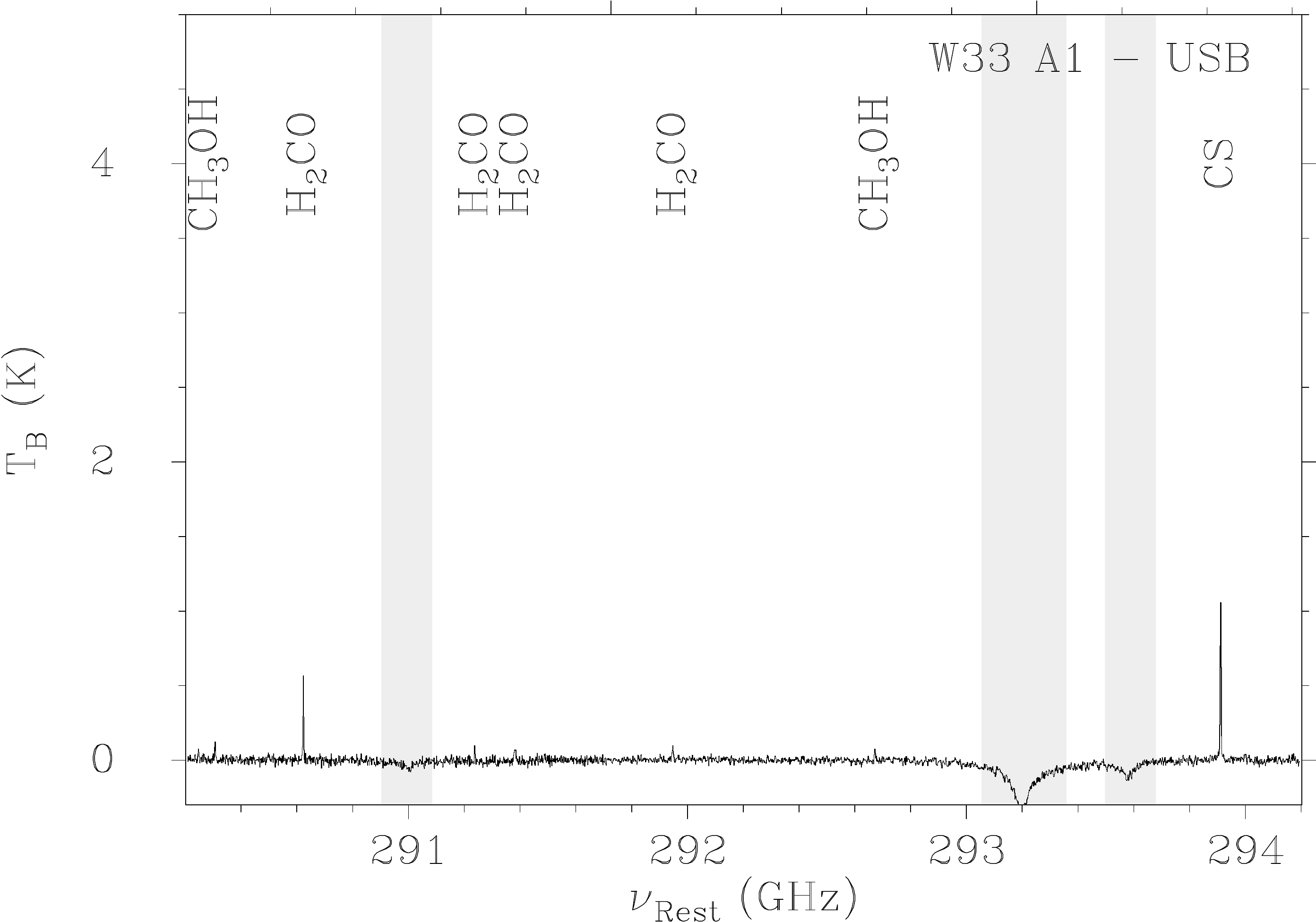}}\\	
	\subfloat{\includegraphics[width=9cm]{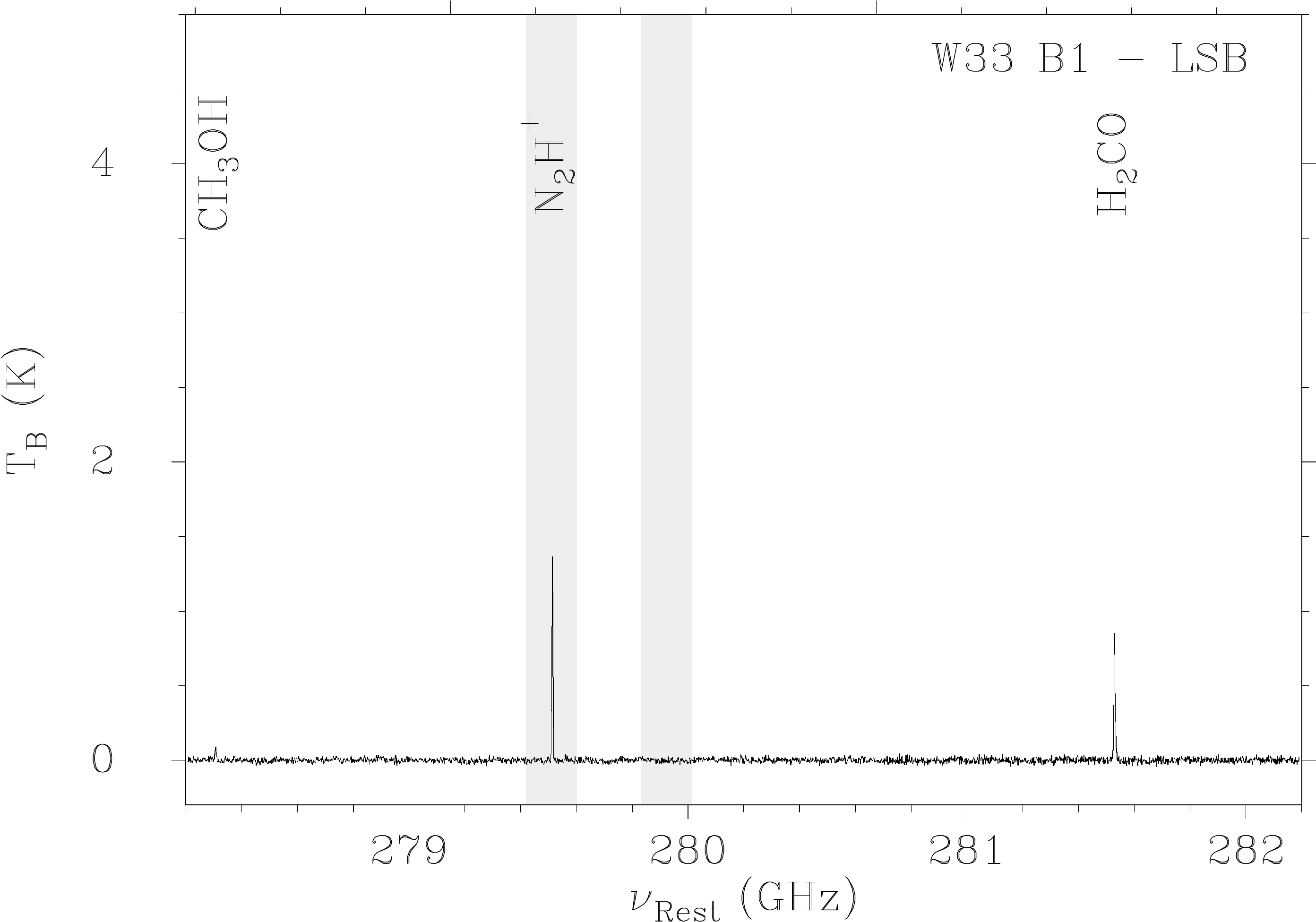}}\hspace{0.2cm} 
	\subfloat{\includegraphics[width=9cm]{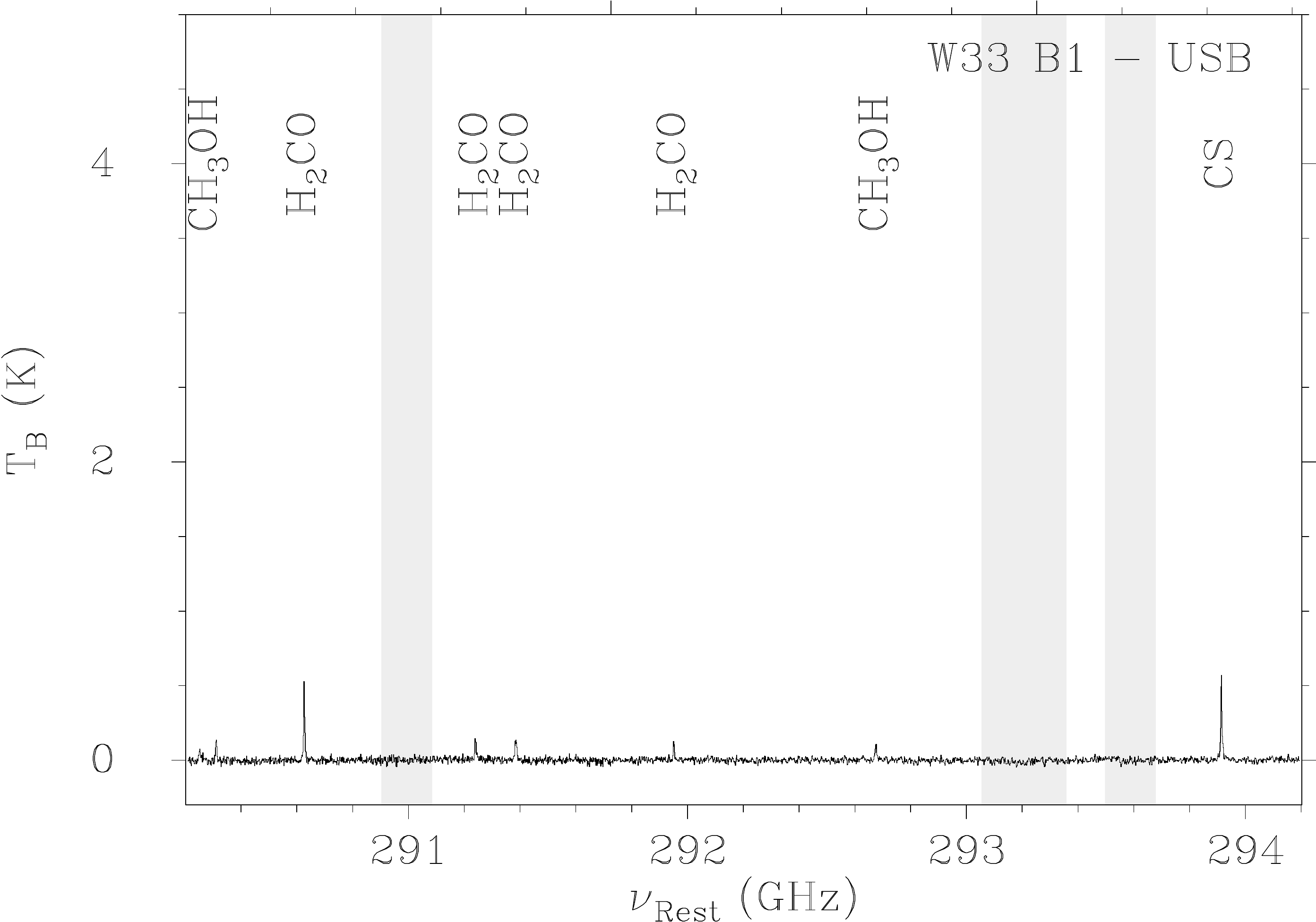}}	
	\label{W33_APEX_Spectra1}
\end{figure*}

\begin{figure*}
	\caption{APEX spectra generated over an area of 24$\arcsec$~x~24$\arcsec$ in W33\,B, W33\,A, and W33\,Main. The grey areas mark atmospheric bands in the APEX spectra that were not completely removed by our calibration.}
	\centering
	\subfloat{\includegraphics[width=9cm]{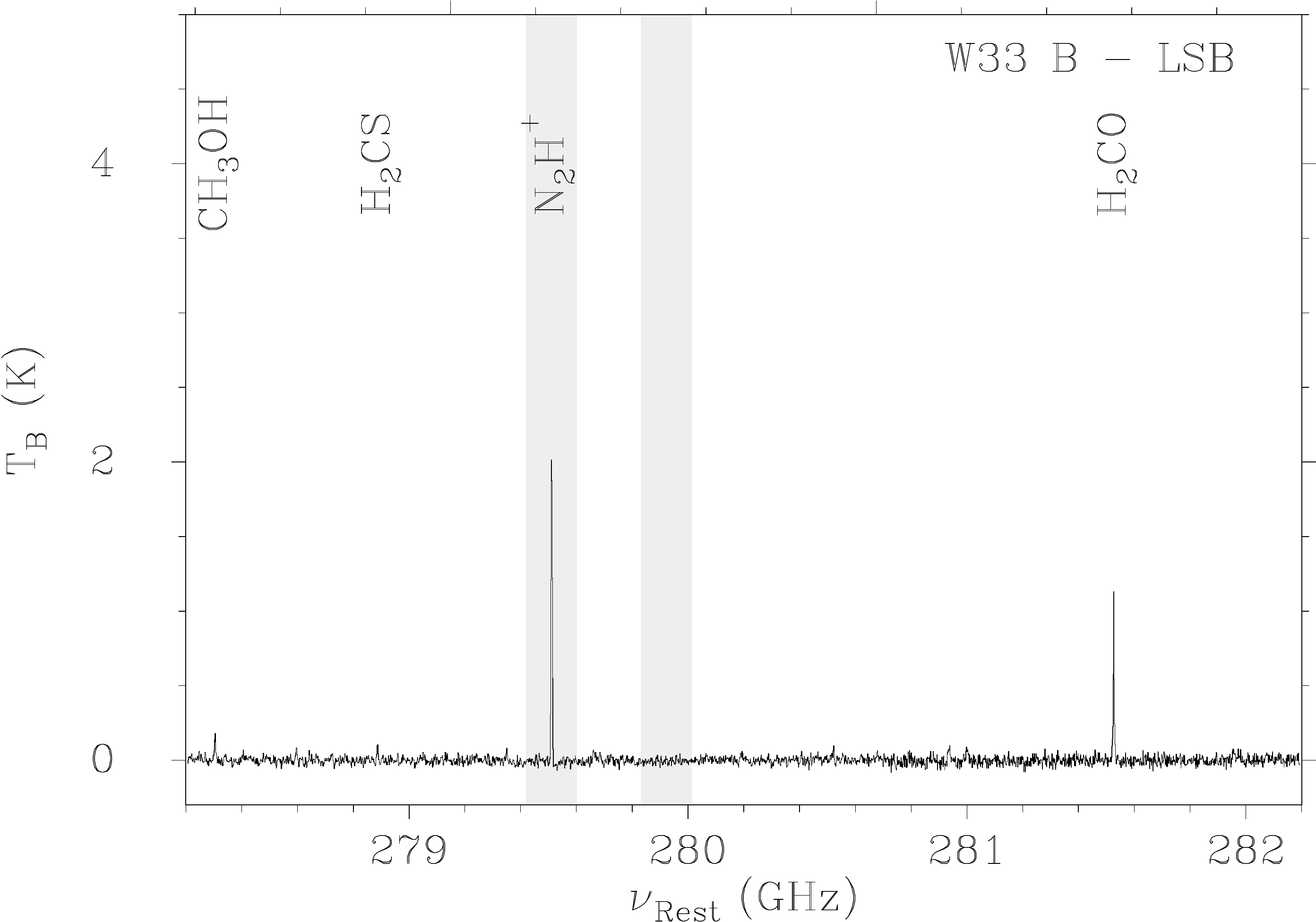}}\hspace{0.2cm}	
	\subfloat{\includegraphics[width=9cm]{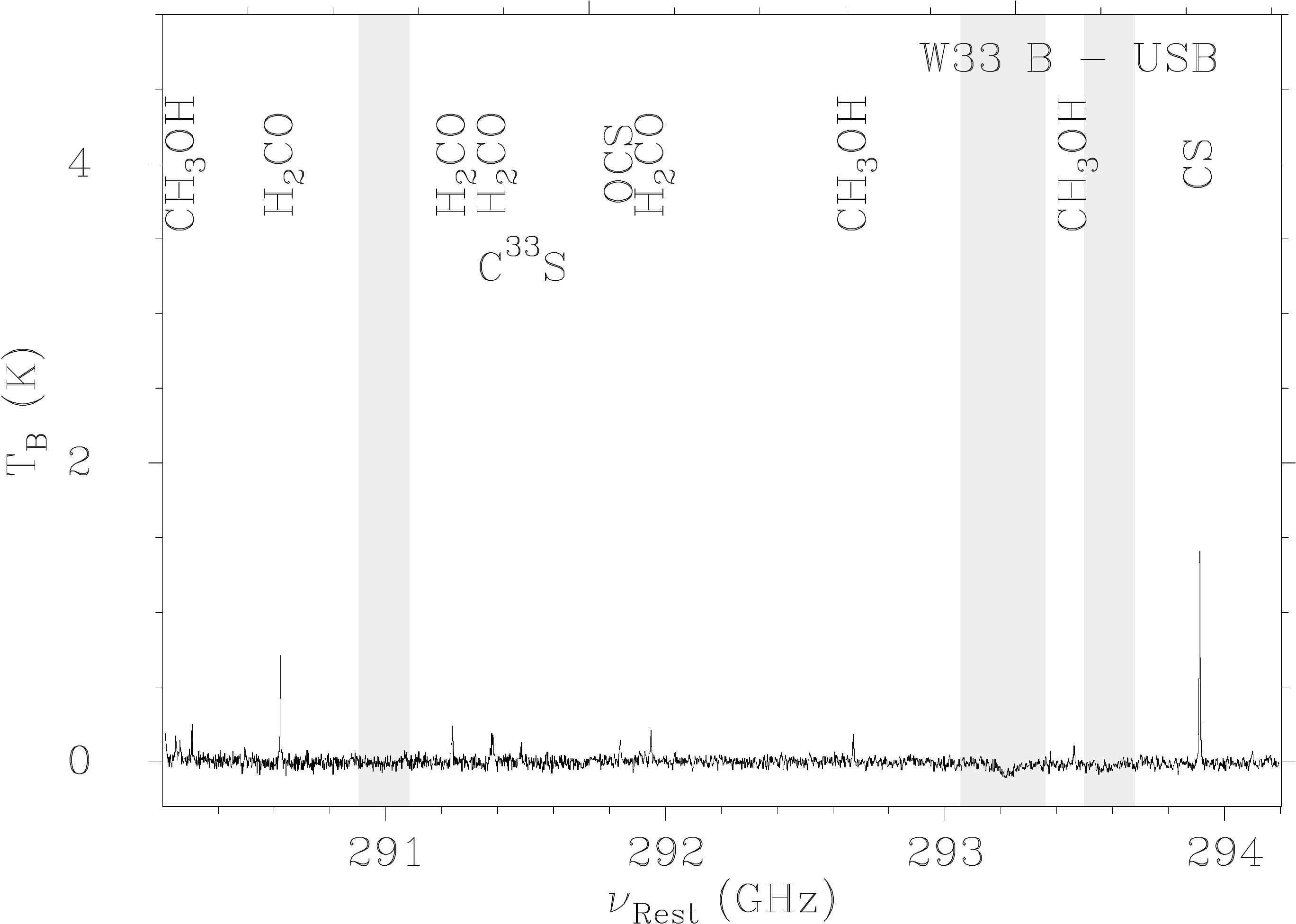}}\\
	\subfloat{\includegraphics[width=9cm]{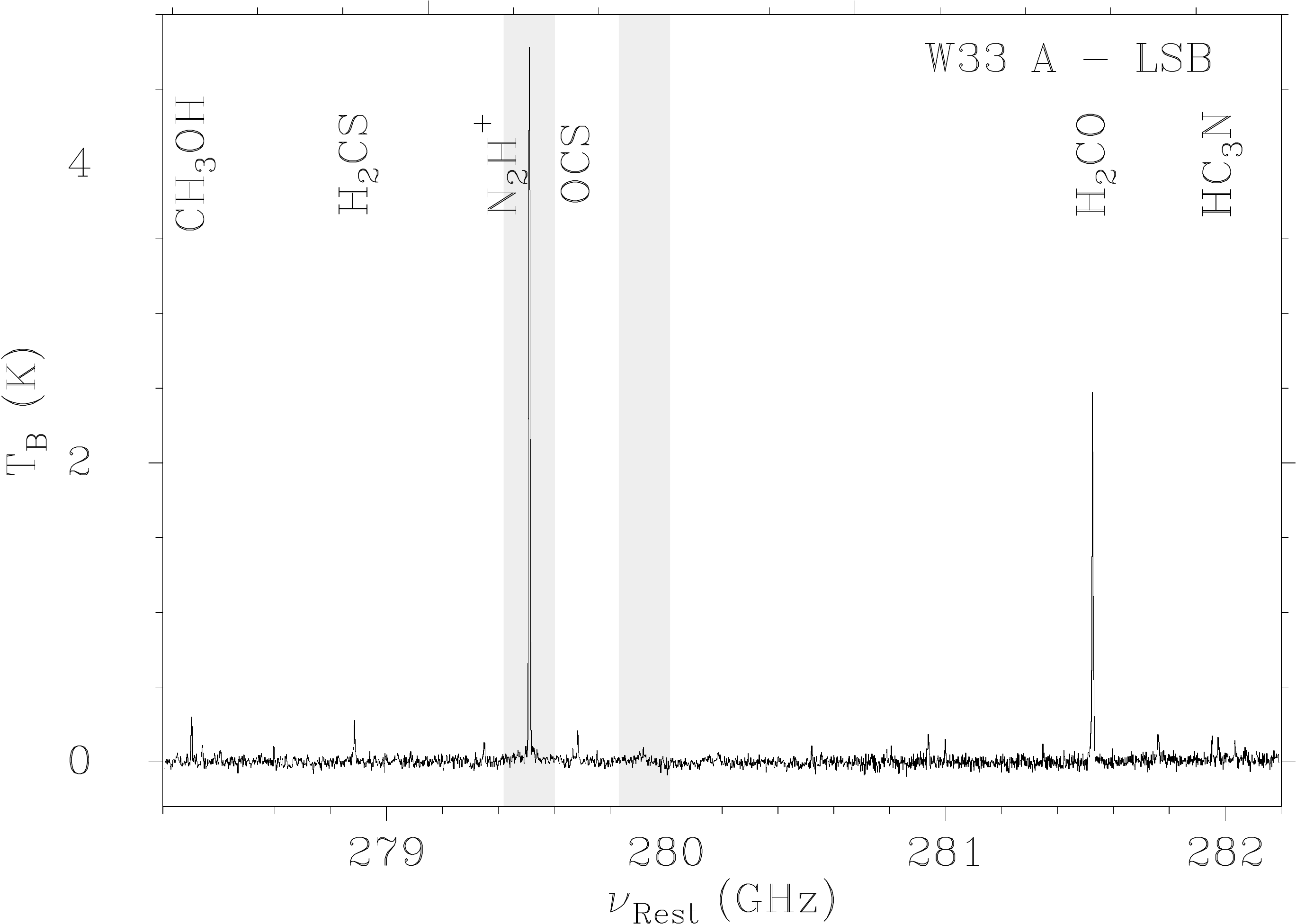}}\hspace{0.2cm} 	
         \subfloat{\includegraphics[width=9cm]{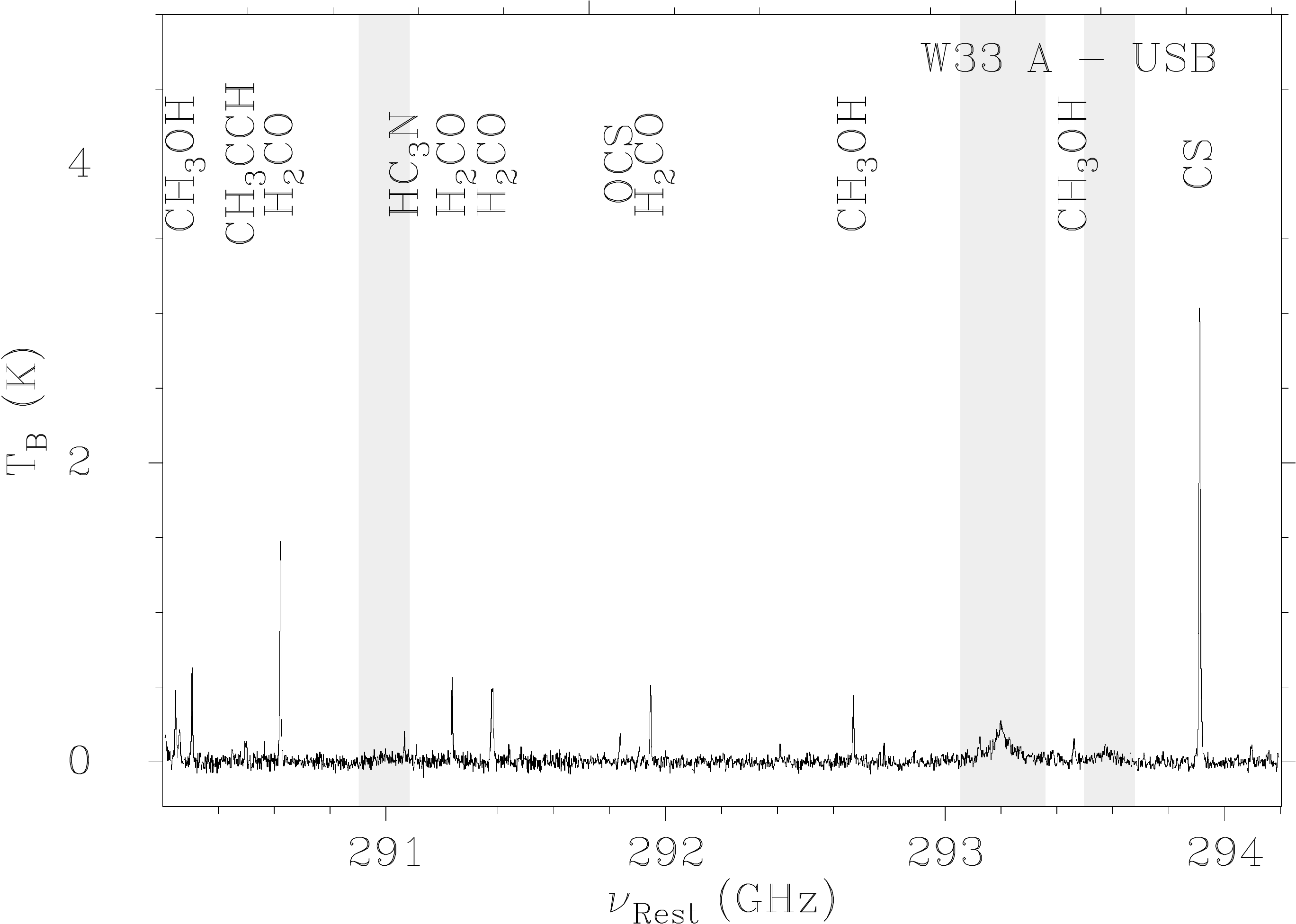}}\\	
	\subfloat{\includegraphics[width=9cm]{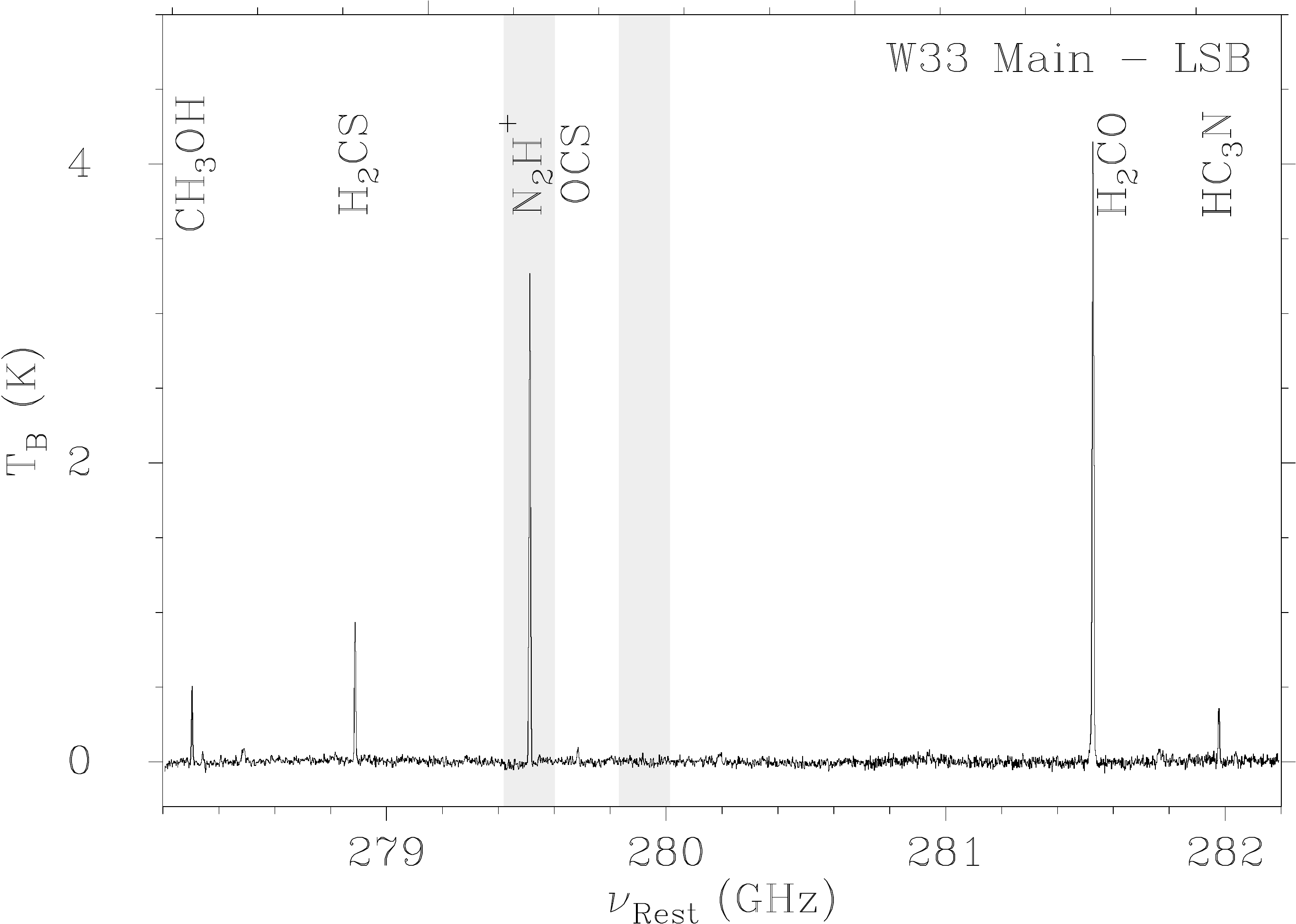}}\hspace{0.2cm} 	
	\subfloat{\includegraphics[width=9cm]{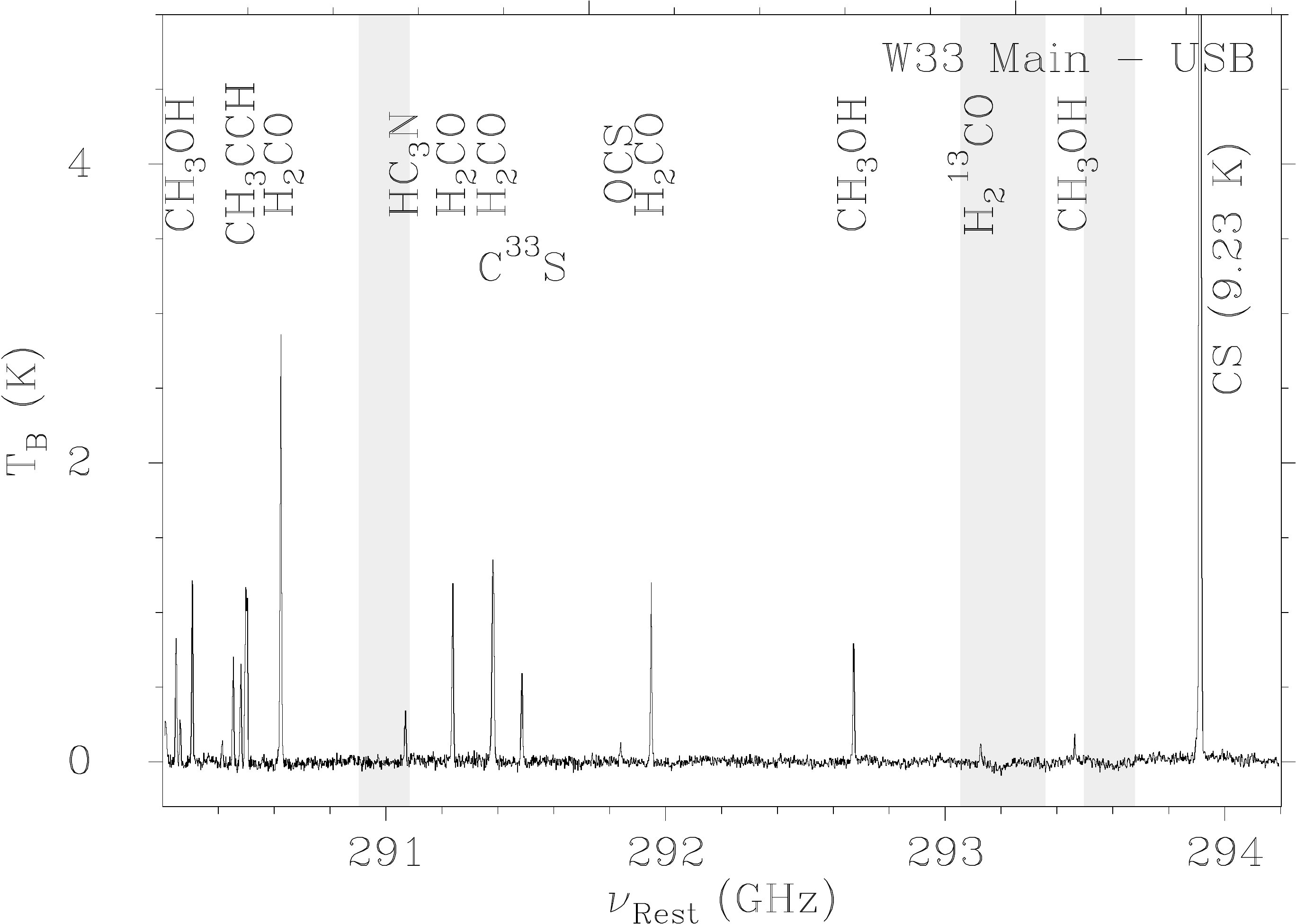}}	
	\label{W33_APEX_Spectra2}
\end{figure*}

The spectral lines in the APEX spectra (Figs. \ref{W33_APEX_Spectra1} and \ref{W33_APEX_Spectra2}) were identified, 
using the Cologne Database for Molecular Spectroscopy catalog \citep[CDMS,][]{Mueller2001}, 
the Jet Propulsion Laboratory spectral line catalog \citep[JPL,][]{Pickett1998}, and the splatalogue 
catalog \citep{Remijan2007}. In Table \ref{LinesTabAPEX} in the online Appendix, the rest frequency (Column 1), the 
transition (Column 2), and the upper energy level E$_{u}$ (Column 3) of each detected spectral line are 
listed. In total, we observed 27 transitions of the ten molecules CH$_{3}$OH, H$_{2}$CS, 
N$_{2}$H$^{+}$, OCS, H$_{2}$CO, H$_{2}^{13}$CO, HC$_{3}$N, CH$_{3}$CCH, C$^{33}$S, and 
CS. The upper energy levels E$_{u}$ of these transitions range between 27 and 241~K. All 
detected spectral lines show Gaussian line profiles. We fitted them with single Gaussians 
to obtain the velocity integrated intensity F$_{Int.}$, the peak intensity F$_{Peak}$, the central 
velocity v$_{central}$, and the FWHM of each line in each source. The fitting results are 
shown in Table \ref{LinesTabAPEX}. We made moment 0 maps of 
each line in each source (Figs. \ref{W33M1-APEX-IntInt} $-$ \ref{W33M-APEX-IntInt} in the online appendix) to study the distribution of the line emission over the entire 
OTF maps and compare them to the location of the ATLASGAL continuum peaks at 870 $\mu$m.
The spectra and moment maps of the W33 sources are discussed in detail in online Appendix \ref{ResultsAPEXApp}. In the next paragraph, we will summarize the source-specific APEX results.
In some of the APEX spectra, absorption features are present (see for example the spectra of W33\,A1 in Fig. \ref{W33_APEX_Spectra1}). These are not ``real'' features but are caused by poorly calibrated atmospheric absorption. We marked the atmospheric bands within our frequency range in the spectra of Figs. \ref{W33_APEX_Spectra1} and \ref{W33_APEX_Spectra2}.

The transitions N$_{2}$H$^{+}$(3$-$2), 
H$_{2}$CO(4$_{\textnormal{1,4}}$$-$3$_{\textnormal{1,3}}$), 
H$_{2}$CO(4$_{\textnormal{0,4}}$$-$3$_{\textnormal{0,3}}$), 
H$_{2}$CO(4$_{\textnormal{2,3}}$$-$3$_{\textnormal{2,2}}$), 
H$_{2}$CO(4$_{\textnormal{3,2}}$$-$3$_{\textnormal{3,1}}$) and 
H$_{2}$CO(4$_{\textnormal{3,1}}$$-$3$_{\textnormal{3,0}}$) (which are blended), 
H$_{2}$CO(4$_{\textnormal{2,2}}$$-$3$_{\textnormal{2,1}}$), 
CH$_{3}$OH(6$_{\textnormal{-2,5}}$--5$_{\textnormal{-2,4}}$) 
and CH$_{3}$OH(6$_{\textnormal{2,4}}$--5$_{\textnormal{2,4}}$) (which are blended), 
CH$_{3}$OH(6$_{\textnormal{1,5}}$$-$5$_{\textnormal{1,4}}$), and CS(6$-$5) are 
detected in all sources. CS and N$_{2}$H$^{+}$ trace cold and dense gas in the clumps.
The strongest lines in all sources are the two low-excitation 
transitions of H$_{2}$CO at 35 and 46 K, N$_{2}$H$^{+}$(3$-$2), and CS(6$-$5).
The least number of transitions (see above) are detected in W33\,Main1 (Fig. \ref{W33_APEX_Spectra1}). The spatial distribution of the spectral line emission hints to a cold interior of W33\,Main1 and possibly an external heating source in the vicinity of W33\,Main1. In W33\,A1 and W33\,B1, two additional CH$_{3}$OH transitions, compared to the lines detected in W33\,Main1, are observed (Fig. \ref{W33_APEX_Spectra1}). The 
spatial distribution of the emission of the high-excitation lines of H$_{2}$CO and CH$_{3}$OH
indicate the presence of heating sources in W33\,A1 and W33\,B1. The object W33\,A1, however, seems to be less developed than W33\,B1. Besides the molecules detected in W33\,A1 and W33\,B1, we 
observe emission from H$_{2}$CS, C$^{33}$S, and OCS in W33\,B and W33\,A and additionally  HC$_{3}$N and CH$_{3}$CCH in W33\,A (Fig. \ref{W33_APEX_Spectra2}). The detection of transitions at higher excitation energies and stronger spectral lines compared to W33\,A1 and W33\,B1 suggests that W33\,B and W33\,A are more evolved than W33\,B1 and W33\,A1. The spectra of W33\,Main are the most line-rich of the six sources (Fig. \ref{W33_APEX_Spectra2}). The emission of the cold gas tracers CS and N$_{2}$H$^{+}$ peaks offset the continuum emission, indicating that the gas towards the center of W33\,Main is not cold anymore. 

\paragraph{}
Following these results, we sort the W33 sources along an evolutionary sequence: W33\,Main1, W33\,A1, W33\,B1, W33\,B, W33\,A, and W33\,Main. The more evolved sources emit transitions with higher excitation energies, which indicate that the temperature in our sources increases along this sequence. Furthermore, the detected lines tend to get stronger and broader along this sequence. 
\citet{Sanhueza2012} showed that the median values of the N$_{2}$H$^{+}$ line widths for their four classes of objects increases from 2.7 to 3.4 km s$^{-1}$ along their established evolutionary sequence from IRDCs to \ion{H}{ii} regions \citep[also seen in][]{Sakai2008, Vasyunina2011} due to the increase of turbulence during the star formation process. We see a similar trend in our observations, but the line widths of the N$_{2}$H$^{+}$ transition are larger than the values of \citet{Sanhueza2012} ranging from 4.7 to 6.4 km s$^{-1}$ compared to 1.6 to 4.6 km s$^{-1}$ in the observations of \citet{Sanhueza2012}.

\paragraph{Chemical clocks in the APEX data?}

\begin{figure*}
	\caption{Integrated intensity ratios N$_{2}$H$^{+}$(3$-$2)/CS(6$-$5) and N$_{2}$H$^{+}$(3$-$2)/H$_{2}$CO(4$_{2,2}$$-$3$_{2,1}$) plotted versus (a) the evolutionary sequence of the six W33 source, (b) their bolometric luminosities, (c) their total masses, (d) their bolometric luminosity to total mass ratio L$_{bol}$/M, and (e) their H$_{2}$ peak column densities. The latter ratio is divided by 10 to be plotted in the same range as the former ratio.}
	\centering
	\subfloat[Evolutionary State\label{LR-EvolState}]{\includegraphics[width=9cm]{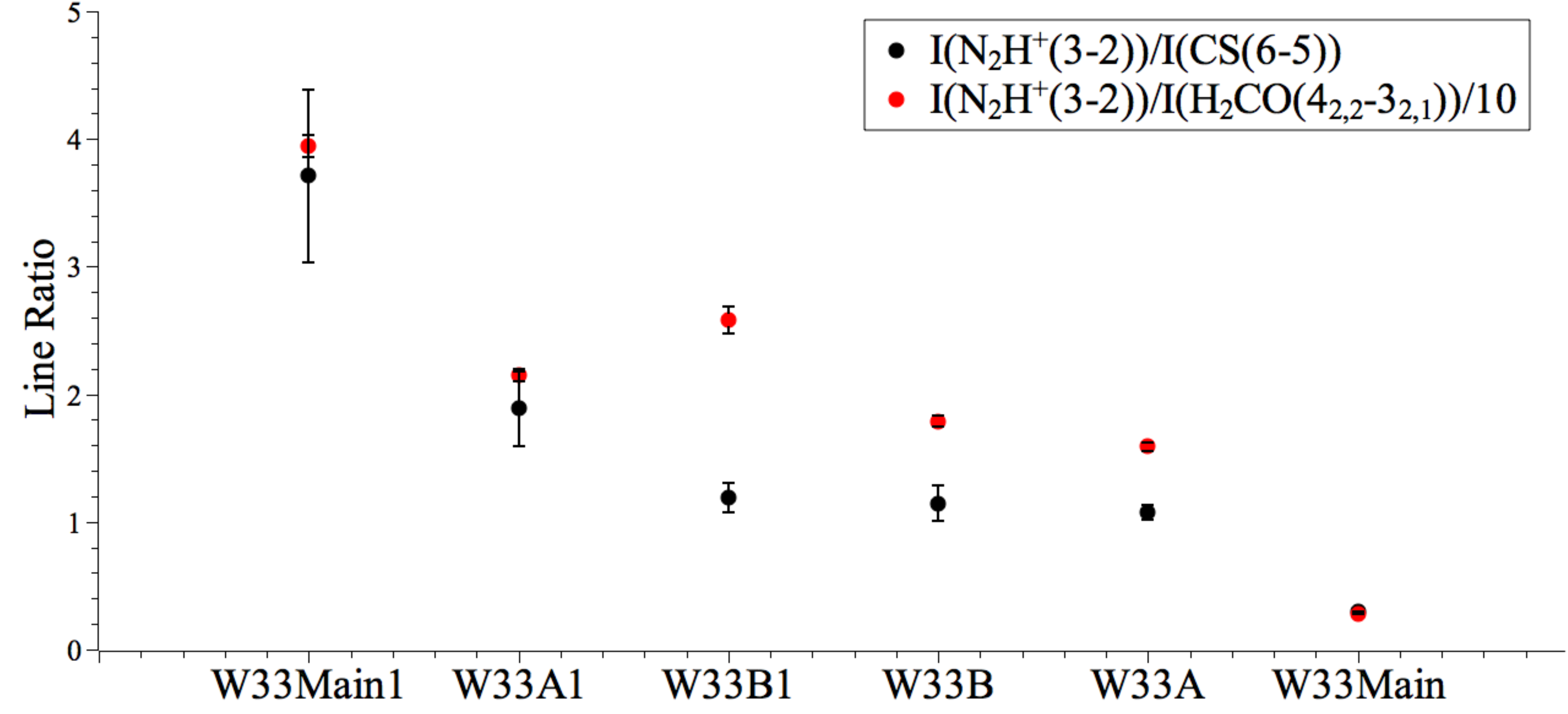}}
	\vspace{3pt}
	\subfloat[Bolometric luminosity L$_{bol}$\label{LR-Luminosity}]{\includegraphics[width=9cm]{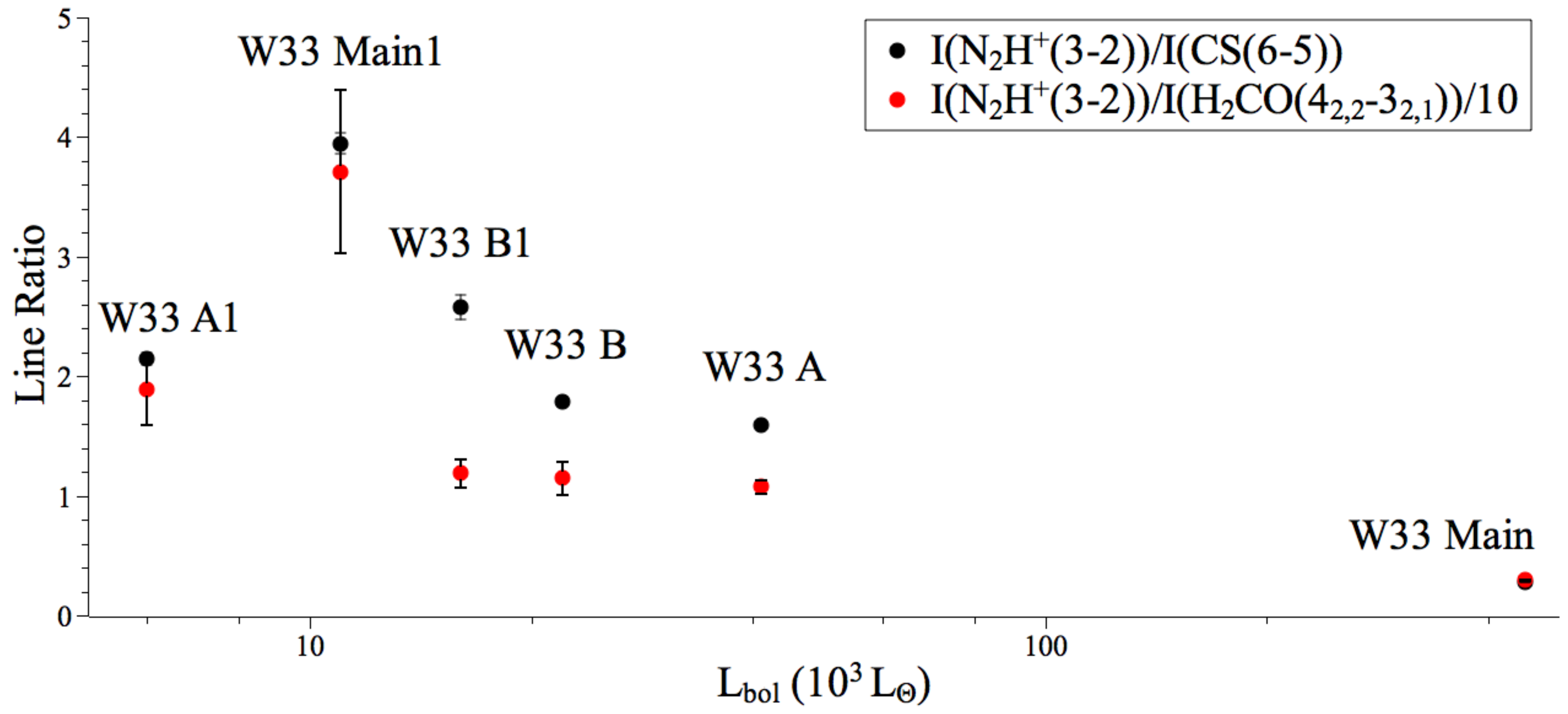}}\\
	\vspace{3pt}
	\subfloat[Total mass M\label{LR-Mass}]{\includegraphics[width=9cm]{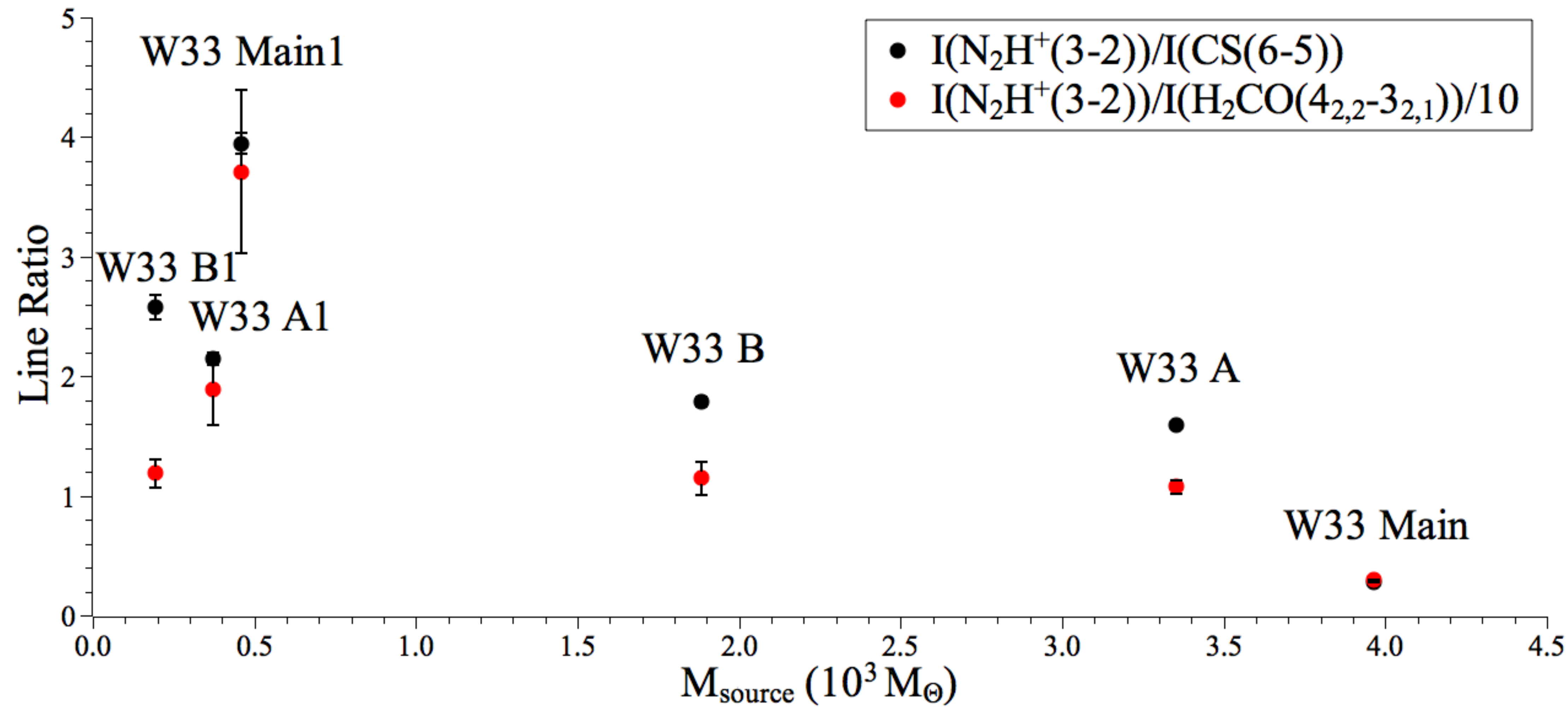}}
	\vspace{3pt}
	\subfloat[L$_{bol}$/M\label{LR-LM}]{\includegraphics[width=9cm]{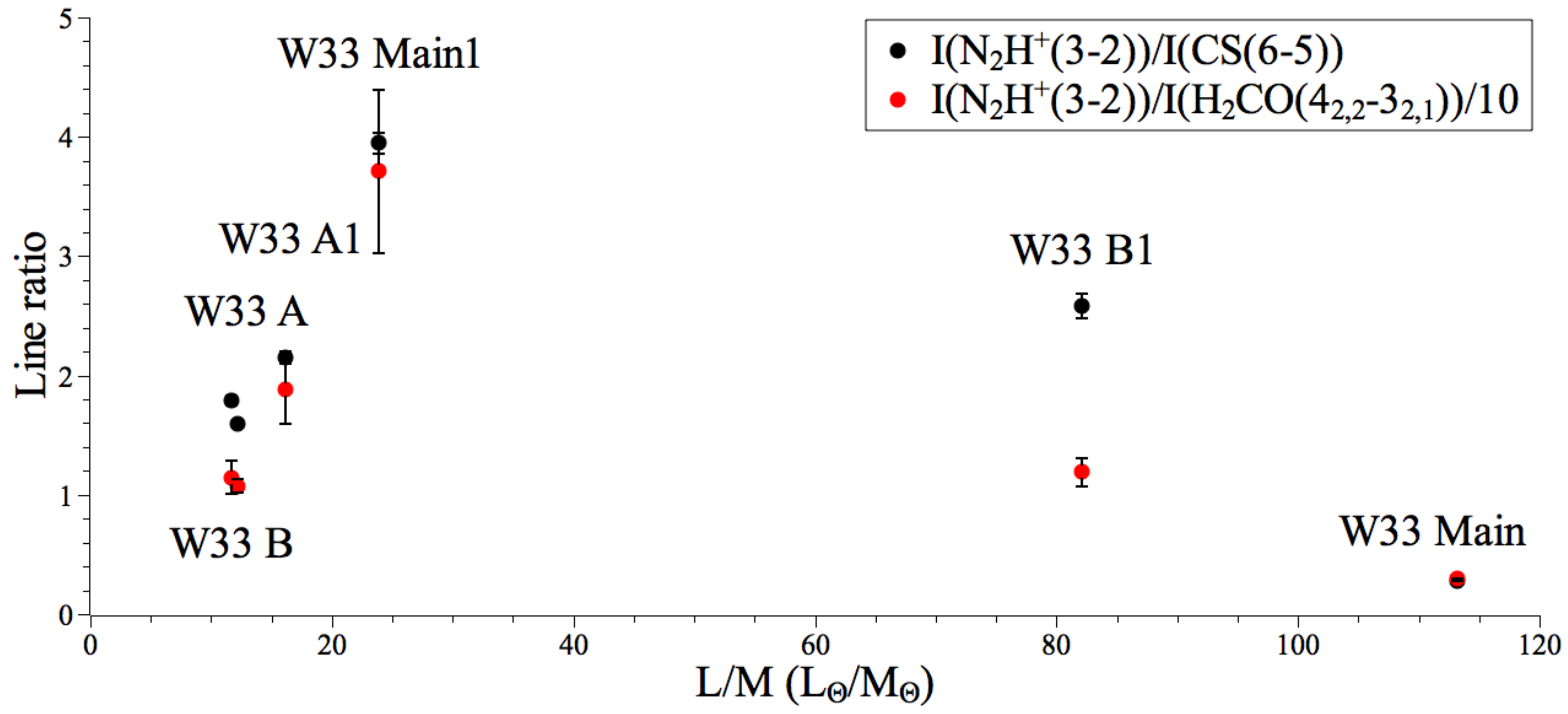}}\\
	\vspace{3pt}
	\subfloat[H$_{2}$ peak column density\label{LR-Peak-Col-Dens}]{\includegraphics[width=9cm]{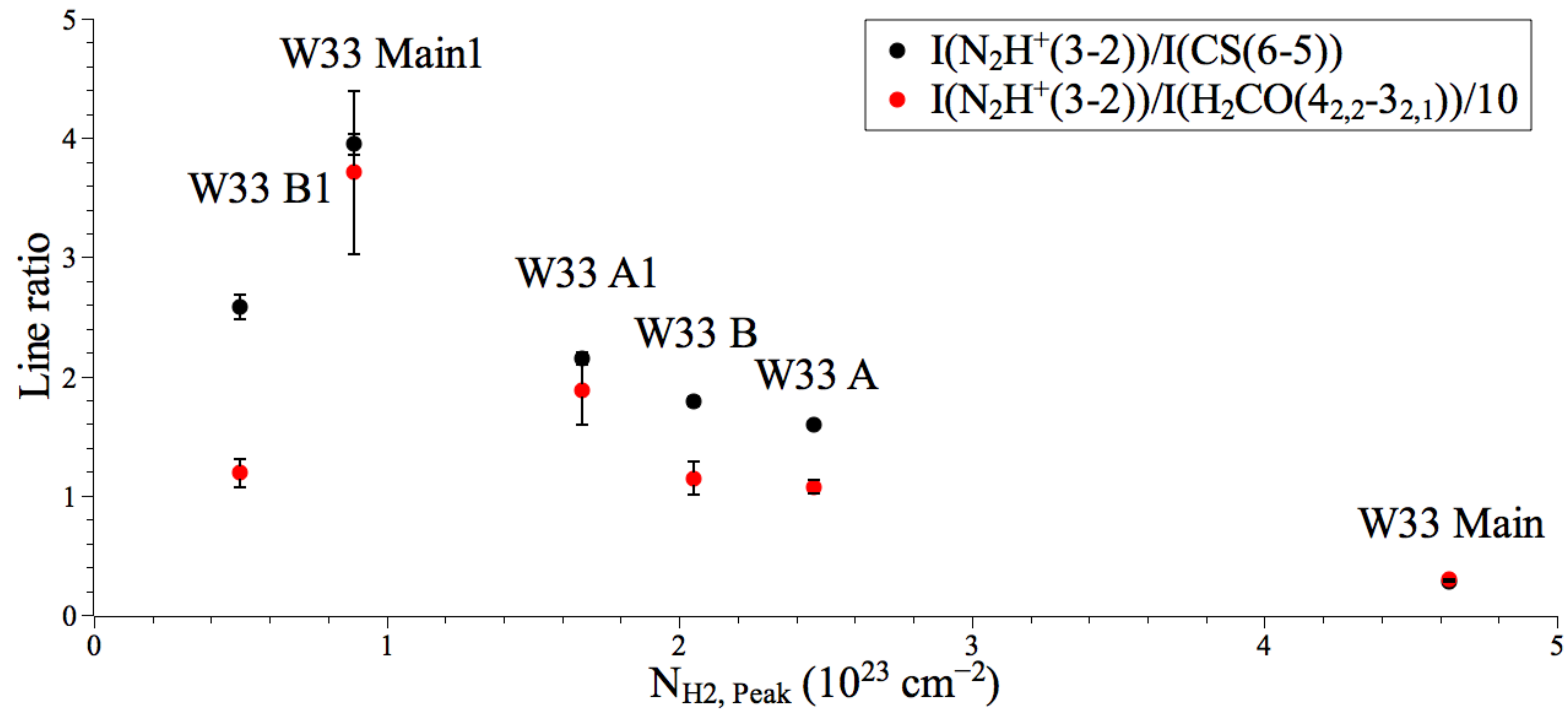}}
	\label{IntIntRatio}
\end{figure*}

A total of 11 transitions of the molecules N$_{2}$H$^{+}$, CS, H$_{2}$CO, and CH$_{3}$OH are observed in all six W33 clumps. While the detection of single transitions does not yield 
information about the evolutionary state of a clump, abundance or integrated intensity ratios of molecules \citep[e.g. SiO/CS, HCN/N$_{2}$H$^{+}$, or NH$_{3}$/N$_{2}$H$^{+}$, ][]{Fuente2005,Palau2007} have proven to be usable as ``chemical clocks'' for evolutionary stages. Since the emission of the above mentioned molecules has a similar spatial distribution in each of the six clumps, a comparison of the integrated intensity of the different transitions is reasonable. To test if a combination of the above mentioned molecules could be used as a tracer of evolutionary state, we calculated the integrated intensity ratios of all combinations of transitions that are detected towards all W33 clumps and checked their correlations with the evolutionary sequence that we established. 

We found clear trends as a function of evolutionary stage in the ratios N$_{2}$H$^{+}$(3$-$2)/CS(6$-$5) and N$_{2}$H$^{+}$(3$-$2)/H$_{2}$CO(4$_{2,2}$$-$3$_{2,1}$) (Fig. \ref{LR-EvolState}). The N$_{2}$H$^{+}$ and CS molecules both trace cold and dense gas in the clumps while the H$_{2}$CO molecule is released from the dust grains, once the protostars starts to warm up the material. However, to determine if these two integrated intensity ratios can be reliably used as chemical clocks, they have to be tested on a larger sample of star forming regions in different stages of evolution. 

Both integrated intensity ratios are the largest in W33\,Main1. Plotting the two integrated intensity ratios against the bolometric luminosities (Fig. \ref{LR-Luminosity}), the total masses (Fig. \ref{LR-Mass}), the bolometric luminosity to total mass ratio (Fig. \ref{LR-LM}), and the H$_{2}$ peak column densities (Fig. \ref{LR-Peak-Col-Dens}) of our six targets, we see clear trends of first increasing and then decreasing integrated intensity ratios with increasing luminosity, luminosity to mass ratio, and peak column density. Especially for sources with total clump masses $\geq$ 500 M$_{\sun}$, the integrated intensity ratios consistently decrease with increasing luminosity, mass, and peak column density. Comparing Fig. \ref{LR-EvolState} with Fig. \ref{LR-Peak-Col-Dens} shows that sources with higher peak column densities also look more chemically and physically evolved. This supports the assumption that sources with high peak column densities form stars earlier (see Section \ref{TempCloudMass}). The clump W33\,B1 is an exception; however, its evolutionary state may be influenced by the exterior illumination from the star cluster SC-1. In spite of its low luminosity, W33\,A1 appears more chemically evolved than W33\,Main1 and W33\,B1, which may be explained by the higher peak column density of W33\,A1 and thus a faster/earlier formation of OB stars in this clump.

\section{Results of the SMA observations}
\label{ResultsSMA}

\subsection{230 GHz continuum emission}

\begin{figure*}
	\caption{230 GHz continuum maps of W33\,Main1, W33\,A1, W33\,B1, W33\,B, W33\,A, and W33\,Main. The black contours are at $-$4, $-$2, 2, 4, 8, ... $\cdot$ $\sigma$ (1$\sigma$ = 6,  5, 5, 10, 5, and 70 mJy beam$^{-1}$). Negative contours are shown with a dashed line. The map of W33\,A was obtained from \citet{GalvanMadrid2010}. The synthesised beams are shown in the lower left corners of the images. In the upper left corners, a 0.1 pc scale is indicated. For comparison, the white contours show the ATLASGAL continuum emission at 345 GHz (same levels as in Figs. \ref{W33M1-APEX-IntInt} $-$ \ref{W33M-APEX-IntInt}).}
	\centering
	\subfloat{\includegraphics[width=8.5cm]{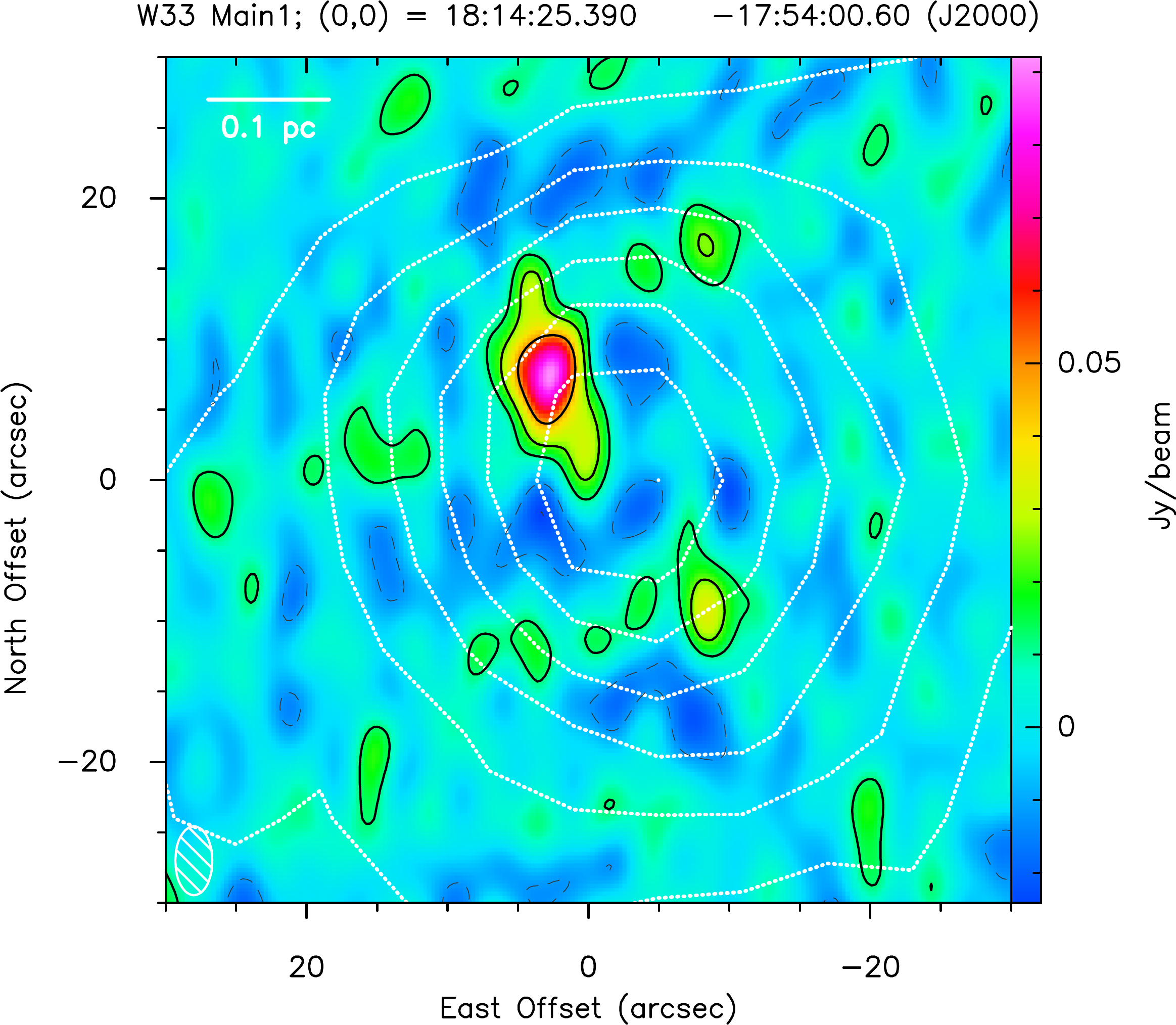}}\hspace{0.2cm}	
	\subfloat{\includegraphics[width=8.5cm]{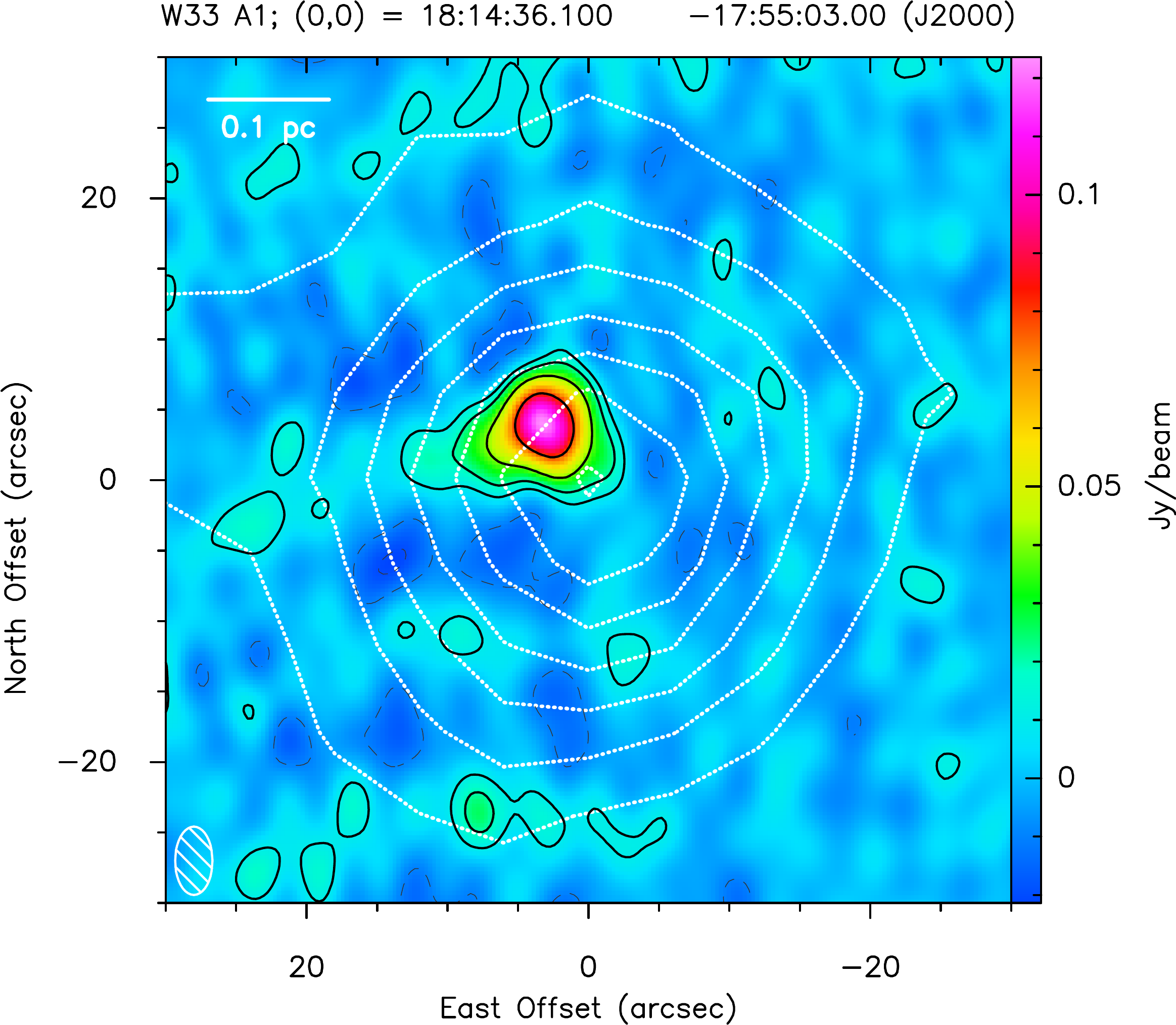}}	\\
	\subfloat{\includegraphics[width=8.5cm]{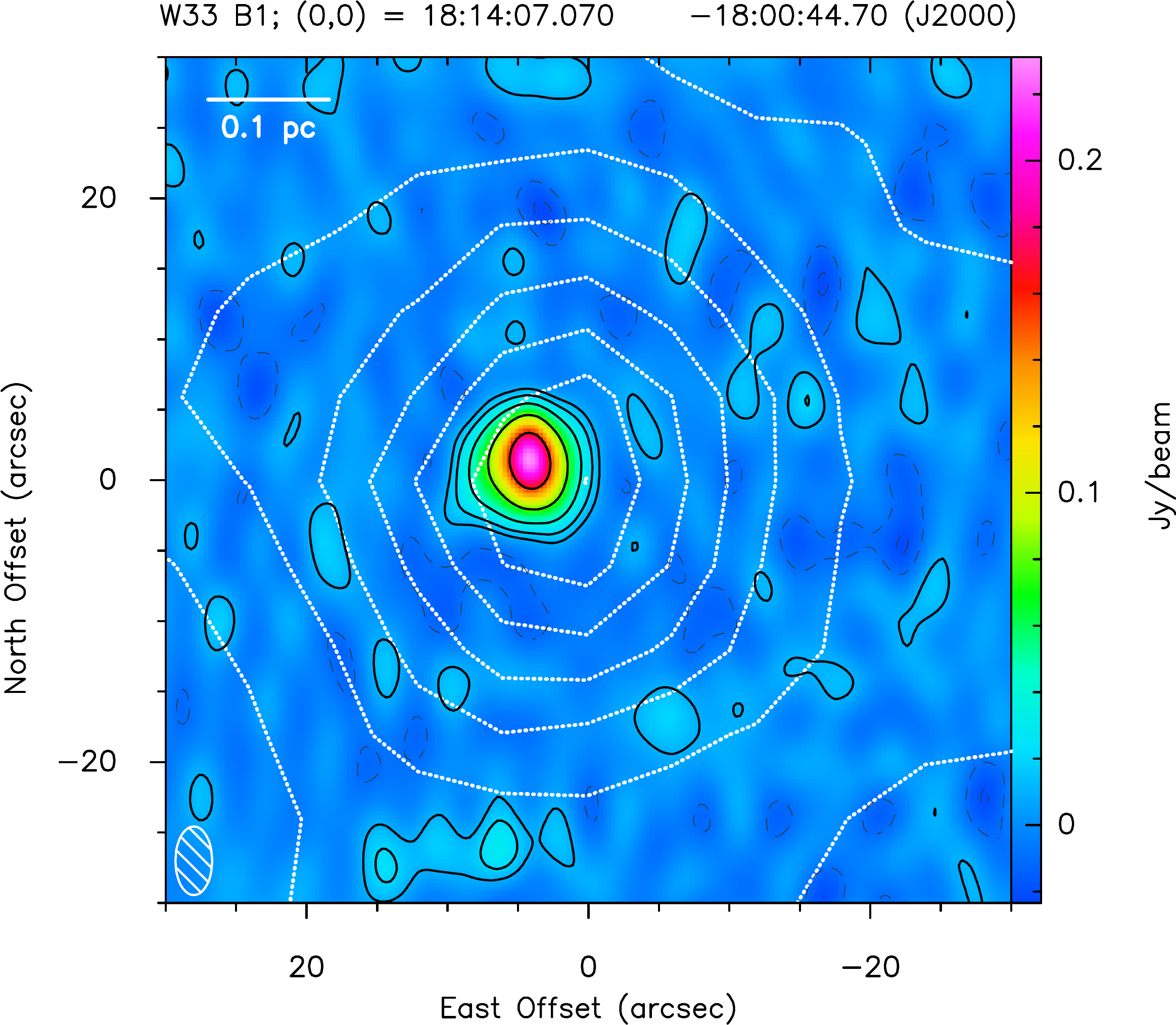}} \hspace{0.2cm}
	\subfloat{\includegraphics[width=8.5cm]{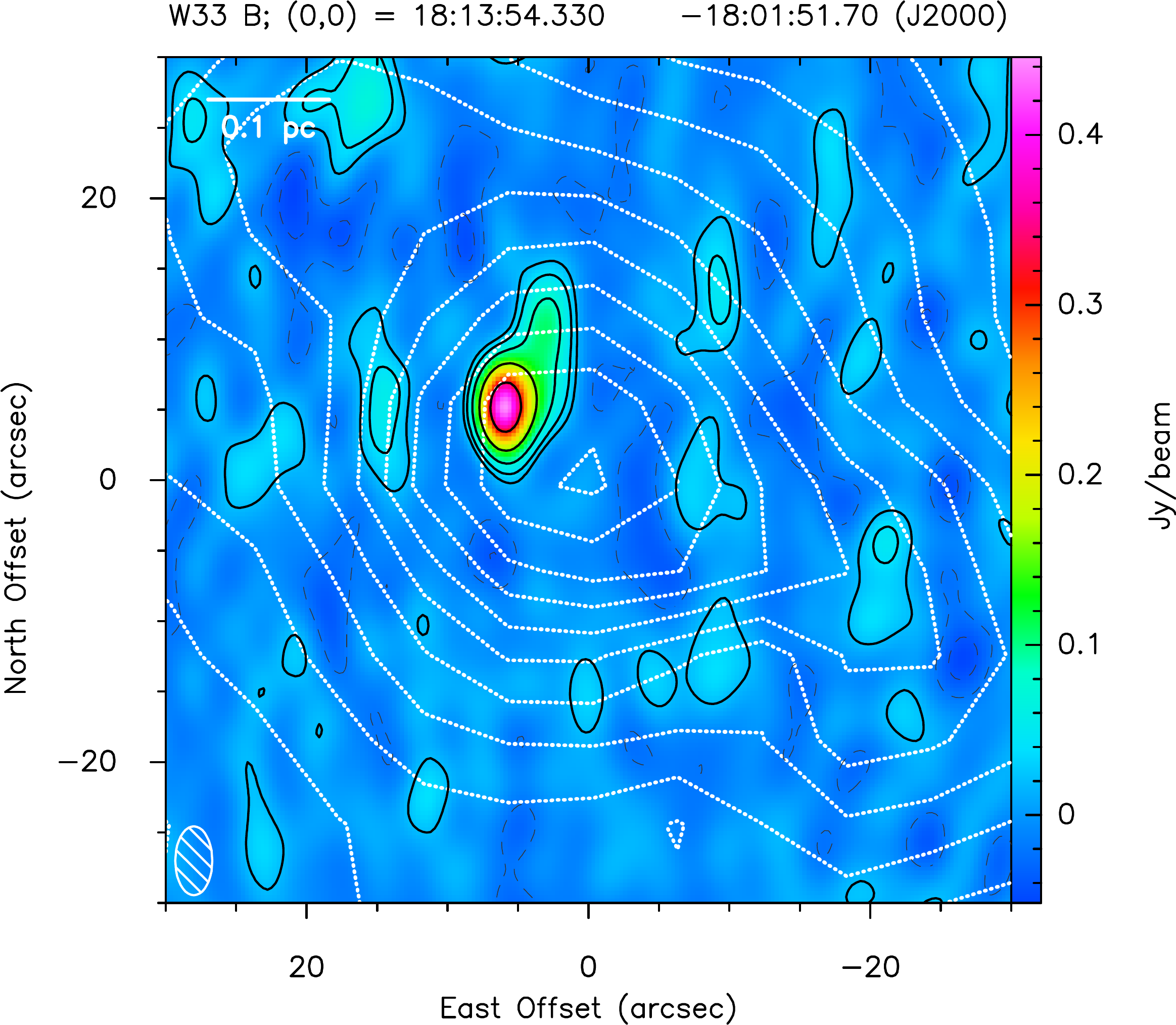}}\\ 
	\subfloat{\includegraphics[width=8.5cm]{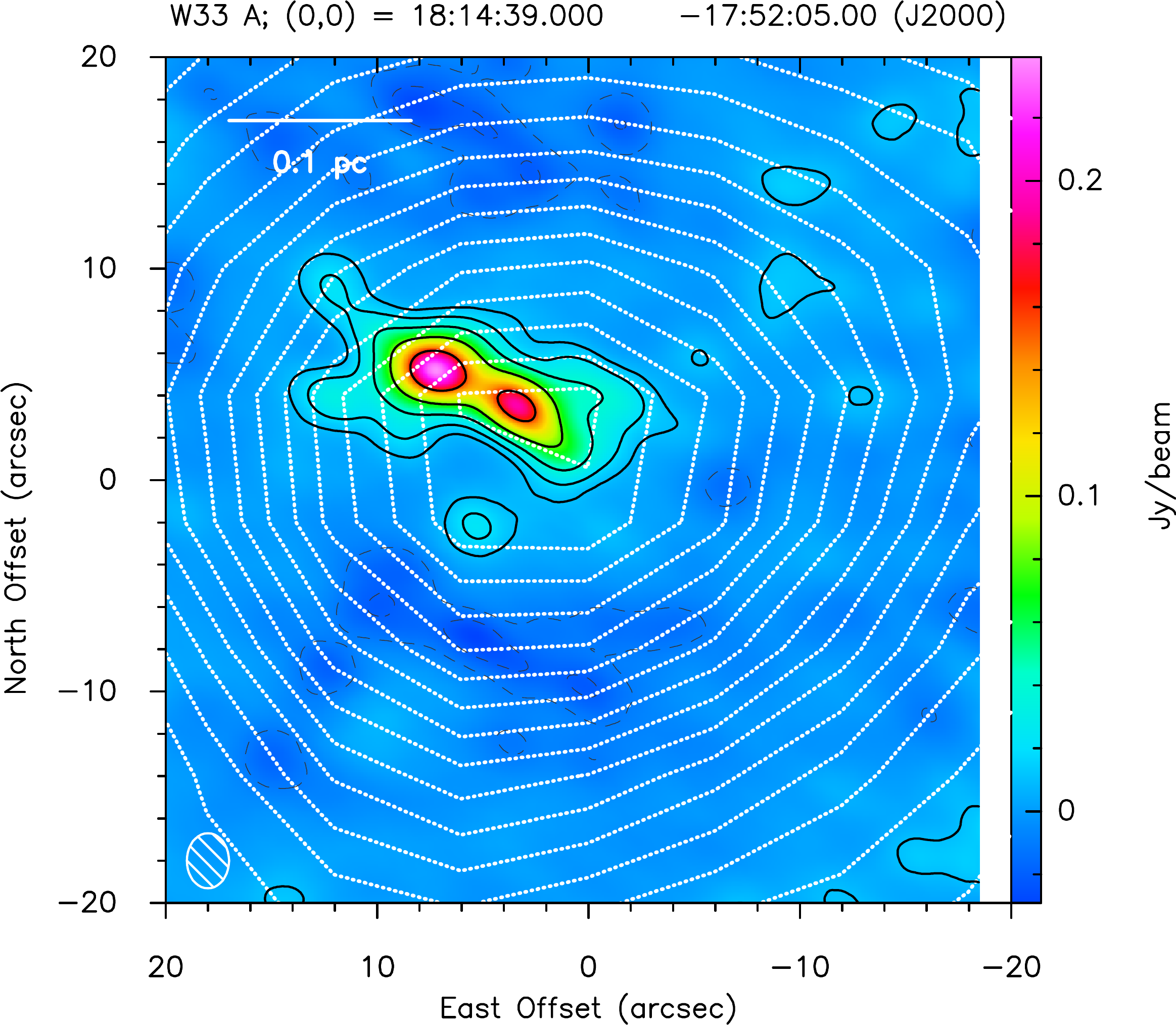}} \hspace{0.2cm}			
         \subfloat{\includegraphics[width=8.5cm]{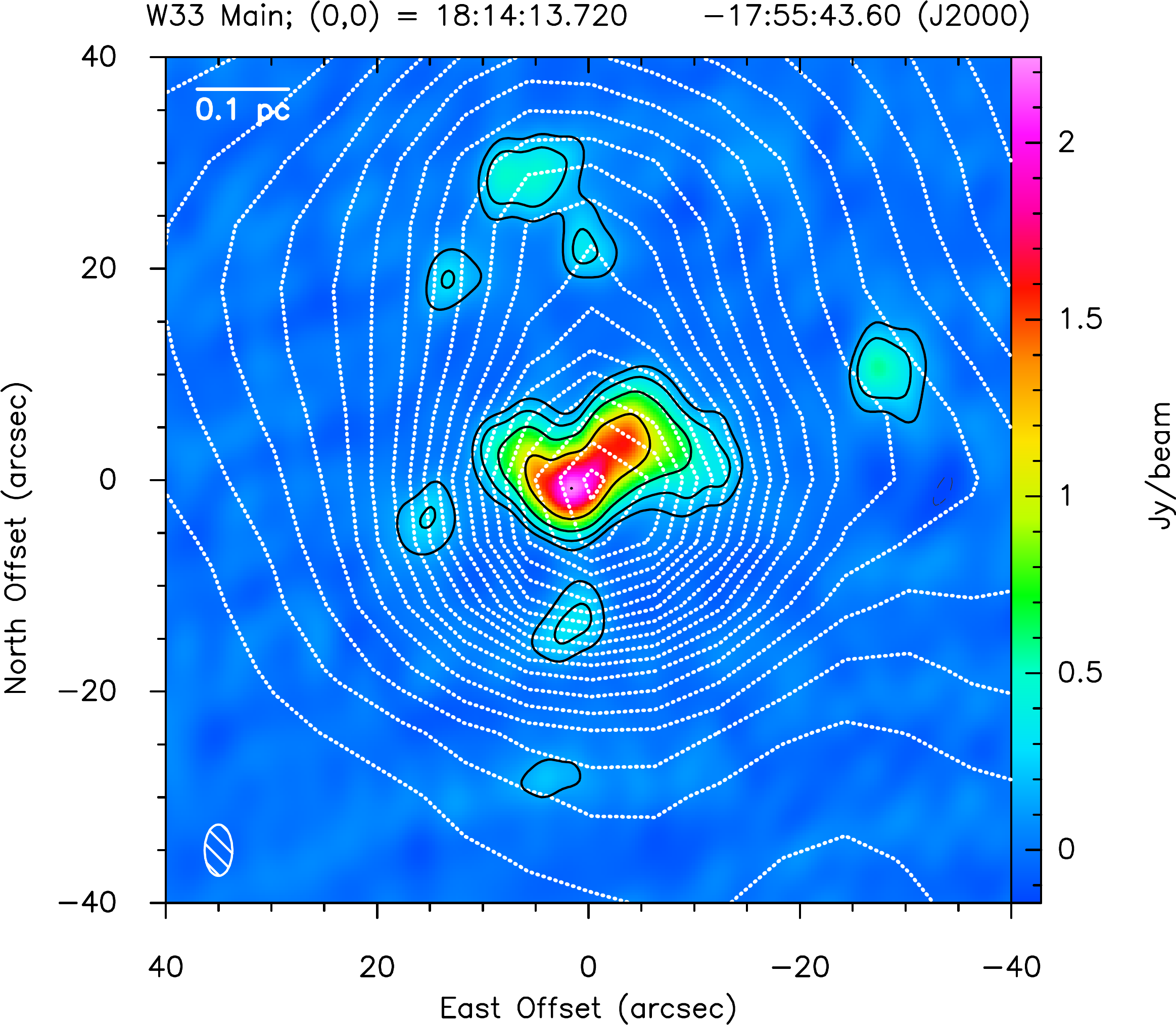}}	
	\label{W33_SMA_CH0}
\end{figure*}

\begin{figure*}
\caption{230 (left) and 345 (right) GHz continuum maps of W33\,Main. The black contours show the 230 GHz emission (same contour levels as in Fig. \ref{W33_SMA_CH0}, sixth panel). The synthesised beams are indicated by the ellipses in the lower left corners. In the upper left corners, a 0.1 pc scale is indicated. As in Fig. \ref{W33_SMA_CH0}, the white contours show the ATLASGAL continuum emission at 345 GHz.}
\centering
	\subfloat{\includegraphics[width=8.5cm]{W33Main_SMA_CH0-eps-converted-to.pdf}}\hspace{0.2cm}			
         \subfloat{\includegraphics[width=8.5cm]{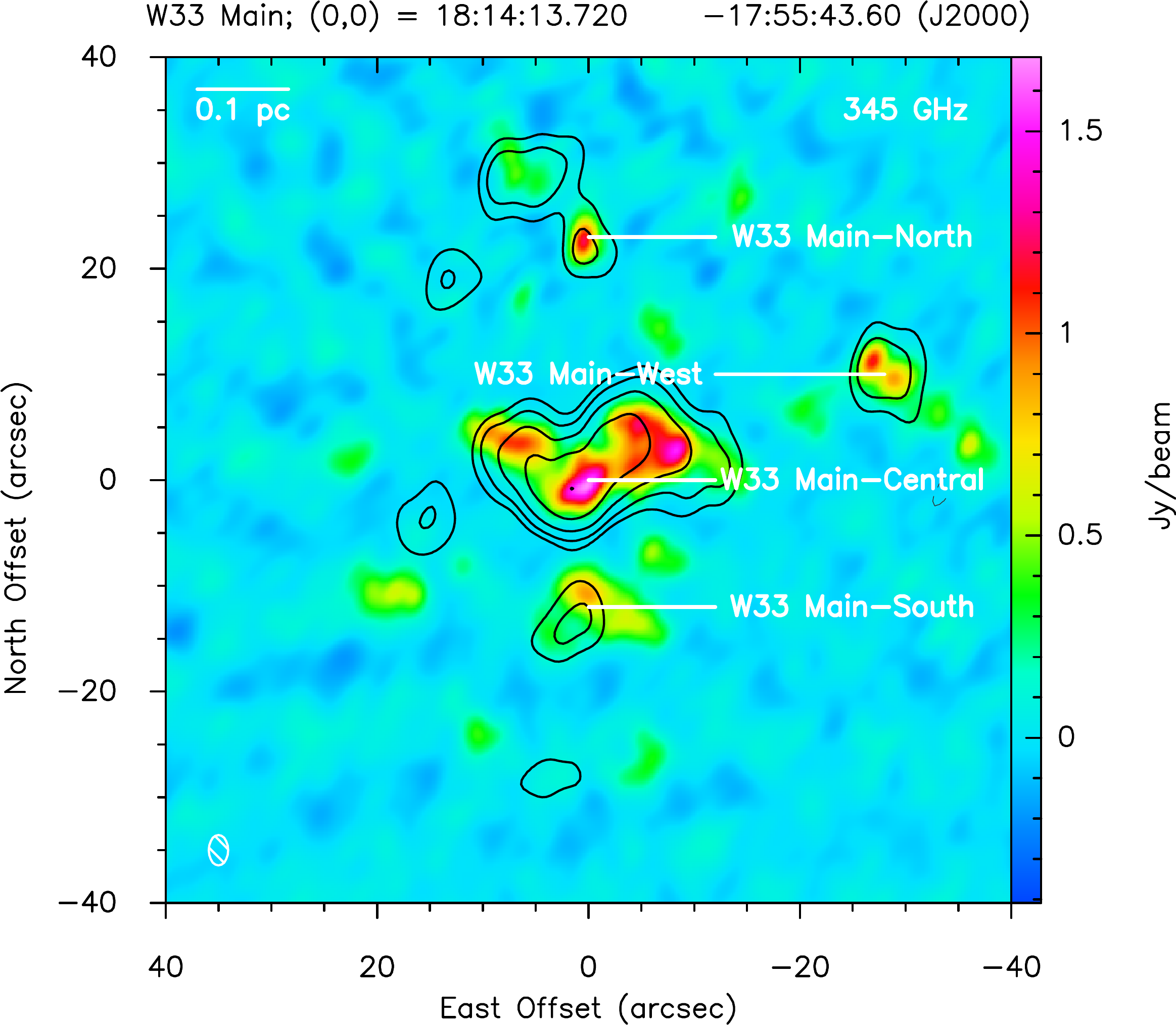}}
\label{W33M_SMA_CH0}
\end{figure*}

Figure \ref{W33_SMA_CH0} shows the 230 GHz continuum emission of the six W33 sources W33\,Main1, W33\,A1, W33\,B1, W33\,B, W33\,A, and W33\,Main, as observed 
with the SMA. Except for W33\,A and W33\,Main, 
the dust cores are single-peaked (at physical scales of 0.1 pc). The object W33\,B1 is the only source that is nearly circular. The other sources are elongated along one axis. The cores have sizes of about 0.1 pc.
The dust emission of W33\,Main shows two peaks in the main core, which has a size of about 0.25 pc. Surrounding the main core, several smaller cores with weaker emission are observed. 
The image of W33\,A was obtained from a combination of the SMA observations in compact and very extended configuration of \citet{GalvanMadrid2010}. For a better comparison with our SMA observations, the image was cleaned with natural weighting, which yields a synthesised beam size of 2.6$\arcsec$~x~2.0$\arcsec$. The continuum emission shows two peaks within one core. The core has a size of $\sim$0.15 pc.

The left and right panels of Fig. \ref{W33M_SMA_CH0} show the continuum emission of W33\,Main at 230 and 345 GHz. The contours correspond to the dust emission at 230 GHz with the same levels as in the sixth panel of Fig. \ref{W33_SMA_CH0}. Several emission peaks are detected in the main core of W33\,Main in the 345 GHz map. Outside of the main core, three more cores are detected at both frequencies, which we name W33\,Main-North, W33\,Main-West, and W33\,Main-South. In the following text, we refer to the main core as W33\,Main-Central.

\subsection{230 GHz line emission}

\begin{figure*}
	\caption{SMA spectra generated at the continuum peak positions of W33\,Main1, W33\,A1, and W33\,B1.}
	\centering
	\subfloat{\includegraphics[width=9cm]{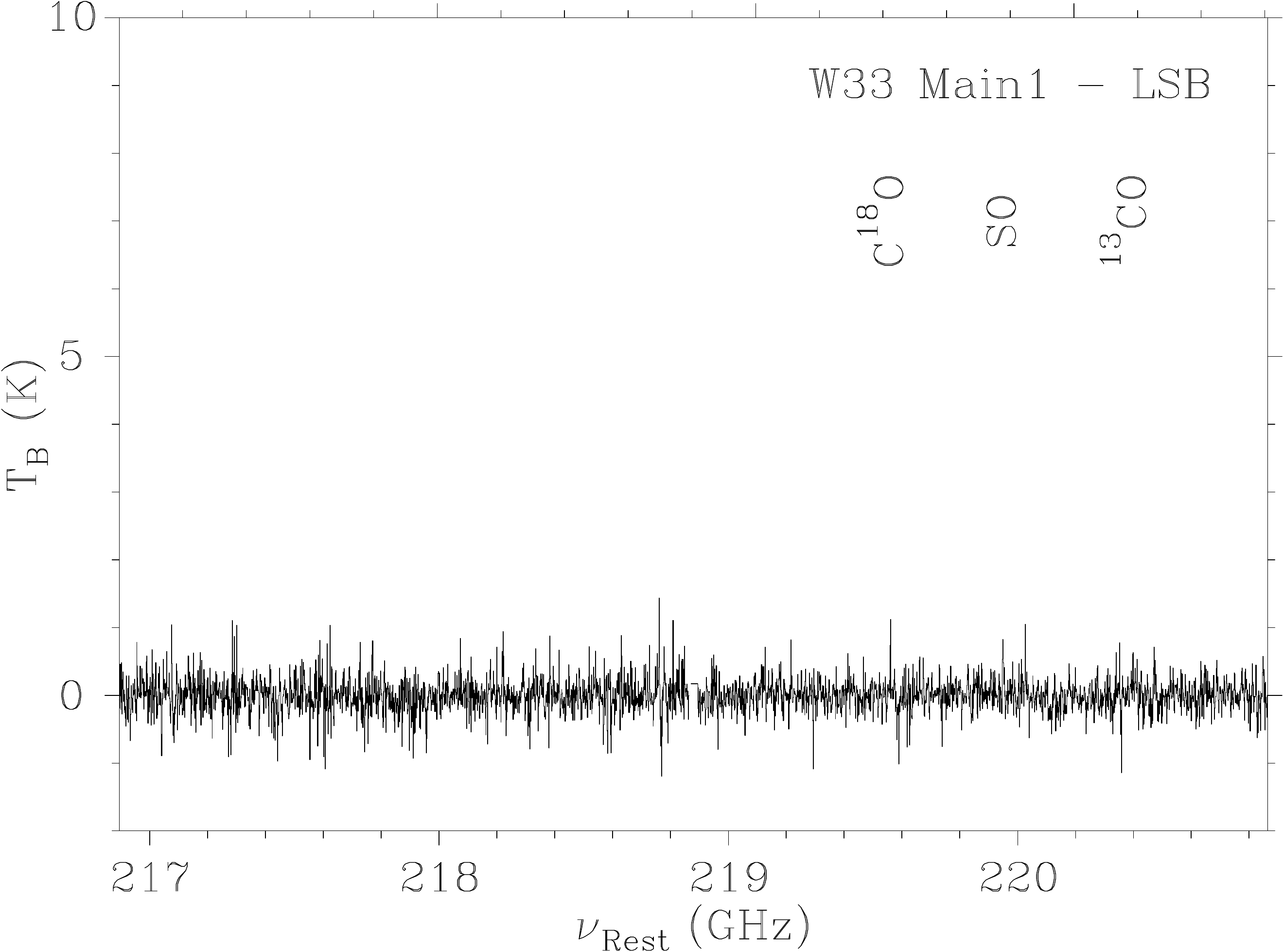}}\hspace{0.2cm}
	\subfloat{\includegraphics[width=9cm]{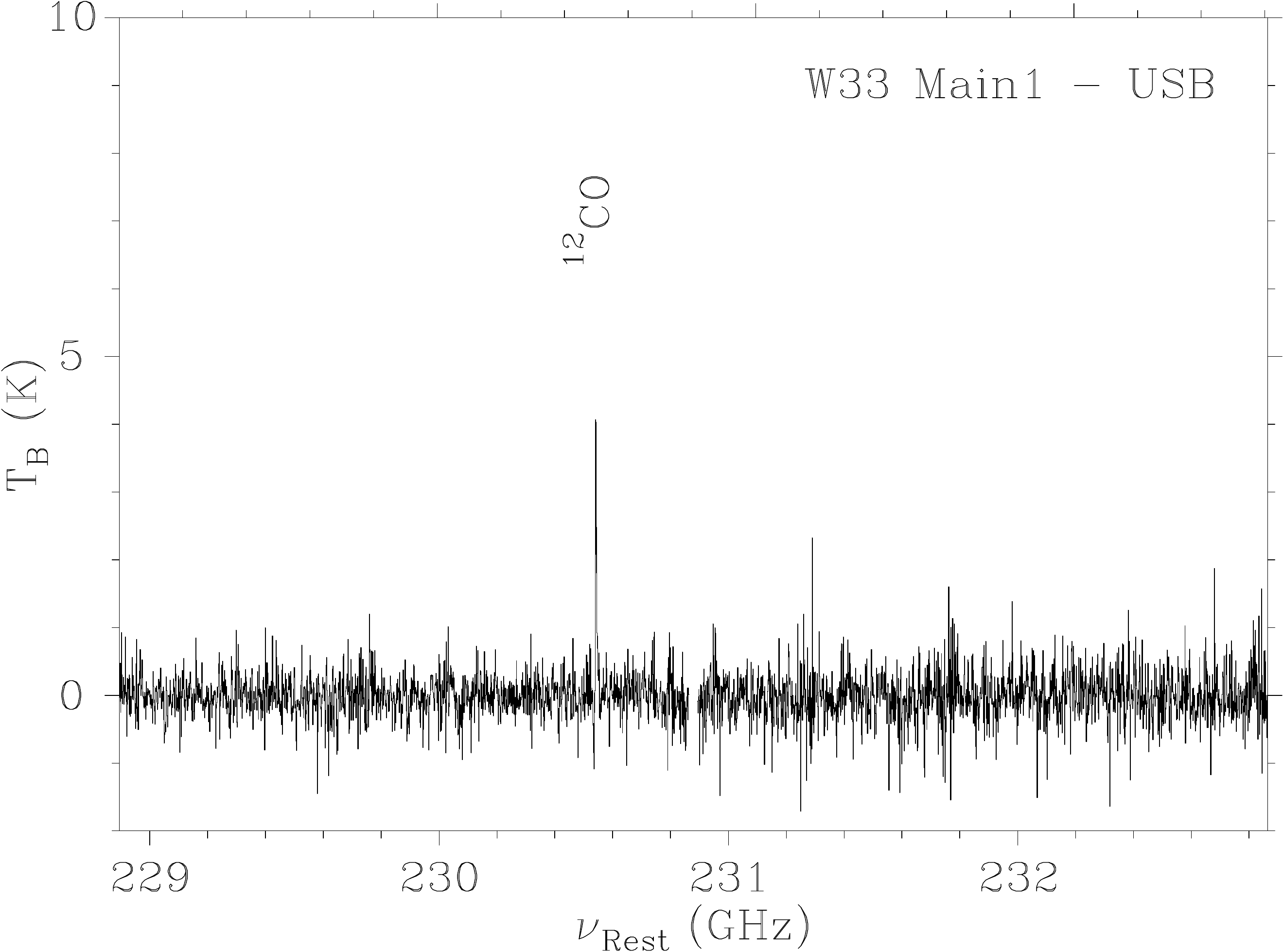}}\\	
	\subfloat{\includegraphics[width=9cm]{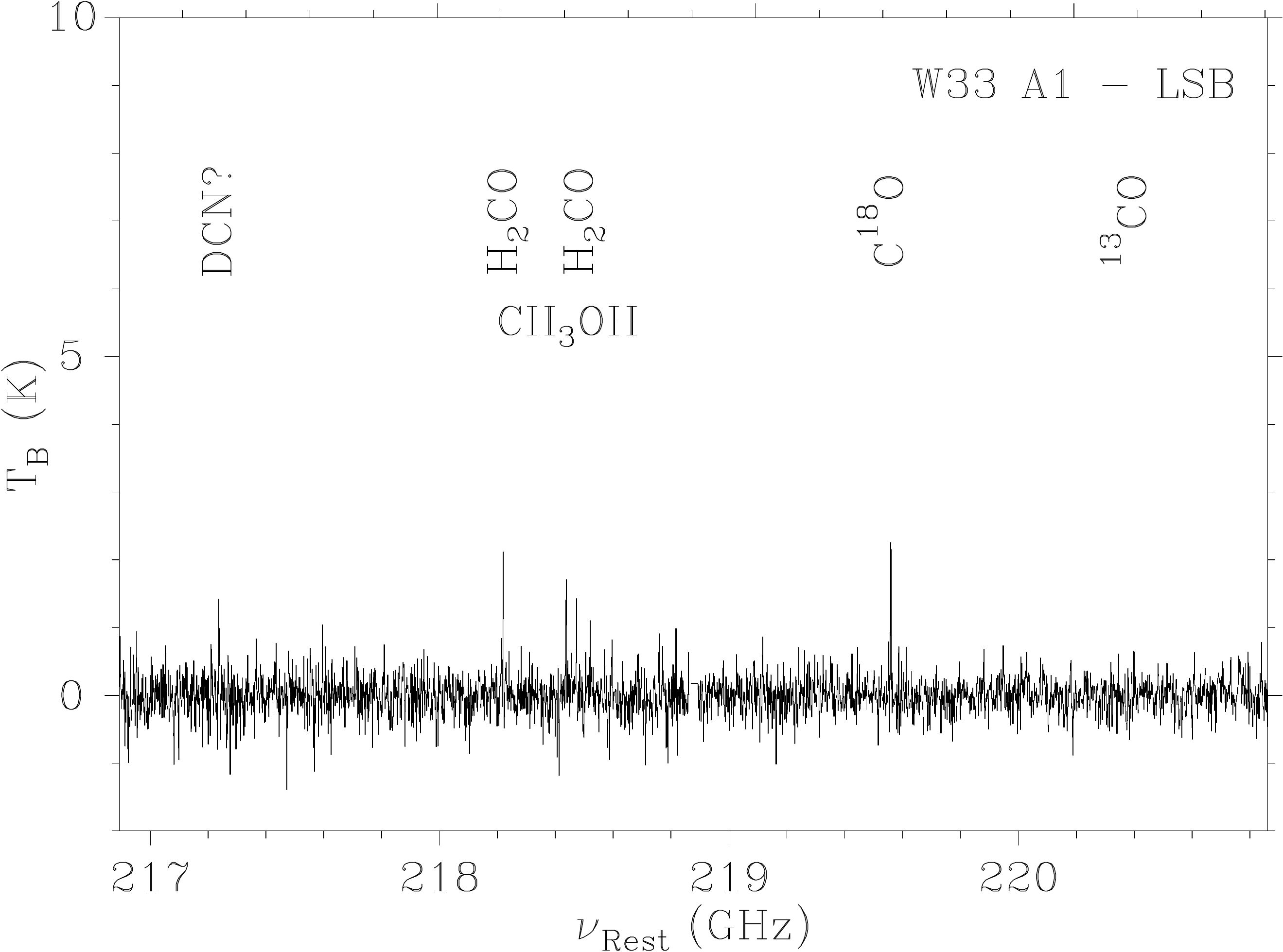}}\hspace{0.2cm}
	\subfloat{\includegraphics[width=9cm]{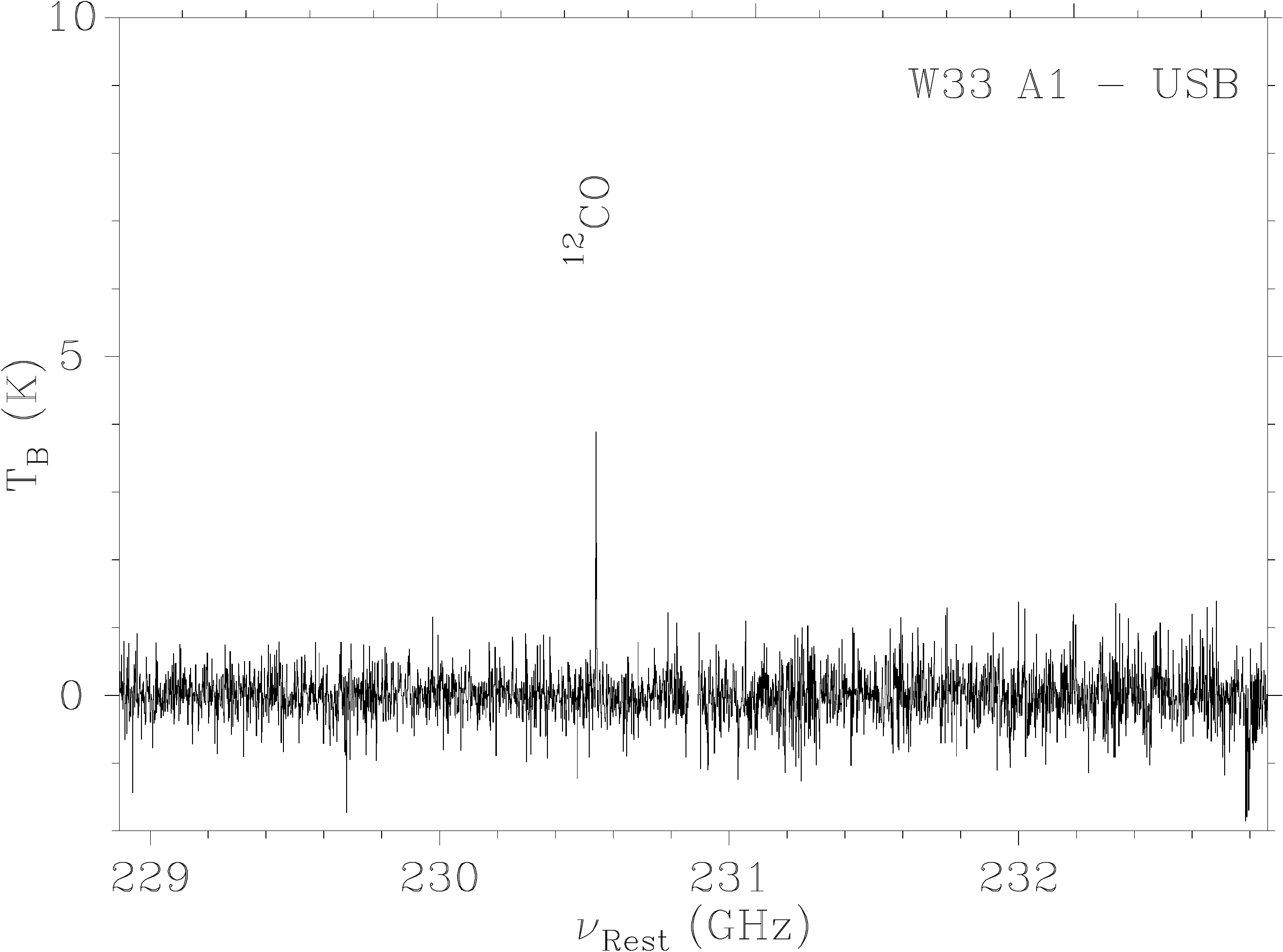}}	\\
	\subfloat{\includegraphics[width=9cm]{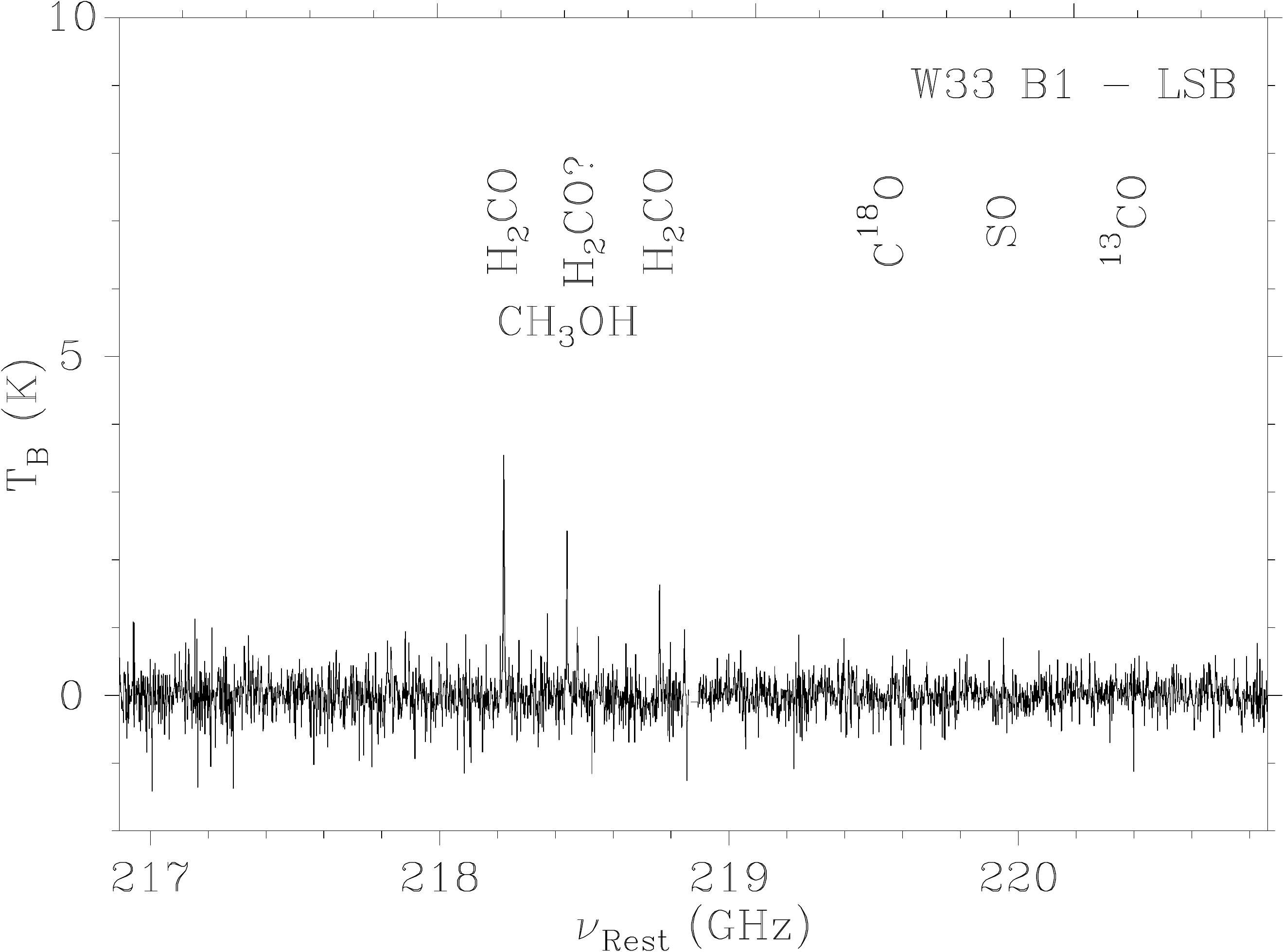}} \hspace{0.2cm}		
	\subfloat{\includegraphics[width=9cm]{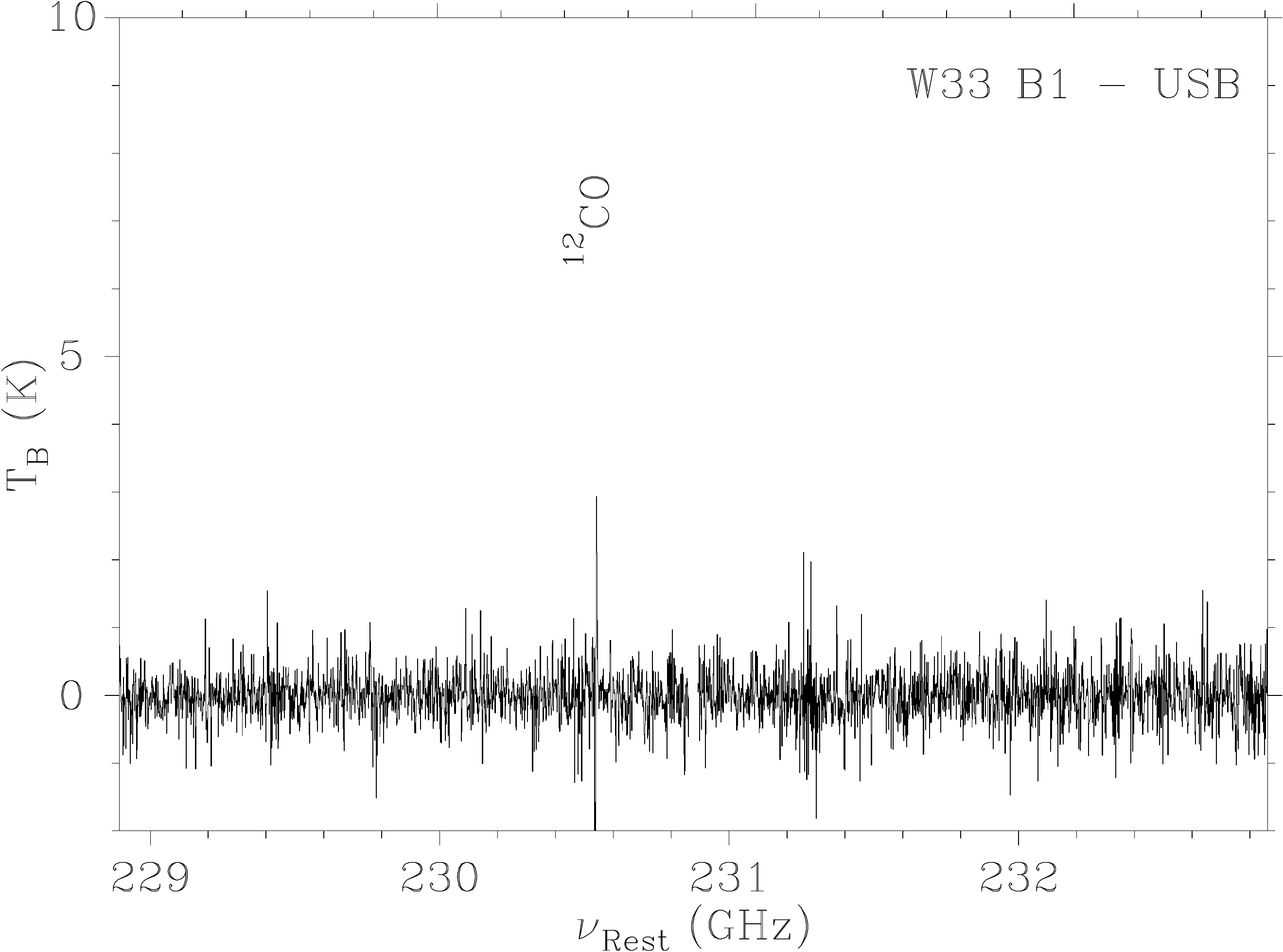}}	
	\label{W33_SMA_Spectra1}
\end{figure*}

\begin{figure*}
	\caption{SMA spectra generated at the continuum peak positions of W33\,B, W33\,A, and W33\,Main.}
	\centering
	\subfloat{\includegraphics[width=9cm]{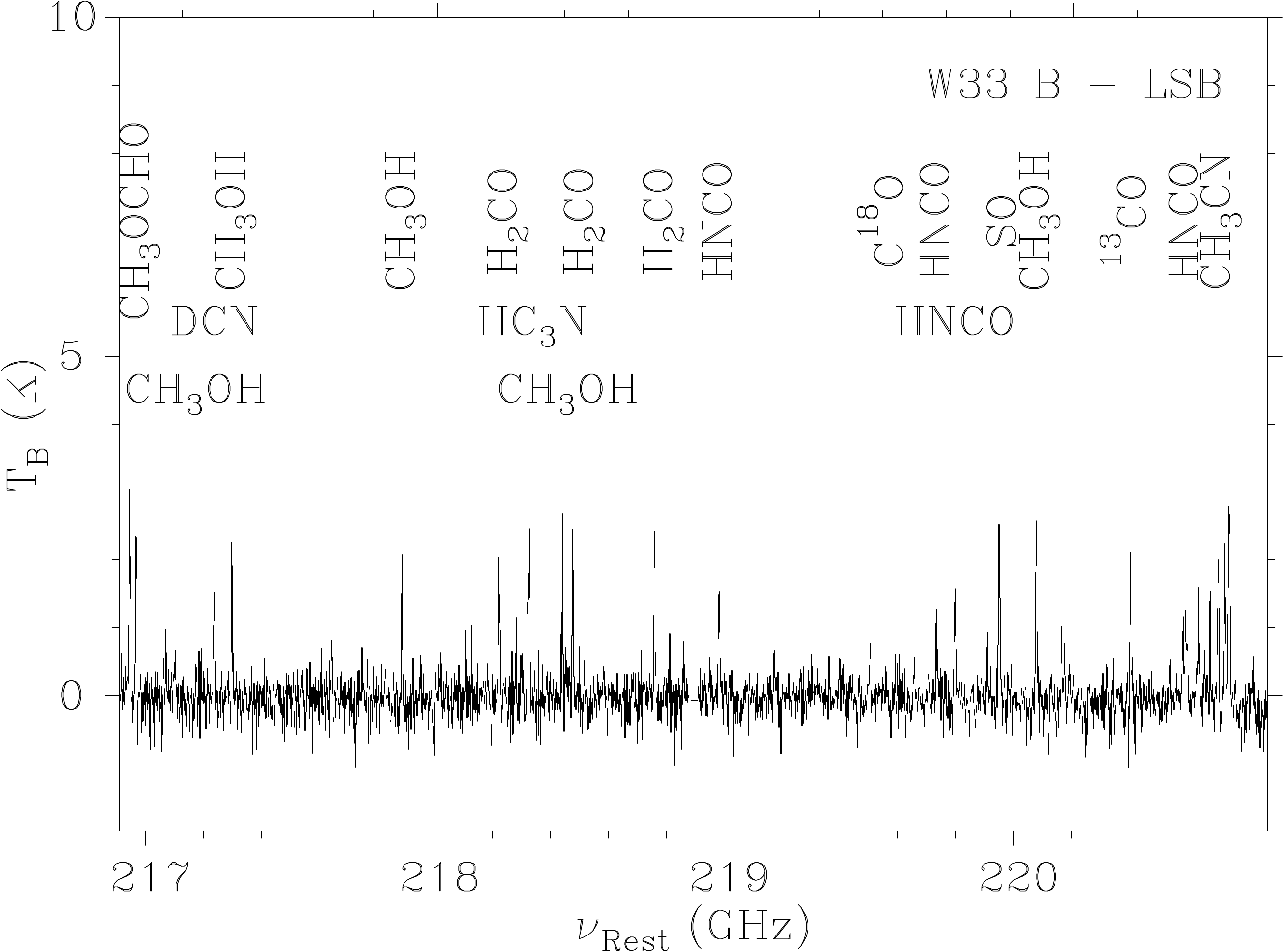}}\hspace{0.2cm} 		
	\subfloat{\includegraphics[width=9cm]{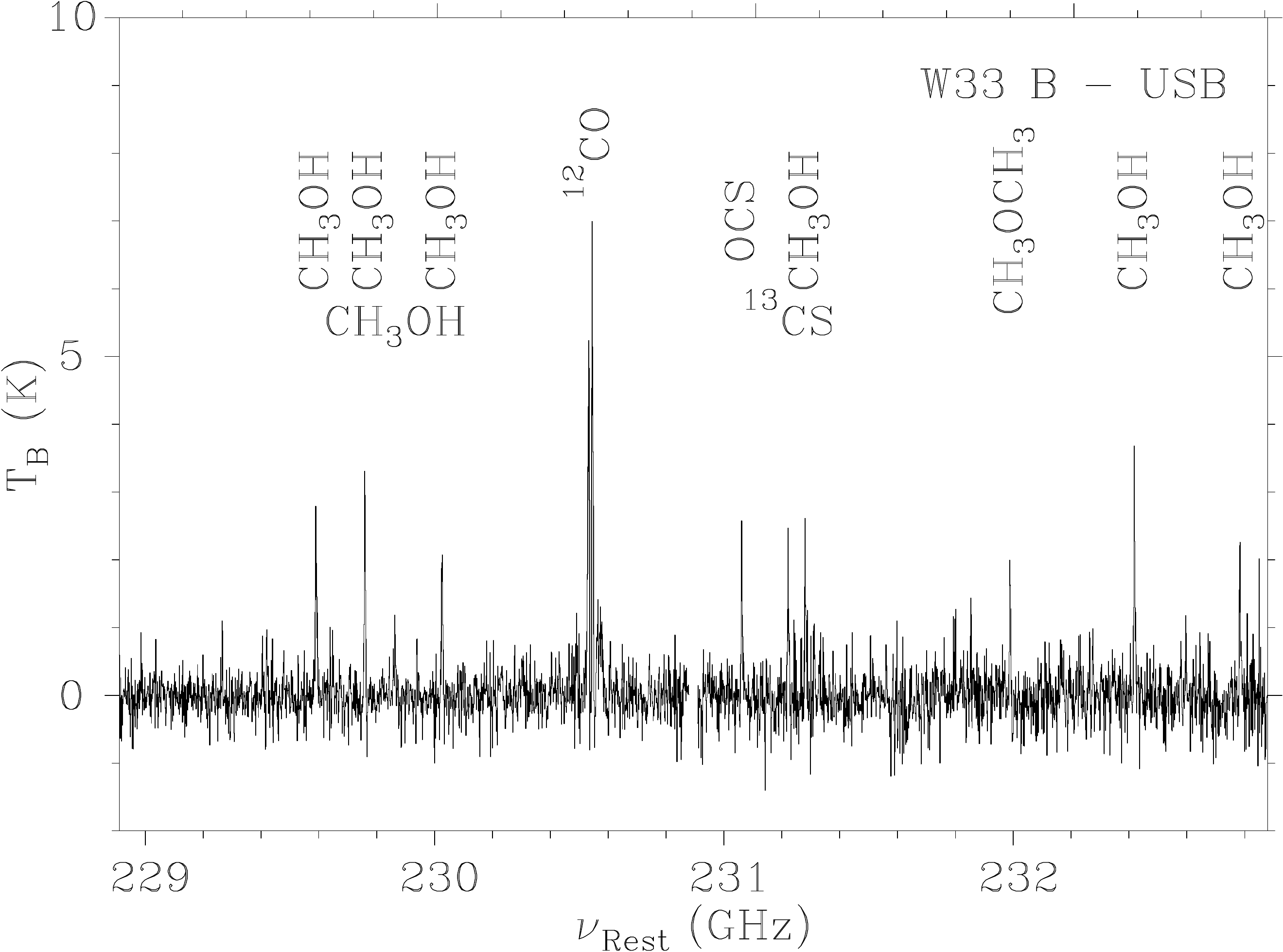}}\\	
	\subfloat{\includegraphics[width=9cm]{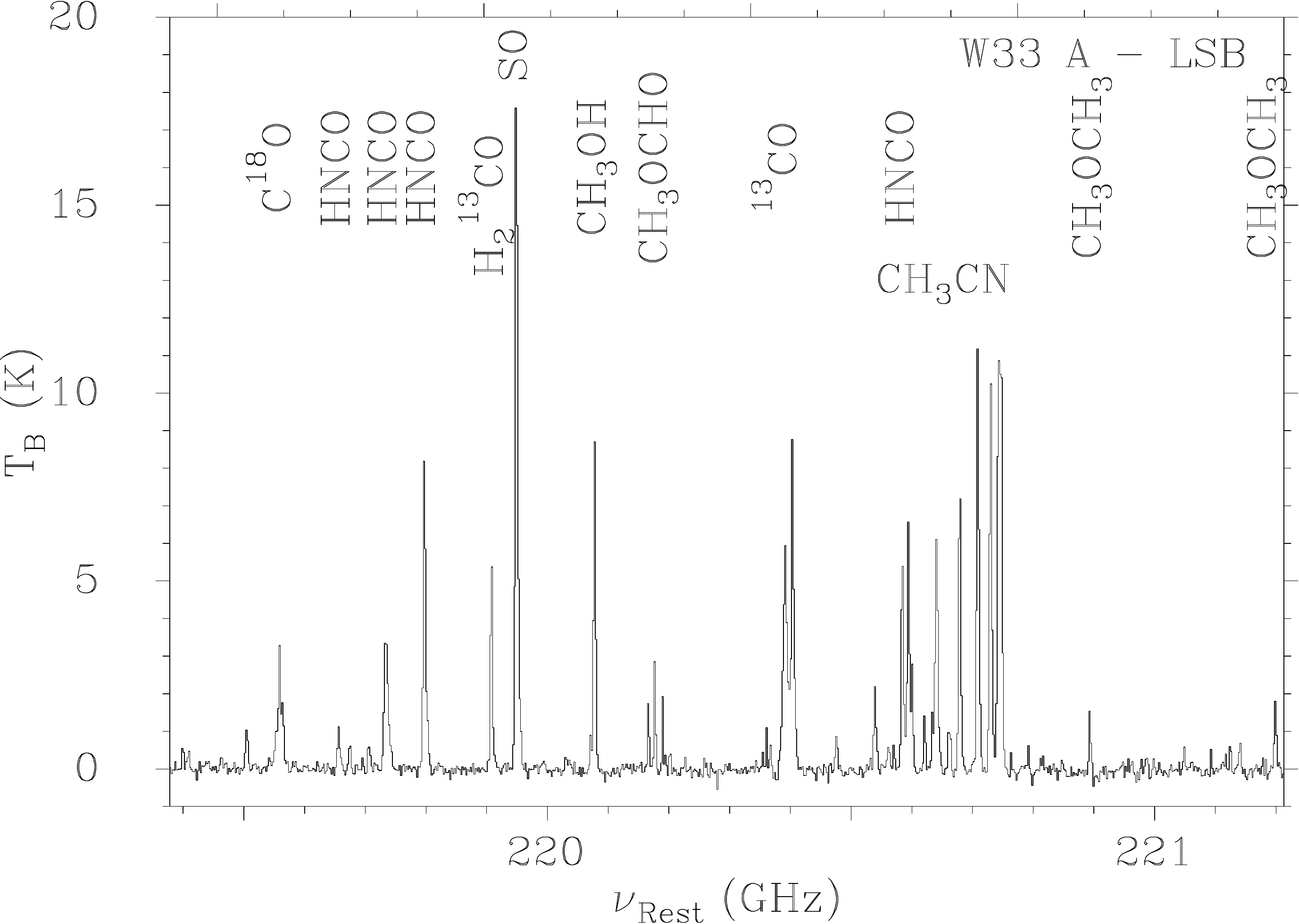}}\hspace{0.2cm} 	
	\subfloat{\includegraphics[width=9cm]{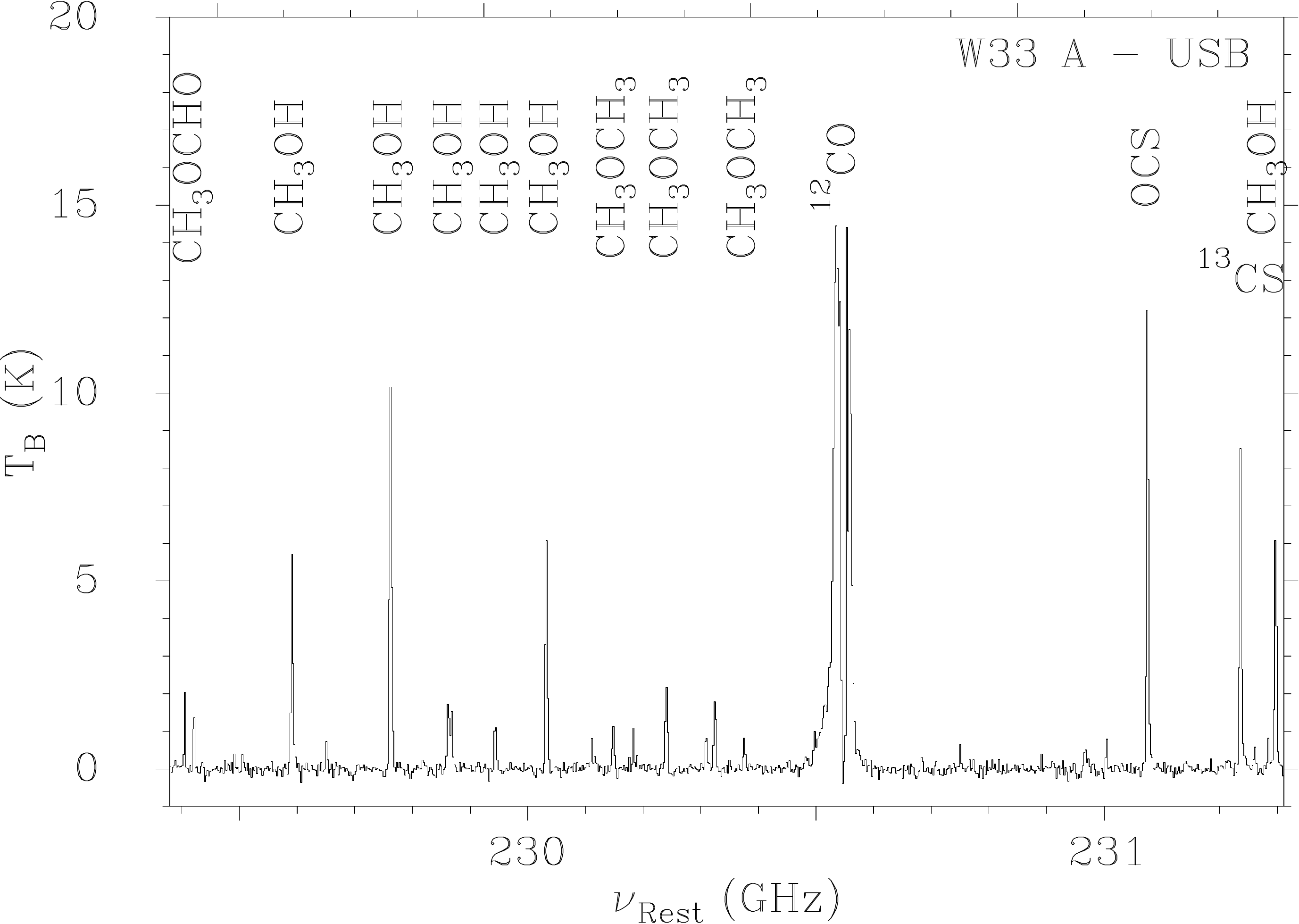}}\\	
	\subfloat{\includegraphics[width=9cm]{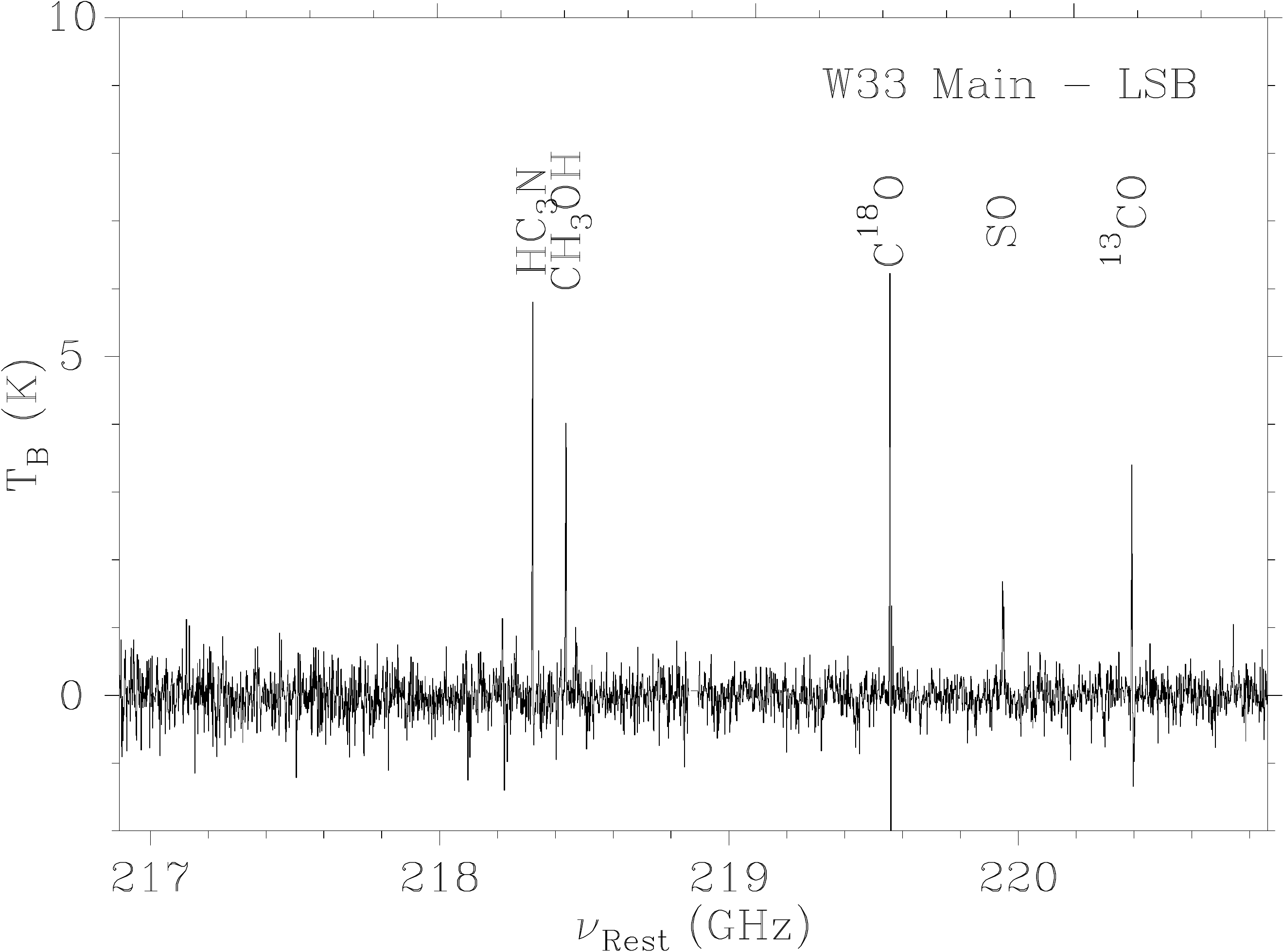}}\hspace{0.2cm} 	
	\subfloat{\includegraphics[width=9cm]{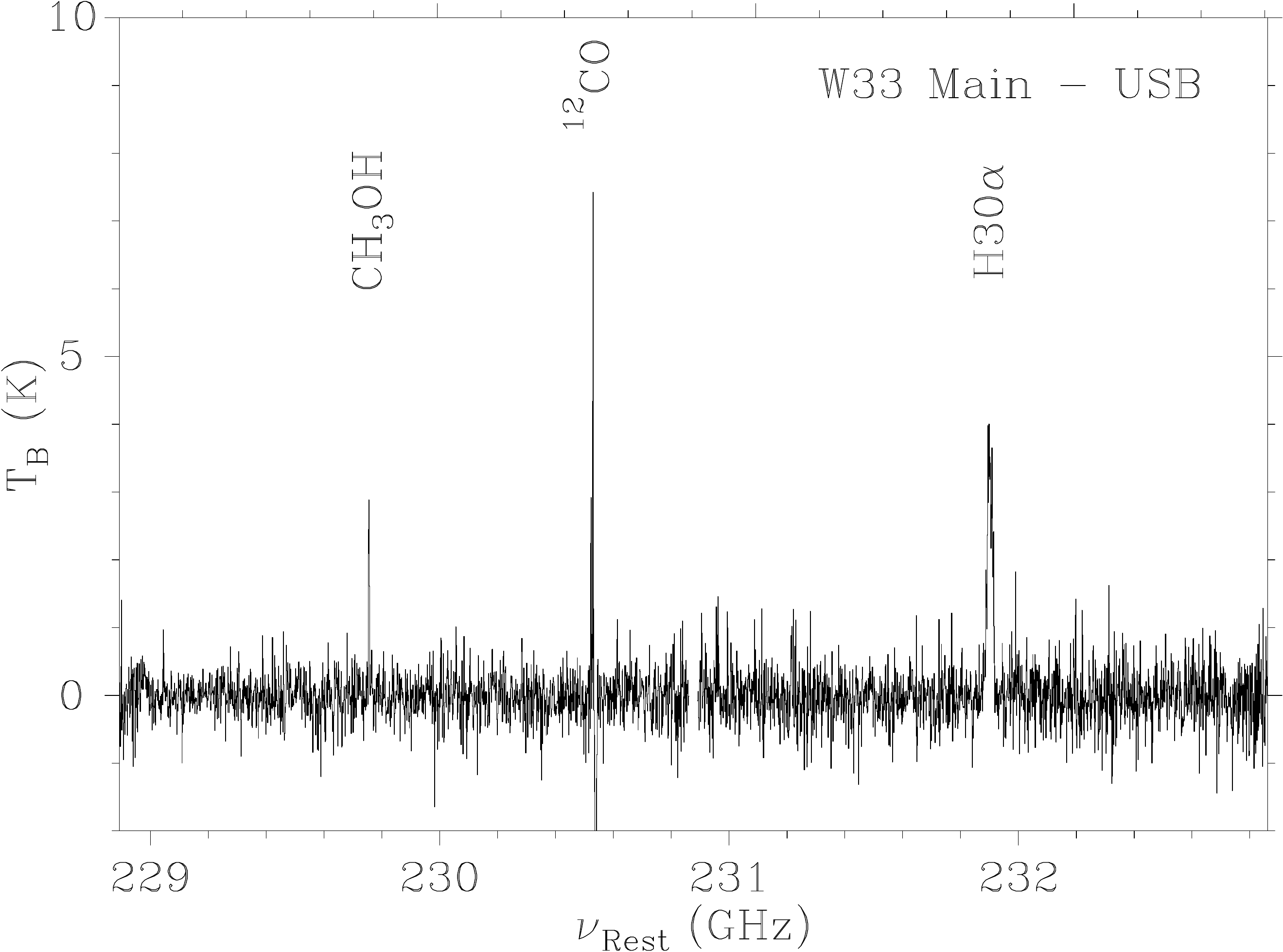}}\\	
	\label{W33_SMA_Spectra2}
\end{figure*}

We generated spectra of all six cores by integrating the emission over one synthesised beam at the 230 GHz continuum peak positions (Figs. \ref{W33_SMA_Spectra1} and \ref{W33_SMA_Spectra2}). In W33\,Main, we obtained the spectrum at the position of the 
stronger peak in W33\,Main$-$Central. 
We again used the CDMS, JPL, and splatalogue catalogs for the spectral line identification. The detected transitions are listed in Table \ref{LinesTabSMA} in the online appendix. In total, we observed 52 transitions of the 16 molecules SiO, SO, \textsuperscript{12}CO and its isotopologues \textsuperscript{13}CO and C\textsuperscript{18}O; the deuterated species DCN; the dense gas tracer \textsuperscript{13}CS; the dense and hot gas tracers CH\textsubscript{3}OH, H\textsubscript{2}CO and its isotopologue H\textsubscript{2}\textsuperscript{13}CO, CH\textsubscript{3}CN, HNCO, OCS; and HC\textsubscript{3}N, CH\textsubscript{3}OCHO, and CH\textsubscript{3}OCH\textsubscript{3}. 
The detected transitions have upper energy levels E\textsubscript{u} between 16 and 579~K. Again, the lines with a Gaussian line profile are fitted with single Gaussians to obtain the same line parameters as in Sect. \ref{ResultsAPEX}. Moment 0 maps (Figs. \ref{W33M1-SMA-IntInt} $-$ \ref{W33M-SMA-IntInt} in the online appendix) were generated for all transitions. Moment 1 maps, showing a velocity gradient, are presented in Fig. \ref{W33_SMA_VelGrad_All} in the online appendix.

As described in Section \ref{Intro}, accretion disks and outflows are found during a certain period of the star formation process. Since the spatial resolution of our observations is too coarse to resolve the accretion disk, we are looking for outflow signatures to characterise these evolutionary stages in our sample. Furthermore, velocity gradients, detected in the moment 1 maps of the spectral emission, can give hints to rotation, infall or expansion of material (Fig. \ref{W33_SMA_VelGrad_All}).
Since the emission of the CO lines is the most widespread, it is affected the most by the lack of short-spacing information. Thus, the identification and interpretation of outflows and velocity gradients in the moment 0 and 1 maps of the CO lines is complicated (Fig. \ref{W33_SMA_12CO}).

Comparing the SMA and the IRAM30m+SMA data of W33\,Main (see Fig. \ref{W33M_COMB_CO} and the associated discussion in Online-Appendix \ref{SMAResApp}), we estimated the fraction of the missing flux for the $^{13}$CO and C$^{18}$O molecules due to the spatial filtering of the interferometer. We integrated the moment 0 maps in Fig. \ref{W33M_COMB_CO} over one beam at the continuum peak position of W33\,Main-Central and then took the ratio of the obtained values, yielding fractions of interferometer flux to total flux of $\sim$20\% and $\sim$40\% for $^{13}$CO and C$^{18}$O, respectively.

In the following paragraph, we summarise the source-specific SMA results. A detailed description of the detected molecules in the W33 sources is given in online Appendix \ref{SMAResApp}.
The object W33\,Main1 is again the source with the least number of detected spectral lines (Fig. \ref{W33_SMA_Spectra1}). The lack of spectral lines besides CO and SO transitions either indicates low temperatures in the dust core or the emission of more complex molecules is not compact enough to be detected with the SMA, which both hint to a very early evolutionary age of W33\,Main1. We conclude that W33\,Main1 is probably in an early protostellar phase before the protostar strongly influences the surrounding material and strong emission of primary molecules like H$_{2}$CO and CH$_{3}$OH is detected.
Besides CO and SO transitions, emission of H$_{2}$CO and CH$_{3}$OH is observed in W33\,A1 and W33\,B1 (Fig. \ref{W33_SMA_Spectra1}). The higher-energy transitions of these two molecules hint to the presence of heating sources in both clumps. However, since we do not see the line-richness of hot cores in these cores, we conclude that both sources are young protostellar cores. The object W33\,B is one of the most line-rich sources of the six W33 cores (Fig. \ref{W33_SMA_Spectra2}). The detection of complex molecules like CH$_{3}$CN, HNCO, HC$_{3}$CN, CH$_{3}$OCHO or CH$_{3}$OCH$_{3}$ indicates that W33\,B is in the hot core stage. In W33\,A, we observe  CH$_{3}$OH, CH$_{3}$OCHO, and CH$_{3}$OCH$_{3}$ transitions at higher excitation energies as in W33\,B (Fig. \ref{W33_SMA_Spectra2}), indicating that W33\,A is more evolved than W33\,B. The variety of detected molecules (including complex molecules) and the high temperature needed for the excitation of the high-energy transitions supports the identification of W33\,A as a hot core. Compared to W33\,B and W33\,A, the spectrum of W33\,Main shows significantly fewer spectral lines (Fig. \ref{W33_SMA_Spectra2}). The object W33\,Main is the only source in our SMA sample that shows emission of the radio recombination line H30$\alpha$, supporting the identification of W33\,Main as a more evolved object where ionised emission of \ion{H}{ii} regions(s) is observed.

\section{Gas temperatures and column densities}
\label{TempDens}

In this section, we determine gas temperatures and column densities of molecules for 
which several transitions are detected in the six W33 clumps. We first determine these 
parameters with the rotational temperature diagram method and then 
test and refine the results by constructing synthetic spectra for the molecules and comparing 
them with the observed spectra.

\subsection{Rotational temperature diagrams and Weeds modelling}
\label{RTD}

From molecules for which multiple transitions can be observed, the rotational temperature and beam-averaged column density of the molecular material can be estimated using the rotational temperature diagram (RTD) method.
Assuming optically thin emission and local thermodynamic equilibrium (LTE), the observed line intensities are proportional to the level populations and the level populations are determined by a single temperature $T_{rot}$. The upper level population is given by 
\begin{eqnarray}
N_{u} = \frac{8\pi^{3}k\nu^2}{hc^3A_{ul}}\int{T_{b}dv},
\label{EQNNu}
\end{eqnarray}
where $A_{ul}$ is the Einstein coefficient of the transition \citep[see the detailed description of this method in][]{Goldsmith1999}.

In LTE, 
\begin{eqnarray}
\ln \frac{N_{u}}{g_{u}} = \ln \frac{N_{tot}}{Q(T_{rot})} - \frac{E_{u}}{k}\frac{1}{T_{rot}},
\label{EQNRTD}
\end{eqnarray}
where $g_{u}$ is the statistical weight of the upper level, $Q(T_{rot})$ the partition function, and $N_{tot}$ the total number column density.\\

Weeds is an extension of the CLASS software, which belongs to the GILDAS package \citep{Maret2011}. It is mostly written in Python. If several transitions of a given species are detected, Weeds permits the computation of a synthetic spectrum to inspect if the relative intensities of the transitions agree with a single excitation temperature. The synthesised spectrum is constructed under the assumption of LTE conditions in the gas. The advantage of Weeds compared to the rotational temperature diagram method is that the modelling directly includes opacity effects and that the synthetic spectrum can directly be compared to the observed spectrum. 
To compute a synthetic spectrum, the user has to provide a text file including the name of the modeled species, the column density, the kinetic temperature, source size, and line width. In the optically thin case, the column density and the source size are degenerate. 

If the source size is smaller than the beam, beam dilution occurs, which affects the line intensity and thus the estimated column densities. Since the transitions of the molecules for which 
we determined the column densities are detected at different frequencies, the beam size of the observations changes slightly (e.g. from 21.3$\arcsec$ to 22.2$\arcsec$ for the APEX observations of H$_{2}$CO). Thus, the column densities that are determined from the synthetic spectra are beam-averaged over the smallest observed beam size 
per molecule (21.3$\arcsec$ for H$_{2}$CO and CH$_{3}$OH, 21.5$\arcsec$ for CH$_{3}$CCH, 5$\arcsec$ for the SMA observations). For the transitions that are observed with a larger beam, a beam dilution factor is taken into account. A smaller beam dilution factor results in a higher column density. Thus, the estimated column densities are only lower limits.

As initial parameters for the construction of the synthetic spectra, we used the temperatures and column densities, determined from the RTD fitting. The line width was determined from the average of the FWHMs of the detected transitions. If the synthetic spectra, computed from the RTD results, did not yield a good fit of the observed spectra, we changed the temperature and column density until we reached a good fit to the strongest transitions by eye. Since the column densities from the RTD fitting have larger uncertainties than the determined temperatures, we tried to keep the input temperature for the synthetic spectrum close to the RTD temperature. Due to uncertainties in the source size and possible temperature substructures, the synthetic spectra only give an estimate of the temperature and column density of the material but do not yield a perfect fit of the observed spectra. A detailed description of the Weeds modelling of the APEX and SMA spectra of the W33 sources is given in online Appendix \ref{WeedsApp}.

\paragraph{APEX observations}

\begin{figure*}
\centering
\caption{Rotational temperature diagrams (RTDs) of the six W33 sources from the APEX observations.}
\subfloat[RTD of H$_{2}$CO in W33\,Main1]{\includegraphics[width=9cm]{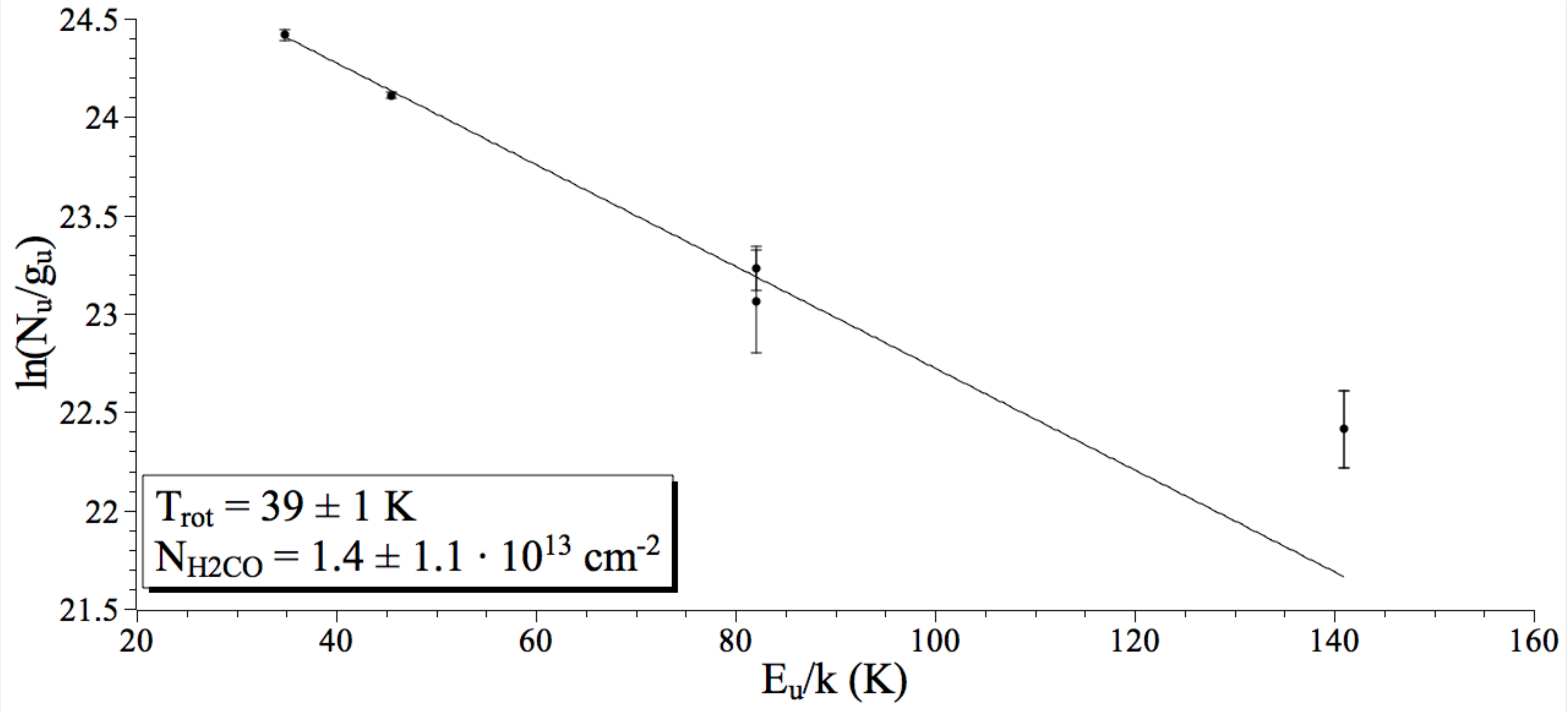}}\hspace{0.1cm}
\subfloat[RTD of H$_{2}$CO in W33\,A1]{\includegraphics[width=9cm]{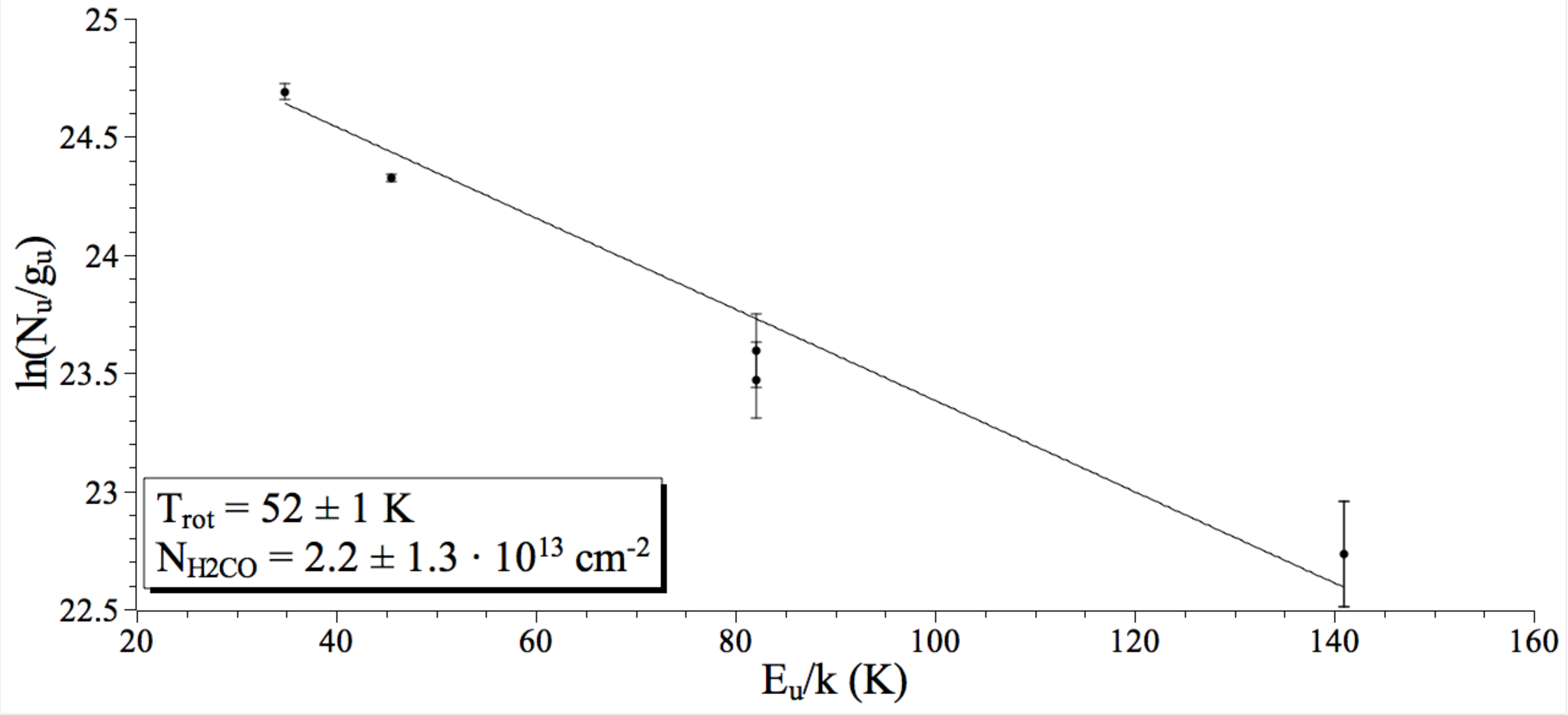}}\\
\subfloat[RTD of H$_{2}$CO in W33\,B1]{\includegraphics[width=9cm]{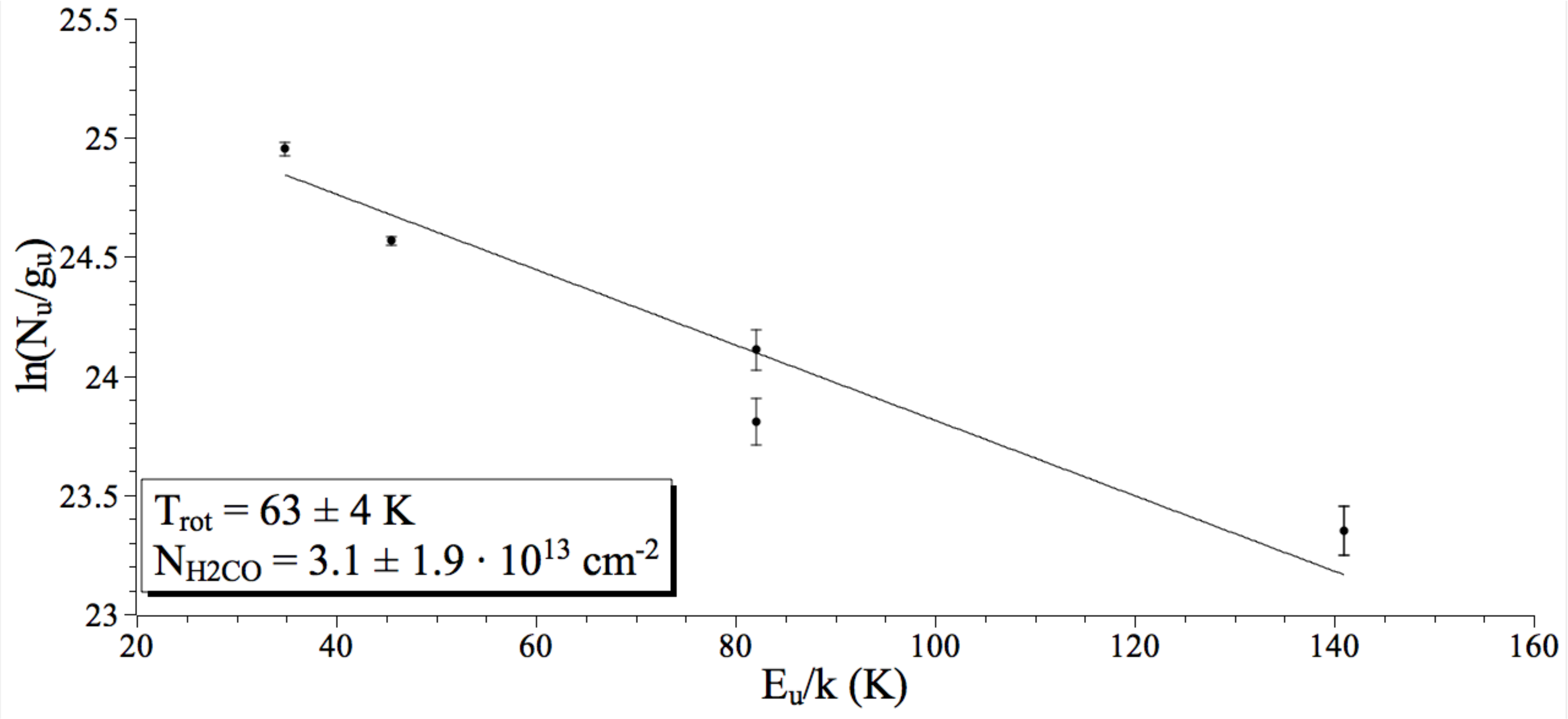}}\hspace{0.1cm}
\subfloat[RTD of H$_{2}$CO in W33\,B]{\includegraphics[width=9cm]{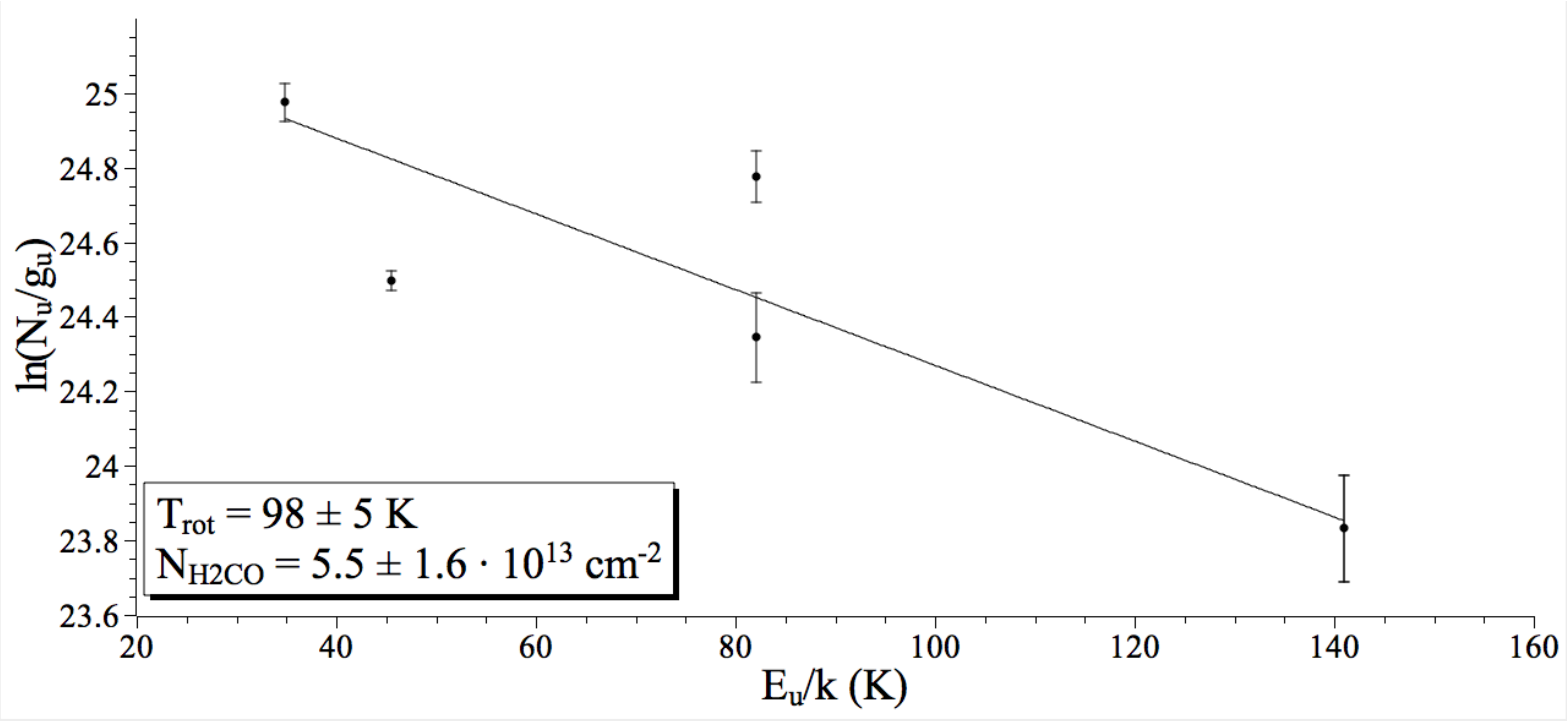}}\\
\subfloat[RTD of H$_{2}$CO in W33\,A]{\includegraphics[width=9cm]{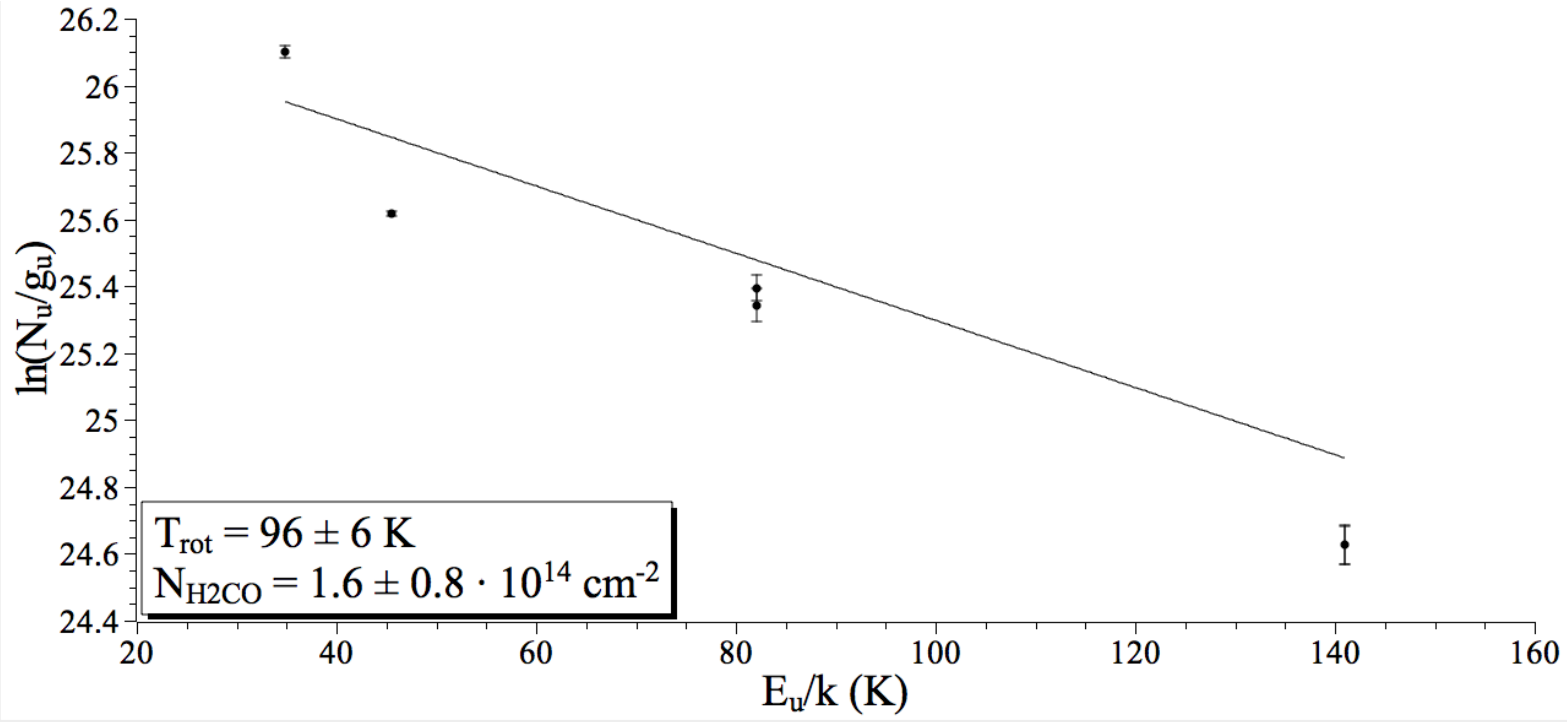}}\hspace{0.1cm}
\subfloat[RTD of H$_{2}$CO in W33\,Main]{\includegraphics[width=9cm]{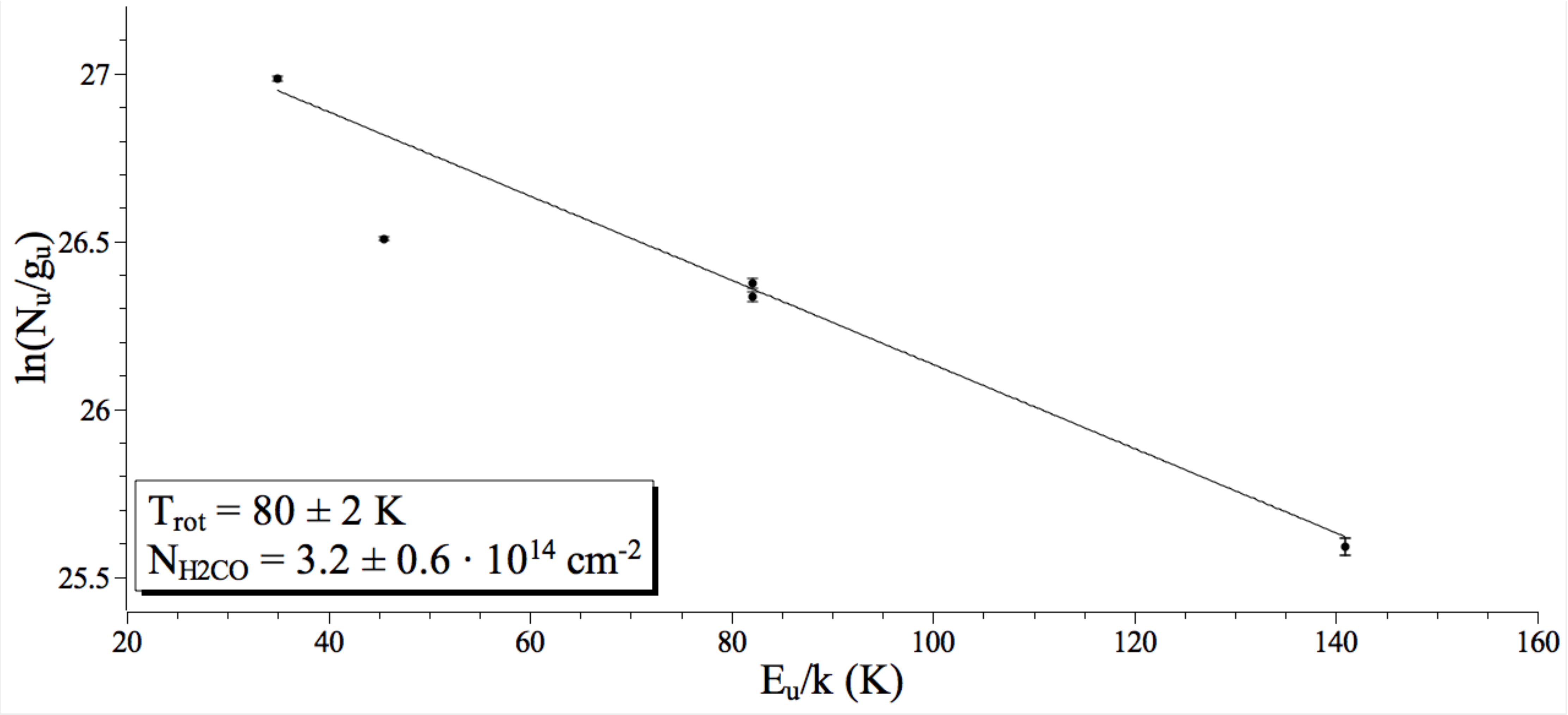}}\\
\subfloat[RTD of CH$_{3}$OH in W33\,Main \label{CH3OHW33Main}]{\includegraphics[width=9cm]{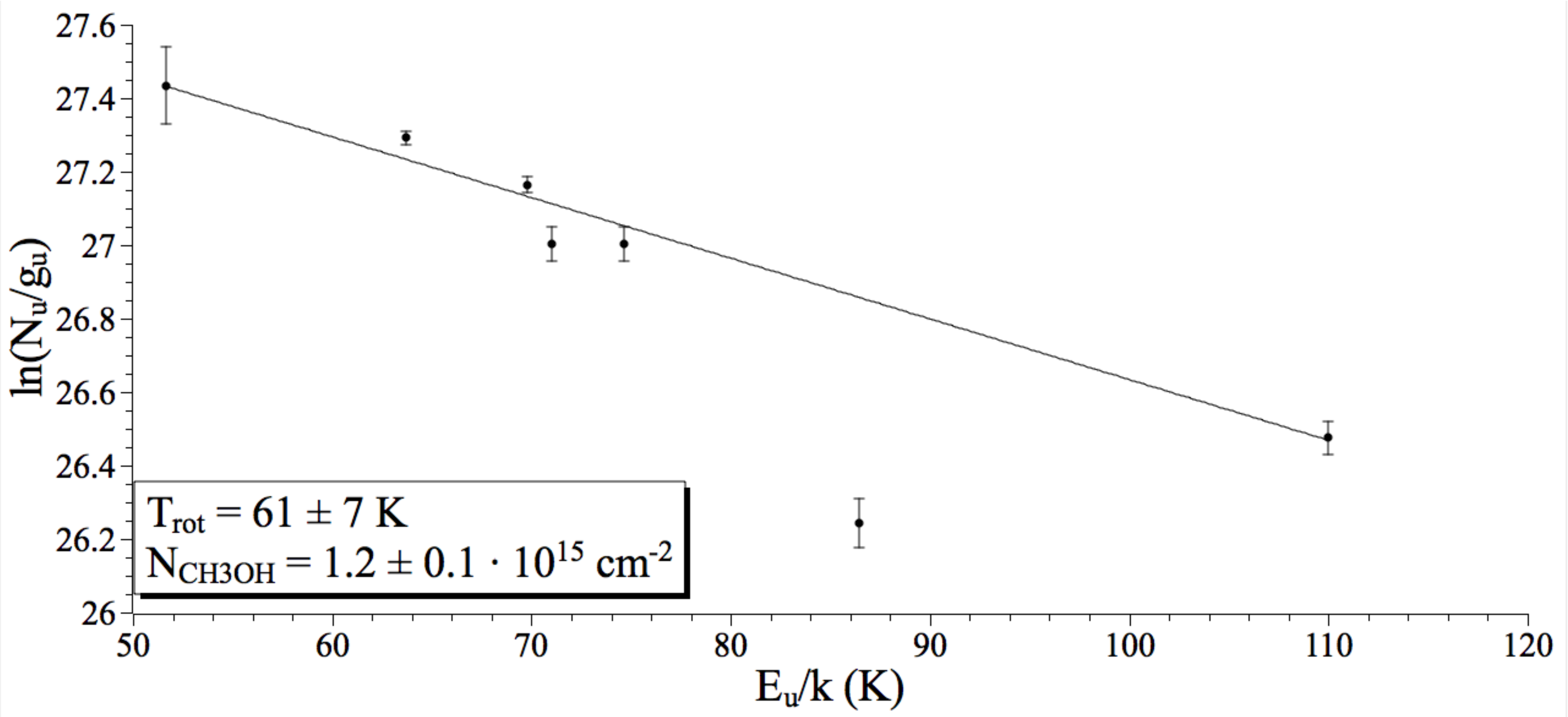}}\hspace{0.1cm}
\subfloat[RTD of CH$_{3}$CCH in W33\,Main \label{CH3CCHW33Main}]{\includegraphics[width=9cm]{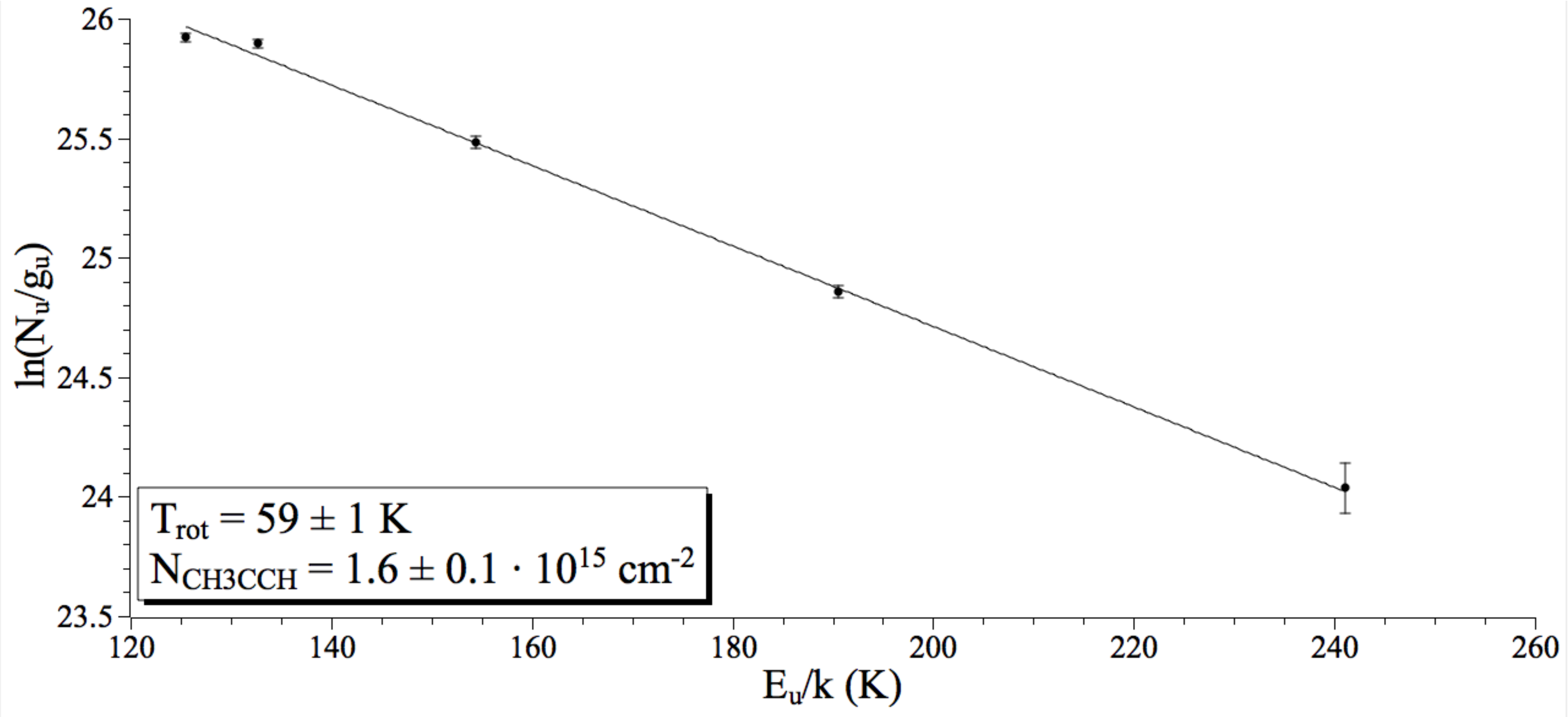}}
\label{APEXRTD}
\end{figure*}

\begin{table*}
\centering
\caption{Gas temperatures T and column densities N$_{tot}$ from RTDs and Weeds modelling of the APEX (upper part) and SMA (lower part) spectra. The molecular abundances were calculated from the ratio of the column densities of the molecule and the H$_{2}$ column densities N$_{H_{2}, 24\arcsec}$.}
\label{TempDenseRTD}
\begin{tabular}{lcccccc}\\\hline
& & \multicolumn{3}{c}{RTD} & \multicolumn{2}{c}{Weeds} \\
Source & Molecule & T & N$_{tot}$ & Abundance & T & N$_{tot}$ \\
& & (K) & (10$^{13}$ cm$^{-2}$) & (10$^{-10}$) & (K) & (10$^{13}$ cm$^{-2}$)\\ \hline 
W33\,Main1 & H$_{2}$CO & 39 & 1.4 & 2.0 & 40 & 3.4 \\
W33\,A1 & H$_{2}$CO & 52 & 2.2 & 1.9 & 30 & 2 \\
W33\,A1 & H$_{2}$CO &  &  & & 55 & 2.5 \\
W33\,B1 & H$_{2}$CO & 63 & 3.1 & 8.7 & 30 & 2\\
W33\,B1 & H$_{2}$CO & & & & 60 & 2.5 \\
W33\,B & H$_{2}$CO &98 & 5.5 & 3.4 & 30 & 2.3\\
W33\,B & H$_{2}$CO & & & & 100 & 5.9 \\
W33\,A & H$_{2}$CO & 96 & 16.1 & 8.2 & 40 & 9 \\
W33\,A & H$_{2}$CO &  &  & & 100 & 16 \\
W33\,Main & H$_{2}$CO & 80 & 31.8 & 8.7 & 50 & 20.0 \\
W33\,Main & H$_{2}$CO &  & & & 100 & 42.0 \\
W33\,Main & CH$_{3}$OH & 61 & 116 & 31.9 & 40 & 130 \\
W33\,Main & CH$_{3}$CCH & 59 & 159.6 & 43.9 & 59 & 340 \\ \hline\hline
W33\,B & HNCO & 351 & 161.3 & & 280 & 250 \\
W33\,B & CH$_{3}$CN & 335 & 83.1&  & 350 & 300 \\
W33\,B & CH$_{3}$OH & 219 & 4194.3 & & 350 & 300 \\ \hline
\end{tabular}
\end{table*}

Since we observe the H$_{2}$CO transitions in the APEX spectra of all six W33 sources and they span a broad range of upper energy levels from 35 to 141~K, we first calculated the level populations for these transitions from the observed integrated intensities, using equation \ref{EQNNu} and assuming that the beam filling factor for the APEX data is unity. Then, we plotted the results in RTDs (Fig. \ref{APEXRTD}). We also tried to fit the CH$_{3}$OH transitions in RTDs but the level populations in W33\,B, W33\,B1, W33\,A1, and W33\,A did not fall along a line in the RTDs. This indicates that either the assumption of optically thin gas or LTE does not apply for CH$_{3}$OH or that the CH$_{3}$OH emission has a temperature substructure in these sources. In the CH$_{3}$OH-RTD of W33\,Main (Fig. \ref{CH3OHW33Main}), the transition at 86~K was ignored in the fitting due to the strong difference of the level population compared to the other CH$_{3}$OH transitions. We also generated an RTD for the CH$_{3}$CCH transitions detected in W33\,Main and fitted the level populations (Fig. \ref{CH3CCHW33Main}).The temperatures (Column 3) and total column densities (Column 4) are listed in Table \ref{TempDenseRTD}. 

The Weeds modelling results (Columns 5, 6 in Table \ref{TempDenseRTD}) show that we need a cold and a warm component to fit the H$_{2}$CO spectral lines well for all sources except W33\,Main1. In all five sources, the warm component has a higher column density than the cold component, indicating that it is excited closer to the center of the core assuming a density gradient from the outer area to the center of the core. Since we estimate high temperatures for the warm component in W33\,B1, W33\,B, W33\,A, and W33\,Main, we expect the existence of a heating source in these cores.
The temperatures and column densities of both components increase from W33\,Main1 to W33\,Main (see Table \ref{TempDenseRTD}), supporting the sorting of the sources along the evolutionary sequence that we established in Sects. \ref{ResultsAPEX} and \ref{ResultsSMA}.

To determine the abundances of H$_{2}$CO in all sources as well as CH$_{3}$OH and CH$_{3}$CCH in W33\,Main, we calculated the ratio of the molecular column densities and the H$_{2}$ column densities over an area of 24$\arcsec$~x~24$\arcsec$ (N$_{H_{2}, 24\arcsec}$; see Sect. \ref{TempCloudMass}, Column 5 in Table \ref{TempDenseRTD}). The obtained abundances for H$_{2}$CO are consistent with other chemical studies \citep[e.g.][]{Vasyunina2014,Gerner2014}. The CH$_{3}$CCH and CH$_{3}$OH abundances in W33\,Main are similar to the values determined towards this source \citep{Miettinen2006} and towards other star forming regions \citep[e.g.][]{Alakoz2000,vanderTak2000,Vasyunina2014,Gerner2014}. Except for W33\,B1, the H$_{2}$CO abundances seem to increase along the evolutionary sequence, which would support an increasing release of this molecule from the dust grains during the evolution of the star forming region. A similar trend was seen in the chemical study of \citet{Gerner2014}.

\paragraph{SMA observations}

\begin{figure*}
\caption{CH$_{3}$CN ladder spectrum (continuous line) and Gaussian fits (dashed line) from the SMA observations of W33\,B.}
\centering
\includegraphics[width=8.5cm]{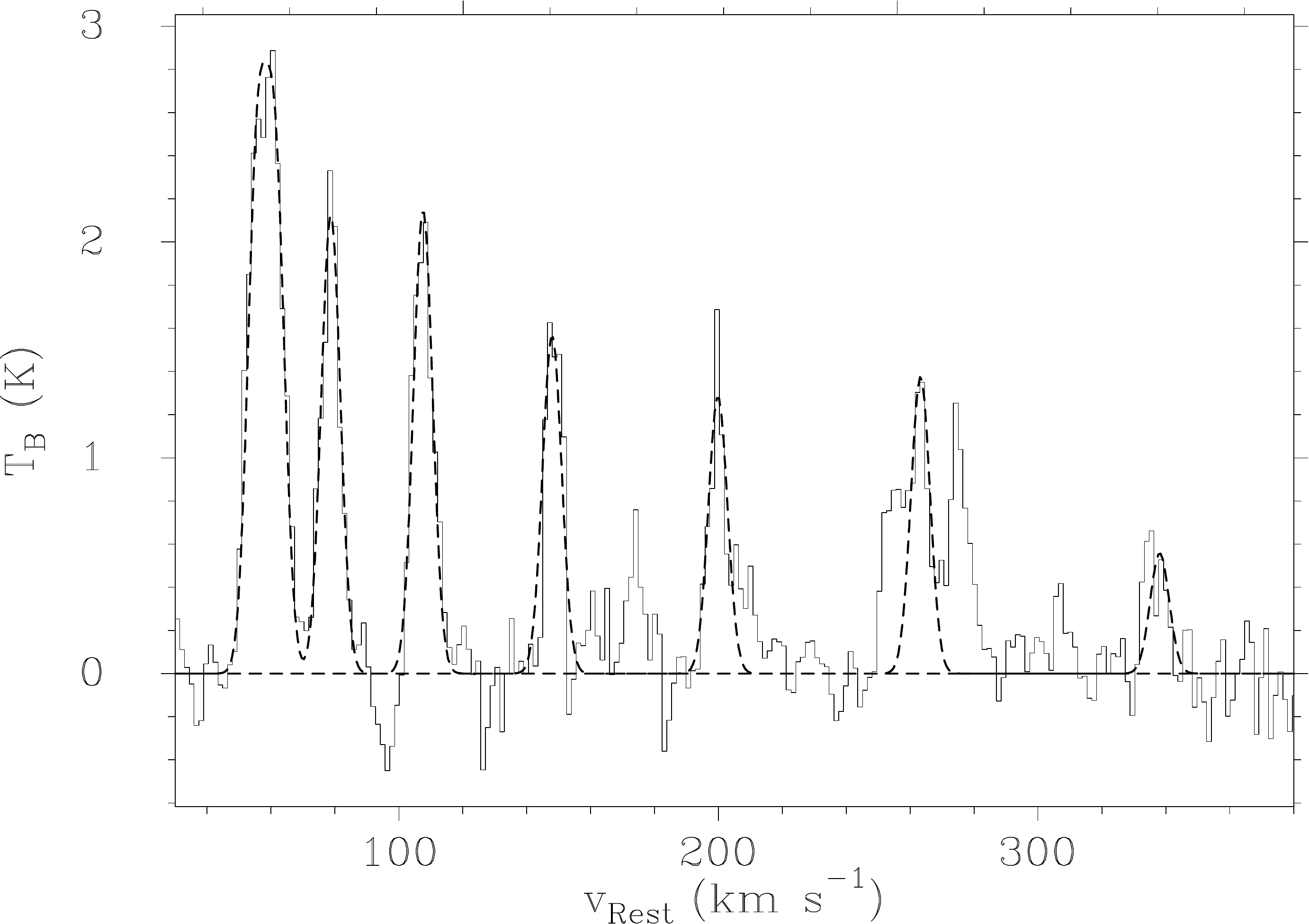}
\label{CH3CNFit}
\end{figure*}

\begin{figure*}
\centering
\caption{Rotational temperature diagrams of CH$_{3}$CN, CH$_{3}$OH, and HNCO from the SMA observations of W33\,B.}
\subfloat[W33B, CH$_{3}$CN]{\includegraphics[width=9cm]{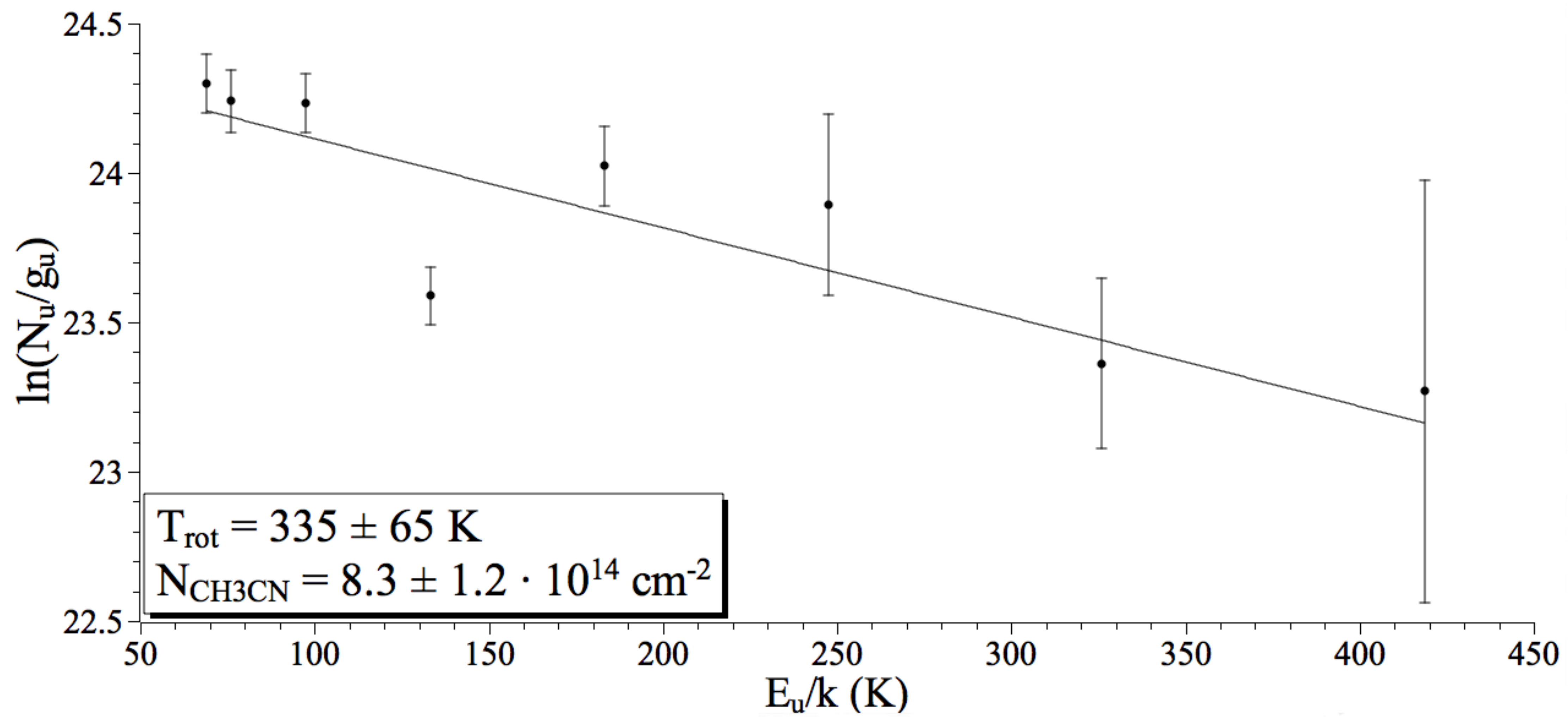}}
\subfloat[W33B, CH$_{3}$OH]{\includegraphics[width=9cm]{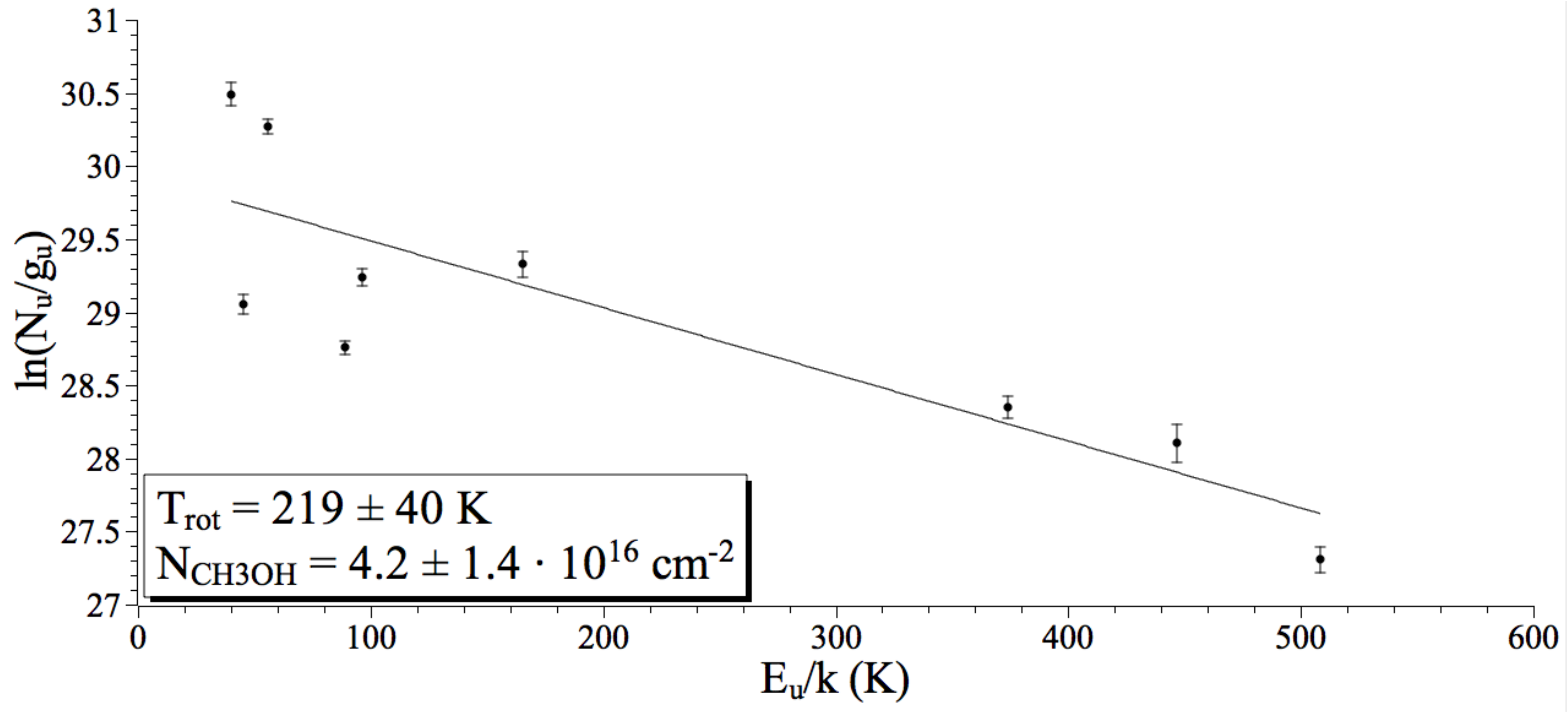}}\\
\subfloat[W33B, HNCO]{\includegraphics[width=9cm]{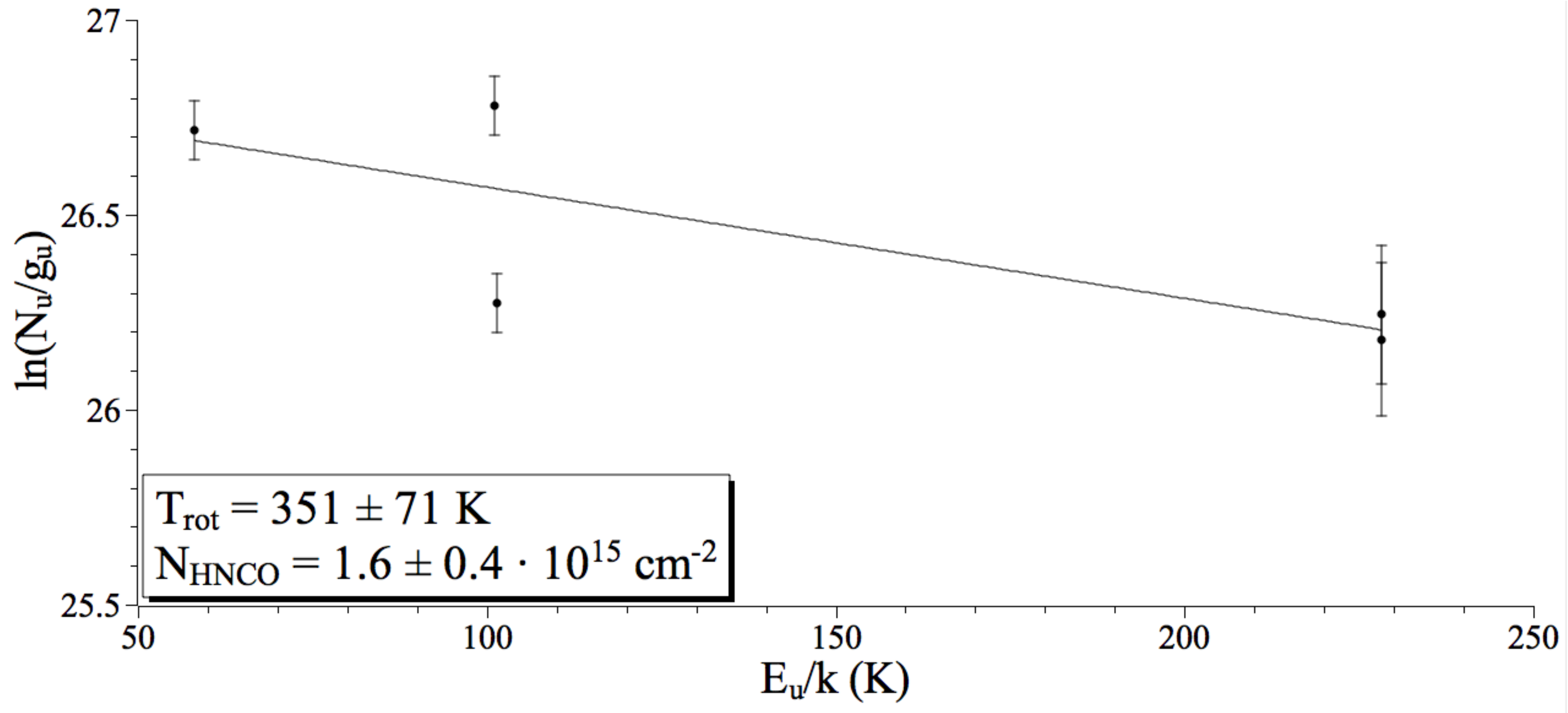}}
\label{SMARTD}
\end{figure*}

In the SMA spectra, H$_{2}$CO is the only molecule with multiple transitions that is detected towards all sources. However, two of these three transitions are at the same upper energy level, and we decided to use the RTD method only for molecules with transitions at more than two different upper energy levels. We tried to compute synthetic spectra from one component for the H$_{2}$CO transitions, but it was not possible to fit all transitions well, indicating that several components are needed for the modelling or that the assumption of LTE is not valid for the H$_{2}$CO transitions.

In W33\,B, we detect multiple transitions of CH$_{3}$OH, HNCO, and CH$_{3}$CN, and we applied the RTD method for these lines.
Figure \ref{CH3CNFit} shows the spectrum of the K~=~0...7 CH$_{3}$CN(12$_{K}$$-$11$_{K}$) ladder in W33\,B. The K~=~0 and K~=~1 transitions are blended together, and K~=~6 and 7 are blended with emission from other lines. Since the K-transitions are so close in frequency, they can be fitted simultaneously. We assume that the K-components have the same line width. To estimate the line width, we first fitted the K-components 2, 3, 4, and 5 with Gaussian line profiles, yielding a line width of 7.0~$\pm$~0.5~km~s$^{-1}$. We then fitted all lines with Gaussian line profiles with fixed frequency spacings and a line width of 7~km~s$^{-1}$ (see Fig. \ref{CH3CNFit}). The integrated intensities of the CH$_{3}$CN transitions are used to calculate the level populations of the different transitions, which are then plotted in an RTD (Fig. \ref{SMARTD}). The least-squares fit to the level populations yields a rotational temperature of 335~$\pm$~65~K and a CH$_{3}$CN column density of 8.3~$\pm$~1.2~$\cdot$~10$^{14}$~cm$^{-2}$.

The RTDs of CH$_{3}$OH and HNCO are also shown in Fig. \ref{SMARTD}. The rotational temperature and column density of HNCO are similar to the values of CH$_{3}$CN. The RTD of CH$_{3}$OH yields a rotational temperature that is much lower than for HNCO. However, the CH$_{3}$OH-RTD shows that a line is not a good fit to the level populations, especially for transitions at lower excitation energies, indicating that either the emission is not optically thin, the CH$_{3}$OH gas has a temperature substructure, or the assumption of LTE does not apply. In the optically thick case, the rotational temperature is overestimated, and the column density underestimated. If the emission is optically thin and the gas has a temperature substructure, the line intensities and, thus, the column densities might be affected by beam dilution and beam filling factors have to be applied (see discussion in Sect. \ref{RTD}). The synthetic spectrum of the CH$_{3}$OH transitions shows that at least two components are necessary to fit the observed spectrum. However, we do not find two combinations of temperatures and column densities that fit all transitions well. Since the opacities of the synthesised spectra are consistent with optically thin emission, the bad fits indicate that the CH$_{3}$OH gas has a more complex temperature substructure or is not in LTE. Comparing the RTD results for the APEX and SMA observations of W33\,B, it is clear that the SMA observations trace a hotter and denser region in the source than the APEX data.

\citet{GalvanMadrid2010} produced an RTD of CH$_{3}$CN for W33\,A, yielding a rotational temperature of 410~K and a column density of 3.4~$\cdot$~10$^{16}$~cm$^{-2}$. Although the rotational temperatures of W33\,B and W33\,A are comparable within the errors, the column density of CH$_{3}$CN in W33\,A is a factor of 40 larger than the column density of CH$_{3}$CN in W33\,B, which might be explained by a disk-like accretion flow in W33\,A, revealed by subarcsecond resolution observations \citep{GalvanMadrid2010}.

\section{Chemical diversity in the W33 complex}
\label{ChemDiv}

Studying the chemistry of high-mass star forming regions in a single star forming complex like W33 has several advantages. Since all star forming regions belong to the same complex and probably evolved from the same giant molecular cloud, the primary material of the birth clouds is similar (w.r.t. metallicity or initial abundance ratios, for example). Furthermore, the star forming regions are at very similar distances. Thus, the observations of each telescope cover the same spatial scales which makes the observations per telescope directly comparable for the different star forming regions within the complex. In addition, missing short spacings in the SMA observations have a similar effect on the data of all star forming regions, assuming the line emission is similarly extended in all the sources.

Due to the different resolutions and the filtering of the interferometer data, the APEX and SMA data sets are sensitive to different physical scales (APEX: >~0.2~pc, SMA: 0.05$-$0.2~pc). 
The APEX observations are sensitive to all the emission from the clump. However, if the emitting gas is confined to a small region, and, thus, the emission is very compact, the large APEX beam dilutes the emission and depending on the strength of the signal compared to the noise level of the observations the emission might be undetected. The SMA observations on the contrary are only sensitive to compact emission while the more extended emission is resolved out. Thus, with these two data sets, we are able to trace the chemical composition on different scales in the star forming regions.

In W33, the molecular clumps cover a range of bolometric luminosities from 6$-$445~$\cdot$~10$^{3}$~L$_{\sun}$, and they seem to be in different evolutionary stages of star formation. Thus, we can compare the observations for different bolometric luminosities and along an evolutionary sequence.
We first discuss each of the W33 sources and then draw conclusions about the change of chemistry along the detected evolutionary sequence.

\subsection{Single sources}

\subsubsection{W33\,Main1} 
W33\,Main1 seems to be in a very early stage of star formation. On larger scales (>~0.2~pc), traced by the APEX data, transitions from H$_{2}$CO and CH$_{3}$OH up to an upper energy level of 141~K are detected. The Weeds modelling showed that only a ``cold'' component of 40 K is needed to fit the H$_{2}$CO transitions. However, H$_{2}$CO and CH$_{3}$OH are not detected on smaller scales, traced by the SMA data. Either the medium is too cold and the molecules are not released yet into the gas phase, or the emission is too diffuse to be detected with the SMA interferometer. If a heating source is present in W33\,Main1, this suggests that it does not strongly influence the chemistry yet. The lack of molecules other than CO (and its isotopologues) and SO suggests that W33\,Main1 is in the collapse phase during the formation of a protostar.

\subsubsection{W33\,A1}
The object W33\,A1 is the source with the lowest luminosity in our sample. The chemistry on larger scales looks similar to W33\,Main1 except that two more CH$_{3}$OH transitions are observed. On smaller scales, we observe low-energy transitions of H$_{2}$CO and CH$_{3}$OH, indicating the presence of a heating source which evaporates these molecules off the dust grains. The necessity of a second warmer component in the LTE modelling supports the existence of a heating source in W33\,A1. We also observe emission of the DCN molecule, which is a tracer of cold and dense gas. Although W33\,A1 is a factor of two less luminous than W33\,Main1, it is probably more evolved. We conclude that W33\,A1 is in the protostellar phase but before the excitation of a hot core.

\subsubsection{W33\,B1}
The object W33\,B1 is the clump with the lowest mass in our sample. The APEX and SMA spectra of W33\,B1 look very similar to the spectra of W33\,A1 but the H$_{2}$CO gas in W33\,B1 seems to have a slightly higher temperature and column density than in W33\,A1. While the average temperature of W33\,B1 measured by the dust SED is consistent with that of the most evolved region W33\,Main (see Section \ref{TempCloudMass}), W33\,B1 shows yet weak emission of hot core tracers, which may imply the dominant heating by the exterior star cluster rather than the (proto)stars embedded in this clump. Although the bolometric luminosity of W33\,B1 is a factor of three larger than the luminosity of W33\,A1, there are no obvious differences in the SMA or APEX observations between these two sources, except that the detected lines tend to be stronger and broader in W33\,B1. We conclude that W33\,B1 is also in the protostellar phase before the excitation of a hot core.

\subsubsection{W33\,B}

Observations of water, methanol, and hydroxyl masers in W33\,B indicate that high-mass stars are forming in this clump. However, the APEX spectra of W33\,B are comparable to the spectra of W33\,B1 or W33\,A1, except that we additionally observe transitions of OCS, H$_{2}$CS, and C$^{33}$S. The compactness of the OCS emission and the high energy needed to excite the detected transition suggest that the emission is excited closer to the forming star at temperatures that are higher than the temperatures traced by H$_{2}$CO and CH$_{3}$OH. The LTE modelling of the APEX data of W33\,B shows that column density and temperature of the envelope are higher than for the previous sources with the temperature of the warm component reaching $\sim$100~K. 

On smaller scales, W33\,B shows a forest of lines. Especially, nitrogen-bearing molecules like HNCO, CH$_{3}$CN or HC$_{3}$N are detected in the SMA spectra, indicating that W33\,B is already in the hot core phase. However, from more complex molecules, only one transition of CH$_{3}$OCHO and one of CH$_{3}$OCH$_{3}$ are observed. Other complex molecules like CH$_{3}$CH$_{2}$CN, CH$_{2}$CHCN, or C$_{2}$H$_{5}$OH, as seen in typical hot cores like Orion-KL, G29.96$-$0.02, G33.92+0.11 or W49A \citep[e.g.][]{Beuther2009, Liu2012, GalvanMadrid2013} are not detected in W33\,B. 

The temperatures and column densities of W33\,B on smaller scales, inferred from LTE modelling of the SMA data, are much higher than on larger scales, as deduced from the APEX data, indicating temperature and column density gradients in W33\,B.

\subsubsection{W33\,A}

As in W33\,B, methanol, water, and hydroxyl masers are detected in W33\,A indicating that high-mass star formation is taking place in this clump. The high-energy transitions of HC$_{3}$N and CH$_{3}$CCH in the APEX data of W33\,A reveal that more complex molecules are detected on larger scales. The LTE modelling shows that the temperatures of the two components are comparable to the temperatures in W33\,B, but the column densities are a factor of three higher. In the SMA spectra, many transitions of a large number of simple and complex molecules are detected with evidence for W33\,A being in the hot core phase. Compared to W33\,B, several transitions of complex molecules like CH$_{3}$OCHO and CH$_{3}$OCH$_{3}$ are observed. As in \citet{Gerner2014}, CH$_{3}$OCHO is only detected in the hot core phase. However, also the spectra of W33\,A lack emission of complex molecules that are observed in typical hot cores. Although W33\,A and W33\,B have very similar masses, W33\,A is twice as luminous as W33\,B. The object W33\,A is probably more evolved than W33\,B. Radio continuum observations of W33\,A by \citet{Rengarajan1996} and \citet{vanderTak2005} with the VLA suggest that the faint emission either comes from an ionised wind or a hypercompact 
\ion{H}{ii} region in W33\,A. This is evidence for W33\,A being more evolved than W33\,B, perhaps in transition from the pure hot core phase to the more evolved \ion{H}{ii} region phase.

\subsubsection{W33\,Main}

The object W33\,Main is the only source in the W33 complex that shows strong radio emission, indicating that this source is in the \ion{H}{ii} region phase. Water and Class I methanol masers are observed in W33\,Main. The mass and luminosity of W33\,Main are a factor of three and ten higher than for W33\,A, respectively. The APEX spectra of W33\,Main are similar to the spectra of W33\,A, except that we detect higher-energy transitions of CH$_{3}$CCH and emission of C$^{33}$S and H$_{2}^{13}$CO. In addition, the column density of the H$_{2}$CO gas is highest in W33\,Main. However, at smaller scales, we do not detect emission of complex molecules such as in the hot core sources anymore. The SMA spectra of W33\,Main are similar to the spectra of W33\,A1 and W33\,B1 plus emission of HC$_{3}$N and the radio recombination line H30$\alpha$, which traces ionised gas. The molecule HC$_{3}$N is the only complex molecule that is detected in W33\,Main and it peaks at the edge of W33\,Main-Central. Other complex molecules like CH$_{3}$CN, CH$_{3}$OCHO or CH$_{3}$OCH$_{3}$ are either destroyed close to the heating source or their emission is extended without compact peaks and thus resolved out by the interferometer observations. 
We conclude that W33\,Main is already in the \ion{H}{ii} region phase.

\subsection{Chemical complexity -- evolutionary sequence}

From our SMA and APEX observations, we infer an evolutionary sequence of star formation in W33. We sort the sources in different groups: 
\begin{itemize}
\item Early Protostellar Phase (W33\,Main1): Cold interior that is probably still contracting with no strong heating source yet
\item Protostellar Phase (W33\,A1, W33\,B1): Heating source starts to warm up the surrounding material, releasing primary molecules from dust grains into gas phase
\item Hot Cores (W33\,B, W33\,A): Heating source strongly influences the chemistry of the surrounding material with release of primary molecules from dust grains and synthesis of complex secondary molecules in gas phase
\item \ion{H}{ii} regions (W33\,Main): Heating source ionises the surrounding material, strong radio emission is detectable, and complex molecules are probably dissociated 
\end{itemize}

Our observations show that the chemistry changes dramatically along this evolutionary sequence, especially on smaller spatial scales. The detected transitions tend to get stronger and broader along this sequence in both the SMA and APEX data sets. The trend of increasing detection rates for different molecules up to the hot core stage and then decreasing detection rates in the \ion{H}{ii} region phase is also seen in the IRAM 30m line survey of \citet{Gerner2014}.

Our SMA observations suggest that emission of CO and its isotopologues dominates in early phases of star formation. Once a heating source is present, H$_{2}$CO and CH$_{3}$OH are evaporated off dust grains. When the temperature in the core rises, more complex molecules are evaporated off the dust grains or synthesised in the gas phase. Once an \ion{H}{ii} region emerges, the more complex molecules seem to be destroyed by the ionisation or their emission is too diffuse to be detected by the interferometer. On larger scales, traced by the APEX observations, the chemical diversity and complexity increases with the evolution of the star forming region and complex molecules are still detected in the \ion{H}{ii} region phase. \citet{Bisschop2007} observed a set of complex molecules in seven different hot cores and showed that the molecules can be classified as either ``hot'' (e.g. CH$_{3}$OH, H$_{2}$CO, CH$_{3}$CN, ..., T > 100 K) or ``cold'' (e.g. CH$_{3}$CHO, T < 100 K), depending on their rotational temperatures. Our APEX and SMA observations of the hot cores W33\,B and W33\,A support these results for the molecules H$_{2}$CO, CH$_{3}$OH, HNCO, and CH$_{3}$CN. \citet{Bisschop2007} concluded that both types of molecules are probably formed on dust grains, but the cold molecules tend to be dissociated at higher temperatures. Furthermore, CH$_{3}$CCH abundances are well-modeled with gas-phase chemistry only \citep{Bisschop2007}, supporting the identification of this molecule as a tracer for more evolved stages as seen in our APEX observations.

We show the importance of combining spectral information on different scales. While the APEX data show emission of more complex molecules for more evolved star forming regions, the SMA spectra show an increase in chemical complexity up to the hot core stage but then a decrease in detected molecules in the \ion{H}{ii} region phase. Thus, the sources W33\,B and W33\,Main would probably have been classified differently if only one of the data sets would have been considered.

\section{Summary}
\label{Summary} 

Infrared, submillimeter, and radio continuum and maser observations of the high-mass star forming complex W33 show that it contains star forming regions at different stages of formation from quiescent clumps to developed \ion{H}{ii} regions. The star forming regions are located at the same distance of 2.4 kpc and probably consist of similar birth material, making a comparative chemical study along the evolutionary sequence feasible. We conducted SMA and APEX observations of six molecular clumps in the W33 complex at 230 and 280 GHz, respectively. In the APEX and SMA data, 27 transitions of ten different molecules and 52 transitions of 16 different molecules, respectively, were detected from simple molecules, like CO with widespread emission, to complex molecules, like CH$_{3}$CCH, CH$_{3}$OCHO or CH$_{3}$OCH$_{3}$ with compact emission. The two data sets probe different physical scales. While the APEX data are sensitive to emission on scales of >~0.2~pc, the SMA data trace compact emission on smaller scales ($\sim$0.05$-$0.2~pc), allowing us to compare the chemical compositions with different resolutions. We established an evolutionary sequence for the 
observed clumps: W33\,Main1, W33\,A1/W33\,B1, W33\,B, W33\,A, and W33\,Main, where W33\,Main is the most evolved clump.

We constructed SEDs for the six clumps W33\,Main1, W33\,A1, W33\,B1, W33\,B, W33\,A, and W33\,Main, and determined dust temperatures, bolometric luminosities, and spectral emissivity indices, and inferred the total masses of the clumps. Except for W33\,B1, the sources follow the trend of increasing luminosity with increasing total mass. The H$_{2}$ peak column density is correlated with the evolutionary stage of the six clumps, being the largest in the most evolved source W33\,Main.

We plotted the integrated intensity ratios N$_{2}$H$^{+}$(3$-$2)/CS(6$-$5) and N$_{2}$H$^{+}$(3$-$2)/H$_{2}$CO(4$_{2,2}$$-$3$_{2,1}$) against the evolutionary state, the luminosity, the total mass, the luminosity to mass ratio, and the H$_{2}$ peak column density of the clouds. These two ratios are the largest in W33\,Main1 and decrease during the evolution of the star forming regions. With increasing luminosity and peak column density, the ratios first increase and then decrease again. Our plots indicate that sources with higher peak column densities look more chemically and physically evolved.

Generating rotational temperature diagrams and constructing synthetic spectra with the Weeds software, we estimated gas rotational temperatures, column densities, and abundances of the H$_{2}$CO, CH$_{3}$OH, and CH$_{3}$CCH molecules from our APEX observations and HNCO, CH$_{3}$CN, and CH$_{3}$OH from the SMA observations. The synthetic spectra of the H$_{2}$CO transitions required a warm and a cold component in the construction for all clumps except W33\,Main1. The temperatures and column densities of the two components increase along the evolutionary sequence.

In the early protostellar phase (W33\,Main1), the protostar starts to release primary molecules like H$_{2}$CO and CH$_{3}$OH from the dust grains, which are observed in the APEX spectra. However, in the SMA data, only low-excitation transitions of CO, its isotopologues, and SO are detected. Thus, the emission of the more complex molecules is probably still too diffuse to be detected by the interferometer. Once the protostar becomes more powerful, it releases more H$_{2}$CO and CH$_{3}$OH into the gas phase, whose emission can then also be detected on small scales. However, the APEX spectra of the sources in the protostellar phase (W33\,A1, W33\,B1) did not change significantly compared to the spectra of W33\,Main1. In the hot core phase (W33\,A, W33\,B), the diversity and complexity of the chemical composition in the SMA spectra changes strongly. The chemistry on larger scales in the APEX spectra of W33\,B and W33\,A is more evolved than in the protostellar phase but does not show the chemical complexity observed on smaller scales in the SMA data. The object W33\,Main is in the \ion{H}{ii} region phase. We detected the radio recombination line H30$\alpha$ in W33\,Main, which traces ionised gas. 
The SMA spectra of W33\,Main are similar to the spectra of the sources in the protostellar phase, except for the detection of the radio recombination line and a transition of HC$_{3}$N. This indicates that the complex molecules that are observed in the hot core phase are either destroyed by the \ion{H}{ii} region or their emission is too diffuse to be detected by the SMA. The APEX data of W33\,Main show transitions of complex molecules, which indicates that the destruction of complex molecules has not reached the larger scales yet.

A study of the kinematics of the W33 complex shows that W33\,A and W33\,B are not gravitationally bound to W33\,Main and thus, the larger clumps in the W33 complex 
will probably drift apart with time.

\Online

\begin{appendix}
\section{Results of the APEX observations}
\label{ResultsAPEXApp}

\paragraph{W33\,Main1} 
The detected transitions of W33\,Main1 are common to all six sources and are listed in 
Sect. \ref{ResultsAPEX} (Fig. \ref{W33_APEX_Spectra1}). The upper energy levels E$_{u}$ of the observed transitions range from 27 to 141~K. The average central velocity is 36.4 km s$^{-1}$ and the average line width is 3.2 km s$^{-1}$. The detection of high-energy transitions of H$_{2}$CO at 82 and 141~K and CH$_{3}$OH at 64 and 75~K hints to the existence of a heating source within or near W33\,Main1. The moment 0 maps of these lines (Fig. \ref{W33M1-APEX-IntInt}) show only extended emission at the position of the continuum peak, while the peak of the line emission is located close to the edge of the maps, which point to an external rather than an internal heating source. The moment 0 maps of the two low-temperature transitions of H$_{2}$CO at 35 and 46~K, and the N$_{2}$H$^{+}$ and the CS transitions show strong compact emission, which peaks within $\sim$6$\arcsec$ of the continuum peak, indicating that the interior of W33\,Main1 is still cold (Fig. \ref{W33M1-APEX-IntInt}).

\paragraph{W33\,A1 and W33\,B1}
In W33\,A1 and W33\,B1 (Fig. \ref{W33_APEX_Spectra1}), we observe the same transitions 
(CH$_{3}$OH(9$_{\textnormal{$-$1,9}}$$-$8$_{\textnormal{0,8}}$) and 
CH$_{3}$OH(6$_{\textnormal{1,5}}$$-$5$_{\textnormal{1,4}}$), in addition to the spectral lines 
detected in W33\,Main1). The upper energy levels E$_{u}$ of all detected spectral lines range 
between 27 and 141~K. The average central velocity is 33.3~km~s$^{-1}$ and 36.7~km~s$^{-1}$ 
for W33\,B1 and W33\,A1, respectively. The spectral lines have an average line width of 
5.4~km~s$^{-1}$ and 3.9~km~s$^{-1}$ in W33\,B1 and W33\,A1, respectively. In general, the 
detected spectral lines tend to be stronger and broader in W33\,B1 compared to W33\,A1.
The peaks of the compact emission of the four strongest lines, as seen in the moment 0 maps, 
are located within $\sim$6$\arcsec$ of the continuum peaks in both sources (Figs. \ref{W33A1-APEX-IntInt} and \ref{W33B1-APEX-IntInt}). In W33\,B1, the emission 
of the high-excitation transitions of H$_{2}$CO and CH$_{3}$OH is also compact and peaks 
close to the continuum peak (Fig. \ref{W33B1-APEX-IntInt}). This suggests that a heating source is present in W33\,B1. The moment 0 maps of these high-temperature lines in W33\,A1 show more extended emission and less isolated peaks (Fig. \ref{W33A1-APEX-IntInt}), which indicates that W33\,A1 already contains a heating source but is probably less developed than W33\,B1.

\paragraph{W33\,B}
Besides the spectral lines detected in W33\,B1 and W33\,A1, we observe emission from 
H$_{2}$CS(8$_{\textnormal{1,7}}$$-$7$_{\textnormal{1,6}}$), 
CH$_{3}$OH(6$_{\textnormal{2,4}}$$-$5$_{\textnormal{2,3}}$), 
CH$_{3}$OH(3$_{\textnormal{2,1}}$$-$4$_{\textnormal{1,4}}$), C$^{33}$S(6$-$5), and 
OCS(24$-$23) in W33\,B (Fig. \ref{W33_APEX_Spectra2}). The spectral lines have upper energy levels E$_{u}$ between 27 and 175~K.
The average central velocity and line width of W33\,B are 55.4~km~s$^{-1}$ and 5.3~km~s$^{-1}$, respectively. Although the radial velocity of W33\,B differs by $\sim$20 km s$^{-1}$ from the radial velocities of the other clumps in W33, \citet{Immer2013} have shown that this clump is located at a similar distance as the other clumps. However, it is still unclear what the reason for this large radial velocity difference is.
While the integrated emission of CS and the two low-excitation transitions of H$_{2}$CO is compact and peaks close to the continuum peak, the emission of N$_{2}$H$^{+}$ is extended in south-east direction and peaks $\sim$12$\arcsec$ east of the continuum peak (Fig. \ref{W33B-APEX-IntInt}). The moment 0 map of H$_{2}$CO(4$_{\textnormal{2,2}}$$-$3$_{\textnormal{2,1}}$) shows extended emission in the north-south direction. The peak of the emission is offset by $\sim$12$\arcsec$ in north-west direction from the continuum peak (Fig. \ref{W33B-APEX-IntInt}). The integrated emission of C$^{33}$S(6$-$5) is extended to the west, but its maximum is located close to the continuum peak (Fig. \ref{W33B-APEX-IntInt}). The emission of the remaining lines is mostly compact and peaks within $\sim$6$\arcsec$ of the continuum peak (Fig. \ref{W33B-APEX-IntInt}). The stronger lines and the detection of transitions at higher excitation energies compared to W33\,B1 and W33\,A1 suggests that W33\,B is even more evolved than W33\,B1 and W33\,A1.

\paragraph{W33\,A}
In addition to the lines found in W33\,B, we also detect emission of OCS(23$-$22), HC$_{3}$N(31$-$30), HC$_{3}$N(32$-$31), and CH$_{3}$CCH(17$_{\textnormal{0}}$$-$16$_{\textnormal{0}}$) in W33\,A (Fig. \ref{W33_APEX_Spectra2}). However, we do not observe emission of C$^{33}$S(6$-$5) or CH$_{3}$OH(3$_{\textnormal{2,1}}$$-$4$_{\textnormal{1,4}}$). The upper energy levels of the detected lines are between 27 and 231~K. The average central velocity of all transitions is 37.6~km~s$^{-1}$, which is close to the systemic velocity of 38.5~km~s$^{-1}$, determined by \citet{GalvanMadrid2010} from their SMA observations of W33\,A. The average line width is 5.4~km~s$^{-1}$. The integrated emission of all lines in W33\,A is compact and their maxima are located close to the continuum peak within $\sim$6$\arcsec-$12$\arcsec$ (Fig. \ref{W33A-APEX-IntInt}). The object W33\,A is probably more evolved than W33\,B.

\paragraph{W33\,Main}
Besides the transitions that we detect in W33\,A and W33\,B, we observe the higher-excitation transitions of CH\textsubscript{3}CCH (CH\textsubscript{3}CCH(17\textsubscript{1}$-$16\textsubscript{1}), CH\textsubscript{3}CCH(17\textsubscript{2}$-$16\textsubscript{2}), CH\textsubscript{3}CCH(17\textsubscript{3}$-$16\textsubscript{3}), CH\textsubscript{3}CCH(17\textsubscript{4}$-$16\textsubscript{4})), and H\textsubscript{2}\textsuperscript{13}CO(4\textsubscript{1,3}$-$3\textsubscript{1,2}) in W33\,Main (Fig. \ref{W33_APEX_Spectra2}). The transition with the highest upper energy level E\textsubscript{u}~=~241~K is detected in this source. The average central velocity and line width of W33\,Main are 35.6~km~s\textsuperscript{$-$1} and 6.0~km~s\textsuperscript{$-$1}, respectively.
Except for the N\textsubscript{2}H\textsuperscript{+}, CS, and OCS transitions, the emission of all spectral lines peaks close to the continuum peak (Fig. \ref{W33M-APEX-IntInt}). The N\textsubscript{2}H\textsuperscript{+} emission is strongest at the northwestern edge of the map. At the center of W33\,Main, the N\textsubscript{2}H\textsuperscript{+} emission is much weaker (Fig. \ref{W33M-APEX-IntInt}). This shows that N\textsubscript{2}H\textsuperscript{+} is not a good tracer of the dust continuum in evolved sources anymore \citep{Reiter2011}. The CS emission is spread from the center of the map to the north and peaks about $\sim$24$\arcsec$ from the continuum peak to the north (Fig. \ref{W33M-APEX-IntInt}). The emission of C\textsuperscript{33}S, CH\textsubscript{3}CCH (except CH\textsubscript{3}CCH(17\textsubscript{4}$-$16\textsubscript{4})), CH\textsubscript{3}OH (except CH\textsubscript{3}OH(3\textsubscript{2,1}$-$4\textsubscript{1,4}) and CH\textsubscript{3}OH(6\textsubscript{2,4}$-$5\textsubscript{2,3})), H\textsubscript{2}CO, and H\textsubscript{2}CS is also extended to the north but peaks close to the continuum peak (Fig. \ref{W33M-APEX-IntInt}). The OCS emission is extended and the peak close to the center of the map is not very pronounced  (Fig. \ref{W33M-APEX-IntInt}).
The extended emission hints to the existence of another source in the north of the map. Since CS and N\textsubscript{2}H\textsuperscript{+} trace cold gas, the line peak offsets of the CS and N\textsubscript{2}H\textsuperscript{+} emission from the continuum peak indicate that the gas towards the center of the W33\,Main clump is not cold anymore. We conclude that W33\,Main is not in an early stage of star formation anymore.

\onecolumn 

\begin{landscape}
\begin{scriptsize}
\onehalfspacing
\tabcolsep=0.15cm
\begin{longtable}{|clc|cccc|cccc|cccc|}
 \caption{\label{LinesTabAPEX} Transitions, detected in W33 with the APEX telescope.}\\ \hline
 &  &  & \multicolumn{4}{|c|}{W33\,Main1} & \multicolumn{4}{|c|}{W33\,A1} & \multicolumn{4}{|c|}{W33\,B1}\\ \hline
$\nu_{0}$ & Transition & E$_{u}$ & F$_{Int.}$ & F$_{Peak}$ & v$_{central}$ & FWHM & F$_{Int}$ & F$_{Peak}$ & v$_{central}$ & FWHM & F$_{Int}$ & F$_{Peak}$ & v$_{central}$ & FWHM\\
(GHz) & & (K) & (K km s$^{-1}$) & (K) & (km s$^{-1}$) & (km s$^{-1}$) & (K km s$^{-1}$) & (K) & (km s$^{-1}$) & (km s$^{-1}$) & (K km s$^{-1}$) & (K) & (km s$^{-1}$) & (km s$^{-1}$)\\ \hline
\endfirsthead
\caption{Continued.}\\ \hline
 & &  & \multicolumn{4}{|c|}{W33\,B} & \multicolumn{4}{|c|}{W33\,A} & \multicolumn{4}{|c|}{W33\,Main}\\ \hline
$\nu_{0}$ & Transition & E$_{u}$ & F$_{Int.}$ & F$_{Peak}$ & v$_{central}$ & FWHM & F$_{Int}$ & F$_{Peak}$ & v$_{central}$ & FWHM & F$_{Int}$ & F$_{Peak}$ & v$_{central}$ & FWHM\\
(GHz) & & (K) & (K km s$^{-1}$) & (K) & (km s$^{-1}$) & (km s$^{-1}$) & (K km s$^{-1}$) & (K) & (km s$^{-1}$) & (km s$^{-1}$) & (K km s$^{-1}$) & (K) & (km s$^{-1}$) & (km s$^{-1}$)\\ \hline
\endhead
\endfoot
278.3045	&	CH\textsubscript{3}OH(9\textsubscript{$-$1,9}$-$8\textsubscript{0,8})	&	109.97	&		&		&		&		&	0.28	&	0.10	&	36.48	&	2.79	&	0.59	&	0.10	&	33.21	&	5.65	\\
278.8864	&	H\textsubscript{2}CS(8\textsubscript{1,7}$-$7\textsubscript{1,6})	&	73.41	&		&		&		&		&		&		&		&		&		&		&		&		\\
279.5117	&	N\textsubscript{2}H\textsuperscript{+}(3$-$2)	&	26.83	&	11.13	&	2.16	&	36.61	&	4.85	&	9.43	&	1.83	&	36.74	&	4.85	&	7.38	&	1.47	&	33.57	&	4.71	\\
279.6853	&	OCS(23$-$22)	&	161.09	&		&		&		&		&		&		&		&		&		&		&		&		\\
281.5269	&	H\textsubscript{2}CO(4\textsubscript{1,4}$-$3\textsubscript{1,3})	&	45.57	&	3.05	&	1.00	&	36.53	&	2.87	&	3.78	&	0.98	&	36.68	&	3.62	&	4.82	&	0.79	&	33.69	&	5.73	\\
281.9768	&	HC\textsubscript{3}N(31$-$30)	&	215.55	&		&		&		&		&		&		&		&		&		&		&		&		\\
290.2487	&	CH\textsubscript{3}OH(6\textsubscript{1,5}$-$5\textsubscript{1,4})	&	69.80	&		&		&		&		&	0.28	&	0.07	&	37.08	&	3.78	&	0.56	&	0.07	&	32.65	&	8.10	\\
290.2641	&	CH\textsubscript{3}OH(6\textsubscript{2,4}$-$5\textsubscript{2,3})	&	86.46	&		&		&		&		&		&		&		&		&		&		&		&		\\
290.3073	&	CH\textsubscript{3}OH(6\textsubscript{$-$2,5}$-$5\textsubscript{$-$2,4})\tablefootmark{a}	&	74.66	&	\multirow{2}{*}{0.19}	&	\multirow{2}{*}{0.08}	&	\multirow{2}{*}{35.96}	&	\multirow{2}{*}{2.39}	&	\multirow{2}{*}{0.48}	&	\multirow{2}{*}{0.12}	&	\multirow{2}{*}{36.75}	&	\multirow{2}{*}{3.78}	&	\multirow{2}{*}{0.88}	&	\multirow{2}{*}{0.15}	&	\multirow{2}{*}{32.74}	&	\multirow{2}{*}{5.58}	\\
290.3076	&	CH\textsubscript{3}OH(6\textsubscript{2,4}$-$5\textsubscript{2,3})\tablefootmark{a}	&	71.00	&		&		&		&		&		&		&		&		&		&		&		&		\\
290.4135	&	CH\textsubscript{3}CCH(17\textsubscript{4}$-$16\textsubscript{4})	&	241.05	&		&		&		&		&		&		&		&		&		&		&		&		\\
290.4523	&	CH\textsubscript{3}CCH(17\textsubscript{3}$-$16\textsubscript{3})	&	190.50	&		&		&		&		&		&		&		&		&		&		&		&		\\
290.4799	&	CH\textsubscript{3}CCH(17\textsubscript{2}$-$16\textsubscript{2})	&	154.39	&		&		&		&		&		&		&		&		&		&		&		&		\\
290.4965	&	CH\textsubscript{3}CCH(17\textsubscript{1}$-$16\textsubscript{1})	&	132.71	&		&		&		&		&		&		&		&		&		&		&		&		\\
290.5021	&	CH\textsubscript{3}CCH(17\textsubscript{0}$-$16\textsubscript{0})	&	125.49	&		&		&		&		&		&		&		&		&		&		&		&		\\
290.6234	&	H\textsubscript{2}CO(4\textsubscript{0,4}$-$3\textsubscript{0,3})	&	34.90	&	1.52	&	0.53	&	36.54	&	2.67	&	2.00	&	0.55	&	36.58	&	3.40	&	2.60	&	0.49	&	33.63	&	5.00	\\
291.0684	&	HC\textsubscript{3}N(32$-$31)	&	230.52	&		&		&		&		&		&		&		&		&		&		&		&		\\
291.2378	&	H\textsubscript{2}CO(4\textsubscript{2,3}$-$3\textsubscript{2,2})	&	82.07	&	0.35	&	0.10	&	36.46	&	3.25	&	0.44	&	0.13	&	36.33	&	3.26	&	0.84	&	0.14	&	33.18	&	5.78	\\
291.3805	&	H\textsubscript{2}CO(4\textsubscript{3,2}$-$3\textsubscript{3,1})\tablefootmark{a}	&	140.94	&	\multirow{2}{*}{0.54}	&	\multirow{2}{*}{0.07}	&	\multirow{2}{*}{36.14}	&	\multirow{2}{*}{7.72}	&	\multirow{2}{*}{0.74}	&	\multirow{2}{*}{0.08}	&	\multirow{2}{*}{36.49}	&	\multirow{2}{*}{8.43}	&	\multirow{2}{*}{1.38}	&	\multirow{2}{*}{0.14}	&	\multirow{2}{*}{32.92}	&	\multirow{2}{*}{9.36}	\\
291.3843	&	H\textsubscript{2}CO(4\textsubscript{3,1}$-$3\textsubscript{3,0})\tablefootmark{a}	&	140.94	&		&		&		&		&		&		&		&		&		&		&		&		\\
291.4859	&	C\textsuperscript{33}S(6$-$5)	&	38.66	&		&		&		&		&		&		&		&		&		&		&		&		\\
291.8397	&	OCS(24$-$23)	&	175.10	&		&		&		&		&		&		&		&		&		&		&		&		\\
291.9481	&	H\textsubscript{2}CO(4\textsubscript{2,2}$-$3\textsubscript{2,1})	&	82.12	&	0.30	&	0.09	&	36.34	&	2.93	&	0.50	&	0.09	&	36.88	&	5.48	&	0.62	&	0.14	&	33.34	&	4.32	\\
292.6729	&	CH\textsubscript{3}OH(6\textsubscript{1,5}$-$5\textsubscript{1,4})	&	63.71	&	0.16	&	0.06	&	36.13	&	2.46	&	0.42	&	0.09	&	36.77	&	4.51	&	0.80	&	0.12	&	32.92	&	6.17	\\
293.1265	&	H\textsubscript{2}\textsuperscript{13}CO(4\textsubscript{1,3}$-$3\textsubscript{1,2})	&	47.01	&		&		&		&		&		&		&		&		&		&		&		&		\\
293.4640	&	CH\textsubscript{3}OH(3\textsubscript{2,1}$-$4\textsubscript{1,4})	&	51.64	&		&		&		&		&		&		&		&		&		&		&		&		\\
293.9122	&	CS(6$-$5)	&	49.37	&	2.82	&	0.89	&	36.56	&	2.97	&	4.39	&	1.22	&	36.96	&	3.38	&	2.86	&	0.62	&	34.16	&	4.34	\\ \hline
\multicolumn{15}{c}{}\\
\multicolumn{15}{c}{}\\
\multicolumn{15}{c}{}\\
\multicolumn{15}{c}{}\\
\multicolumn{15}{c}{}\\
\multicolumn{15}{c}{}\\
\multicolumn{15}{c}{}\\
\multicolumn{15}{c}{}\\
\multicolumn{15}{c}{}\\
\multicolumn{15}{c}{}\\
\multicolumn{15}{c}{}\\
\multicolumn{15}{c}{}\\
\multicolumn{15}{c}{}\\
\multicolumn{15}{c}{}\\
\multicolumn{15}{c}{}\\
\multicolumn{15}{c}{}\\
\multicolumn{15}{c}{}\\
\multicolumn{15}{c}{}\\
\multicolumn{15}{c}{}\\
\multicolumn{15}{c}{}\\
278.3045	&	CH\textsubscript{3}OH(9\textsubscript{$-$1,9}$-$8\textsubscript{0,8})	&	109.97	&	0.98	&	0.19	&	54.73	&	4.88	&	2.24	&	0.34	&	37.52	&	6.12	&	3.05	&	0.56	&	36.07	&	5.12	\\
278.8864	&	H\textsubscript{2}CS(8\textsubscript{1,7}$-$7\textsubscript{1,6})	&	73.41	&	0.49	&	0.13	&	54.12	&	3.69	&	1.62	&	0.28	&	36.63	&	5.40	&	5.90	&	0.99	&	34.59	&	5.58	\\
279.5117	&	N\textsubscript{2}H\textsuperscript{+}(3$-$2)	&	26.83	&	12.26	&	2.24	&	55.39	&	5.16	&	31.07	&	5.11	&	37.28	&	5.71	&	22.91	&	3.35	&	35.40	&	6.43	\\
279.6853	&	OCS(23$-$22)	&	161.09	&		&		&		&		&	1.05	&	0.21	&	37.81	&	4.69	&	0.81	&	0.11	&	36.45	&	7.20	\\
281.5269	&	H\textsubscript{2}CO(4\textsubscript{1,4}$-$3\textsubscript{1,3})	&	45.57	&	4.49	&	1.04	&	55.44	&	4.07	&	13.74	&	2.35	&	37.56	&	5.50	&	33.50	&	4.11	&	35.87	&	7.66	\\
281.9768	&	HC\textsubscript{3}N(31$-$30)	&	215.55	&		&		&		&		&	0.62	&	0.14	&	38.23	&	4.04	&	2.21	&	0.37	&	35.27	&	5.59	\\
290.2487	&	CH\textsubscript{3}OH(6\textsubscript{1,5}$-$5\textsubscript{1,4})	&	69.80	&	0.97	&	0.13	&	55.39	&	6.88	&	2.54	&	0.41	&	37.51	&	5.77	&	5.27	&	0.86	&	35.97	&	5.78	\\
290.2641	&	CH\textsubscript{3}OH(6\textsubscript{2,4}$-$5\textsubscript{2,3})	&	86.46	&	0.87	&	0.12	&	55.39	&	6.72	&	1.50	&	0.18	&	37.51	&	7.68	&	1.75	&	0.30	&	36.32	&	5.54	\\
290.3073	&	CH\textsubscript{3}OH(6\textsubscript{$-$2,5}$-$5\textsubscript{$-$2,4})\tablefootmark{a}	&	74.66	&	\multirow{2}{*}{1.08}	&	\multirow{2}{*}{0.24}	&	\multirow{2}{*}{55.38}	&	\multirow{2}{*}{4.22}	&	\multirow{2}{*}{3.51}	&	\multirow{2}{*}{0.63}	&	\multirow{2}{*}{37.19}	&	\multirow{2}{*}{5.25}	&	\multirow{2}{*}{7.96}	&	\multirow{2}{*}{1.27}	&	\multirow{2}{*}{36.08}	&	\multirow{2}{*}{5.90}	\\
290.3076	&	CH\textsubscript{3}OH(6\textsubscript{2,4}$-$5\textsubscript{2,3})\tablefootmark{a}	&	71.00	&		&		&		&		&		&		&		&		&		&		&		&		\\
290.4135	&	CH\textsubscript{3}CCH(17\textsubscript{4}$-$16\textsubscript{4})	&	241.05	&		&		&		&		&		&		&		&		&	0.94	&	0.15	&	35.24	&	5.98	\\
290.4523	&	CH\textsubscript{3}CCH(17\textsubscript{3}$-$16\textsubscript{3})	&	190.50	&		&		&		&		&		&		&		&		&	4.40	&	0.69	&	35.24	&	5.98	\\
290.4799	&	CH\textsubscript{3}CCH(17\textsubscript{2}$-$16\textsubscript{2})	&	154.39	&		&		&		&		&		&		&		&		&	4.18	&	0.66	&	35.24	&	5.98	\\
290.4965	&	CH\textsubscript{3}CCH(17\textsubscript{1}$-$16\textsubscript{1})	&	132.71	&		&		&		&		&		&		&		&		&	6.40	&	1.00	&	35.24	&	5.98	\\
290.5021	&	CH\textsubscript{3}CCH(17\textsubscript{0}$-$16\textsubscript{0})	&	125.49	&		&		&		&		&	1.07	&	0.13	&	37.02	&	7.68	&	6.58	&	1.03	&	35.24	&	5.98	\\
290.6234	&	H\textsubscript{2}CO(4\textsubscript{0,4}$-$3\textsubscript{0,3})	&	34.90	&	2.66	&	0.69	&	55.20	&	3.65	&	8.19	&	1.48	&	37.50	&	5.19	&	19.83	&	2.83	&	35.56	&	6.57	\\
291.0684	&	HC\textsubscript{3}N(32$-$31)	&	230.52	&		&		&		&		&	0.90	&	0.18	&	38.65	&	4.64	&	2.00	&	0.32	&	35.06	&	5.97	\\
291.2378	&	H\textsubscript{2}CO(4\textsubscript{2,3}$-$3\textsubscript{2,2})	&	82.07	&	1.64	&	0.20	&	55.69	&	7.63	&	3.04	&	0.56	&	37.52	&	5.08	&	8.10	&	1.22	&	35.39	&	6.25	\\
291.3805	&	H\textsubscript{2}CO(4\textsubscript{3,2}$-$3\textsubscript{3,1})\tablefootmark{a}	&	140.94	&	\multirow{2}{*}{2.23}	&	\multirow{2}{*}{0.17}	&	\multirow{2}{*}{55.86}	&	\multirow{2}{*}{12.48}	&	\multirow{2}{*}{4.93}	&	\multirow{2}{*}{0.52}	&	\multirow{2}{*}{37.76}	&	\multirow{2}{*}{8.99}	&	\multirow{2}{*}{12.93}	&	\multirow{2}{*}{1.39}	&	\multirow{2}{*}{34.97}	&	\multirow{2}{*}{8.75}	\\
291.3843	&	H\textsubscript{2}CO(4\textsubscript{3,1}$-$3\textsubscript{3,0})\tablefootmark{a}	&	140.94	&		&		&		&		&		&		&		&		&		&		&		&		\\
291.4859	&	C\textsuperscript{33}S(6$-$5)	&	38.66	&	0.49	&	0.13	&	55.67	&	3.51	&		&		&		&		&	3.91	&	0.57	&	35.65	&	6.46	\\
291.8397	&	OCS(24$-$23)	&	175.10	&	1.06	&	0.14	&	55.97	&	7.30	&	1.41	&	0.18	&	38.58	&	5.49	&	0.90	&	0.12	&	36.41	&	7.23	\\
291.9481	&	H\textsubscript{2}CO(4\textsubscript{2,2}$-$3\textsubscript{2,1})	&	82.12	&	1.07	&	0.17	&	54.98	&	5.84	&	2.89	&	0.54	&	37.82	&	5.05	&	7.80	&	1.23	&	35.40	&	5.97	\\
292.6729	&	CH\textsubscript{3}OH(6\textsubscript{1,5}$-$5\textsubscript{1,4})	&	63.71	&	1.08	&	0.22	&	55.66	&	4.59	&	2.76	&	0.55	&	37.68	&	4.70	&	5.90	&	0.96	&	36.11	&	5.77	\\
293.1265	&	H\textsubscript{2}\textsuperscript{13}CO(4\textsubscript{1,3}$-$3\textsubscript{1,2})	&	47.01	&		&		&		&		&		&		&		&		&	1.00	&	0.14	&	34.95	&	6.58	\\
293.4640	&	CH\textsubscript{3}OH(3\textsubscript{2,1}$-$4\textsubscript{1,4})	&	51.64	&	0.79	&	0.14	&	56.62	&	5.34	&		&		&		&		&	0.98	&	0.20	&	36.32	&	4.64	\\
293.9122	&	CS(6$-$5)	&	49.37	&	6.86	&	1.71	&	55.58	&	3.76	&	19.52	&	3.22	&	37.58	&	5.70	&	81.86	&	9.23	&	36.39	&	8.33	\\ \hline
\end{longtable}
\tablefoot{
\tablefoottext{a}{Blended transitions.}
}
\end{scriptsize}
\end{landscape}

\twocolumn

\begin{figure*}
	\caption{Line emission of detected transitions in W33\,Main1. The contours show the ATLASGAL continuum emission at 345 GHz (levels in steps of 5$\sigma$, starting at 6$\sigma$ ($\sigma$ = 0.081 Jy beam$^{-1}$). The name of the each transition is shown in the upper right corner. A scale of 0.5 pc is marked in the upper left corner, and the synthesised beam is shown in the lower left corner.}
	\centering
	\subfloat{\includegraphics[width=9cm]{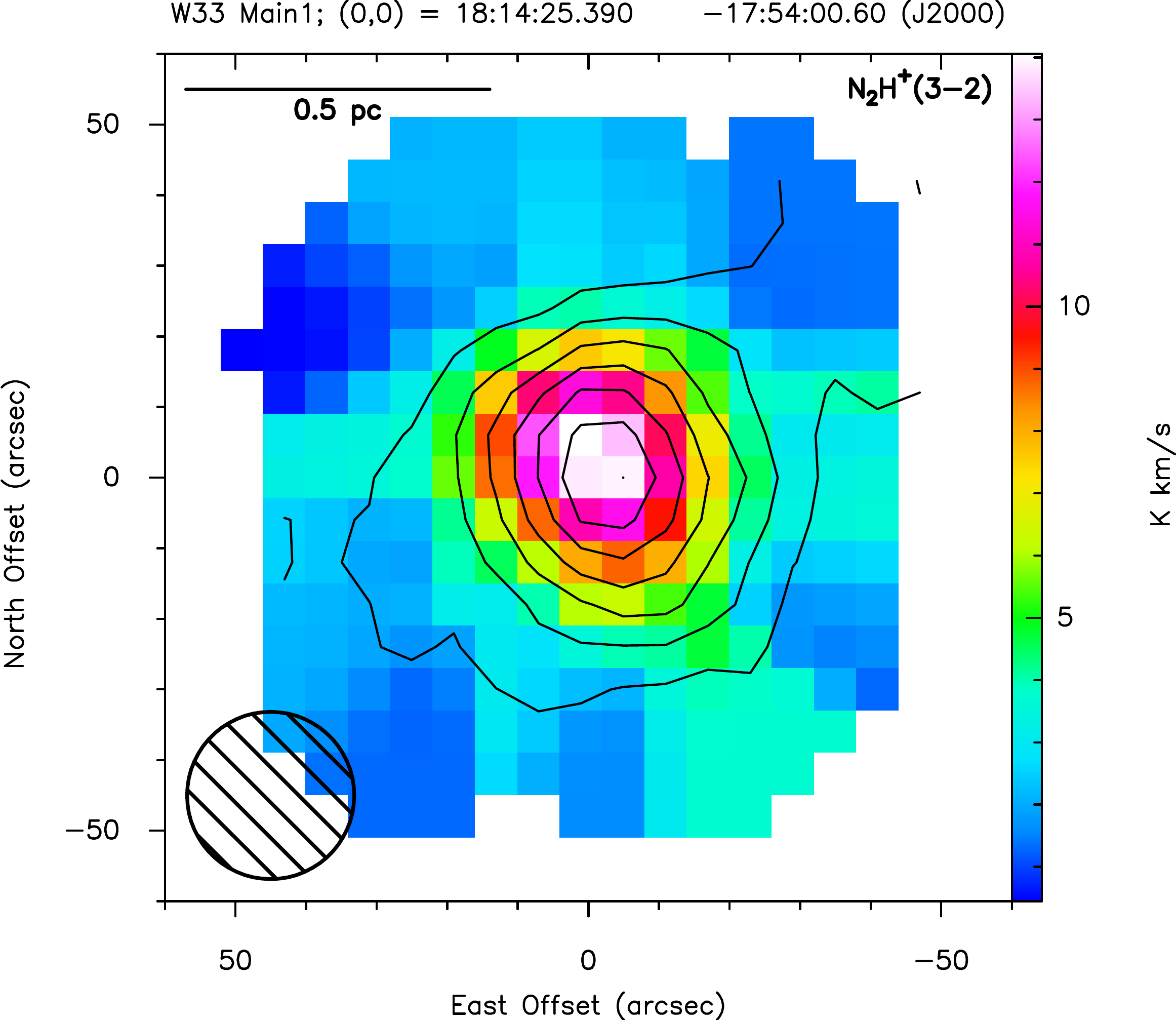}}\hspace{0.2cm}	
	\subfloat{\includegraphics[width=9cm]{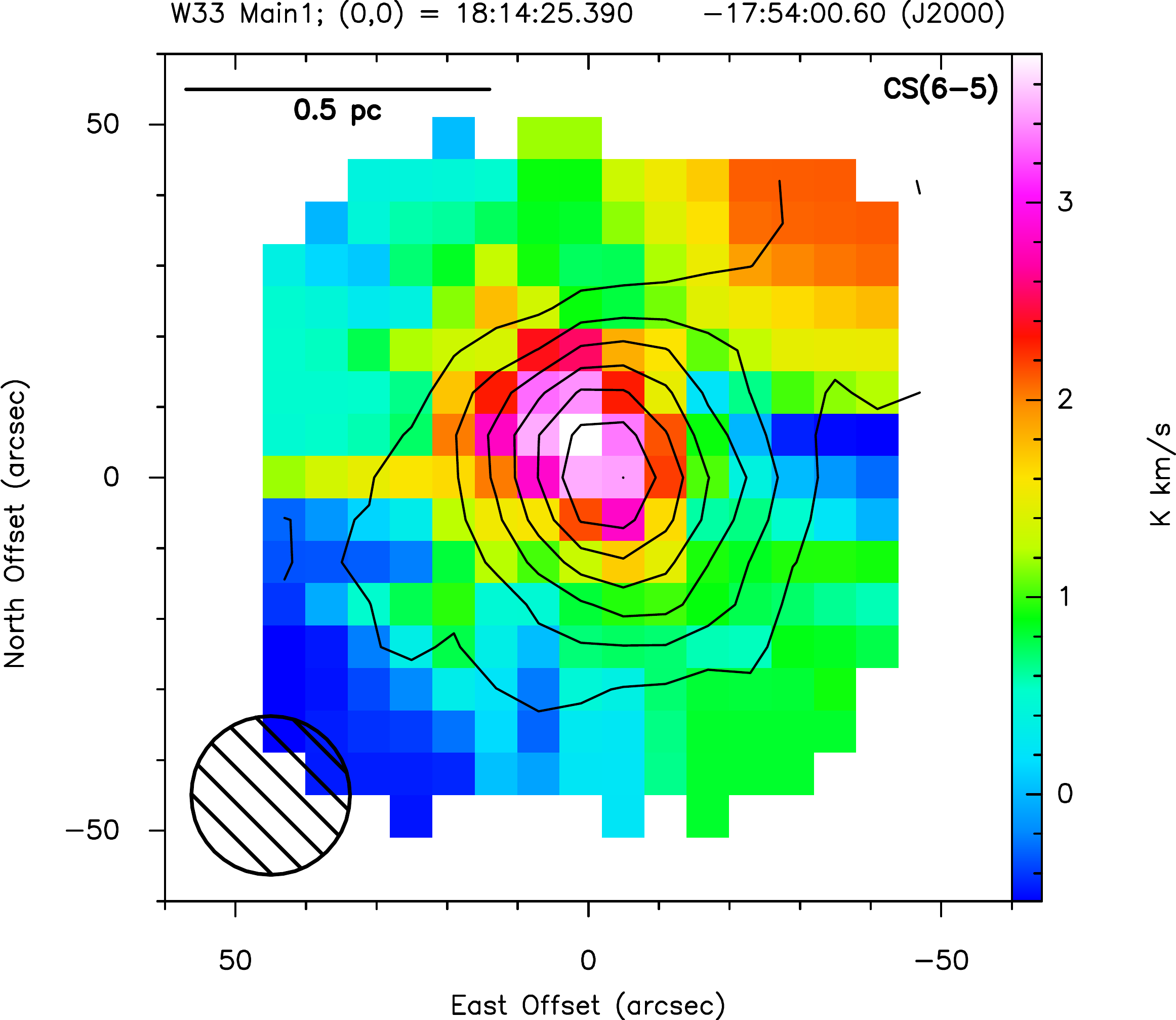}}	\\
	\subfloat{\includegraphics[width=9cm]{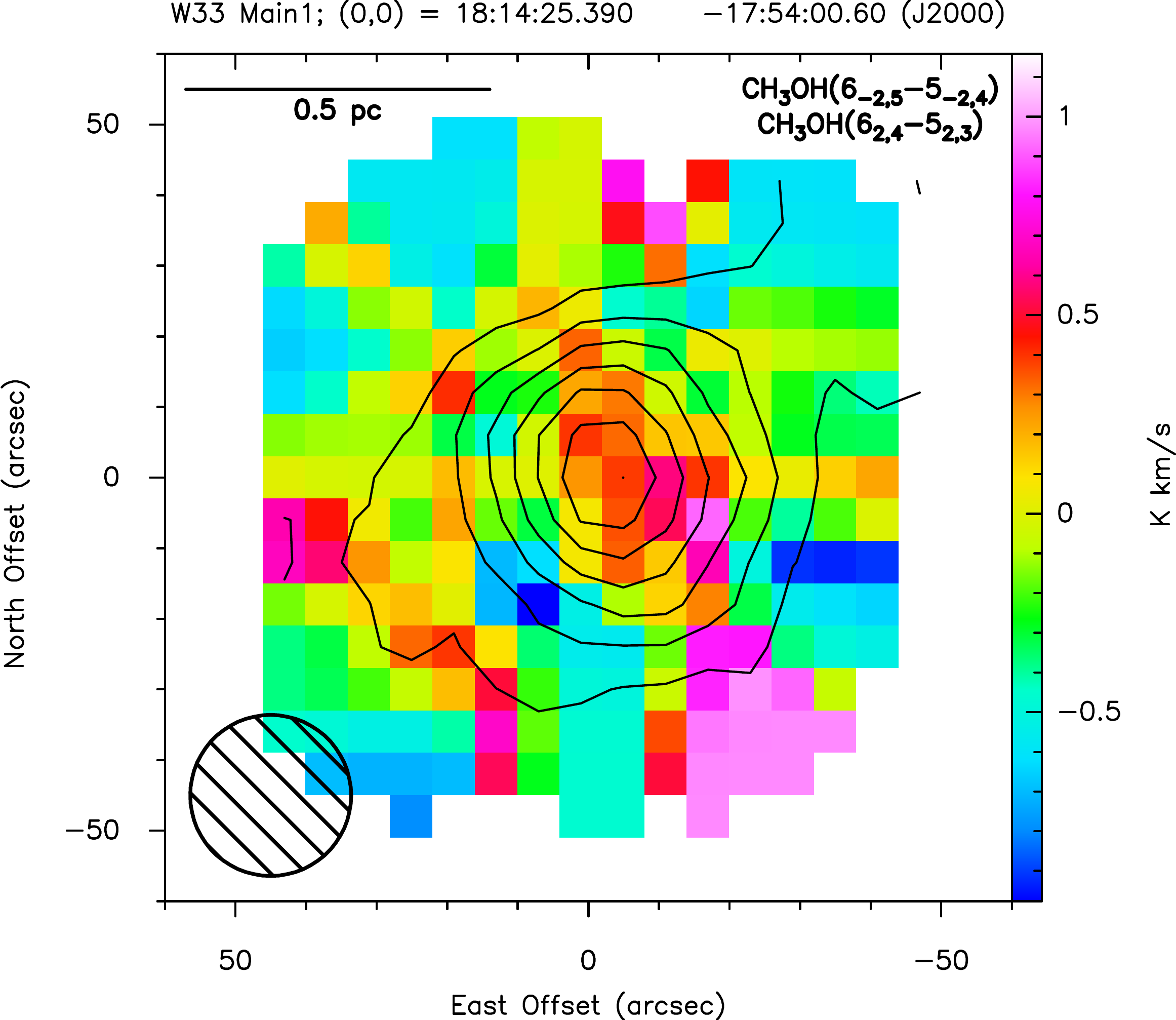}}\hspace{0.2cm}	
	\subfloat{\includegraphics[width=9cm]{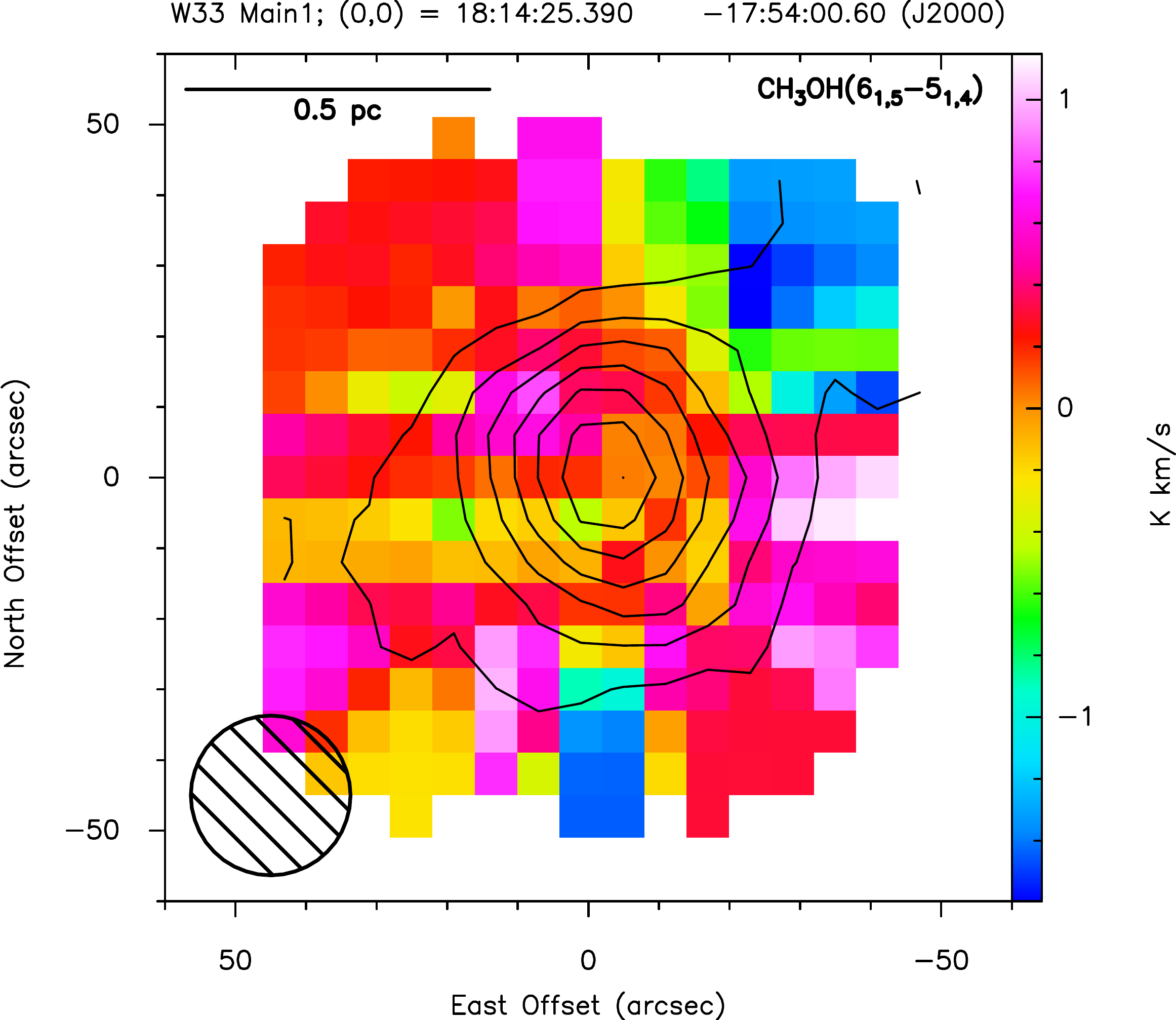}}	
	\label{W33M1-APEX-IntInt}
\end{figure*}

\addtocounter{figure}{-1}
\begin{figure*}
	\caption{Continued.}
	\centering
	\subfloat{\includegraphics[width=9cm]{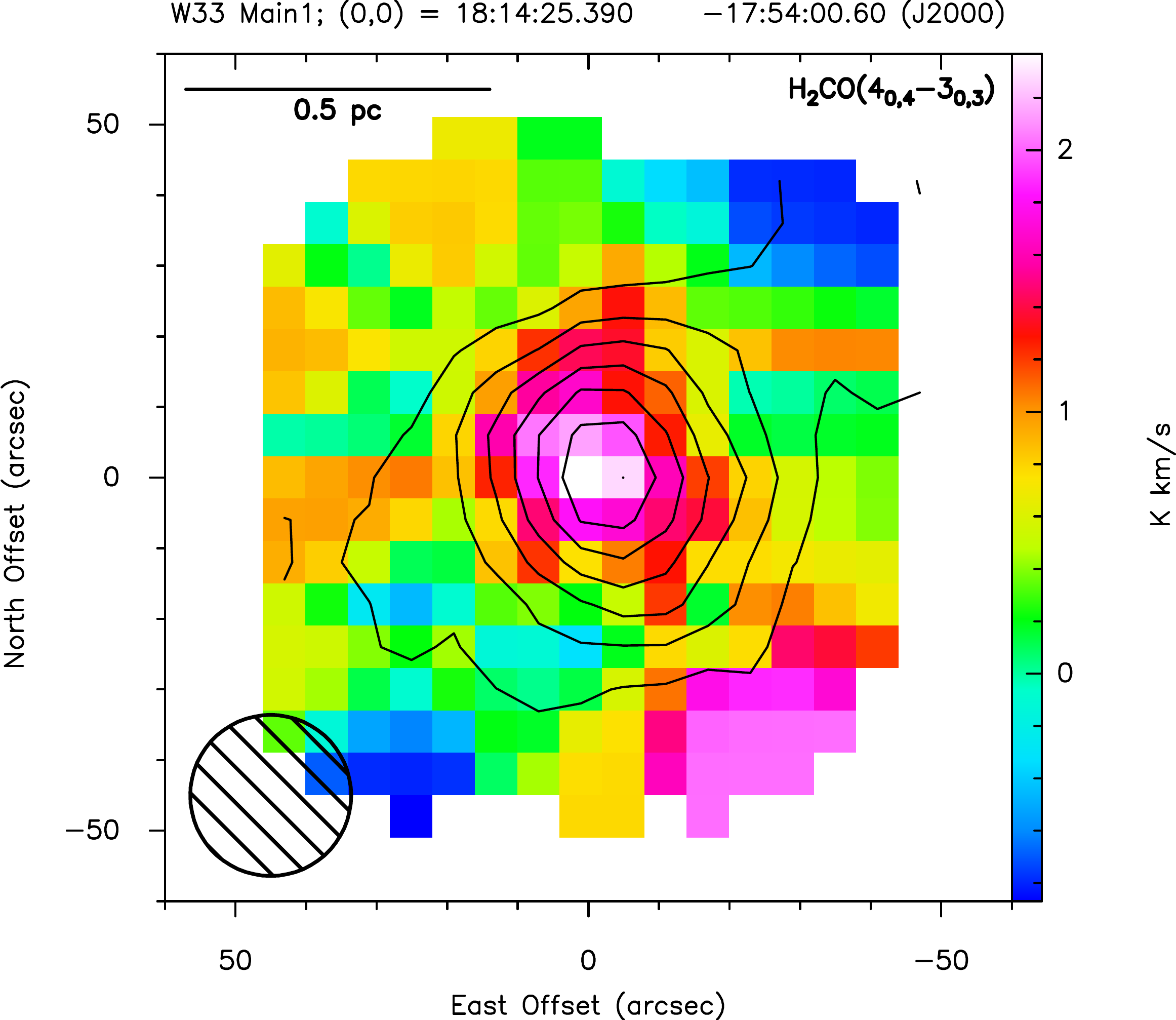}}\hspace{0.2cm}	
	\subfloat{\includegraphics[width=9cm]{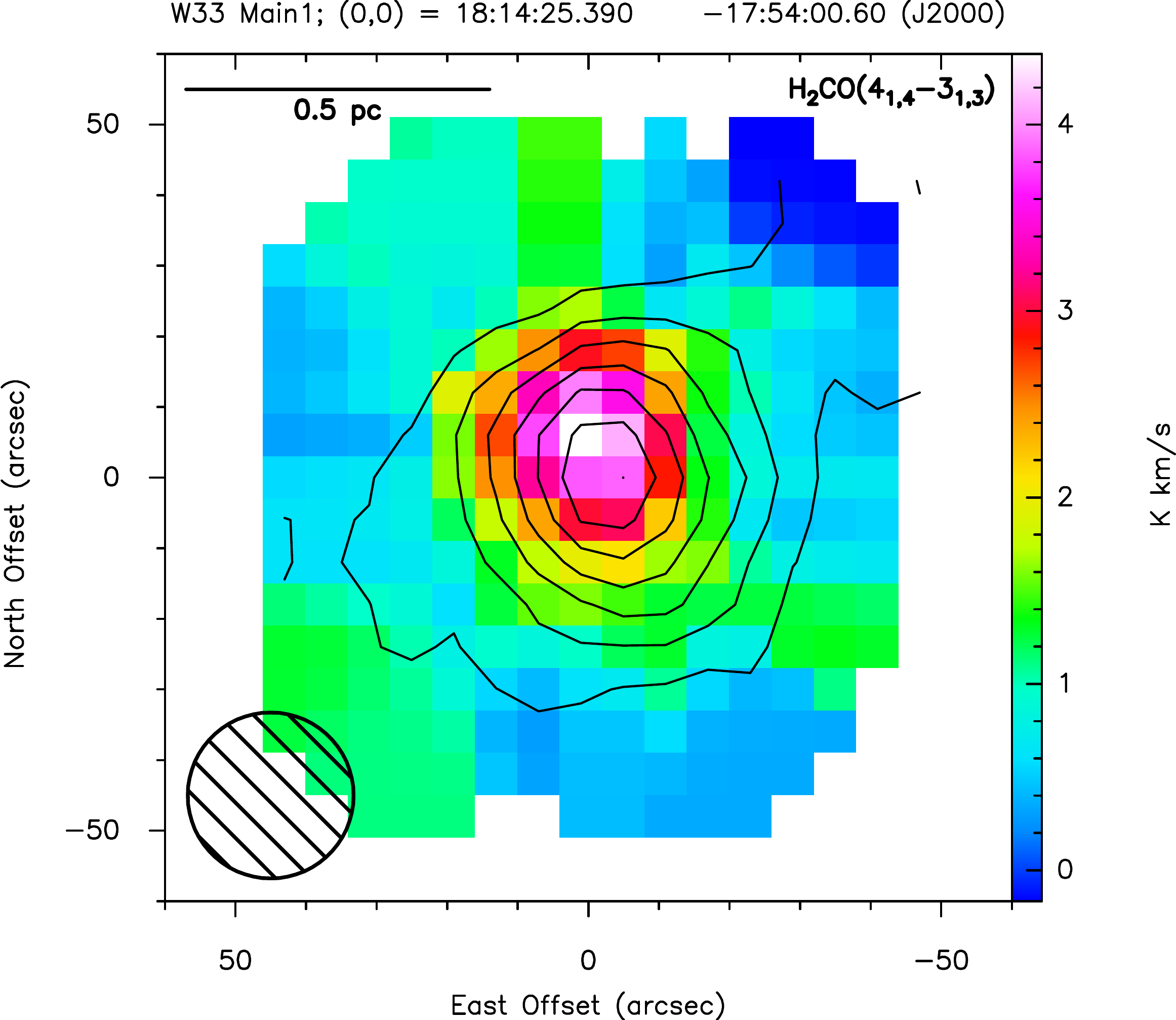}}\\	
	\subfloat{\includegraphics[width=9cm]{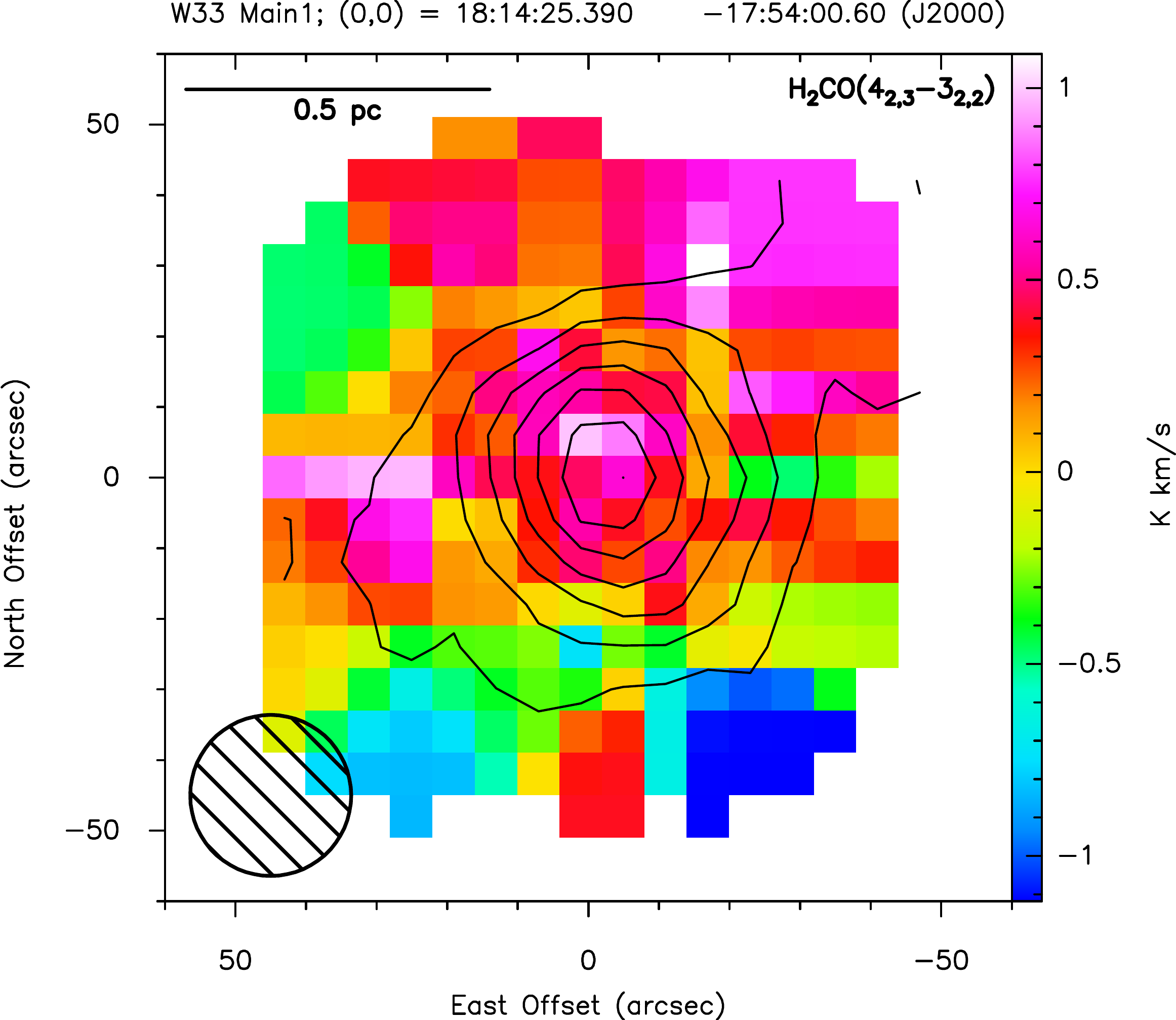}}\hspace{0.2cm}	
	\subfloat{\includegraphics[width=9cm]{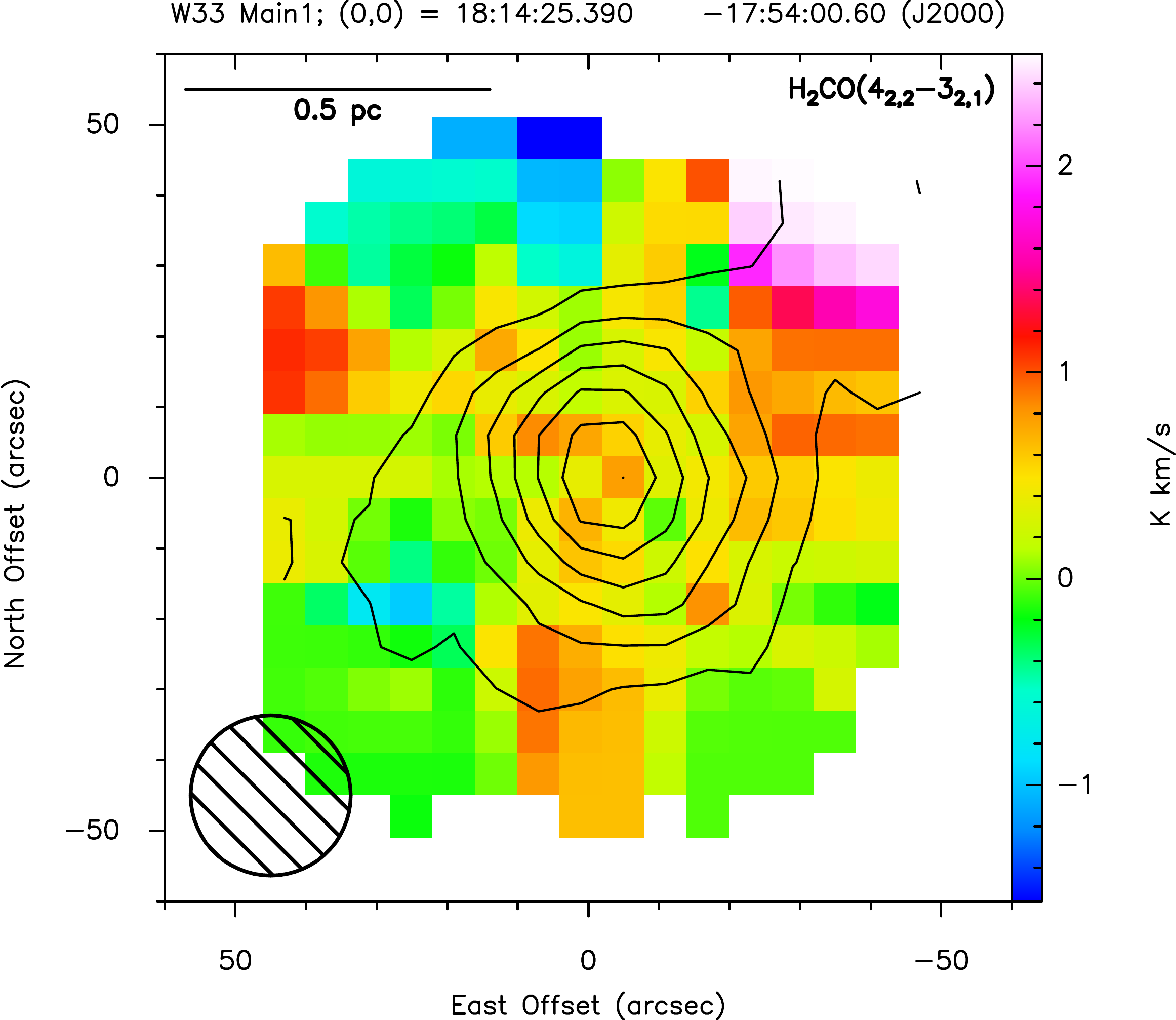}}\\	
	\subfloat{\includegraphics[width=9cm]{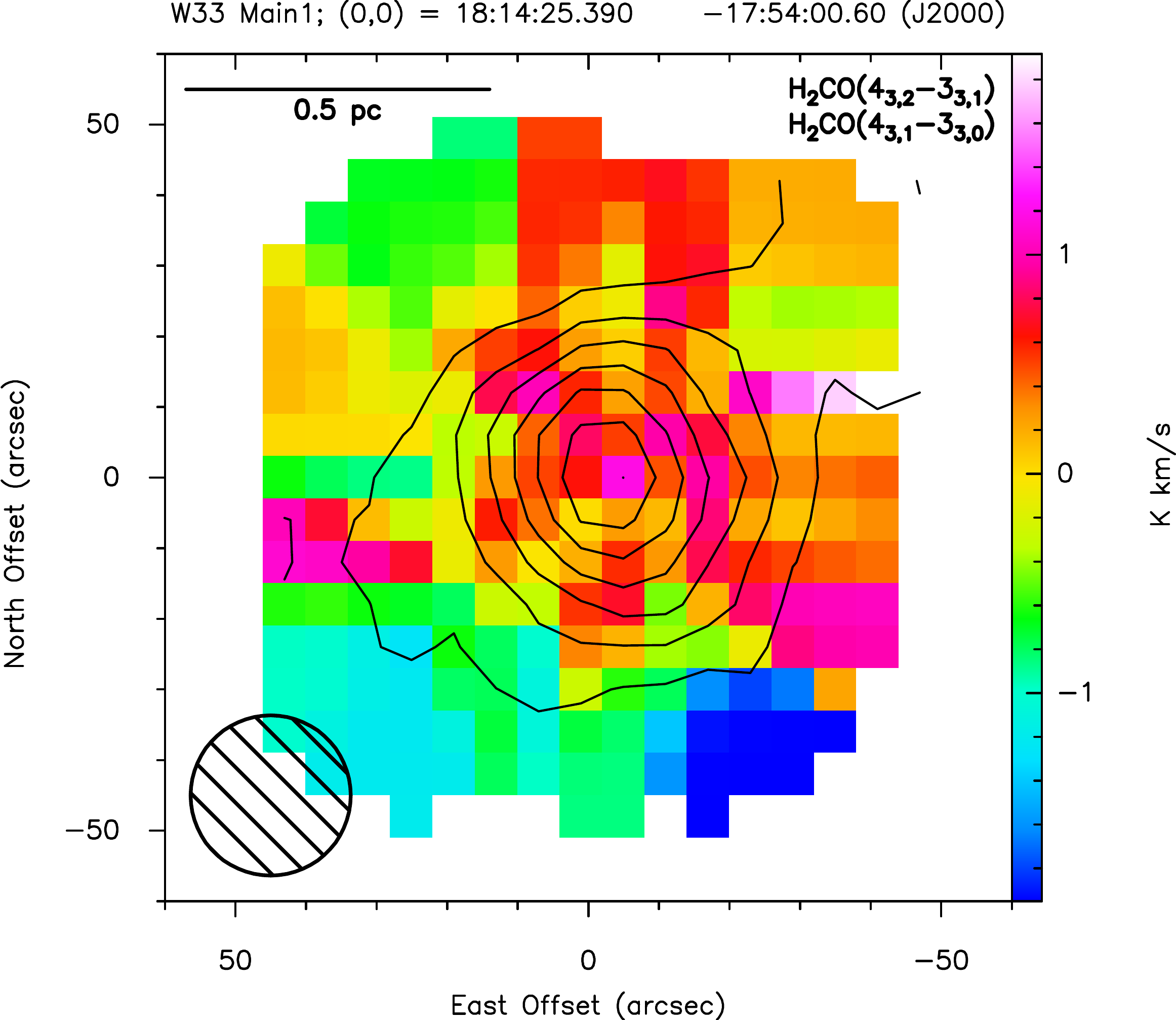}}	
\end{figure*}

\begin{figure*}
	\caption{Line emission of detected transitions in W33\,A1. The contours show the ATLASGAL continuum emission at 345 GHz (levels in steps of 5$\sigma$, starting at 5$\sigma$ ($\sigma$ = 0.081 Jy beam$^{-1}$). The name of the each transition is shown in the upper right corner. A scale of 0.5 pc is marked in the upper left corner, and the synthesised beam is shown in the lower left corner.}
	\centering
	\subfloat{\includegraphics[width=9cm]{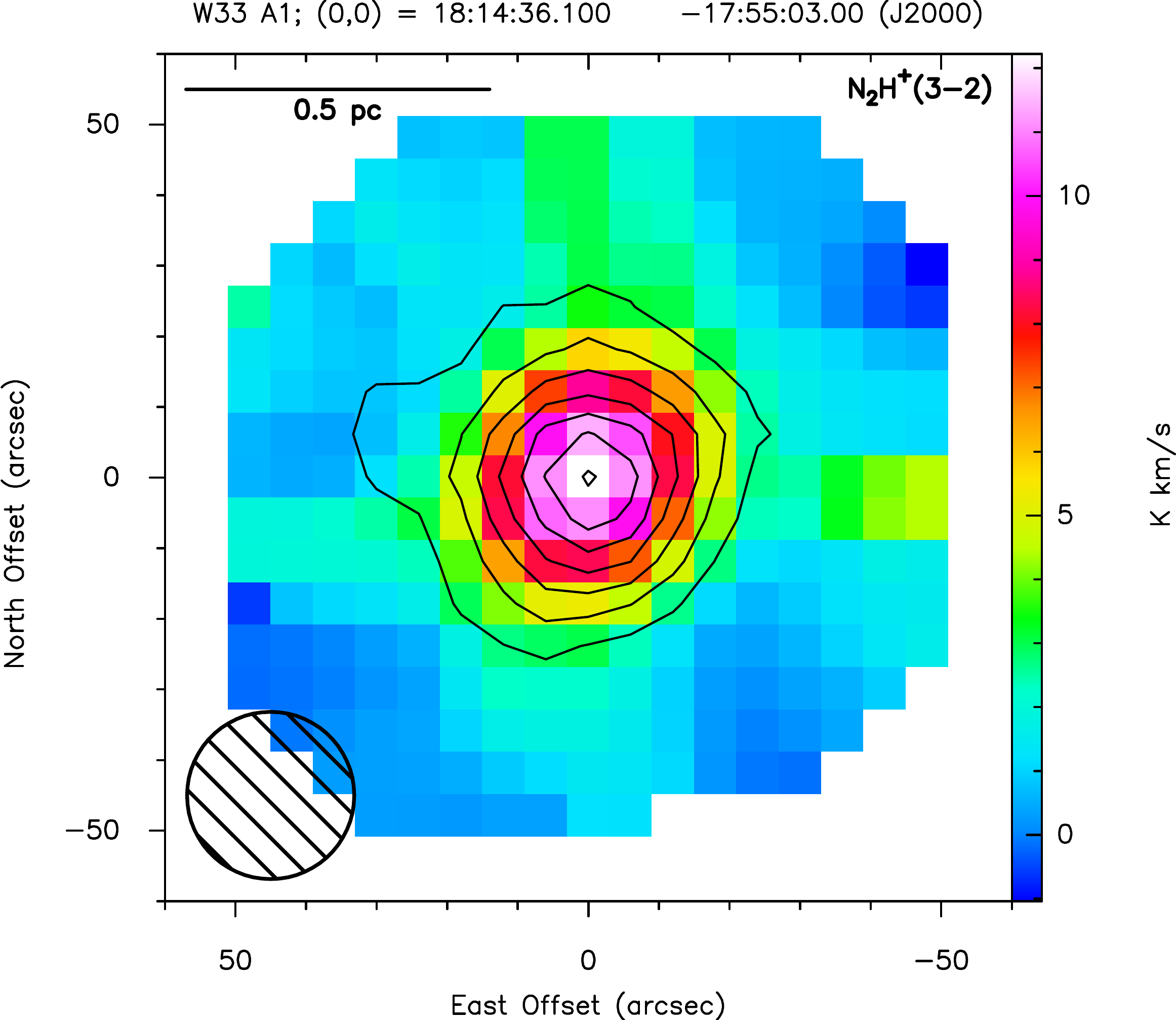}}\hspace{0.2cm}	
	\subfloat{\includegraphics[width=9cm]{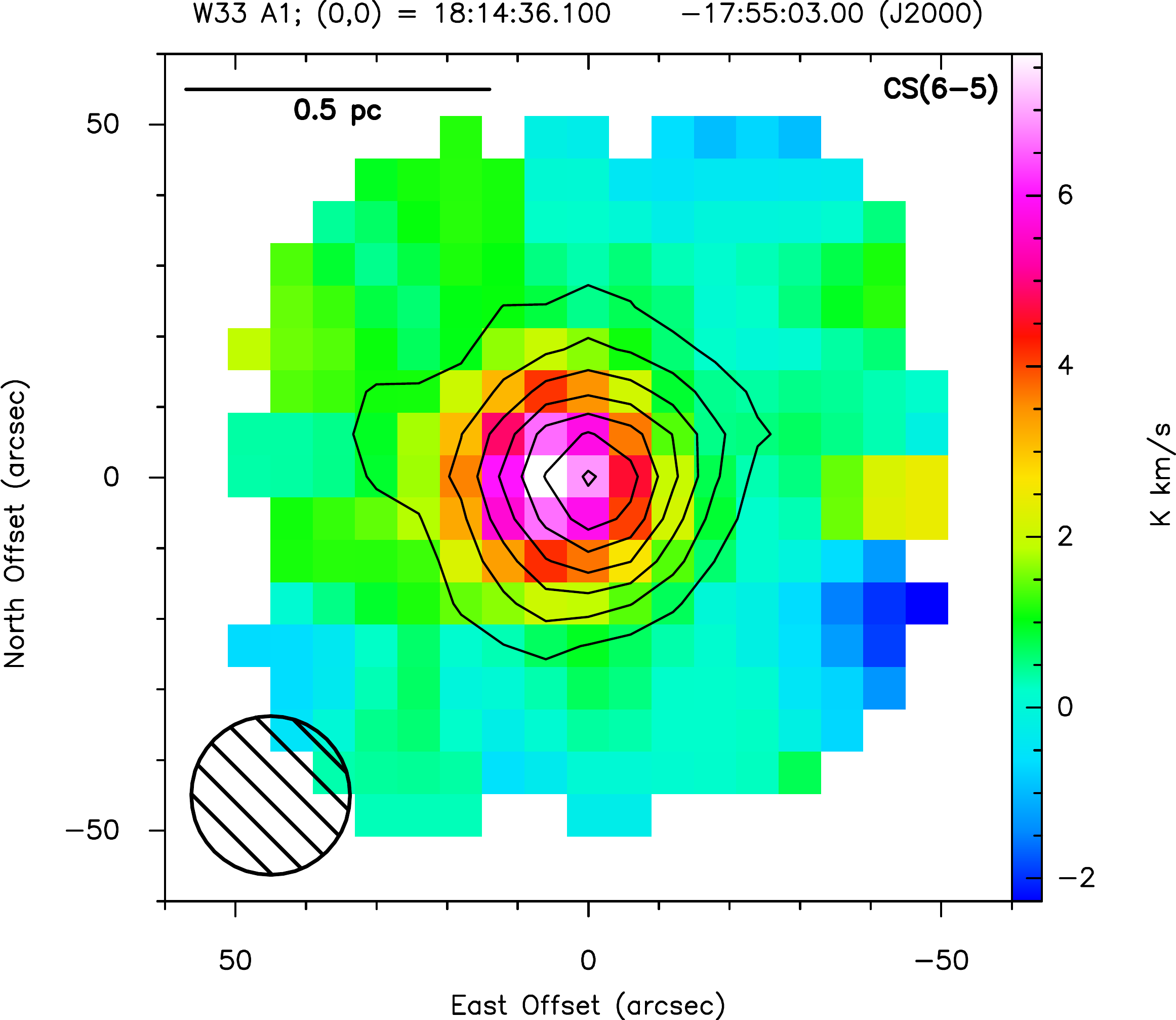}}\\	
	\subfloat{\includegraphics[width=9cm]{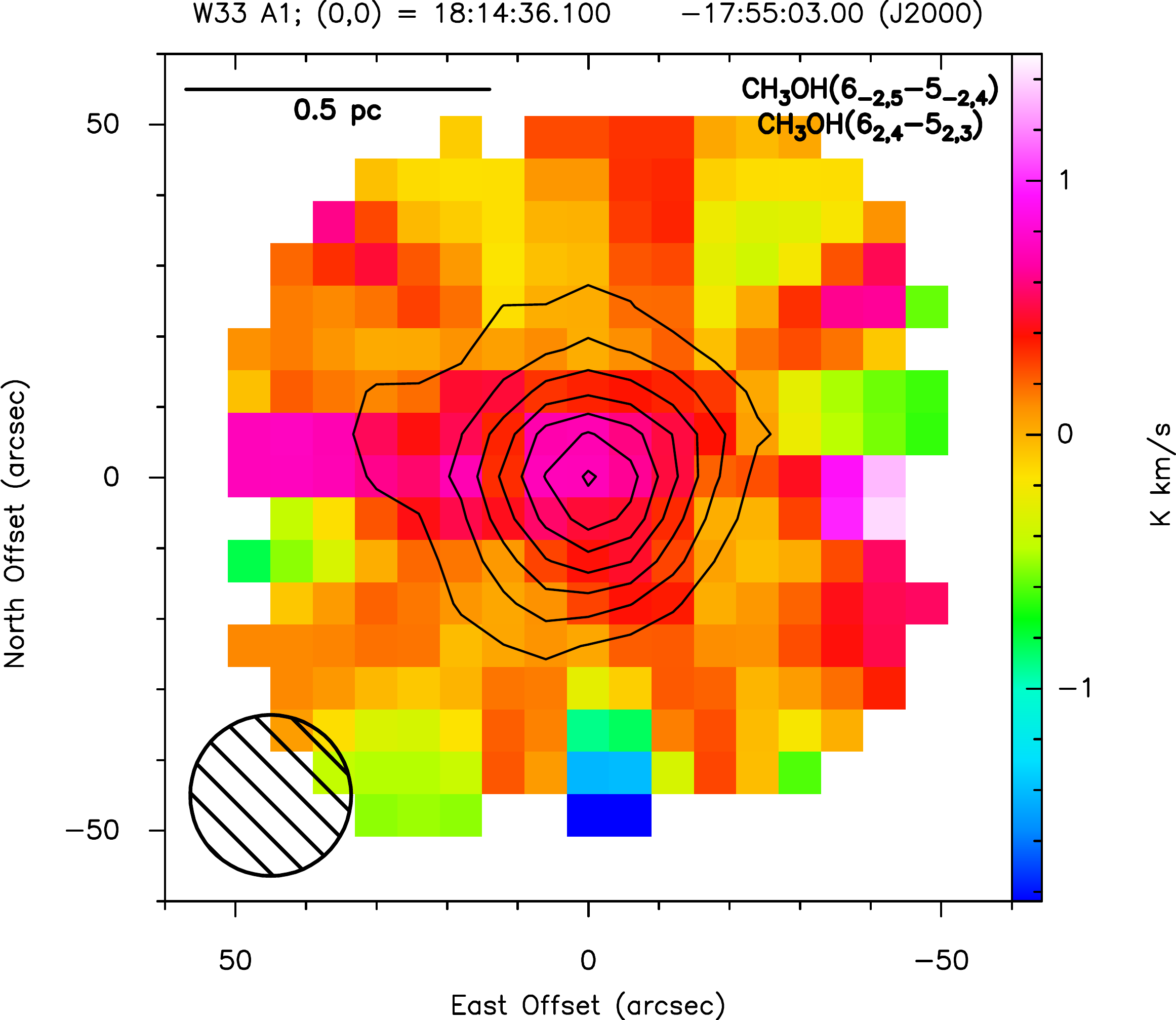}}\hspace{0.2cm}	
	\subfloat{\includegraphics[width=9cm]{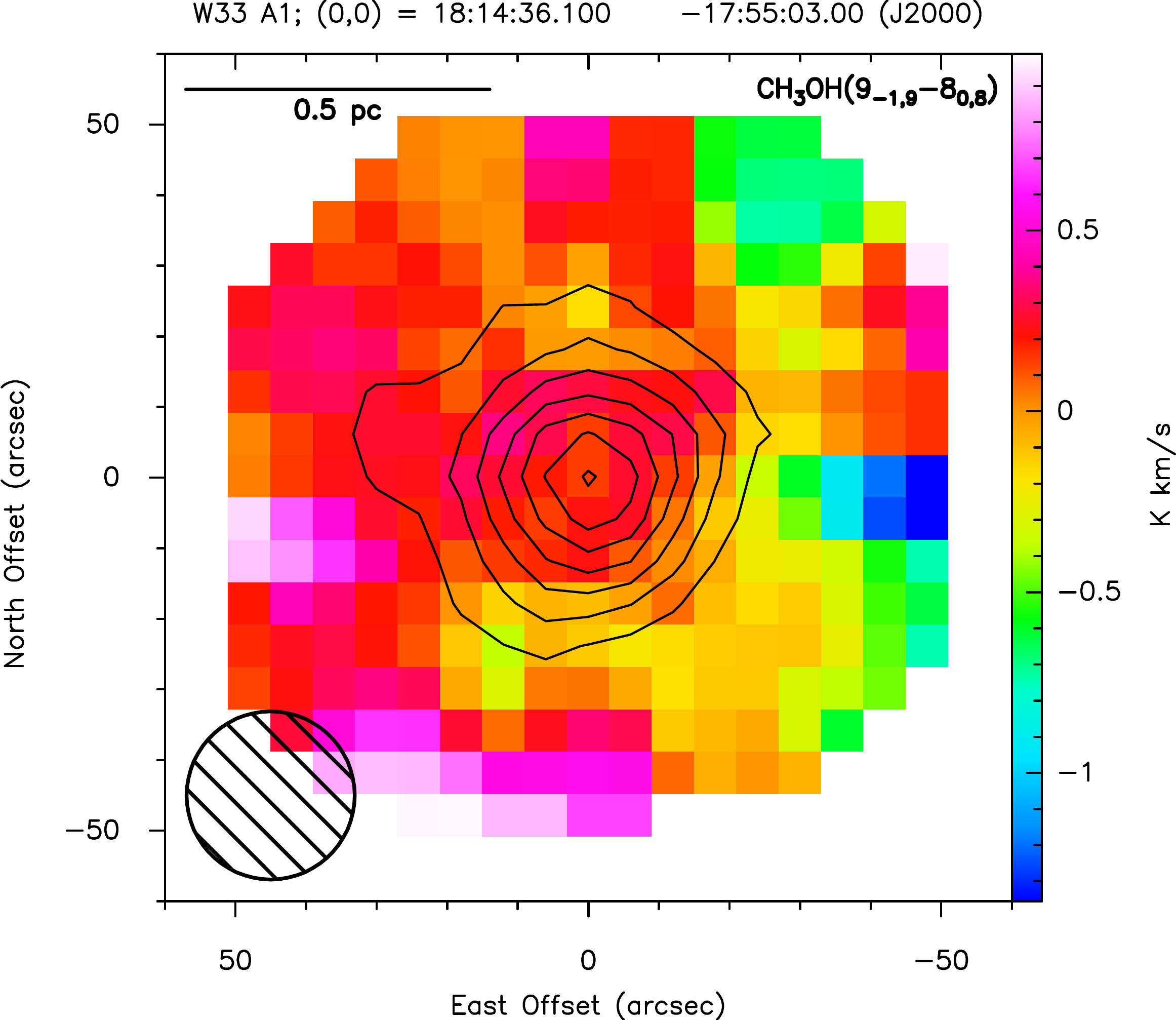}}	\\
	\subfloat{\includegraphics[width=9cm]{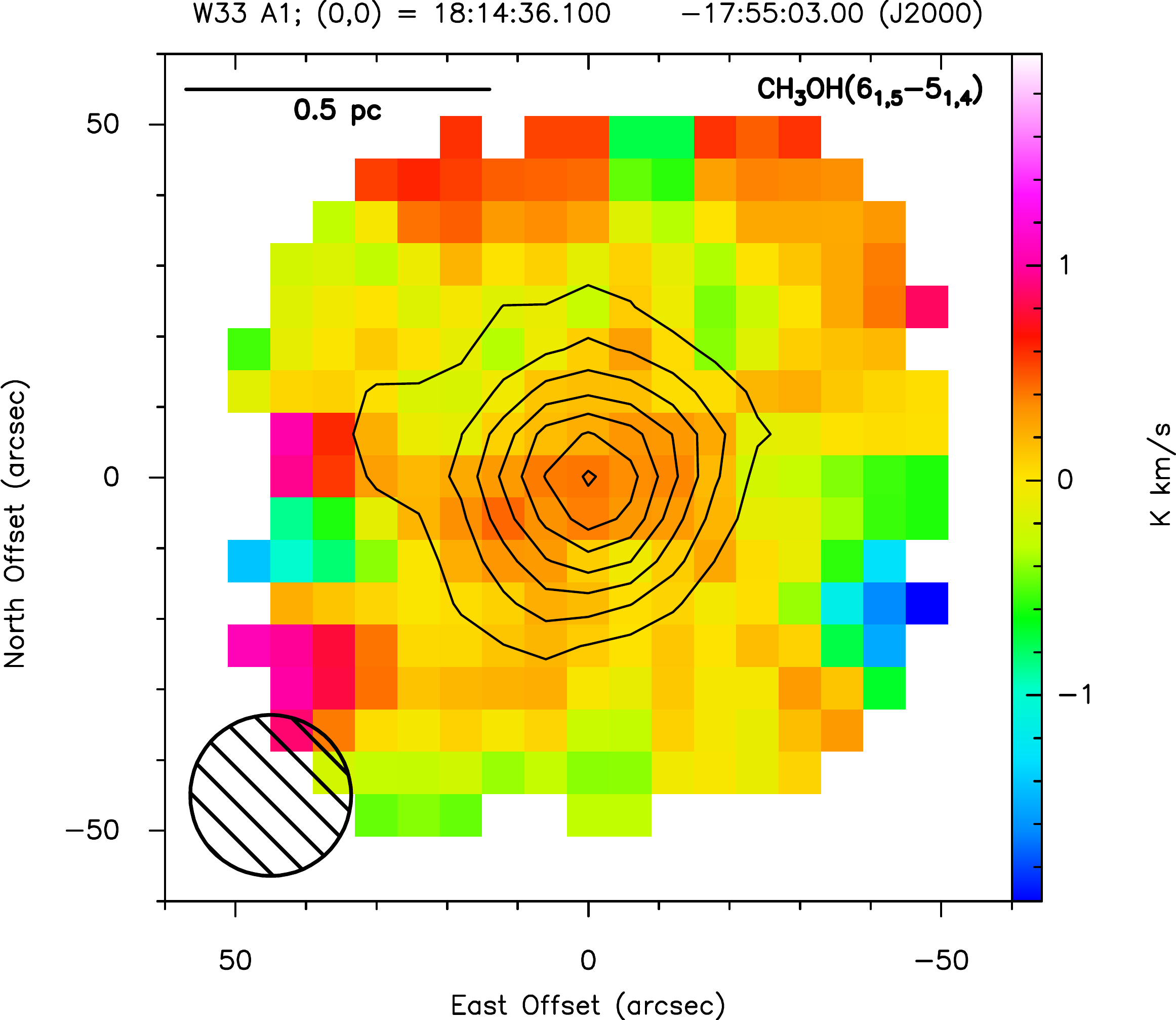}}\hspace{0.2cm}	
	\subfloat{\includegraphics[width=9cm]{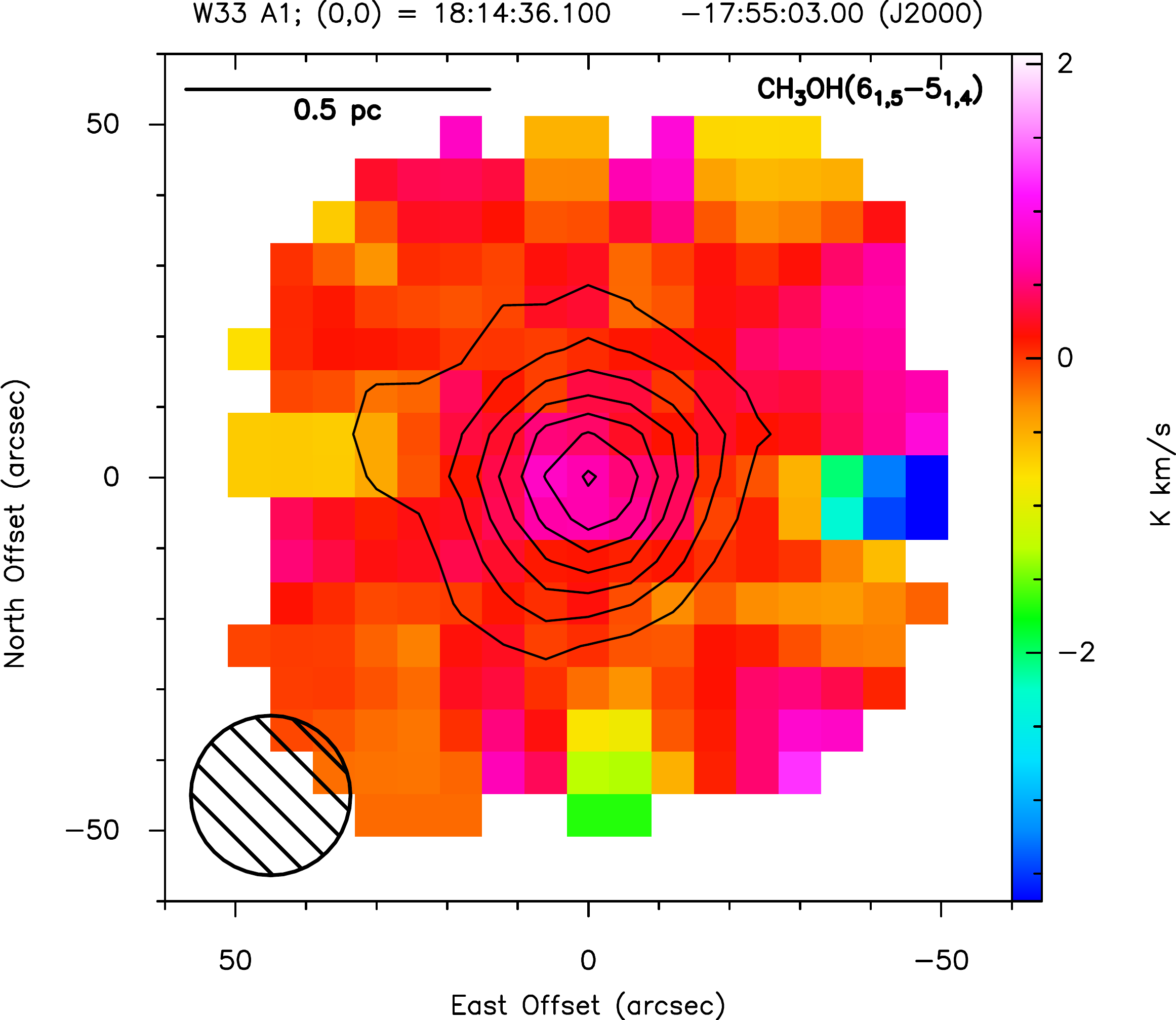}}		
	\label{W33A1-APEX-IntInt}
\end{figure*}

\addtocounter{figure}{-1}
\begin{figure*}
	\centering
	\caption{Continued.}
	\subfloat{\includegraphics[width=9cm]{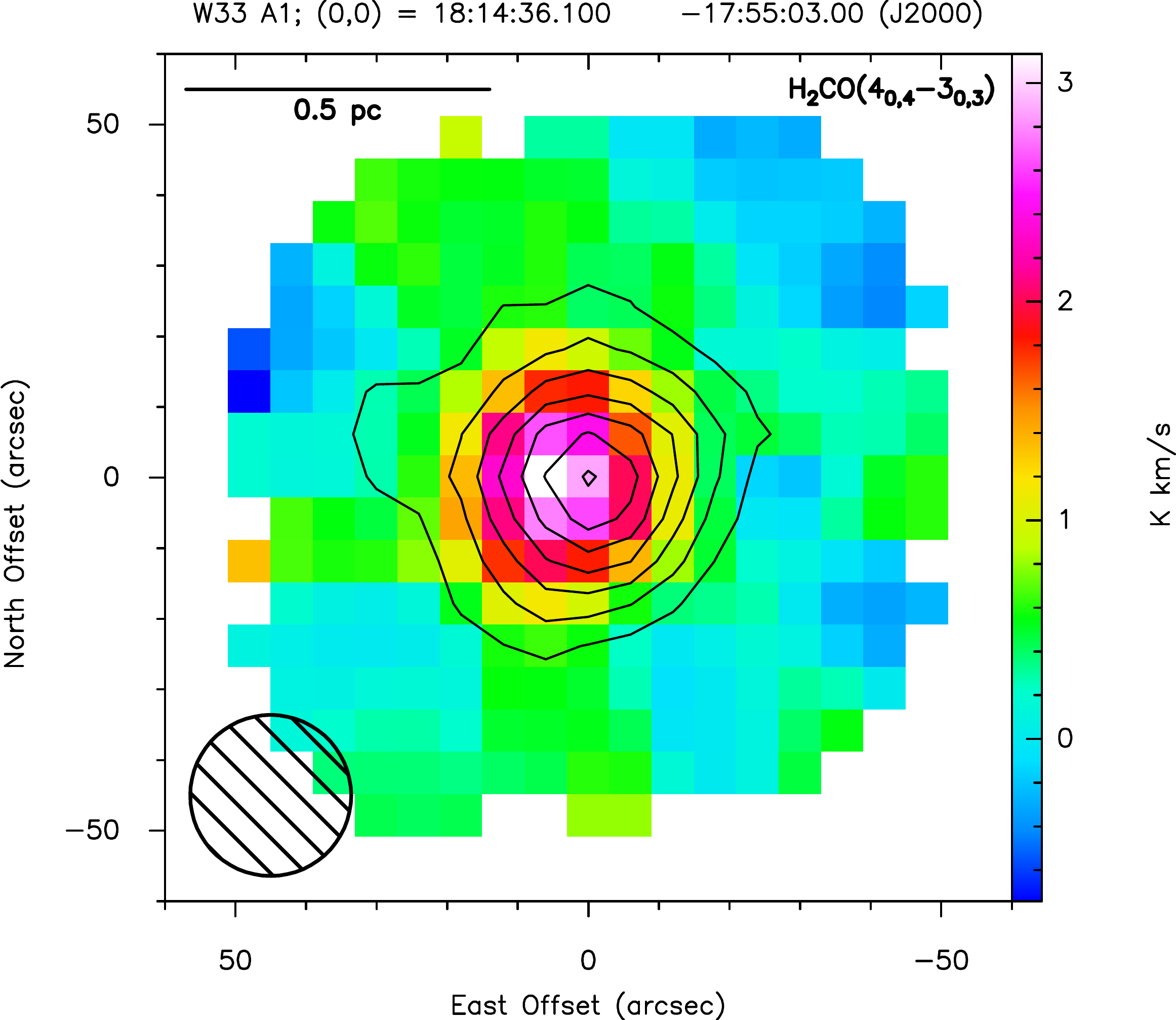}}\hspace{0.2cm}	
	\subfloat{\includegraphics[width=9cm]{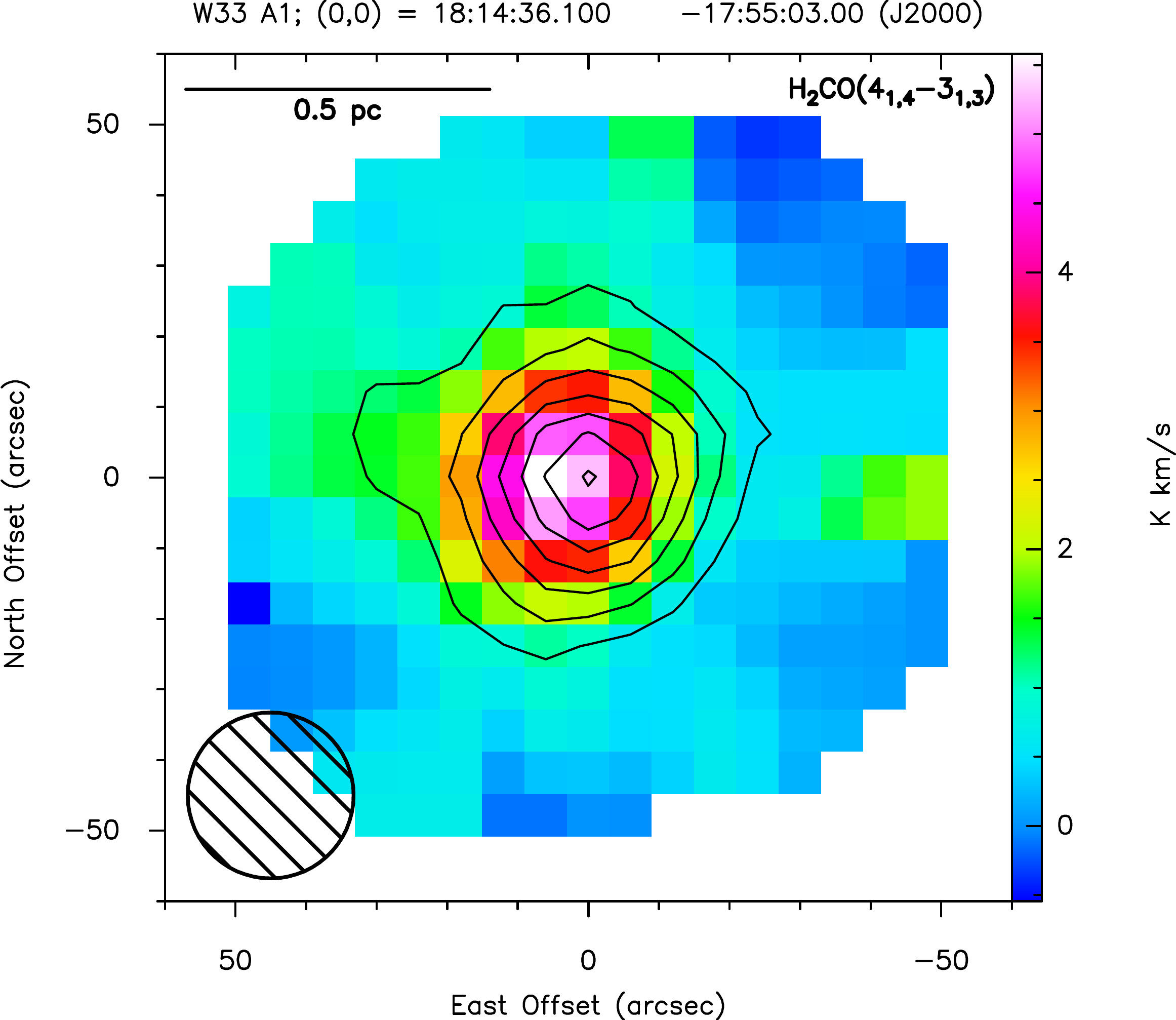}}\\	
	\subfloat{\includegraphics[width=9cm]{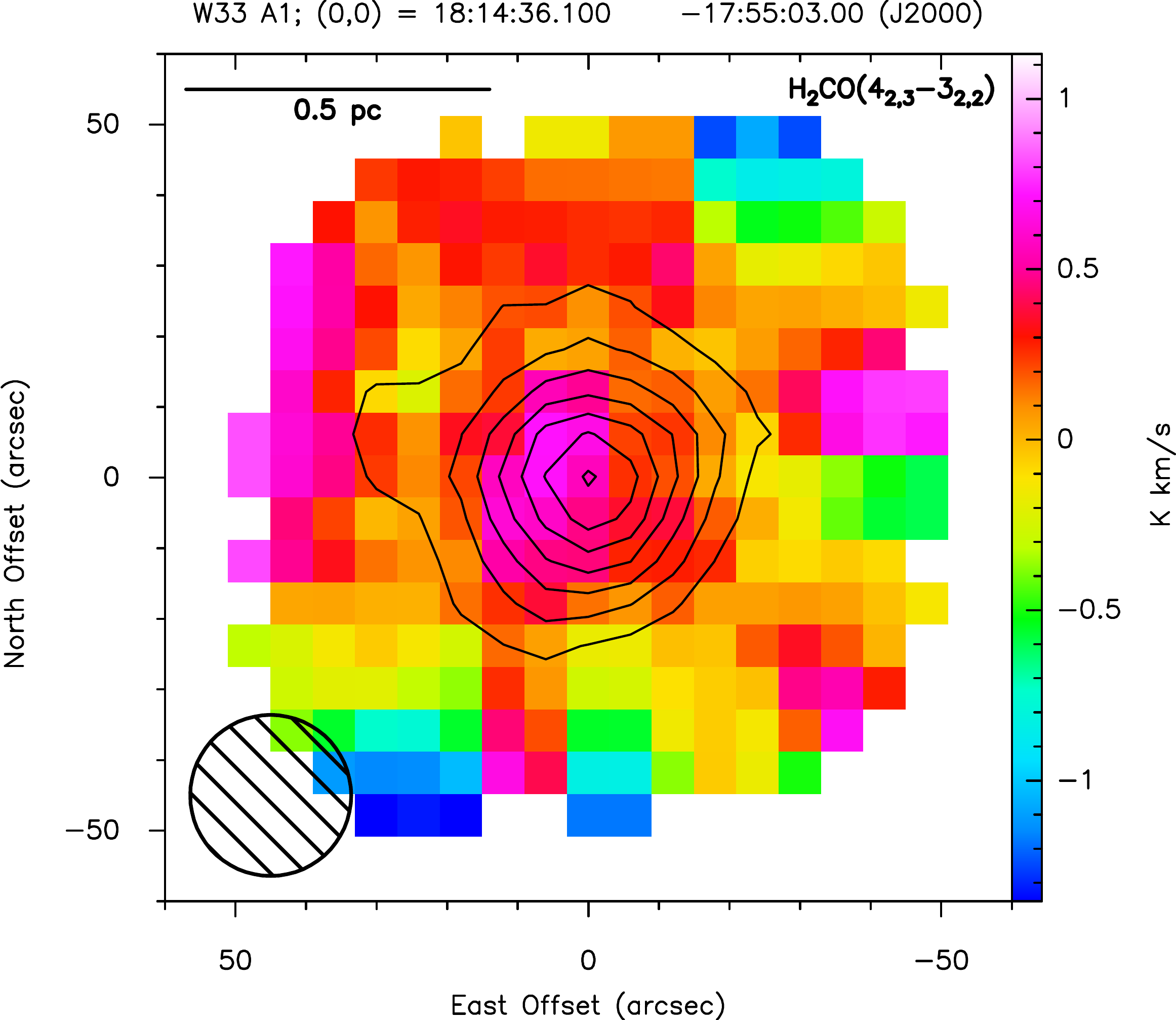}}\hspace{0.2cm}	
	\subfloat{\includegraphics[width=9cm]{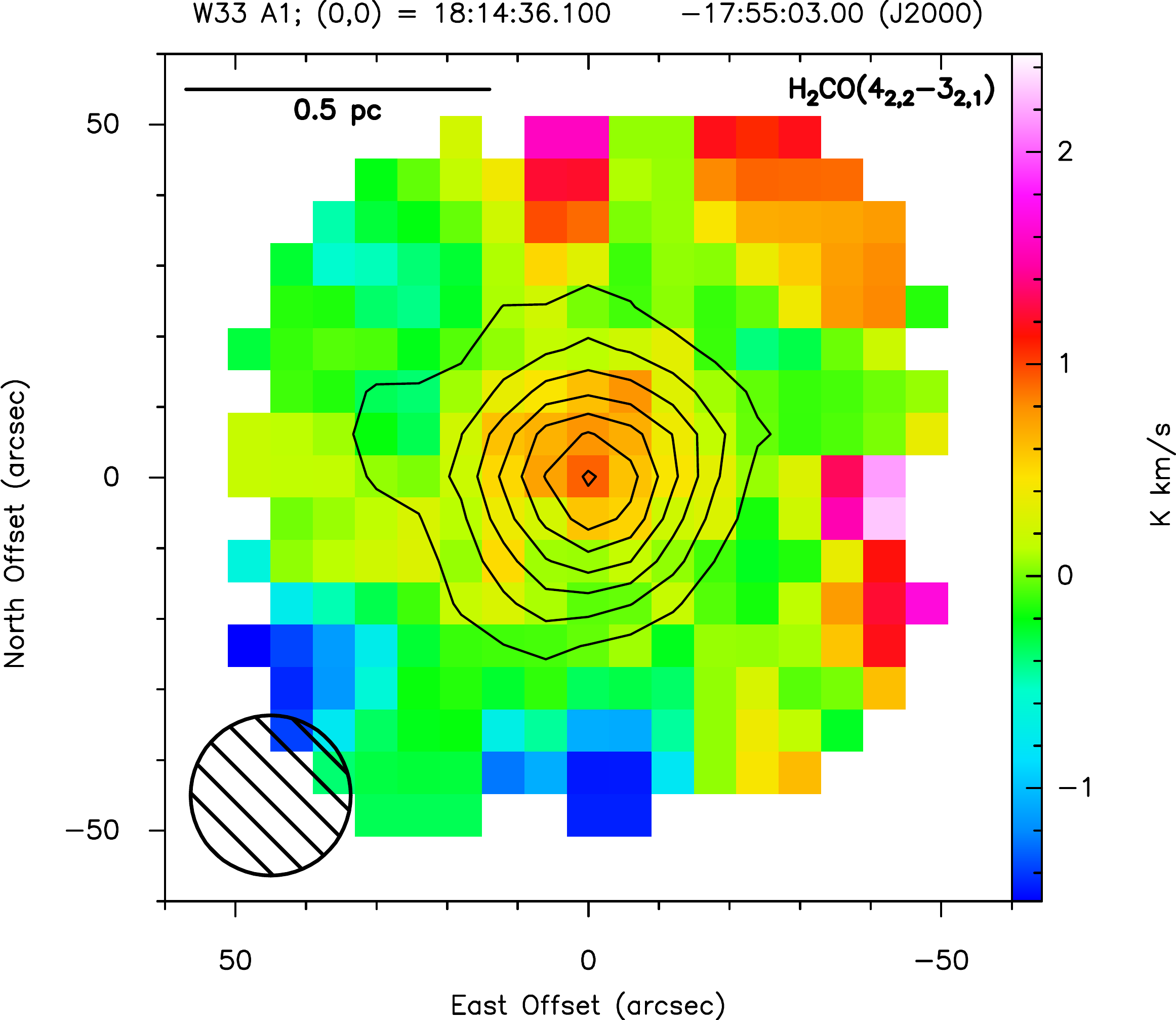}}\\	
	\subfloat{\includegraphics[width=9cm]{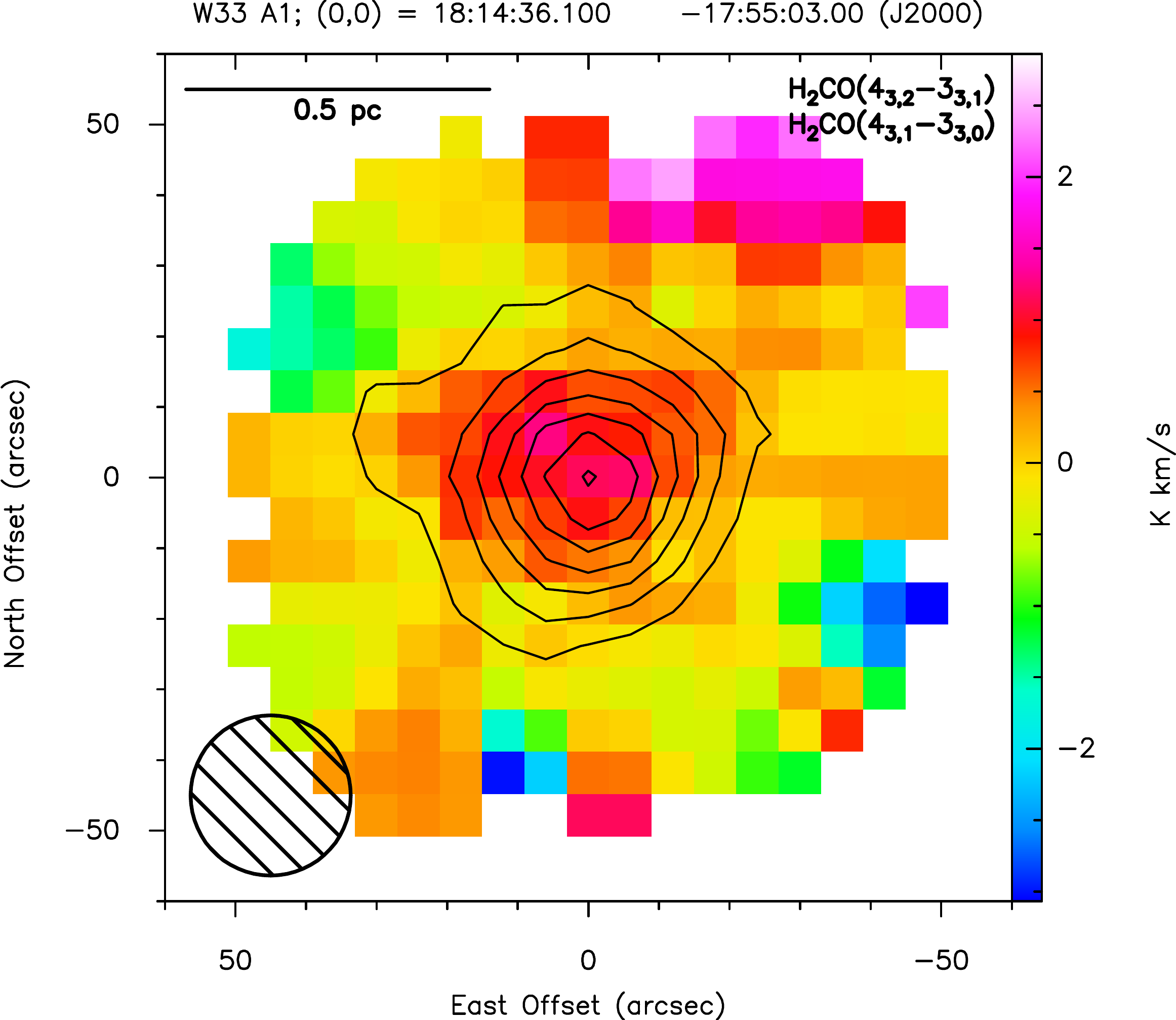}}		
\end{figure*}

\begin{figure*}
	\caption{Line emission of detected transitions in W33\,B1. The contours show the ATLASGAL continuum emission at 345 GHz (levels in steps of 5$\sigma$, starting at 5$\sigma$ ($\sigma$ = 0.081 Jy beam$^{-1}$). The name of the each transition is shown in the upper right corner. A scale of 0.5 pc is marked in the upper left corner, and the synthesised beam is shown in the lower left corner.}
	\centering
	\subfloat{\includegraphics[width=9cm]{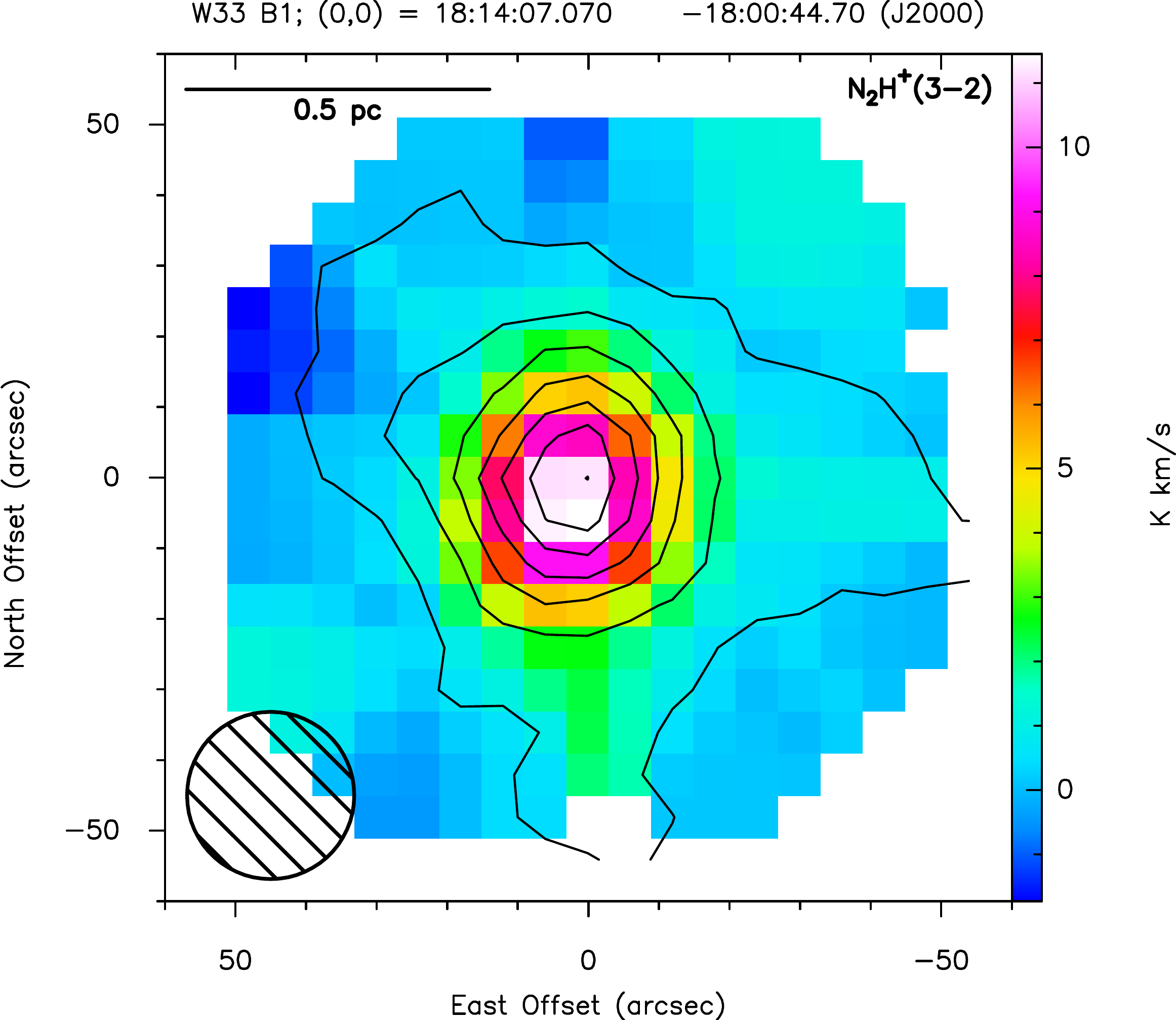}}\hspace{0.2cm}	
	\subfloat{\includegraphics[width=9cm]{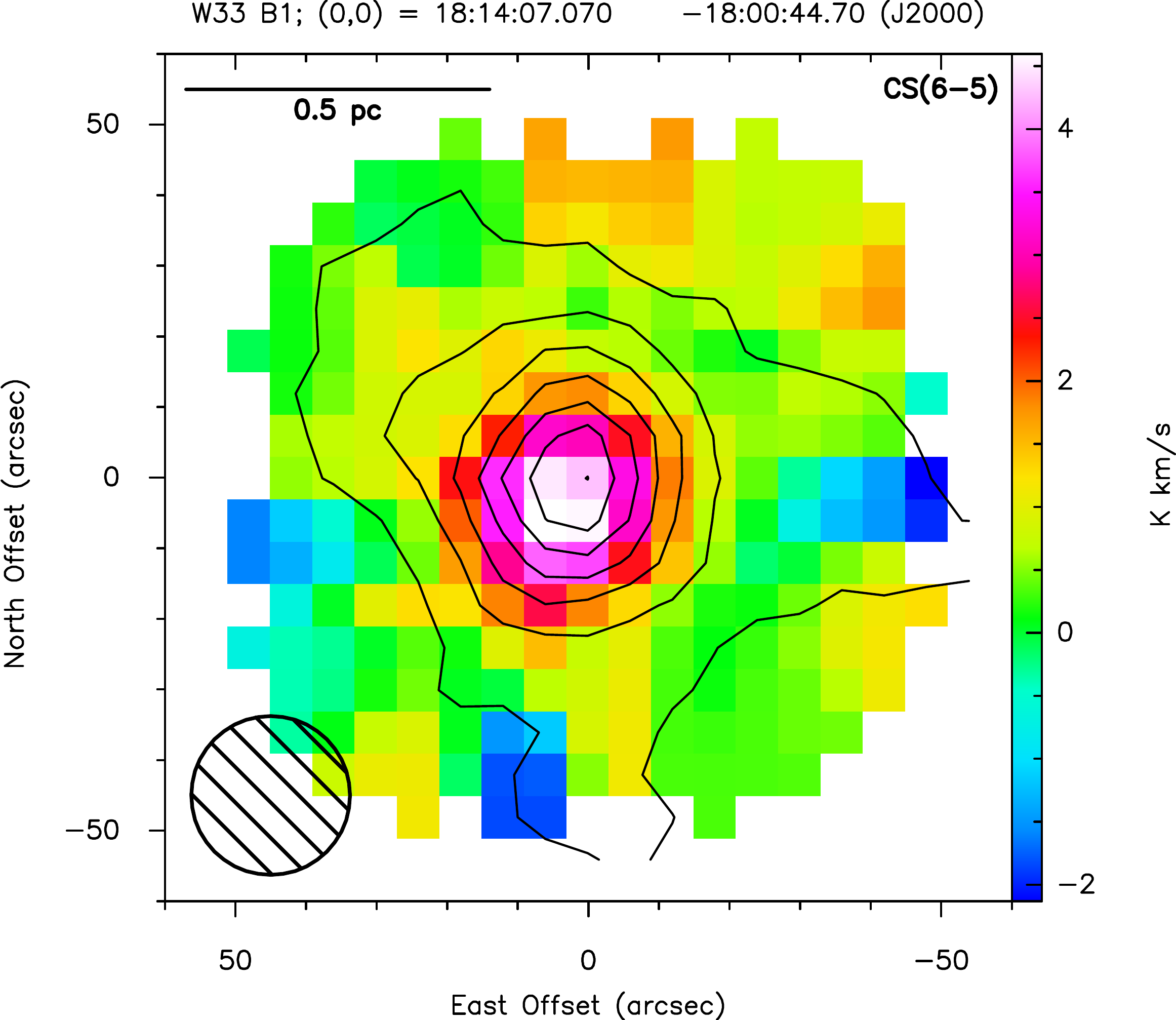}}\\	
	\subfloat{\includegraphics[width=9cm]{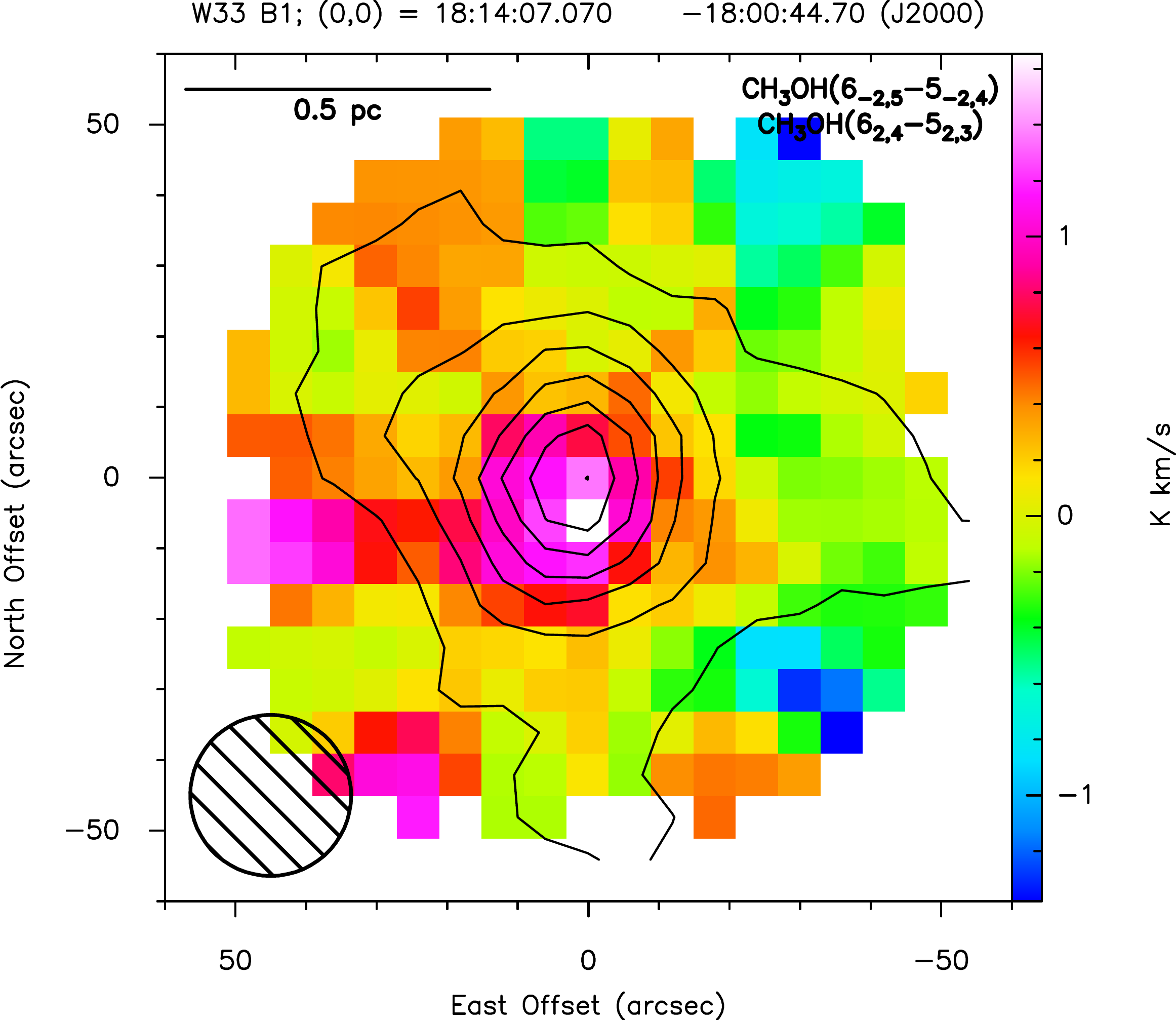}}\hspace{0.2cm}	
	\subfloat{\includegraphics[width=9cm]{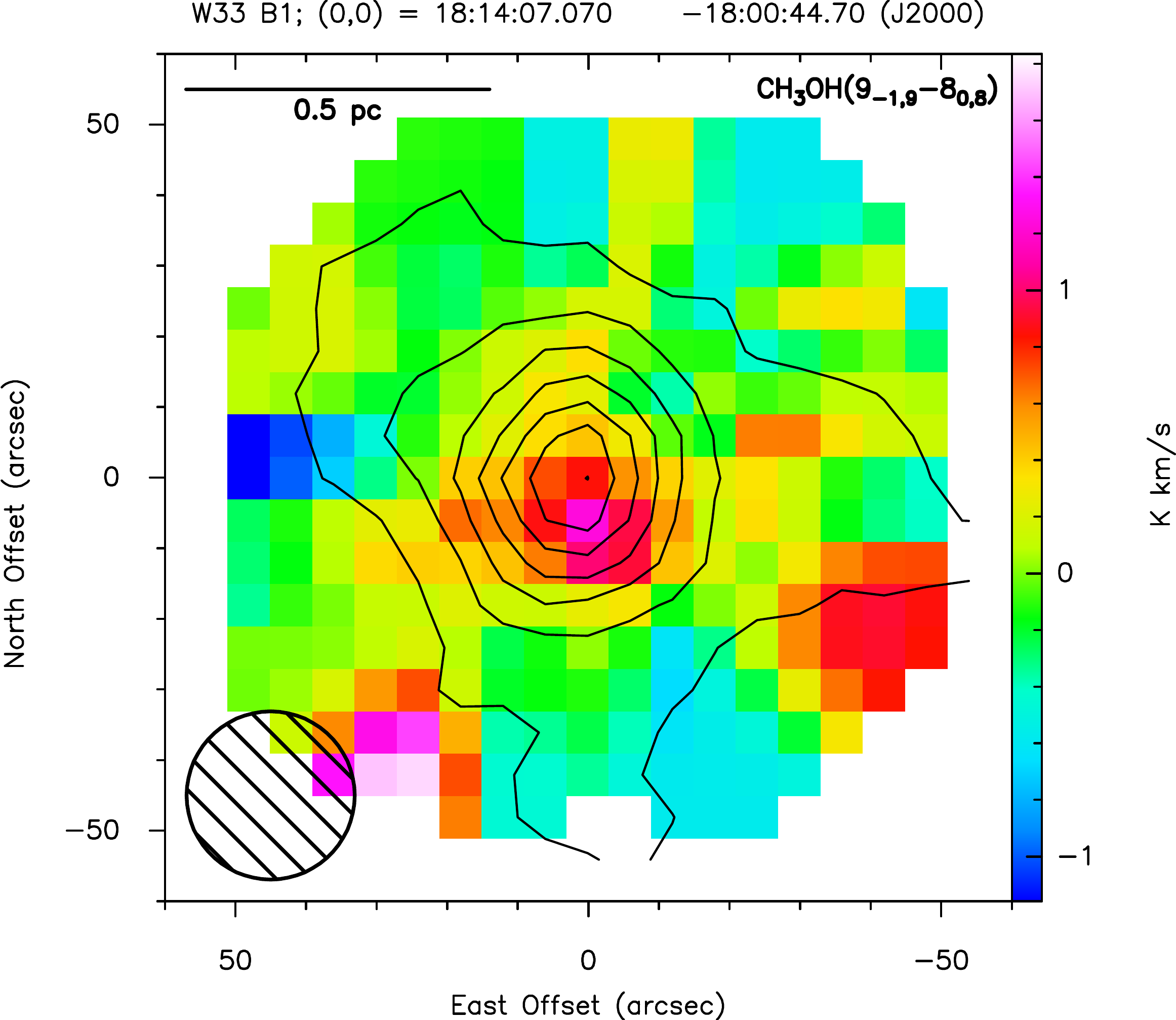}}	\\
	\subfloat{\includegraphics[width=9cm]{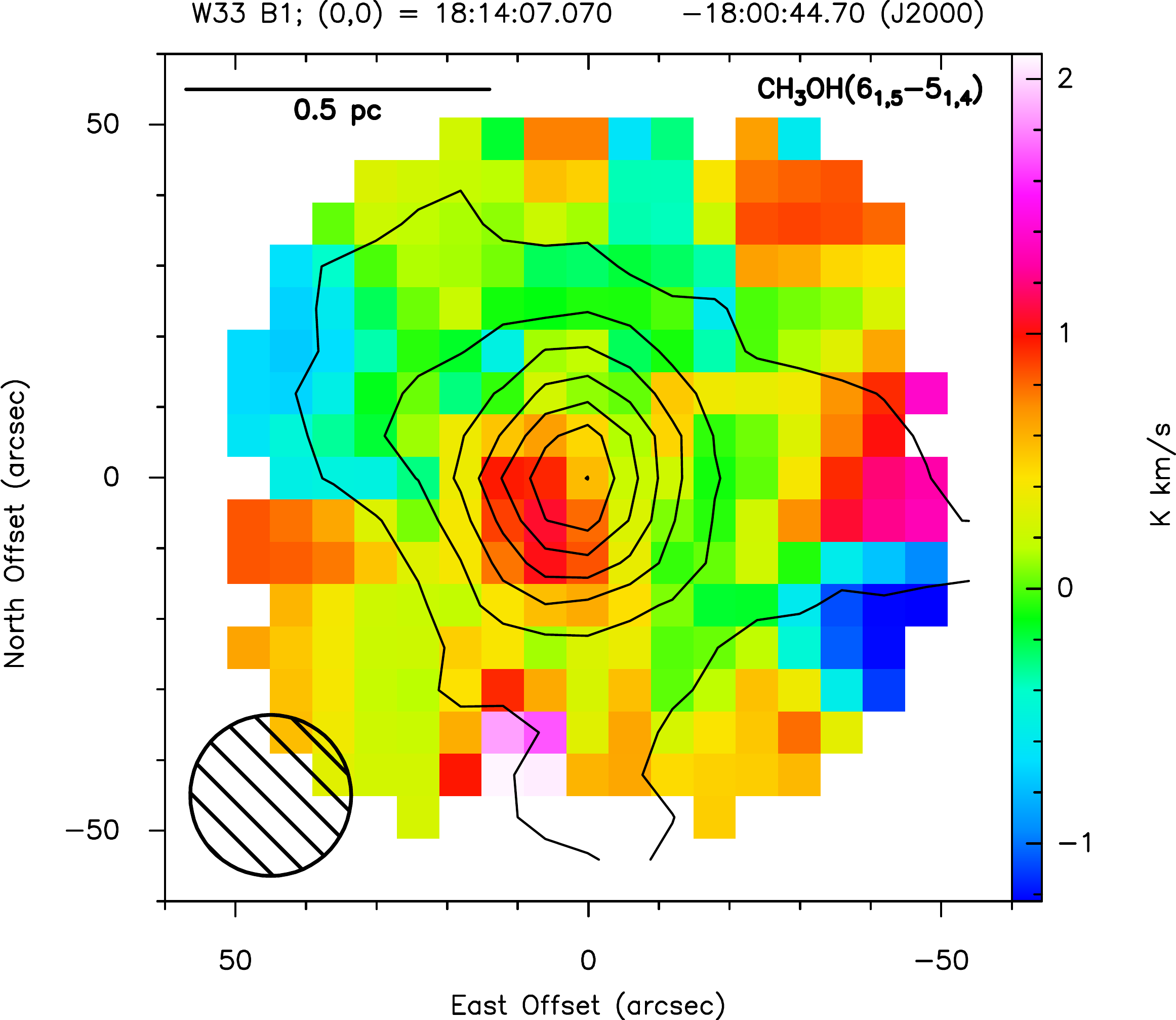}}\hspace{0.2cm}	
	\subfloat{\includegraphics[width=9cm]{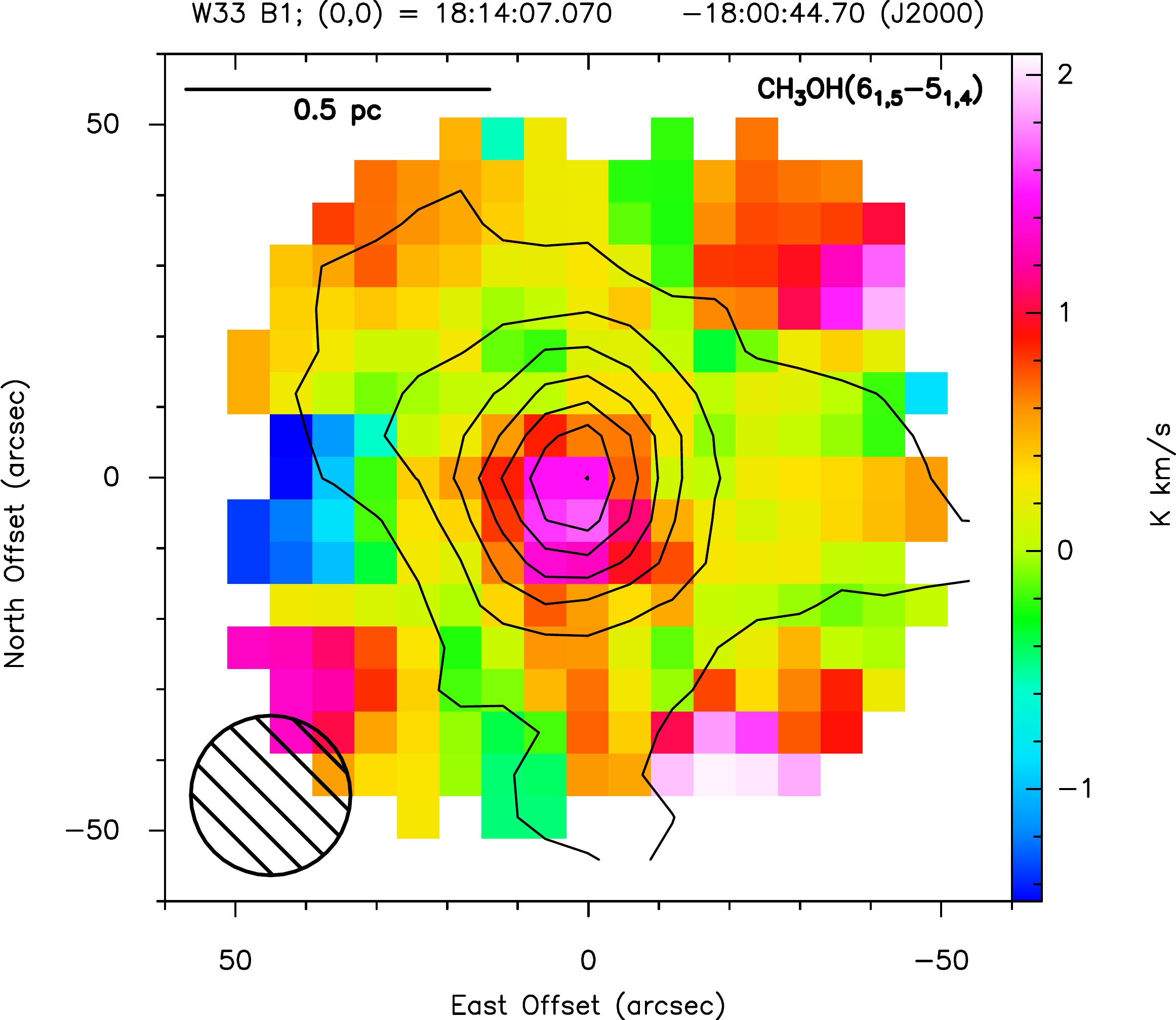}}		
	\label{W33B1-APEX-IntInt}
\end{figure*}

\addtocounter{figure}{-1}
\begin{figure*}
	\centering
	\caption{Continued.}
	\subfloat{\includegraphics[width=9cm]{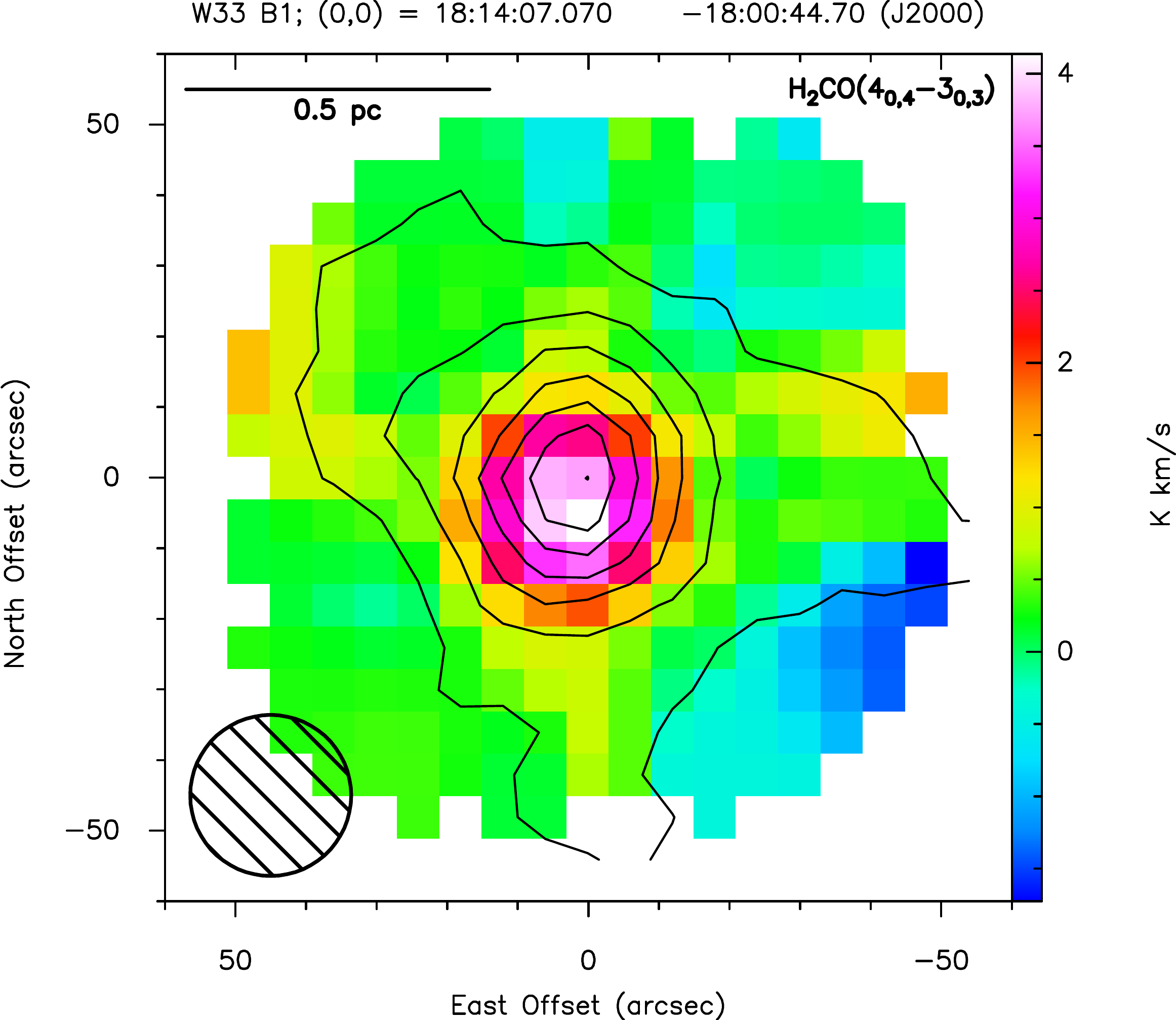}}\hspace{0.2cm}	
	\subfloat{\includegraphics[width=9cm]{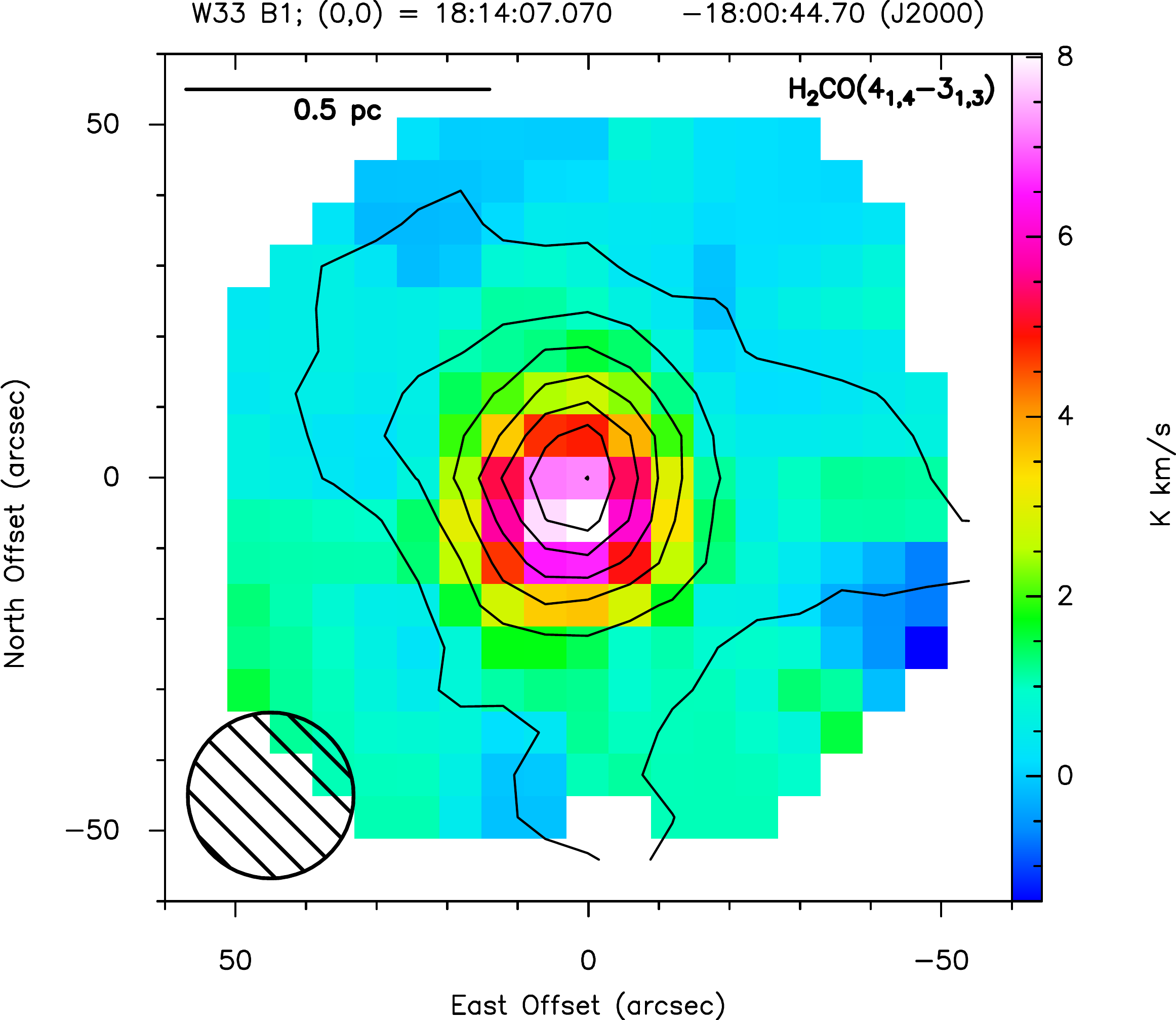}}\\	
	\subfloat{\includegraphics[width=9cm]{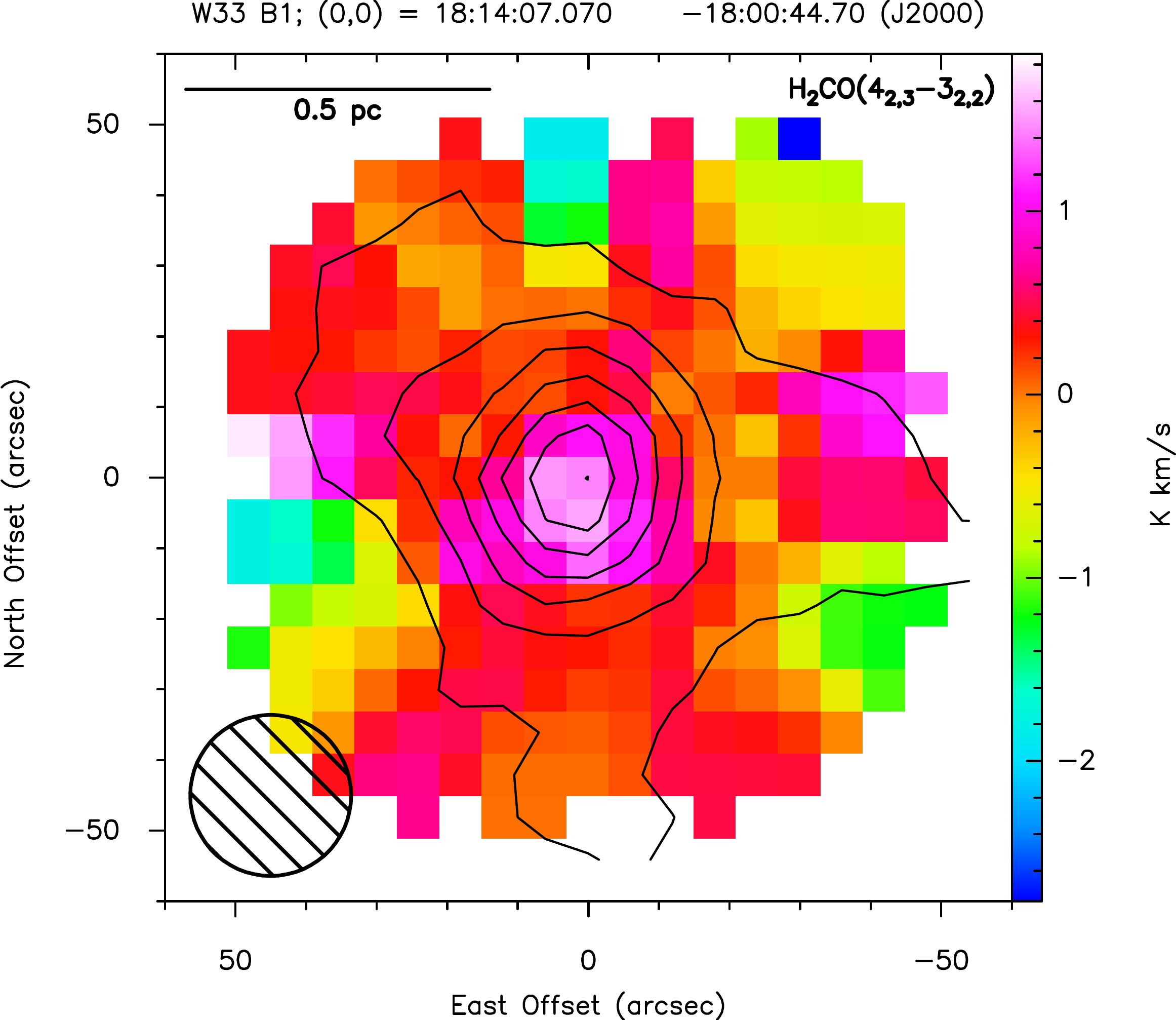}}\hspace{0.2cm}	
	\subfloat{\includegraphics[width=9cm]{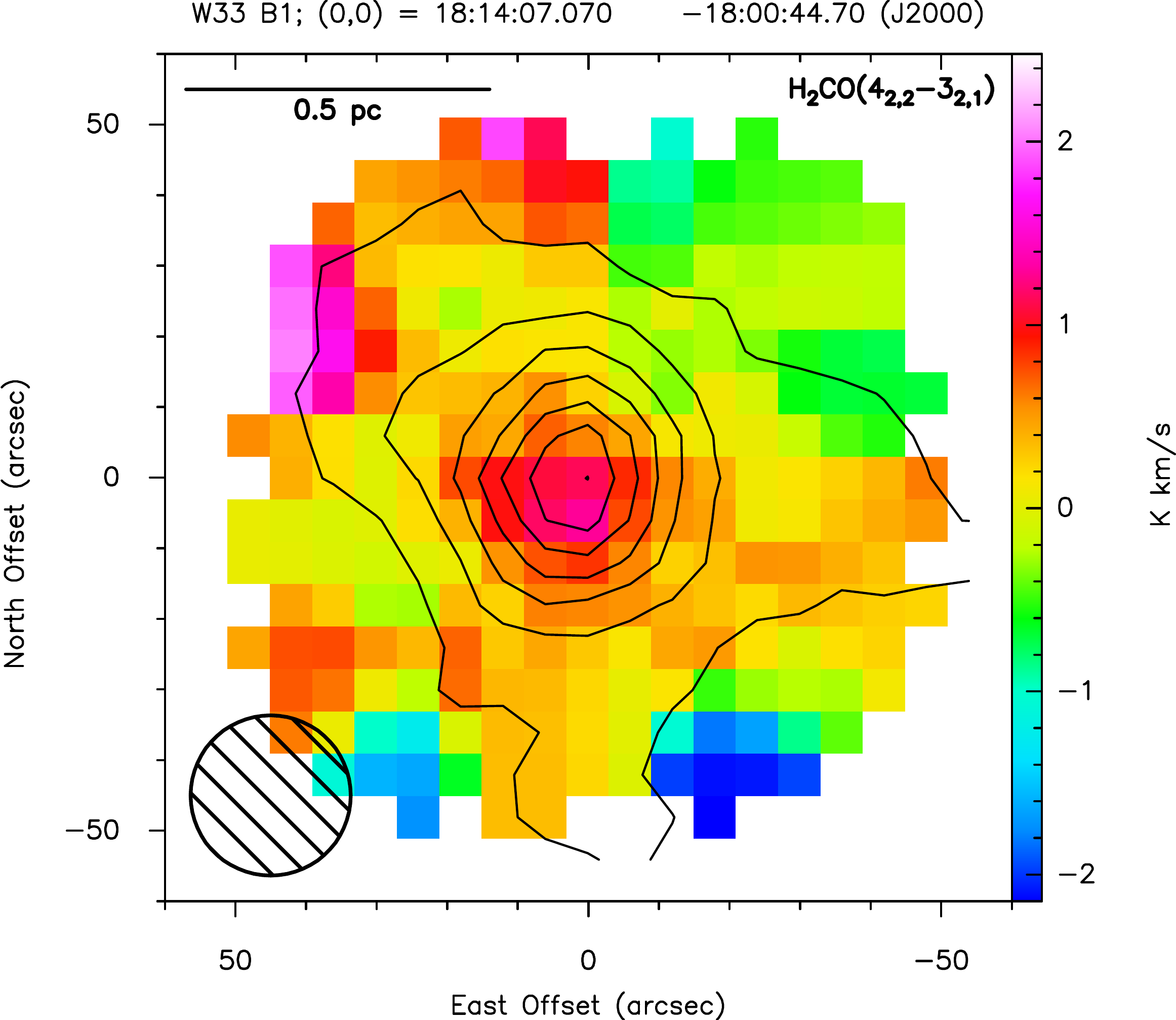}}\\
	\subfloat{\includegraphics[width=9cm]{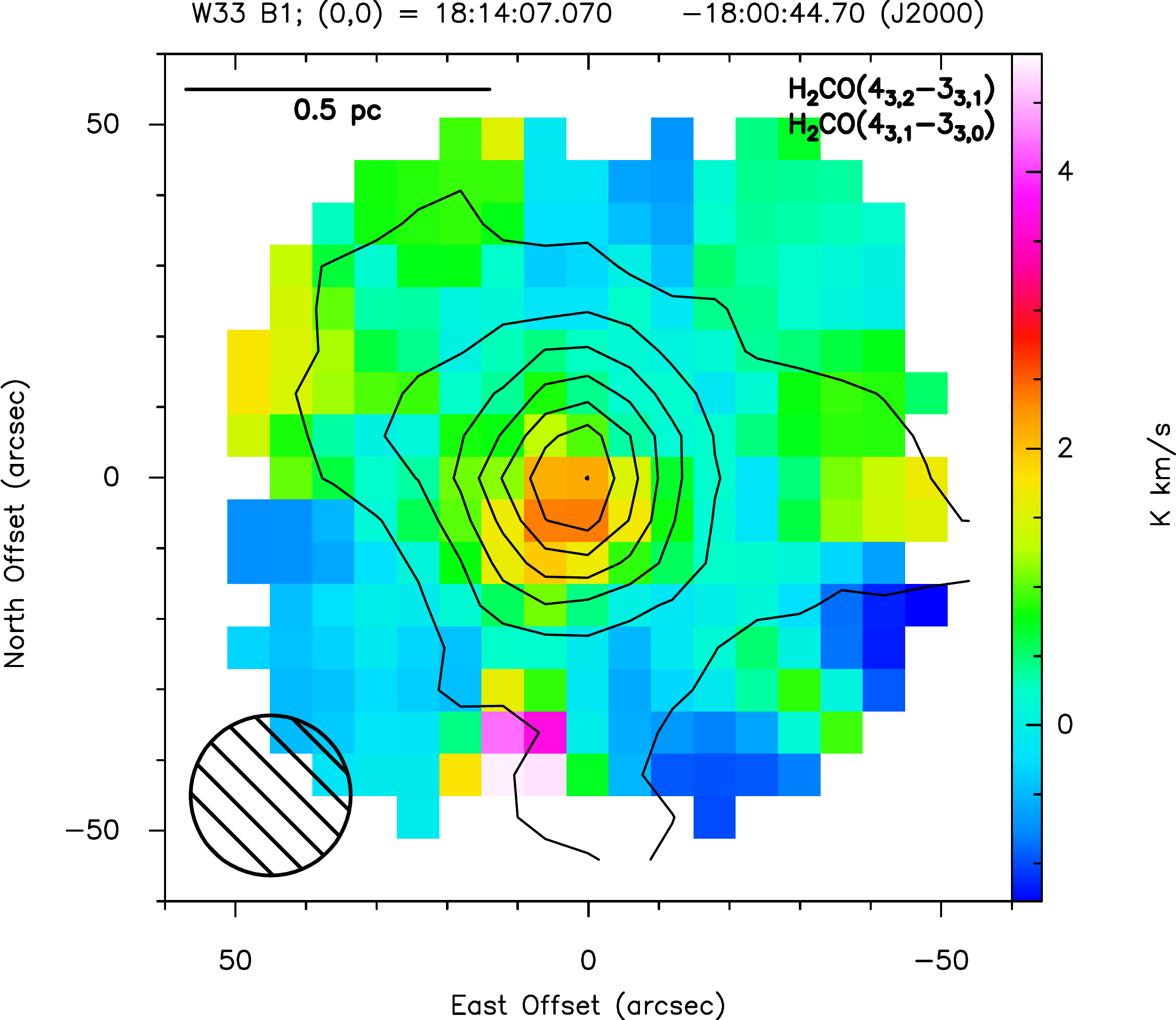}}		
\end{figure*}

\begin{figure*}
	\caption{Line emission of detected transitions in W33\,B. The contours show the ATLASGAL continuum emission at 345 GHz (levels in steps of 5$\sigma$, starting at 10$\sigma$ ($\sigma$ = 0.081 Jy beam$^{-1}$). The name of the each transition is shown in the upper right corner. A scale of 0.5 pc is marked in the upper left corner, and the synthesised beam is shown in the lower left corner.}
	\centering
	\subfloat{\includegraphics[width=9cm]{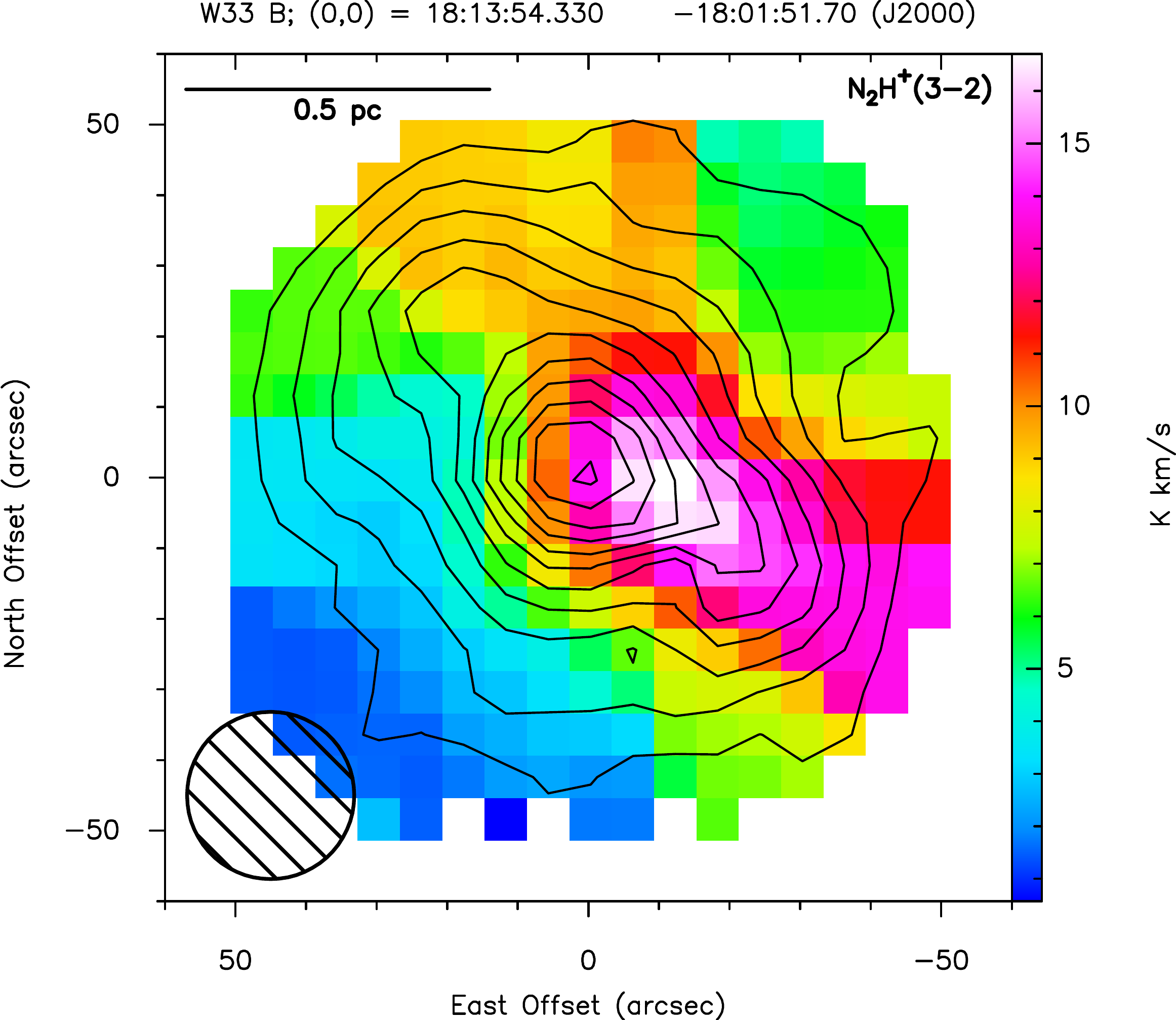}}\hspace{0.2cm}	
	\subfloat{\includegraphics[width=9cm]{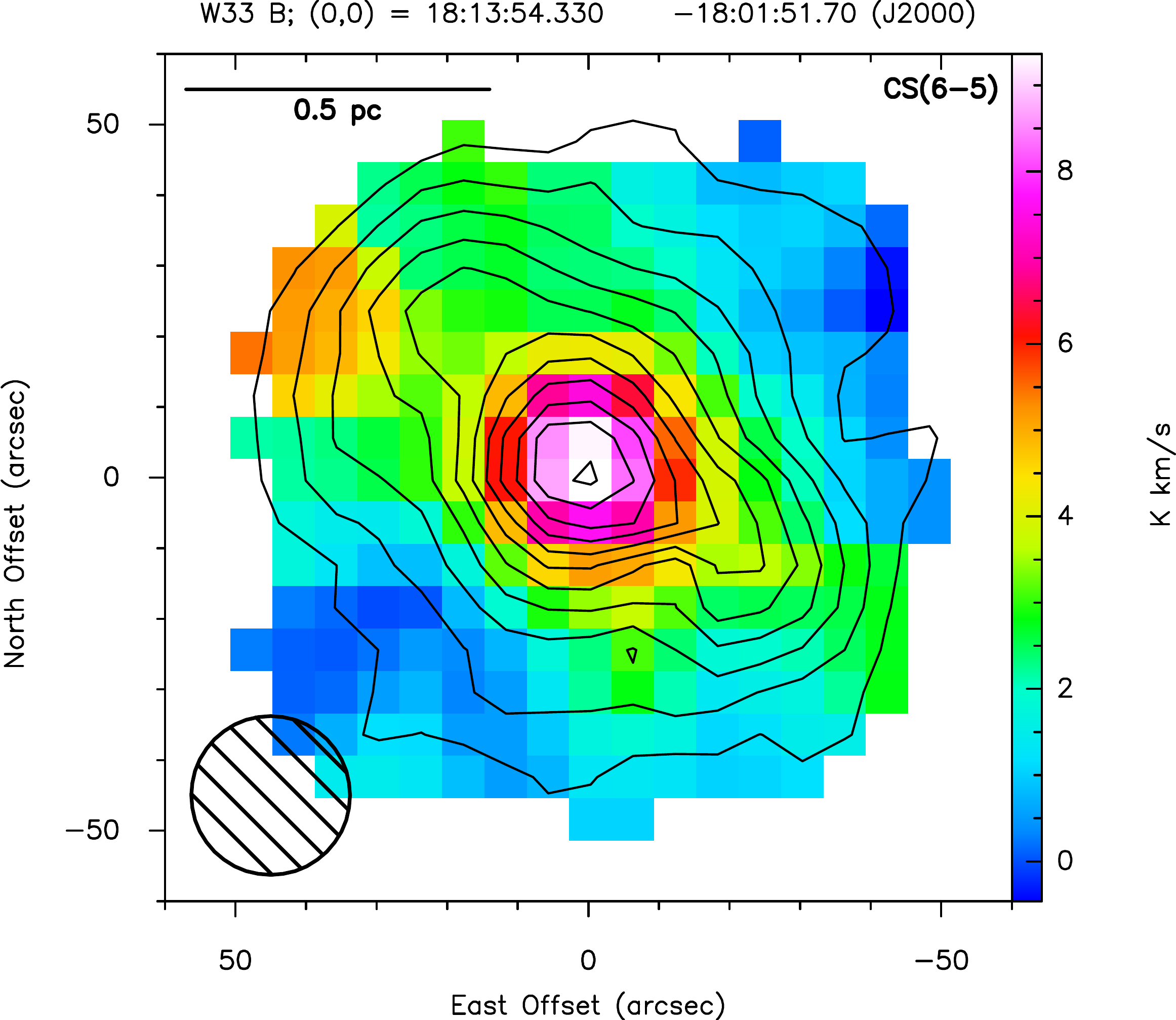}}\\		
	\subfloat{\includegraphics[width=9cm]{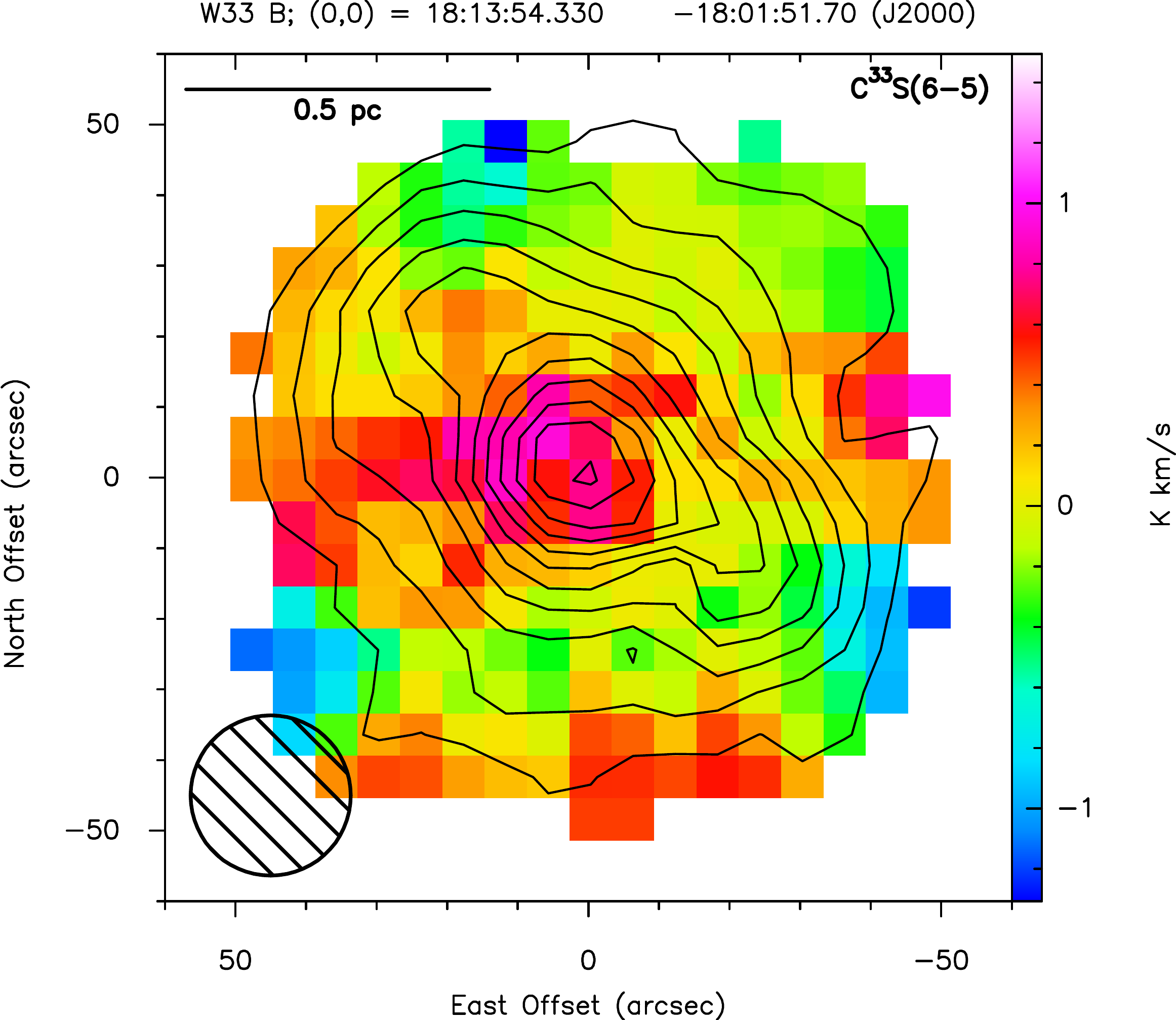}}\hspace{0.2cm}		
	\subfloat{\includegraphics[width=9cm]{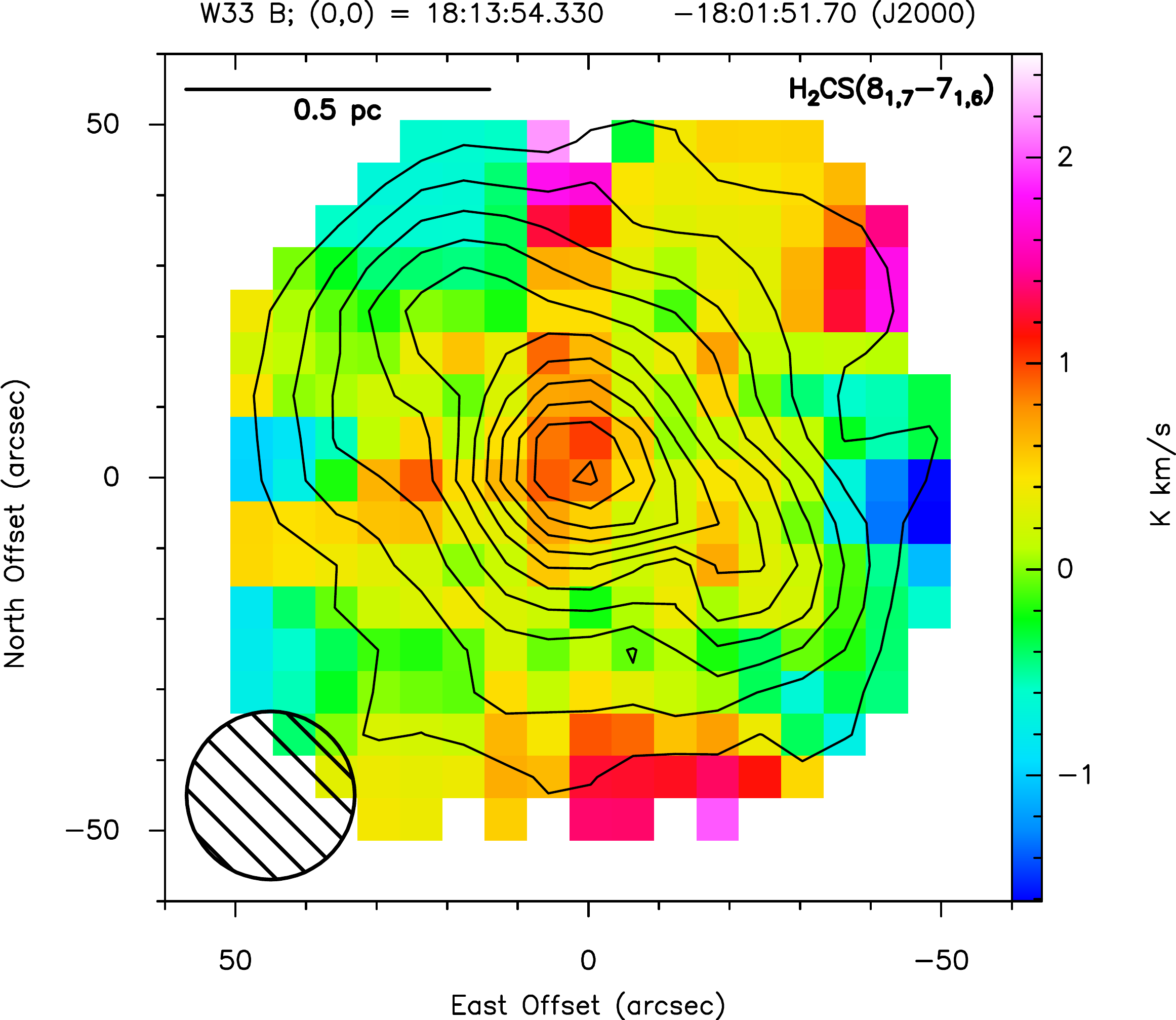}}\\		
	\subfloat{\includegraphics[width=9cm]{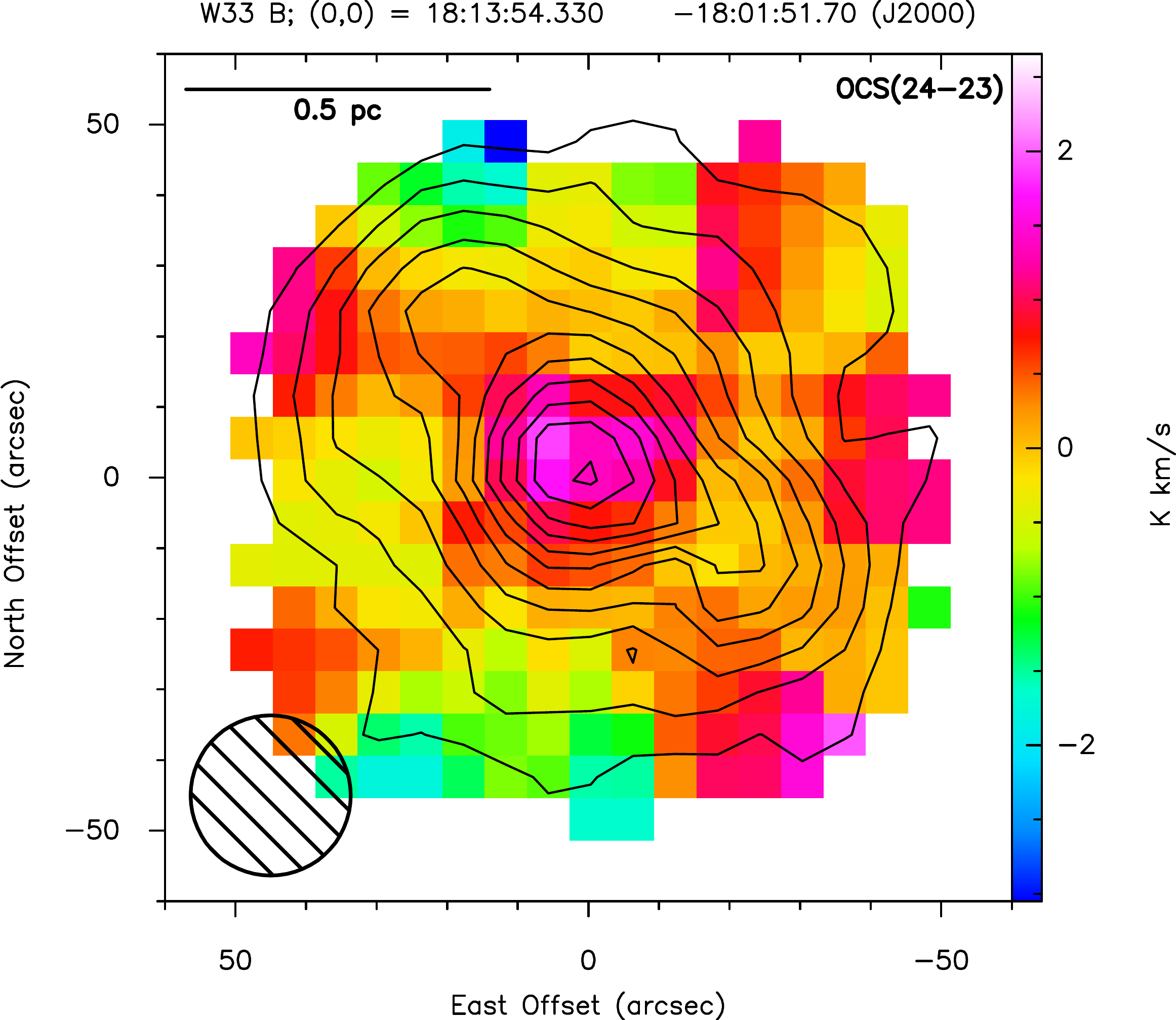}}		
	\label{W33B-APEX-IntInt}
\end{figure*}

\addtocounter{figure}{-1}
\begin{figure*}
	\centering
	\caption{Continued.}
	\subfloat{\includegraphics[width=9cm]{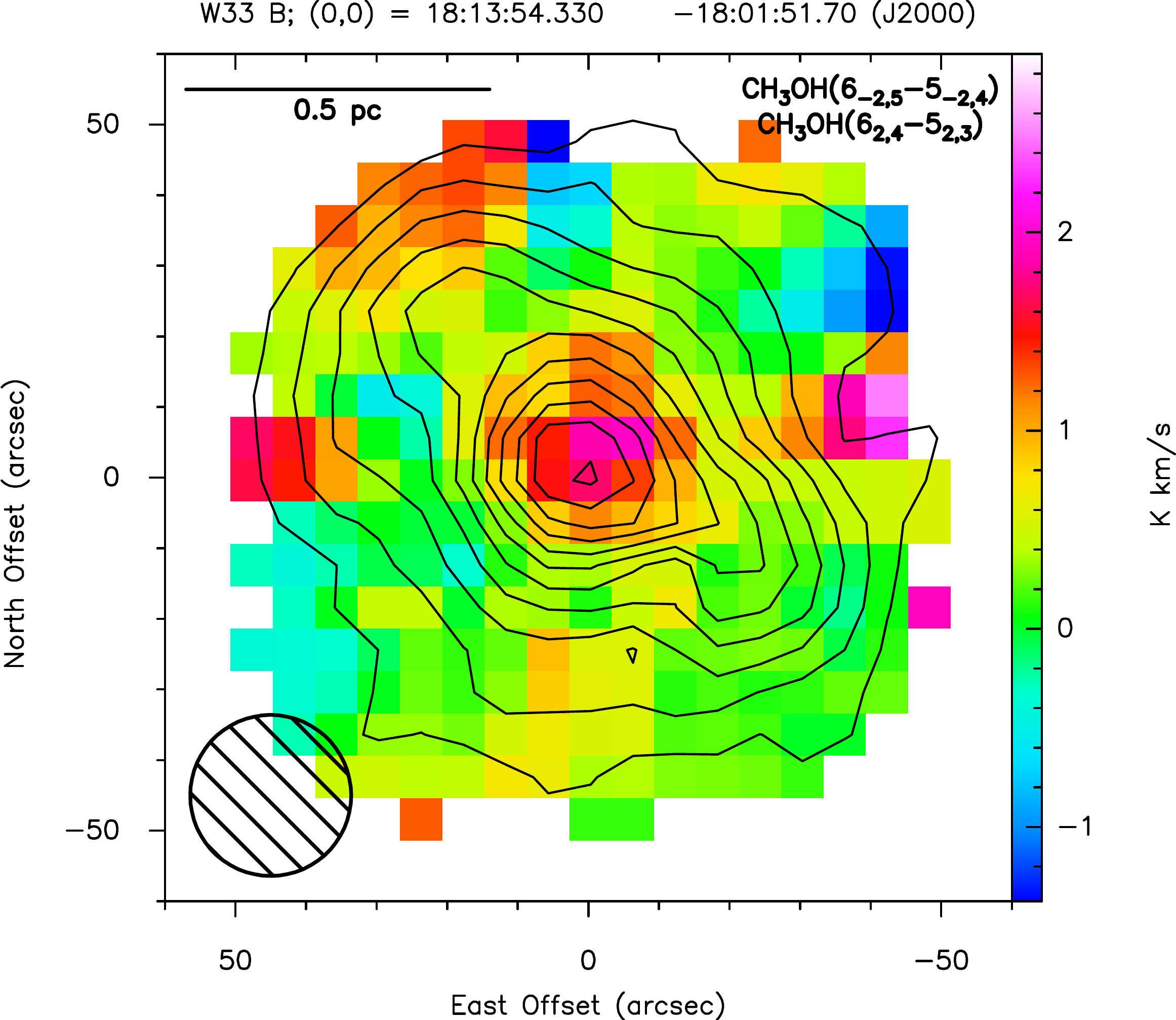}}\hspace{0.2cm}	
	\subfloat{\includegraphics[width=9cm]{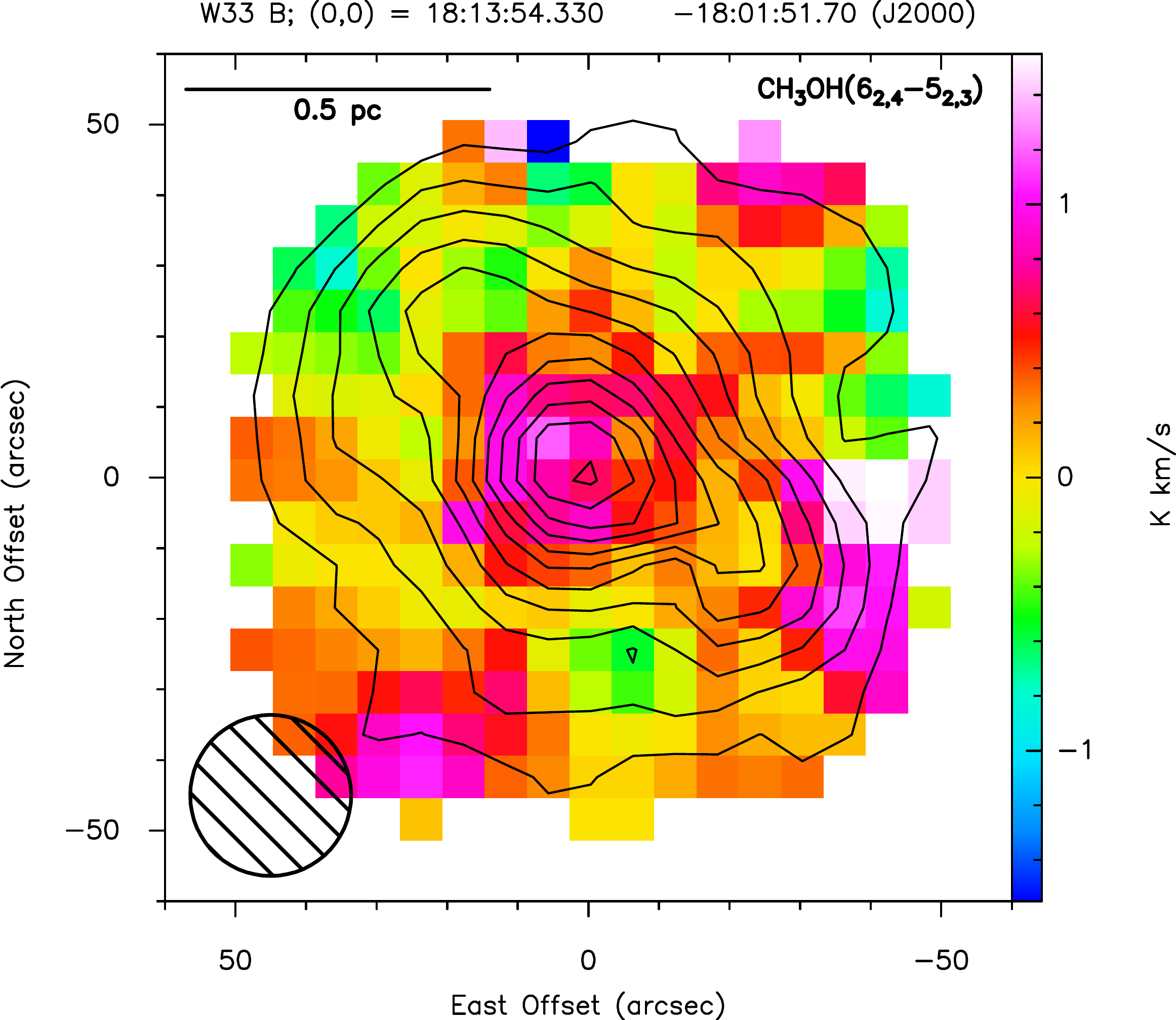}}\\	
	\subfloat{\includegraphics[width=9cm]{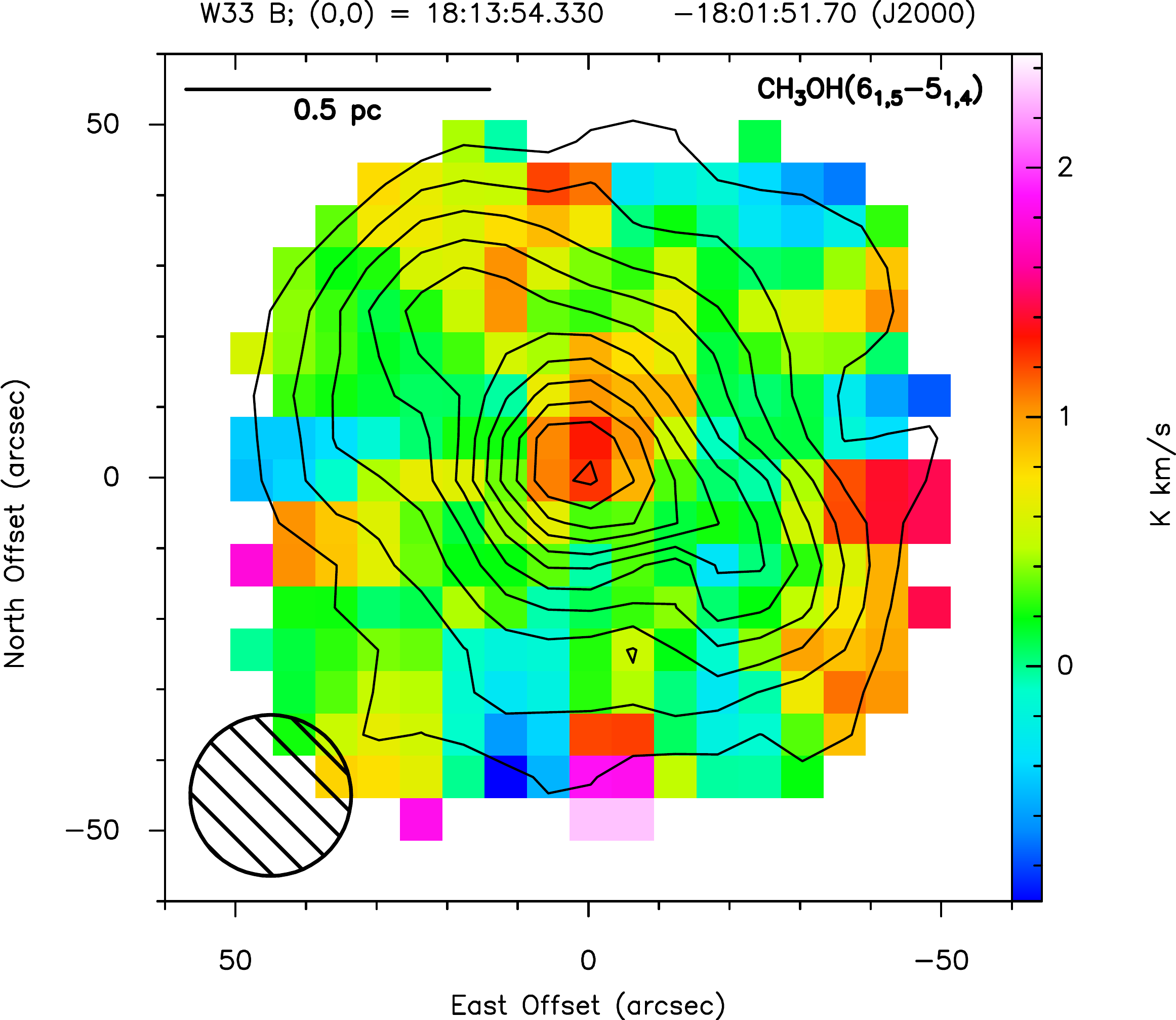}}\hspace{0.2cm}	
	\subfloat{\includegraphics[width=9cm]{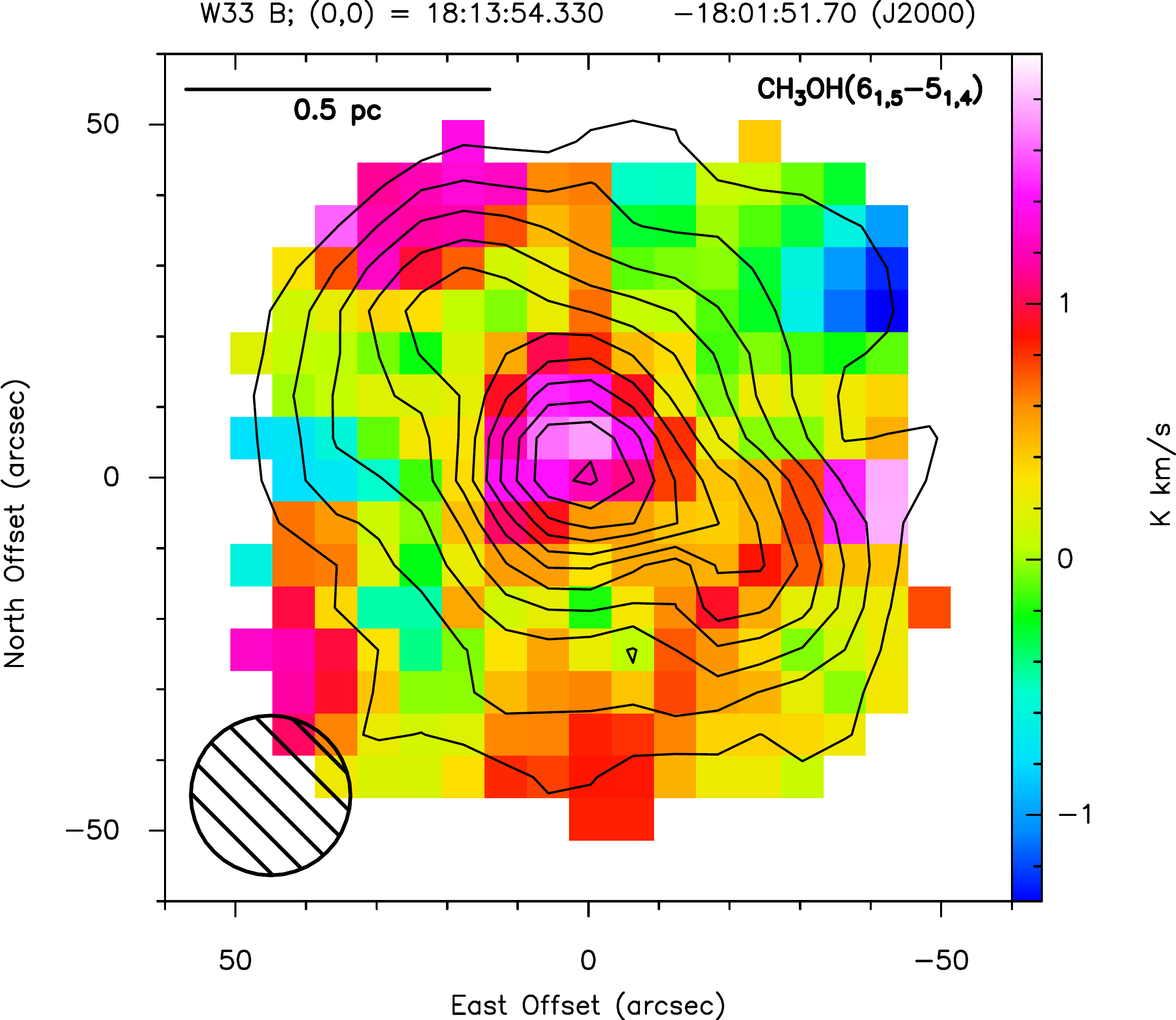}}\\		
	\subfloat{\includegraphics[width=9cm]{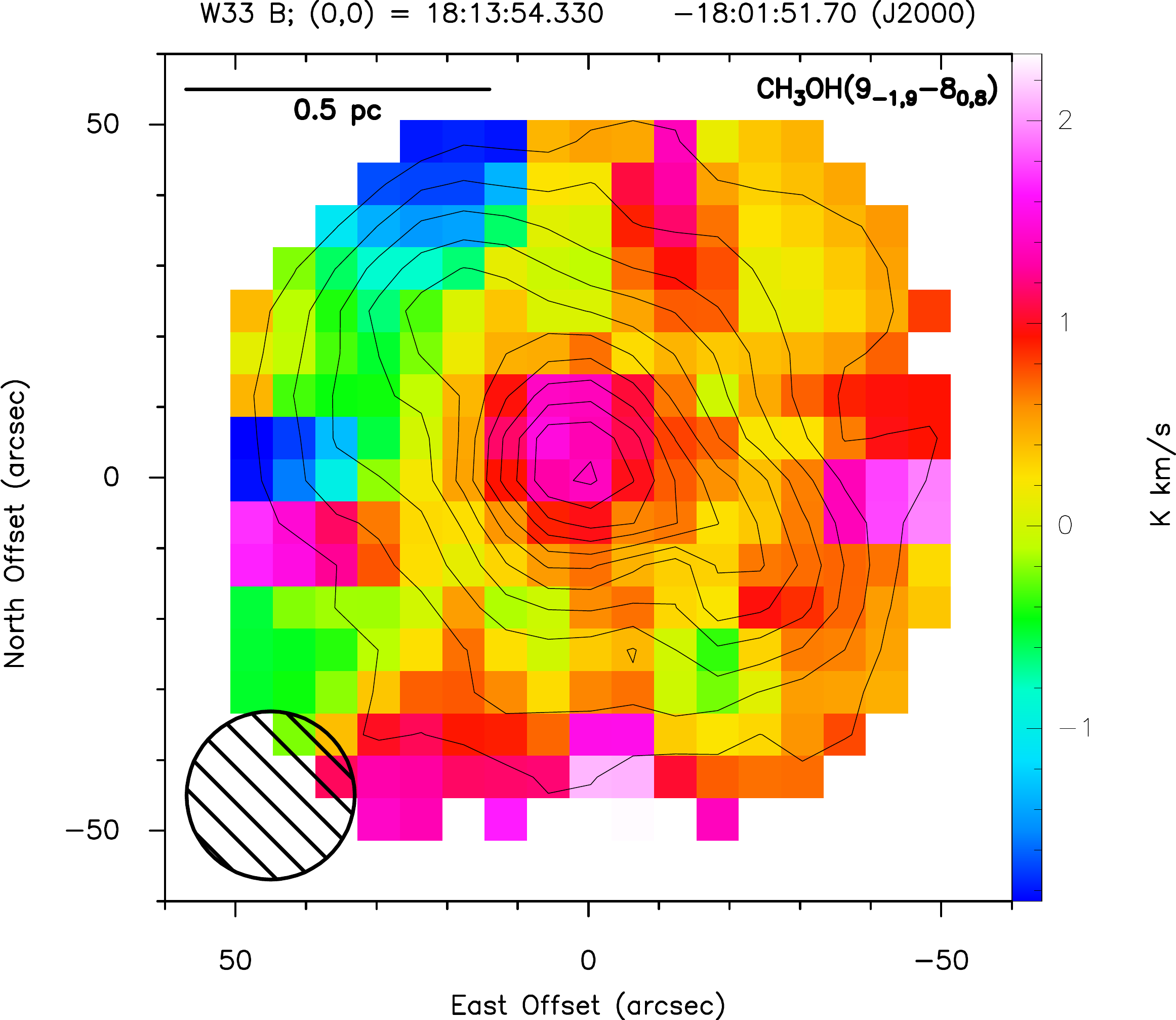}}\hspace{0.2cm}	
	\subfloat{\includegraphics[width=9cm]{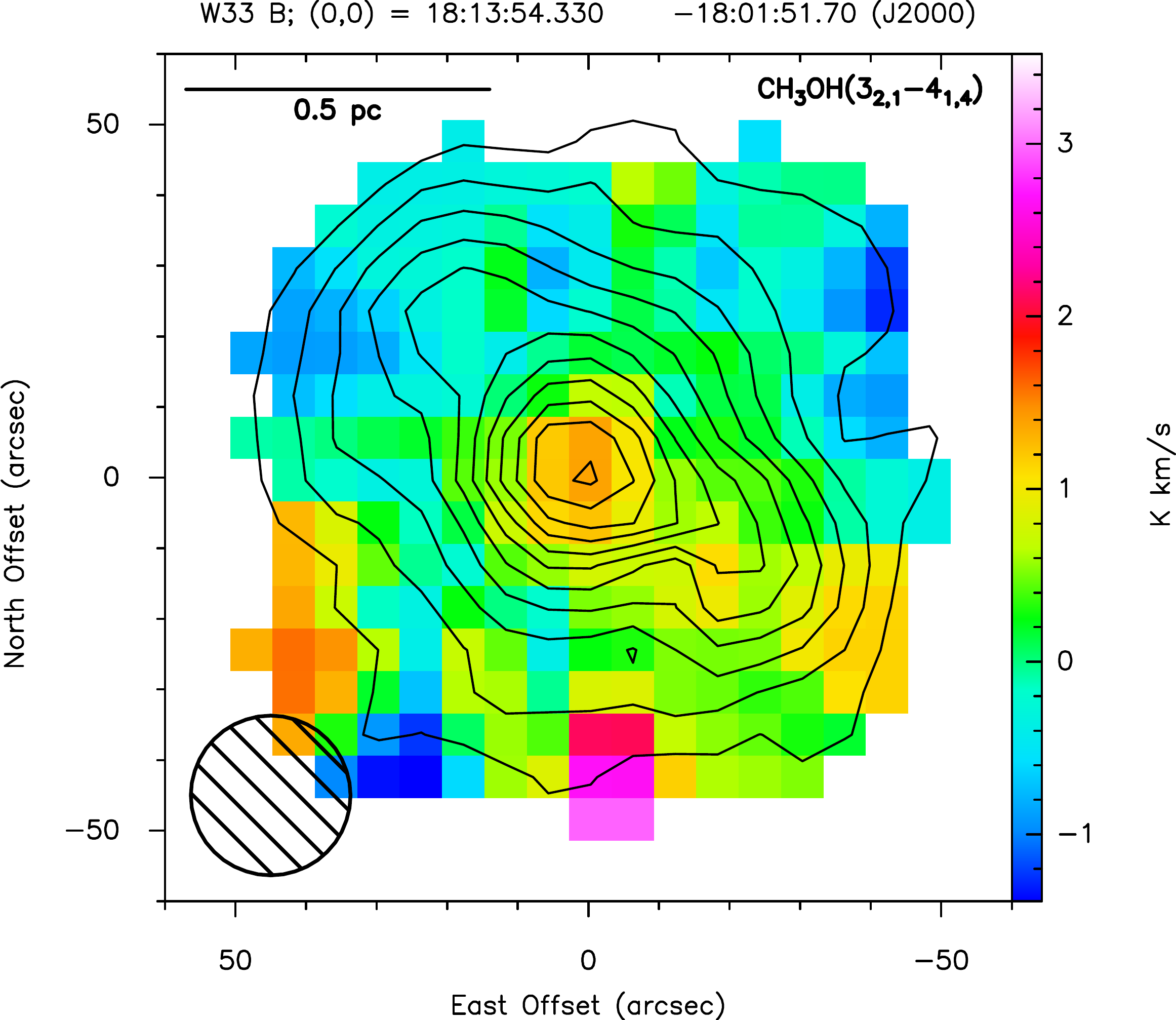}}	
\end{figure*}

\addtocounter{figure}{-1}
\begin{figure*}
	\centering
	\caption{Continued.}
	\subfloat{\includegraphics[width=9cm]{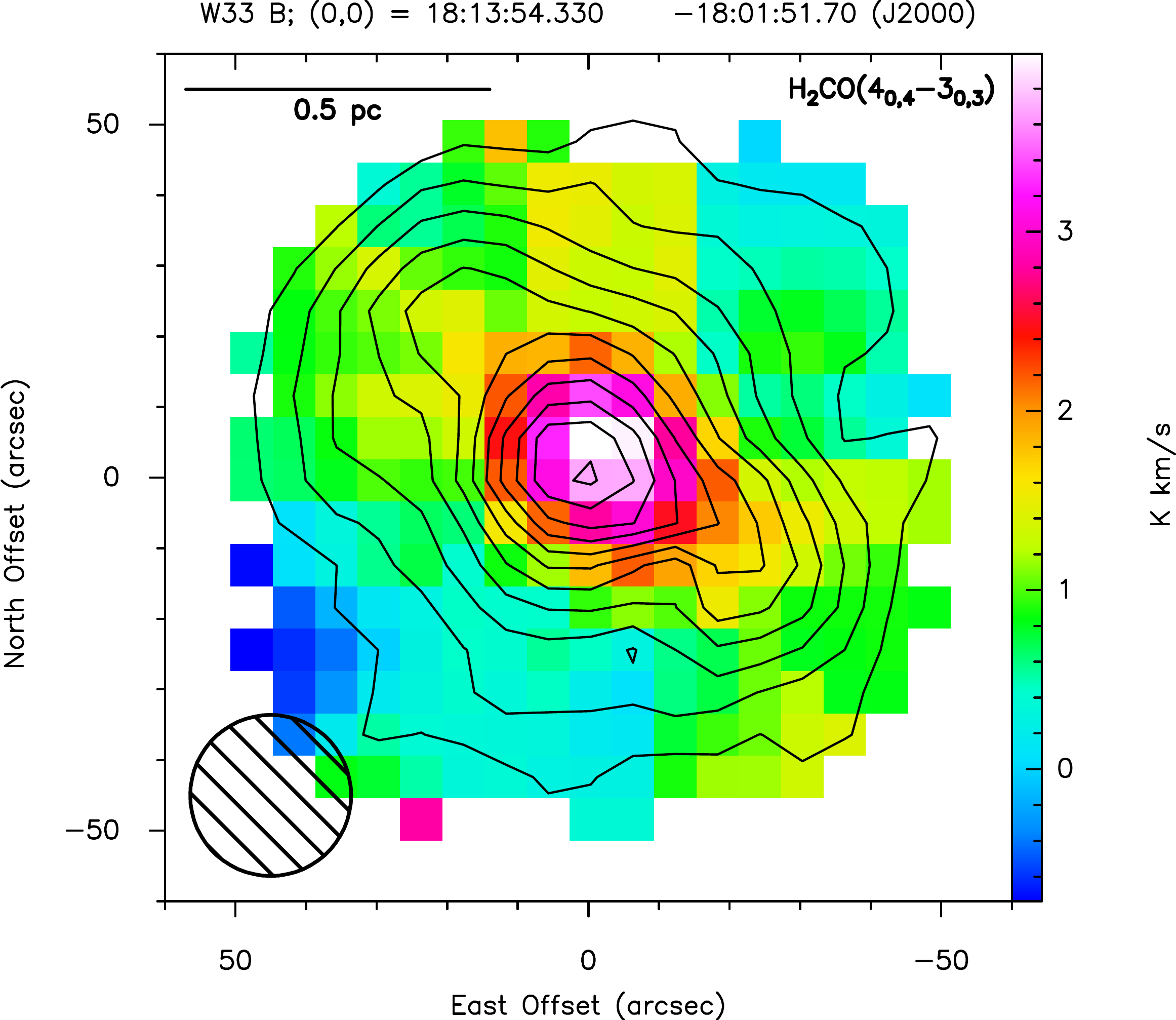}}\hspace{0.2cm}	
	\subfloat{\includegraphics[width=9cm]{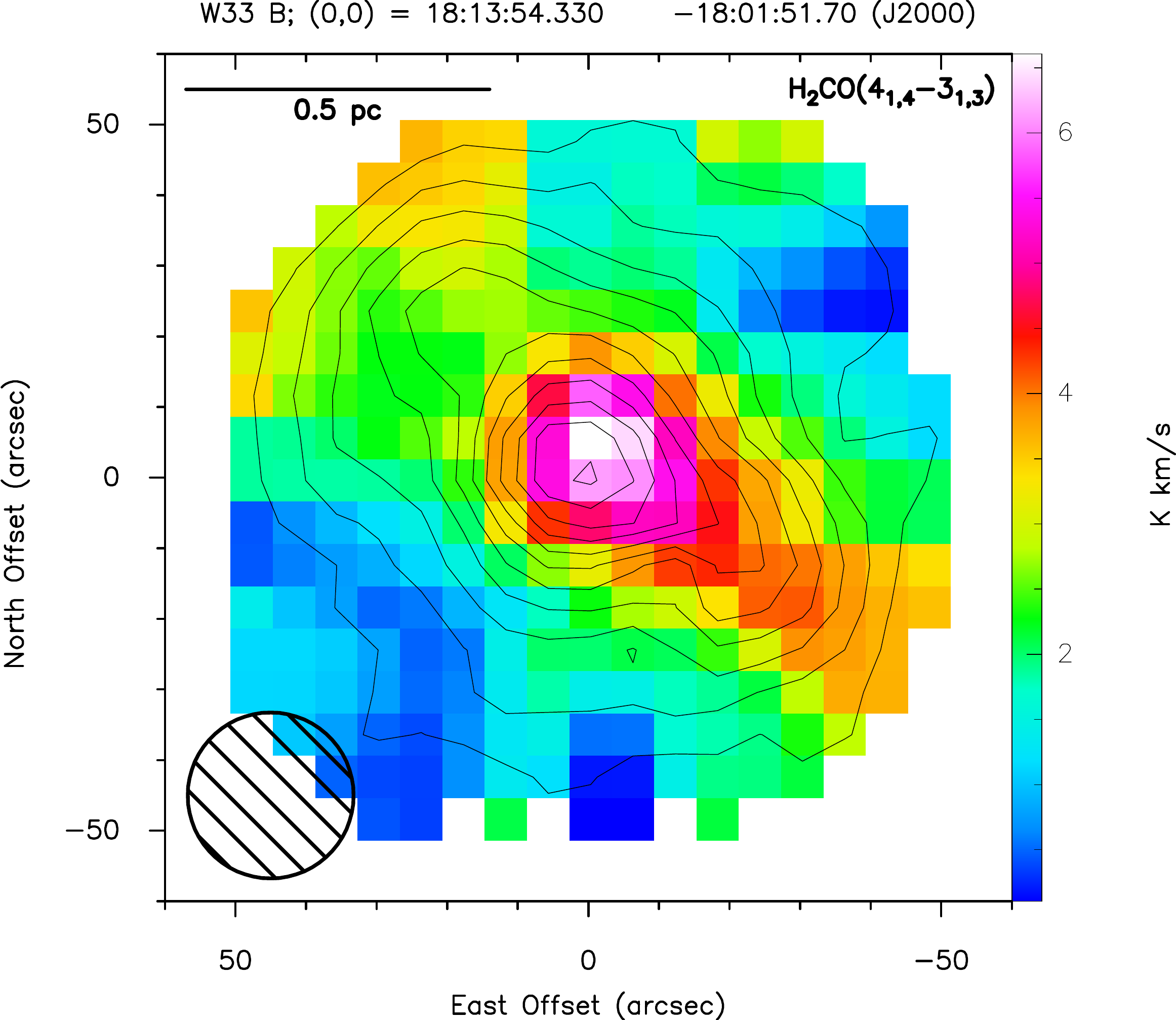}}\\	
	\subfloat{\includegraphics[width=9cm]{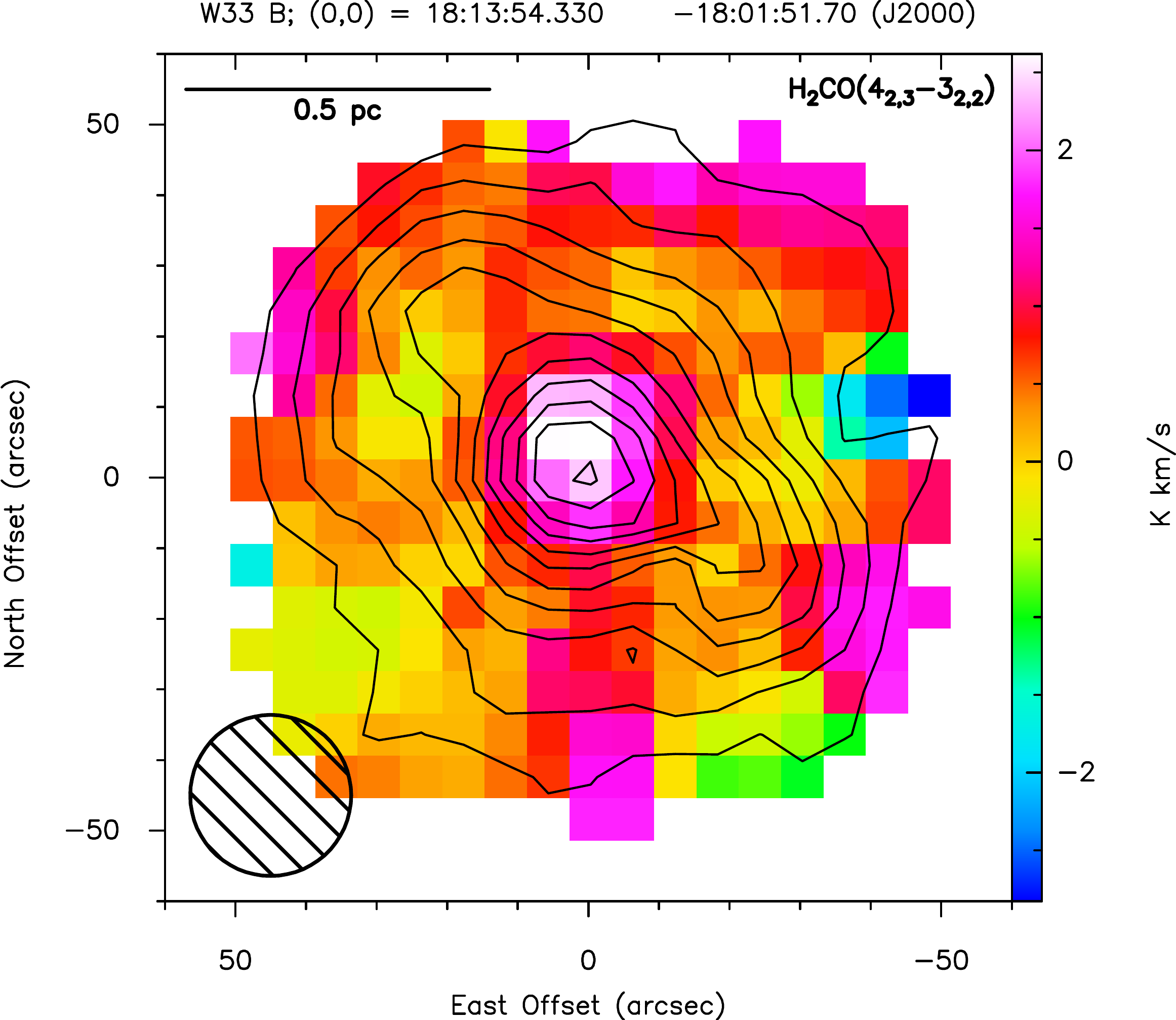}}\hspace{0.2cm}	
	\subfloat{\includegraphics[width=9cm]{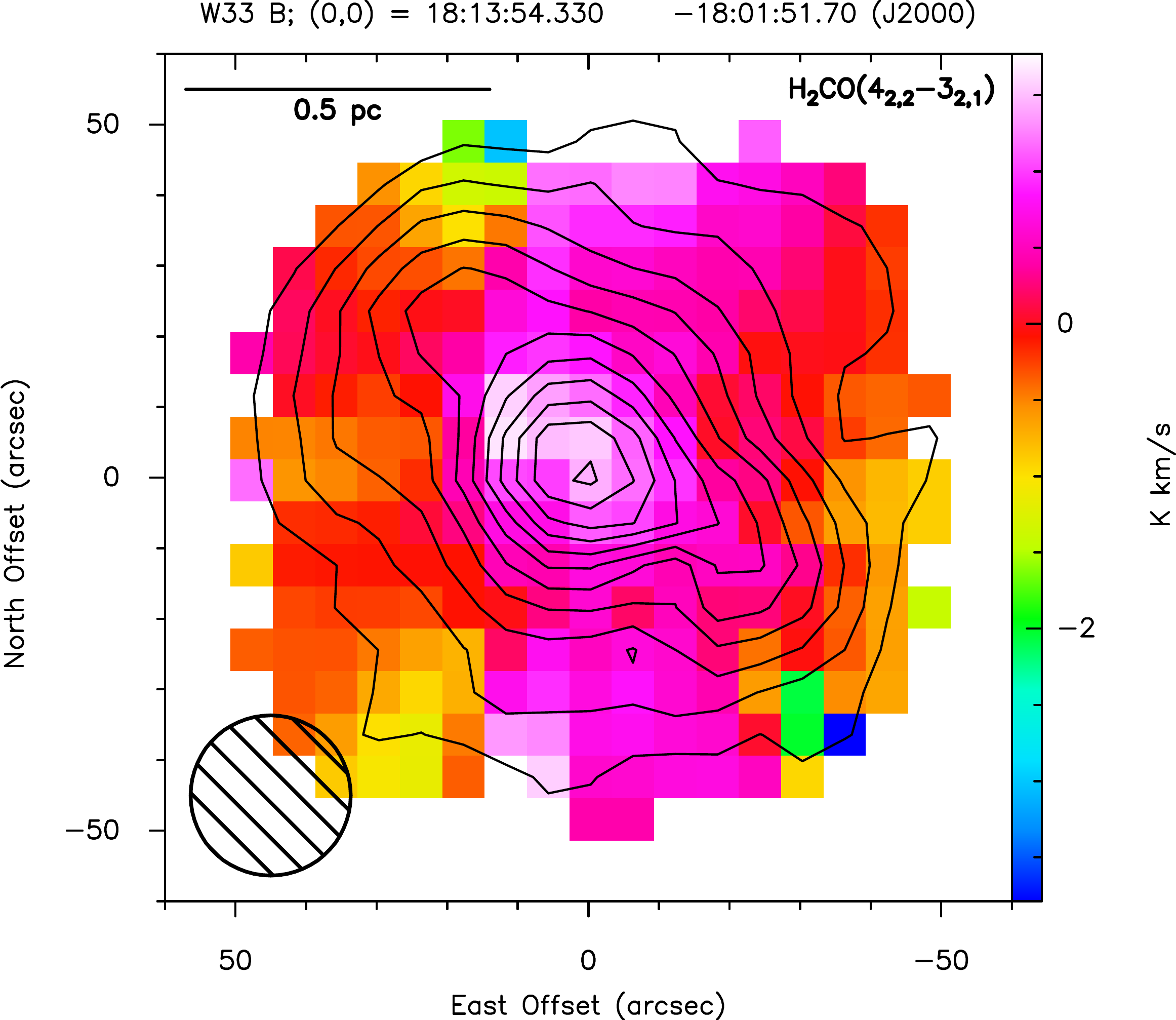}}\\	
	\subfloat{\includegraphics[width=9cm]{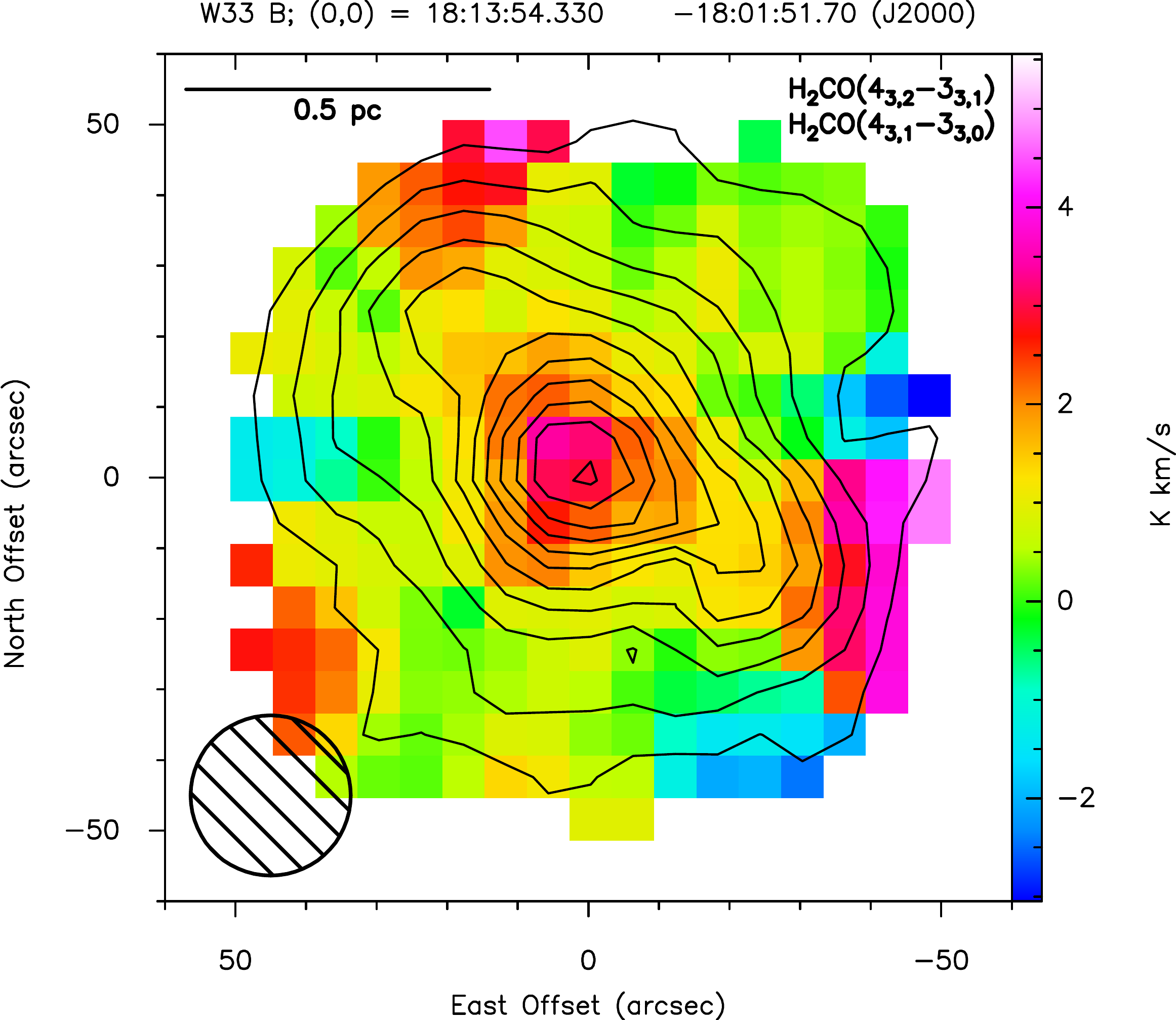}}	
\end{figure*}

\clearpage

\begin{figure*}
	\caption{Line emission of detected transitions in W33\,A. The contours show the ATLASGAL continuum emission at 345 GHz (levels in steps of 10$\sigma$, starting at 20$\sigma$ ($\sigma$ = 0.081 Jy beam$^{-1}$). The name of the each transition is shown in the upper right corner. A scale of 0.5 pc is marked in the upper left corner, and the synthesised beam is shown in the lower left corner.}
	\centering
	\subfloat{\includegraphics[width=9cm]{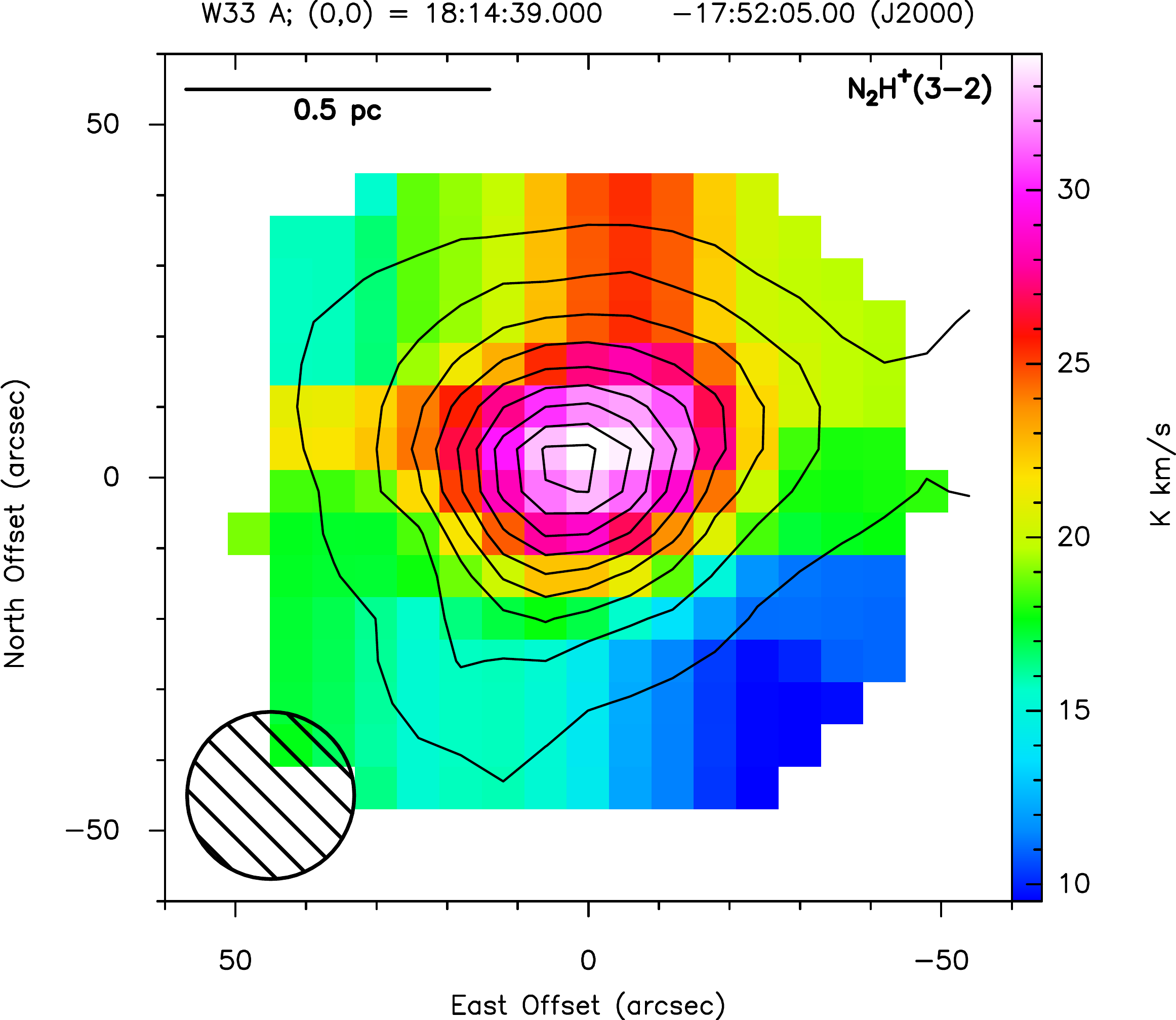}}\hspace{0.2cm}	
	\subfloat{\includegraphics[width=9cm]{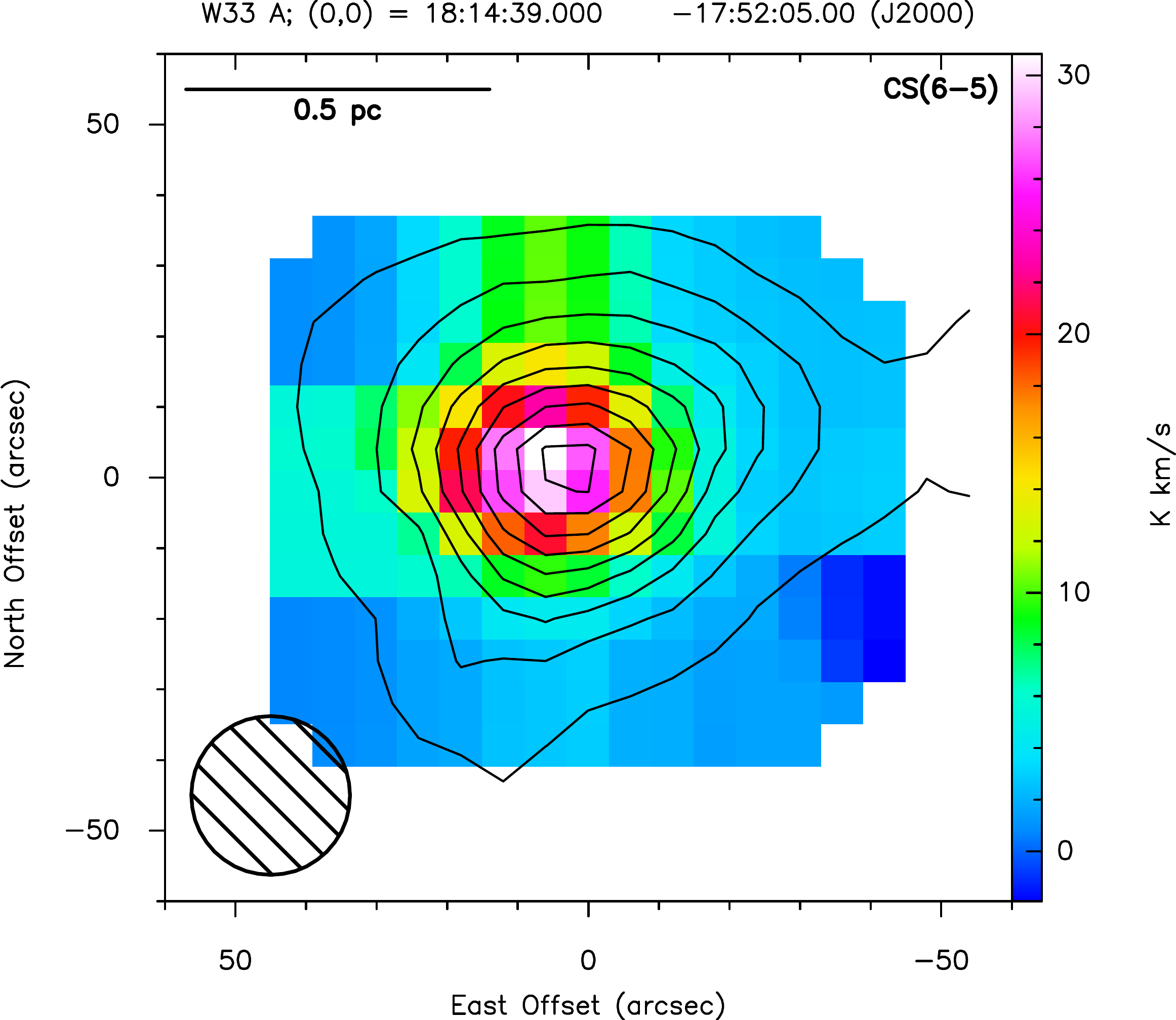}}\\		
	\subfloat{\includegraphics[width=9cm]{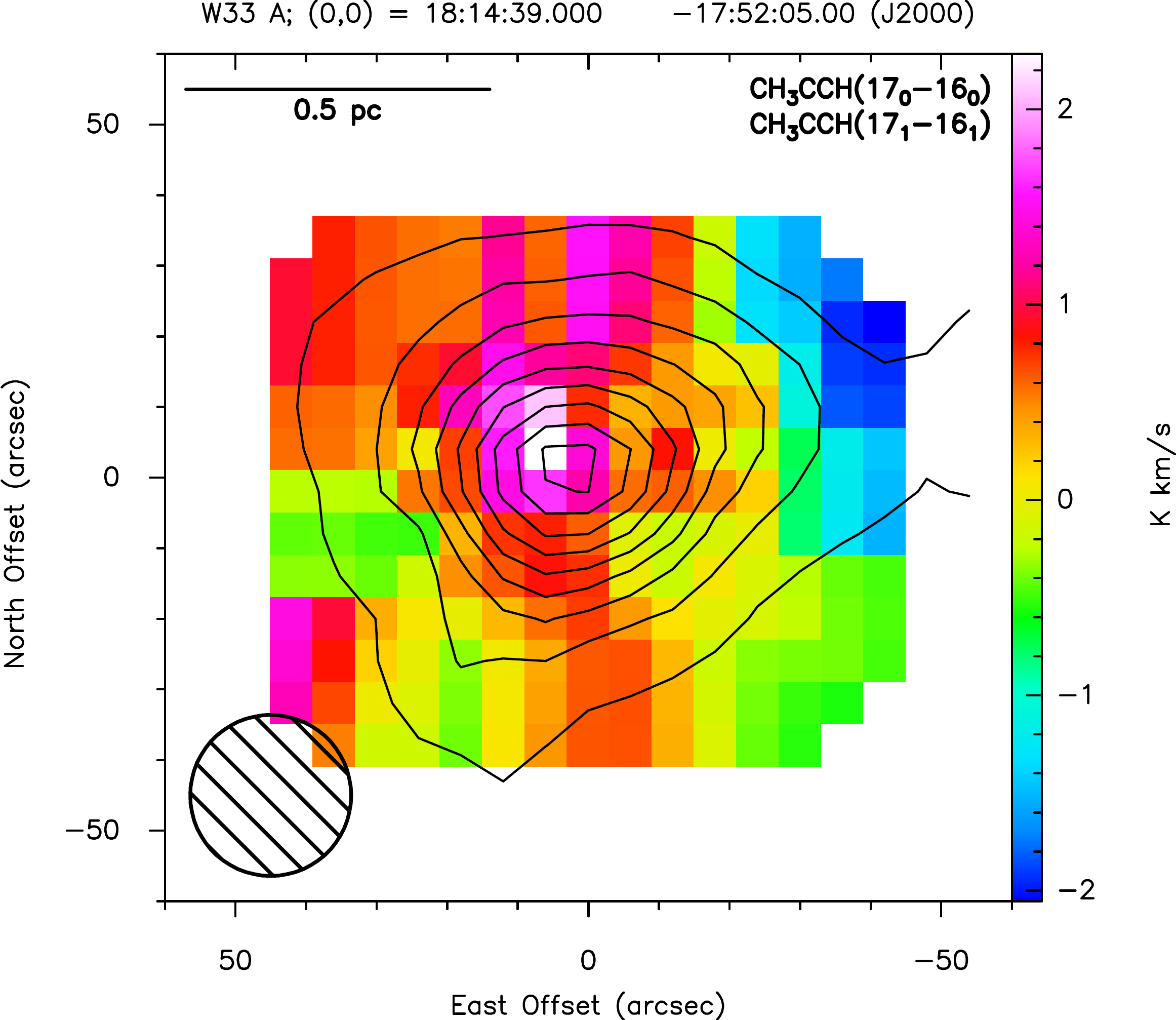}}\hspace{0.2cm}		
	\subfloat{\includegraphics[width=9cm]{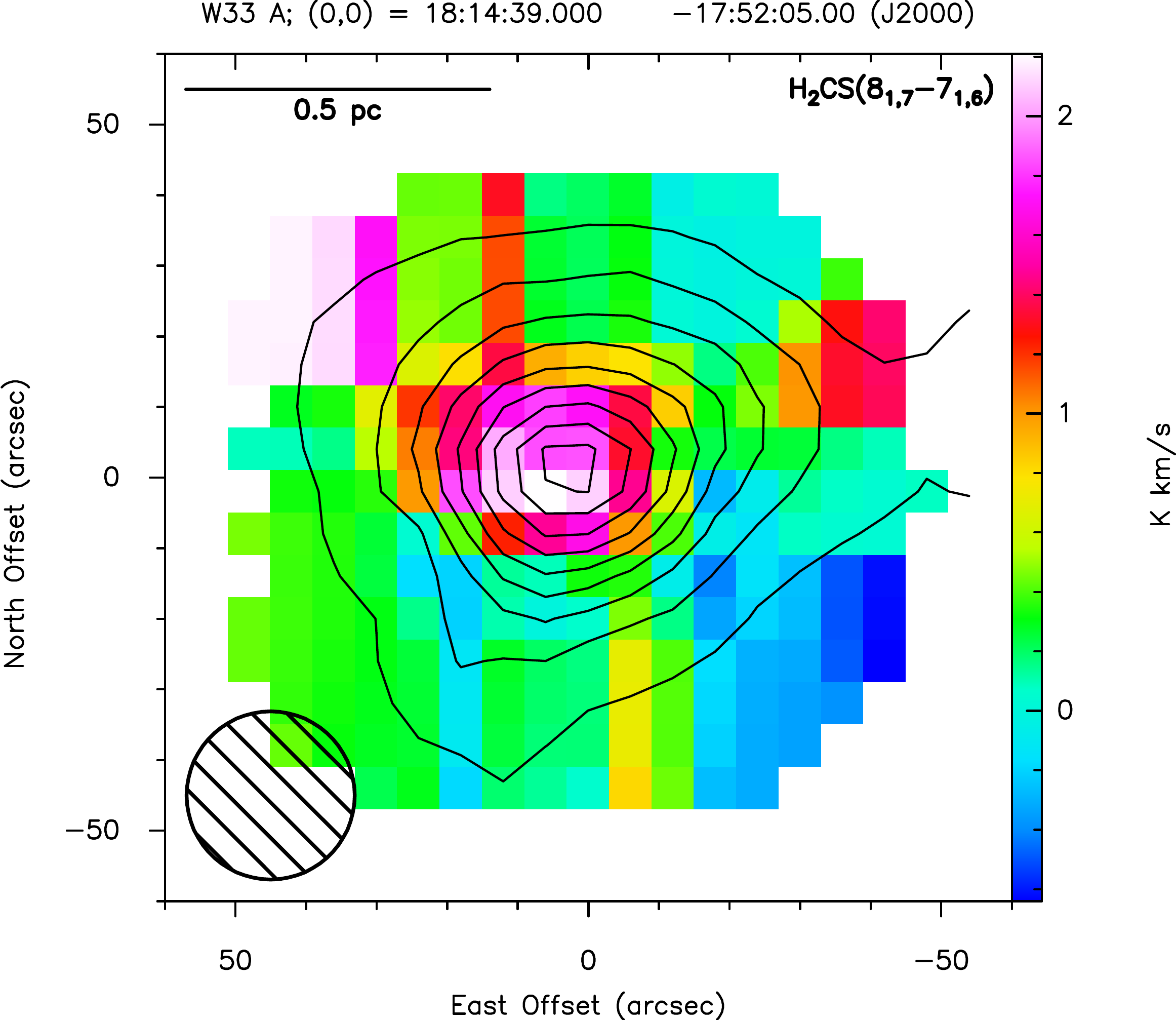}}	
	\label{W33A-APEX-IntInt}
\end{figure*}

\addtocounter{figure}{-1}
\begin{figure*}
	\caption{Continued.}
	\centering	
	\subfloat{\includegraphics[width=9cm]{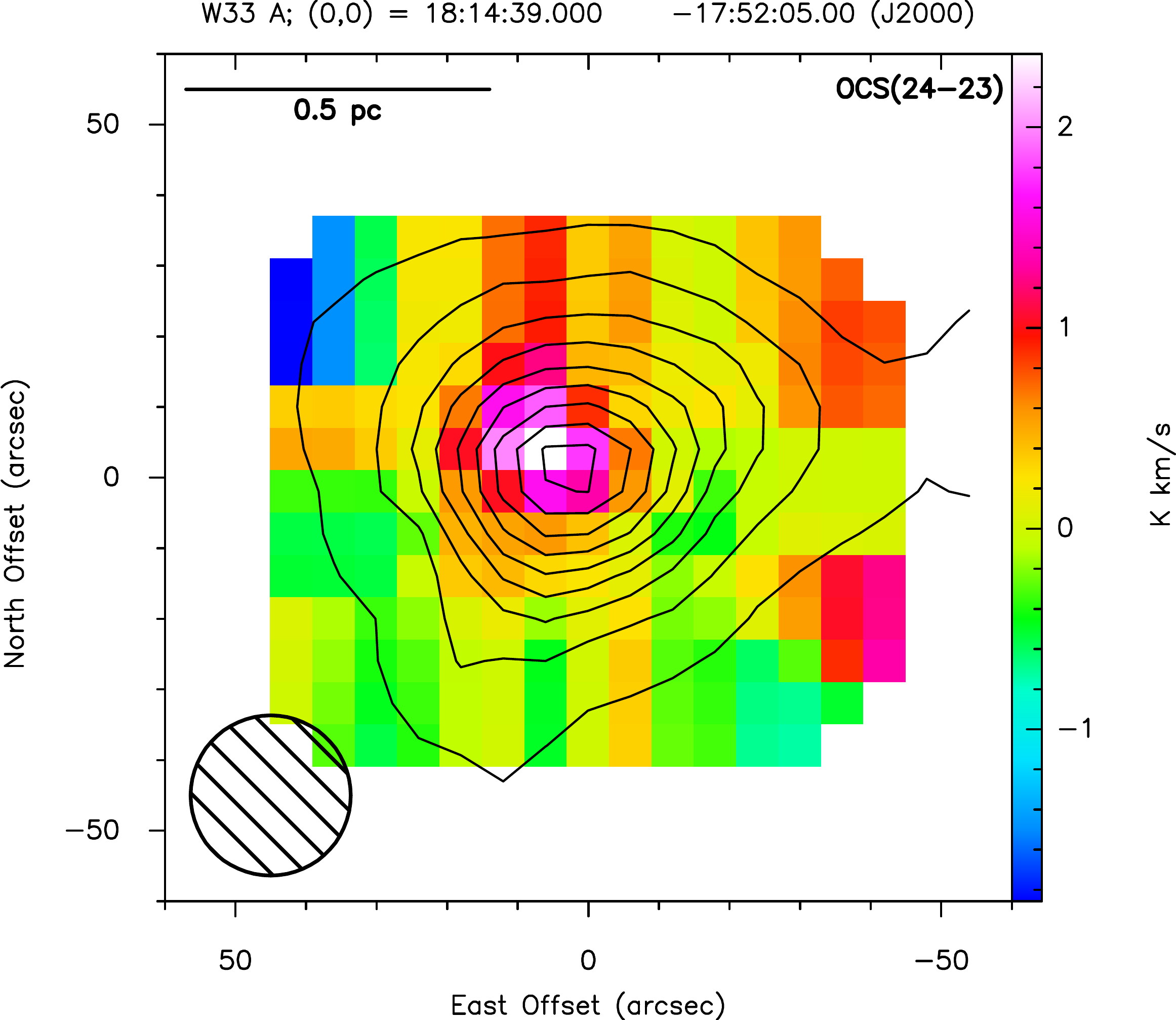}}\hspace{0.2cm}					
\subfloat{\includegraphics[width=9cm]{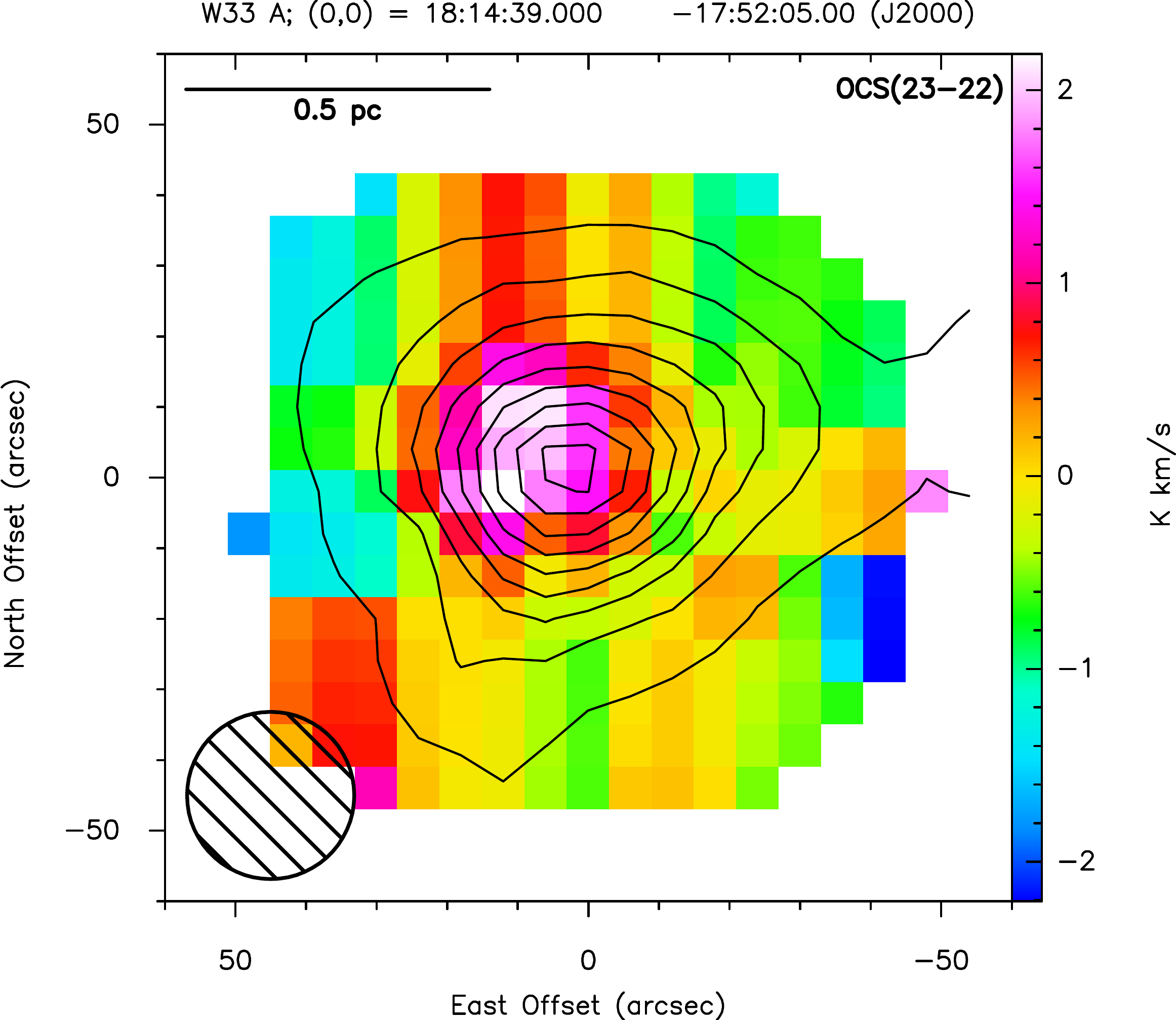}}\\		
	\subfloat{\includegraphics[width=9cm]{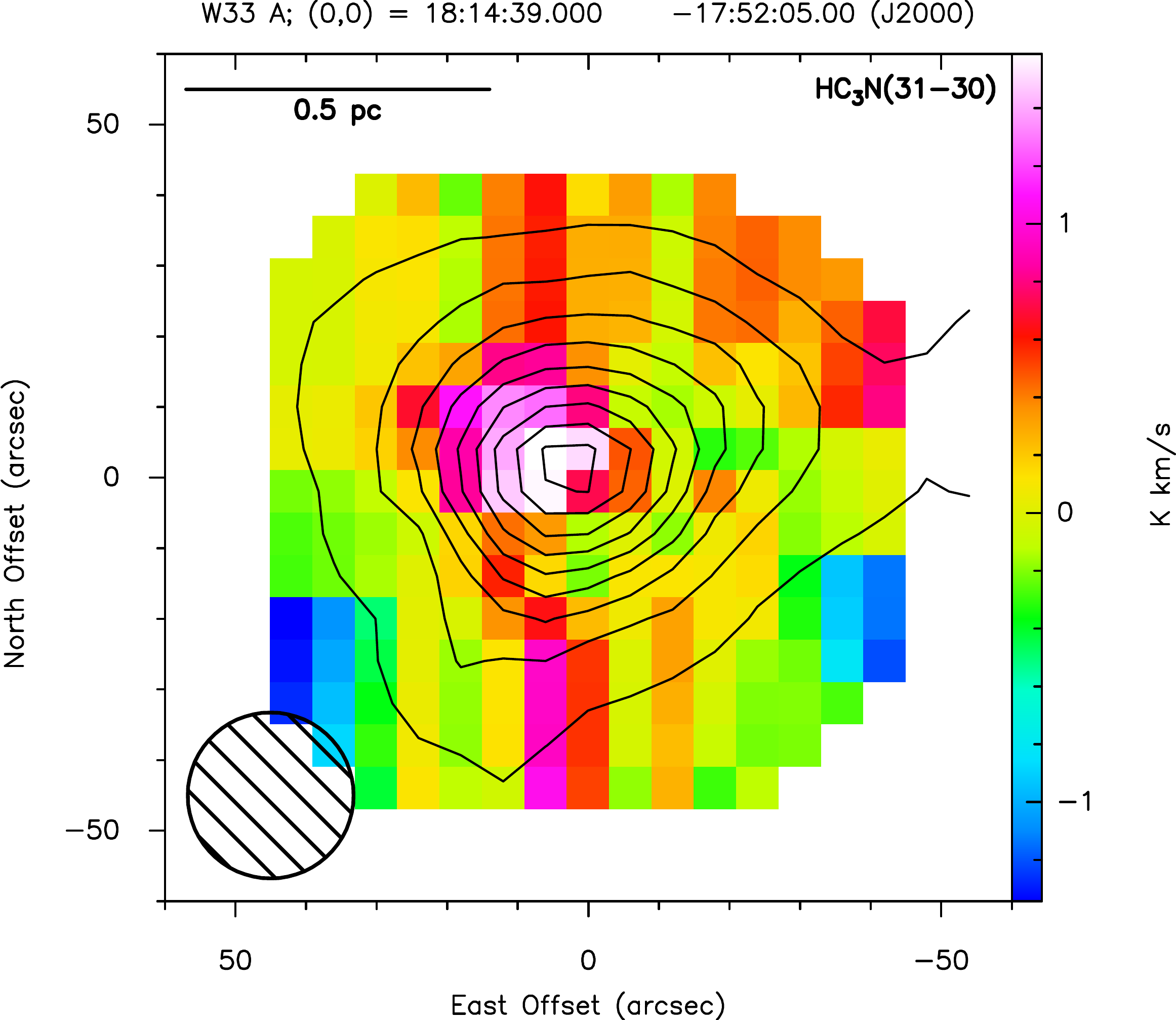}}\hspace{0.2cm}
	\subfloat{\includegraphics[width=9cm]{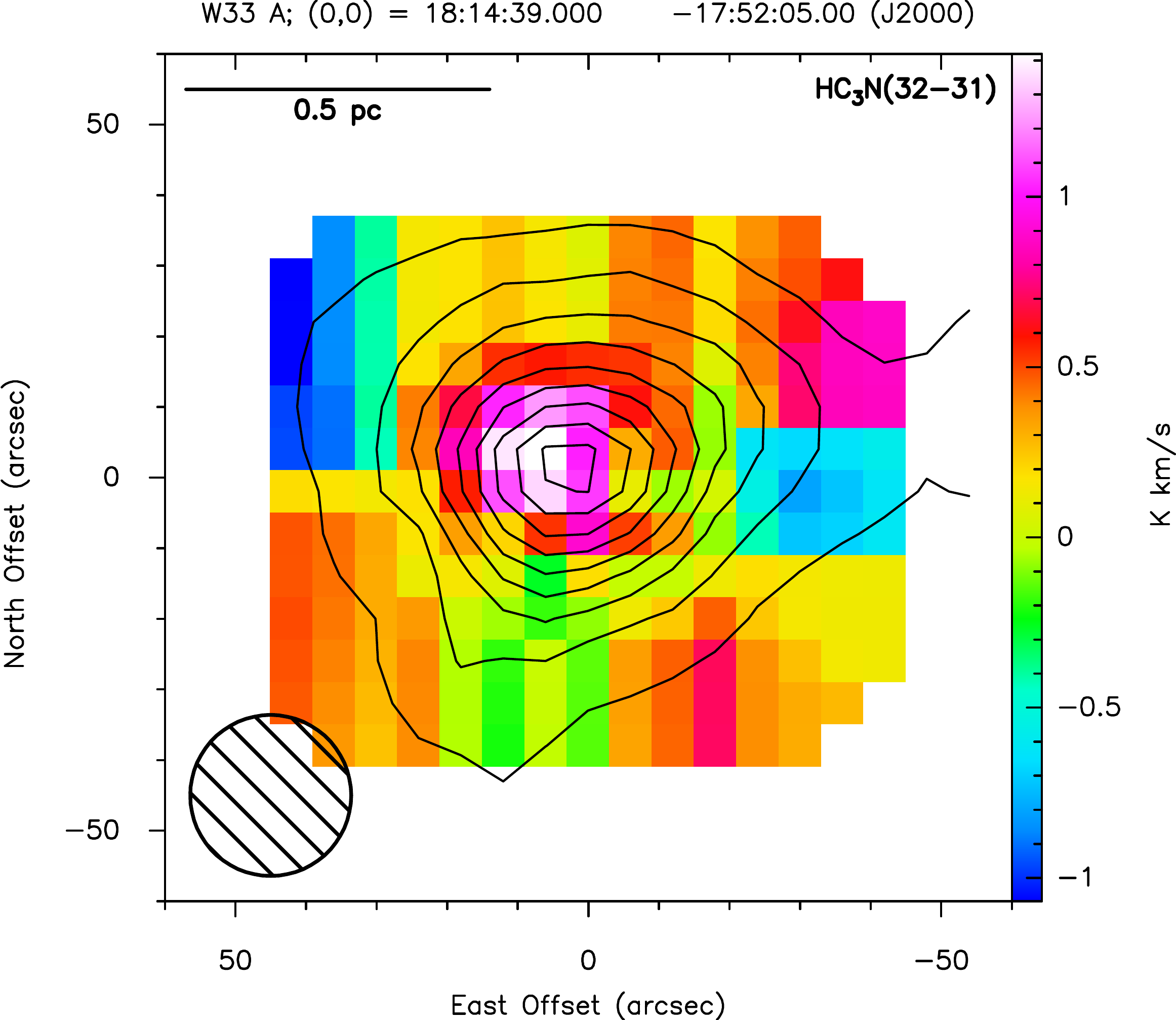}}
\end{figure*}

\addtocounter{figure}{-1}
\begin{figure*}
	\centering
	\caption{Continued.}
	\subfloat{\includegraphics[width=9cm]{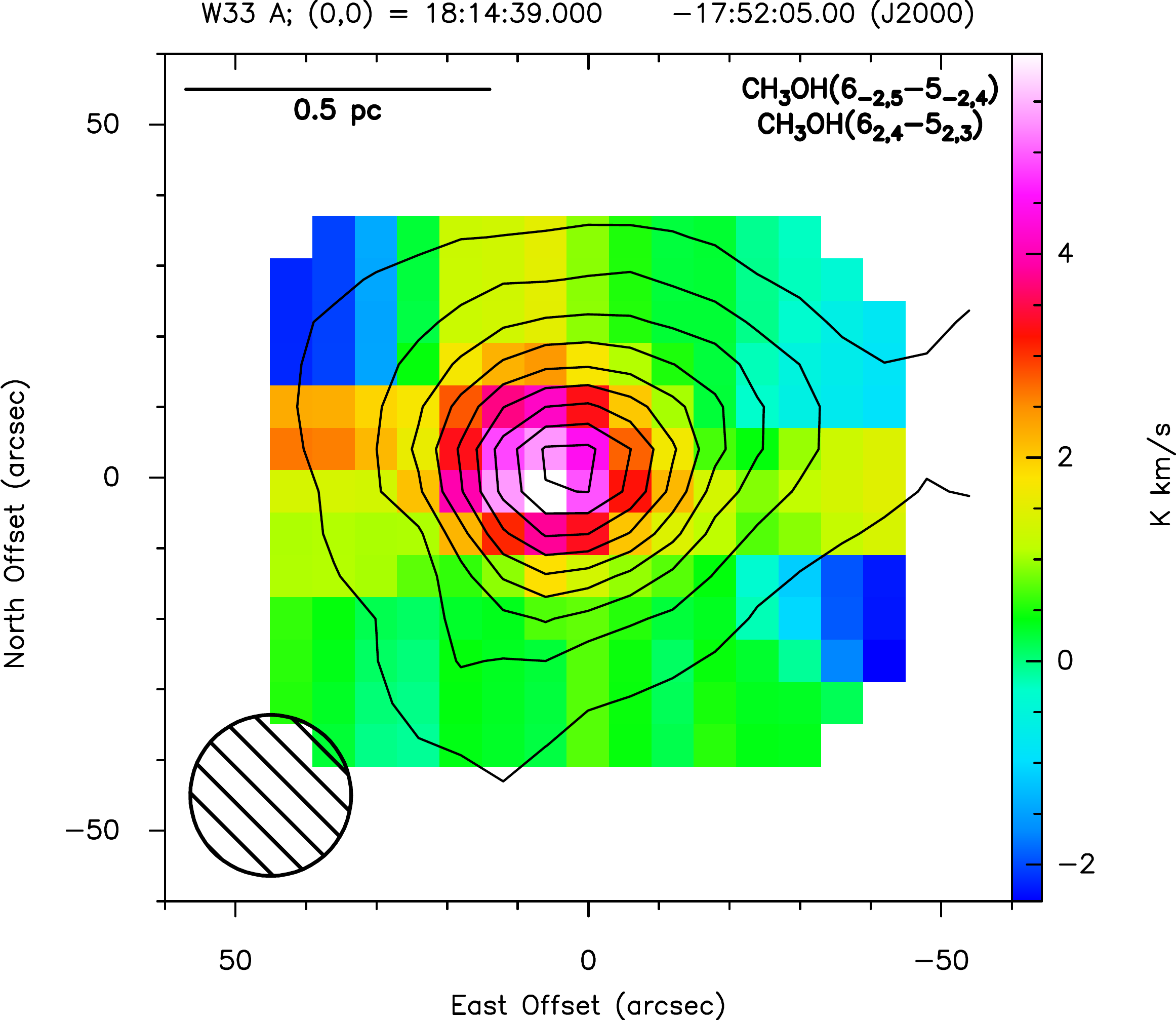}}\hspace{0.2cm}	
	\subfloat{\includegraphics[width=9cm]{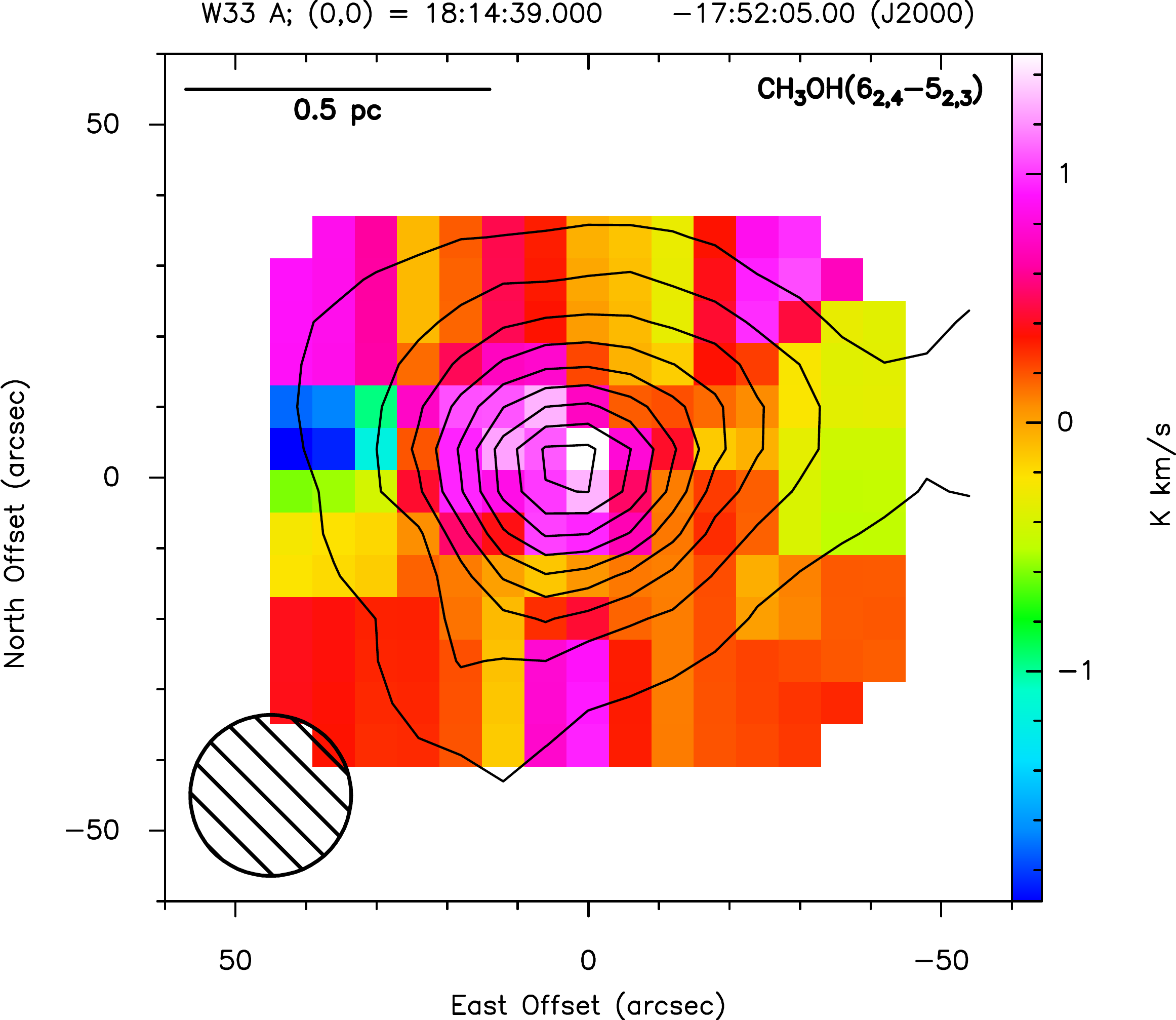}}\\
	\subfloat{\includegraphics[width=9cm]{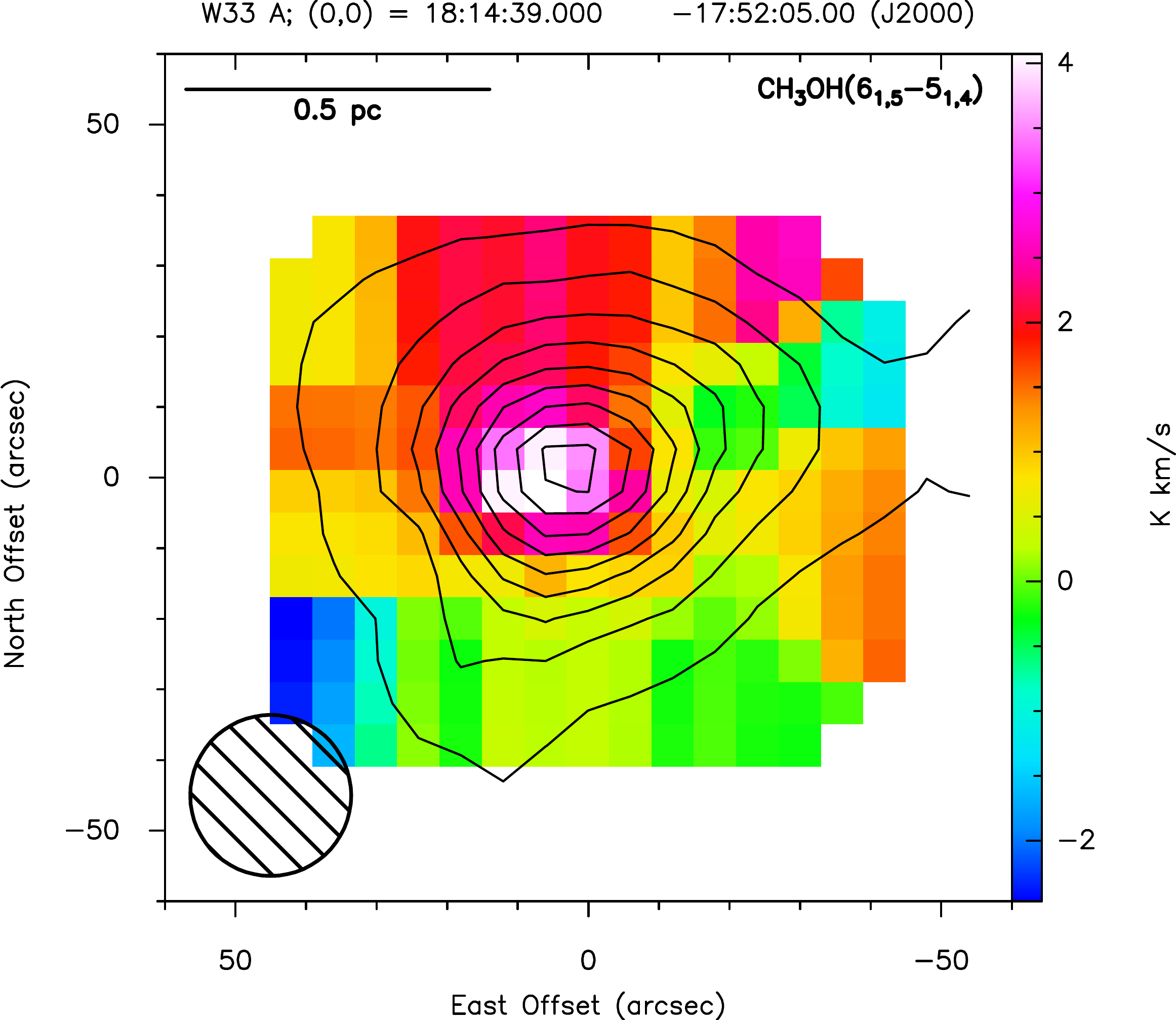}}\hspace{0.2cm}	
	\subfloat{\includegraphics[width=9cm]{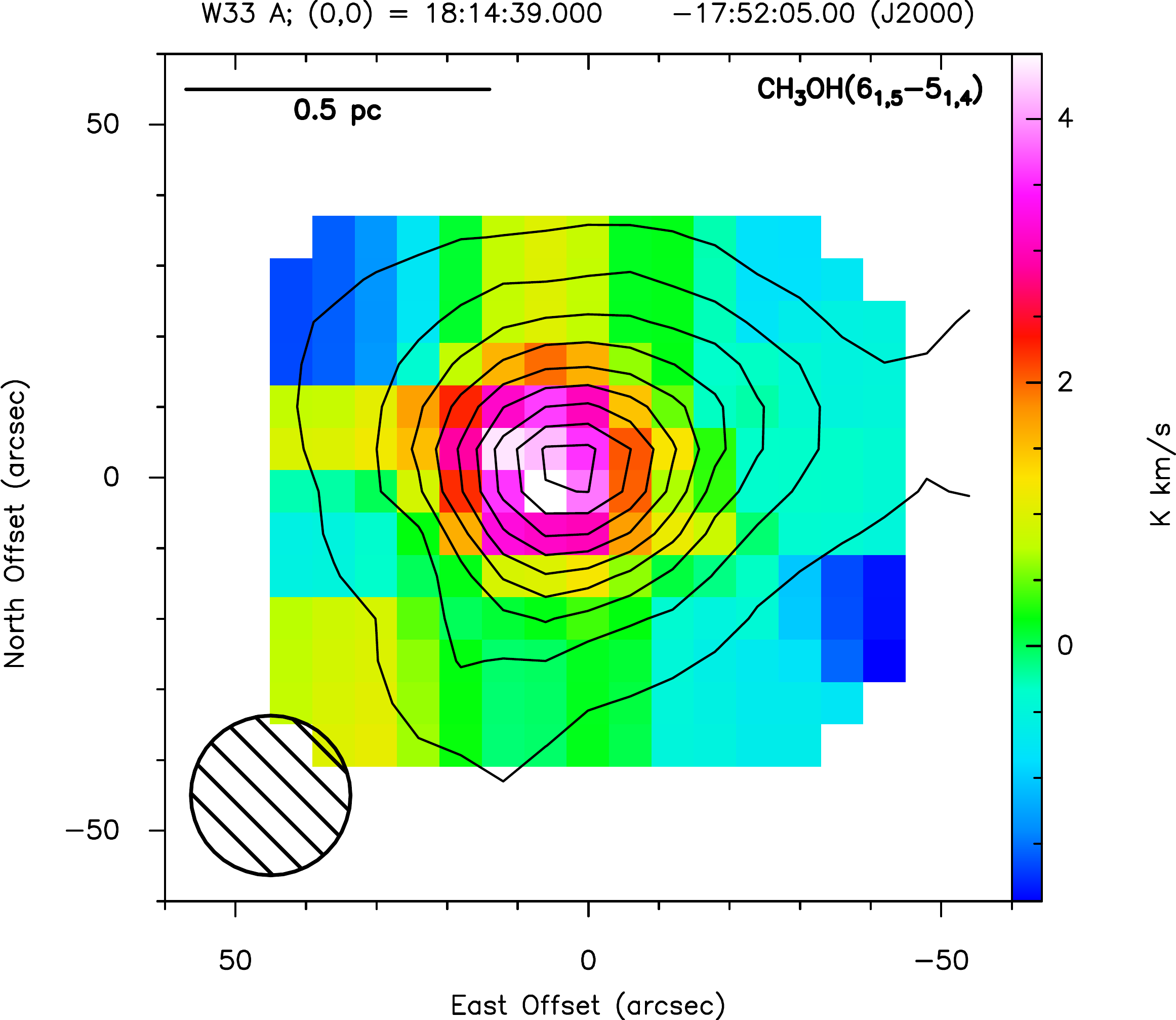}}\\		
	\subfloat{\includegraphics[width=9cm]{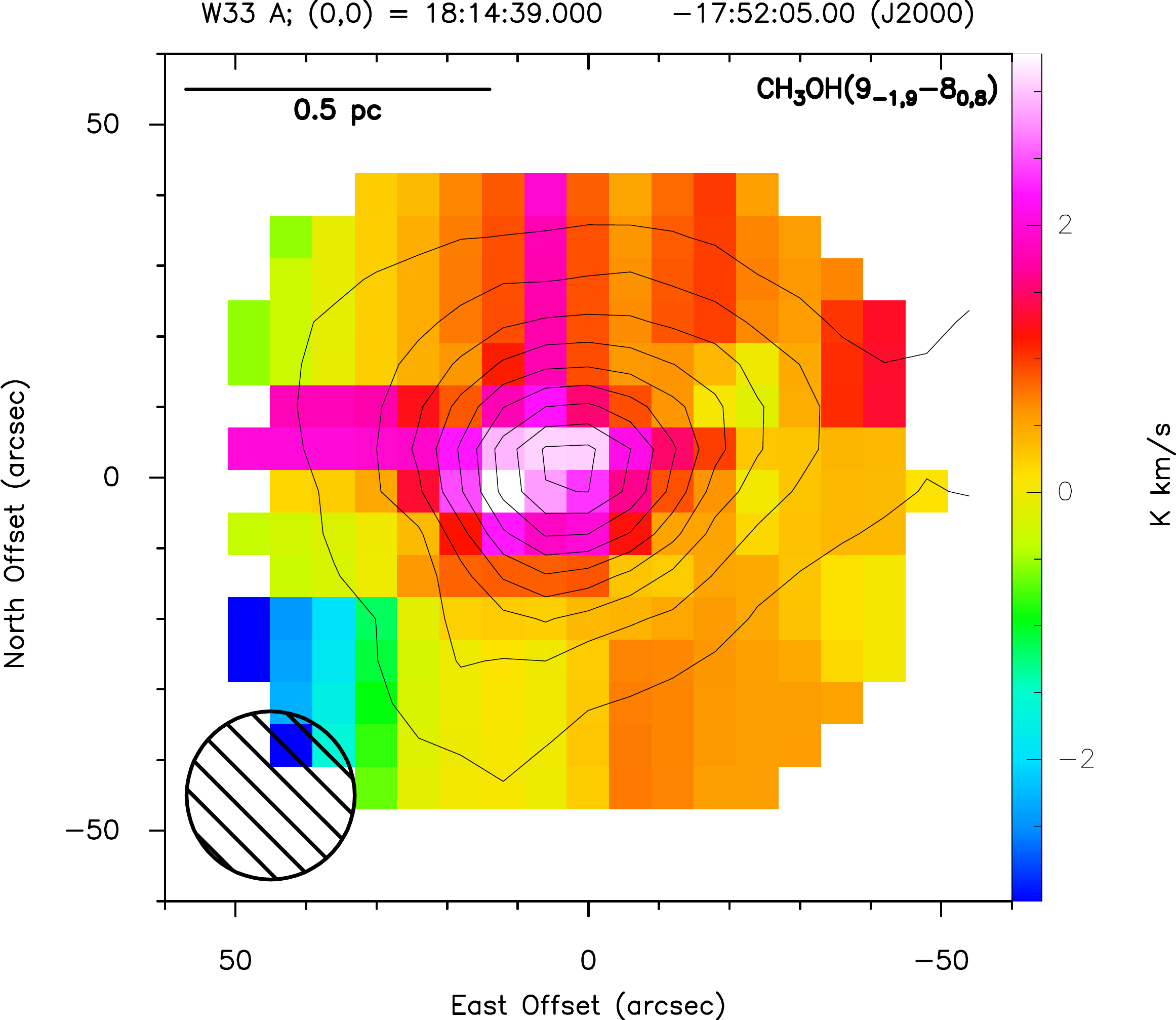}}\hspace{0.2cm}	
	\subfloat{\includegraphics[width=9cm]{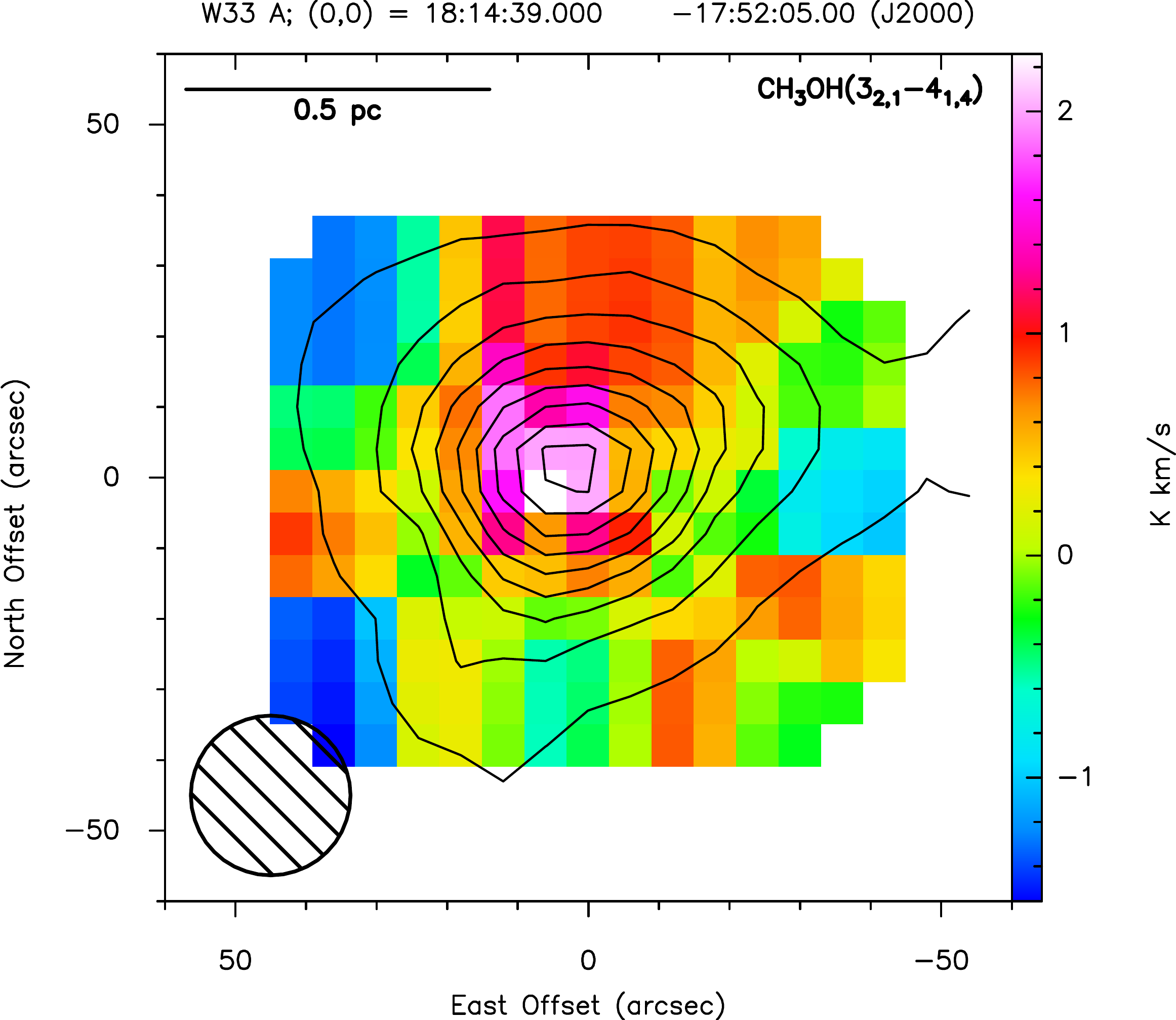}}	
\end{figure*}

\addtocounter{figure}{-1}
\begin{figure*}
	\centering
	\caption{Continued.}
	\subfloat{\includegraphics[width=9cm]{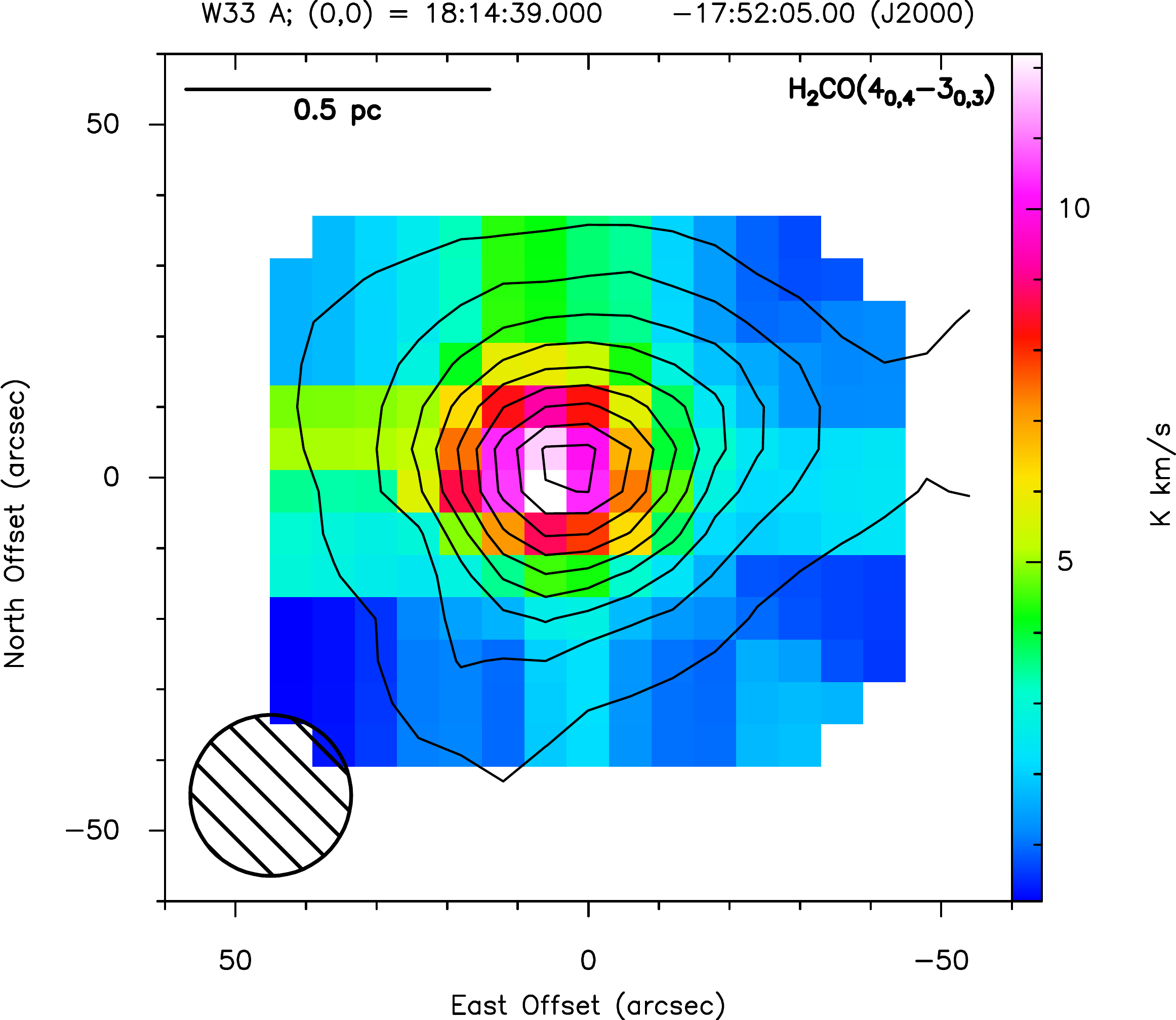}}\hspace{0.2cm}	
	\subfloat{\includegraphics[width=9cm]{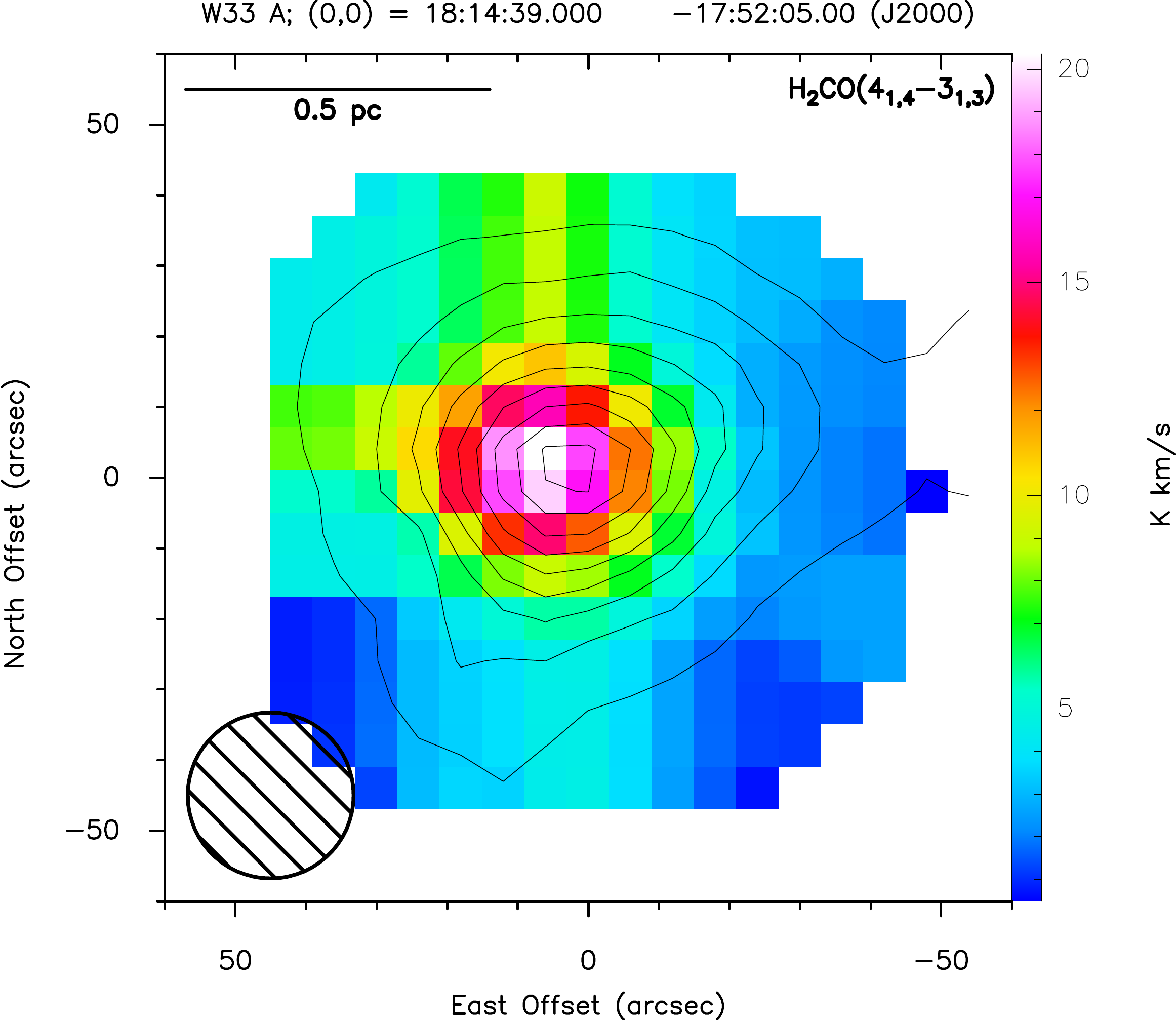}}\\	
	\subfloat{\includegraphics[width=9cm]{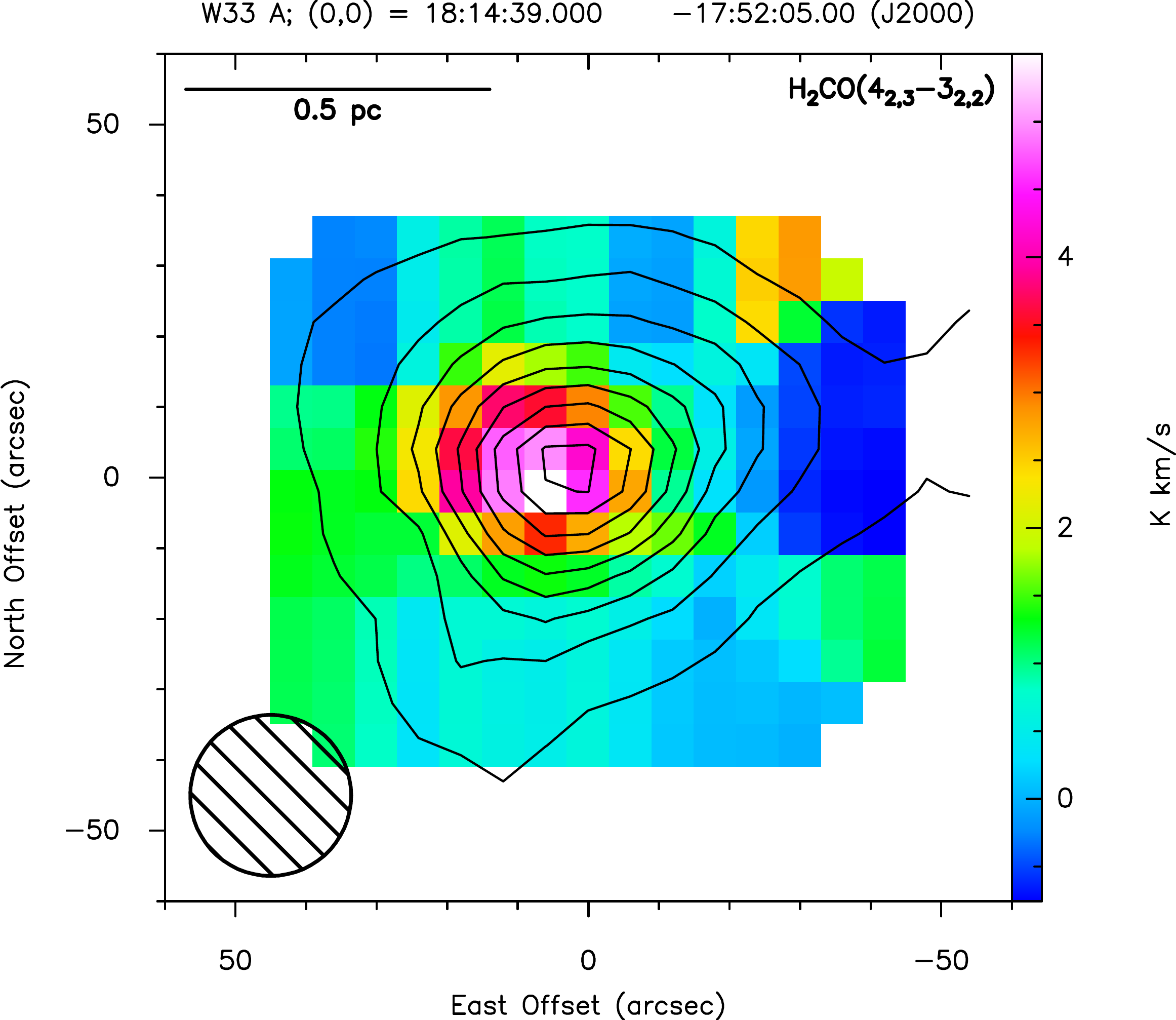}}\hspace{0.2cm}	
	\subfloat{\includegraphics[width=9cm]{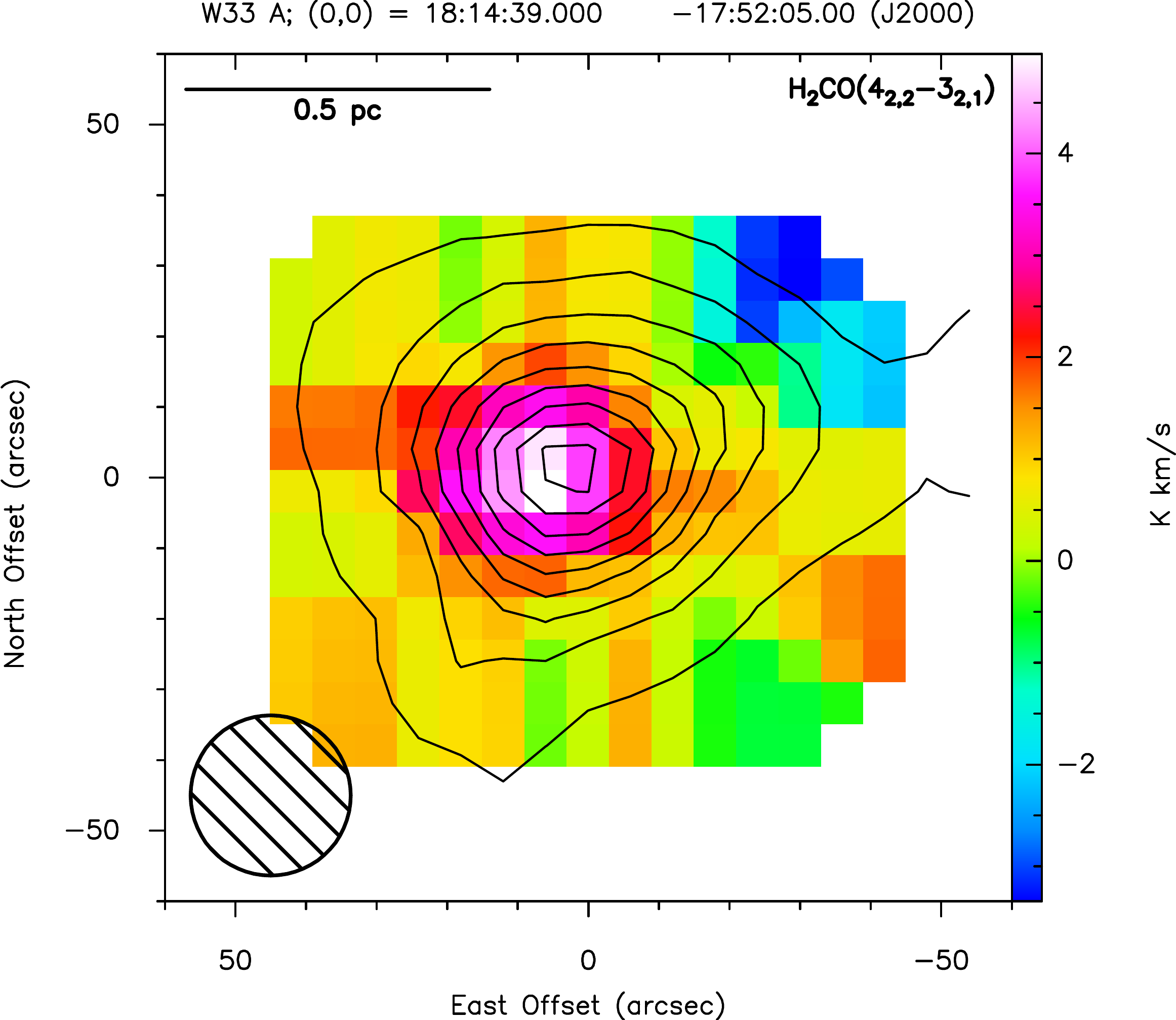}}\\
	\subfloat{\includegraphics[width=9cm]{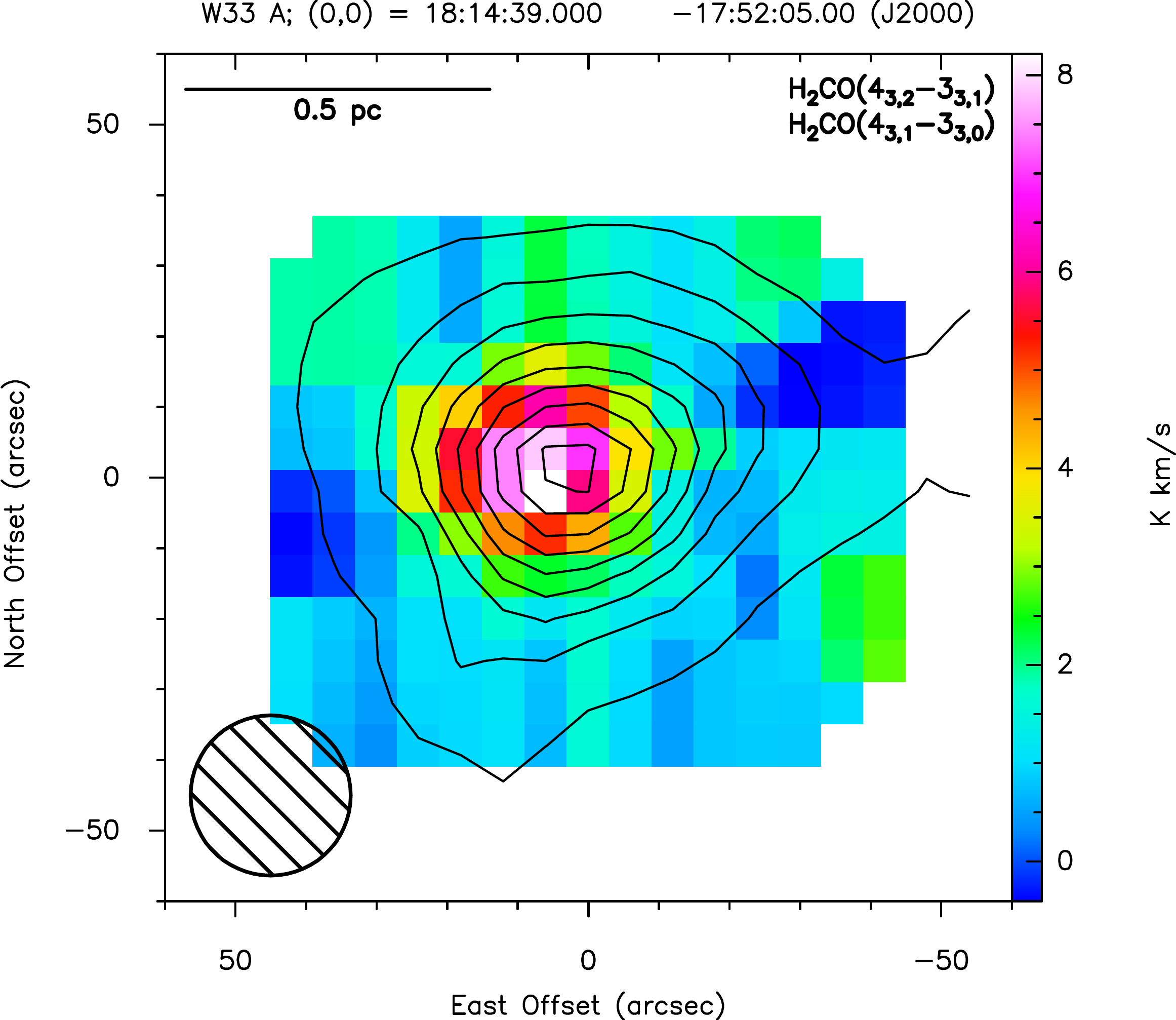}}	
\end{figure*}

\clearpage

\begin{figure*}
	\caption{Line emission of detected transitions in W33\,Main. The contours show the ATLASGAL continuum emission at 345 GHz (levels in steps of 40$\sigma$, starting at 50$\sigma$ ($\sigma$ = 0.081 Jy beam$^{-1}$). The name of the each transition is shown in the upper right corner. A scale of 0.5 pc is marked in the upper left corner, and the synthesised beam is shown in the lower left corner.}
	\centering
	\subfloat{\includegraphics[width=9cm]{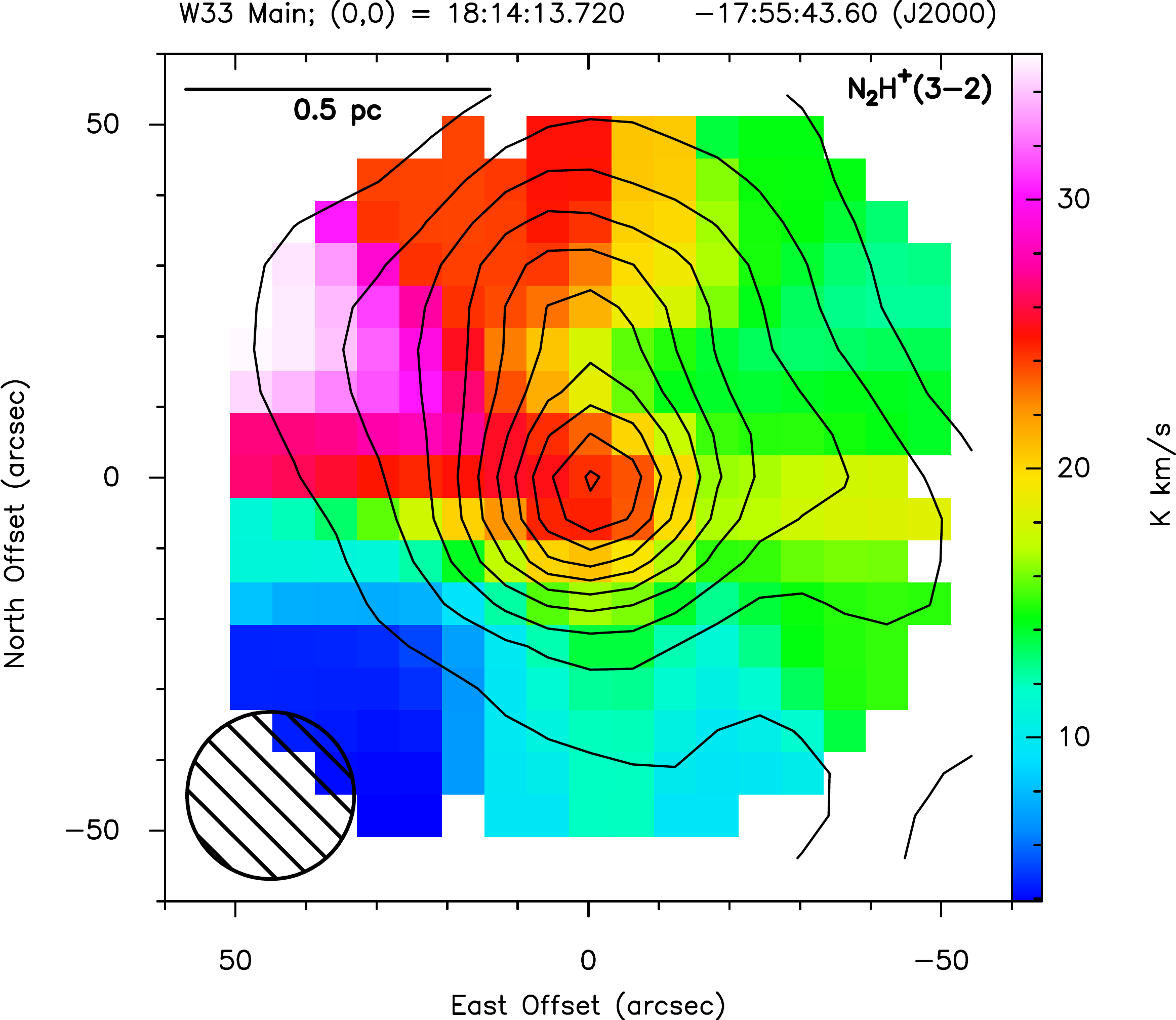}}\hspace{0.2cm}	
	\subfloat{\includegraphics[width=9cm]{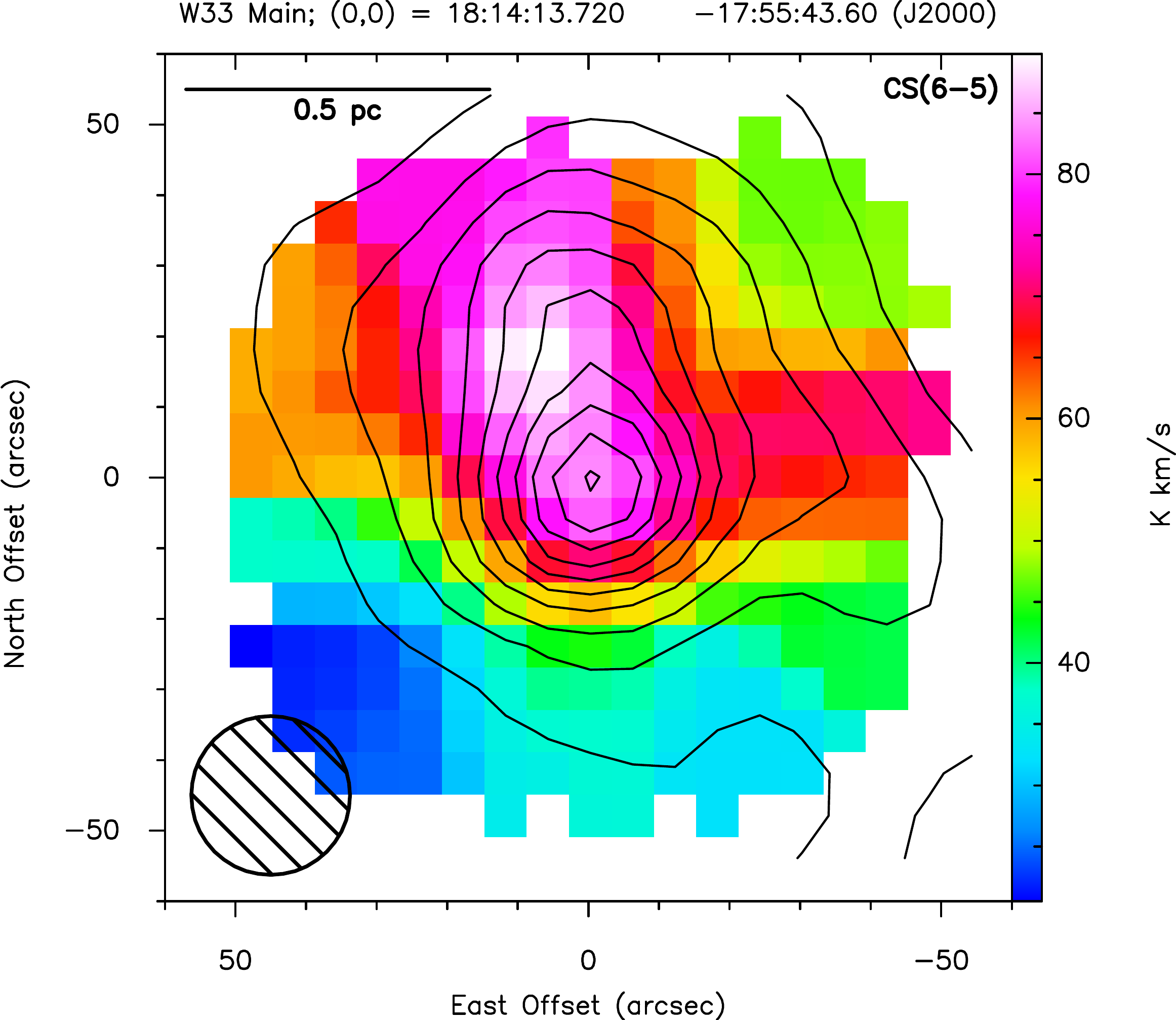}}\\		
	\subfloat{\includegraphics[width=9cm]{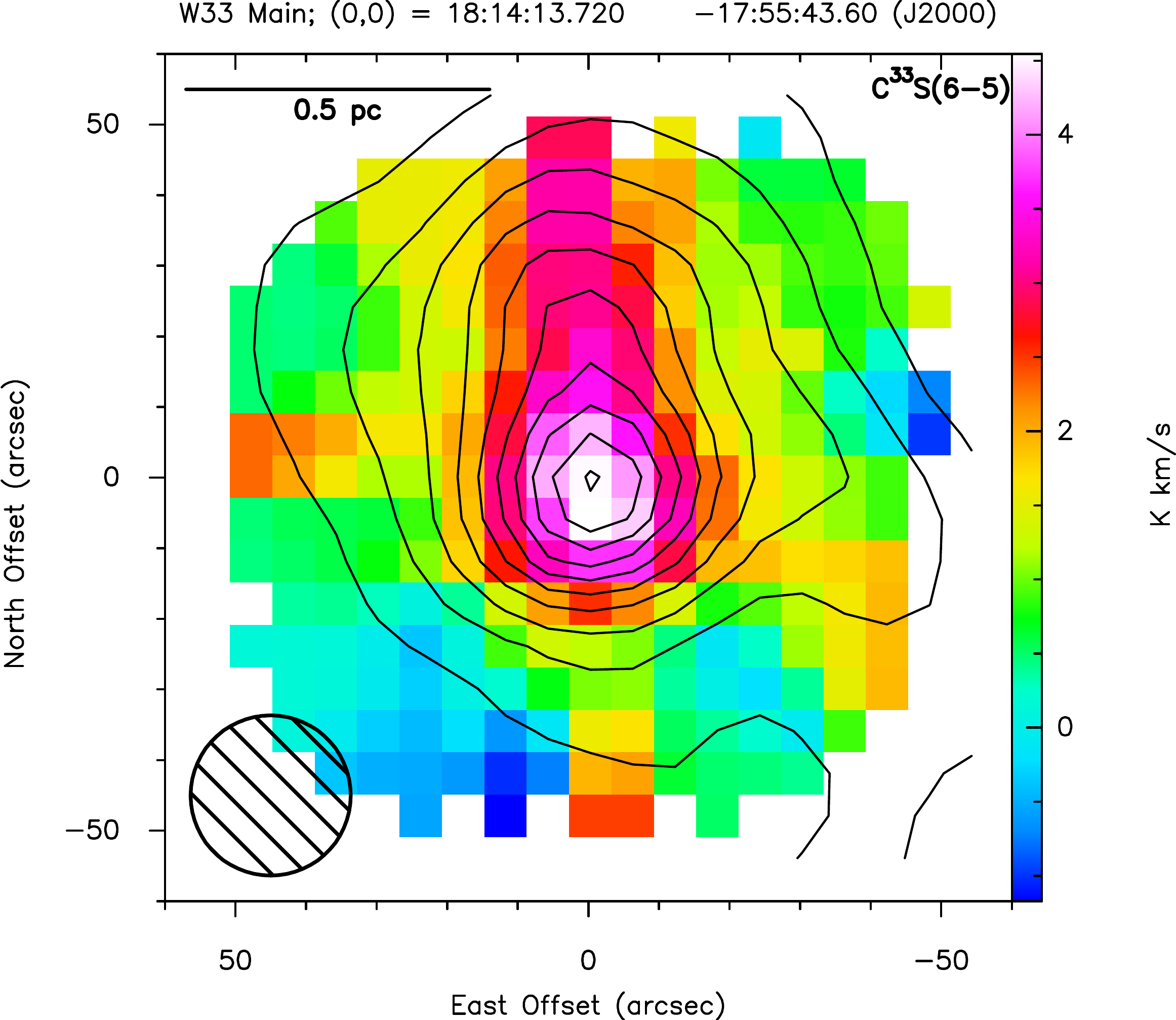}}\hspace{0.2cm}		
	\subfloat{\includegraphics[width=9cm]{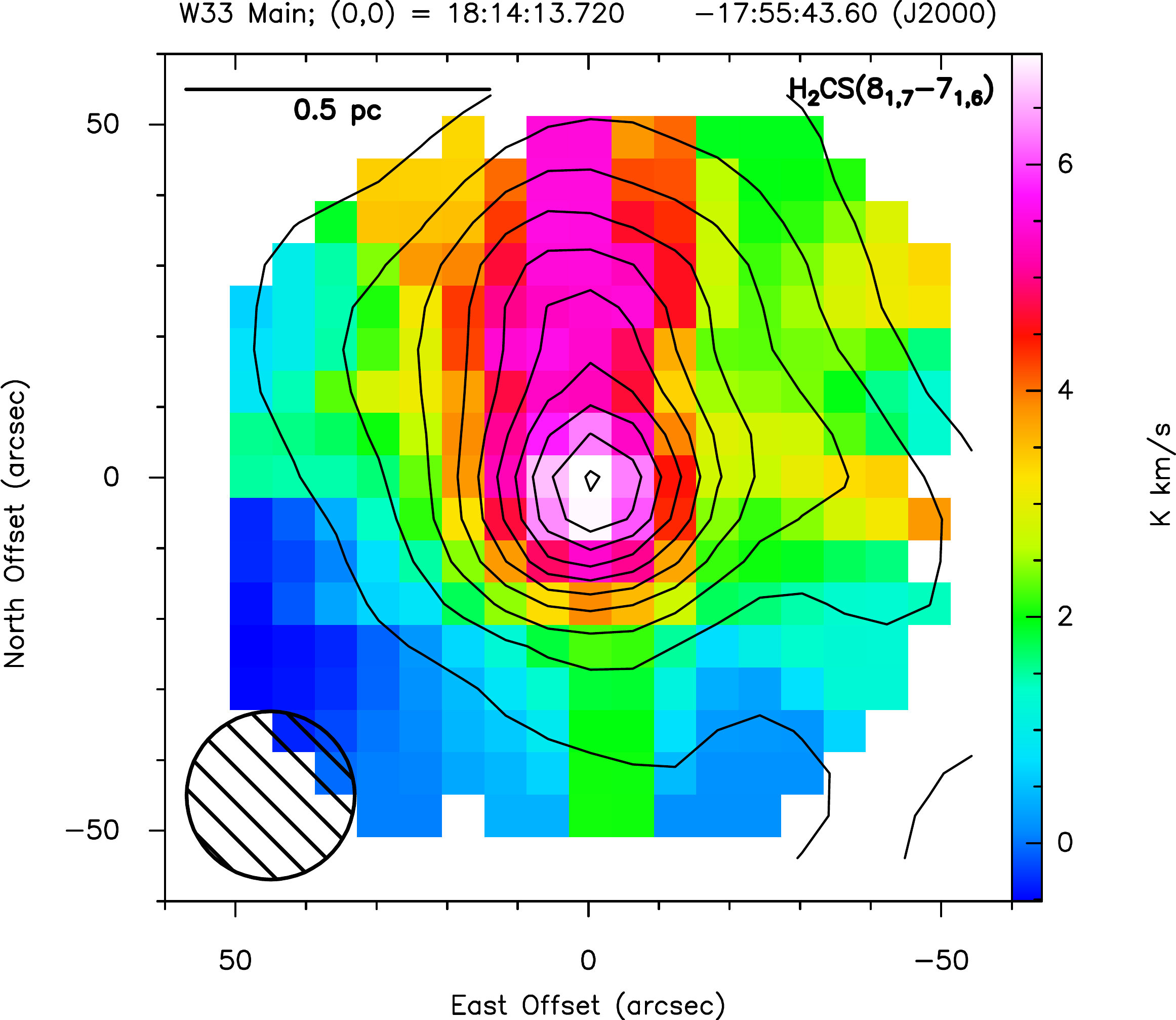}}\\		
	\subfloat{\includegraphics[width=9cm]{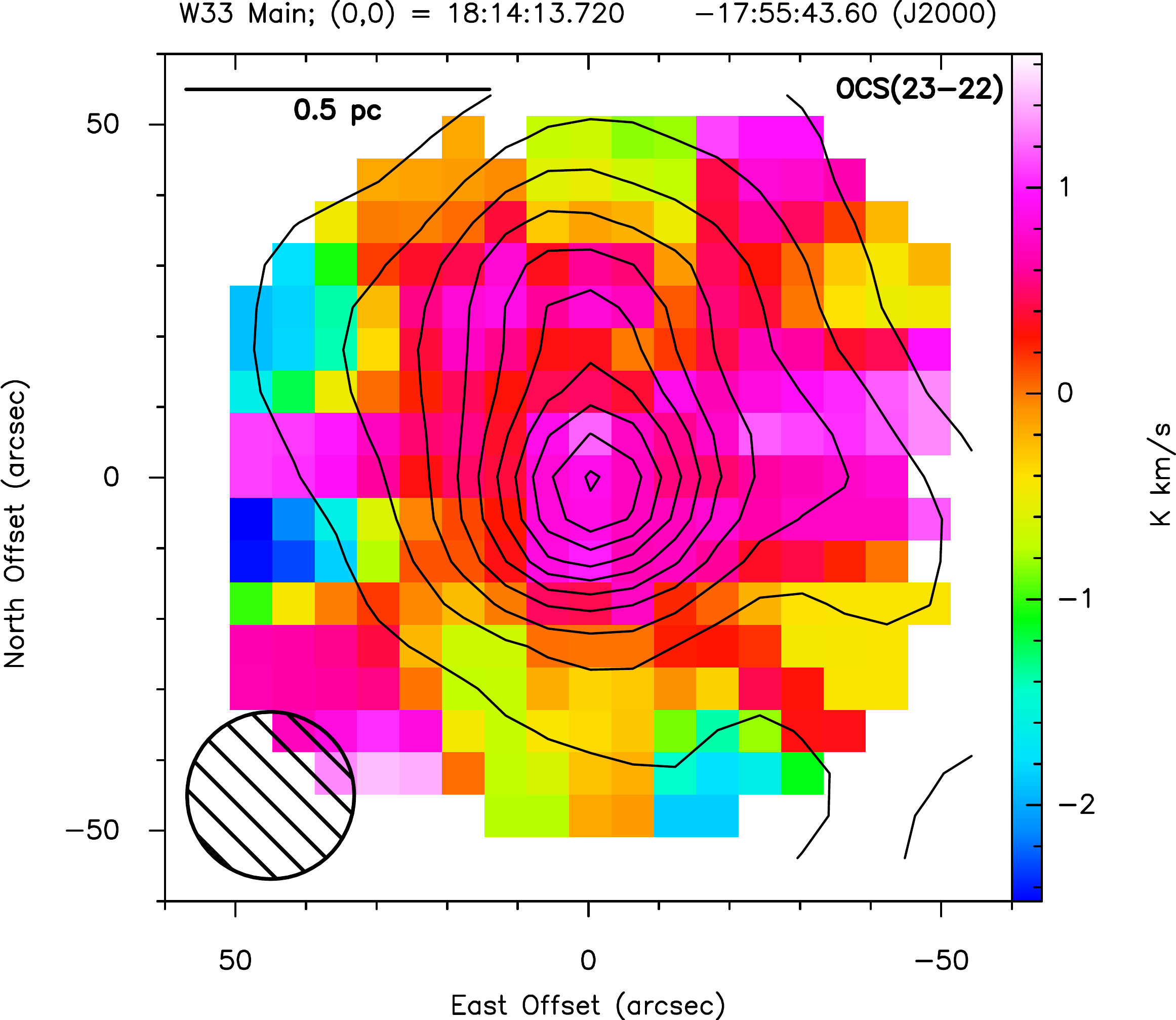}}\hspace{0.2cm}		
	\subfloat{\includegraphics[width=9cm]{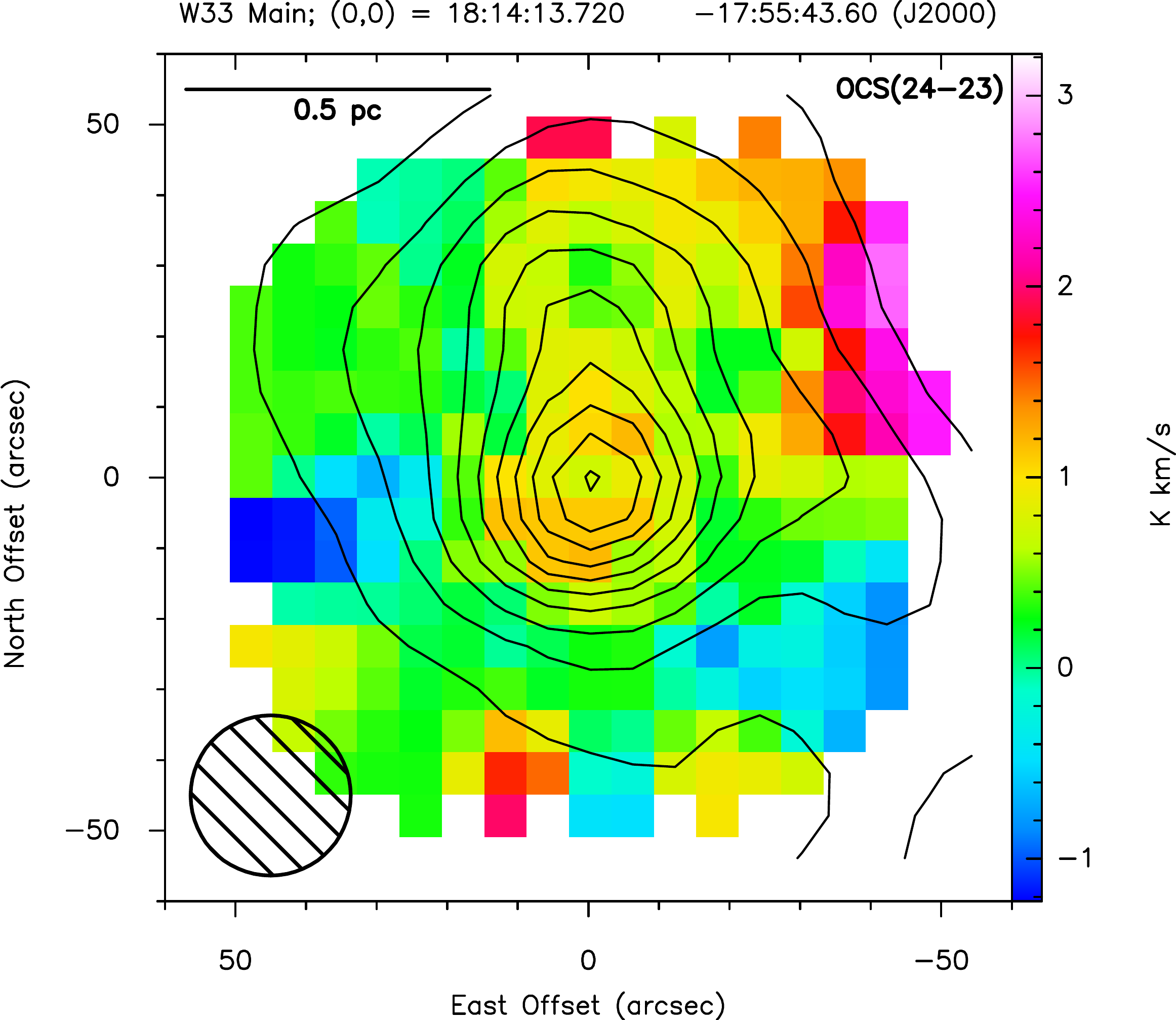}}		
	\label{W33M-APEX-IntInt}
\end{figure*}

\clearpage

\addtocounter{figure}{-1}
\begin{figure*}
	\centering
	\caption{Continued.}
	\subfloat{\includegraphics[width=9cm]{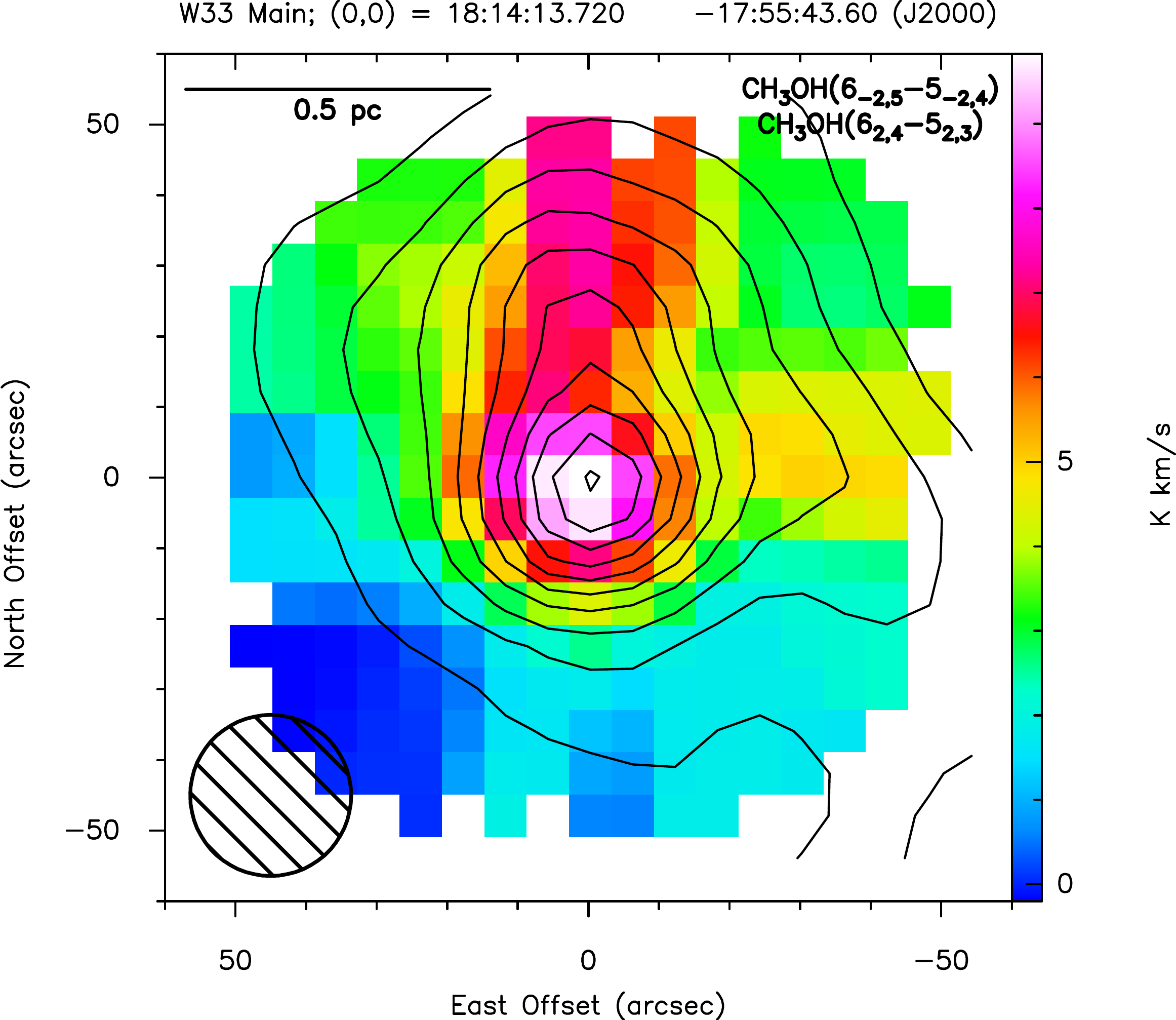}}\hspace{0.2cm}	
	\subfloat{\includegraphics[width=9cm]{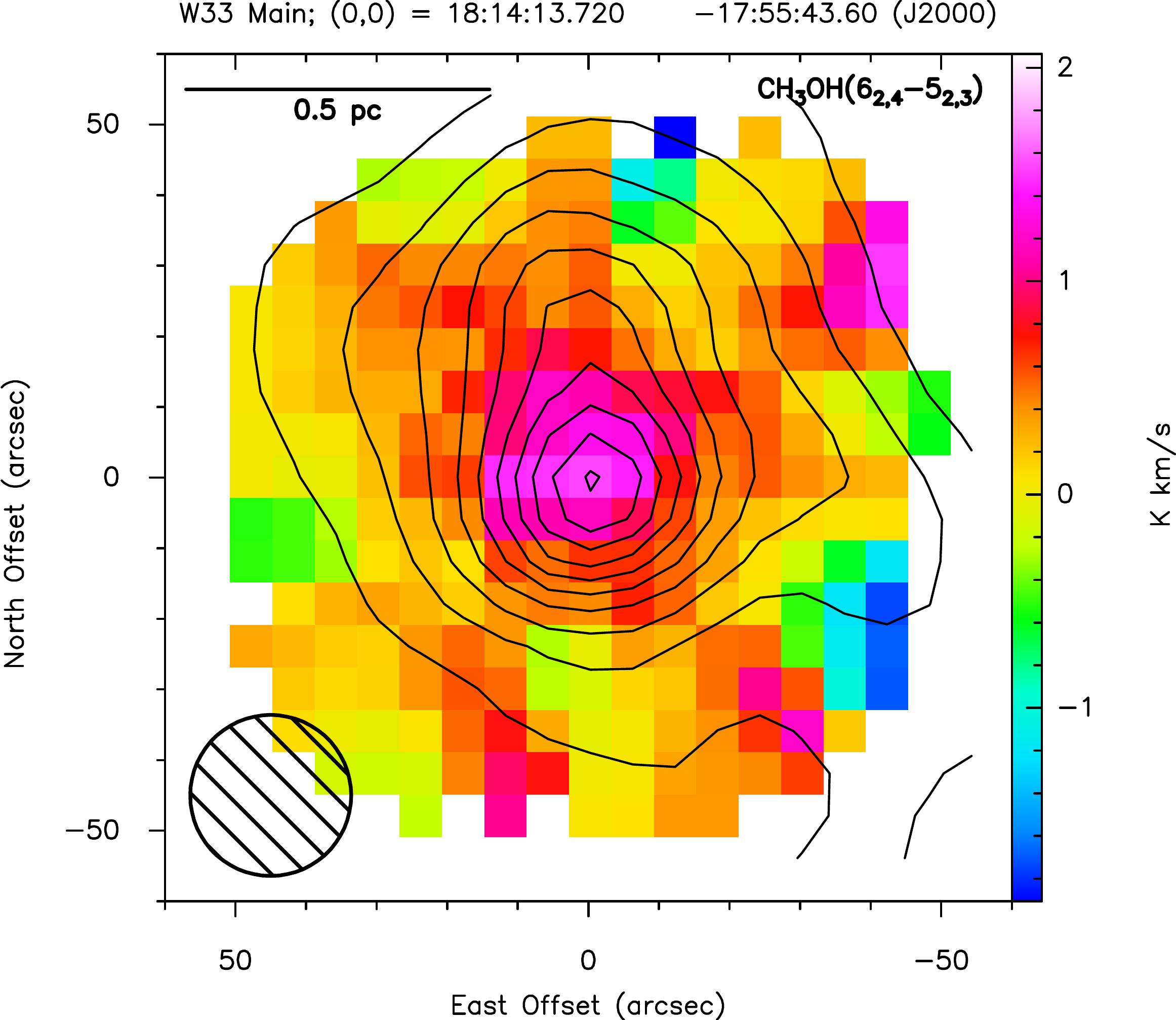}}\\
	\subfloat{\includegraphics[width=9cm]{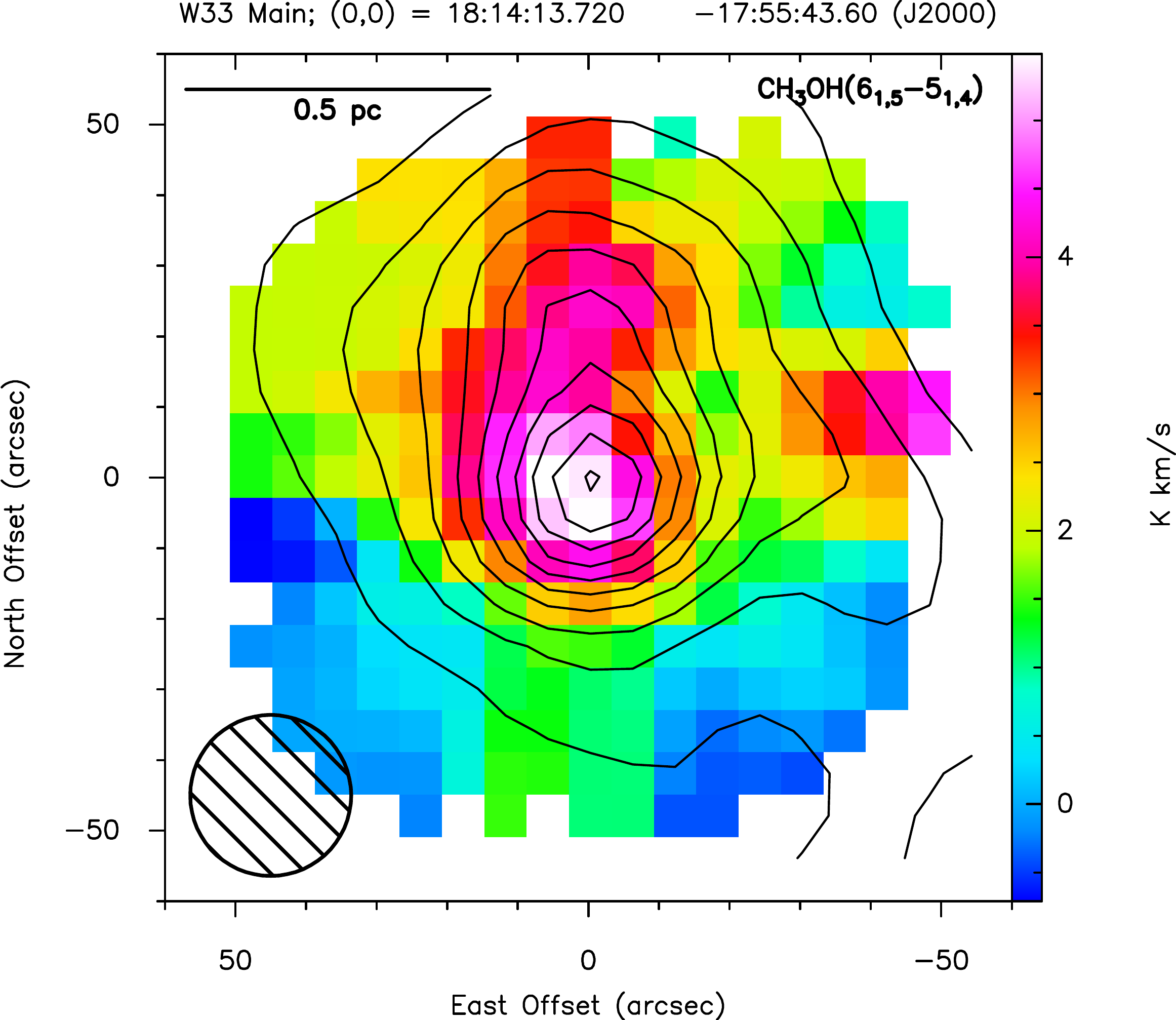}}\hspace{0.2cm}	
	\subfloat{\includegraphics[width=9cm]{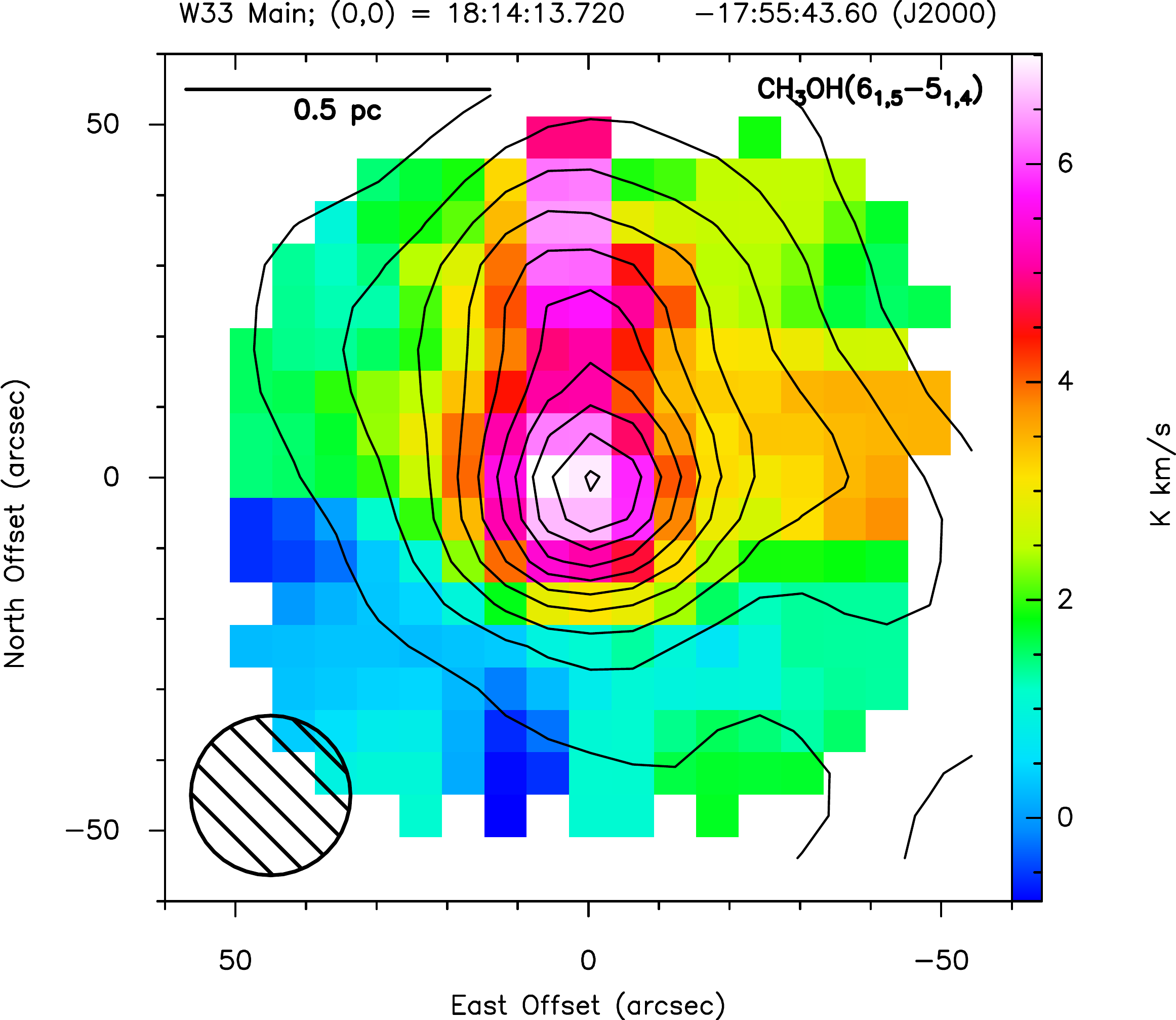}}\\		
	\subfloat{\includegraphics[width=9cm]{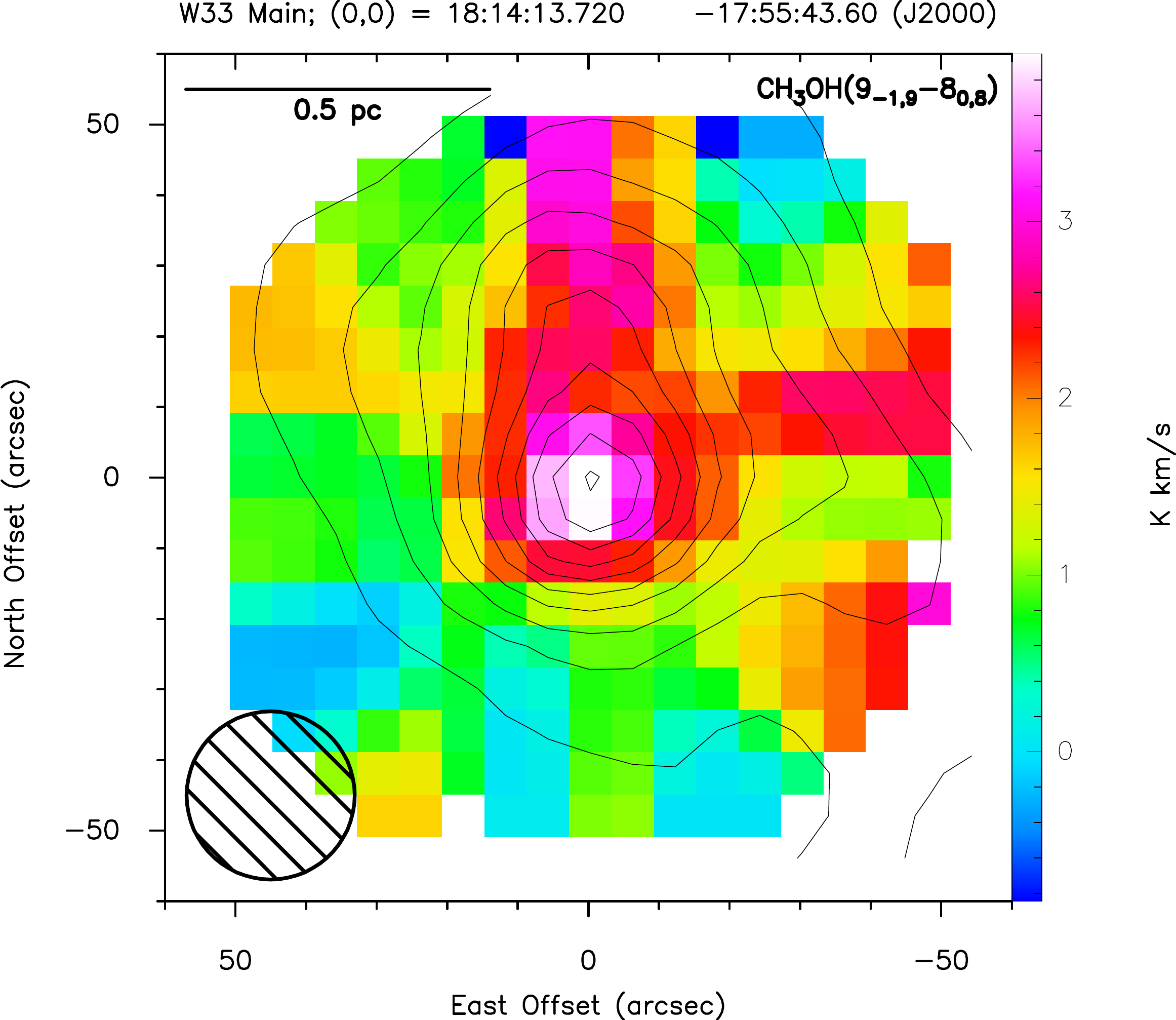}}\hspace{0.2cm}	
	\subfloat{\includegraphics[width=9cm]{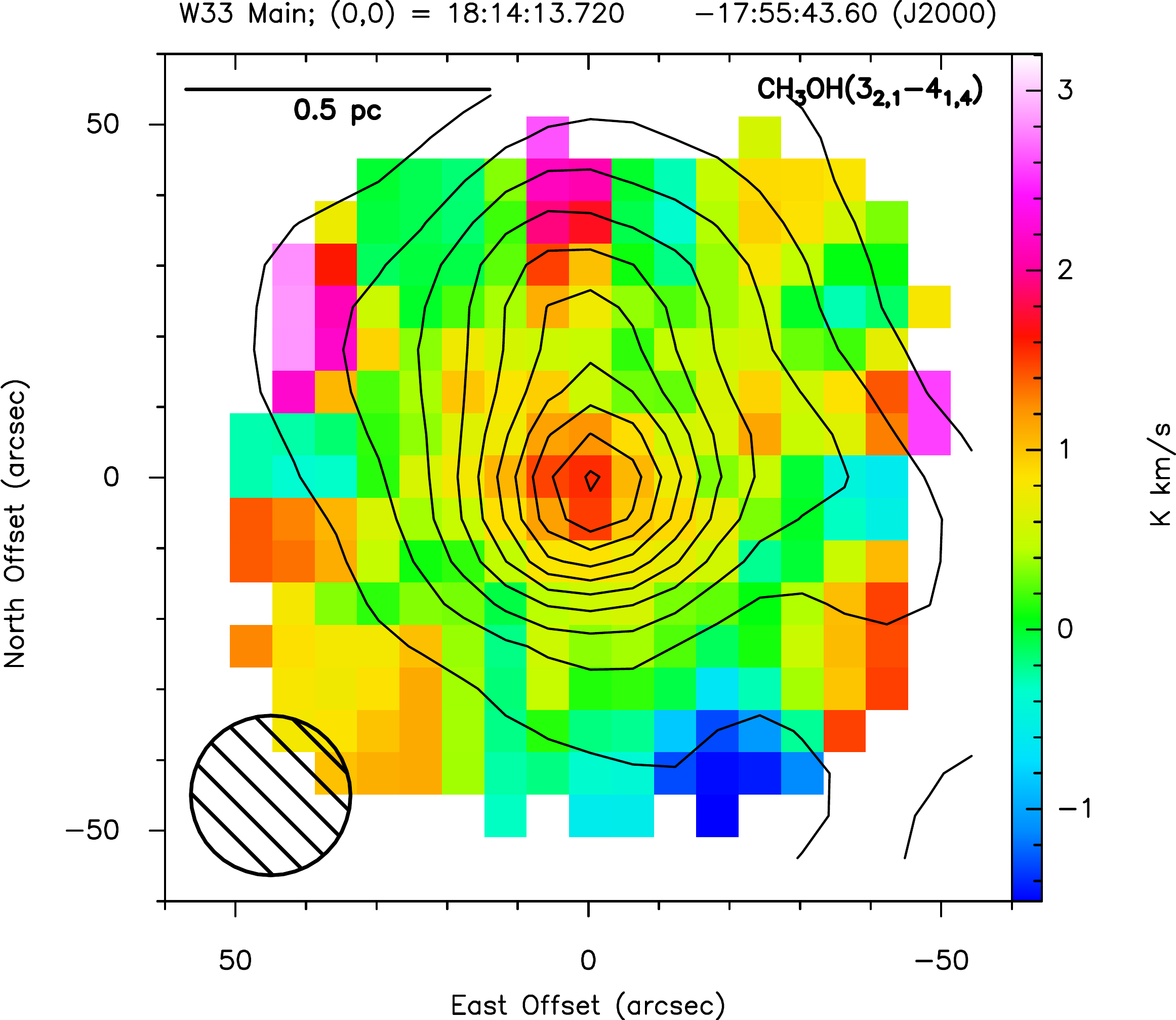}}	
\end{figure*}

\addtocounter{figure}{-1}
\begin{figure*}
	\centering
	\caption{Continued.}
	\subfloat{\includegraphics[width=9cm]{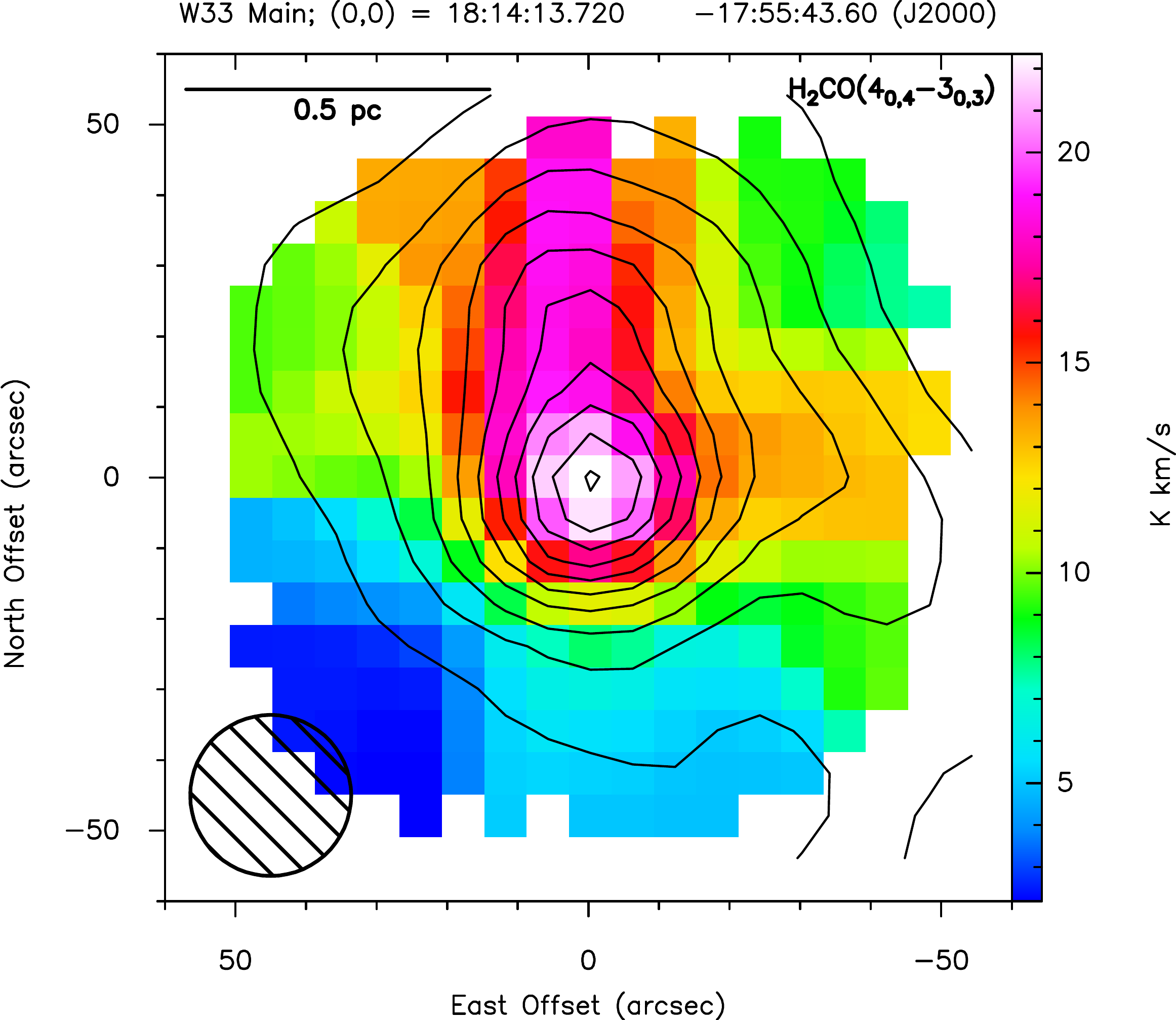}}\hspace{0.2cm}	
	\subfloat{\includegraphics[width=9cm]{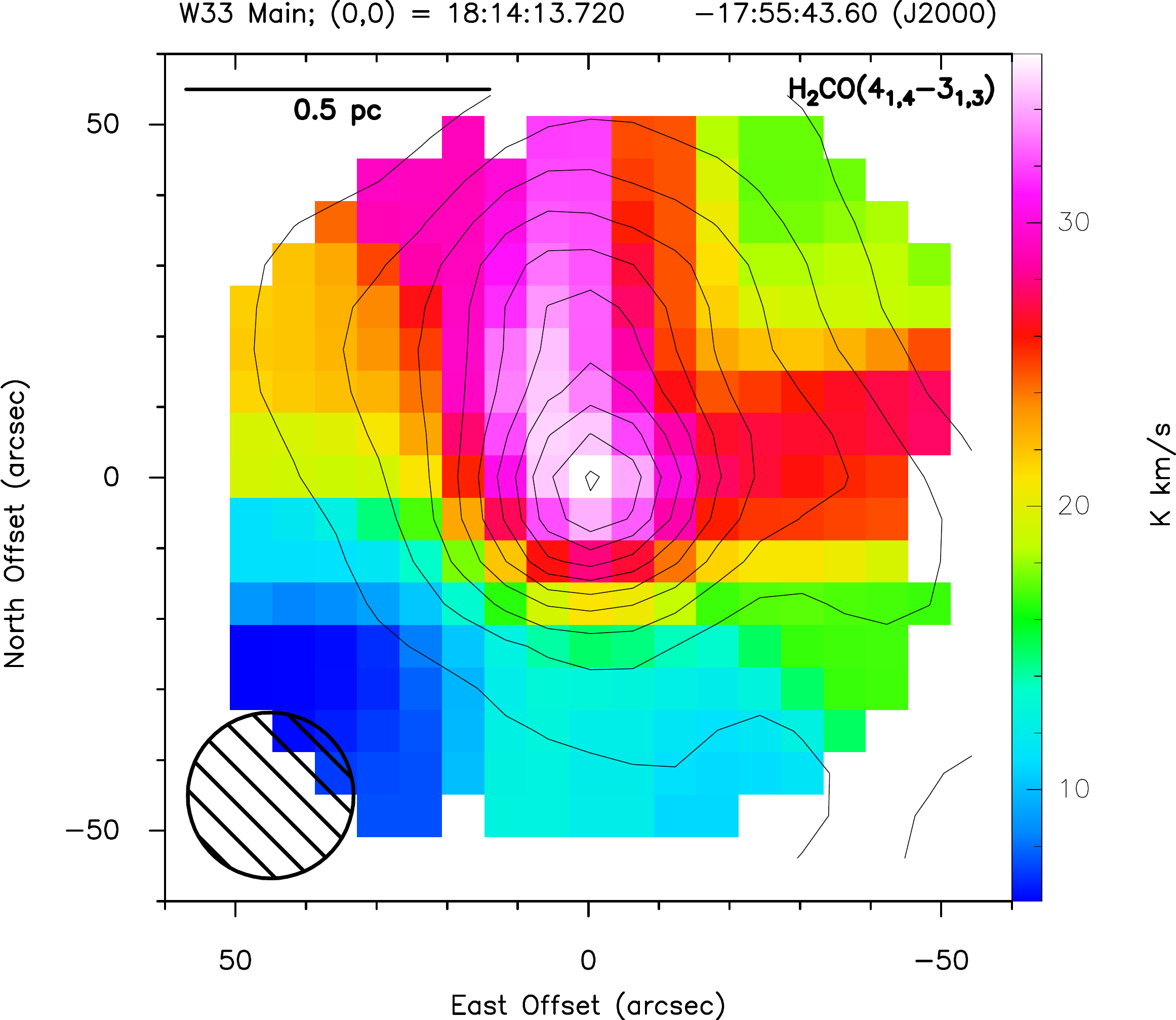}}\\	
	\subfloat{\includegraphics[width=9cm]{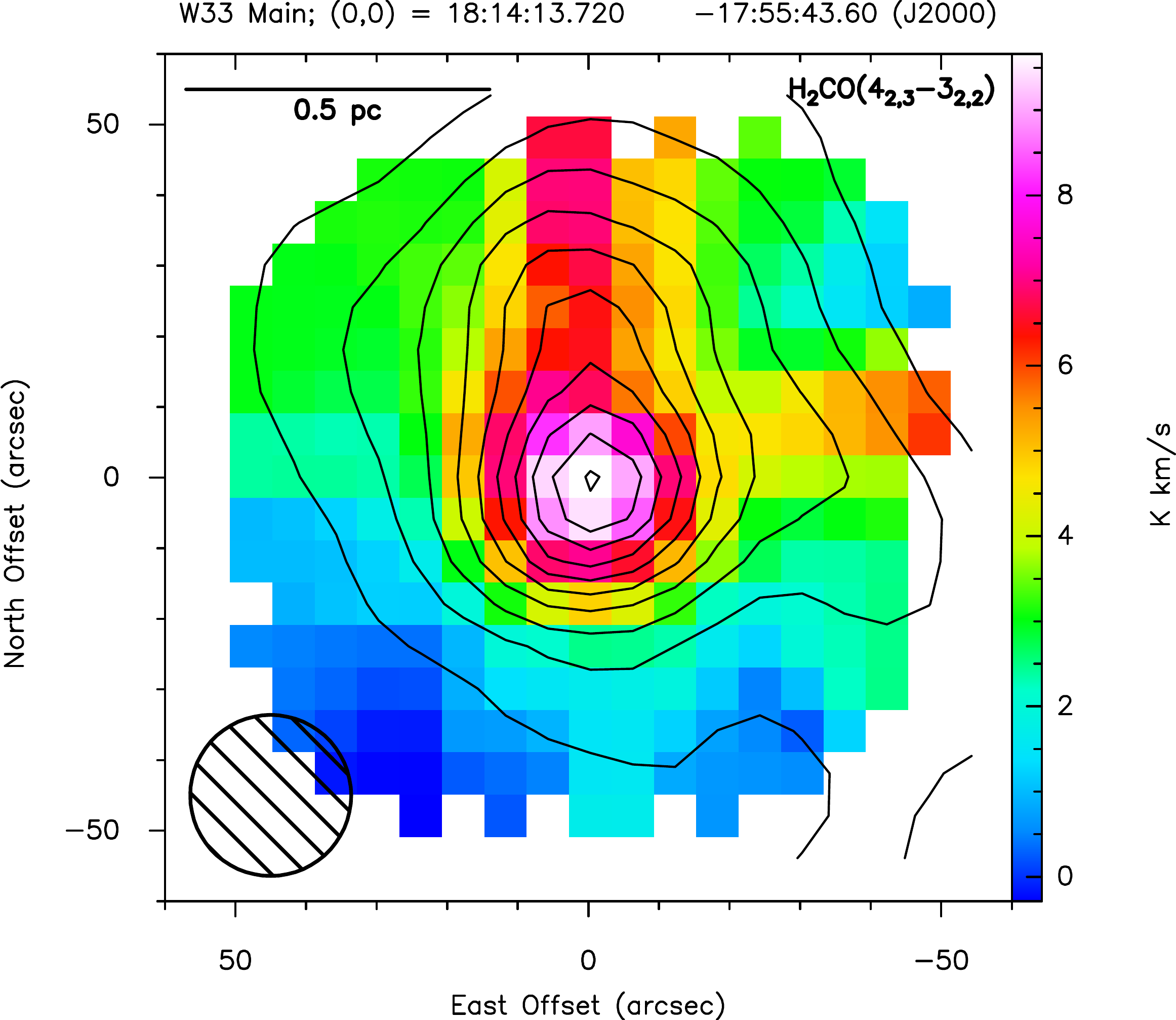}}\hspace{0.2cm}	
	\subfloat{\includegraphics[width=9cm]{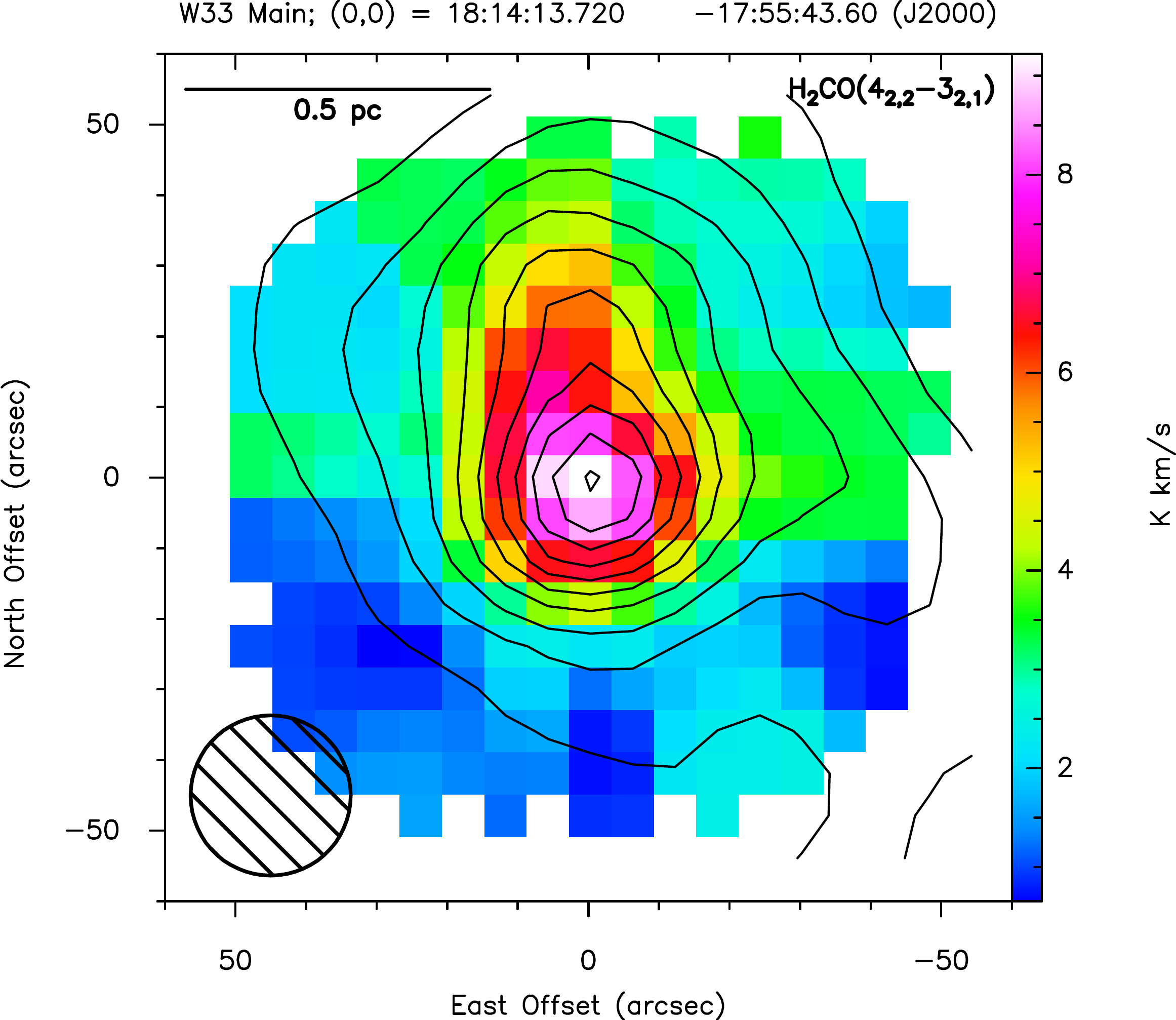}}\\	
	\subfloat{\includegraphics[width=9cm]{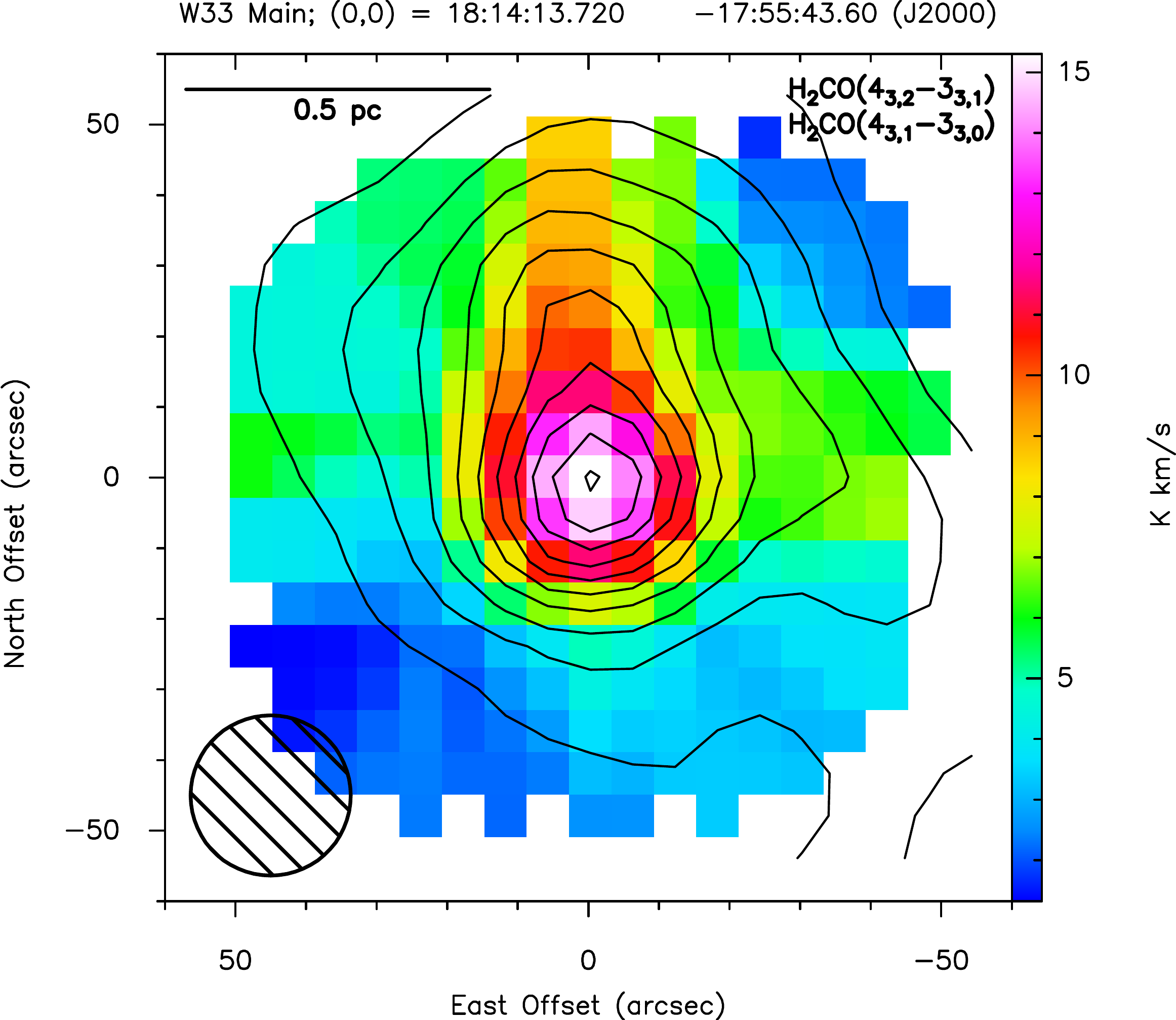}}\hspace{0.2cm}
	\subfloat{\includegraphics[width=9cm]{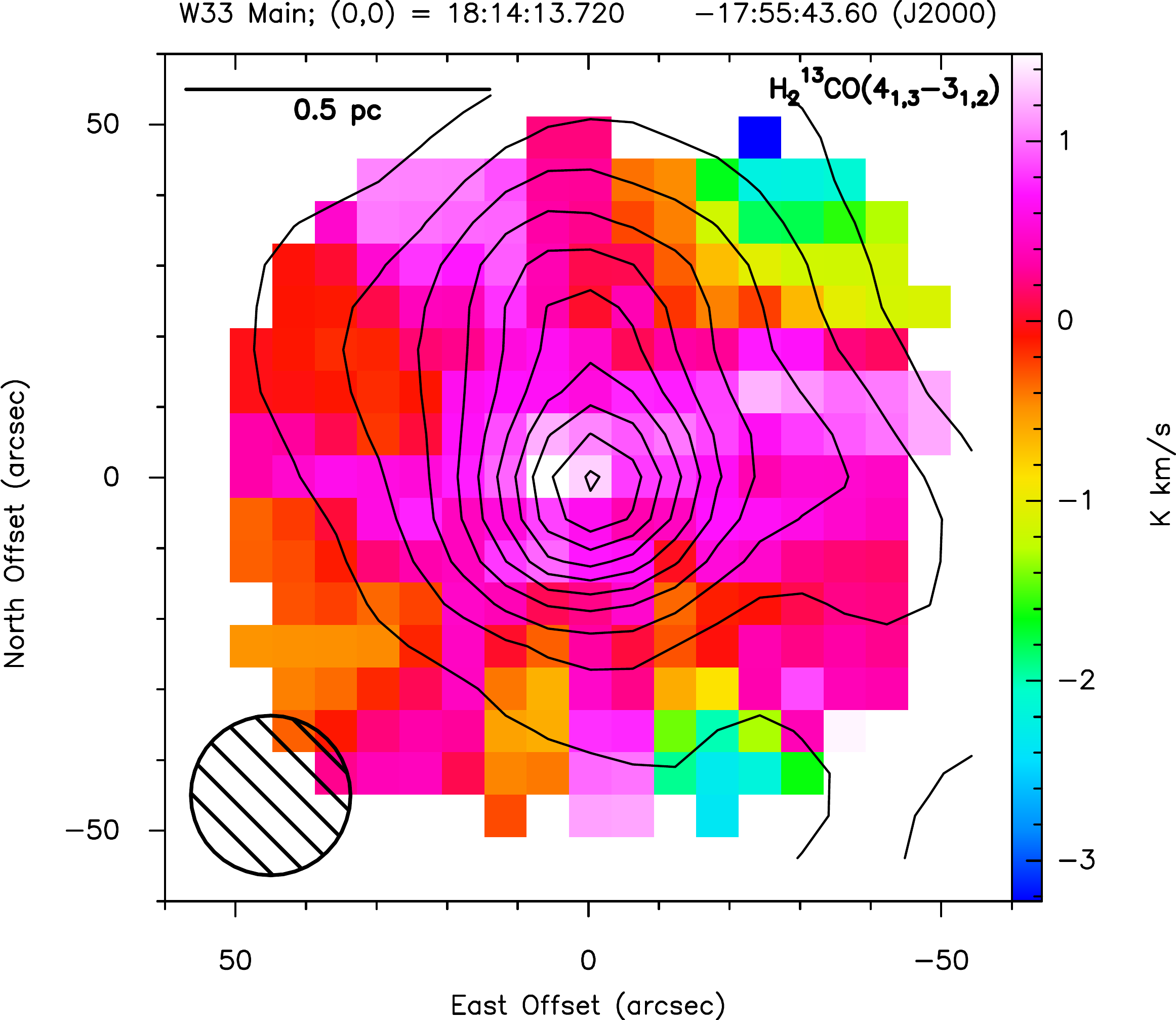}}			
\end{figure*}

\addtocounter{figure}{-1}
\begin{figure*}
	\centering
	\caption{Continued.}
	\subfloat{\includegraphics[width=9cm]{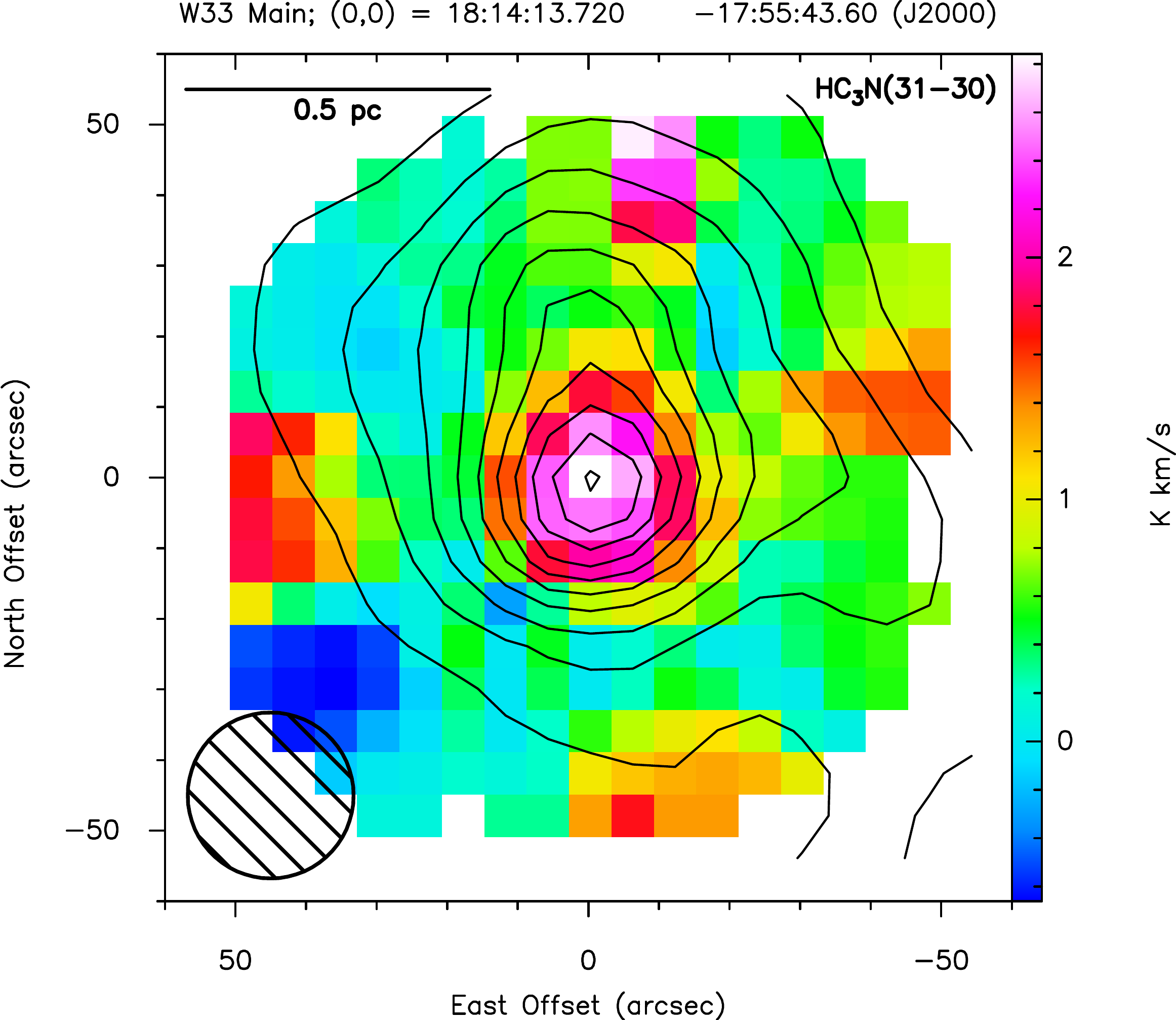}}\hspace{0.2cm}
	\subfloat{\includegraphics[width=9cm]{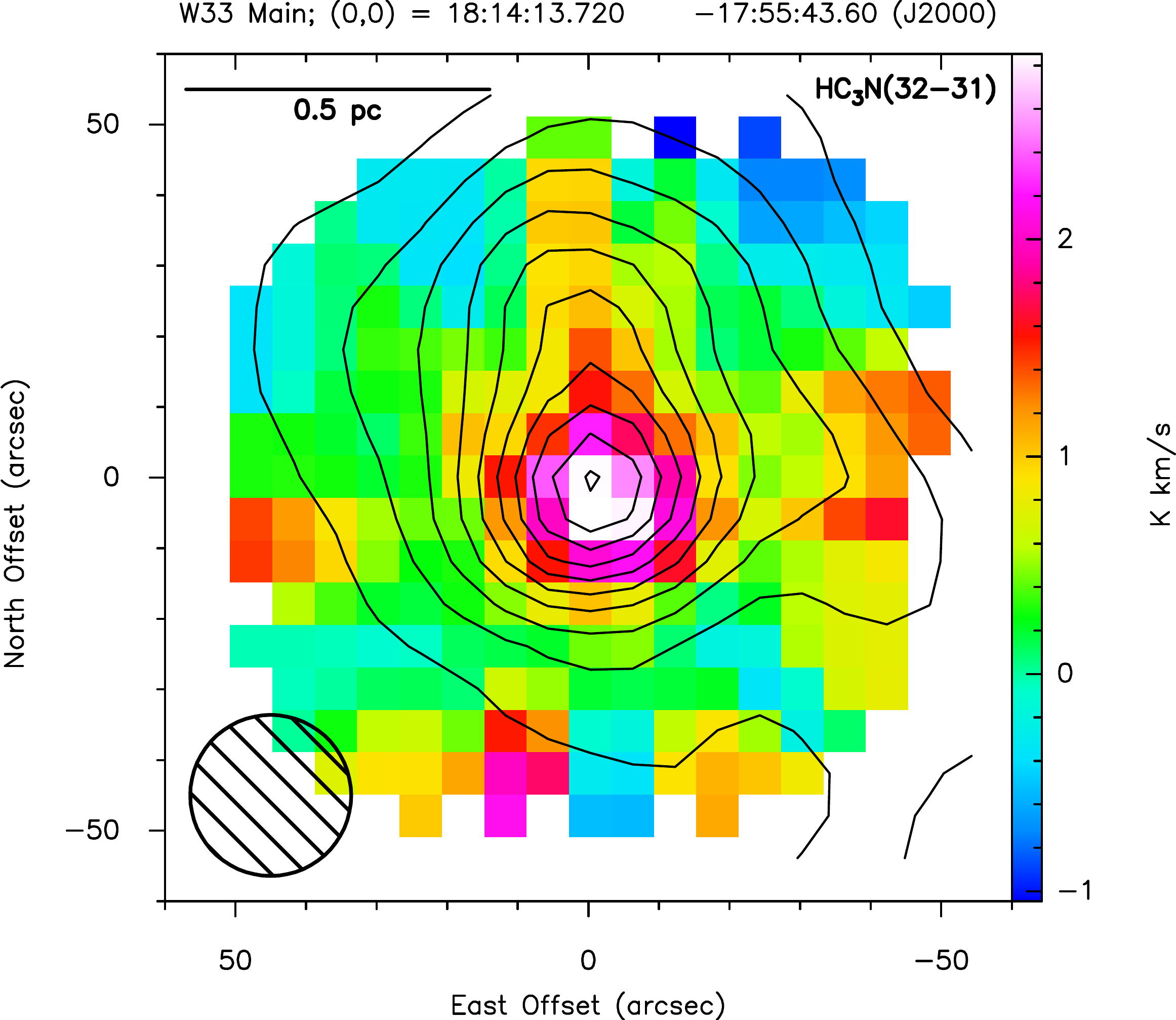}}\\
	\subfloat{\includegraphics[width=9cm]{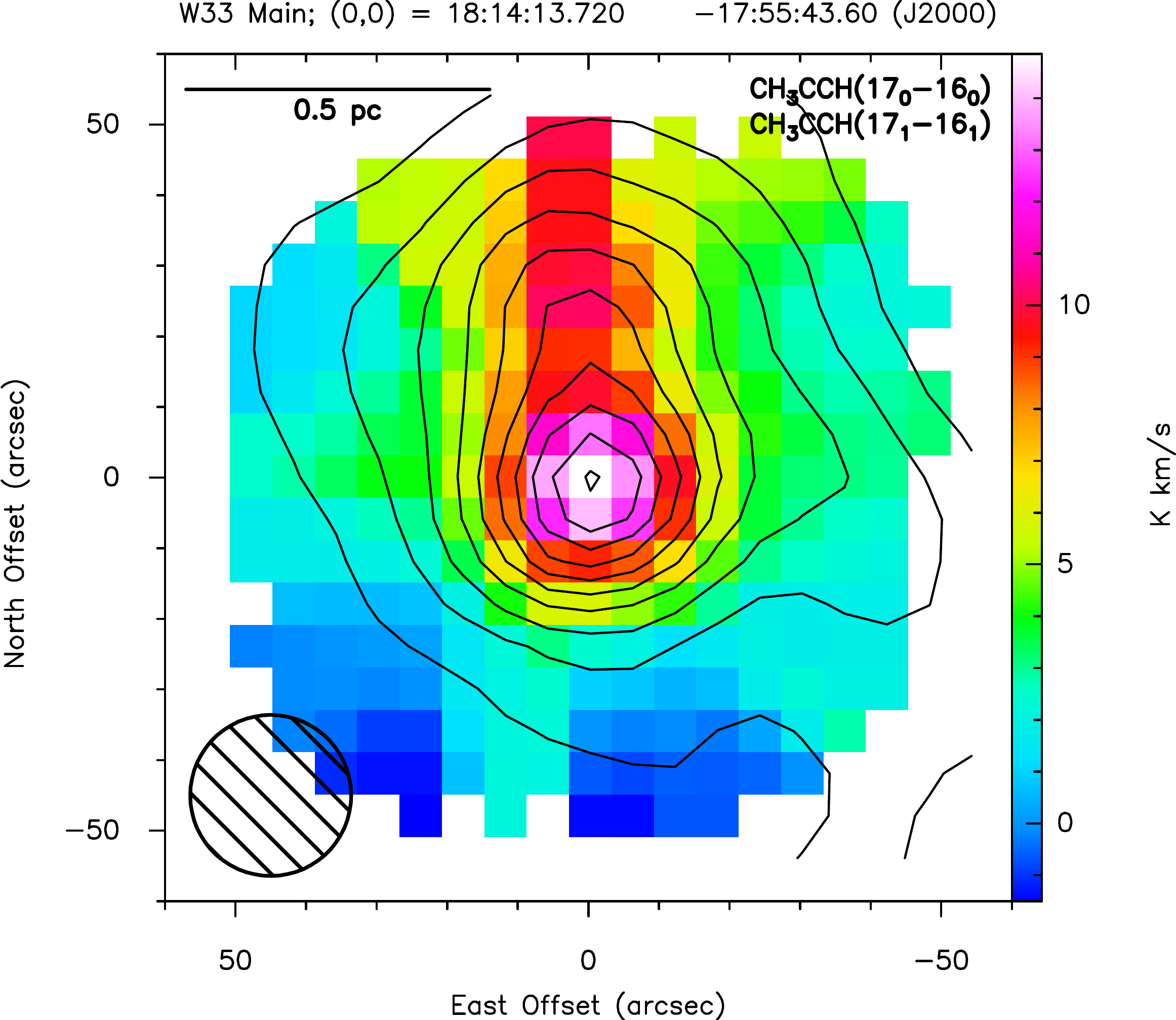}}\hspace{0.2cm}		
	\subfloat{\includegraphics[width=9cm]{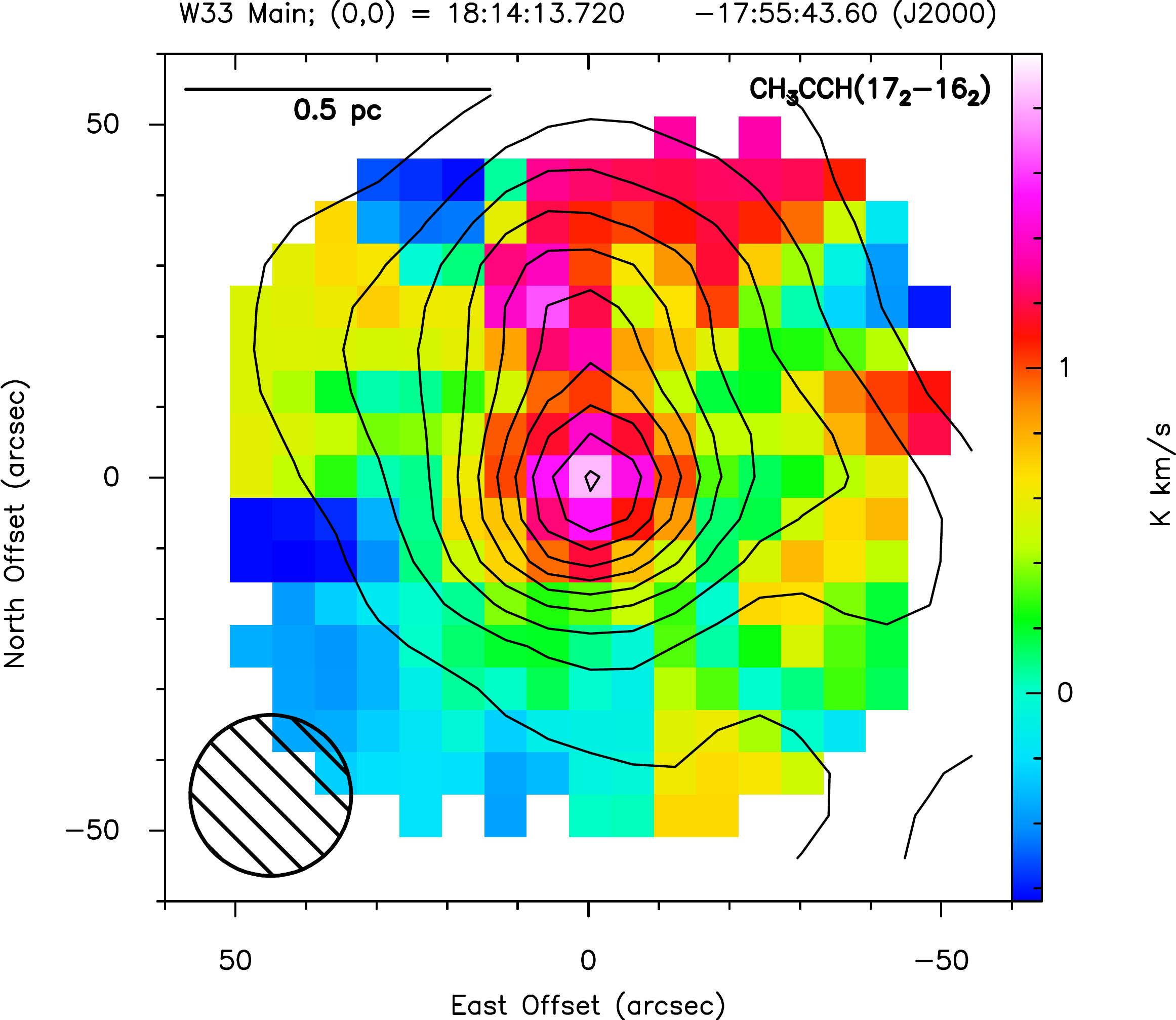}}\\		
	\subfloat{\includegraphics[width=9cm]{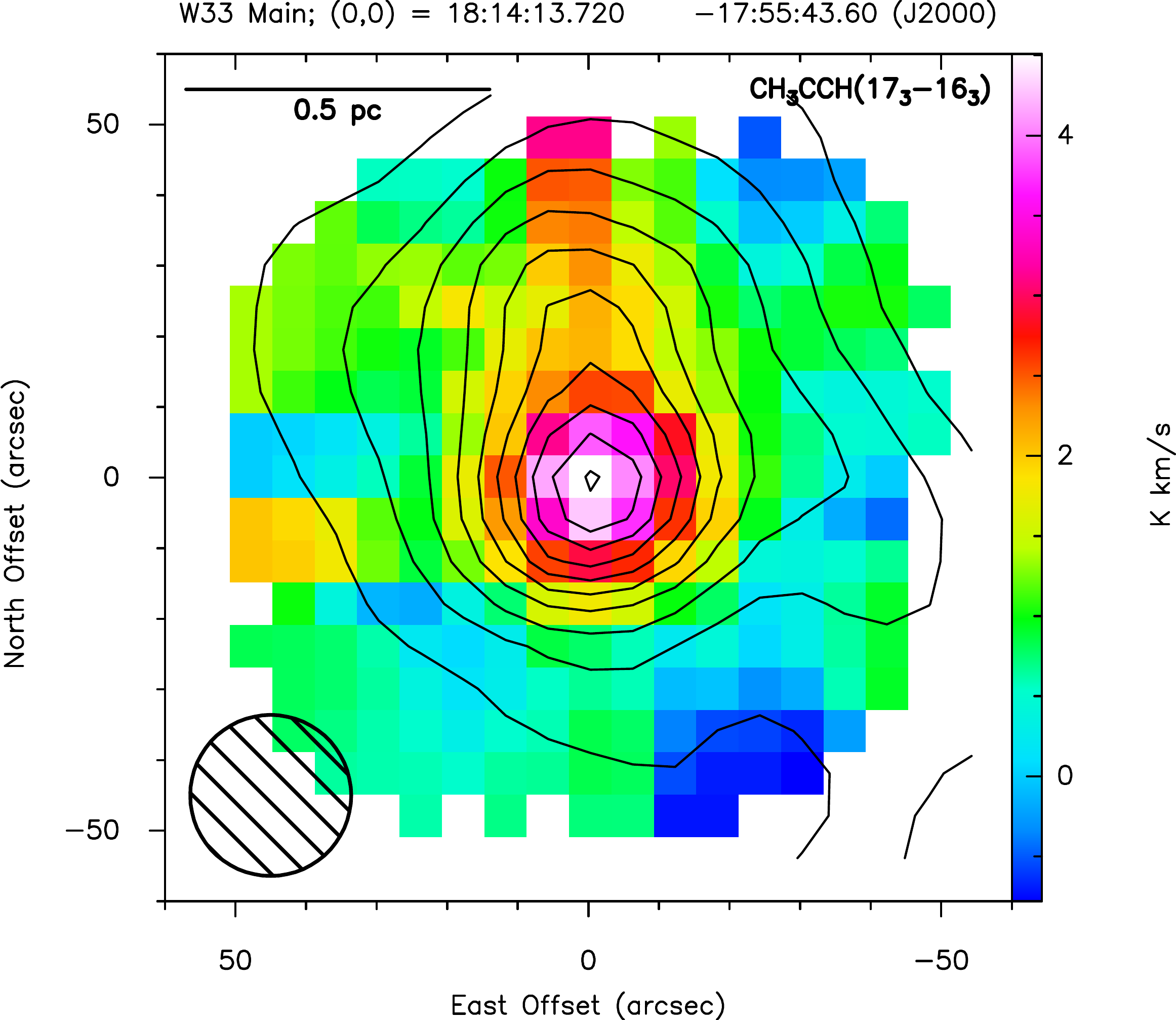}}\hspace{0.2cm}		
	\subfloat{\includegraphics[width=9cm]{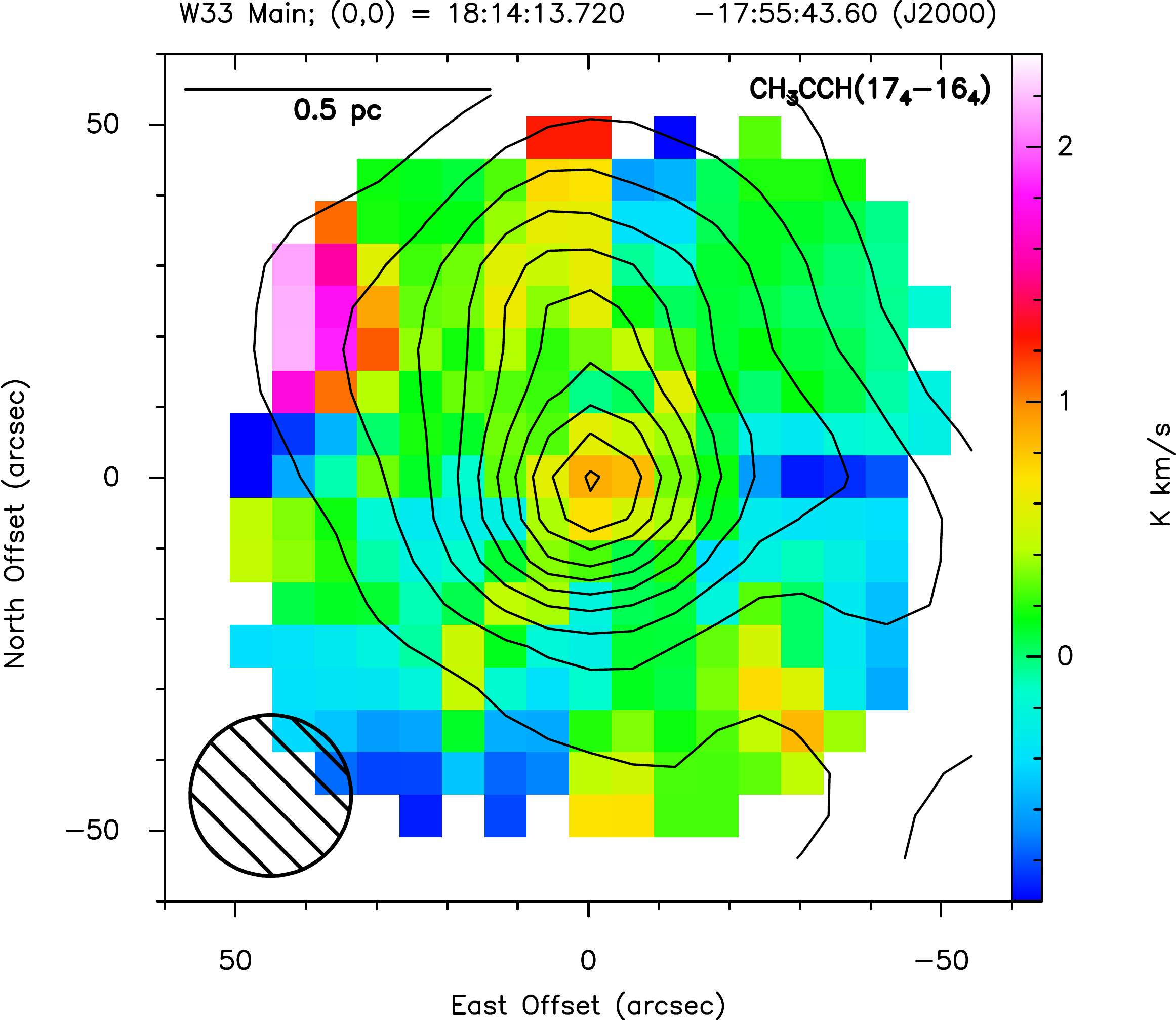}}		
\end{figure*}

\clearpage

\section{Results of the SMA line observations}
\label{SMAResApp}

\begin{figure*}
	\caption{Line emission of detected transitions in W33\,Main1. The contours show the SMA continuum emission at 230 GHz (same contour levels as in Fig. \ref{W33_SMA_CH0}). The name of the each transition is shown in the upper right corner. A scale of 0.1 pc is marked in the upper left corner, and the synthesised beam is shown in the lower left corner.}
	\centering
	\subfloat{\includegraphics[width=9cm]{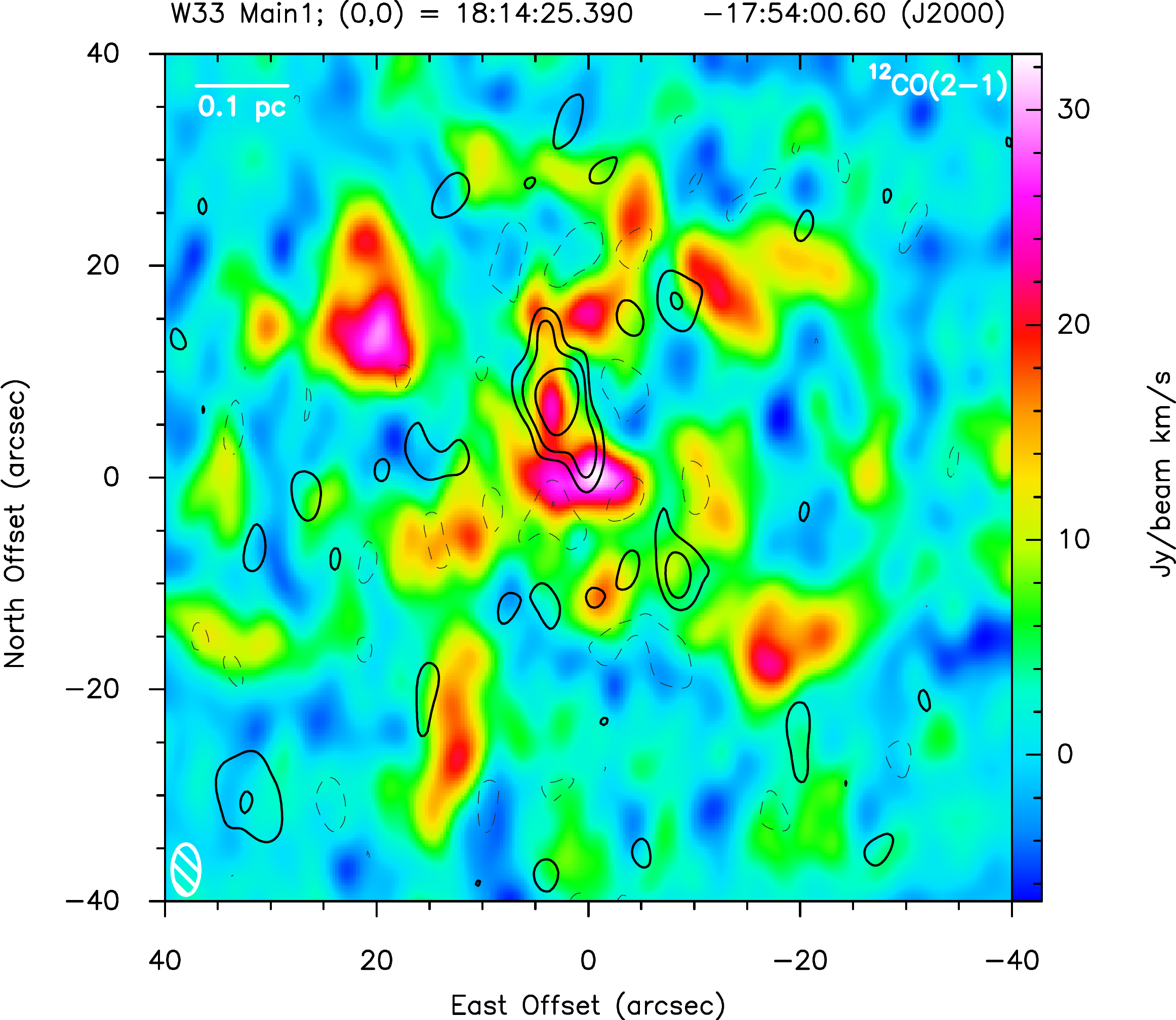}}\hspace{0.2cm}	
	\subfloat{\includegraphics[width=9cm]{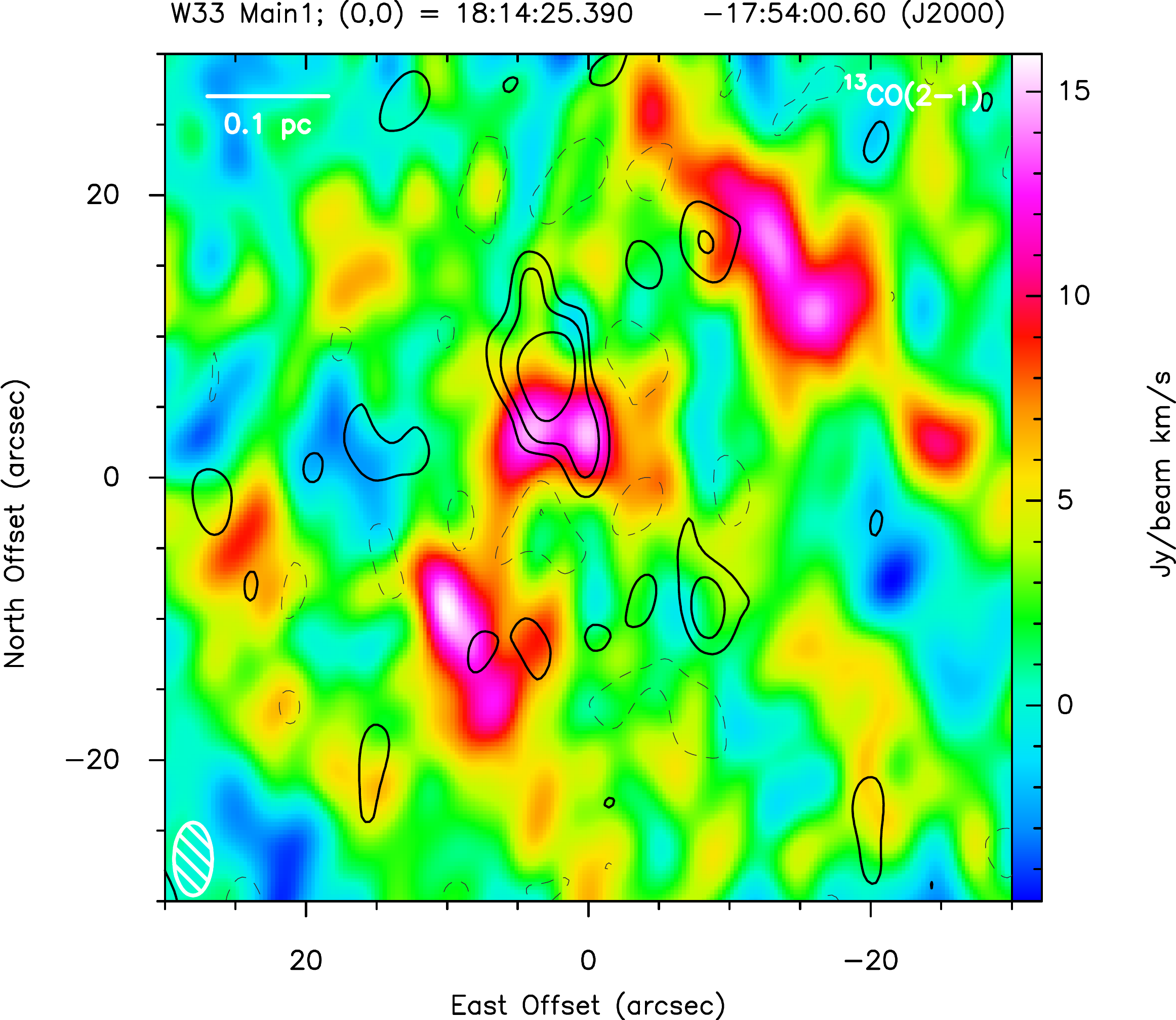}}\\	
	\subfloat{\includegraphics[width=9cm]{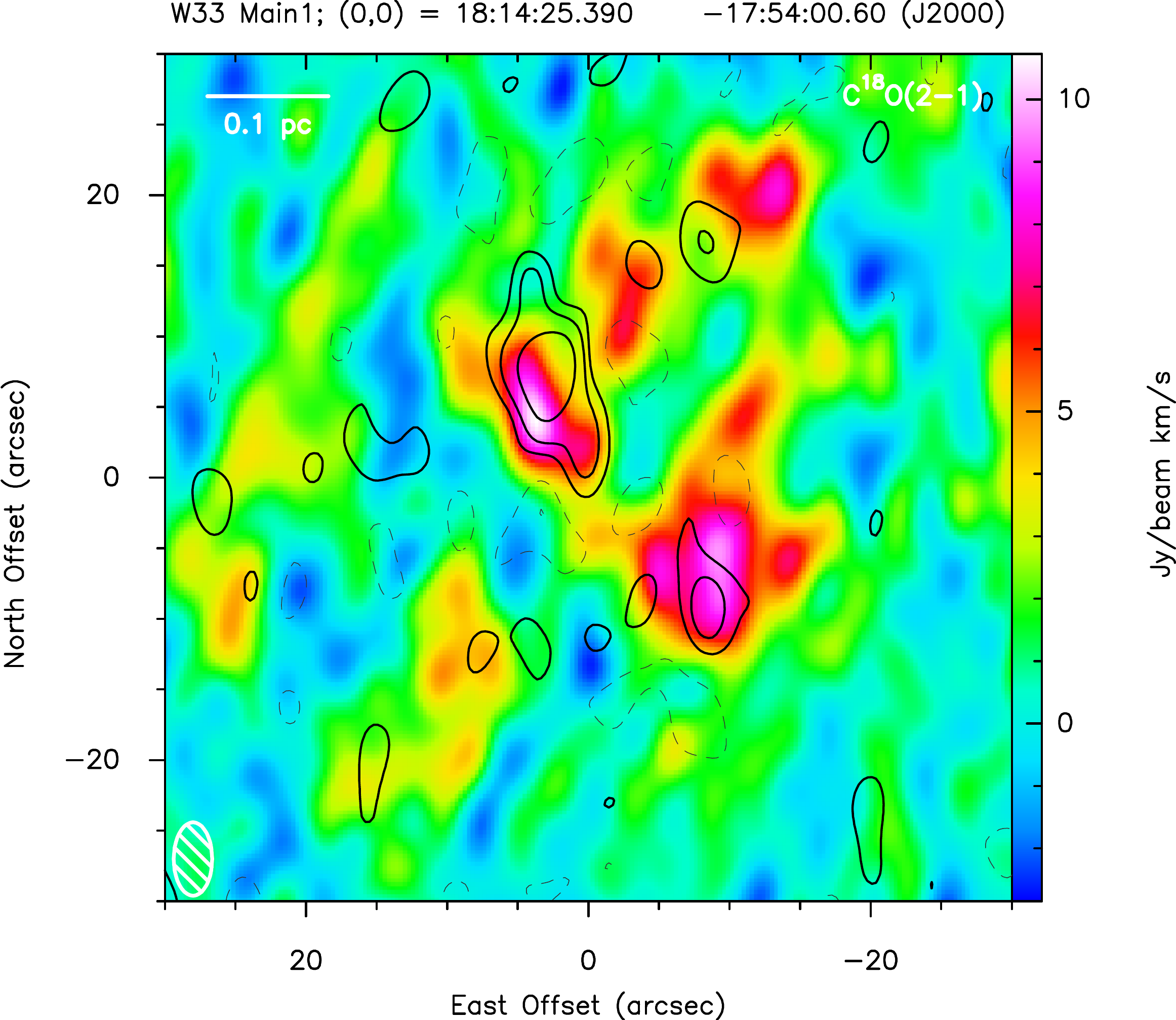}}\hspace{0.2cm}	
	\subfloat{\includegraphics[width=9cm]{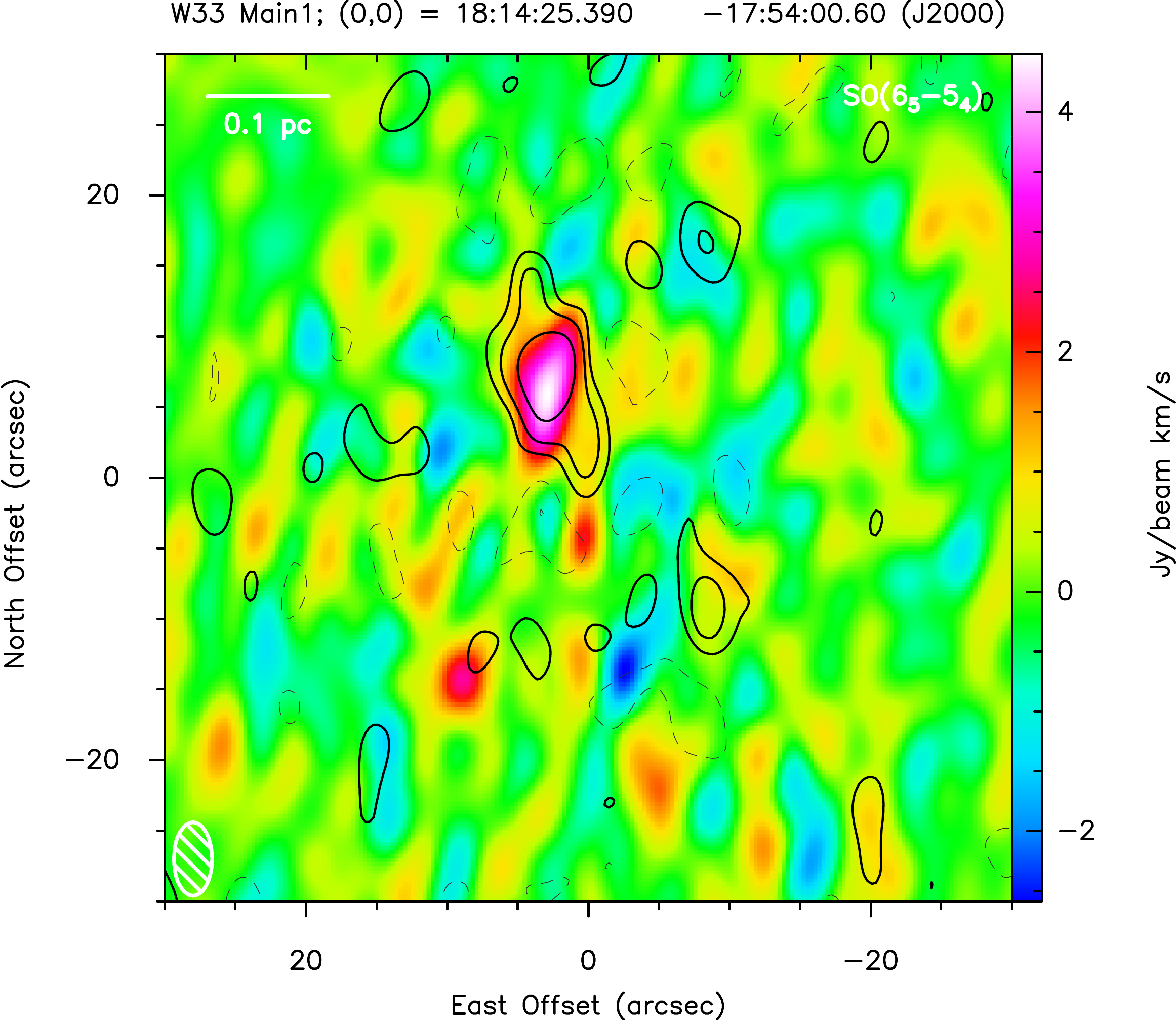}}	
	\label{W33M1-SMA-IntInt}
\end{figure*}

\paragraph{W33\,Main1} 
Besides the two CO lines $^{12}$CO and C$^{18}$O, we only observe emission of SO(6$_{5}$$-$5$_{4}$) in the spectrum of W33\,Main1 (Fig. \ref{W33_SMA_Spectra1}). $^{13}$CO is not detected at the position of the dust continuum peak, but extended emission of $^{13}$CO is observed in the W33\,Main1 mosaic (Fig. \ref{W33M1-SMA-IntInt}). The upper energy levels E$_{u}$ of the observed transitions are 16$-$35~K. The average central velocity and the average line width are 35.9~km~s$^{-1}$ and 4.0~km~s$^{-1}$, which are similar to the results of the APEX observations. 

The blueshifted and redshifted $^{12}$CO(2$-$1) emission of W33\,Main1 shows several peaks at and around the dust continuum peak (Fig. \ref{W33M1_12CO}). The distribution of the emission does not look ordered and might be excited by turbulences in this source.

The moment 0 map of the SO transition shows compact emission, whose peak is offset by $\sim$2$\arcsec$ from the dust continuum peak (Fig. \ref{W33M1-SMA-IntInt}). The moment 1 map of the SO transition shows a velocity gradient over the central part of the source (Fig. \ref{W33_SMA_VelGrad_All}).

The lack of spectral lines besides the CO and SO transitions either indicates low temperatures in the dust core or the emission of more complex molecules is not compact enough to be detected with the SMA. Both hint to a very early evolutionary age of W33\,Main1. We conclude that W33\,Main1 is probably in an early protostellar phase before the protostar strongly influences the surrounding material, and strong emission of primary molecules like H$_{2}$CO and CH$_{3}$OH is detected.

\paragraph{W33\,B1} In the spectrum of W33\,B1, we detect the three transitions of H$_{2}$CO, CH$_{3}$OH(4$_{\textnormal{2,2}}$$-$3$_{\textnormal{1,2}}$), SO(6$_{5}$$-$5$_{4}$), and $^{12}$CO (Fig. \ref{W33_SMA_Spectra1}). Extended emission of $^{13}$CO and C$^{18}$O is observed in the W33\,B1 map but not at the center of the core (Fig. \ref{W33B1-SMA-IntInt}). The detected transitions have upper energy levels between 17 and 68~K, indicating that this source is still relatively cold. The average central velocity and the average line width of W33\,B1 are 32.7~km~s$^{-1}$ and 5.0~km~s$^{-1}$, respectively.
The $^{12}$CO(2$-$1) emission shows a preferred direction in W33\,B1 (north-west to south-east). Most of the emission is detected close to the dust continuum peak (Fig. \ref{W33B1_12CO}).

The moment 0 maps of CH$_{3}$OH and H$_{2}$CO show compact emission, which peaks close to the dust continuum peak (Fig. \ref{W33B1-SMA-IntInt}). The emission of SO is also compact but peaks at the edge of the core (Fig. \ref{W33B1-SMA-IntInt}). The moment 1 maps of H$_{2}$CO(3$_{\textnormal{0,3}}$$-$2$_{\textnormal{0,2}}$) and CH$_{3}$OH(4$_{\textnormal{2,2}}$$-$3$_{\textnormal{1,2}}$) show similar velocity gradients across the source (Fig. \ref{W33_SMA_VelGrad_All}).

The higher-energy transitions of H$_{2}$CO and CH$_{3}$OH hint to the presence of a heating source in W33\,B1. However, since we do not see the line-richness of hot cores in this core, we conclude that W33\,B1 is a young protostellar core.

\paragraph{W33\,A1} In addition to the lines detected in W33\,B1, we observe emission of DCN(3$-$2) and C$^{18}$O at the peak of W33\,A1 (Fig. \ref{W33_SMA_Spectra1}). Again, extended emission of $^{13}$CO is observed in the map but not at the position of the continuum peak (Fig. \ref{W33A1-SMA-IntInt}). The upper energy levels of the detected transitions are between 16 and 68~K. The average central velocity of W33\,A1 is 35.5~km~s$^{-1}$. The average line width is 3.7~km~s$^{-1}$. 

Two streams of $^{12}$CO(2$-$1) emission are detected in W33\,A1 that are almost perpendicular to each other (Fig. \ref{W33A1_12CO}). Blueshifted emission is observed north of the dust continuum peak, while redshifted emission is detected west of the continuum peak.

The emission of DCN is compact and peaks close to the dust continuum peak (Fig. \ref{W33A1-SMA-IntInt}). The moment 0 map of SO shows two peaks at the edges of the dust core with the main peak of the SO emission being offset from the dust continuum peak by $\sim$2$\arcsec$ (Fig. \ref{W33A1-SMA-IntInt}). In the moment 1 map, a small velocity gradient is visible over both SO emission peaks (Fig. \ref{W33_SMA_VelGrad_All}). The integrated emission of H$_{2}$CO and CH$_{3}$OH is compact and peaks within 1$\arcsec$ of the dust continuum peak (Fig. \ref{W33A1-SMA-IntInt}). The low-energy transitions of H$_{2}$CO and CH$_{3}$OH both show a velocity gradient over the source (Fig. \ref{W33_SMA_VelGrad_All}).
For the same reasons as for W33\,B1, we conclude that W33\,A1 is a young protostellar core.

 \paragraph{W33\,B} The object W33\,B is one of the most line-rich sources of the six W33 cores. We observed 37 transitions of 14 molecules with the SMA (Fig. \ref{W33_SMA_Spectra2}). Emission of the CO isotopologue C$^{18}$O is not observed towards the center of the molecular core but is detected as extended emission in the map (Fig. \ref{W33B-SMA-IntInt}). The ionised gas tracer H30$\alpha$ is not detected in W33\,B, indicating that an UC \ion{H}{ii} region is not yet present. The strongest line in the spectrum is $^{12}$CO with a peak brightness temperature of $\sim$6 K. The average central velocity is 55.6~km~s$^{-1}$. The average line width is 6.8~km~s$^{-1}$. 

In W33\,B, we detect an outflow in the $^{12}$CO emission in almost north-south direction. Another stream of $^{12}$CO emission at blueshifted velocities points in the south-west direction and probably marks another outflow of which we only see the blueshifted side (Fig. \ref{W33B_12CO}).

The integrated emission of most of the detected lines is concentrated within the boundaries of the dust emission and only barely resolved (Fig. \ref{W33B-SMA-IntInt}). The lines that show more extended emission are $^{13}$CS, CH$_{3}$OH(4$_{\textnormal{2,2}}$$-$3$_{\textnormal{1,2}}$), H$_{2}$CO(3$_{\textnormal{0,3}}$$-$2$_{\textnormal{0,2}}$), SO(6$_{5}$$-$5$_{4}$), as well as the CO lines, which are spread over the whole field of view (Fig. \ref{W33B-SMA-IntInt}). Velocity gradients are detected in the moment 1 maps of the H$_{2}$CO and CH$_{3}$OH transitions as well as the OCS(19$-$18) and HC$_{3}$N(24$-$23) transitions (Fig. \ref{W33_SMA_VelGrad_All}).

The detection of complex molecules like CH$_{3}$CN, HNCO, HC$_{3}$N, CH$_{3}$OCHO, or CH$_{3}$OCH$_{3}$ indicates that W33\,B is in the hot core stage.

\paragraph{W33\,A}
For comparison with our sample of W33 sources, we obtained the spectrum of W33\,A by integrating the emission over one synthesised beam at the position of the stronger continuum peak \citep[MM1 in][]{GalvanMadrid2010}. The spectra of the two sidebands are shown in Fig. \ref{W33_SMA_Spectra2}. In the covered frequency range, we detect the same transitions as in W33\,B. In addition, high-energy CH$_{3}$OH transitions at E$_{u}$ = 579~K and emission of more complex molecules (CH$_{3}$OCHO, CH$_{3}$OCH$_{3}$) at higher excitation energies are observed. All transitions have upper energy levels E$_{u}$ between 16 and 579~K. Unfortunately, the frequency of the radio recombination line (RRL) H30$\alpha$ is located outside the frequency range of the W33\,A spectra. The variety of detected molecules (including complex molecules) and the high temperature needed for the excitation of the high-energy transitions supports the identification of W33\,A as a hot core. Since W33\,A shows emission of complex molecules at higher excitation energies, we conclude that W33\,A is probably more evolved than W33\,B.\\

\paragraph{W33\,Main} Compared to W33\,B and W33\,A, the spectrum of W33\,Main shows significantly fewer spectral lines (Fig. \ref{W33_SMA_Spectra2}). The dust continuum peak of W33\,Main-Central is almost devoid of H$_{2}$CO emission (Fig. \ref{W33M-SMA-IntInt}). In the spectrum, we detect H$_{2}$CO(3$_{\textnormal{0,3}}$$-$2$_{\textnormal{0,2}}$) and H$_{2}$CO(3$_{\textnormal{2,2}}$$-$2$_{\textnormal{2,1}}$). However, we observe emission of the 68~K H$_{2}$CO transitions (3$_{\textnormal{2,2}}$$-$2$_{\textnormal{2,1}}$) and (3$_{\textnormal{2,1}}$$-$2$_{\textnormal{2,0}}$) in the western part of W33\,Main-Central and at the peak positions of W33\,Main-West and W33\,Main-North (Fig. \ref{W33M-SMA-IntInt}). 
Besides two transitions of CH$_{3}$OH (CH$_{3}$OH(4$_{\textnormal{2,2}}$$-$3$_{\textnormal{1,2}}$), CH$_{3}$OH(8$_{\textnormal{$-$1,8}}$$-$7$_{\textnormal{0,7}}$)), we detect SO, the CO isotopologues $^{12}$CO, $^{13}$CO, and C$^{18}$O, HC$_{3}$N, and the RRL H30$\alpha$ in the spectrum. The detected lines have upper energy levels from 16 to 131~K. The two strongest lines are $^{12}$CO and C$^{18}$O with peak brightness temperatures of $\sim$8 and $\sim$6.5~K, respectively. The average central velocity and the average line width of W33\,Main are 37.1~km~s$^{-1}$ and 4.3~km~s$^{-1}$. 

Figure \ref{W33M_C18O} shows the outflowing gas in the C$^{18}$O emission. While there seem to be two redshifted streams of C$^{18}$O emission, we only see one strong peak of blueshifted emission that is located close to W33\,Main-Central. In Fig. \ref{W33M_COMB_CO}, we show a comparison of the integrated emission of the $^{13}$CO and C$^{18}$O transitions from the SMA data only and from the combination of the SMA and IRAM30m data. In Fig. \ref{W33M_COMB_CO_Spectra}, the $^{13}$CO spectra from the SMA data and the IRAM30m+SMA data, which are integrated over one synthesised beam at the continuum peak of W33\,Main are plotted. While the spectrum of the combined data shows a broad Gaussian with a blue and a red ``shoulder'', the SMA spectrum mimics a P-Cygni profile. These two figures present the importance of zero-spacing information, especially for widespread line emission, and show the strong filtering of large scale emission and the importance of side lobes in spectral features.

The object W33\,Main is the only source in our SMA sample that shows emission of the RRL H30$\alpha$ and the shock tracer SiO (Fig. \ref{W33M-SMA-IntInt}). The integrated emission of the RRL has the same shape as the dust continuum emission of W33\,Main-Central which suggests that part of the continuum emission comes from free-free radiation (Fig. \ref{W33M-SMA-IntInt}, see Sect. \ref{TempCloudMass}). The detection of the RRL supports the identification of W33\,Main as a more evolved object where ionised emission of \ion{H}{ii} region(s) is observed. The integrated emission of CH$_{3}$OH(4$_{\textnormal{2,2}}$$-$3$_{\textnormal{1,2}}$) and SO(6$_{5}$$-$5$_{4}$) also peaks at the center of W33\,Main-Central (Fig. \ref{W33M-SMA-IntInt}). However, the HC$_{3}$N transition peaks in the western part of W33\,Main-Central, which is offset from the main dust peak by $\sim$0.1 pc (Fig. \ref{W33M-SMA-IntInt}). Emission of the SiO(5$-$4) transition is observed at the western and southern edges of W33\,Main-Central (Fig. \ref{W33M-SMA-IntInt}). Velocity gradients are seen in the moment 1 maps of the CH$_{3}$OH and HC$_{3}$N transitions (Fig. \ref{W33_SMA_VelGrad_All}).

The object W33\,Main-North is bright in $^{13}$CS, H$_{2}$CO(3$_{\textnormal{0,3}}$$-$2$_{\textnormal{0,2}}$), SO(6$_{5}$$-$5$_{4}$), and SiO(5$-$4) (Fig. \ref{W33M-SMA-IntInt}). Weaker emission of CH$_{3}$OH(4$_{\textnormal{2,2}}$$-$3$_{\textnormal{1,2}}$) is also observed at this position. The strongest lines in W33\,Main-West are H$_{2}$CO(3$_{\textnormal{0,3}}$$-$2$_{\textnormal{0,2}}$) and CH$_{3}$OH(4$_{\textnormal{2,2}}$$-$3$_{\textnormal{1,2}}$) (Fig. \ref{W33M-SMA-IntInt}). In addition, weaker emission of $^{13}$CS, SO(6$_{5}$$-$5$_{4}$), and SiO(5$-$4) is detected. At the position of W33\,Main-South, we only observed diffuse line emission (Fig. \ref{W33M-SMA-IntInt}).\\
\\
In the sources W33\,Main1, W33\,A1, W33\,B1, and W33\,Main, we detect $^{12}$CO(2$-$1) and $^{13}$CO(2$-$1) emission at velocities of $\sim$60~km~s$^{-1}$. This emission is offset from the dust emission peak by several arcseconds in all sources. This shows that the velocity component of W33\,B (v$_{sys}$~=~56~km~s$^{-1}$) is not unique in the W33 complex but also observed towards the other sources in the low density gas tracers \citep[see also][]{Goldsmith1983,Urquhart2008,Chen2010}.

\onecolumn

\begin{landscape}
\scriptsize
\begin{longtable}{|clc|cccc|cccc|cccc|}
 \caption{\label{LinesTabSMA} Transitions, detected in W33 with the SMA telescope.}\\ \hline
 &  &  & \multicolumn{4}{|c|}{W33\,Main1} & \multicolumn{4}{|c|}{W33\,A1} & \multicolumn{4}{|c|}{W33\,B1}\\ \hline
$\nu_{0}$ & Transition & E$_{u}$ & F$_{Int.}$ & F$_{Peak}$ & v$_{central}$ & FWHM & F$_{Int}$ & F$_{Peak}$ & v$_{central}$ & FWHM & F$_{Int}$ & F$_{Peak}$ & v$_{central}$ & FWHM\\
(GHz) & & (K) & (K km s$^{-1}$) & (K) & (km s$^{-1}$) & (km s$^{-1}$) & (K km s$^{-1}$) & (K) & (km s$^{-1}$) & (km s$^{-1}$) & (K km s$^{-1}$) & (K) & (km s$^{-1}$) & (km s$^{-1}$)\\ \hline
\endfirsthead
\caption{Continued.}\\ \hline
 &  &  & \multicolumn{4}{|c|}{W33\,Main1} & \multicolumn{4}{|c|}{W33\,A1} & \multicolumn{4}{|c|}{W33\,B1}\\ \hline
$\nu_{0}$ & Transition & E$_{u}$ & F$_{Int.}$ & F$_{Peak}$ & v$_{central}$ & FWHM & F$_{Int}$ & F$_{Peak}$ & v$_{central}$ & FWHM & F$_{Int}$ & F$_{Peak}$ & v$_{central}$ & FWHM\\
(GHz) & & (K) & (K km s$^{-1}$) & (K) & (km s$^{-1}$) & (km s$^{-1}$) & (K km s$^{-1}$) & (K) & (km s$^{-1}$) & (km s$^{-1}$) & (K km s$^{-1}$) & (K) & (km s$^{-1}$) & (km s$^{-1}$)\\ \hline
\endhead
\endfoot
216.9456	&	CH\textsubscript{3}OH(5\textsubscript{1,4}$-$4\textsubscript{2,2})	&	55.87	&		&		&		&		&		&		&		&		&		&		&		&		\\
216.9659	&	CH\textsubscript{3}OCHO(20\textsubscript{1,20}$-$19\textsubscript{1,19})A	&	111.48	&		&		&		&		&		&		&		&		&		&		&		&		\\
217.1050	&	SiO(5$-$4)	&	31.26	&		&		&		&		&		&		&		&		&		&		&		&		\\
217.2386	&	DCN(3$-$2)	&	20.85	&		&		&		&		&	5.75	&	1.29	&	34.14	&	4.19	&		&		&		&		\\
217.2992	&	CH\textsubscript{3}OH(6\textsubscript{1,5}$-$7\textsubscript{2,6})	&	373.93	&		&		&		&		&		&		&		&		&		&		&		&		\\
217.8864	&	CH\textsubscript{3}OH(20\textsubscript{1,19}$-$20\textsubscript{0,20})	&	508.38	&		&		&		&		&		&		&		&		&		&		&		&		\\
218.2222	&	H\textsubscript{2}CO(3\textsubscript{0,3}$-$2\textsubscript{0,2})	&	20.96	&		&		&		&		&	8.33	&	2.22	&	35.43	&	3.52	&	21.98	&	3.33	&	32.94	&	6.20	\\
218.3248	&	HC\textsubscript{3}N(24$-$23)	&	130.98	&		&		&		&		&		&		&		&		&		&		&		&		\\
218.4401	&	CH\textsubscript{3}OH(4\textsubscript{2,2}$-$3\textsubscript{1,2})	&	45.46	&		&		&		&		&	7.74	&	1.64	&	36.33	&	4.43	&	13.63	&	2.51	&	32.72	&	5.10	\\
218.4756	&	H\textsubscript{2}CO(3\textsubscript{2,2}$-$2\textsubscript{2,1})	&	68.09	&		&		&		&		&	4.61	&	1.50	&	36.71	&	2.89	&	5.37	&	0.81	&	31.69	&	6.20	\\
218.7601	&	H\textsubscript{2}CO(3\textsubscript{2,1}$-$2\textsubscript{2,0})	&	68.11	&		&		&		&		&	3.44	&	0.96	&	35.43	&	3.38	&	8.73	&	1.64	&	32.78	&	5.02	\\
218.9810	&	HNCO(10\textsubscript{1,10}$-$9\textsubscript{1,9})	&	101.08	&		&		&		&		&		&		&		&		&		&		&		&		\\
219.5604	&	C\textsuperscript{18}O(2$-$1)	&	15.81	&	3.41	&	1.38	&	36.13	&	2.32	&	10.08	&	2.51	&	34.87	&	3.77	&	\multicolumn{4}{c|}{*}							\\
219.7339	&	HNCO(10\textsubscript{2,9}$-$9\textsubscript{2,8})\tablefootmark{a,b}	&	228.28	&		&		&		&		&		&		&		&		&		&		&		&		\\
219.7372	&	HNCO(10\textsubscript{2,8}$-$9\textsubscript{2,7})\tablefootmark{a,b}	&	228.29	&		&		&		&		&		&		&		&		&		&		&		&		\\
219.7982	&	HNCO(10\textsubscript{0,10}$-$9\textsubscript{0,9})	&	58.02	&		&		&		&		&		&		&		&		&		&		&		&		\\
219.9085	&	H\textsubscript{2}\textsuperscript{13}CO(3\textsubscript{1,2}$-$2\textsubscript{1,1})	&	15.56	&		&		&		&		&		&		&		&		&		&		&		&		\\
219.9494	&	SO(6\textsubscript{5}$-$5\textsubscript{4})	&	34.98	&	4.17	&	0.82	&	35.75	&	4.80	&		&		&		&		&	3.37	&	0.9	&	33.35	&	3.53	\\
220.0785	&	CH\textsubscript{3}OH(8\textsubscript{0,8}$-$7\textsubscript{1,6})	&	96.61	&		&		&		&		&		&		&		&		&		&		&		&		\\
220.1669	&	CH\textsubscript{3}OCHO(17\textsubscript{4,13}$-$16\textsubscript{4,12})E	&	103.15	&		&		&		&		&		&		&		&		&		&		&		&		\\
220.1903	&	CH\textsubscript{3}OCHO(17\textsubscript{4,13}$-$16\textsubscript{4,12})E	&	103.15	&		&		&		&		&		&		&		&		&		&		&		&		\\
220.3989	&	\textsuperscript{13}CO(2$-$1)	&	15.89	&	\multicolumn{4}{c|}{*}							&	\multicolumn{4}{c|}{*}							&	\multicolumn{4}{c|}{*}							\\
220.5392	&	CH\textsubscript{3}CN(12\textsubscript{7}$-$11\textsubscript{7})	&	418.63	&		&		&		&		&		&		&		&		&		&		&		&		\\
220.5848	&	HNCO(10\textsubscript{1,9}$-$9\textsubscript{1,8})	&	101.50	&		&		&		&		&		&		&		&		&		&		&		&		\\
220.5944	&	CH\textsubscript{3}CN(12\textsubscript{6}$-$11\textsubscript{6})	&	325.90	&		&		&		&		&		&		&		&		&		&		&		&		\\
220.6410	&	CH\textsubscript{3}CN(12\textsubscript{5}$-$11\textsubscript{5})	&	247.40	&		&		&		&		&		&		&		&		&		&		&		&		\\
220.6792	&	CH\textsubscript{3}CN(12\textsubscript{4}$-$11\textsubscript{4})	&	183.15	&		&		&		&		&		&		&		&		&		&		&		&		\\
220.7090	&	CH\textsubscript{3}CN(12\textsubscript{3}$-$11\textsubscript{3})	&	133.16	&		&		&		&		&		&		&		&		&		&		&		&		\\
220.7302	&	CH\textsubscript{3}CN(12\textsubscript{2}$-$11\textsubscript{2})	&	97.44	&		&		&		&		&		&		&		&		&		&		&		&		\\
220.7430	&	CH\textsubscript{3}CN(12\textsubscript{1}$-$11\textsubscript{1})	&	76.01	&		&		&		&		&		&		&		&		&		&		&		&		\\
220.7472	&	CH\textsubscript{3}CN(12\textsubscript{0}$-$11\textsubscript{0})	&	68.87	&		&		&		&		&		&		&		&		&		&		&		&		\\
220.8934	&	CH\textsubscript{3}OCH\textsubscript{3}(23\textsubscript{4,20}$-$23\textsubscript{3,21})\tablefootmark{c}	&	274.44	&		&		&		&		&		&		&		&		&		&		&		&		\\
221.1984	&	CH\textsubscript{3}OCH\textsubscript{3}(27\textsubscript{5,22}$-$27\textsubscript{4,23})\tablefootmark{c}	&	257.14	&		&		&		&		&		&		&		&		&		&		&		&		\\
229.4050	&	CH\textsubscript{3}OCHO(18\textsubscript{3,15}$-$17\textsubscript{3,14})	&	110.74	&		&		&		&		&		&		&		&		&		&		&		&		\\
229.4203	&	CH\textsubscript{3}OCHO(18\textsubscript{3,15}$-$17\textsubscript{3,14})	&	110.74	&		&		&		&		&		&		&		&		&		&		&		&		\\
229.5891	&	CH\textsubscript{3}OH(15\textsubscript{4,11}$-$16\textsubscript{3,13})	&	374.44	&		&		&		&		&		&		&		&		&		&		&		&		\\
229.7588	&	CH\textsubscript{3}OH(8\textsubscript{$-$1,8}$-$7\textsubscript{0,7})	&	89.10	&		&		&		&		&		&		&		&		&		&		&		&		\\
229.8641	&	CH\textsubscript{3}OH(19\textsubscript{5,15}$-$20\textsubscript{4,16})	&	578.60	&		&		&		&		&		&		&		&		&		&		&		&		\\
229.9392	&	CH\textsubscript{3}OH(19\textsubscript{5,14}$-$20\textsubscript{4,17})	&	578.60	&		&		&		&		&		&		&		&		&		&		&		&		\\
230.0270	&	CH\textsubscript{3}OH(3\textsubscript{$-$2,2}$-$4\textsubscript{$-$1,4})	&	39.83	&		&		&		&		&		&		&		&		&		&		&		&		\\
230.1414	&	CH\textsubscript{3}OCH\textsubscript{3}(25\textsubscript{4,22}$-$25\textsubscript{3,22})	&	319.21	&		&		&		&		&		&		&		&		&		&		&		&		\\
230.2338	&	CH\textsubscript{3}OCH\textsubscript{3}(17\textsubscript{2,15}$-$16\textsubscript{3,14})	&	147.65	&		&		&		&		&		&		&		&		&		&		&		&		\\
230.3682	&	CH\textsubscript{3}OCH\textsubscript{3}(28\textsubscript{5,23}$-$27\textsubscript{6,22})	&	406.35	&		&		&		&		&		&		&		&		&		&		&		&		\\
230.5380	&	\textsuperscript{12}CO(2$-$1)	&	16.60	&	21.51	&	4.08	&	30.01	&	4.95	&	14.19	&	3.56	&	29.44	&	3.74	&	12.76	&	3.16	&	26.40	&	3.79	\\
231.0610	&	OCS(19$-$18)	&	110.90	&		&		&		&		&		&		&		&		&		&		&		&		\\
231.2210	&	\textsuperscript{13}CS(5$-$4)	&	33.29	&		&		&		&		&		&		&		&		&		&		&		&		\\
231.2811	&	CH\textsubscript{3}OH(10\textsubscript{2,9}$-$9\textsubscript{3,6})	&	165.35	&		&		&		&		&		&		&		&		&		&		&		&		\\
231.9009	&	H30$\alpha$	&		&		&		&		&		&		&		&		&		&		&		&		&		\\
231.9879	&	CH\textsubscript{3}OCH\textsubscript{3}(13\textsubscript{0,13}$-$12\textsubscript{1,12})	&	80.92	&		&		&		&		&		&		&		&		&		&		&		&		\\
232.4186	&	CH\textsubscript{3}OH(10\textsubscript{2,8}$-$9\textsubscript{3,7})	&	165.40	&		&		&		&		&		&		&		&		&		&		&		&		\\
232.7836	&	CH\textsubscript{3}OH(18\textsubscript{3,16}$-$17\textsubscript{4,13})	&	446.53	&		&		&		&		&		&		&		&		&		&		&		&		\\ \hline
\end{longtable}
\tablefoot{
\tablefoottext{*}{Emission detected in map but not in the spectrum at the center of the core.}
\tablefoottext{a}{Blended transitions.}
\tablefoottext{b}{Spectral lines fitted with same central velocity and FWHM.}
\tablefoottext{c}{Only in frequency range of W33\,A.}
\tablefoottext{d}{Weak transition (<~3$\sigma$), detected with Weeds (see Sect. \ref{TempDens}).}
}
\end{landscape}

\addtocounter{table}{-1}

\begin{landscape}
\scriptsize
\begin{longtable}{|clc|cccc|cccc|cccc|}
 \caption{Continued.}\\ \hline
 & &  & \multicolumn{4}{|c|}{W33\,B} & \multicolumn{4}{|c|}{W33\,A} & \multicolumn{4}{|c|}{W33\,Main}\\ \hline
$\nu_{0}$ & Transition & E$_{u}$ & F$_{Int.}$ & F$_{Peak}$ & v$_{central}$ & FWHM & \multicolumn{4}{|c|}{} & F$_{Int}$ & F$_{Peak}$ & v$_{central}$ & FWHM\\
(GHz) & & (K) & (K km s$^{-1}$) & (K) & (km s$^{-1}$) & (km s$^{-1}$) & \multicolumn{4}{|c|}{} & (K km s$^{-1}$) & (K) & (km s$^{-1}$) & (km s$^{-1}$) \\ \hline
\endfirsthead
\caption{Continued.}\\ \hline
 & &  & \multicolumn{4}{|c|}{W33\,B} & \multicolumn{4}{|c|}{W33\,A} & \multicolumn{4}{|c|}{W33\,Main}\\ \hline
$\nu_{0}$ & Transition & E$_{u}$ & F$_{Int.}$ & F$_{Peak}$ & v$_{central}$ & FWHM & \multicolumn{4}{|c|}{} & F$_{Int}$ & F$_{Peak}$ & v$_{central}$ & FWHM\\
(GHz) & & (K) & (K km s$^{-1}$) & (K) & (km s$^{-1}$) & (km s$^{-1}$) & \multicolumn{4}{|c|}{} & (K km s$^{-1}$) & (K) & (km s$^{-1}$) & (km s$^{-1}$) \\ \hline
\endhead
\endfoot
216.9456	&	CH\textsubscript{3}OH(5\textsubscript{1,4}$-$4\textsubscript{2,2})	&	55.87	&	20.52	&	2.78	&	55.91	&	6.95	&	\multicolumn{4}{c|}{}	&		&		&		&		\\
216.9659	&	CH\textsubscript{3}OCHO(20\textsubscript{1,20}$-$19\textsubscript{1,19})A	&	111.48	&	19.95	&	2.72	&	55.07	&	6.89	&	\multicolumn{4}{c|}{}	&		&		&		&		\\
217.1050	&	SiO(5$-$4)	&	31.26	&	6.58	&	0.54	&	60.60	&	11.44	&	\multicolumn{4}{c|}{}	&	\multicolumn{4}{c|}{*}							\\
217.2386	&	DCN(3$-$2)	&	20.85	&	10.68	&	1.34	&	55.78	&	7.47	&	\multicolumn{4}{c|}{}	&		&		&		&		\\
217.2992	&	CH\textsubscript{3}OH(6\textsubscript{1,5}$-$7\textsubscript{2,6})	&	373.93	&	12.48	&	2.49	&	56.45	&	4.71	&	\multicolumn{4}{c|}{}	&		&		&		&		\\
217.8864	&	CH\textsubscript{3}OH(20\textsubscript{1,19}$-$20\textsubscript{0,20})	&	508.38	&	10.91	&	2.19	&	55.71	&	4.69	&	\multicolumn{4}{c|}{}	&		&		&		&		\\
218.2222	&	H\textsubscript{2}CO(3\textsubscript{0,3}$-$2\textsubscript{0,2})	&	20.96	&	17.55	&	1.92	&	56.67	&	8.61	&	\multicolumn{4}{c|}{}	&	4.63	&	1.12	&	39.59	&	3.89	\\
218.3248	&	HC\textsubscript{3}N(24$-$23)	&	130.98	&	22.17	&	2.02	&	55.40	&	10.32	&	\multicolumn{4}{c|}{}	&	21.89	&	5.96	&	37.40	&	3.45	\\
218.4401	&	CH\textsubscript{3}OH(4\textsubscript{2,2}$-$3\textsubscript{1,2})	&	45.46	&	18.77	&	2.92	&	55.80	&	6.04	&	\multicolumn{4}{c|}{}	&	19.15	&	4.21	&	37.78	&	4.27	\\
218.4756	&	H\textsubscript{2}CO(3\textsubscript{2,2}$-$2\textsubscript{2,1})	&	68.09	&	12.37	&	2.16	&	55.12	&	5.37	&	\multicolumn{4}{c|}{}	&	4.14	&	0.81	&	36.00	&	4.79	\\
218.7601	&	H\textsubscript{2}CO(3\textsubscript{2,1}$-$2\textsubscript{2,0})	&	68.11	&	14.34	&	2.57	&	56.26	&	5.23	&	\multicolumn{4}{c|}{}	&		&		&		&		\\
218.9810	&	HNCO(10\textsubscript{1,10}$-$9\textsubscript{1,9})	&	101.08	&	13.56	&	1.60	&	54.58	&	7.94	&	\multicolumn{4}{c|}{}	&		&		&		&		\\
219.5604	&	C\textsuperscript{18}O(2$-$1)	&	15.81	&	\multicolumn{4}{c|}{*}							&	\multicolumn{4}{c|}{$\checkmark$}	&	22.99	&	6.46	&	38.57	&	3.34	\\
219.7339	&	HNCO(10\textsubscript{2,9}$-$9\textsubscript{2,8})\tablefootmark{a,b}	&	228.28	&	5.30	&	1.29	&	56.32	&	3.85	&	\multicolumn{4}{c|}{$\checkmark$}	&		&		&		&		\\
219.7372	&	HNCO(10\textsubscript{2,8}$-$9\textsubscript{2,7})\tablefootmark{a,b}	&	228.29	&	2.19	&	0.53	&	56.32	&	3.85	&	\multicolumn{4}{c|}{$\checkmark$}	&		&		&		&		\\
219.7982	&	HNCO(10\textsubscript{0,10}$-$9\textsubscript{0,9})	&	58.02	&	13.99	&	1.56	&	55.82	&	8.44	&	\multicolumn{4}{c|}{$\checkmark$}	&		&		&		&		\\
219.9085	&	H\textsubscript{2}\textsuperscript{13}CO(3\textsubscript{1,2}$-$2\textsubscript{1,1})	&	15.56	&		&		&		&		&	\multicolumn{4}{c|}{$\checkmark$}	&		&		&		&		\\
219.9494	&	SO(6\textsubscript{5}$-$5\textsubscript{4})	&	34.98	&	21.40	&	2.34	&	55.15	&	8.60	&	\multicolumn{4}{c|}{$\checkmark$}	&	14.61	&	1.45	&	36.18	&	9.48	\\
220.0785	&	CH\textsubscript{3}OH(8\textsubscript{0,8}$-$7\textsubscript{1,6})	&	96.61	&	22.74	&	2.34	&	55.22	&	9.13	&	\multicolumn{4}{c|}{$\checkmark$}	&		&		&		&		\\
220.1669	&	CH\textsubscript{3}OCHO(17\textsubscript{4,13}$-$16\textsubscript{4,12})E	&	103.15	&		&		&		&		&	\multicolumn{4}{c|}{$\checkmark$}	&		&		&		&		\\
220.1903	&	CH\textsubscript{3}OCHO(17\textsubscript{4,13}$-$16\textsubscript{4,12})E	&	103.15	&		&		&		&		&	\multicolumn{4}{c|}{$\checkmark$}	&		&		&		&		\\
220.3989	&	\textsuperscript{13}CO(2$-$1)	&	15.89	&	10.18	&	1.98	&	48.46	&	4.82	&	\multicolumn{4}{c|}{$\checkmark$}	&	12.74	&	3.20	&	41.69	&	3.74	\\
220.5392	&	CH\textsubscript{3}CN(12\textsubscript{7}$-$11\textsubscript{7})	&	418.63	&	4.18	&	0.56\tablefootmark{d}	&	55.65\tablefootmark{b}	&	7.04\tablefootmark{b}	&	\multicolumn{4}{c|}{$\checkmark$}	&		&		&		&		\\
220.5848	&	HNCO(10\textsubscript{1,9}$-$9\textsubscript{1,8})	&	101.5	&	5.42	&	0.94	&	54.63	&	5.44	&	\multicolumn{4}{c|}{$\checkmark$}	&		&		&		&		\\
220.5944	&	CH\textsubscript{3}CN(12\textsubscript{6}$-$11\textsubscript{6})	&	325.9	&	10.31	&	1.38	&	55.65\tablefootmark{b}	&	7.04\tablefootmark{b}	&	\multicolumn{4}{c|}{$\checkmark$}	&		&		&		&		\\
220.6410	&	CH\textsubscript{3}CN(12\textsubscript{5}$-$11\textsubscript{5})	&	247.4	&	9.63	&	1.28	&	55.65\tablefootmark{b}	&	7.04\tablefootmark{b}	&	\multicolumn{4}{c|}{$\checkmark$}	&		&		&		&		\\
220.6792	&	CH\textsubscript{3}CN(12\textsubscript{4}$-$11\textsubscript{4})	&	183.15	&	11.74	&	1.57	&	55.65\tablefootmark{b}	&	7.04\tablefootmark{b}	&	\multicolumn{4}{c|}{$\checkmark$}	&		&		&		&		\\
220.7090	&	CH\textsubscript{3}CN(12\textsubscript{3}$-$11\textsubscript{3})	&	133.16	&	16.11	&	2.15	&	55.65\tablefootmark{b}	&	7.04\tablefootmark{b}	&	\multicolumn{4}{c|}{$\checkmark$}	&		&		&		&		\\
220.7302	&	CH\textsubscript{3}CN(12\textsubscript{2}$-$11\textsubscript{2})	&	97.44	&	15.95	&	2.13	&	55.65\tablefootmark{b}	&	7.04\tablefootmark{b}	&	\multicolumn{4}{c|}{$\checkmark$}	&		&		&		&		\\
220.7430	&	CH\textsubscript{3}CN(12\textsubscript{1}$-$11\textsubscript{1})	&	76.01	&	16.72	&	2.23	&	55.65\tablefootmark{b}	&	7.04\tablefootmark{b}	&	\multicolumn{4}{c|}{$\checkmark$}	&		&		&		&		\\
220.7472	&	CH\textsubscript{3}CN(12\textsubscript{0}$-$11\textsubscript{0})	&	68.87	&	17.12	&	2.28	&	55.65\tablefootmark{b}	&	7.04\tablefootmark{b}	&	\multicolumn{4}{c|}{$\checkmark$}	&		&		&		&		\\
220.8934	&	CH\textsubscript{3}OCH\textsubscript{3}(23\textsubscript{4,20}$-$23\textsubscript{3,21})\tablefootmark{c}	&	274.44	&		&		&		&		&	\multicolumn{4}{c|}{$\checkmark$}	&		&		&		&		\\
221.1984	&	CH\textsubscript{3}OCH\textsubscript{3}(27\textsubscript{5,22}$-$27\textsubscript{4,23})\tablefootmark{c}	&	257.14	&		&		&		&		&	\multicolumn{4}{c|}{$\checkmark$}	&		&		&		&		\\
229.4050	&	CH\textsubscript{3}OCHO(18\textsubscript{3,15}$-$17\textsubscript{3,14})	&	110.74	&		&		&		&		&	\multicolumn{4}{c|}{$\checkmark$}	&		&		&		&		\\
229.4203	&	CH\textsubscript{3}OCHO(18\textsubscript{3,15}$-$17\textsubscript{3,14})	&	110.74	&		&		&		&		&	\multicolumn{4}{c|}{$\checkmark$}	&		&		&		&		\\
229.5891	&	CH\textsubscript{3}OH(15\textsubscript{4,11}$-$16\textsubscript{3,13})	&	374.44	&	21.97	&	2.40	&	54.87	&	8.61	&	\multicolumn{4}{c|}{$\checkmark$}	&		&		&		&		\\
229.7588	&	CH\textsubscript{3}OH(8\textsubscript{$-$1,8}$-$7\textsubscript{0,7})	&	89.10	&	21.57	&	3.06	&	55.8	&	6.62	&	\multicolumn{4}{c|}{$\checkmark$}	&	15.01	&	2.95	&	37.67	&	4.78	\\
229.8641	&	CH\textsubscript{3}OH(19\textsubscript{5,15}$-$20\textsubscript{4,16})	&	578.60	&		&		&		&		&	\multicolumn{4}{c|}{$\checkmark$}	&		&		&		&		\\
229.9392	&	CH\textsubscript{3}OH(19\textsubscript{5,14}$-$20\textsubscript{4,17})	&	578.60	&		&		&		&		&	\multicolumn{4}{c|}{$\checkmark$}	&		&		&		&		\\
230.0270	&	CH\textsubscript{3}OH(3\textsubscript{$-$2,2}$-$4\textsubscript{$-$1,4})	&	39.83	&	17.40	&	2.19	&	57.01	&	7.46	&	\multicolumn{4}{c|}{$\checkmark$}	&		&		&		&		\\
230.1414	&	CH\textsubscript{3}OCH\textsubscript{3}(25\textsubscript{4,22}$-$25\textsubscript{3,22})	&	319.21	&		&		&		&		&	\multicolumn{4}{c|}{$\checkmark$}	&		&		&		&		\\
230.2338	&	CH\textsubscript{3}OCH\textsubscript{3}(17\textsubscript{2,15}$-$16\textsubscript{3,14})	&	147.65	&		&		&		&		&	\multicolumn{4}{c|}{$\checkmark$}	&		&		&		&		\\
230.3682	&	CH\textsubscript{3}OCH\textsubscript{3}(28\textsubscript{5,23}$-$27\textsubscript{6,22})	&	406.35	&		&		&		&		&	\multicolumn{4}{c|}{$\checkmark$}	&		&		&		&		\\
230.5380	&	\textsuperscript{12}CO(2$-$1)	&	16.60	&	45.09	&	6.21	&	46.31	&	6.82	&	\multicolumn{4}{c|}{$\checkmark$}	&	18.62	&	7.95	&	42.98	&	2.20	\\
231.0610	&	OCS(19$-$18)	&	110.90	&	14.26	&	2.33	&	55.82	&	5.76	&	\multicolumn{4}{c|}{$\checkmark$}	&		&		&		&		\\
231.2210	&	\textsuperscript{13}CS(5$-$4)	&	33.29	&	13.55	&	1.78	&	54.81	&	7.16	&	\multicolumn{4}{c|}{$\checkmark$}	&	\multicolumn{4}{c|}{*}							\\
231.2811	&	CH\textsubscript{3}OH(10\textsubscript{2,9}$-$9\textsubscript{3,6})	&	165.35	&	16.79	&	1.70	&	54.67	&	9.30	&	\multicolumn{4}{c|}{$\checkmark$}	&		&		&		&		\\
231.9009	&	H30$\alpha$	&		&		&		&		&		&	\multicolumn{4}{c|}{}	&	83.80	&	3.19	&	32.91	&	24.65	\\
231.9879	&	CH\textsubscript{3}OCH\textsubscript{3}(13\textsubscript{0,13}$-$12\textsubscript{1,12})	&	80.92	&	12.04	&	1.93	&	54.72	&	5.87	&	\multicolumn{4}{c|}{}	&		&		&		&		\\
232.4186	&	CH\textsubscript{3}OH(10\textsubscript{2,8}$-$9\textsubscript{3,7})	&	165.40	&	20.10	&	3.12	&	55.86	&	6.05	&	\multicolumn{4}{c|}{}	&		&		&		&		\\
232.7836	&	CH\textsubscript{3}OH(18\textsubscript{3,16}$-$17\textsubscript{4,13})	&	446.53	&	12.13	&	2.32	&	56.07	&	4.92	&	\multicolumn{4}{c|}{}	&		&		&		&		\\ \hline
\end{longtable}
\tablefoot{
\tablefoottext{*}{Emission detected in map but not in the spectrum at the center of the core.}
\tablefoottext{a}{Blended transitions.}
\tablefoottext{b}{Spectral lines fitted with same central velocity and FWHM.}
\tablefoottext{c}{Only in frequency range of W33\,A.}
\tablefoottext{d}{Weak transition (<~3$\sigma$), detected with Weeds (see Sect. \ref{TempDens}).}
}
\end{landscape}

\twocolumn

\begin{figure*}
	\centering
	\caption{Line emission of detected transitions in W33\,A1. The contours show the SMA continuum emission at 230 GHz (same contour levels as in Fig. \ref{W33_SMA_CH0}). The name of the each transition is shown in the upper right corner. A scale of 0.1 pc is marked in the upper left corner, and the synthesised beam is shown in the lower left corner.}
	\subfloat{\includegraphics[width=9cm]{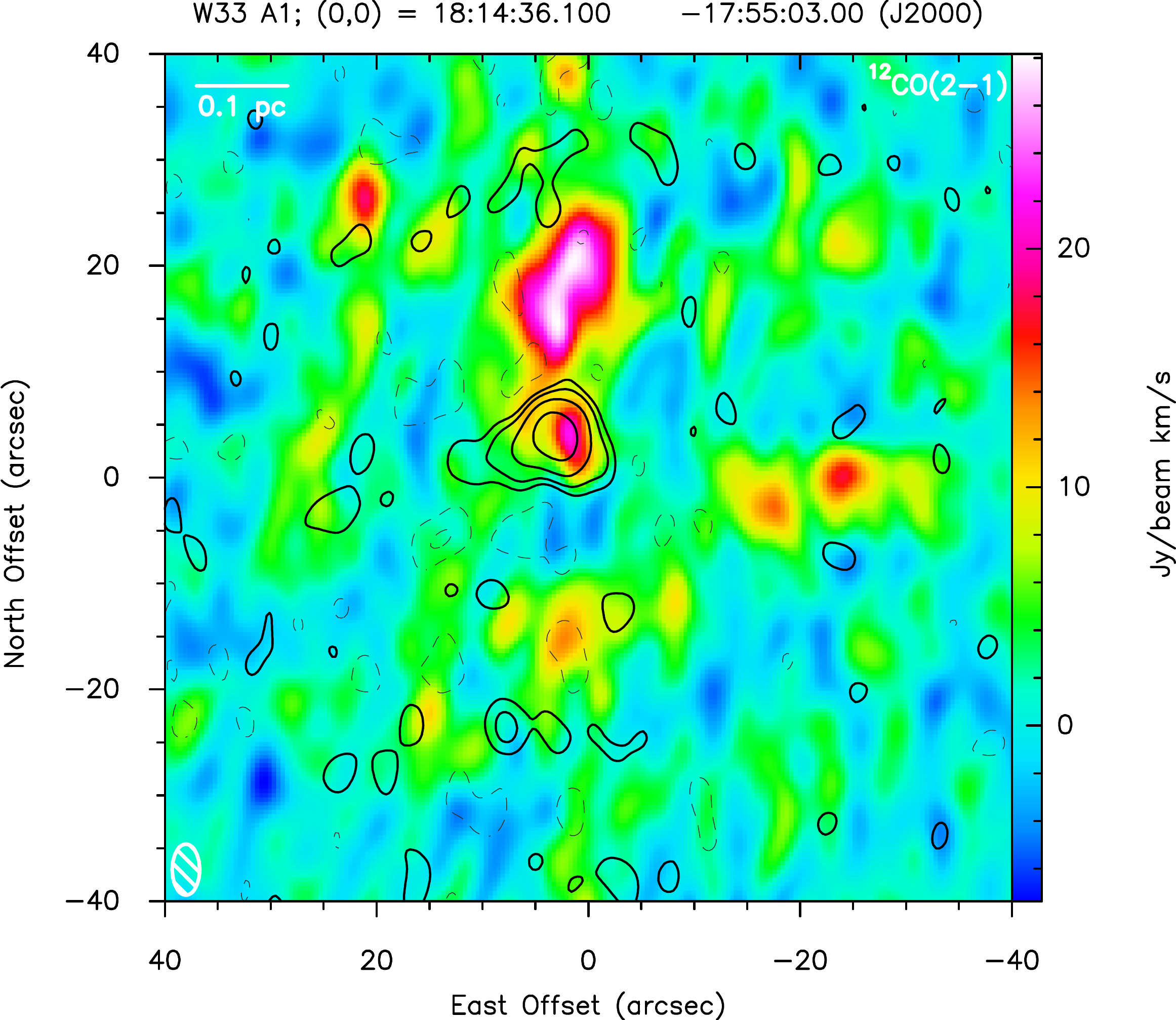}}\hspace{0.2cm}	
	\subfloat{\includegraphics[width=9cm]{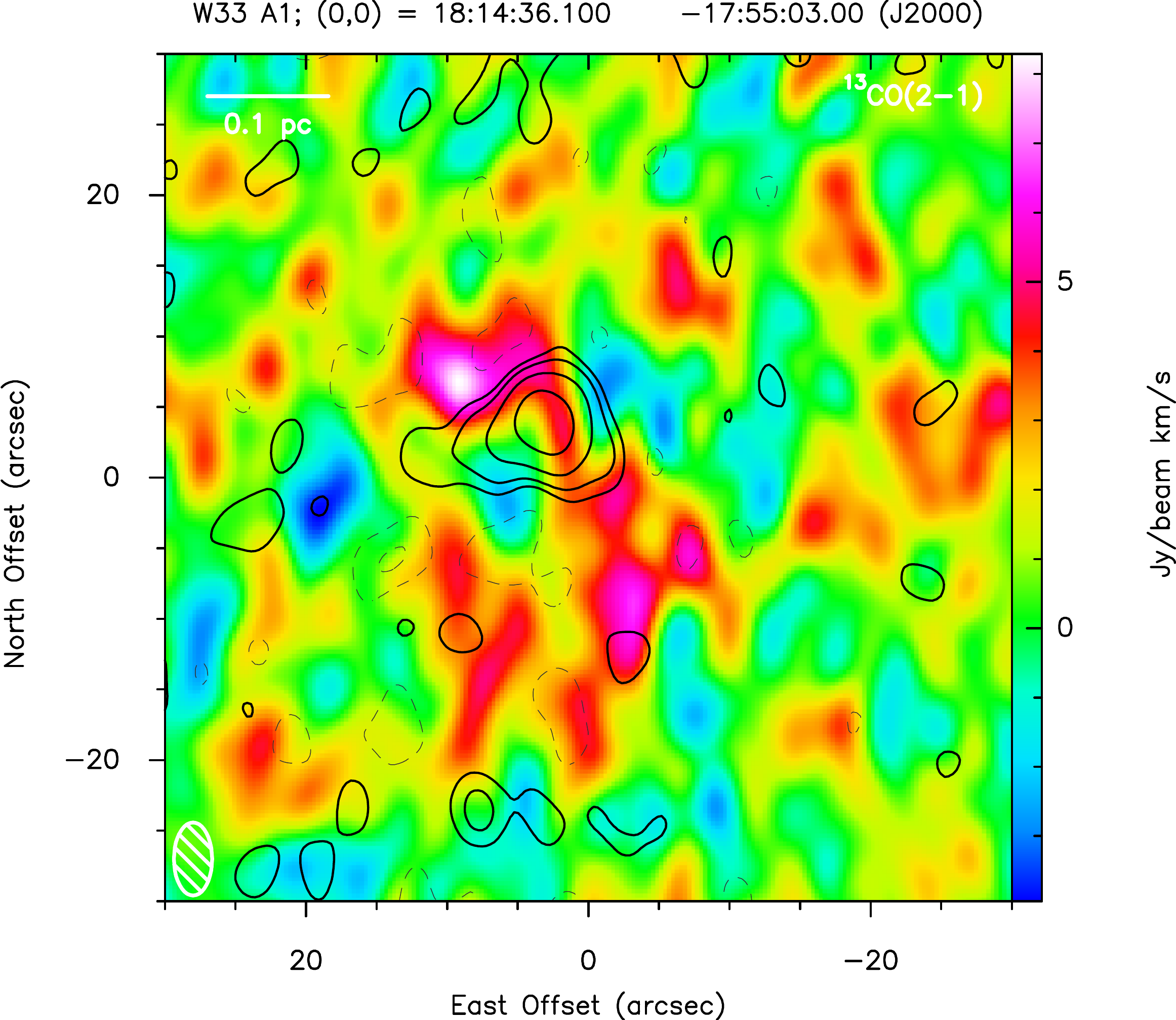}}\\	
	\subfloat{\includegraphics[width=9cm]{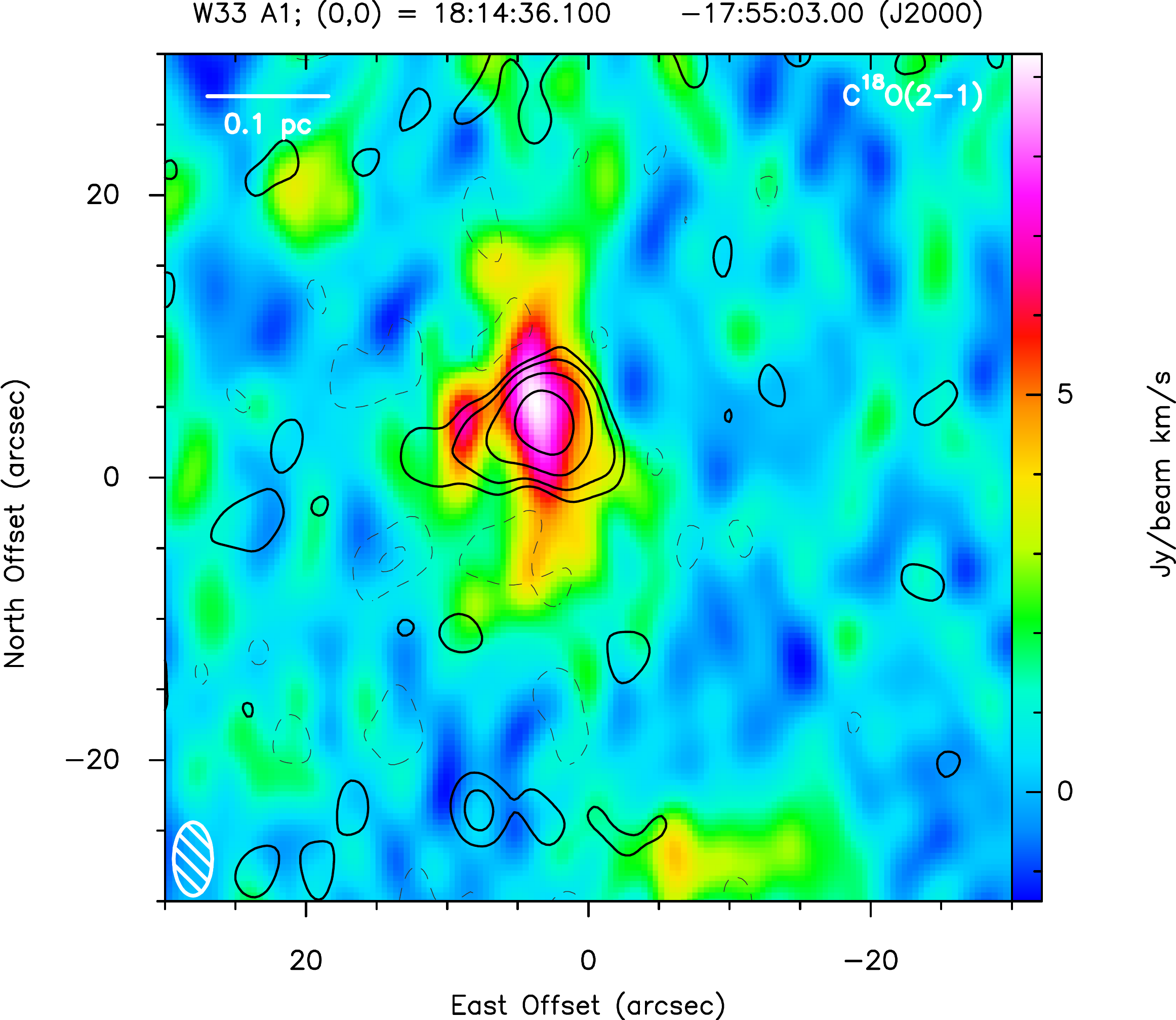}}\hspace{0.2cm}	
	\subfloat{\includegraphics[width=9cm]{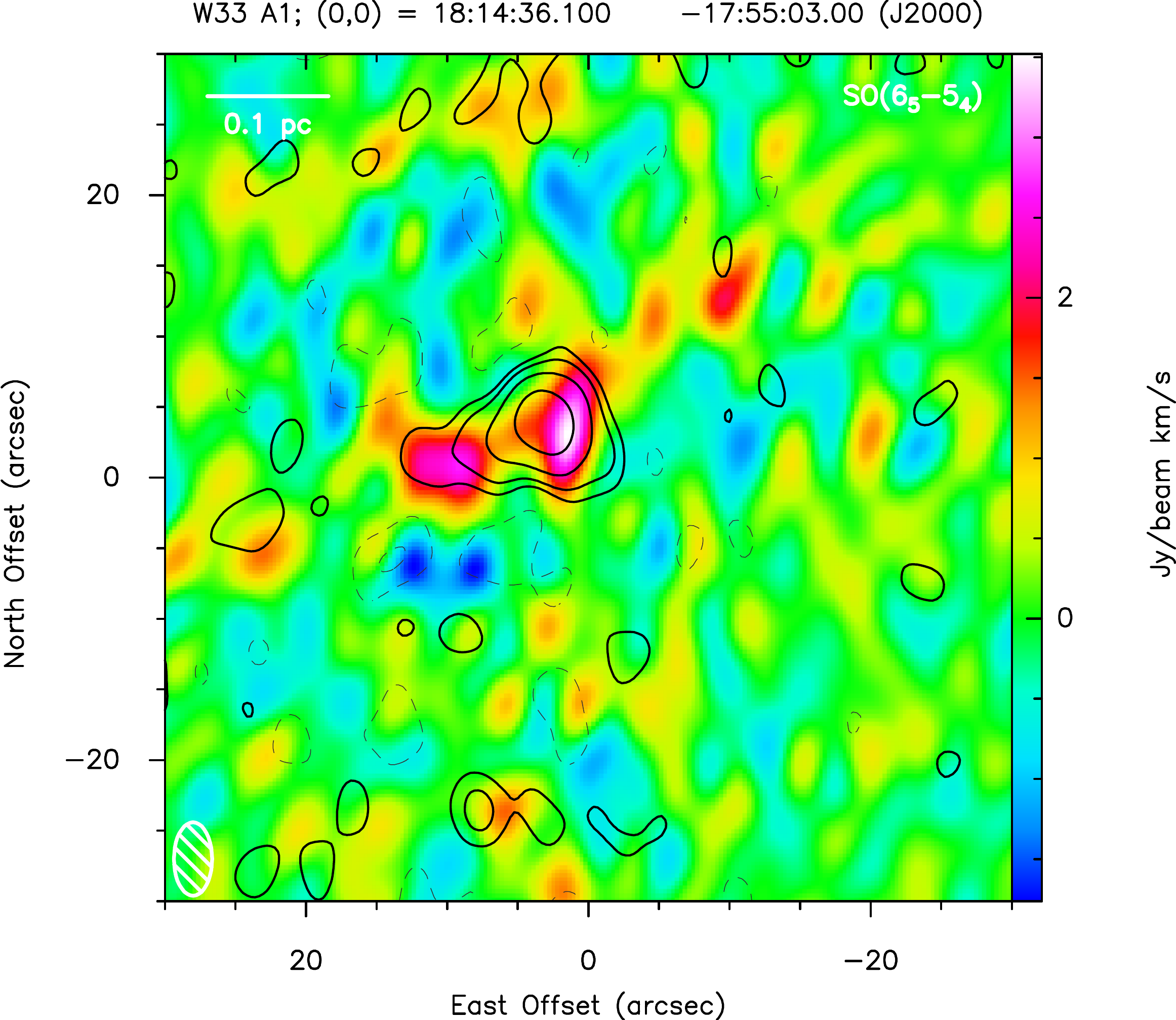}}\\	
	\subfloat{\includegraphics[width=9cm]{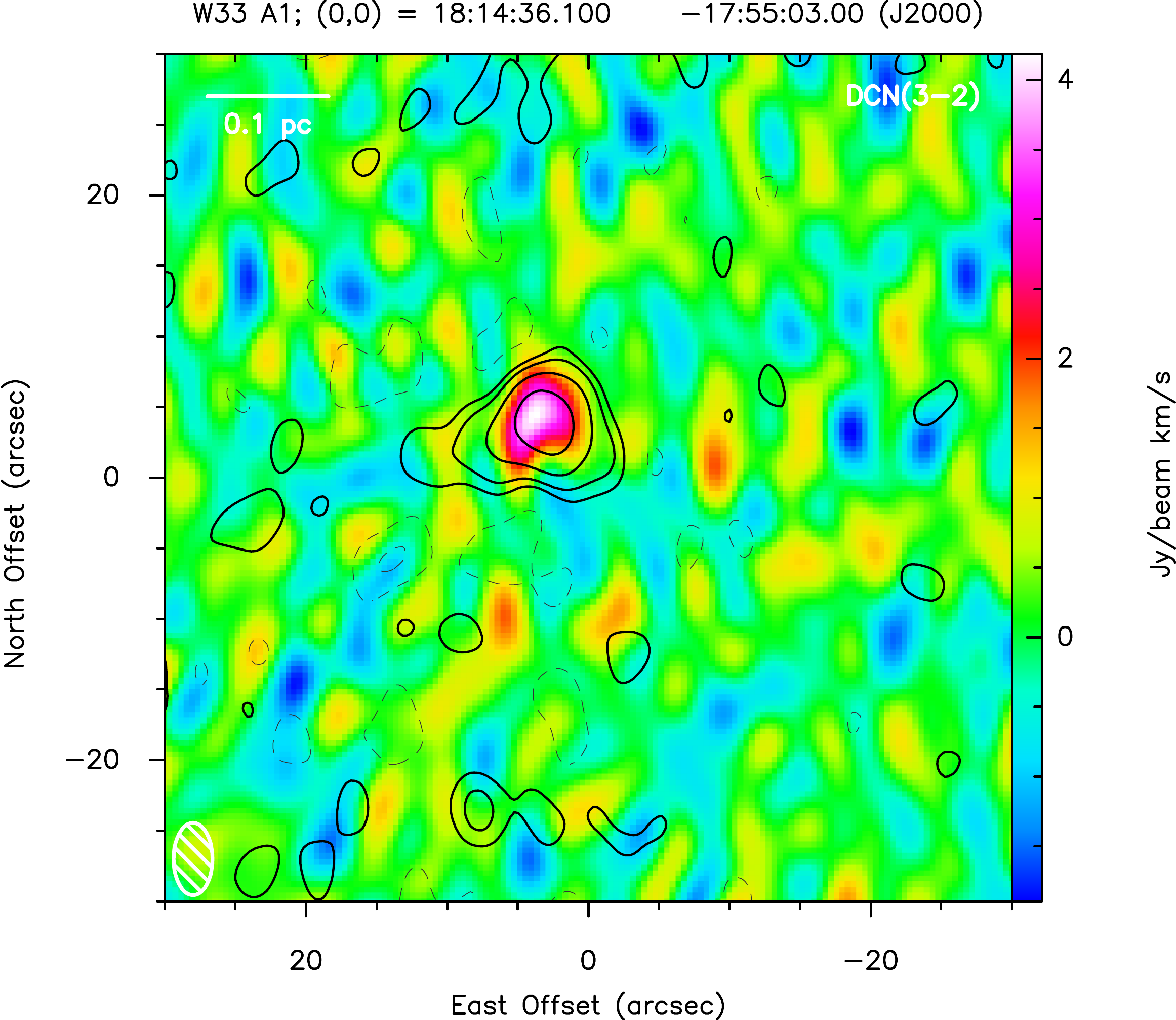}}	
	\label{W33A1-SMA-IntInt}
\end{figure*}

\addtocounter{figure}{-1}
\begin{figure*}
	\centering
	\caption{Continued.}
	\subfloat{\includegraphics[width=9cm]{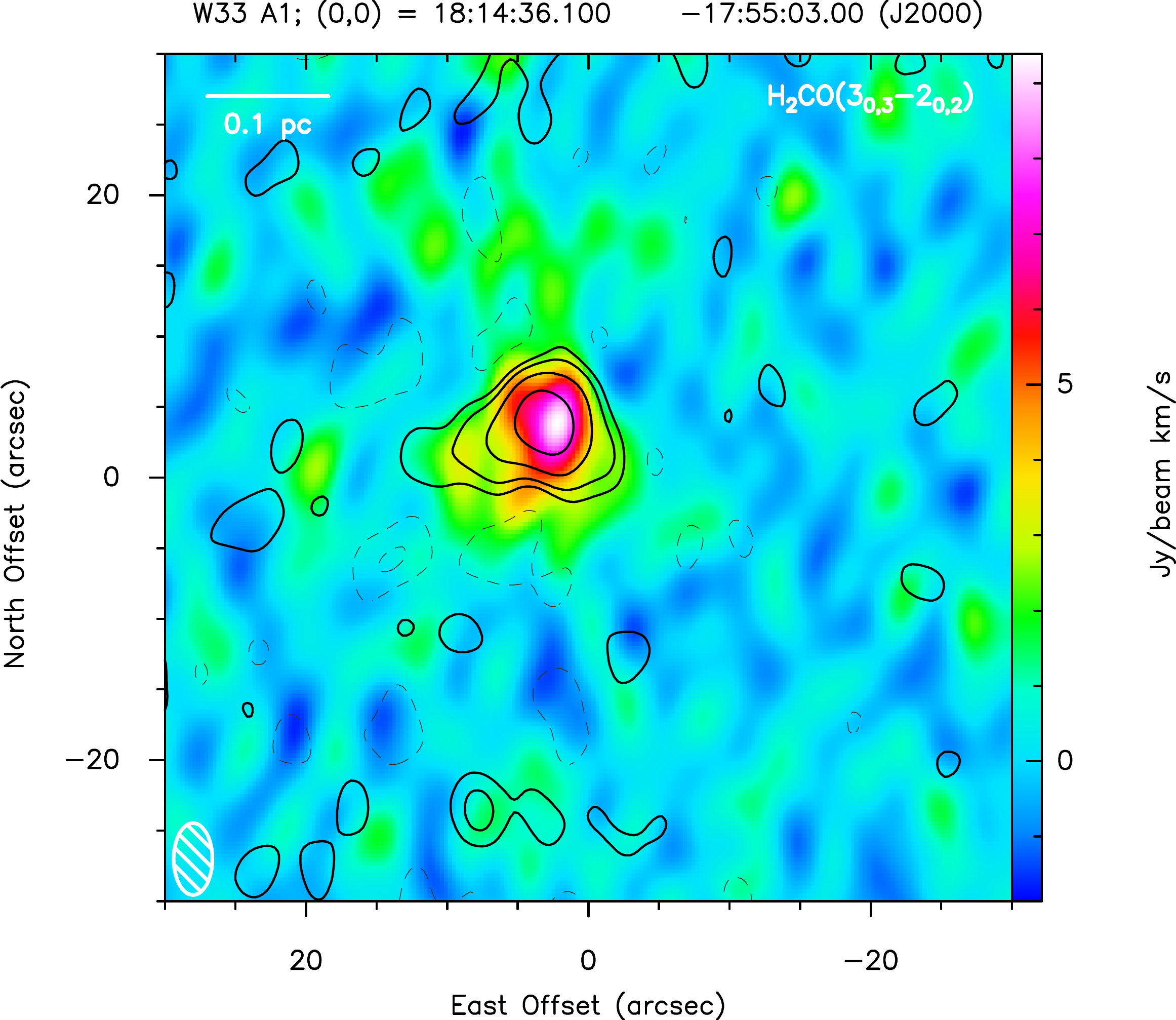}}\hspace{0.2cm}	
	\subfloat{\includegraphics[width=9cm]{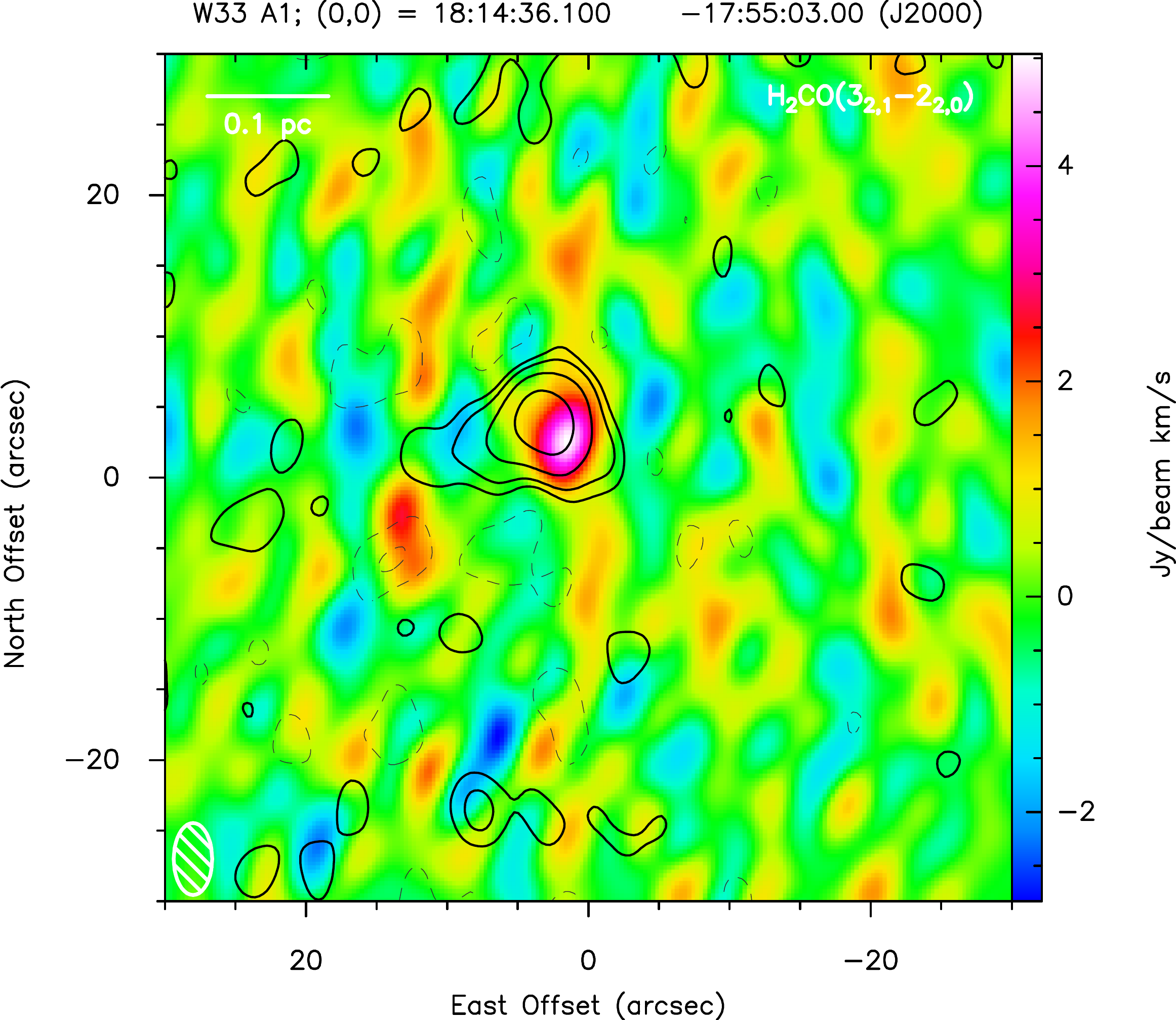}}\\	
	\subfloat{\includegraphics[width=9cm]{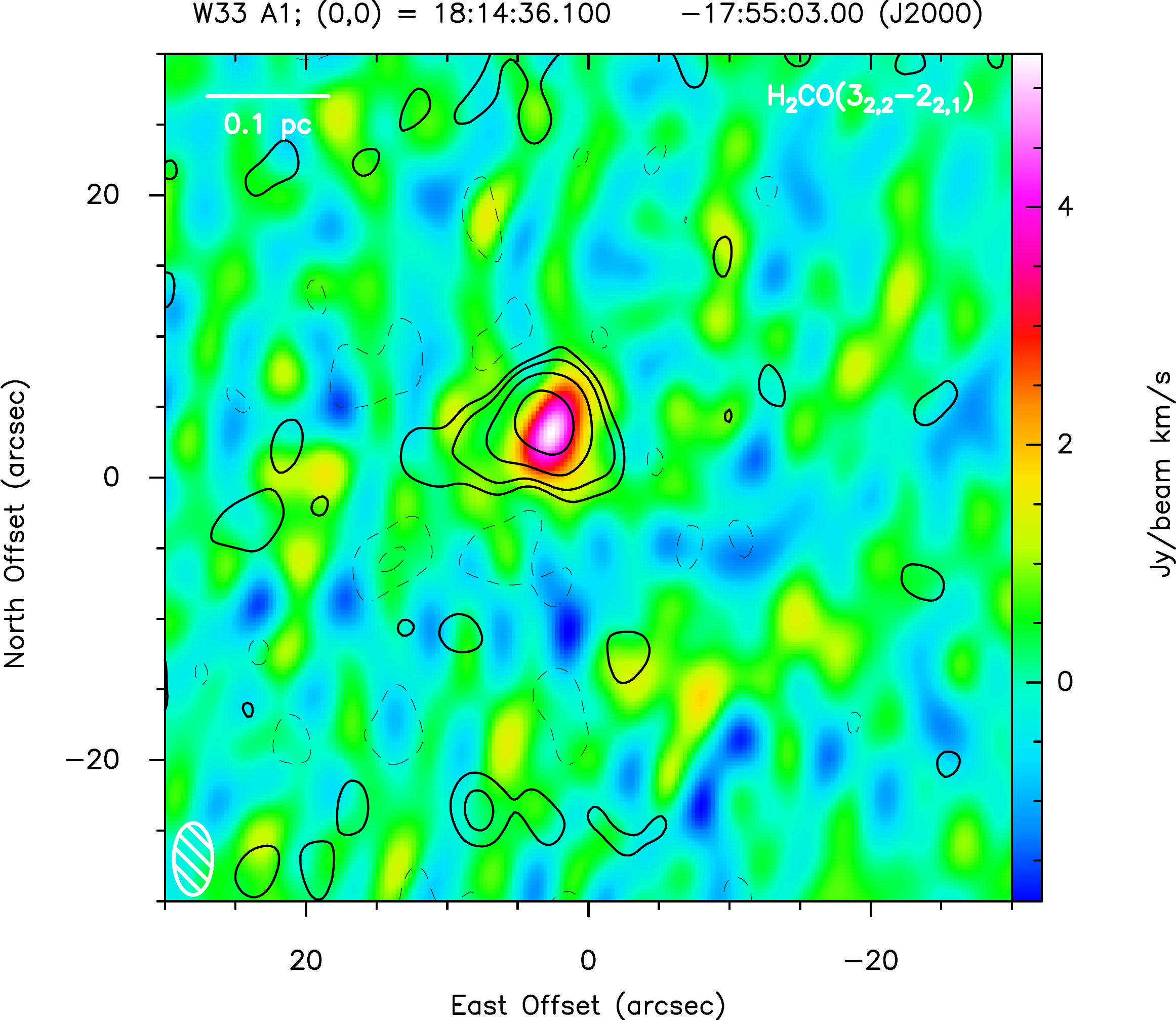}}\hspace{0.2cm}	
	\subfloat{\includegraphics[width=9cm]{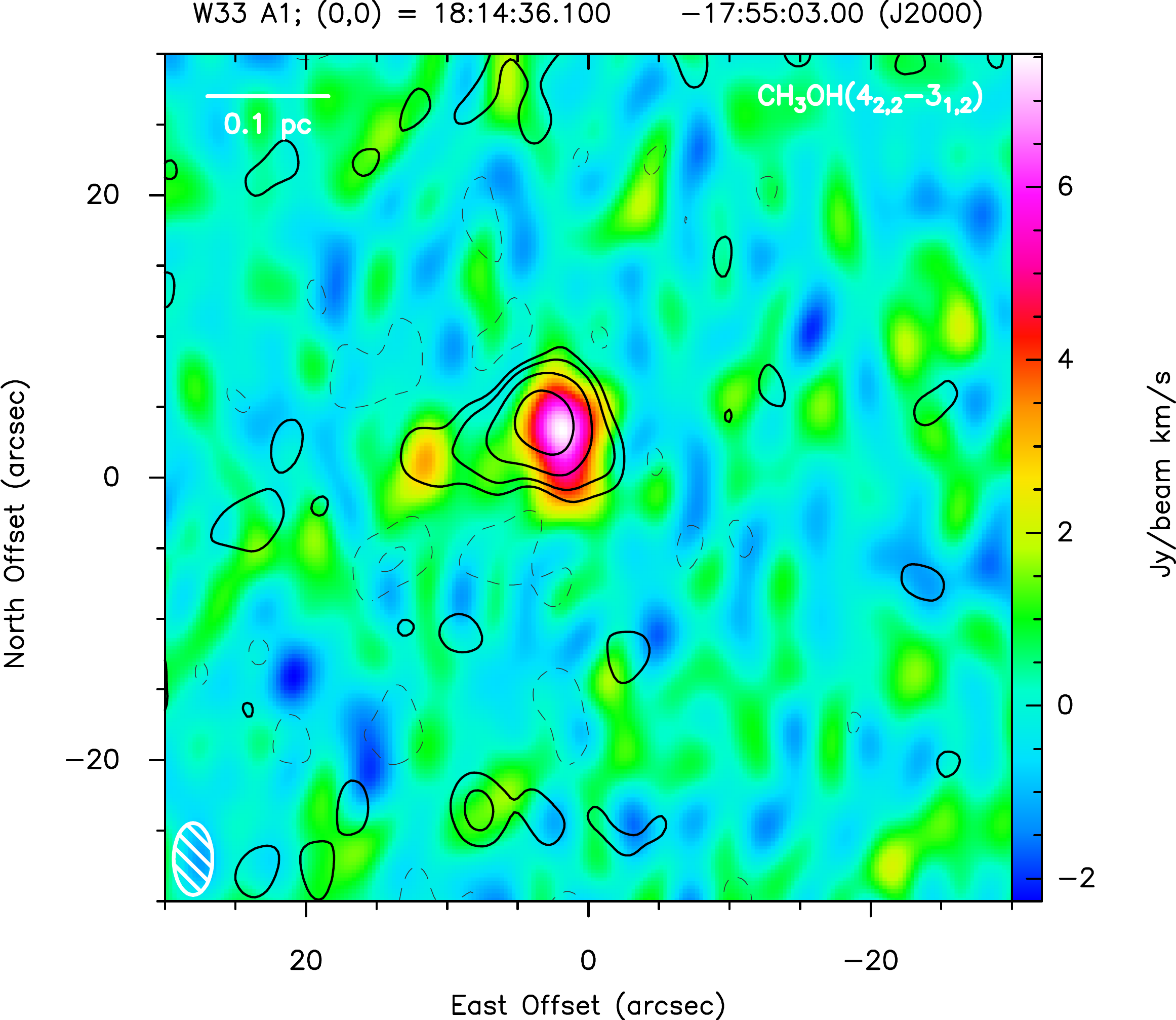}}	
\end{figure*}

\begin{figure*}
	\centering
	\caption{Line emission of detected transitions in W33\,B1. The contours show the SMA continuum emission at 230 GHz (same contour levels as in Fig. \ref{W33_SMA_CH0}). The name of the each transition is shown in the upper right corner. A scale of 0.1 pc is marked in the upper left corner, and the synthesised beam is shown in the lower left corner.}
	\subfloat{\includegraphics[width=9cm]{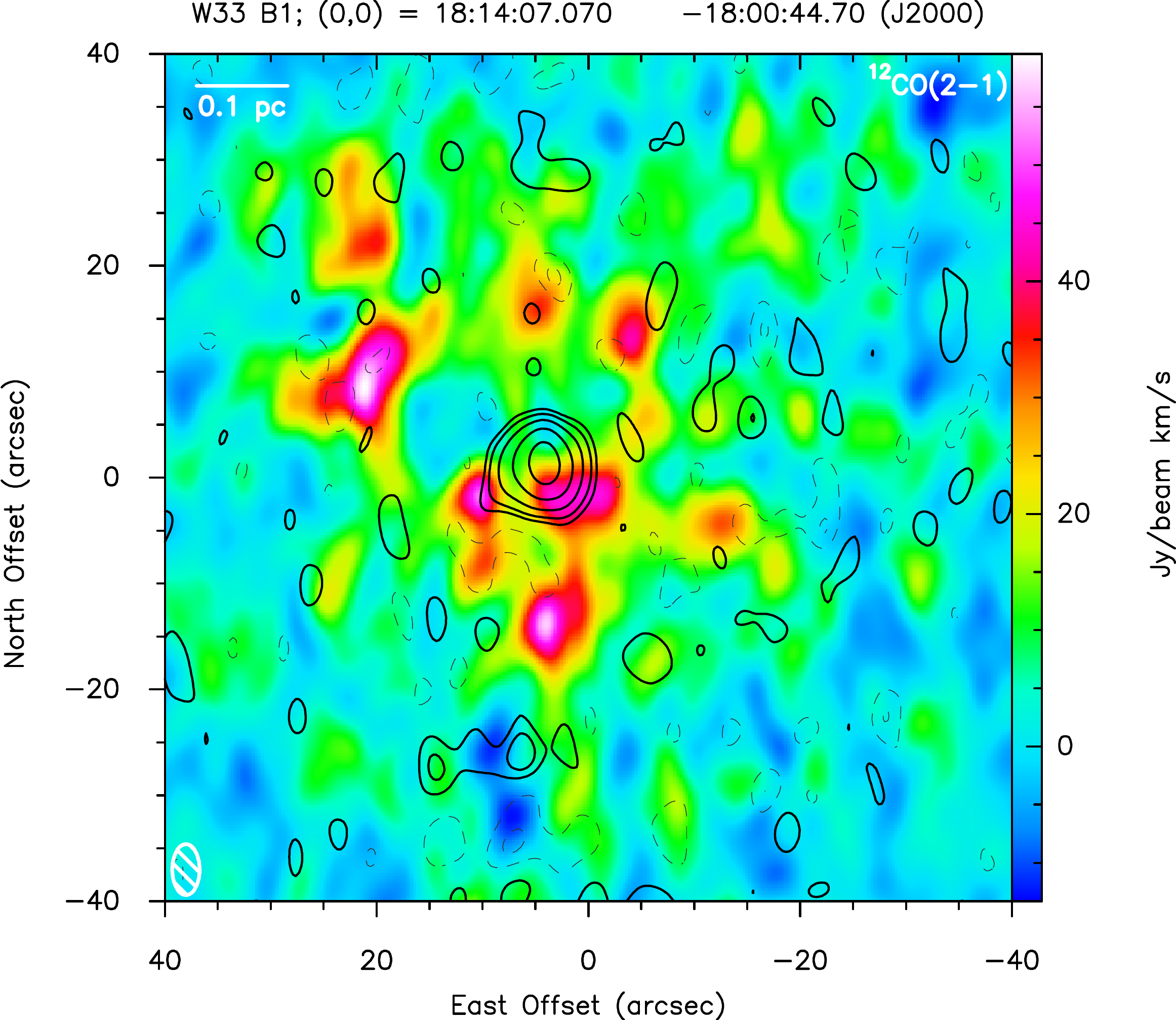}}\hspace{0.2cm}	
	\subfloat{\includegraphics[width=9cm]{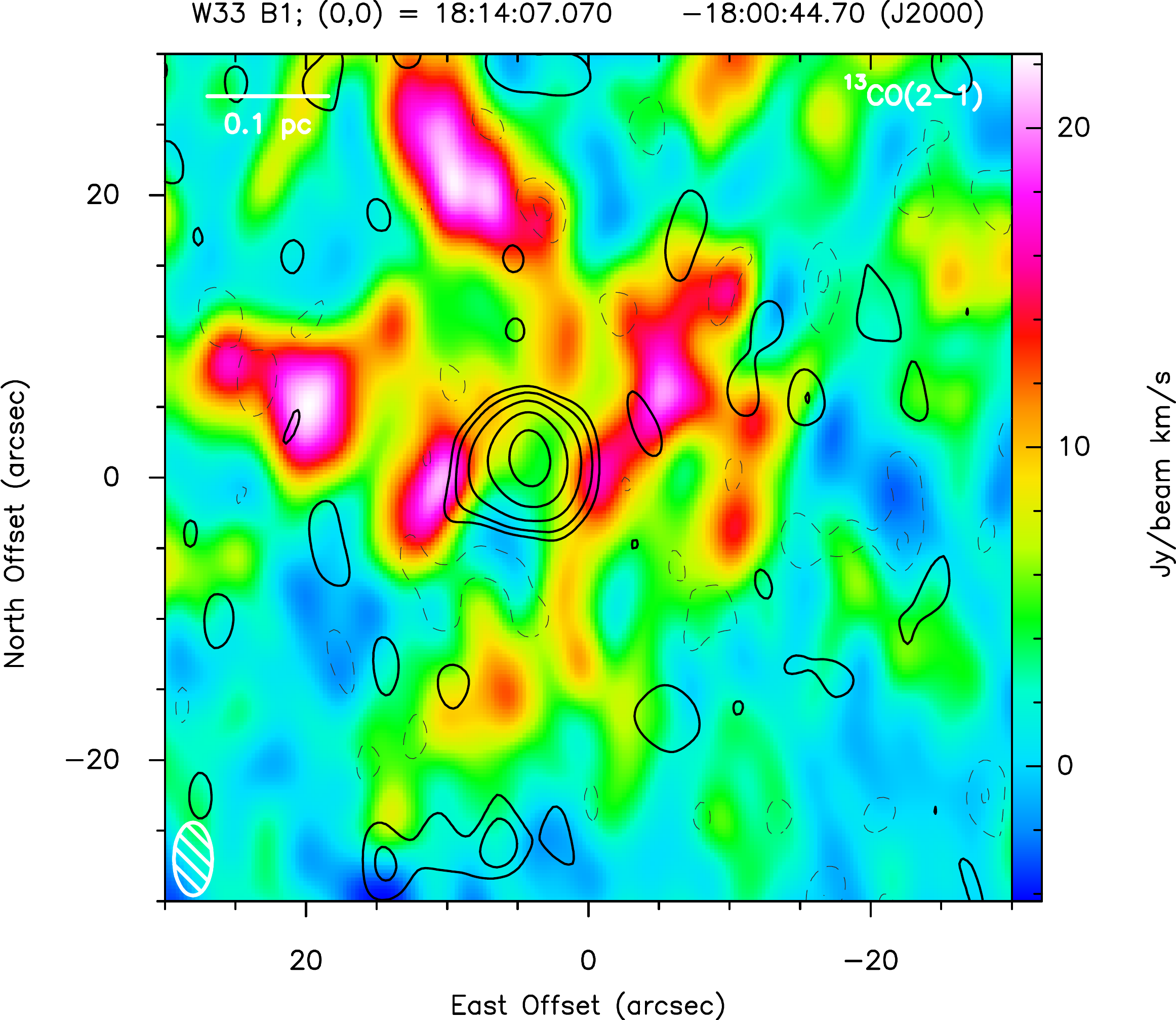}}\\
	\subfloat{\includegraphics[width=9cm]{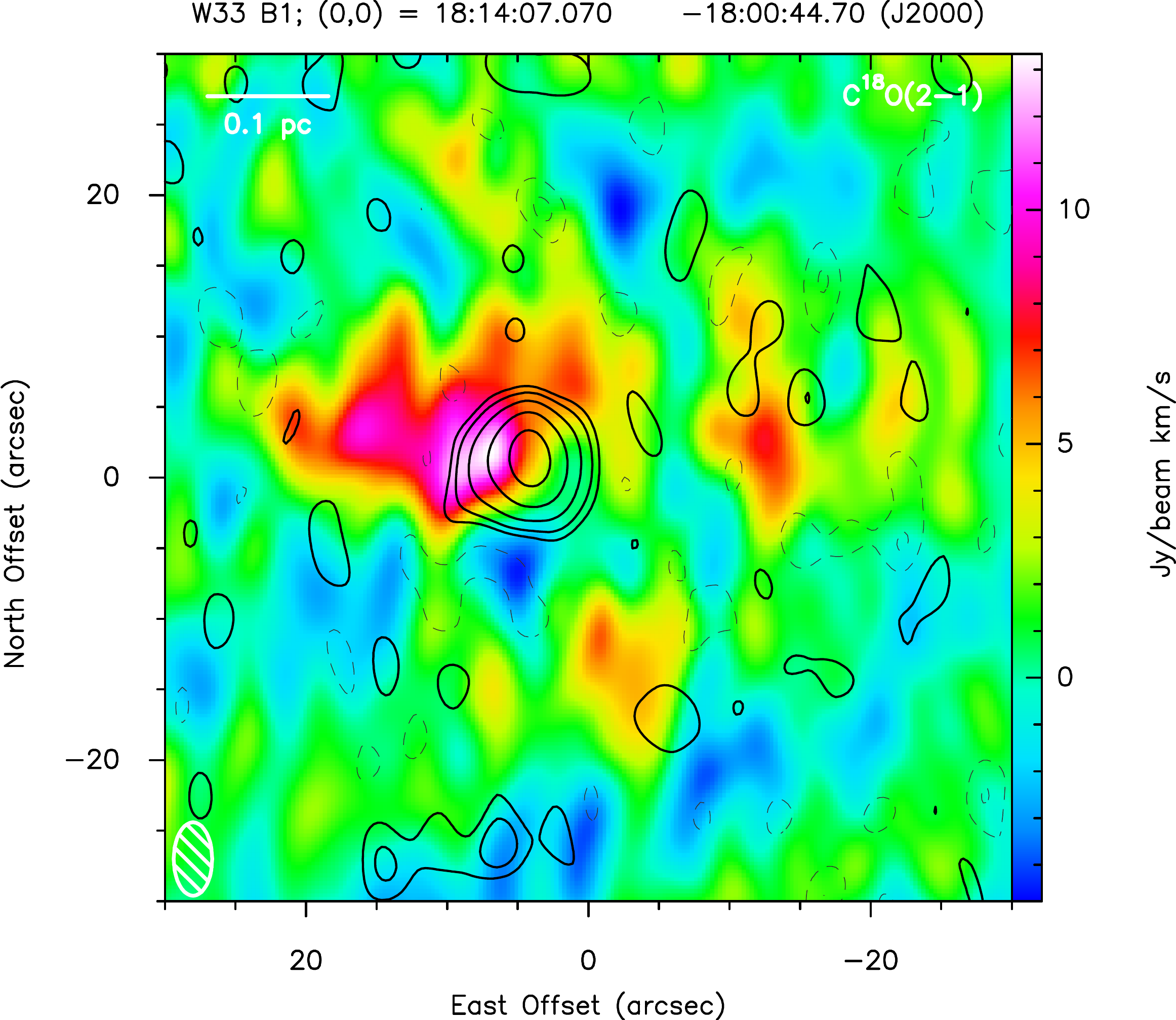}}\hspace{0.2cm}	
	\subfloat{\includegraphics[width=9cm]{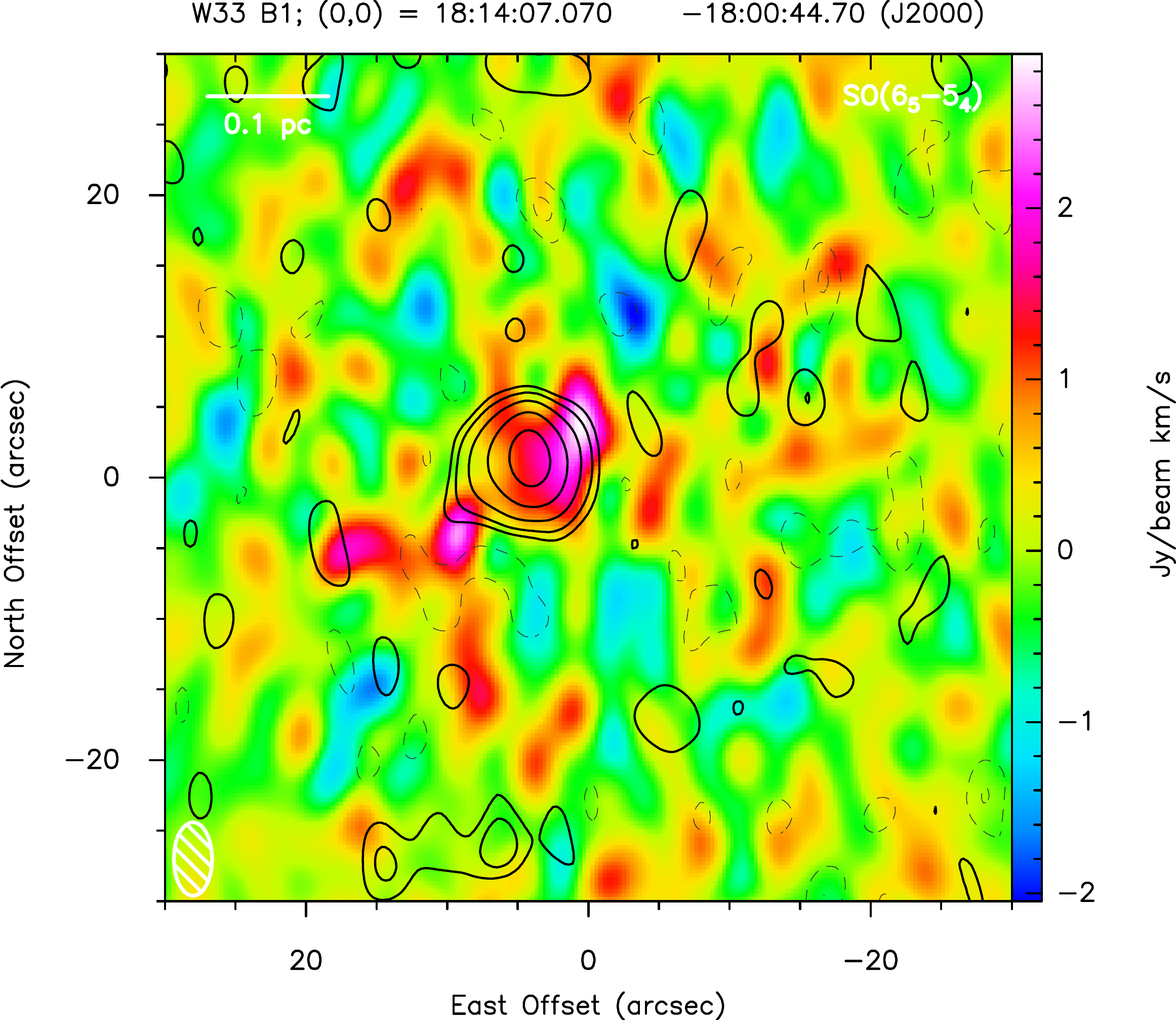}}	
	\label{W33B1-SMA-IntInt}
\end{figure*}

\addtocounter{figure}{-1}
\begin{figure*}
	\centering
	\caption{Continued.}
	\subfloat{\includegraphics[width=9cm]{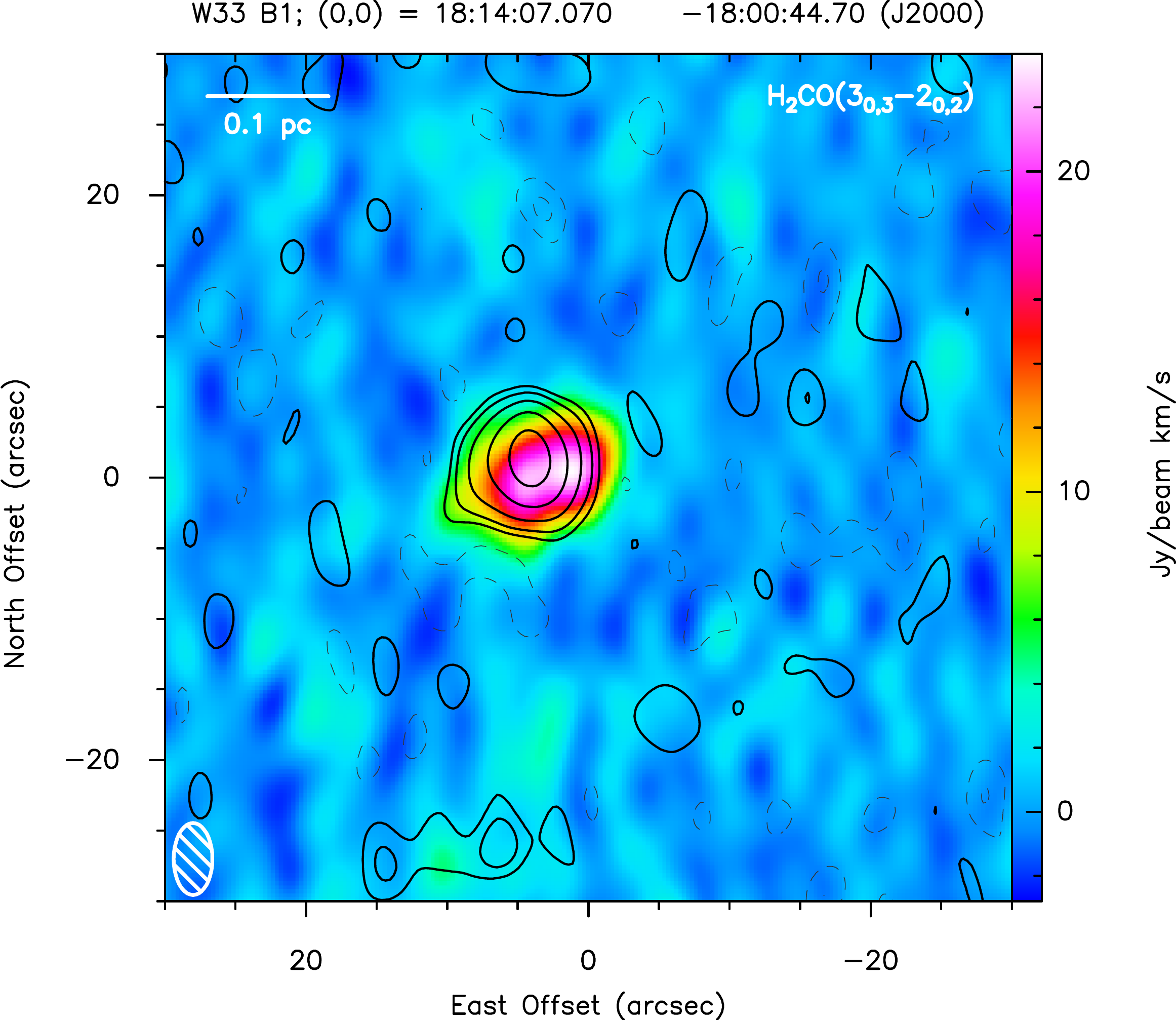}}\hspace{0.2cm}	
	\subfloat{\includegraphics[width=9cm]{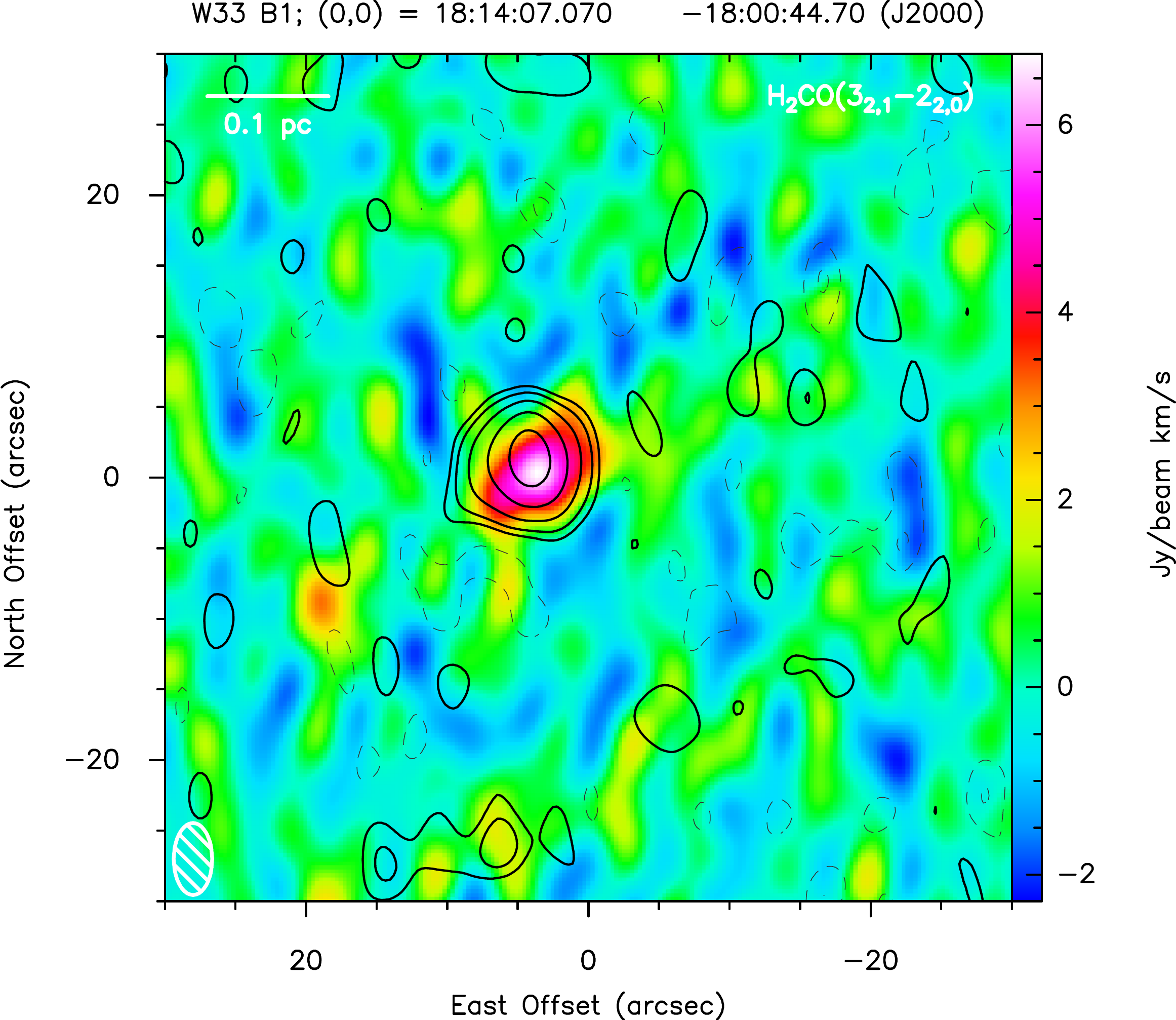}}\\ 	
	\subfloat{\includegraphics[width=9cm]{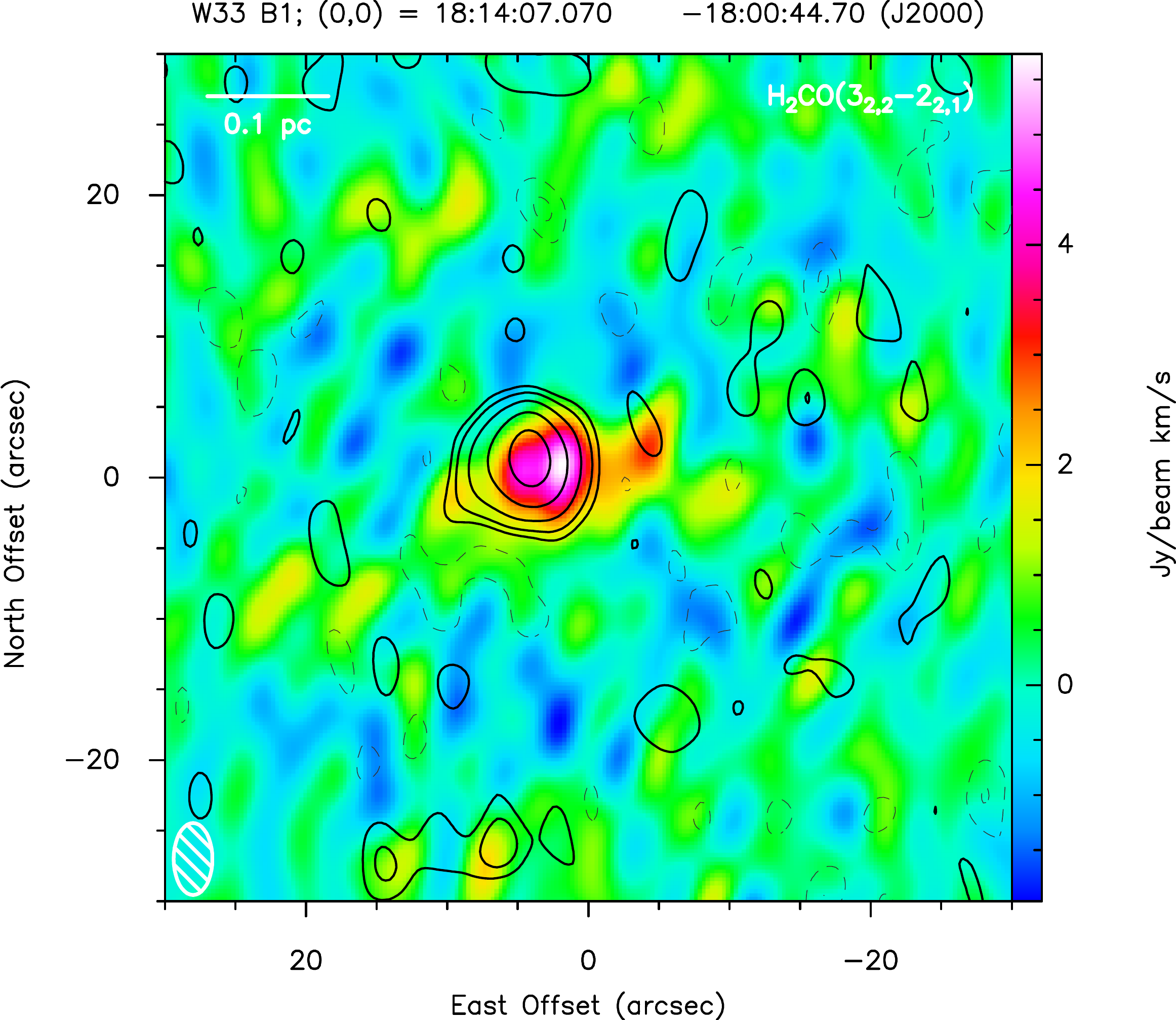}}\hspace{0.2cm}	
	\subfloat{\includegraphics[width=9cm]{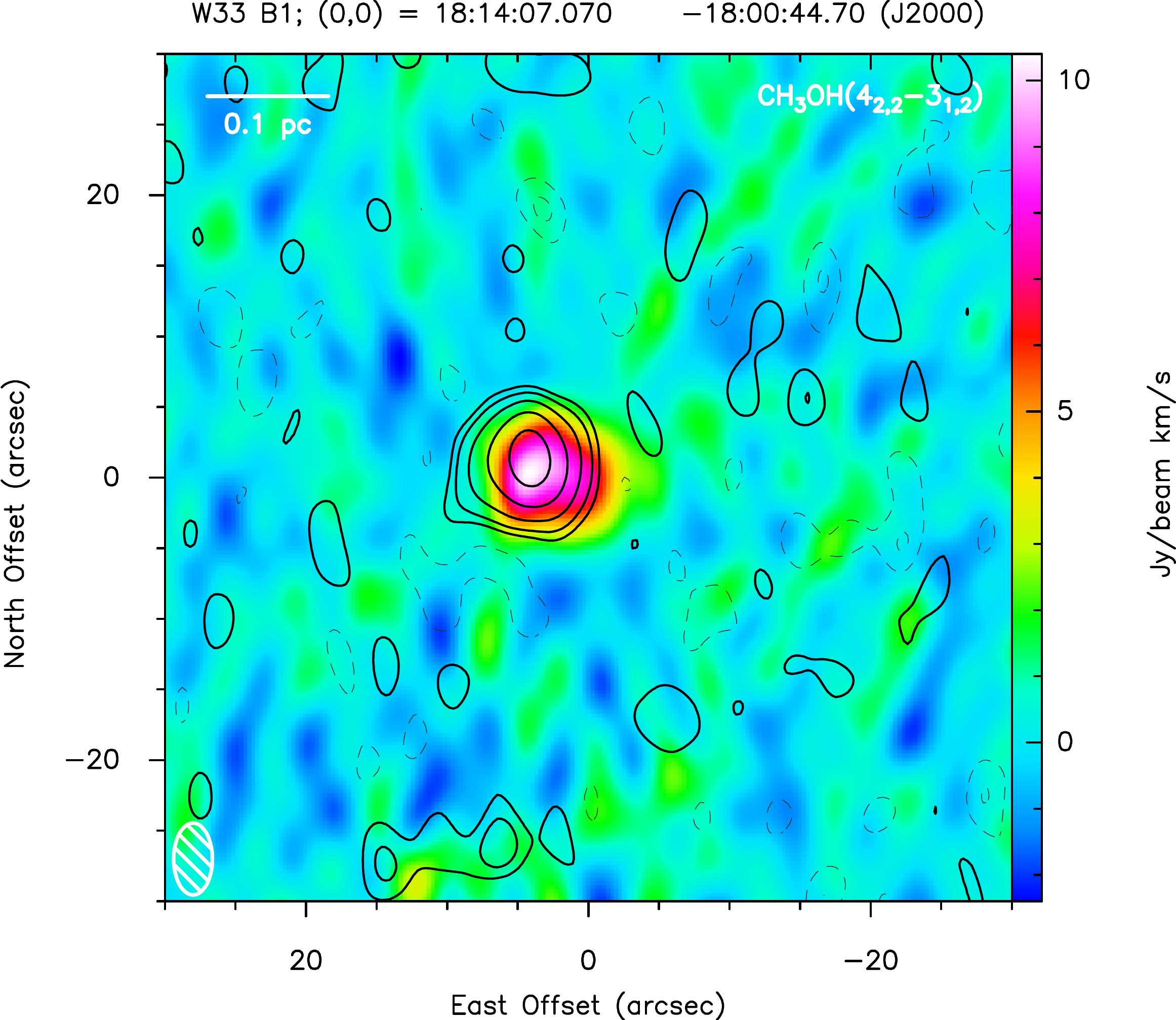}}	
\end{figure*}

\begin{figure*}
	\centering
	\caption{Line emission of detected transitions in W33\,B. The contours show the SMA continuum emission at 230 GHz (same contour levels as in Fig. \ref{W33_SMA_CH0}). The name of the each transition is shown in the upper right corner. A scale of 0.1 pc is marked in the upper left corner, and the synthesised beam is shown in the lower left corner.}
	\subfloat{\includegraphics[width=9cm]{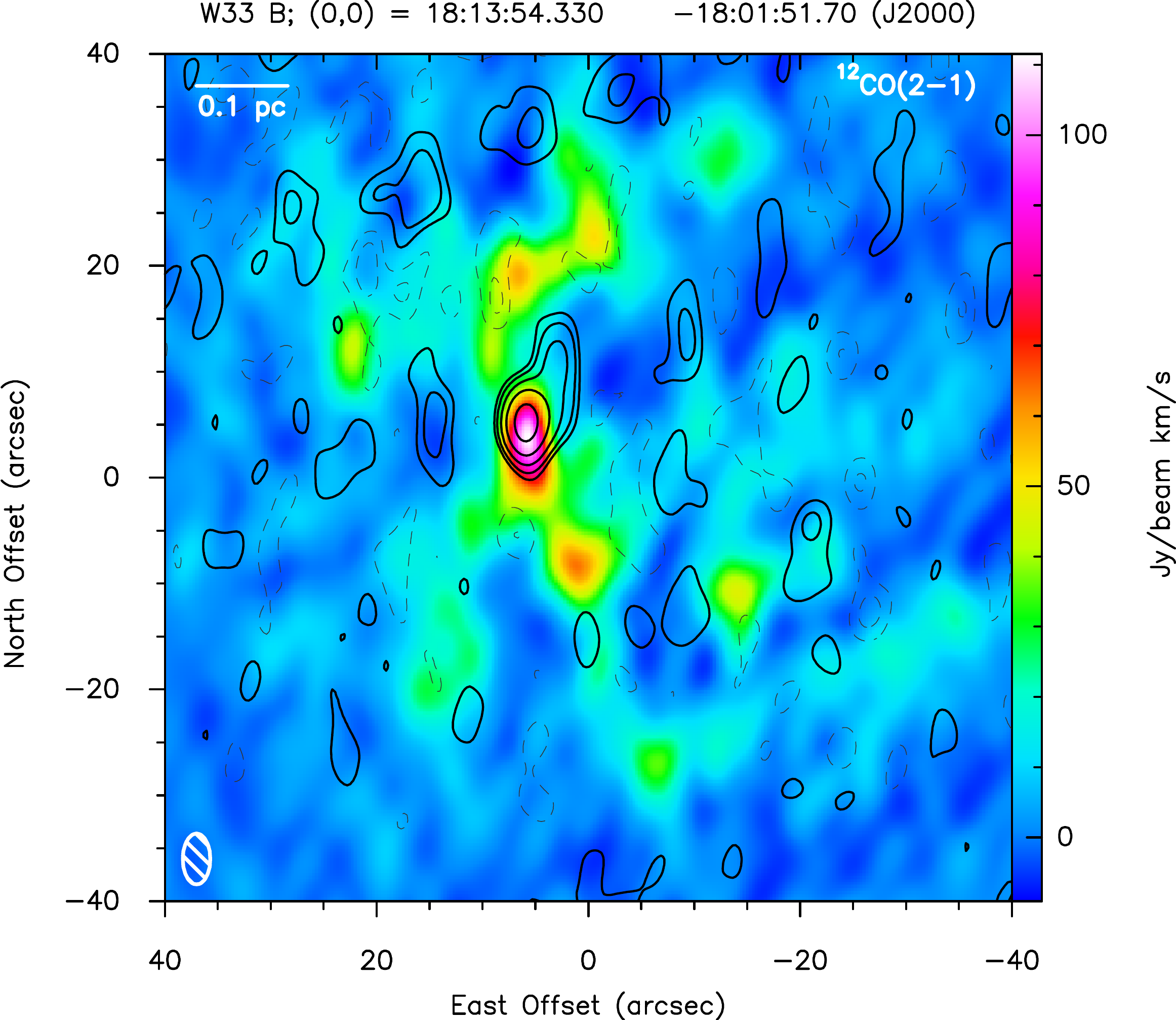}}\hspace{0.2cm}	
	\subfloat{\includegraphics[width=9cm]{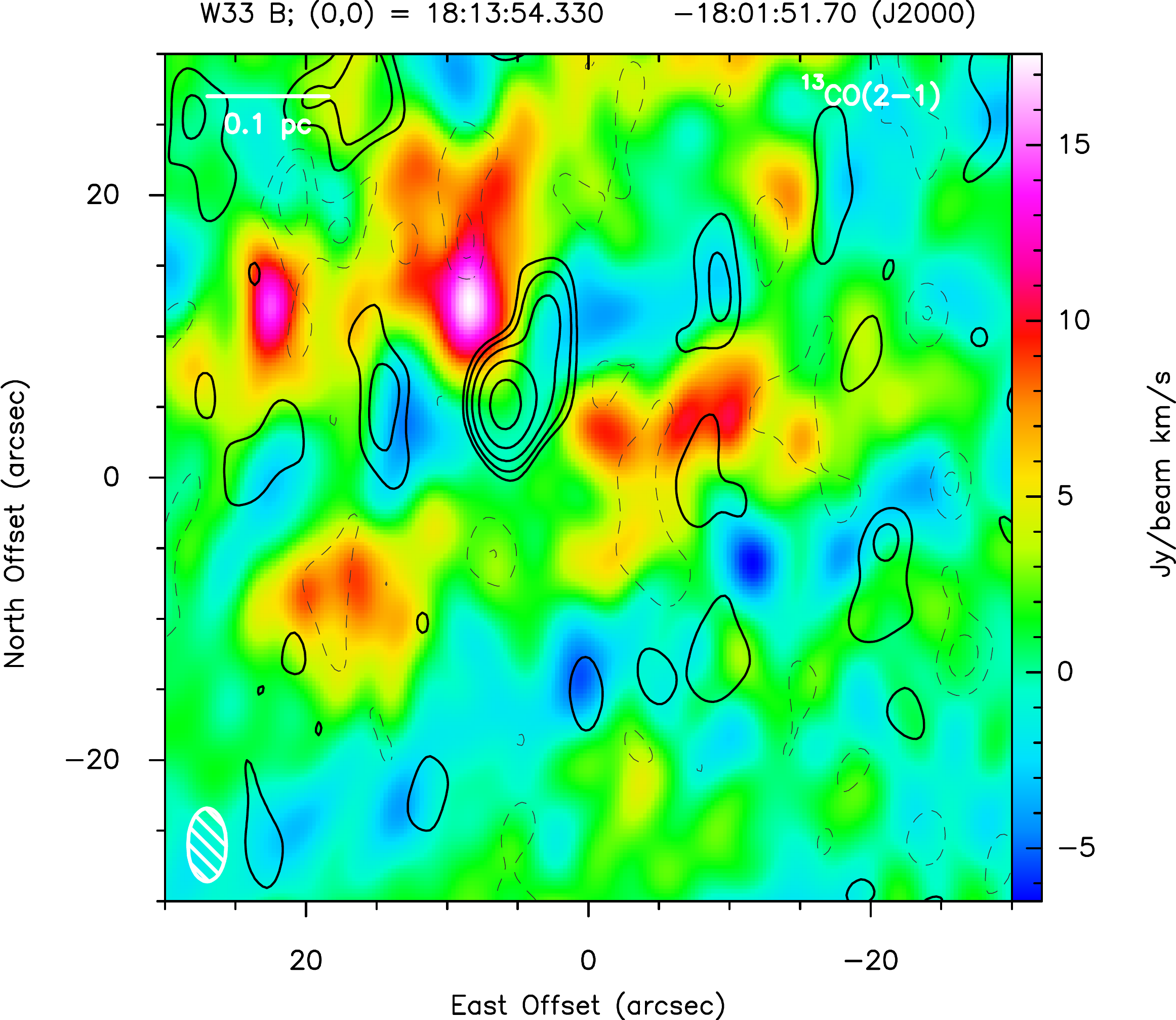}}\\
	\subfloat{\includegraphics[width=9cm]{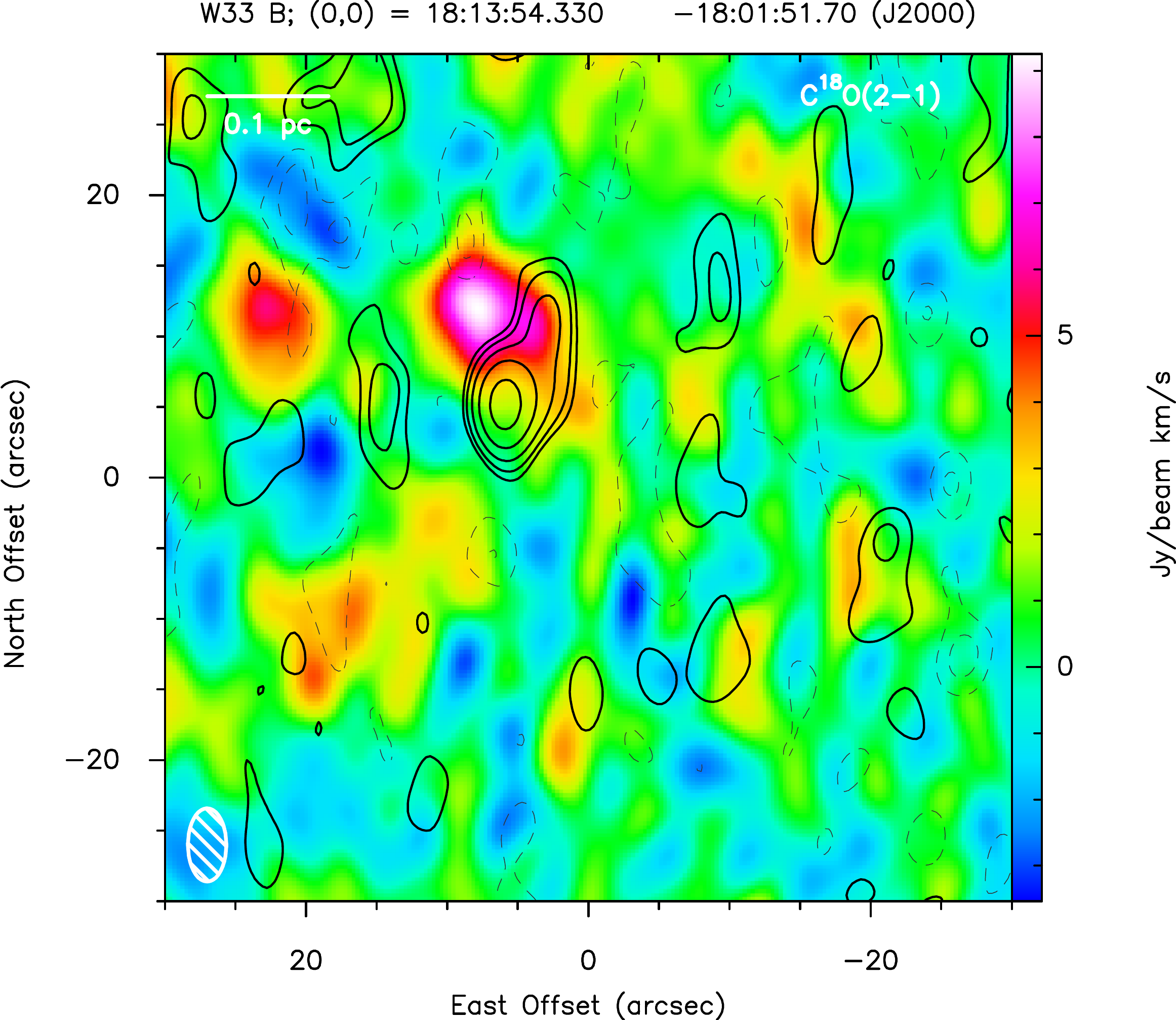}}\hspace{0.2cm}	
	\subfloat{\includegraphics[width=9cm]{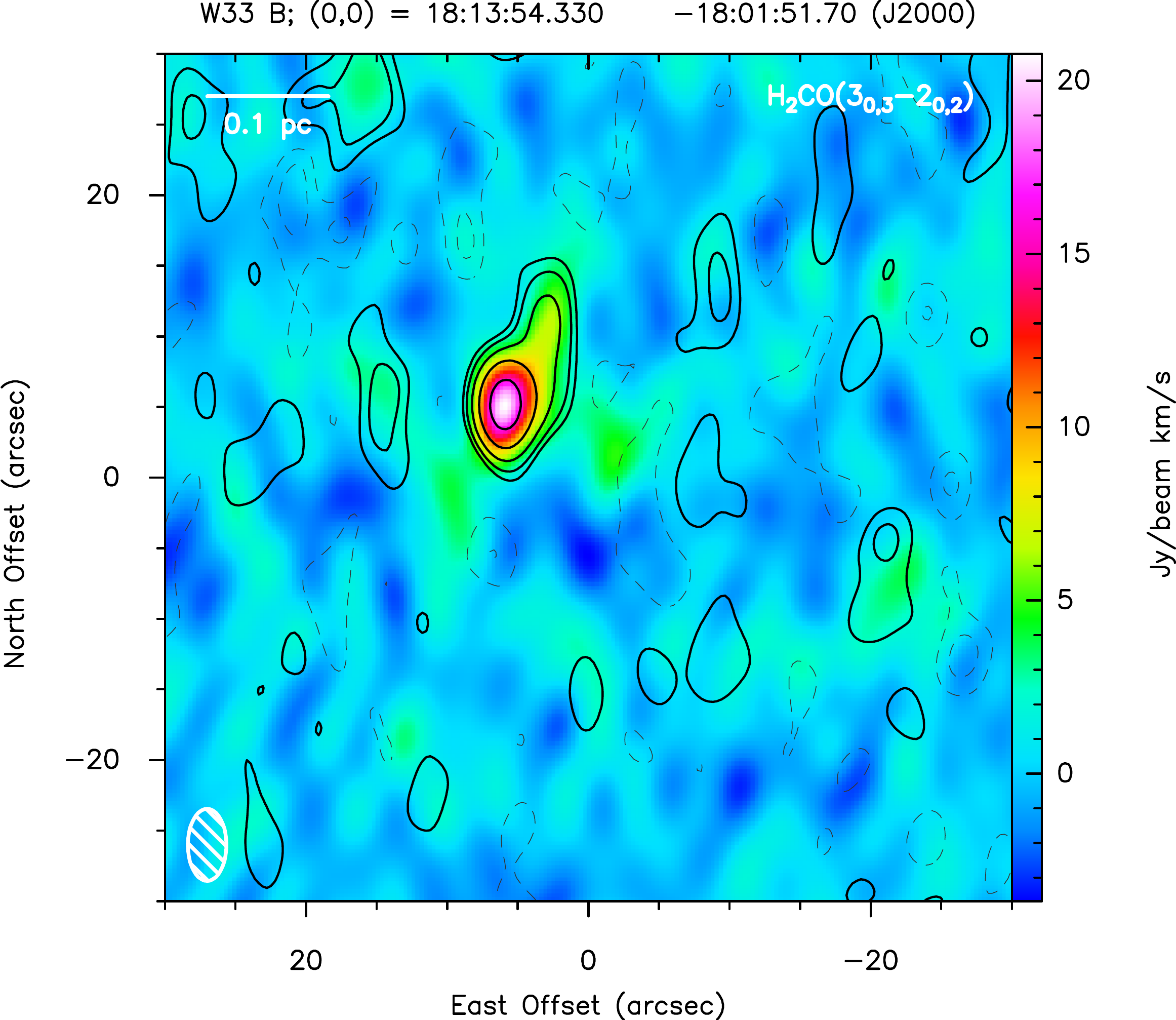}}\\	
	\subfloat{\includegraphics[width=9cm]{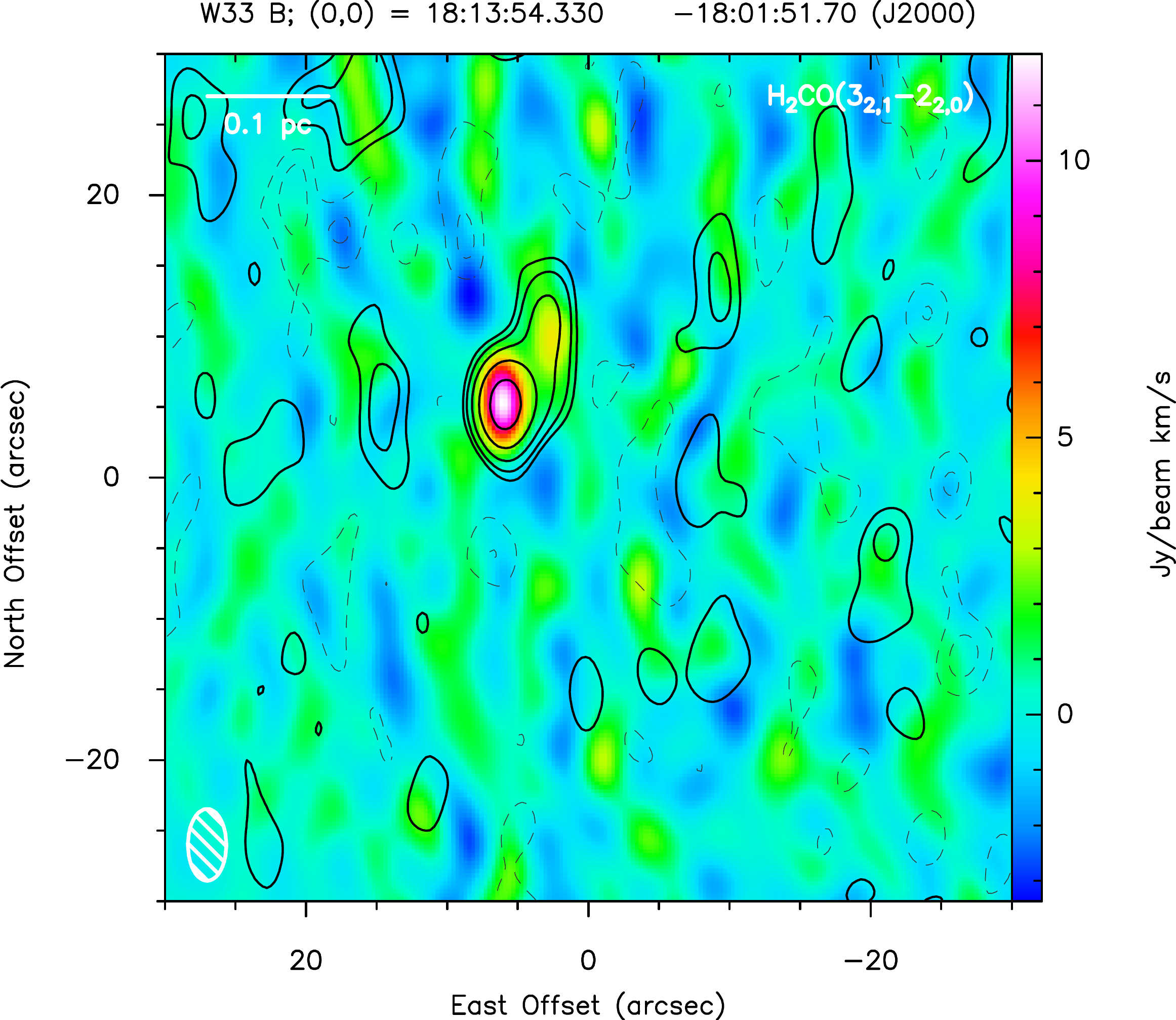}}\hspace{0.2cm} 
	\subfloat{\includegraphics[width=9cm]{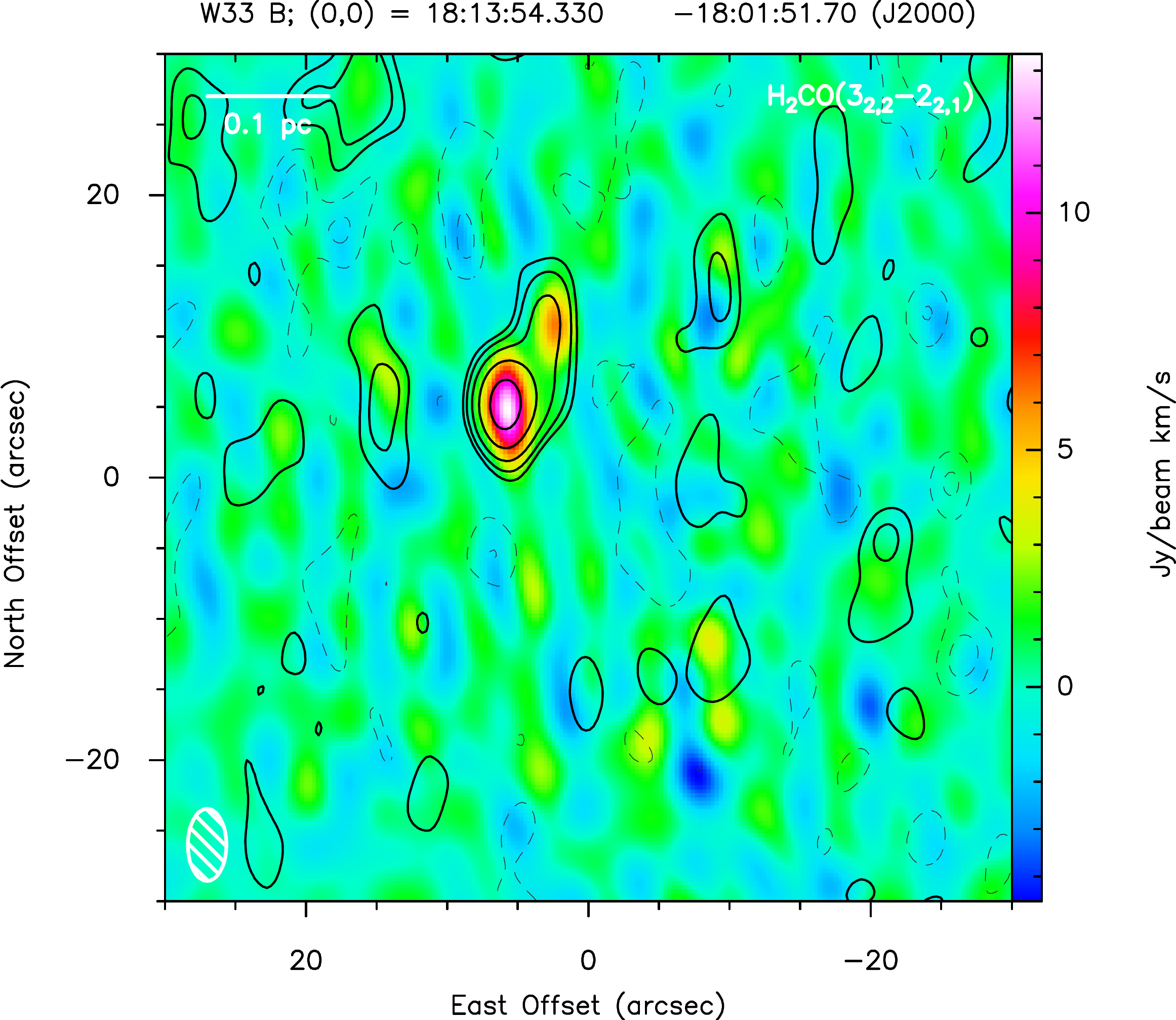}}	
	\label{W33B-SMA-IntInt}
\end{figure*}

\addtocounter{figure}{-1}
\begin{figure*}
	\centering
	\caption{Continued.}
	\subfloat{\includegraphics[width=9cm]{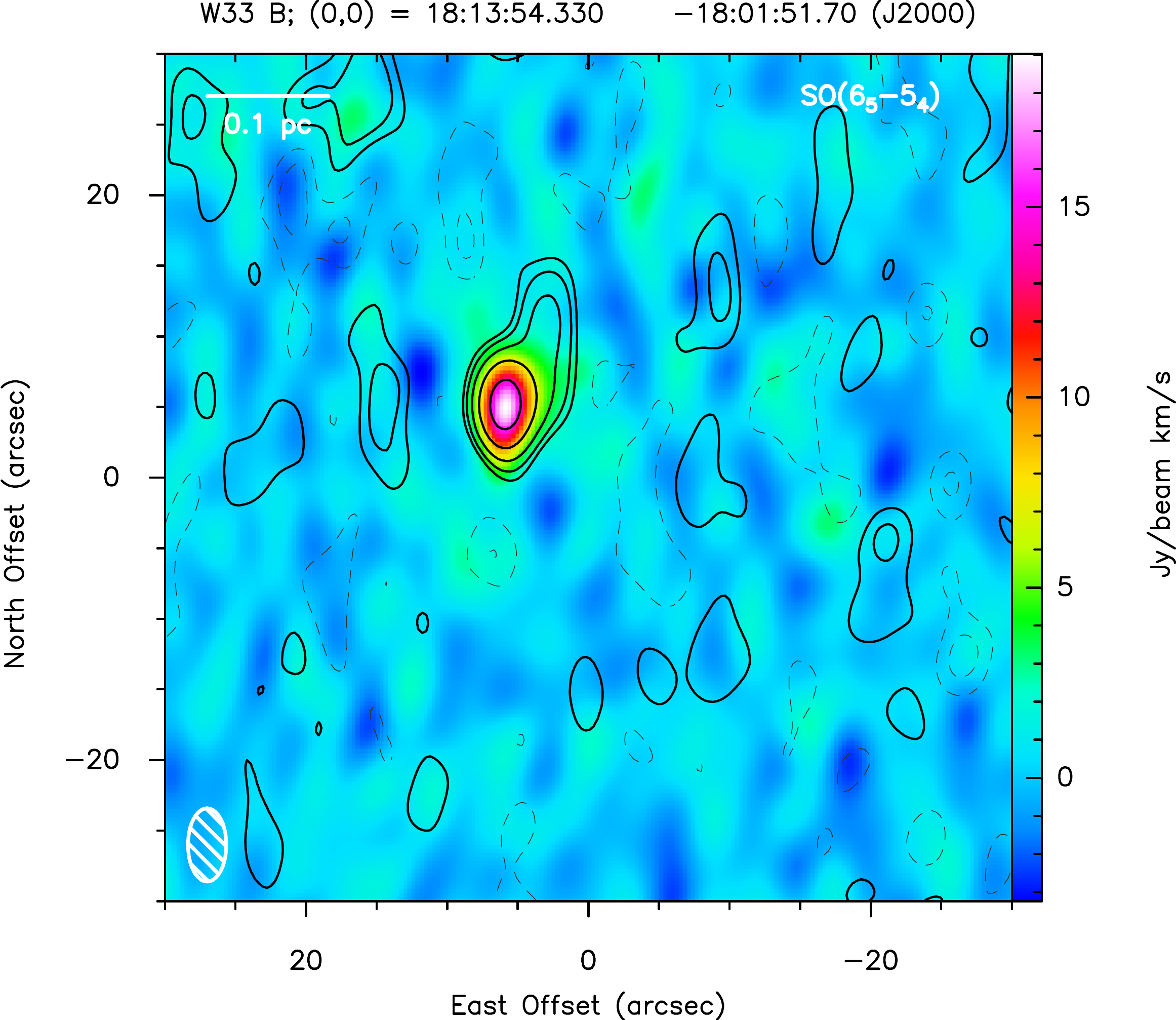}}	\hspace{0.2cm}
	\subfloat{\includegraphics[width=9cm]{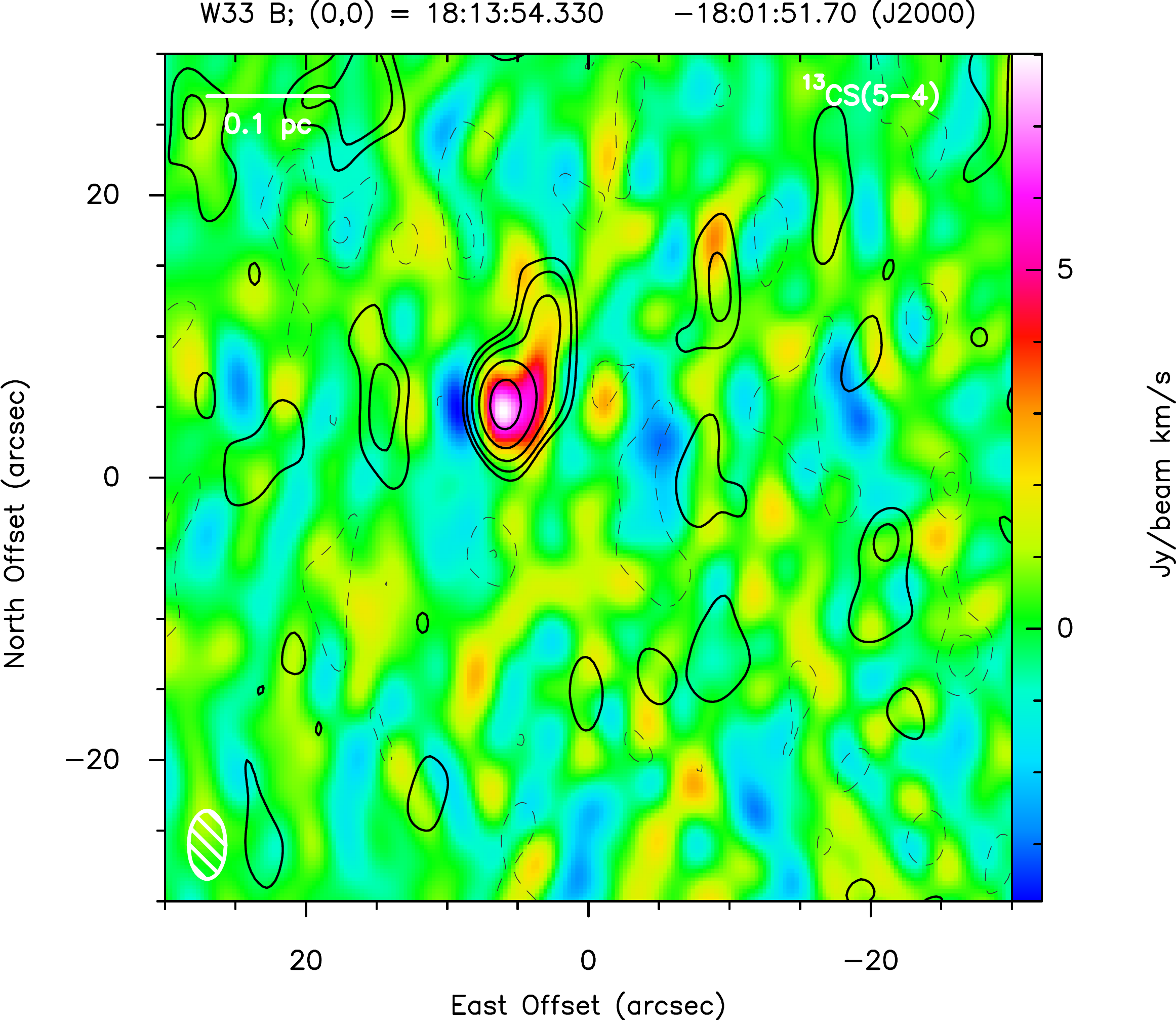}}\\	
	\subfloat{\includegraphics[width=9cm]{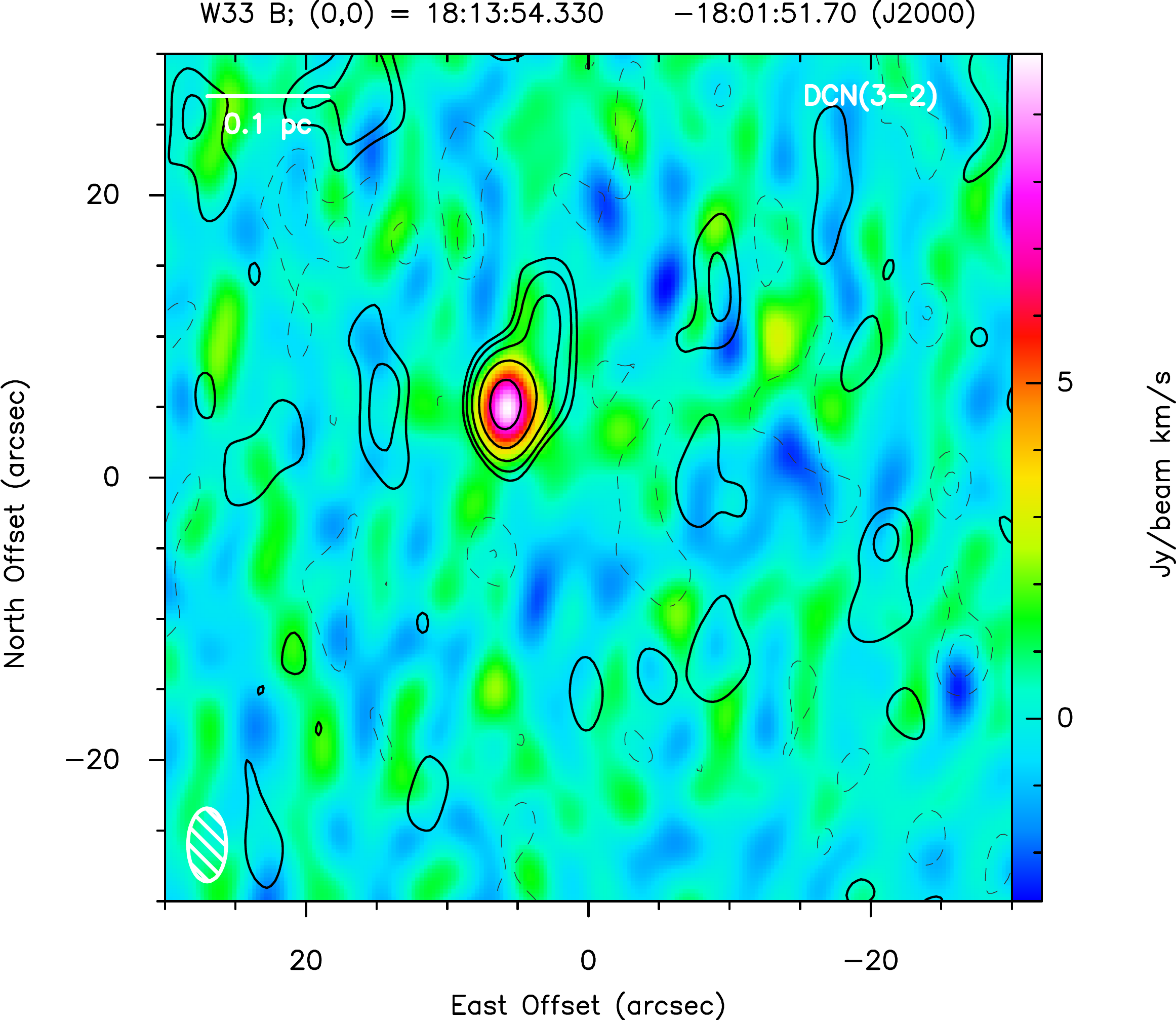}} \hspace{0.2cm}
	\subfloat{\includegraphics[width=9cm]{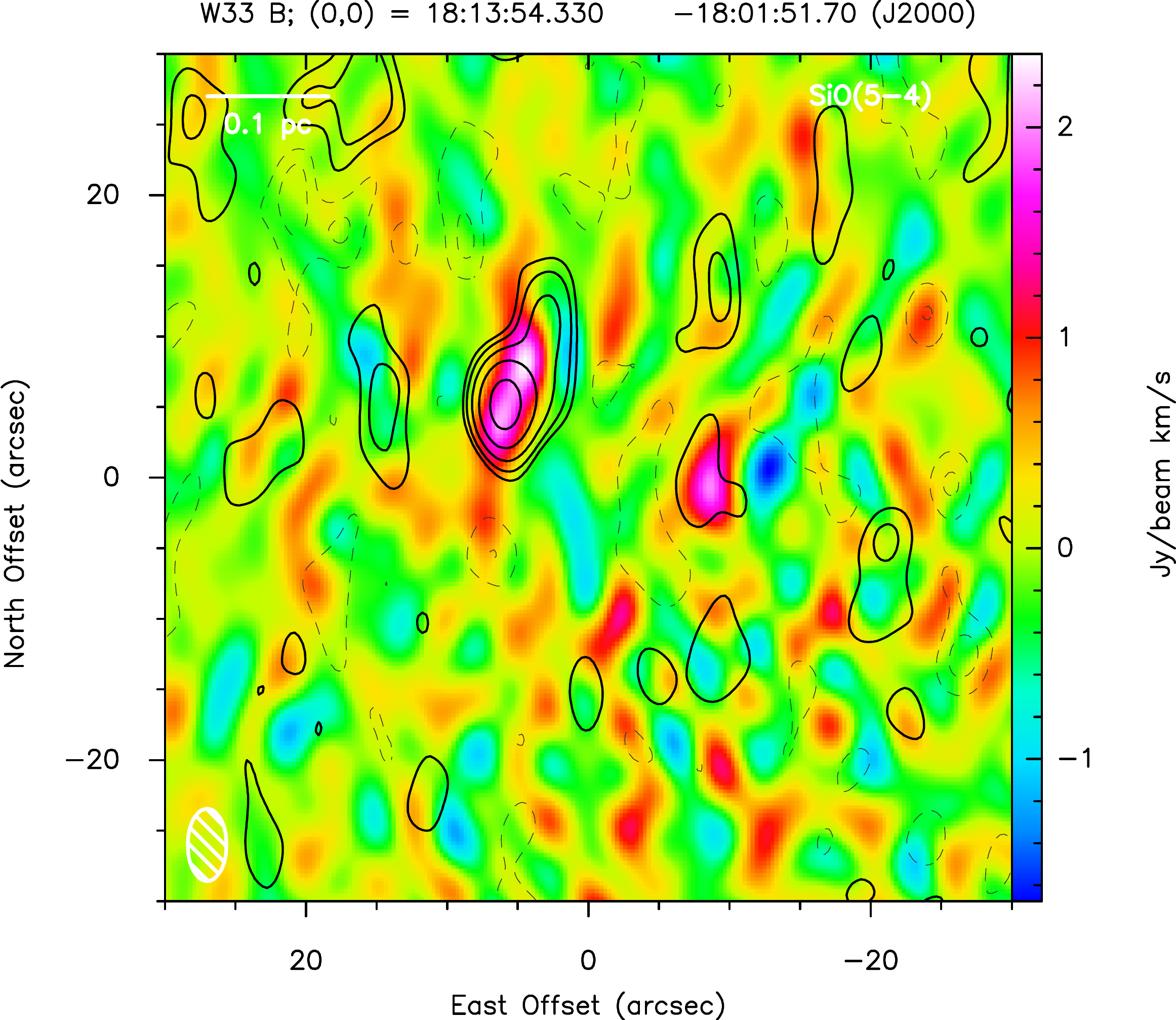}}	
\end{figure*}

\addtocounter{figure}{-1}
\begin{figure*}
	\centering
	\caption{Continued.}
	\subfloat{\includegraphics[width=9cm]{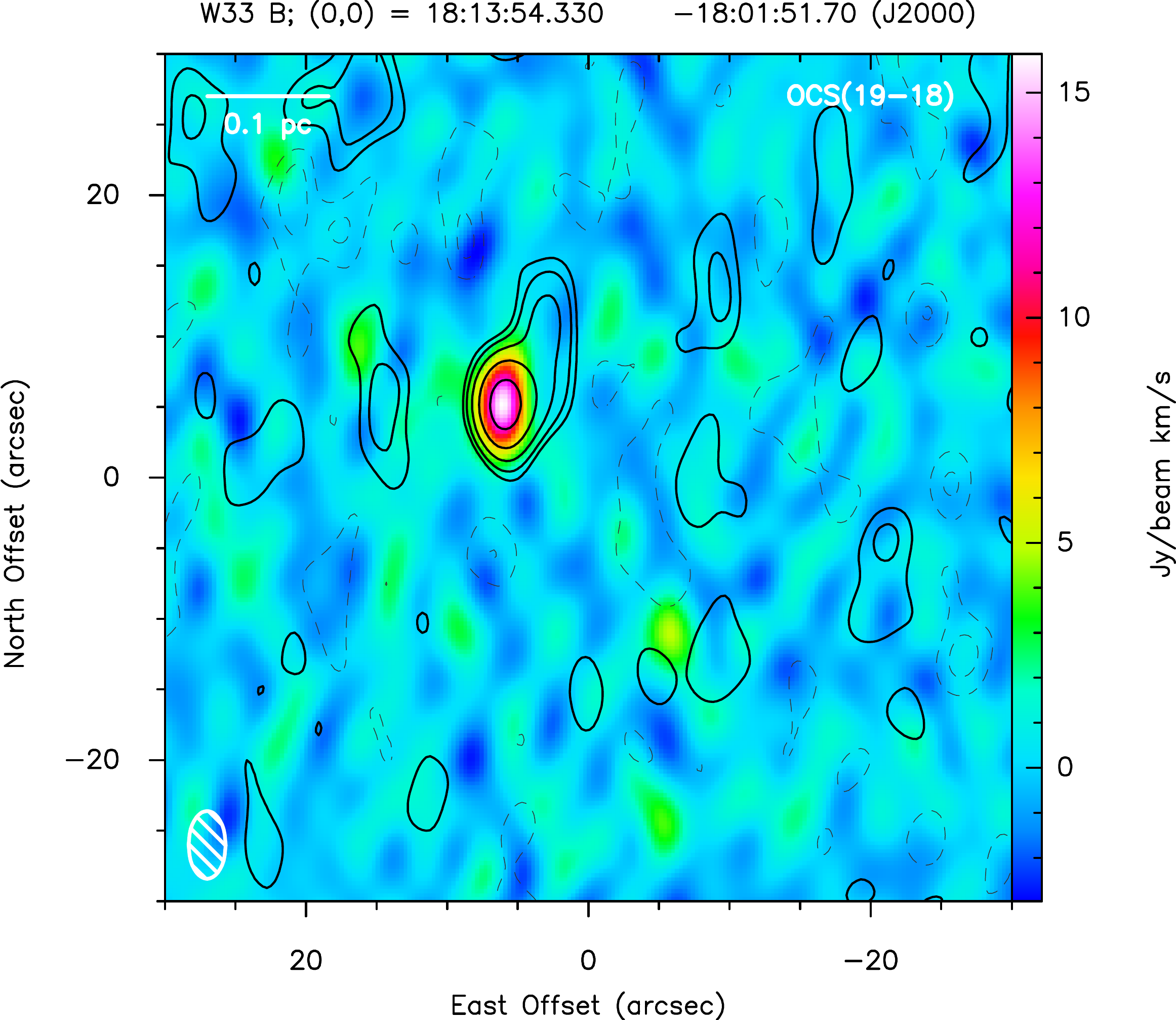}}\hspace{0.2cm}	
	\subfloat{\includegraphics[width=9cm]{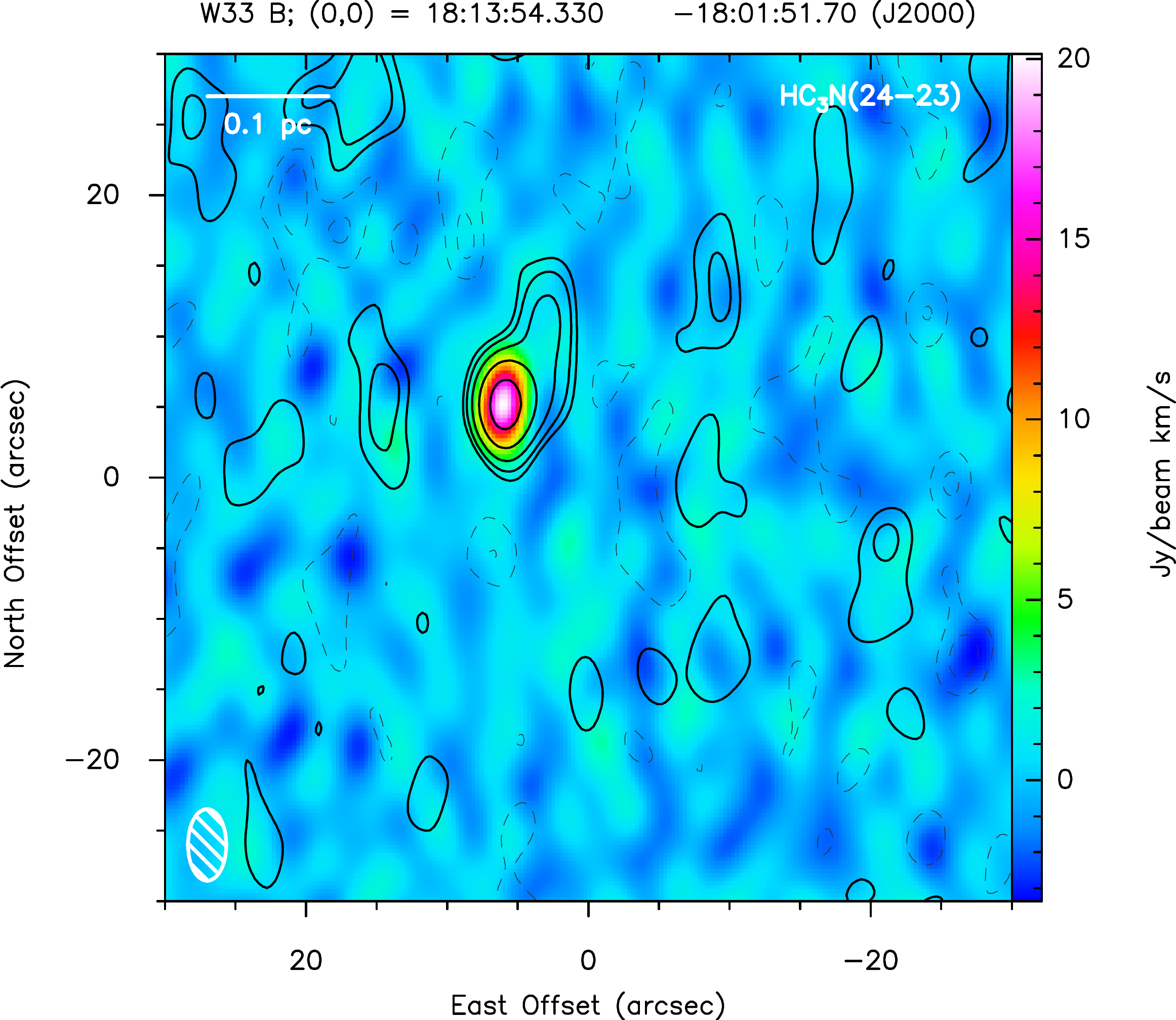}}\\	
	\subfloat{\includegraphics[width=9cm]{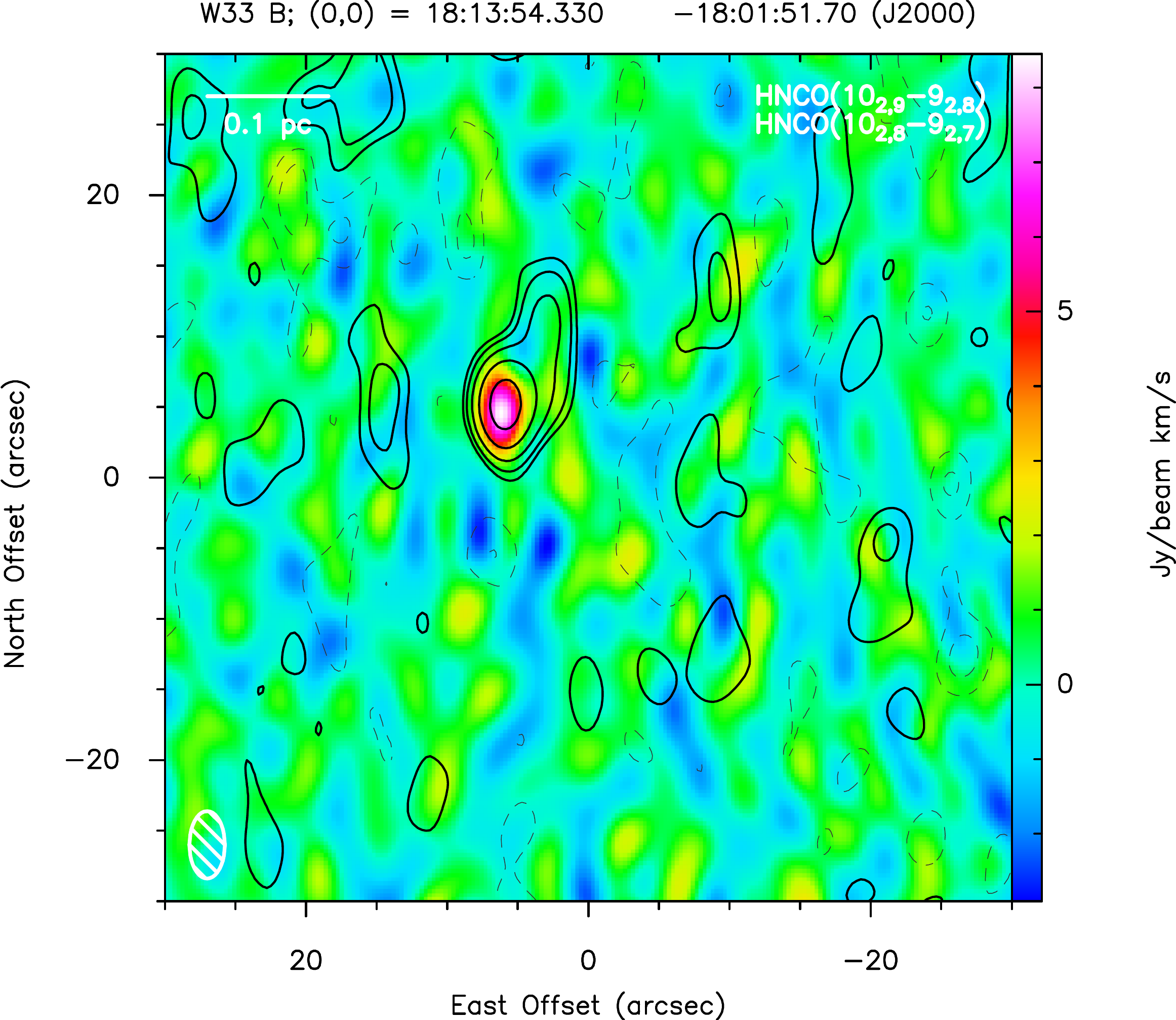}}\hspace{0.2cm}	
	\subfloat{\includegraphics[width=9cm]{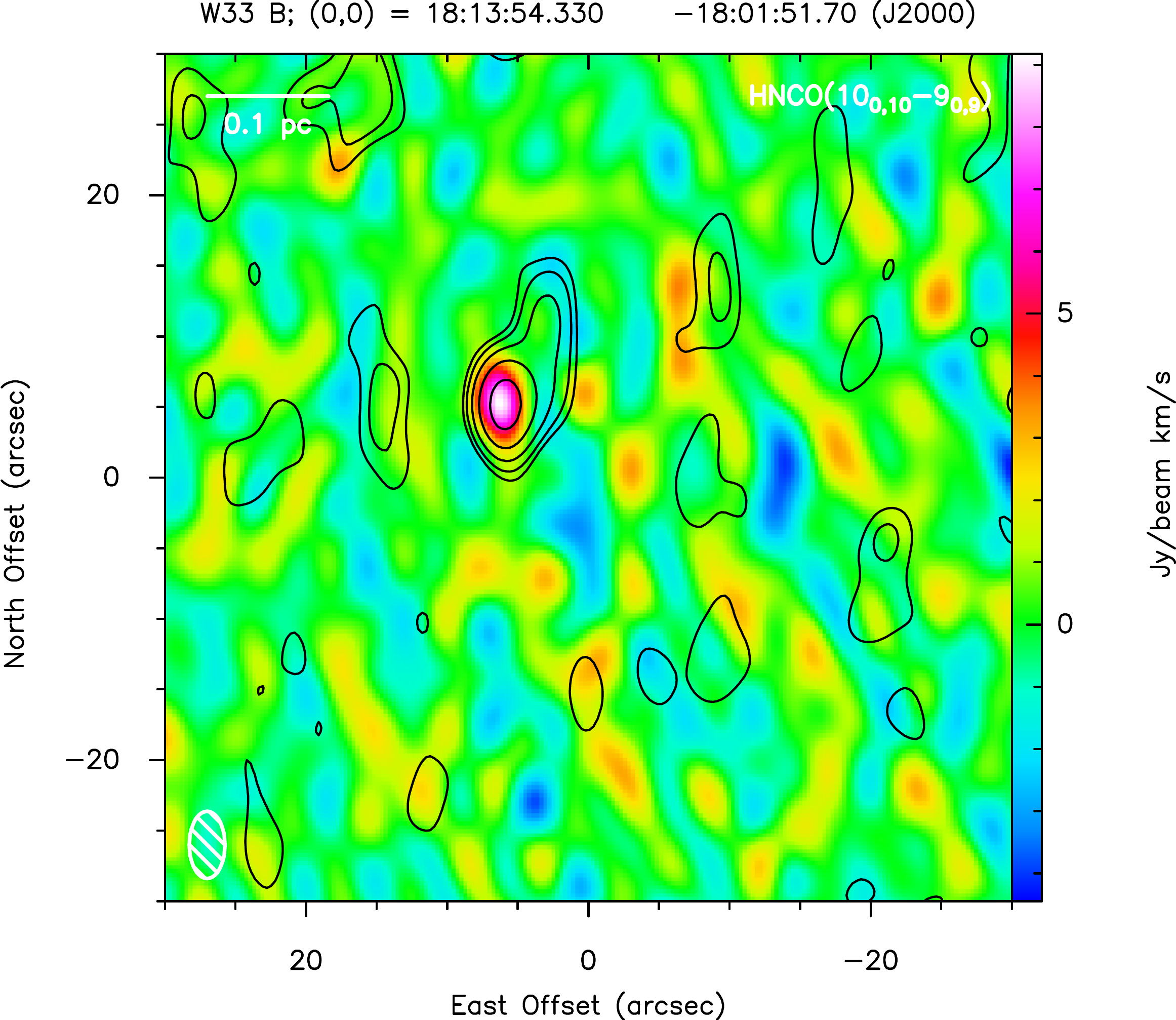}} \\	
	\subfloat{\includegraphics[width=9cm]{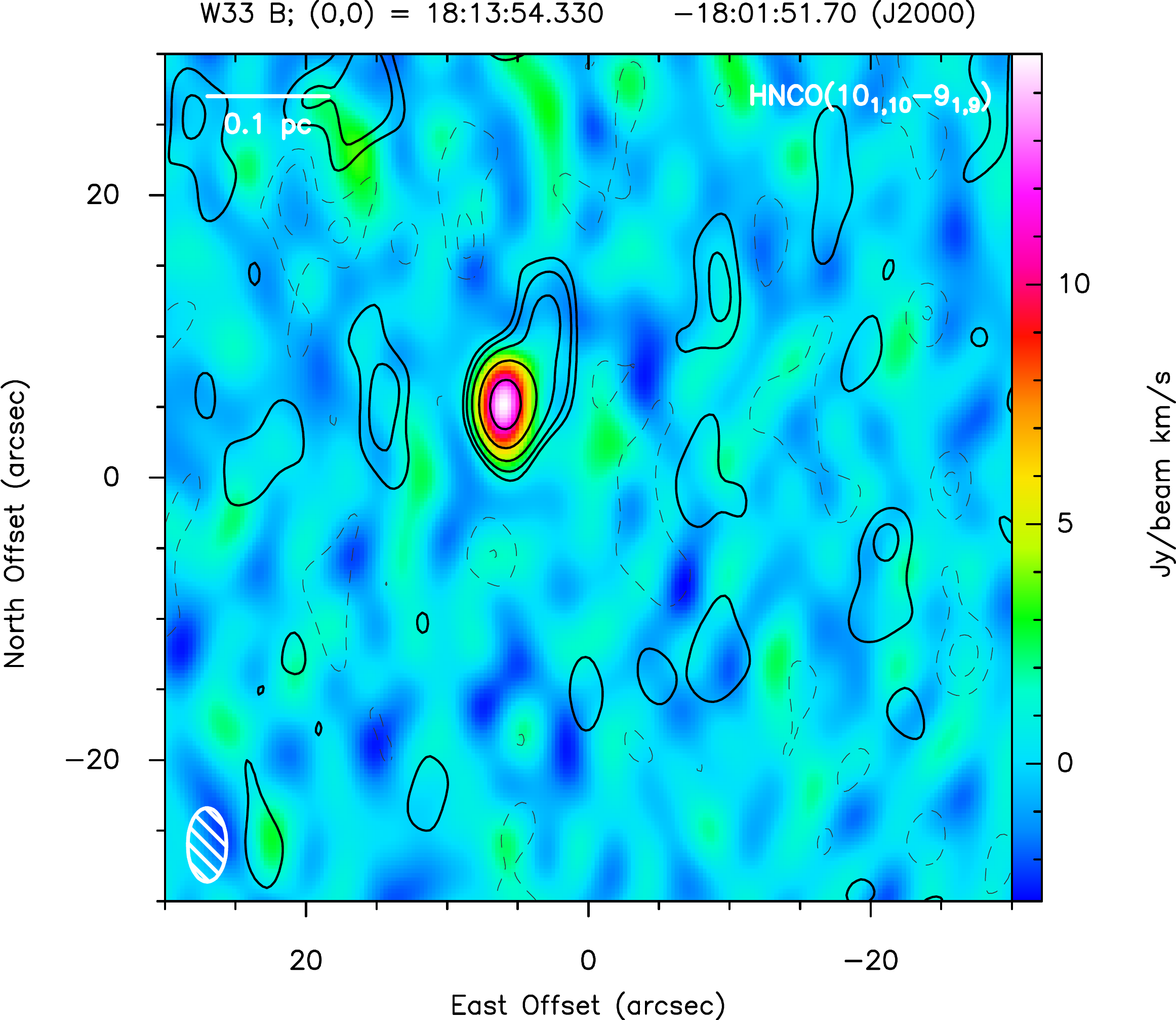}}\hspace{0.2cm}	
	\subfloat{\includegraphics[width=9cm]{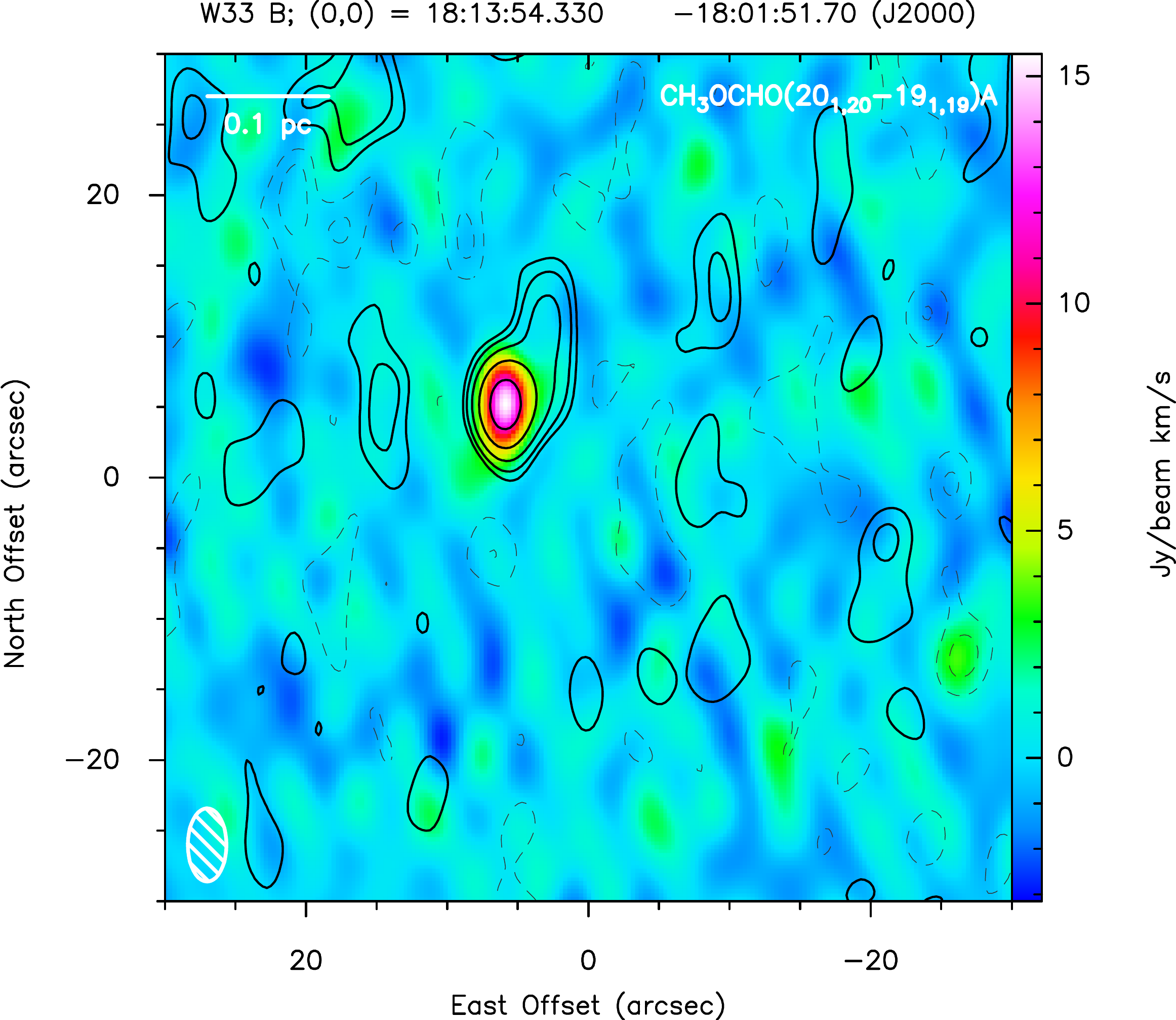}}	
\end{figure*}

\addtocounter{figure}{-1}
\begin{figure*}
	\centering
	\caption{Continued.}
	\subfloat{\includegraphics[width=9cm]{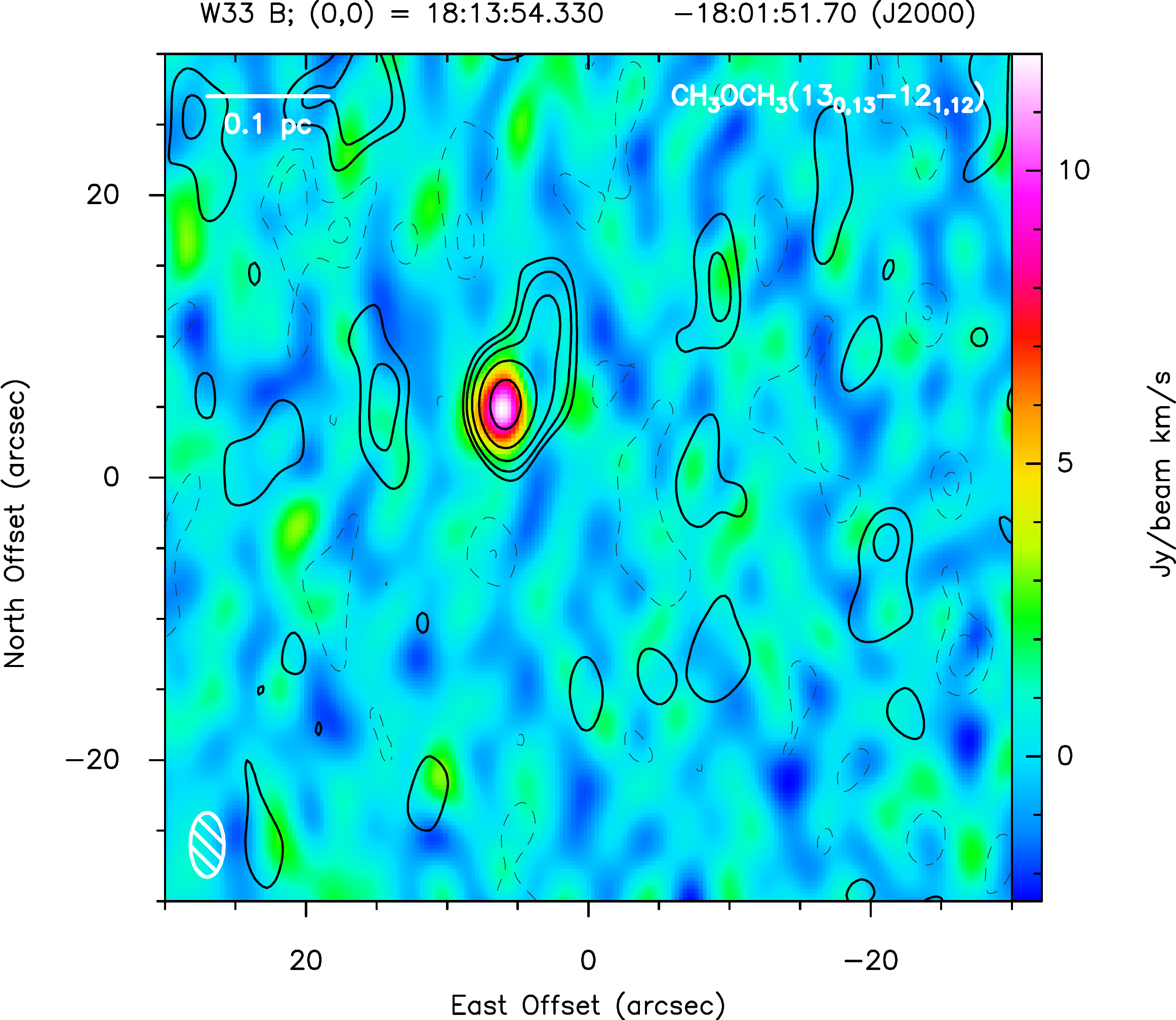}}\hspace{0.2cm}	
	\subfloat{\includegraphics[width=9cm]{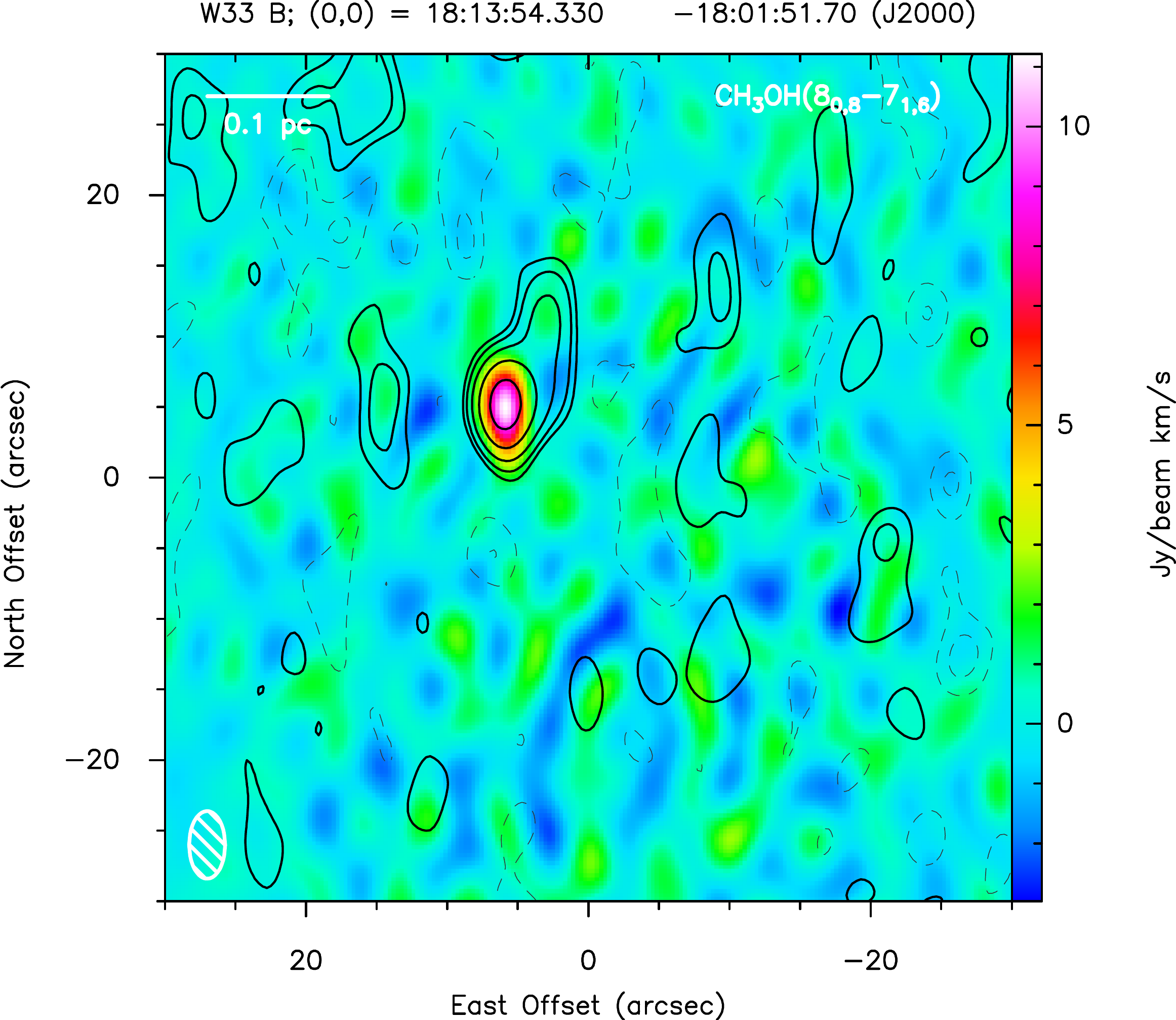}} \\	
	\subfloat{\includegraphics[width=9cm]{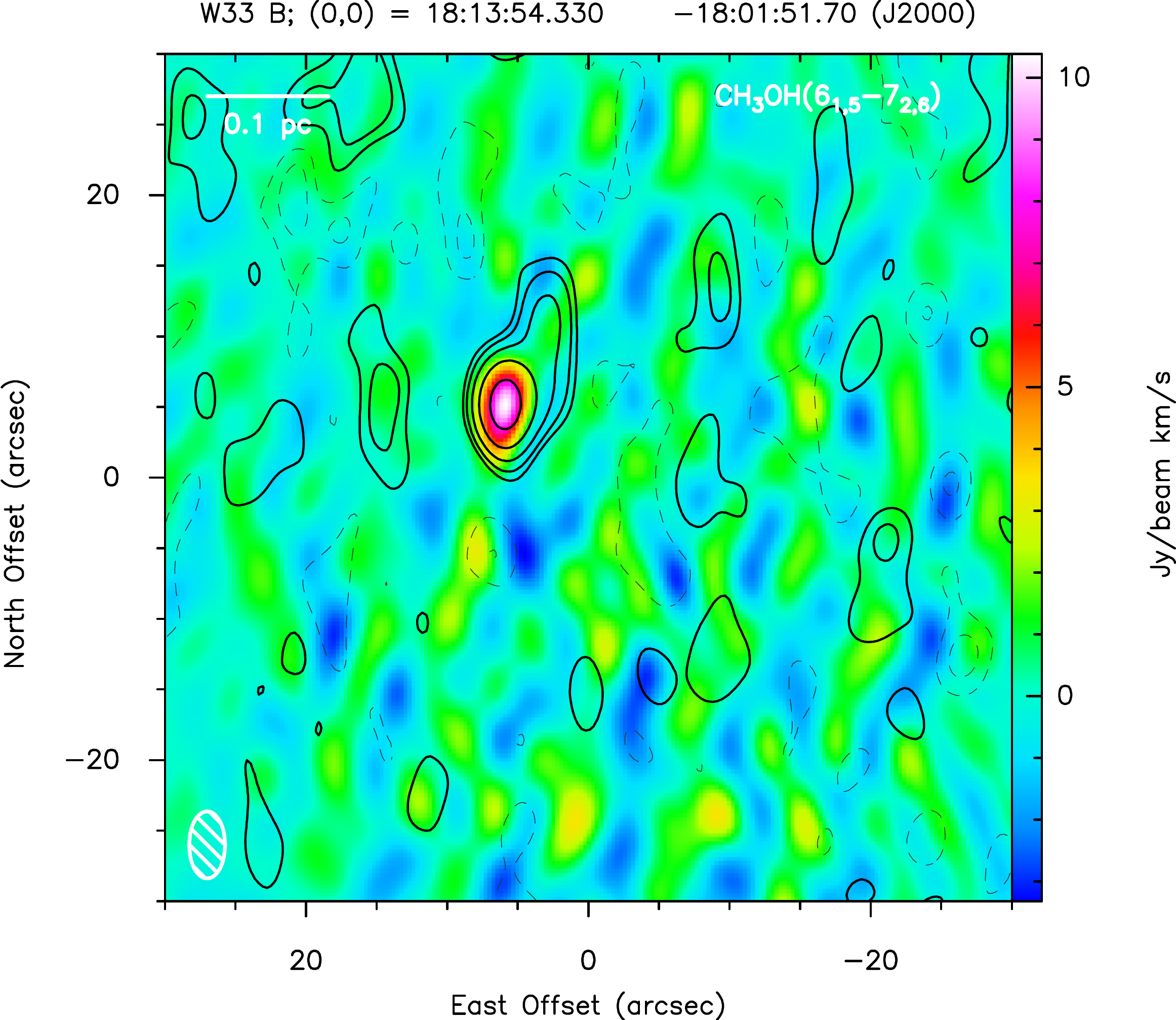}}\hspace{0.2cm}	
	\subfloat{\includegraphics[width=9cm]{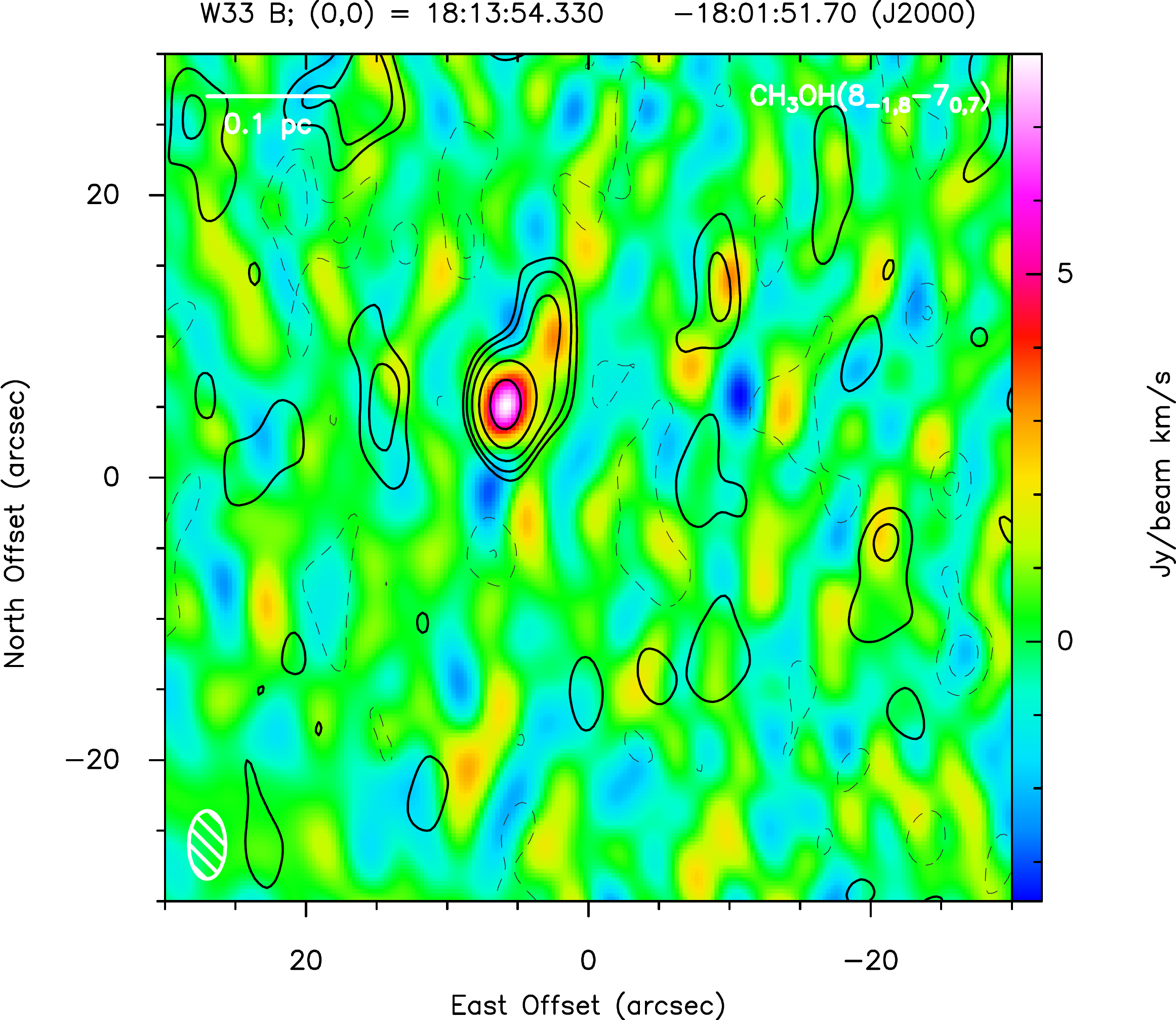}}\\	
	\subfloat{\includegraphics[width=9cm]{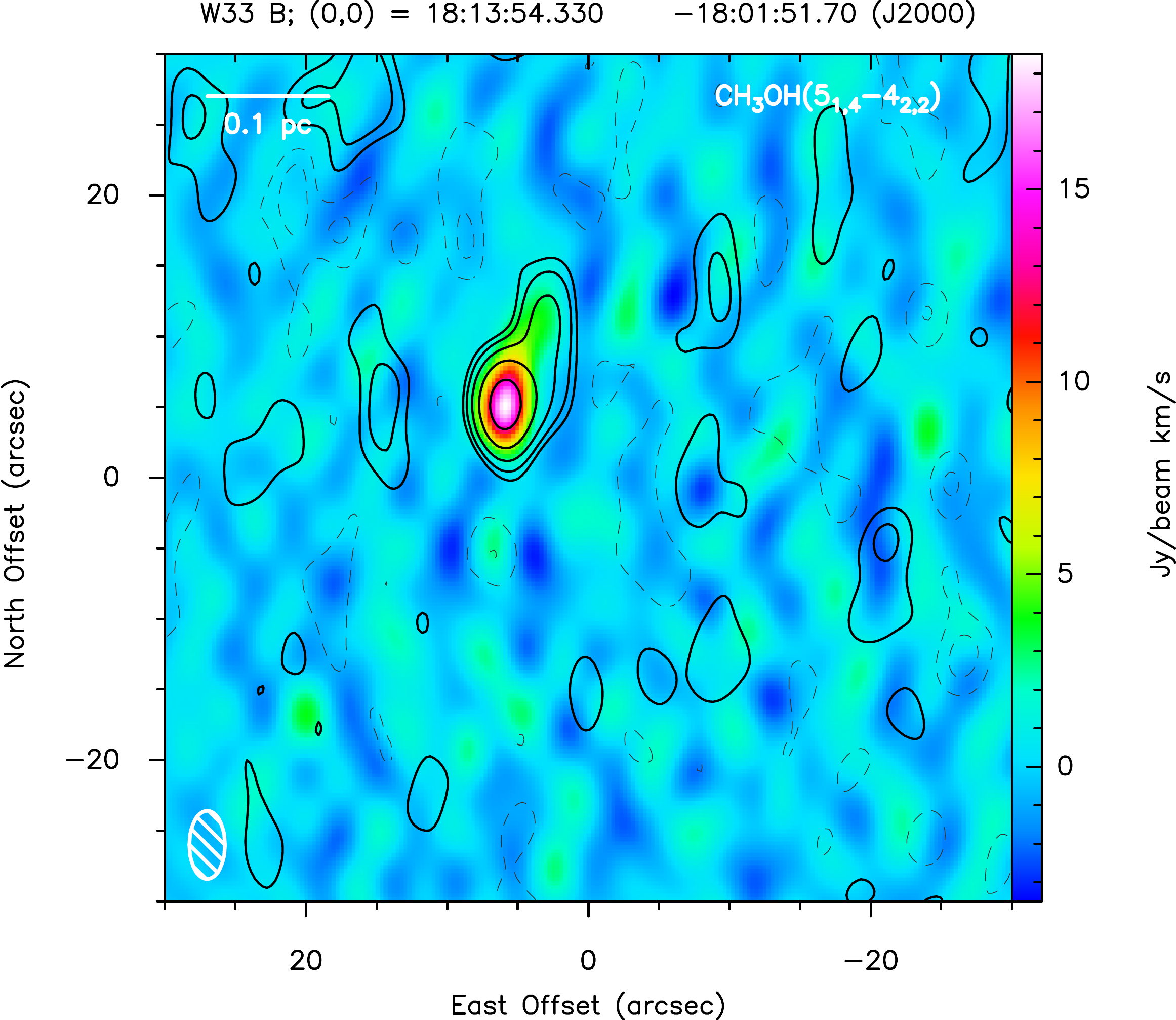}}\hspace{0.2cm}	
	\subfloat{\includegraphics[width=9cm]{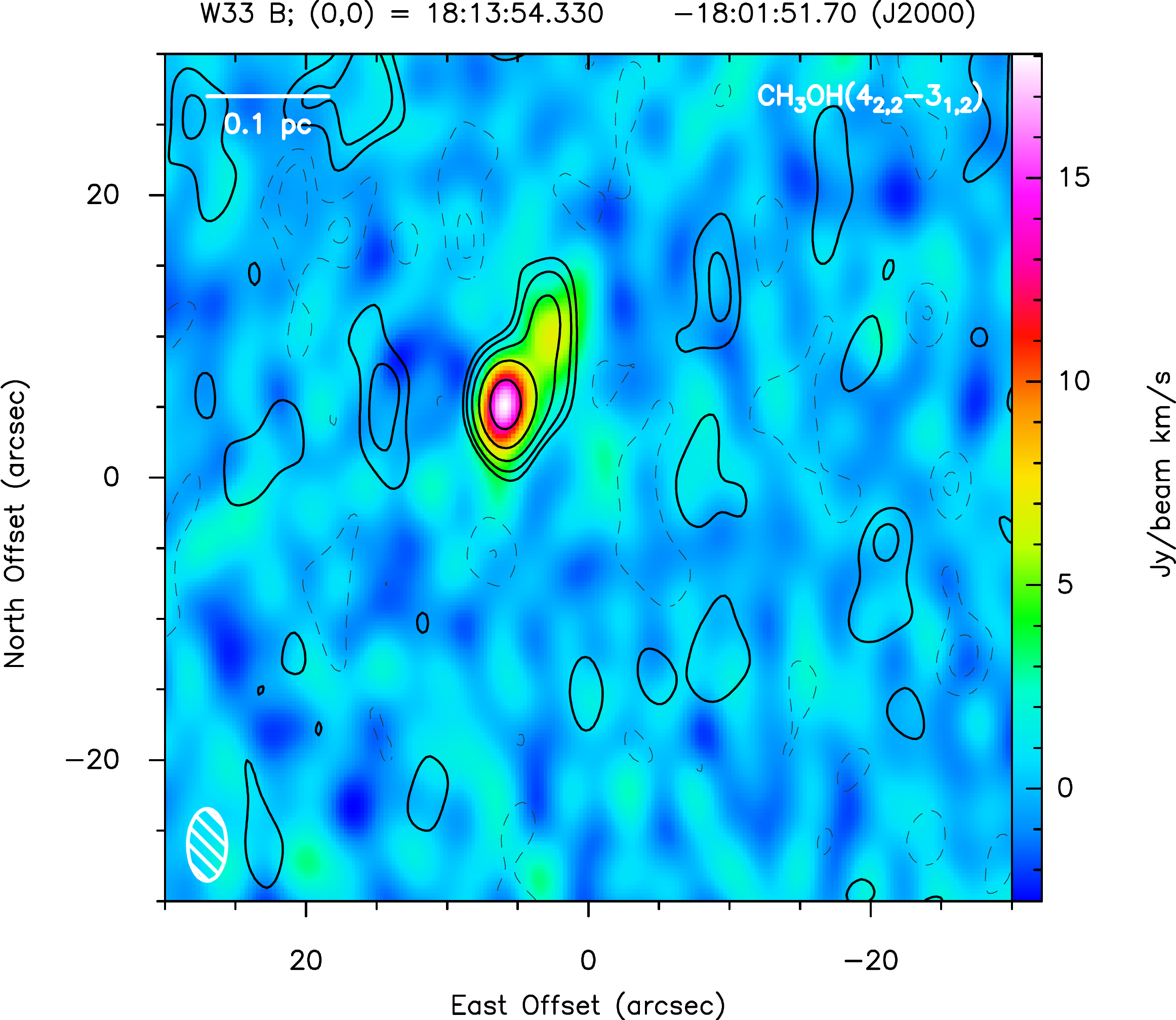}} 	
\end{figure*}

\clearpage

\addtocounter{figure}{-1}
\begin{figure*}
	\centering
	\caption{Continued.}
	\subfloat{\includegraphics[width=9cm]{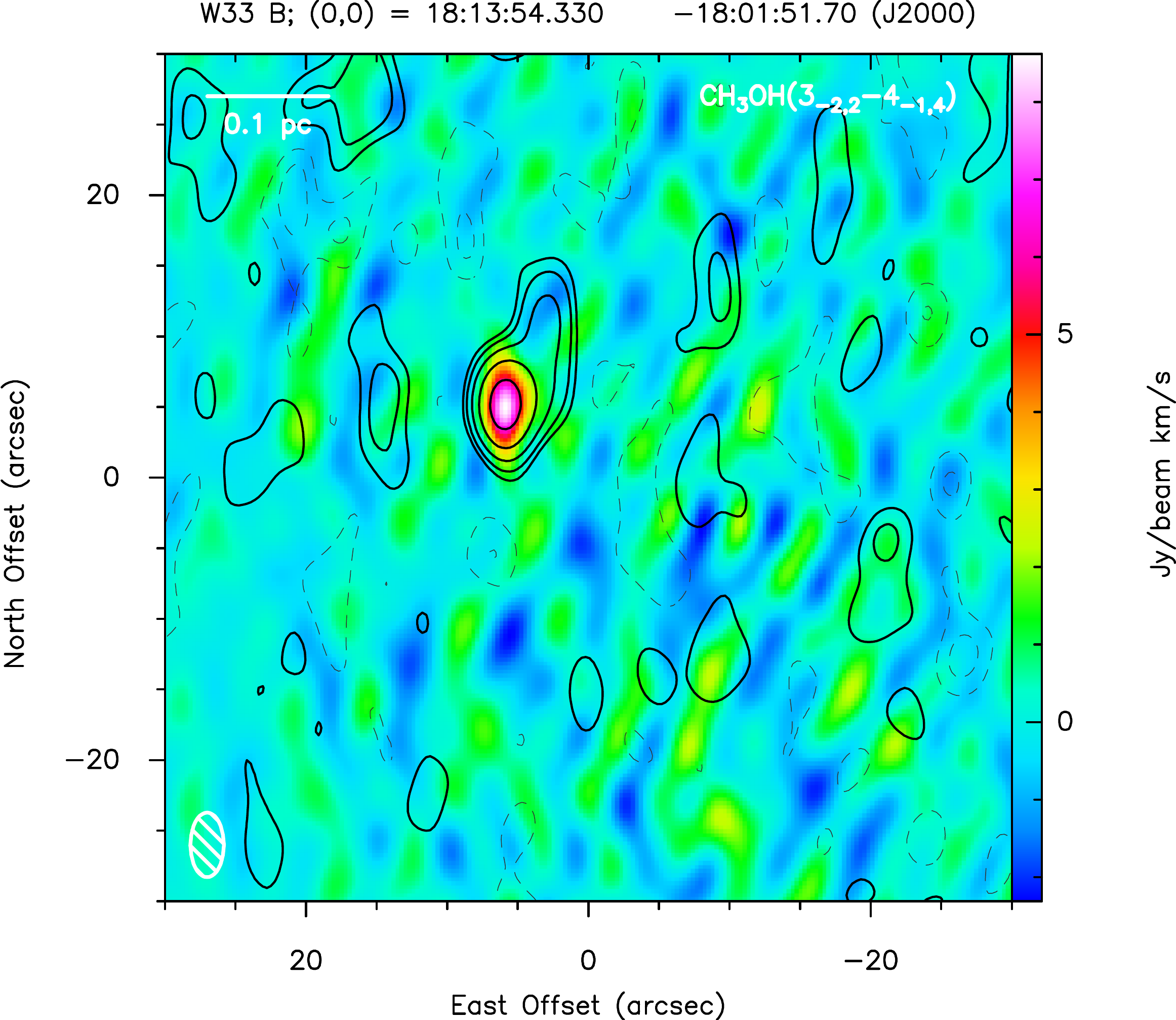}}\hspace{0.2cm}
	\subfloat{\includegraphics[width=9cm]{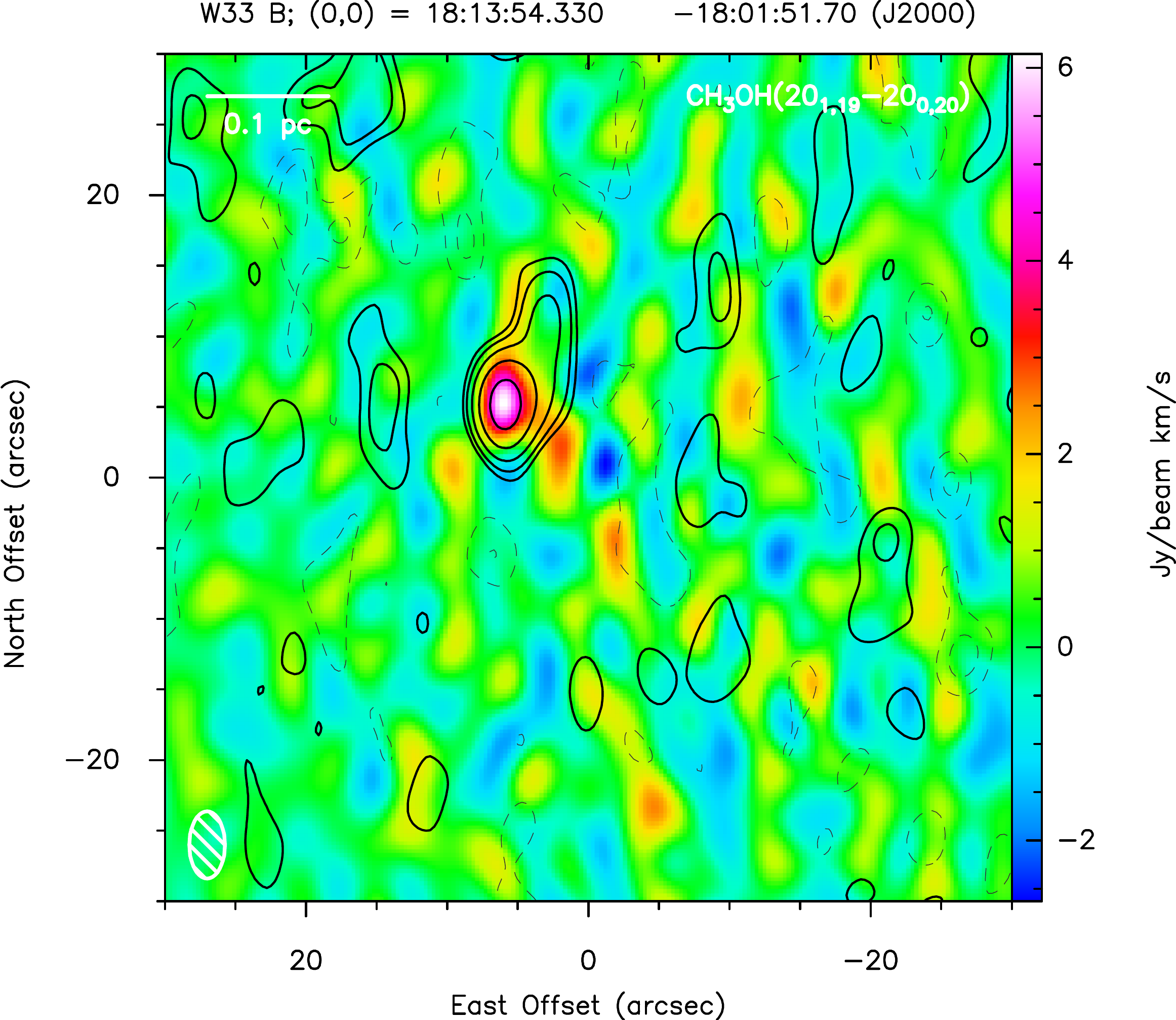}}\\	
	\subfloat{\includegraphics[width=9cm]{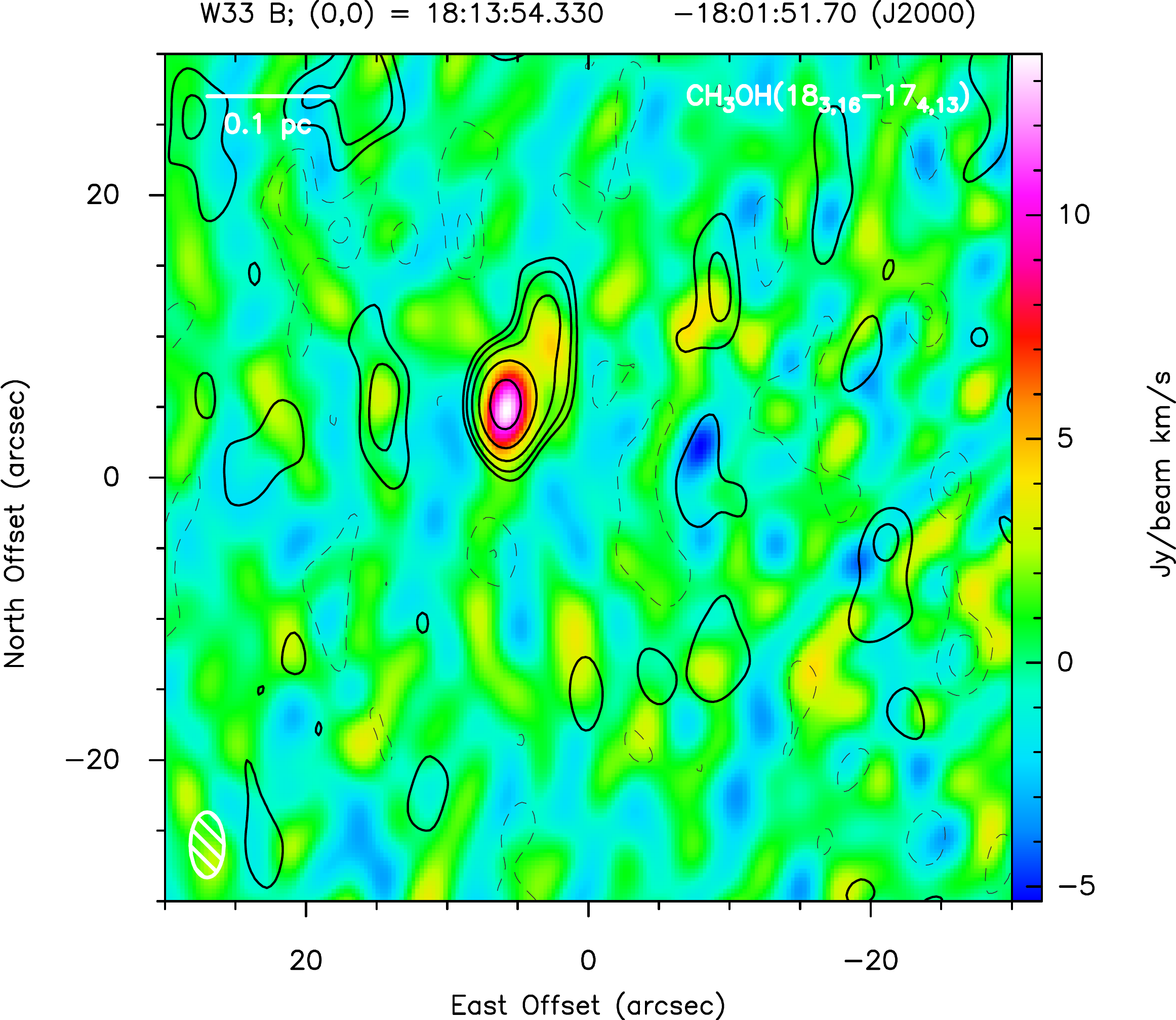}}\hspace{0.2cm}	
	\subfloat{\includegraphics[width=9cm]{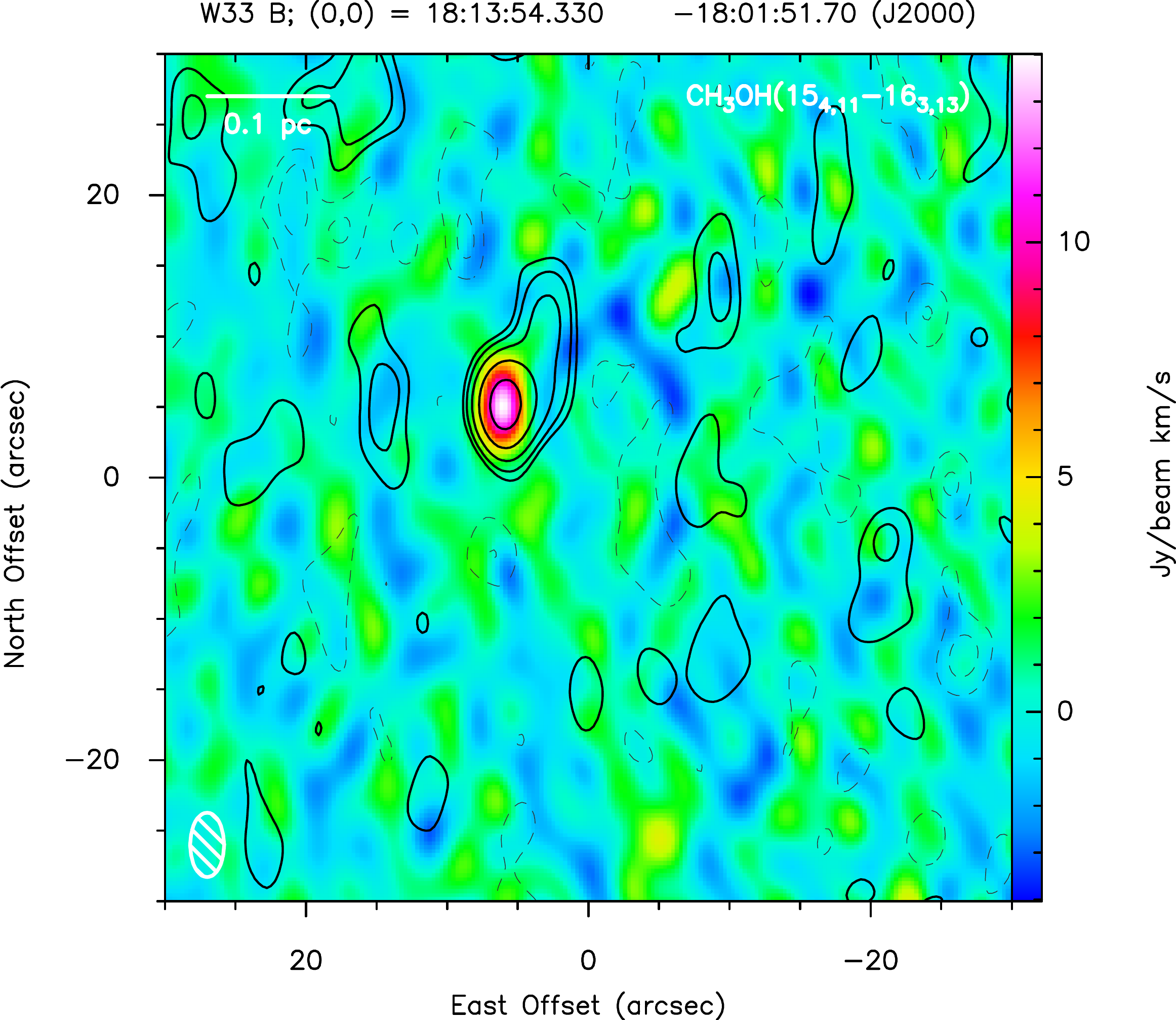}} \\	
	\subfloat{\includegraphics[width=9cm]{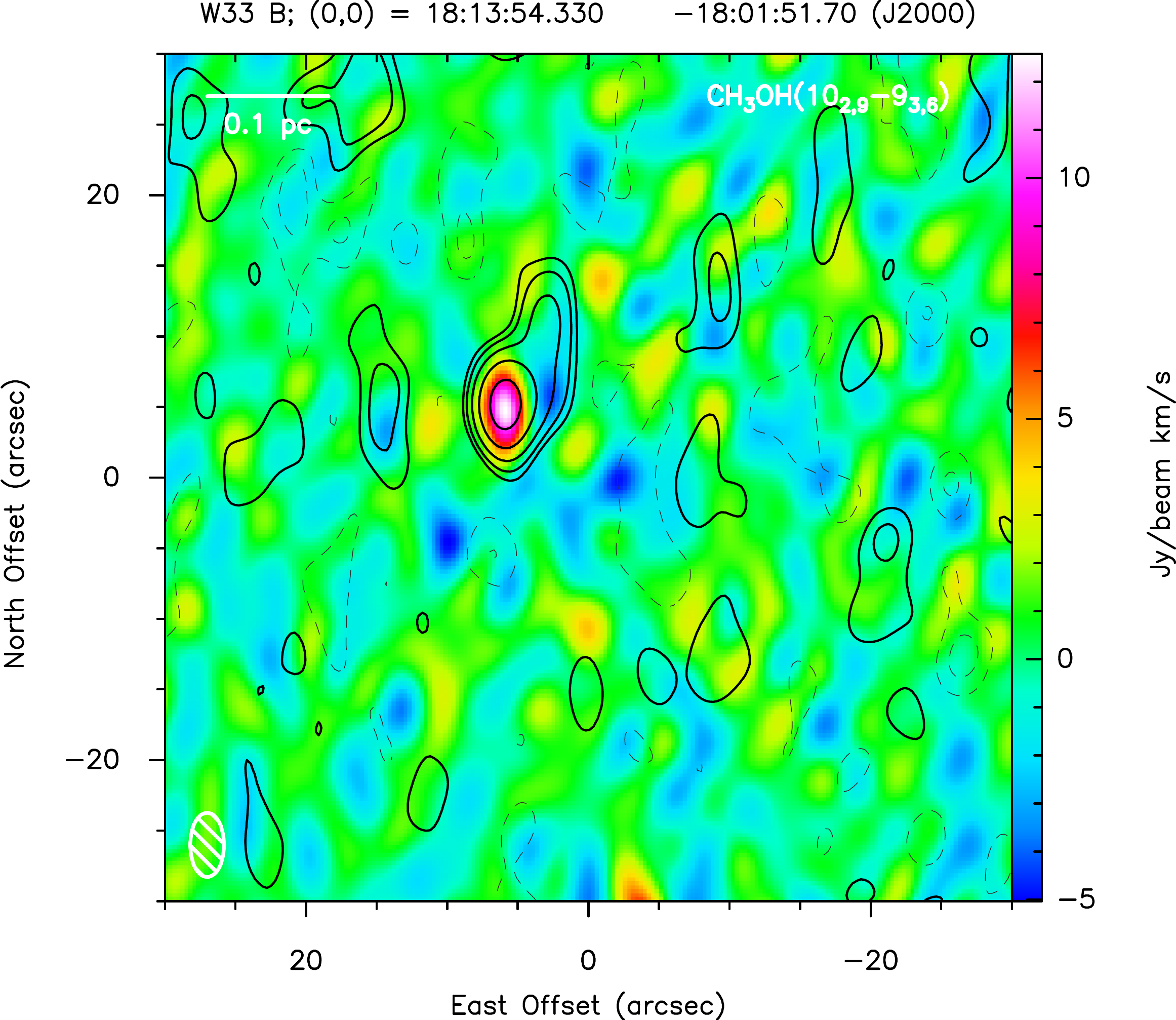}}\hspace{0.2cm}	
	\subfloat{\includegraphics[width=9cm]{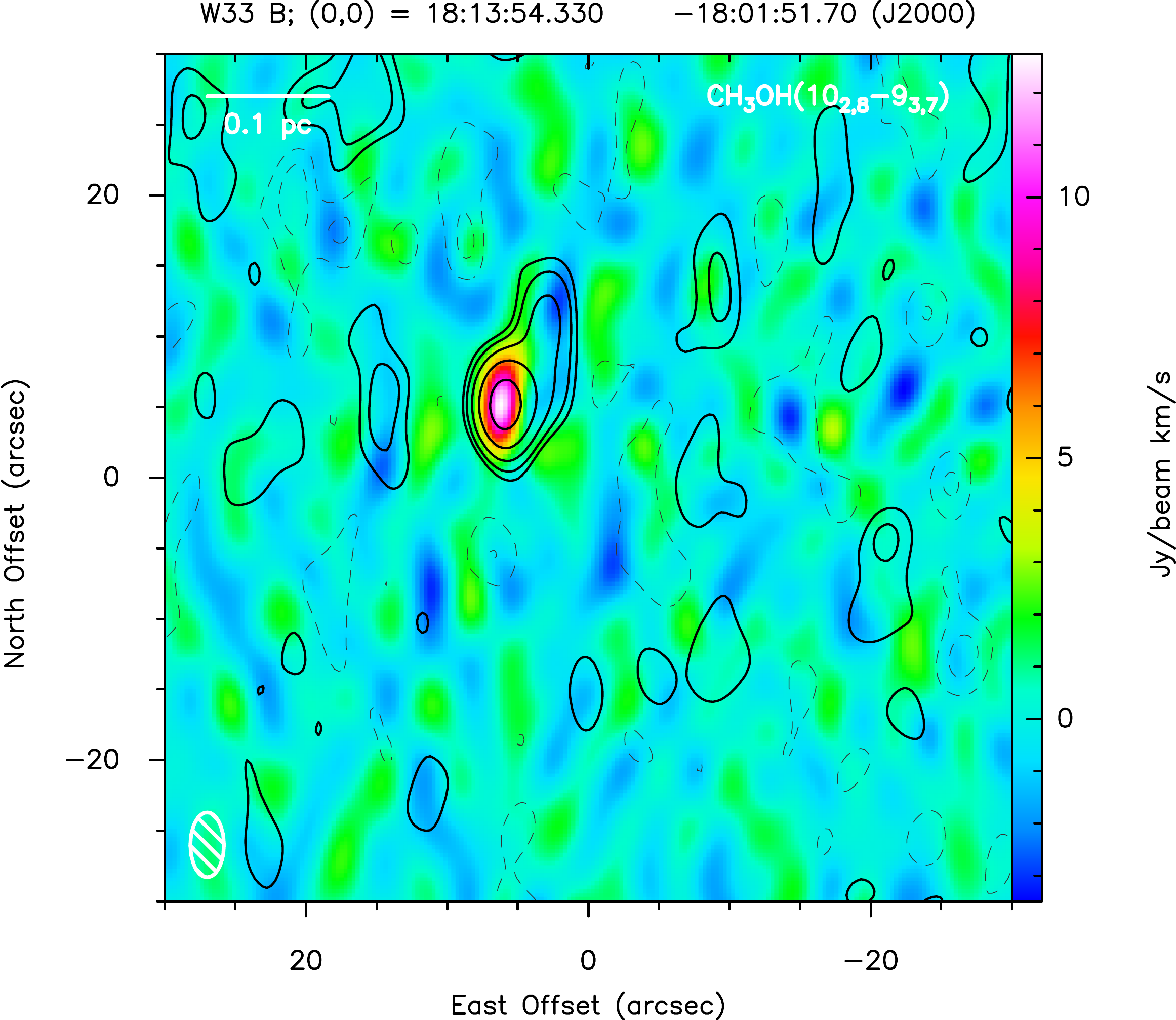}}	
\end{figure*}

\addtocounter{figure}{-1}
\begin{figure*}
	\centering
	\caption{Continued.}
	\subfloat{\includegraphics[width=9cm]{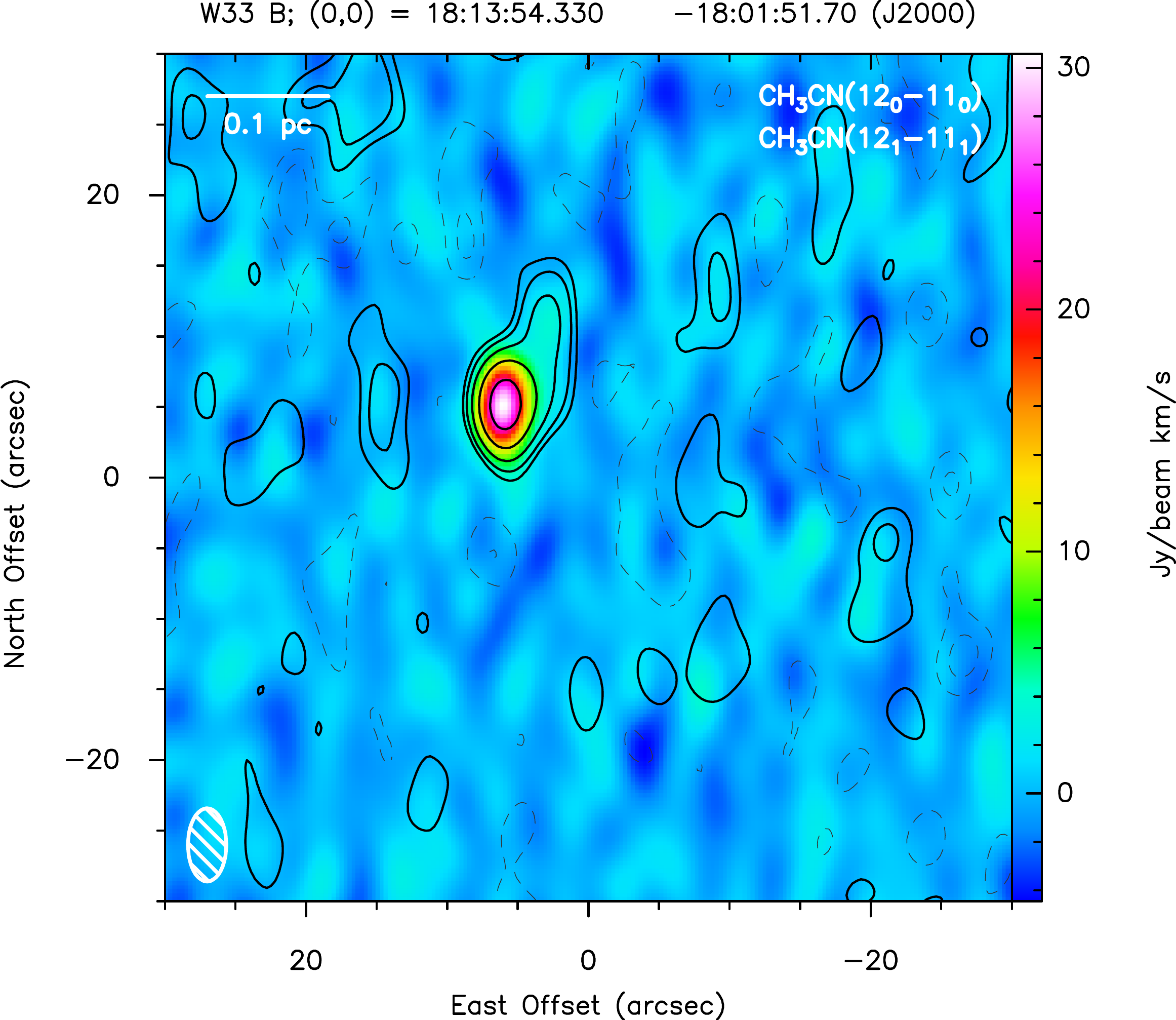}}\hspace{0.2cm}	
	\subfloat{\includegraphics[width=9cm]{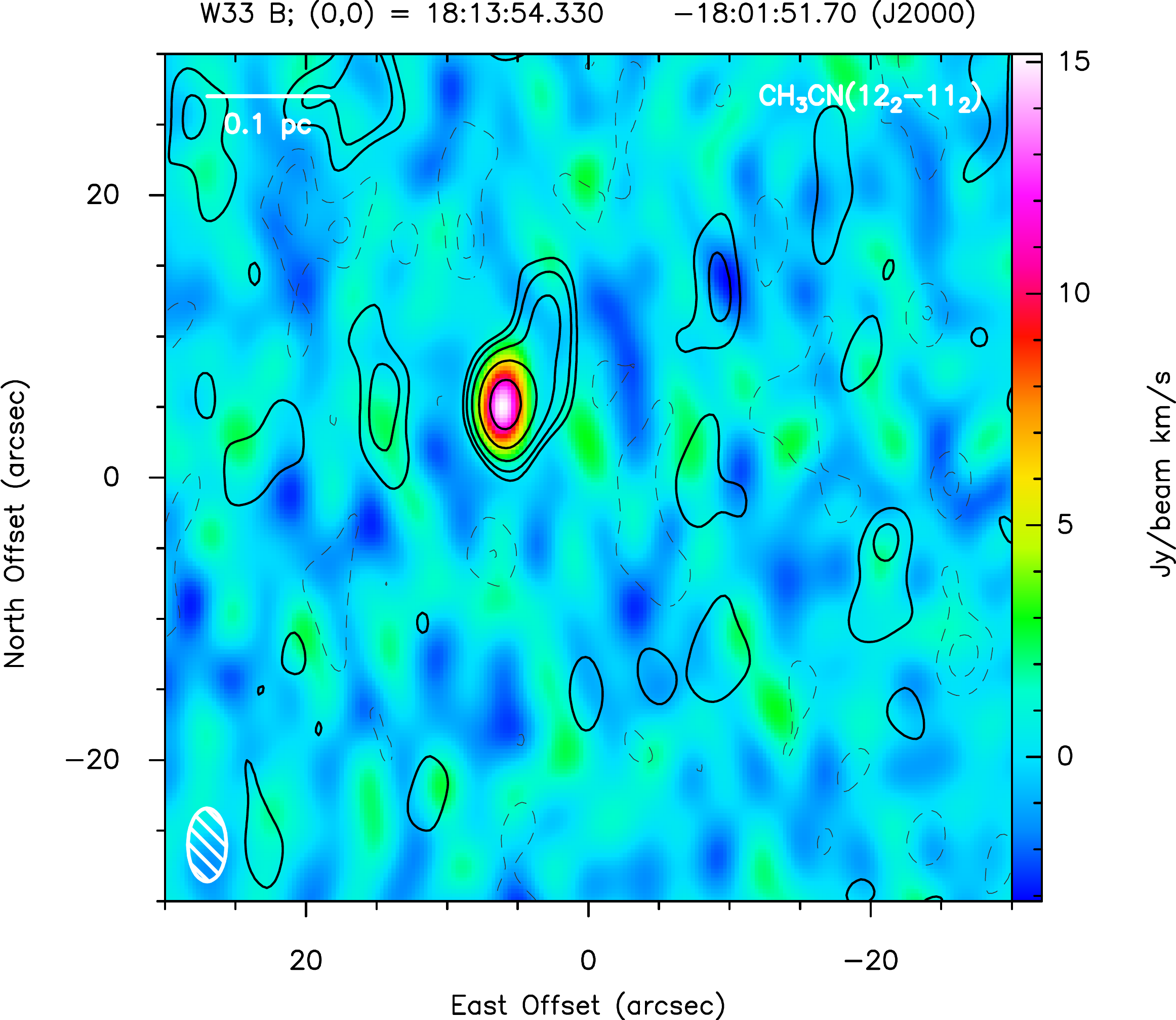}}\\
	\subfloat{\includegraphics[width=9cm]{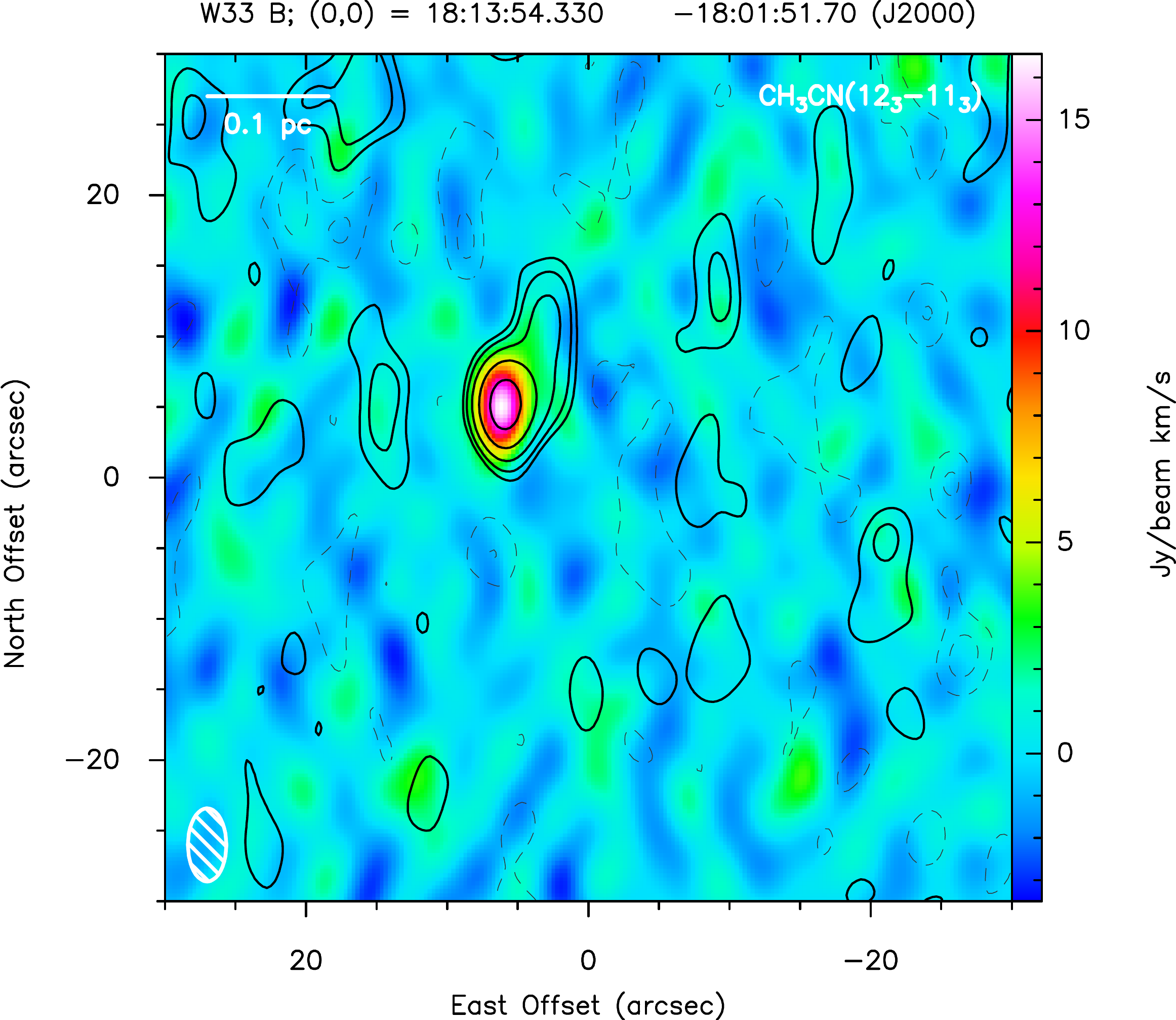}}\hspace{0.2cm}	
	\subfloat{\includegraphics[width=9cm]{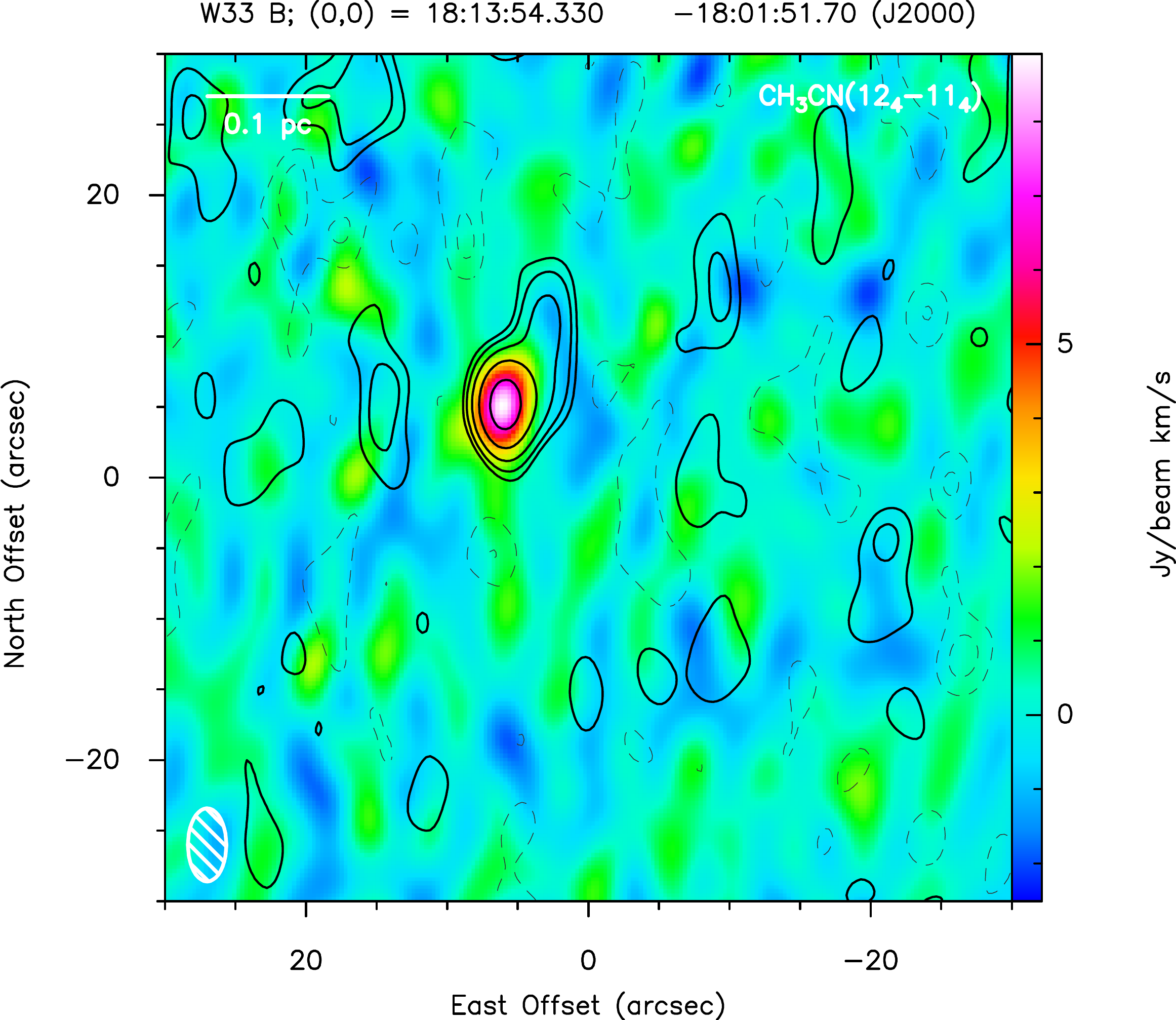}}\\	
	\subfloat{\includegraphics[width=9cm]{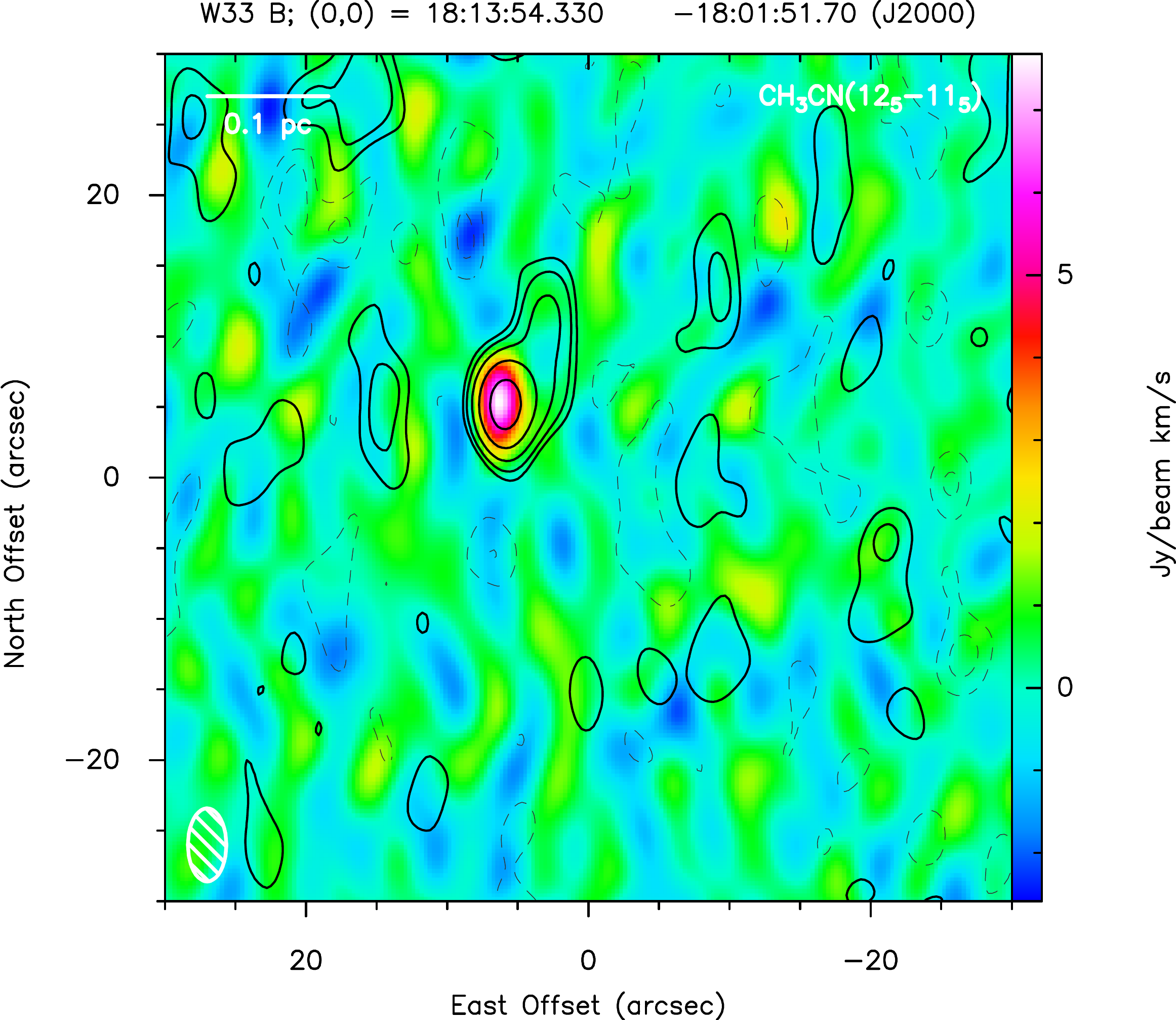}}\hspace{0.2cm}	
	\subfloat{\includegraphics[width=9cm]{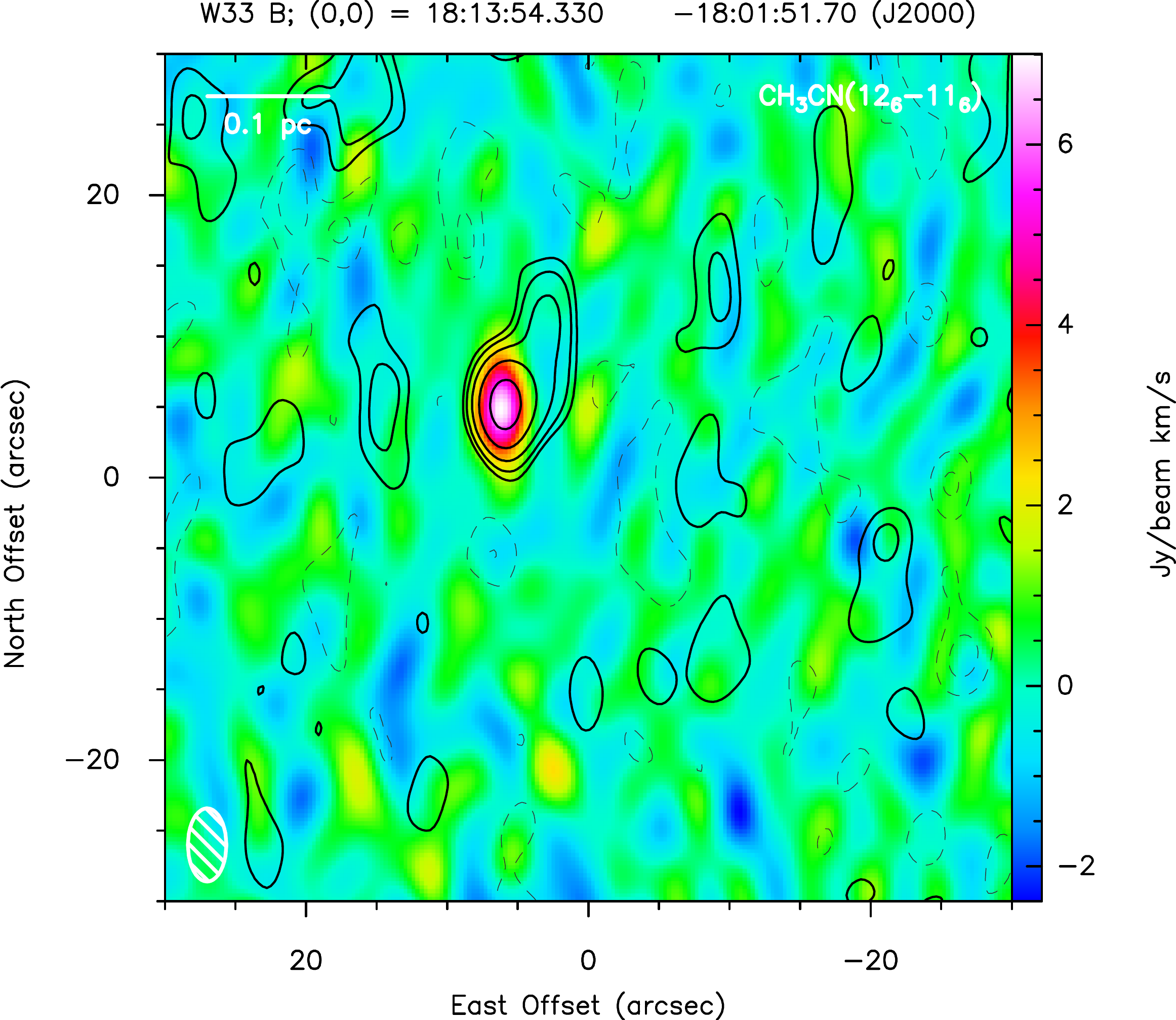}}	
\end{figure*}

\begin{figure*}
	\centering
	\caption{Line emission of detected transitions in W33\,Main. The contours show the SMA continuum emission at 230 GHz (same contour levels as in Fig. \ref{W33_SMA_CH0}). The name of the each transition is shown in the upper right corner. A scale of 0.1 pc is marked in the upper left corner, and the synthesised beam is shown in the lower left corner.}
	\subfloat{\includegraphics[width=9cm]{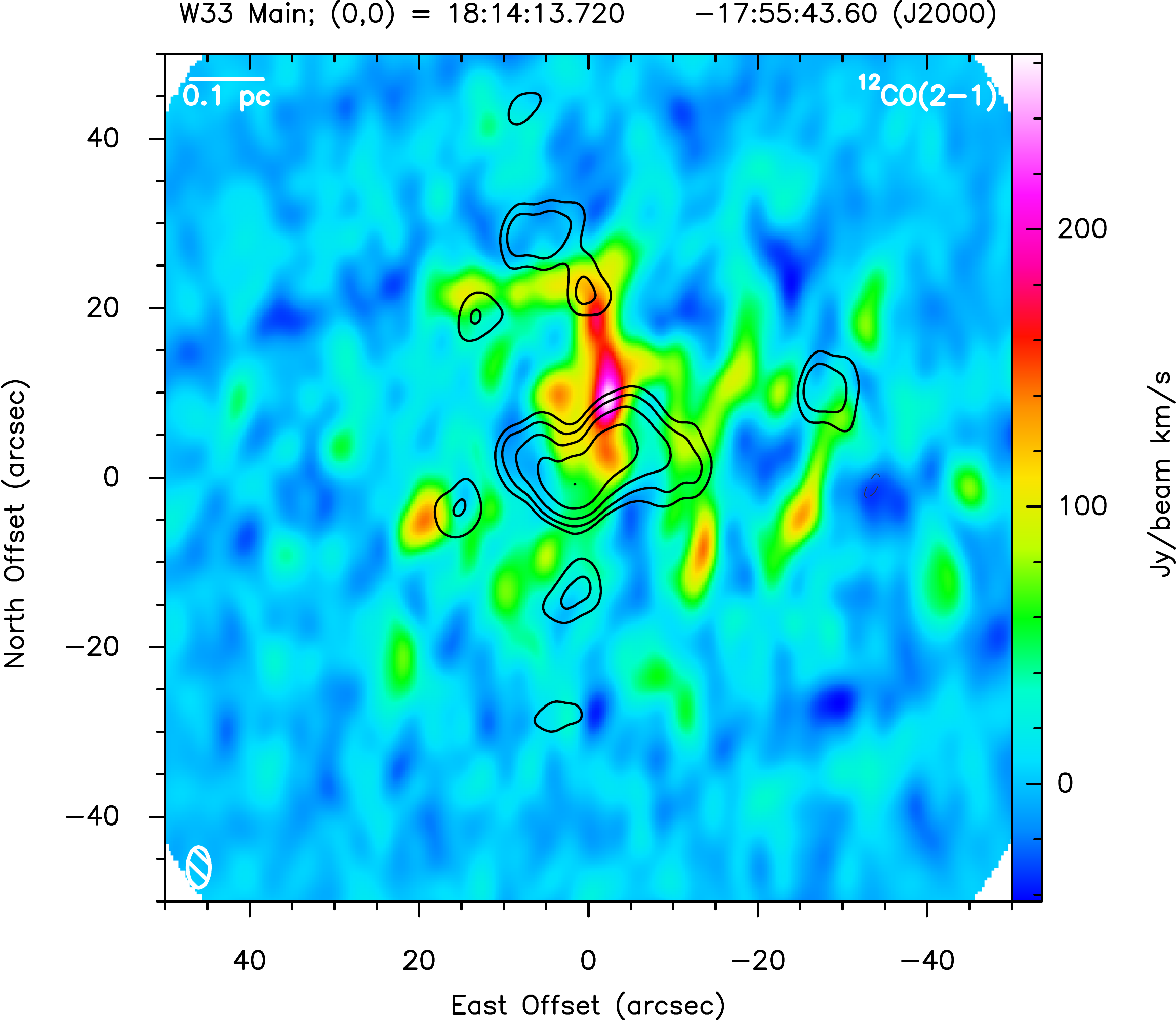}}\hspace{0.2cm}	
	\subfloat{\includegraphics[width=9cm]{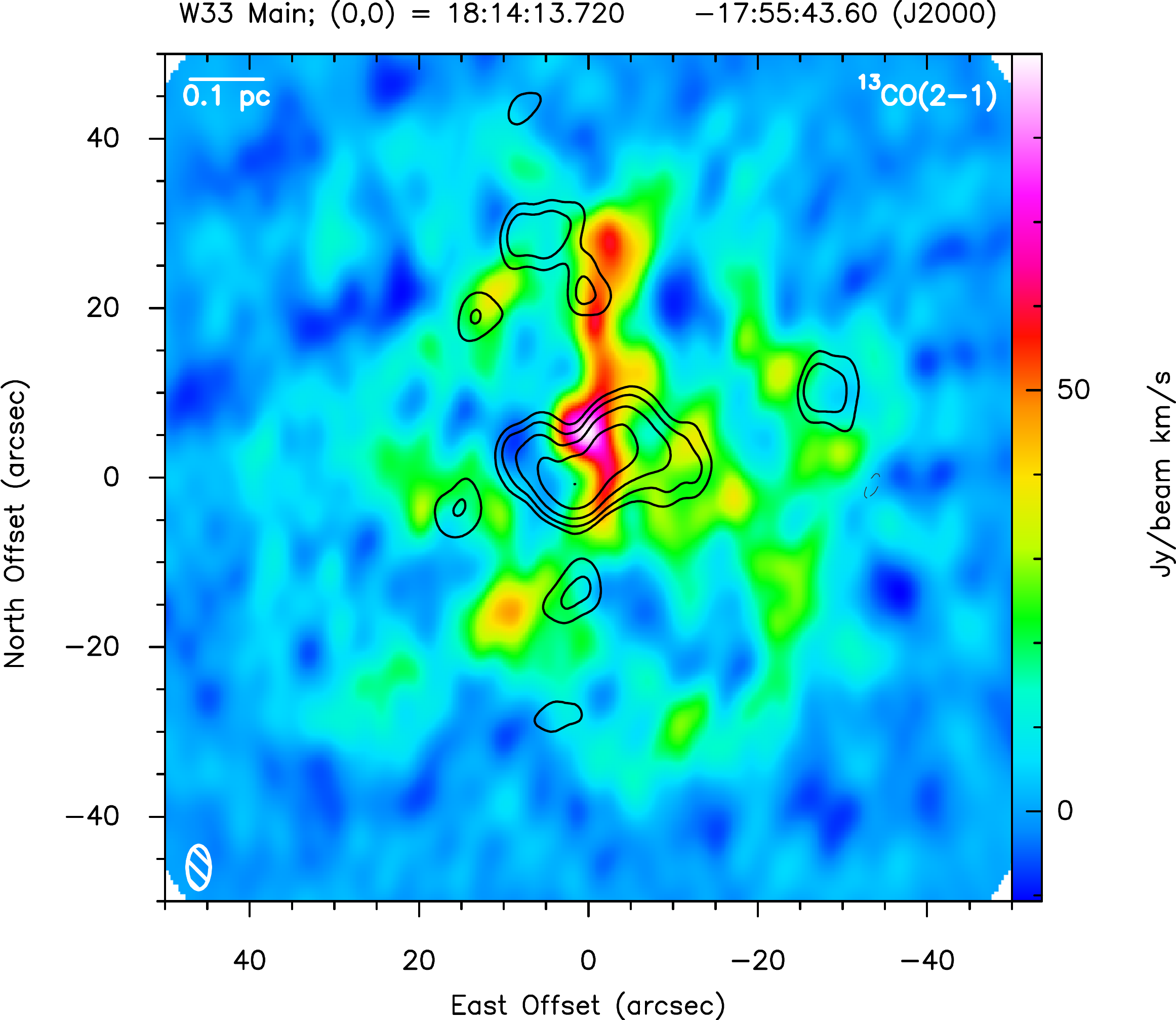}}\\
	\subfloat{\includegraphics[width=9cm]{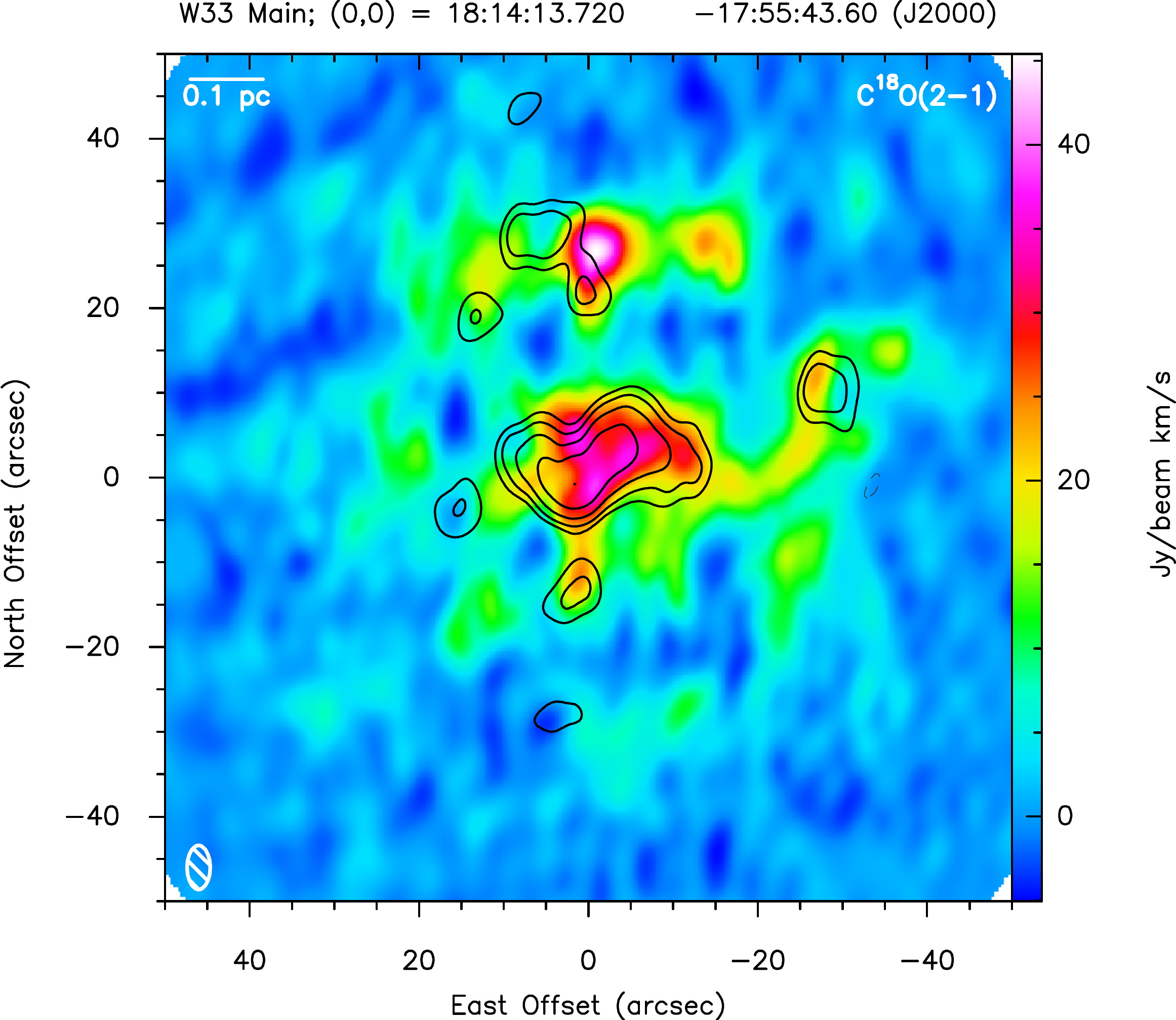}}\hspace{0.2cm}	
	\subfloat{\includegraphics[width=9cm]{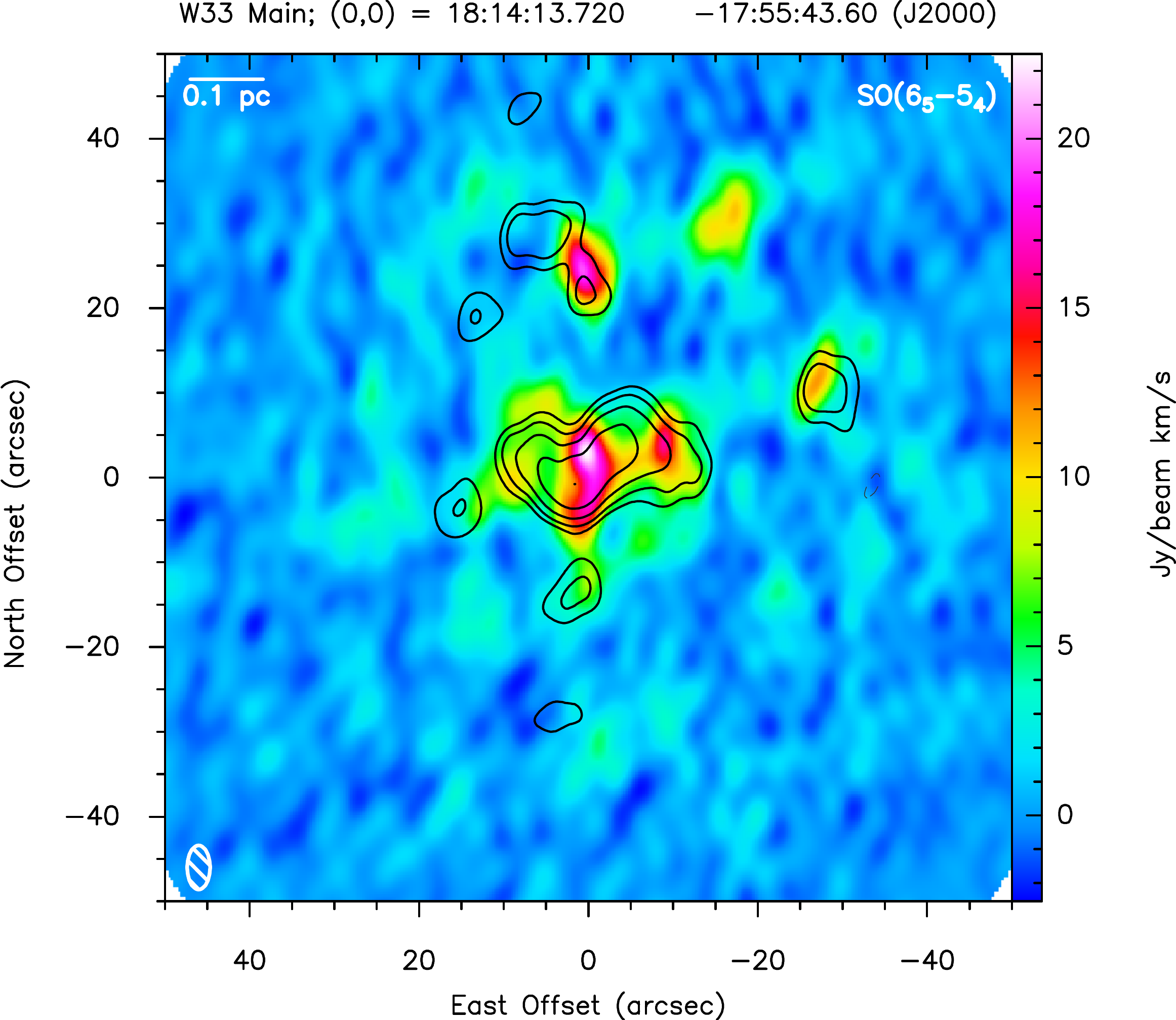}}	
	\label{W33M-SMA-IntInt}
\end{figure*}

\addtocounter{figure}{-1}
\begin{figure*}
	\centering
	\caption{Continued.}
	\subfloat{\includegraphics[width=9cm]{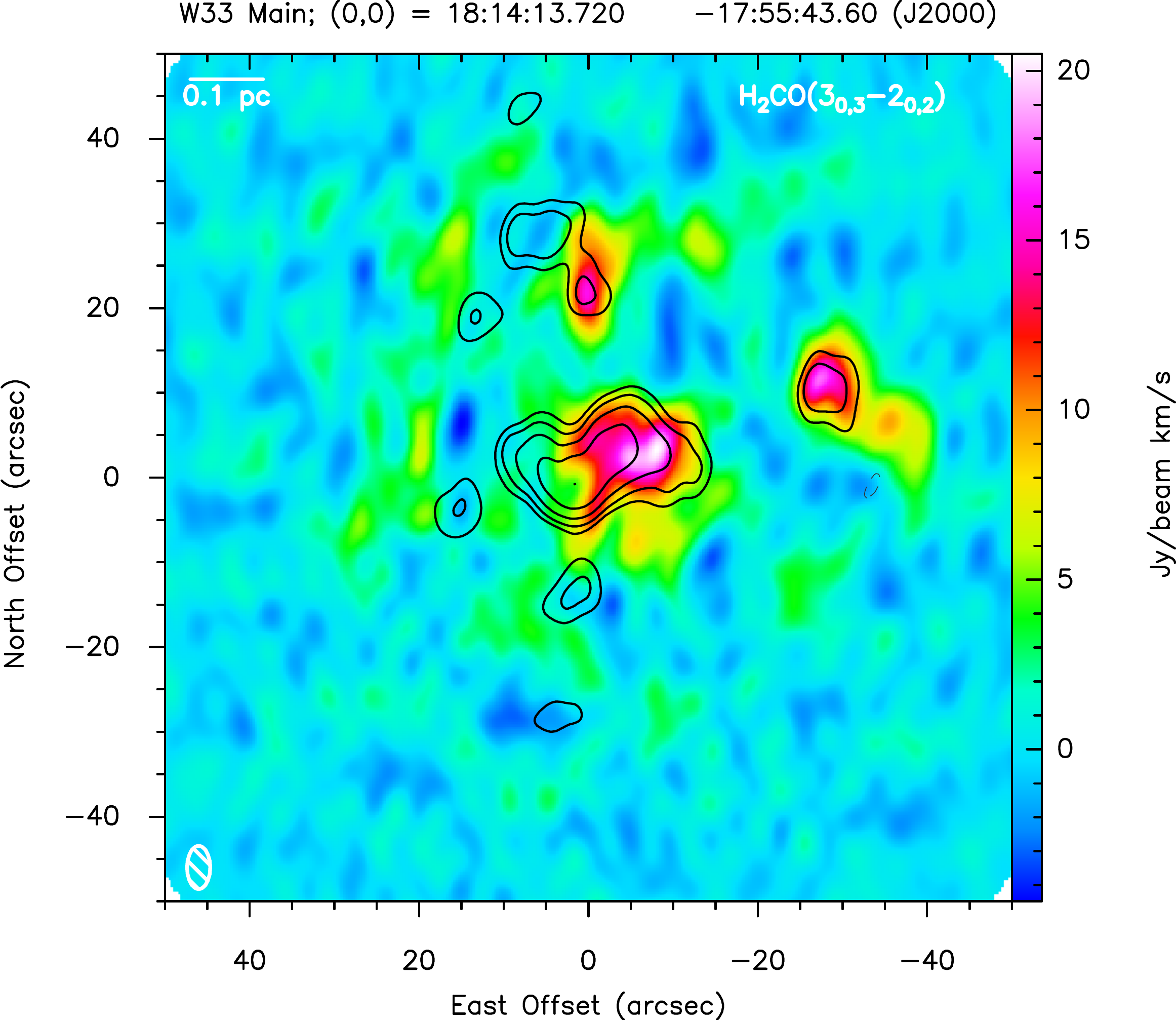}}\hspace{0.2cm}	
	\subfloat{\includegraphics[width=9cm]{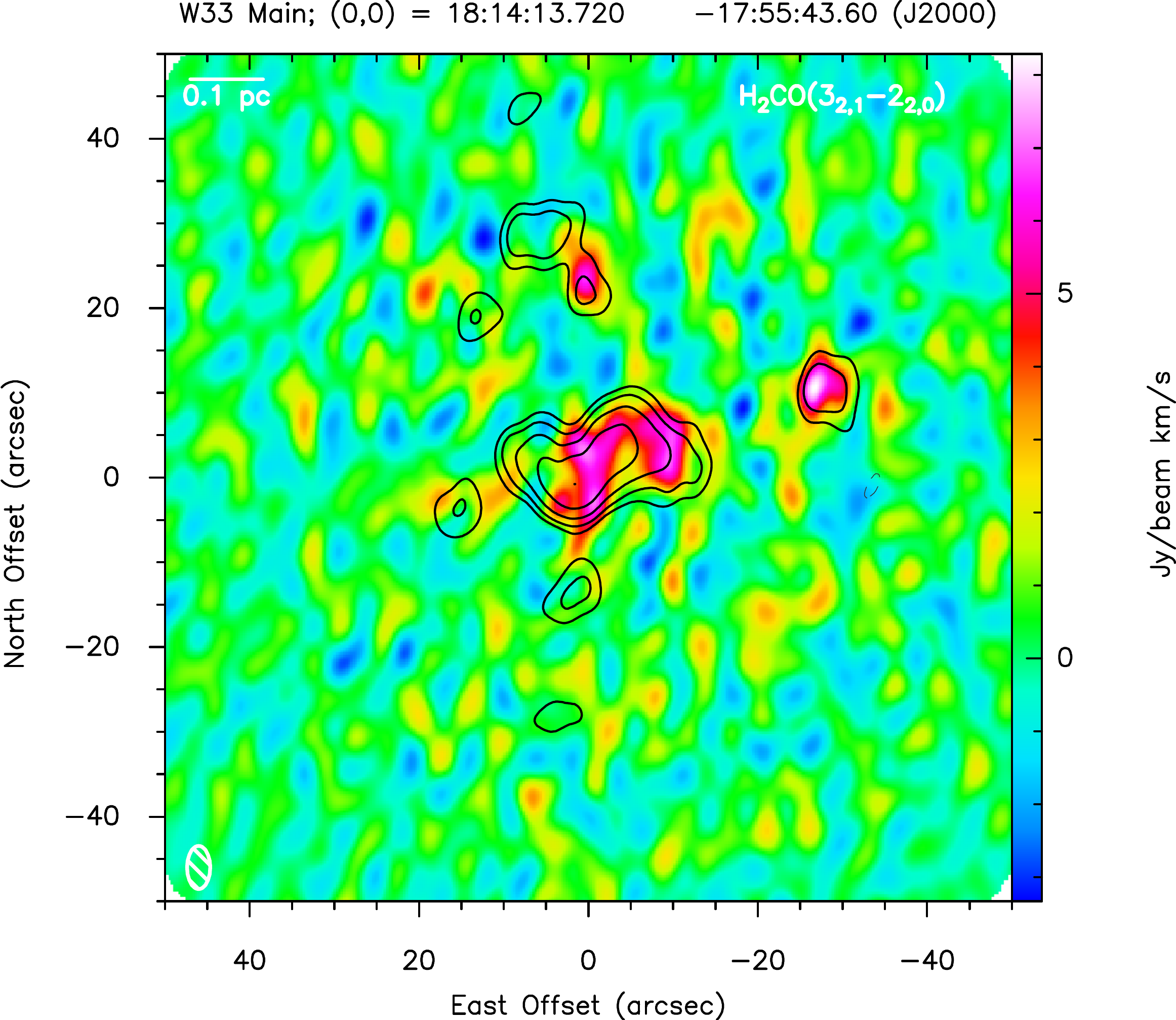}} \\	
	\subfloat{\includegraphics[width=9cm]{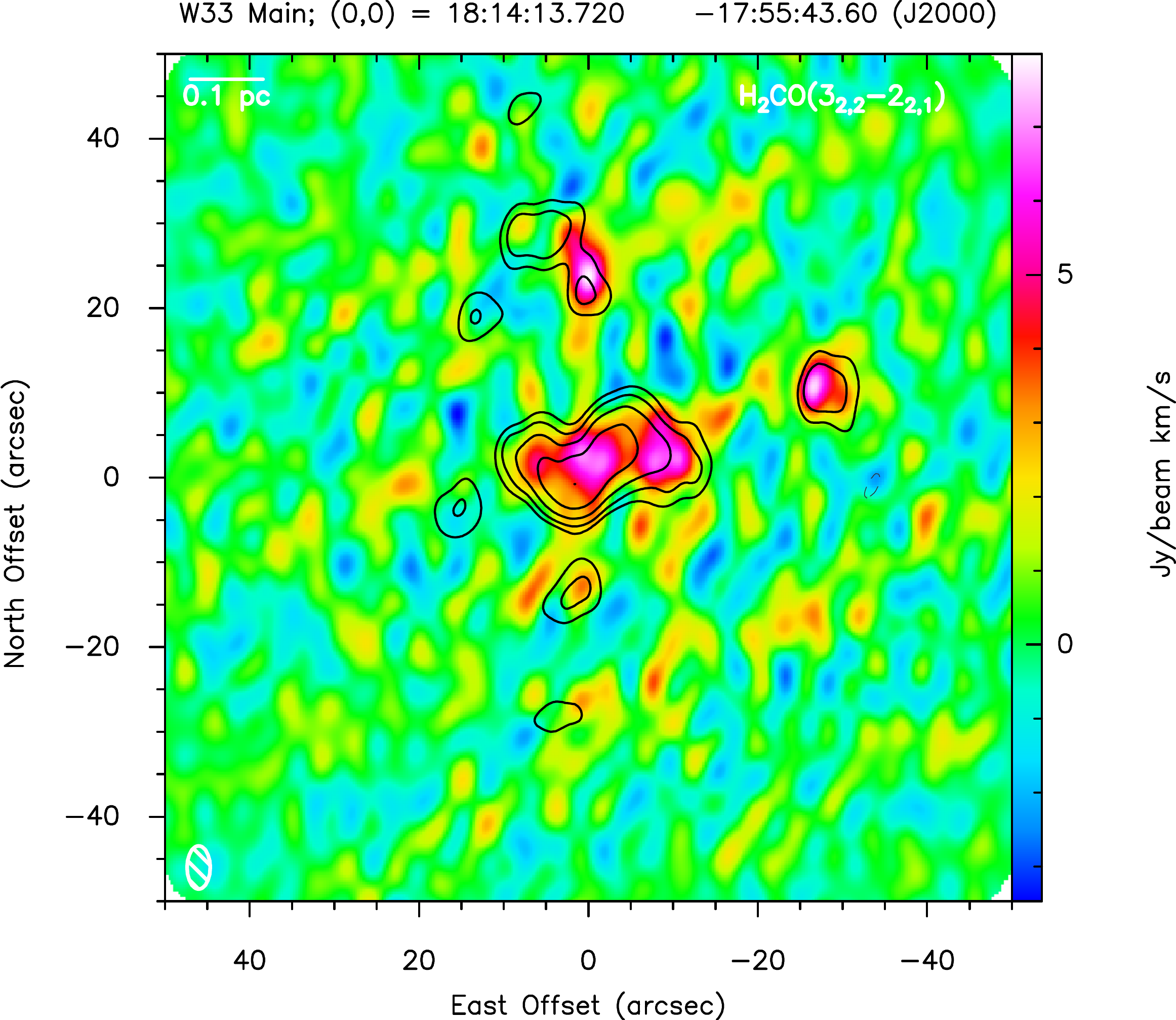}}	
\end{figure*}

\addtocounter{figure}{-1}
\begin{figure*}
	\centering
	\caption{Continued.}
	\subfloat{\includegraphics[width=9cm]{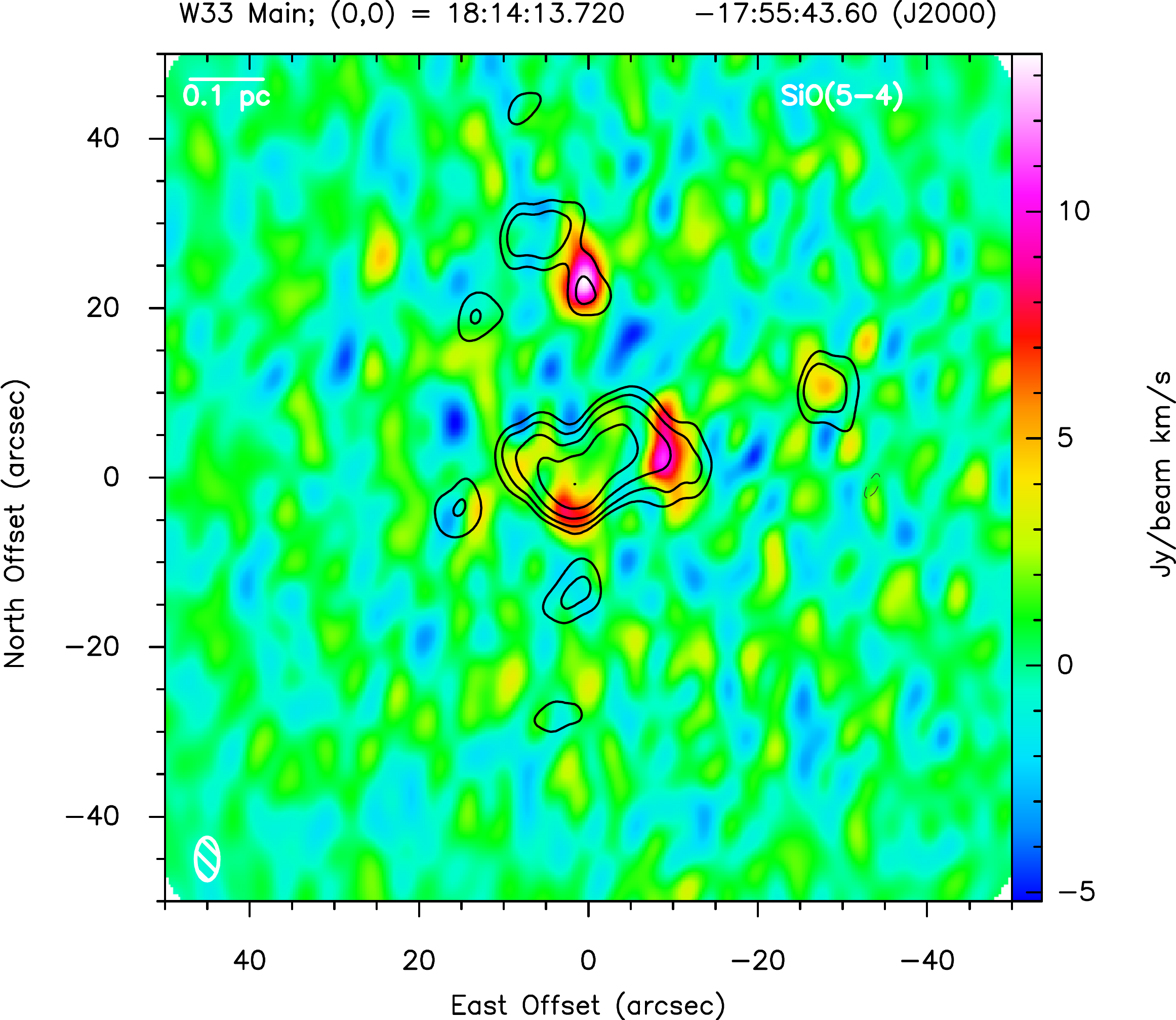}}\hspace{0.2cm}	
	\subfloat{\includegraphics[width=9cm]{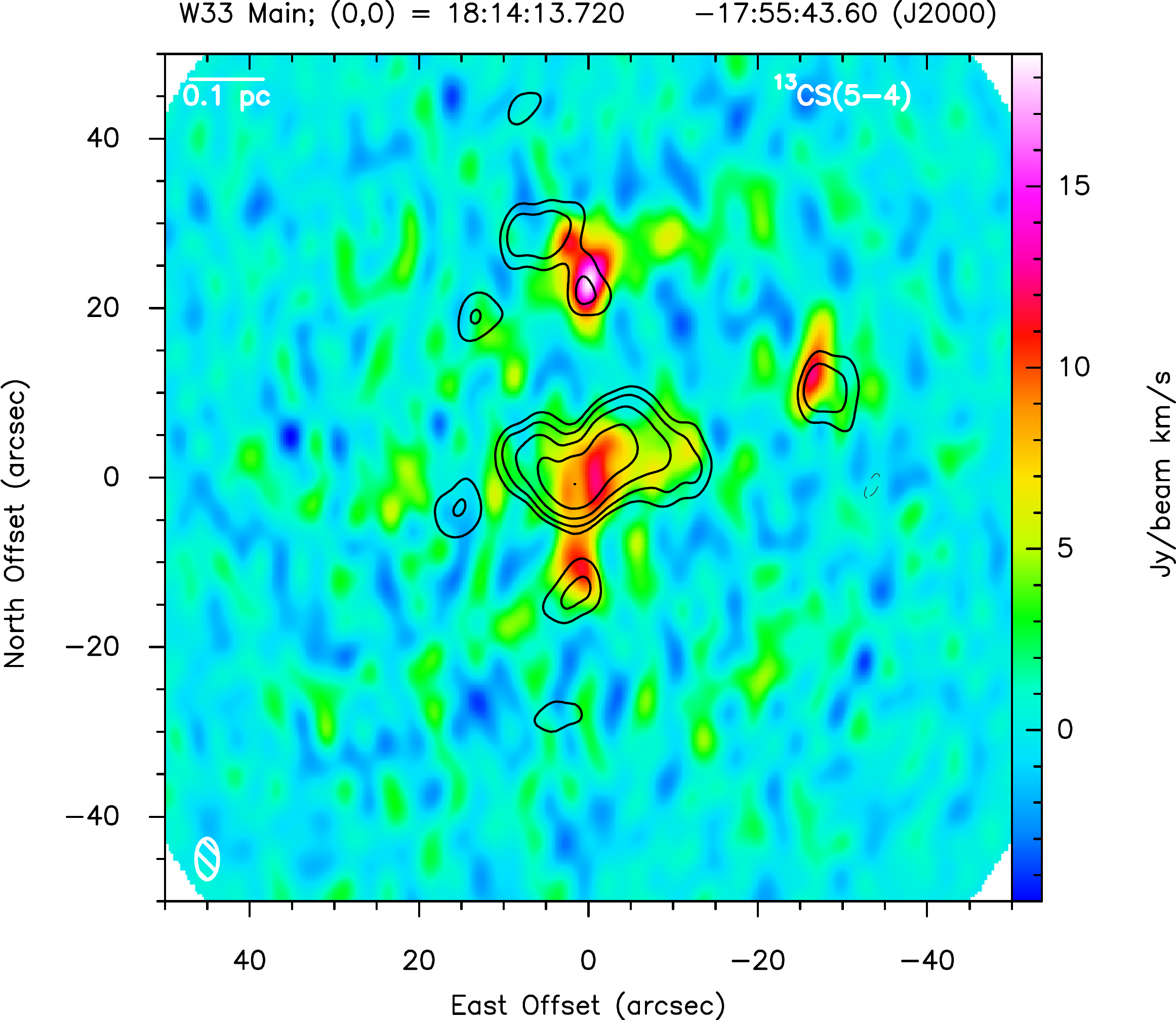}}	\\
	\subfloat{\includegraphics[width=9cm]{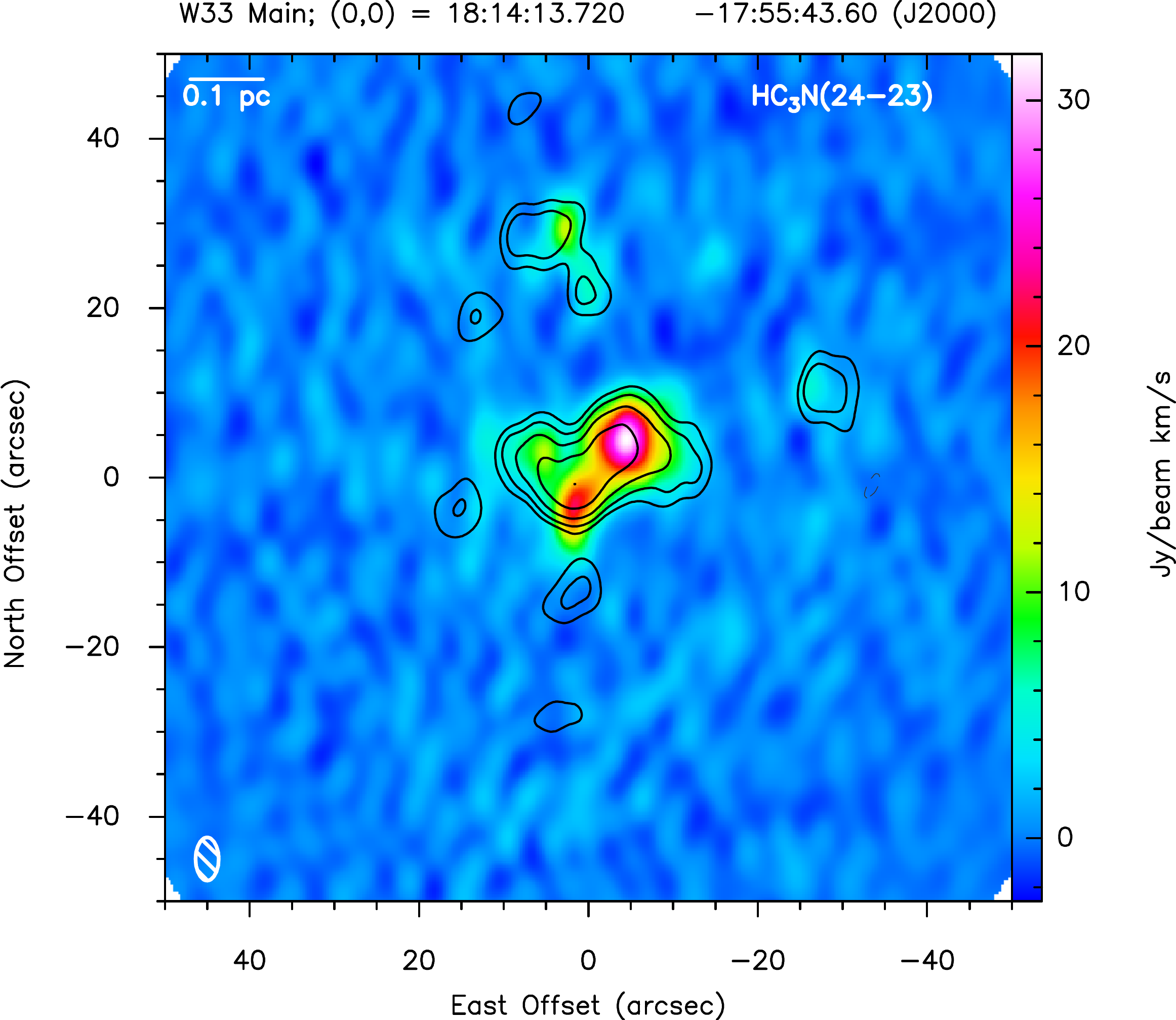}}\hspace{0.2cm}	
	\subfloat{\includegraphics[width=9cm]{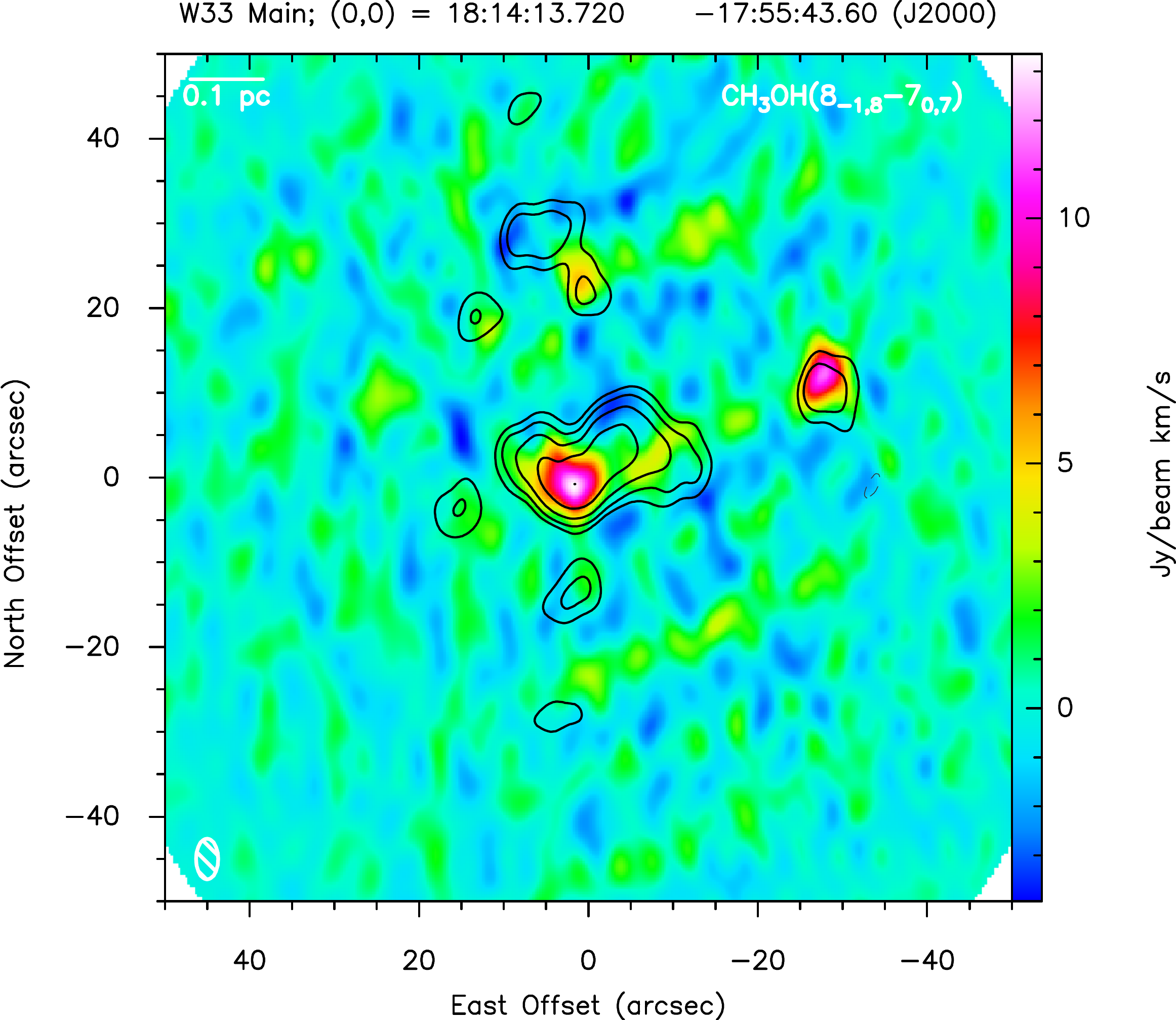}}\\	
	\subfloat{\includegraphics[width=9cm]{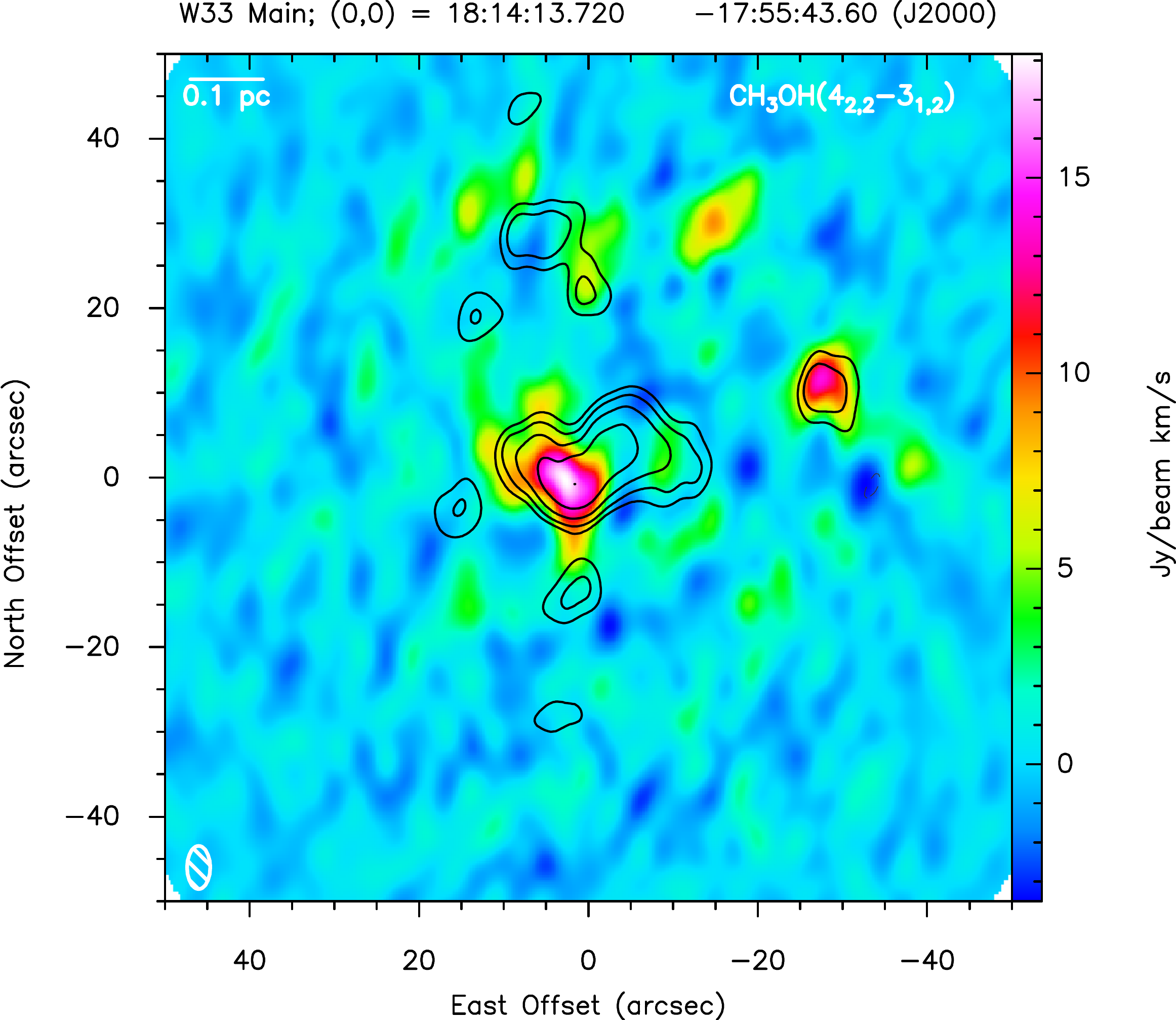}} \hspace{0.2cm}
	\subfloat{\includegraphics[width=9cm]{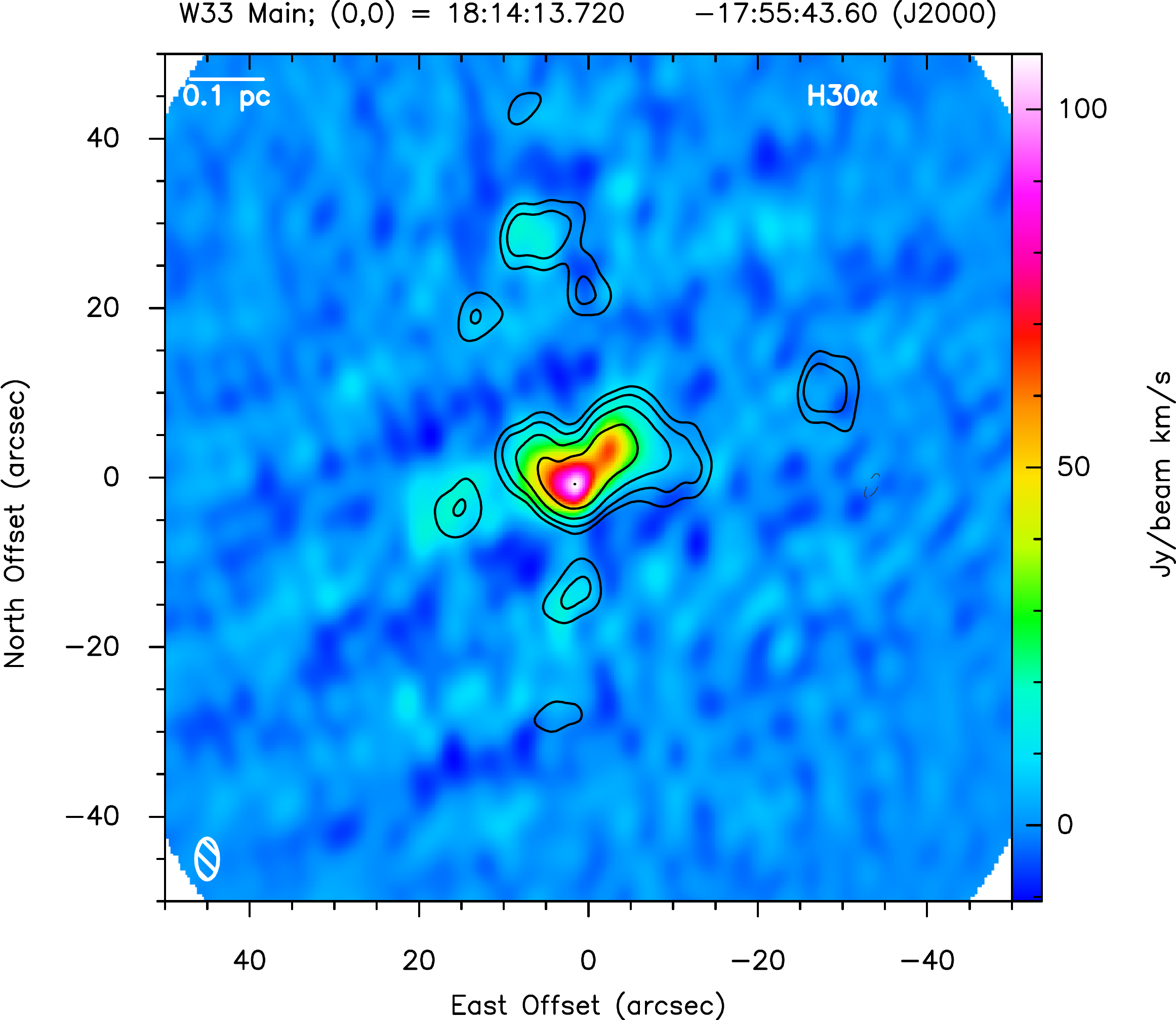}}	
\end{figure*}

\begin{figure*}
\caption{Moment 1 maps of different transitions, showing velocity gradients in the W33 sources. The contours show the continuum emission at 230 GHz (positive contour levels the same as in Fig. \ref{W33_SMA_CH0}).}
	\centering
	\subfloat{\includegraphics[width=9cm]{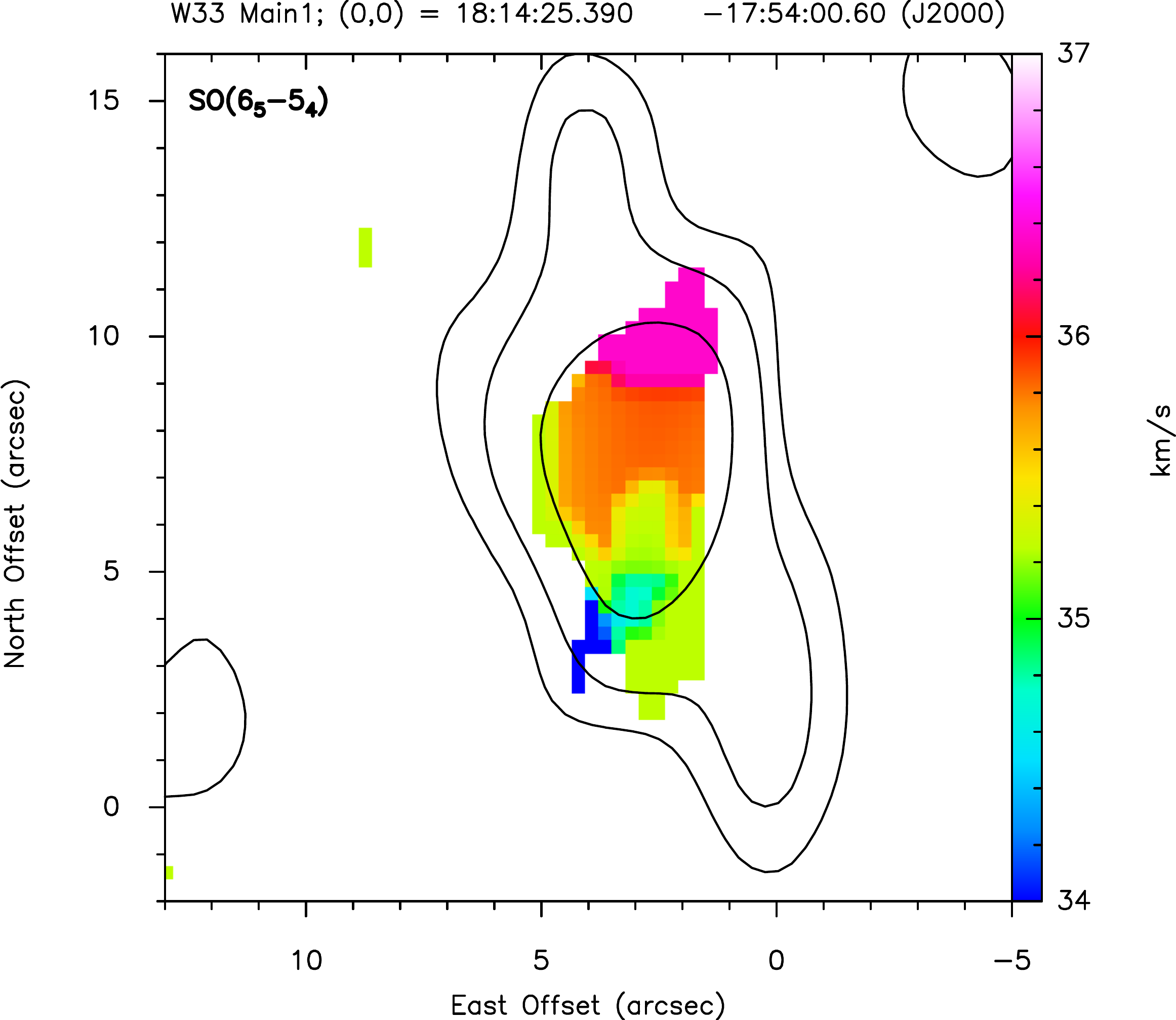}}\hspace{0.2cm}
	\subfloat{\includegraphics[width=9cm]{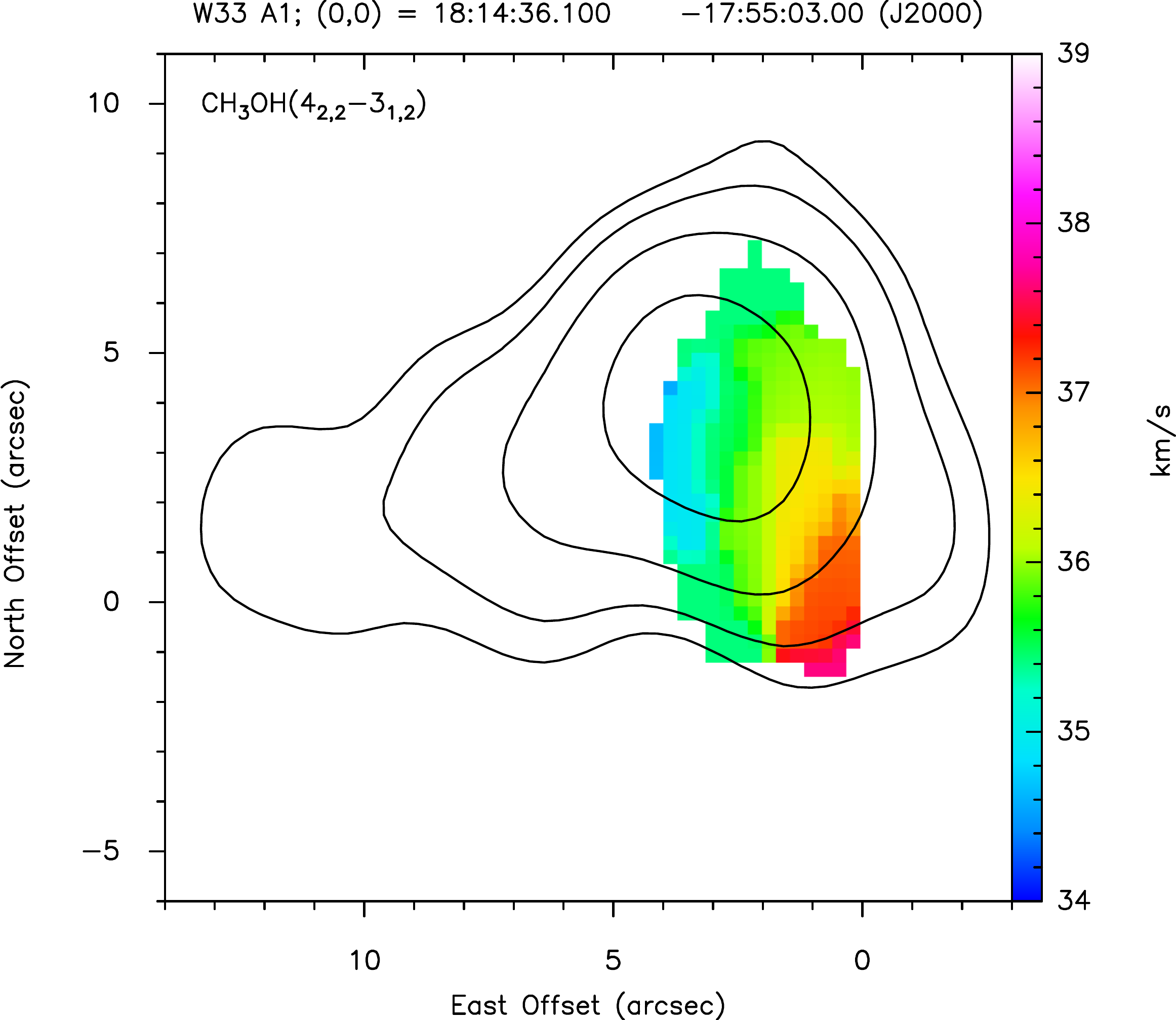}} \\	
	\subfloat{\includegraphics[width=9cm]{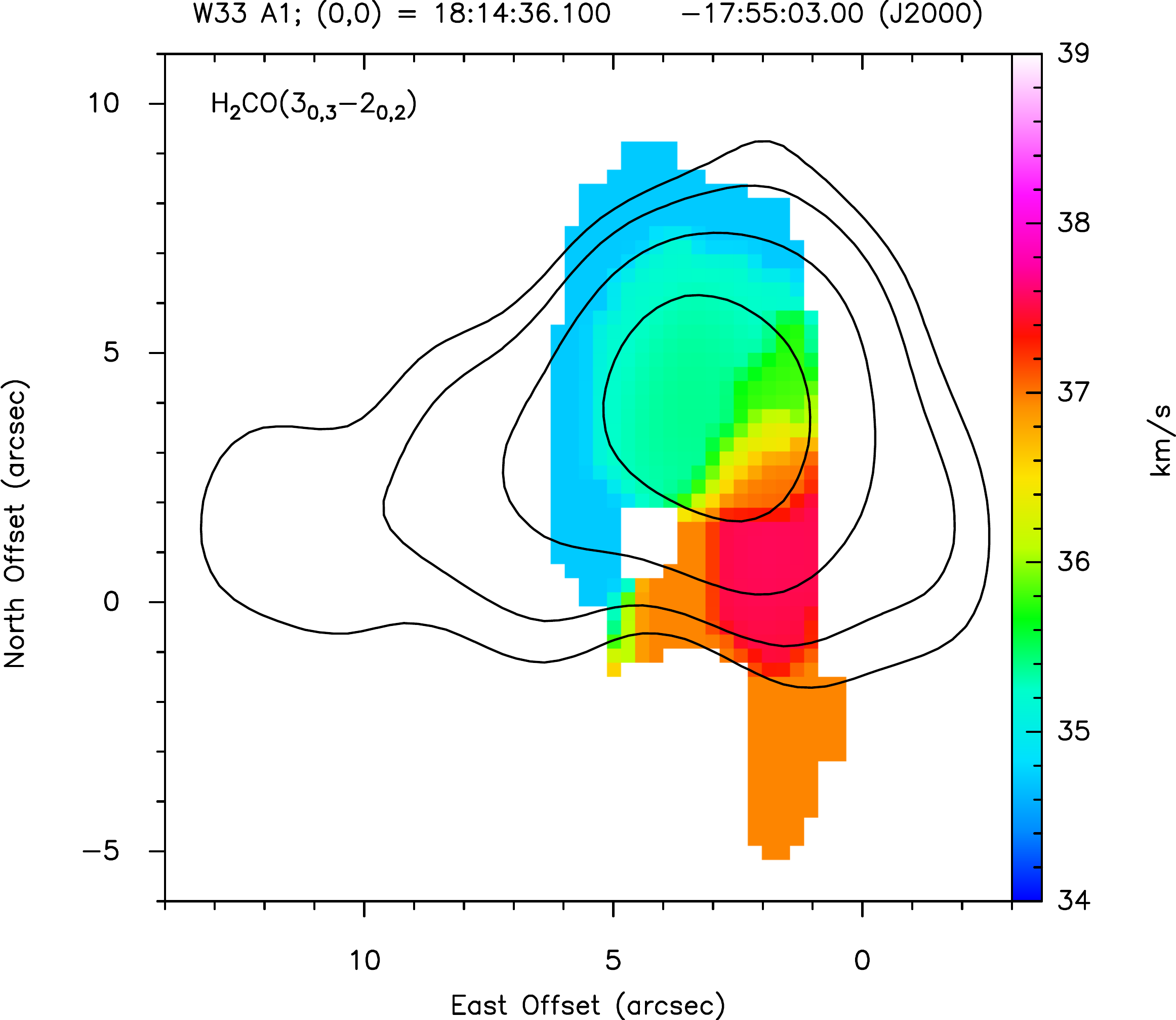}}\hspace{0.2cm} 	
	\subfloat{\includegraphics[width=9cm]{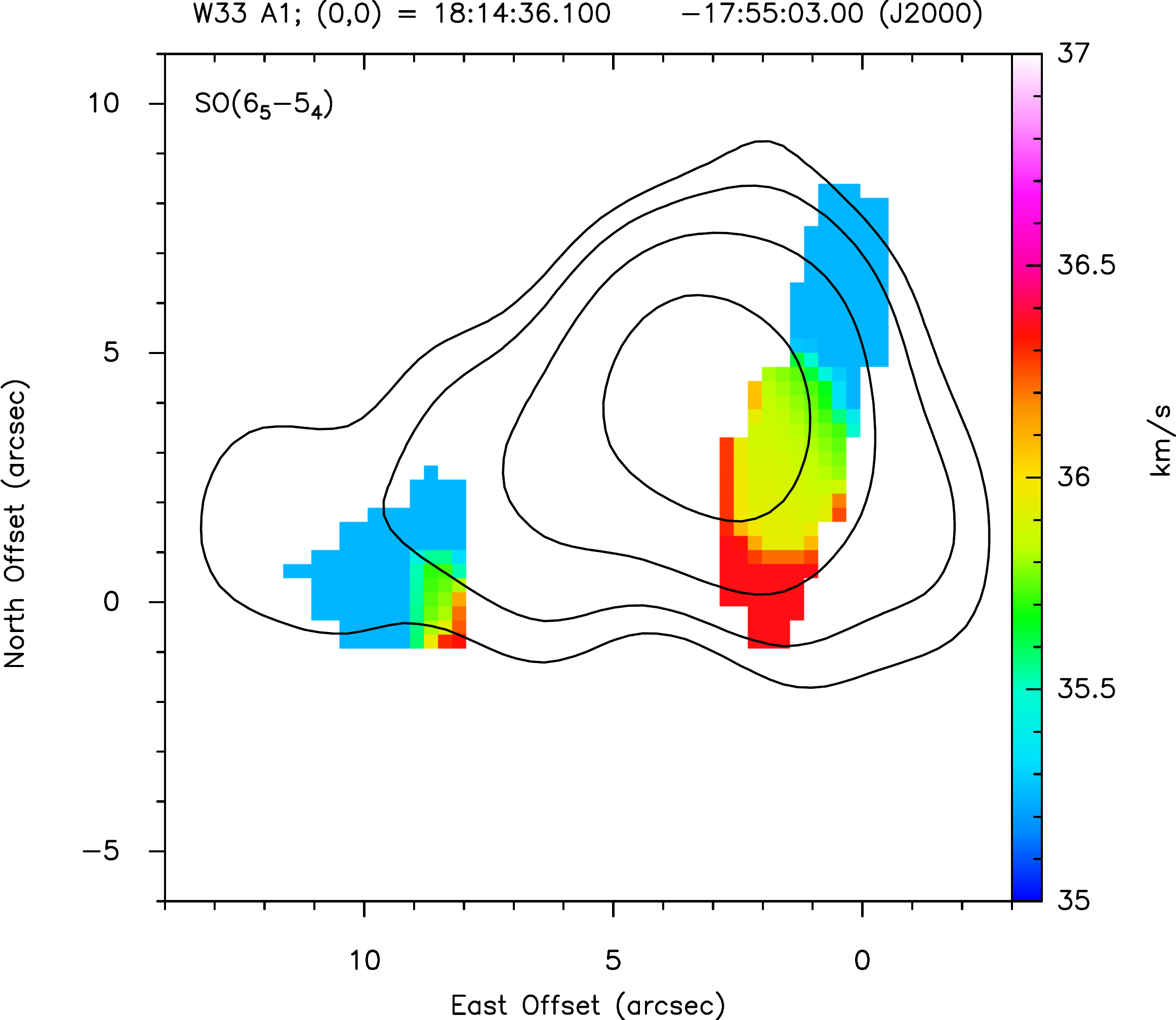}} \\	
	\subfloat{\includegraphics[width=9cm]{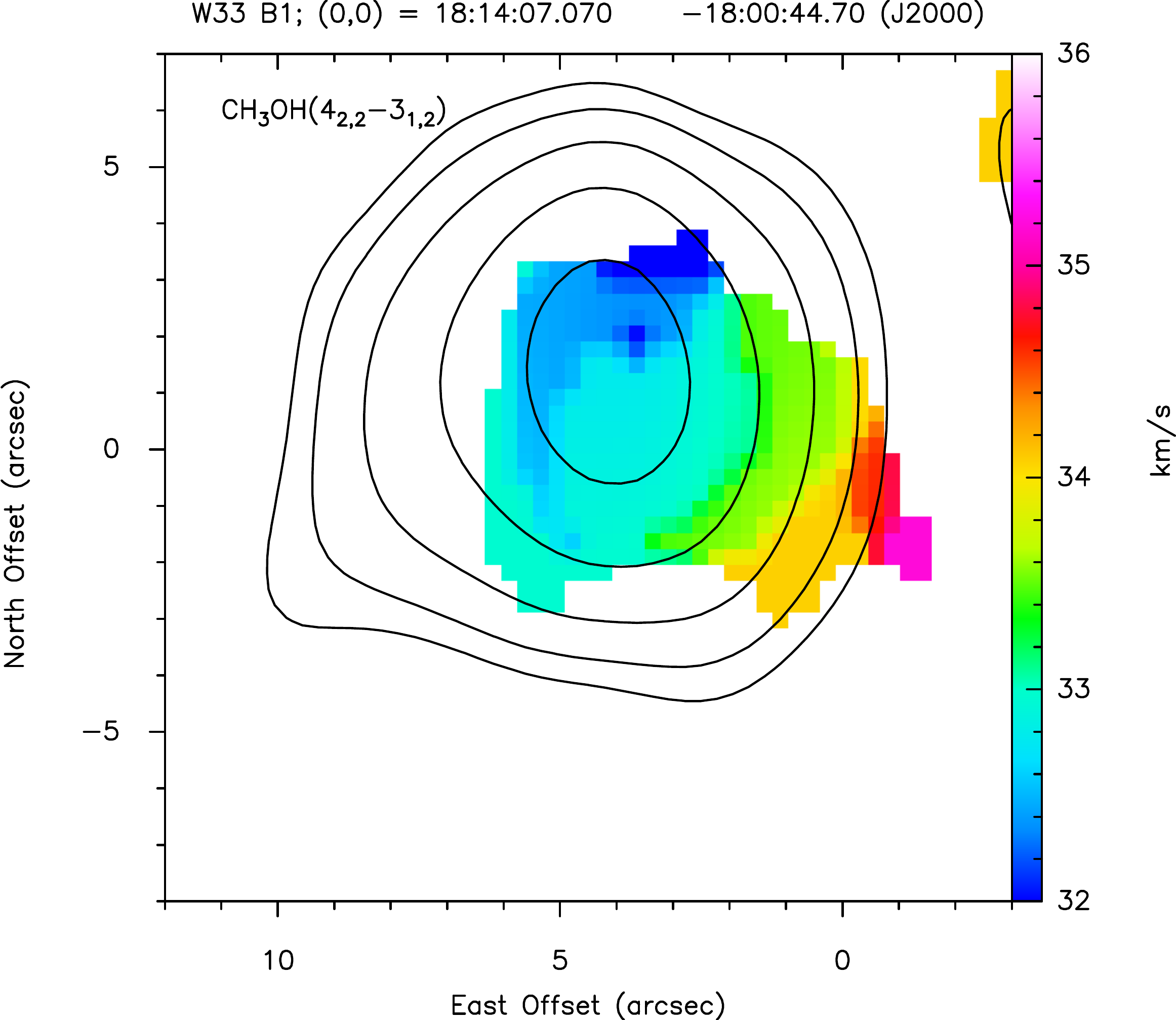}}\hspace{0.2cm}
	\subfloat{\includegraphics[width=9cm]{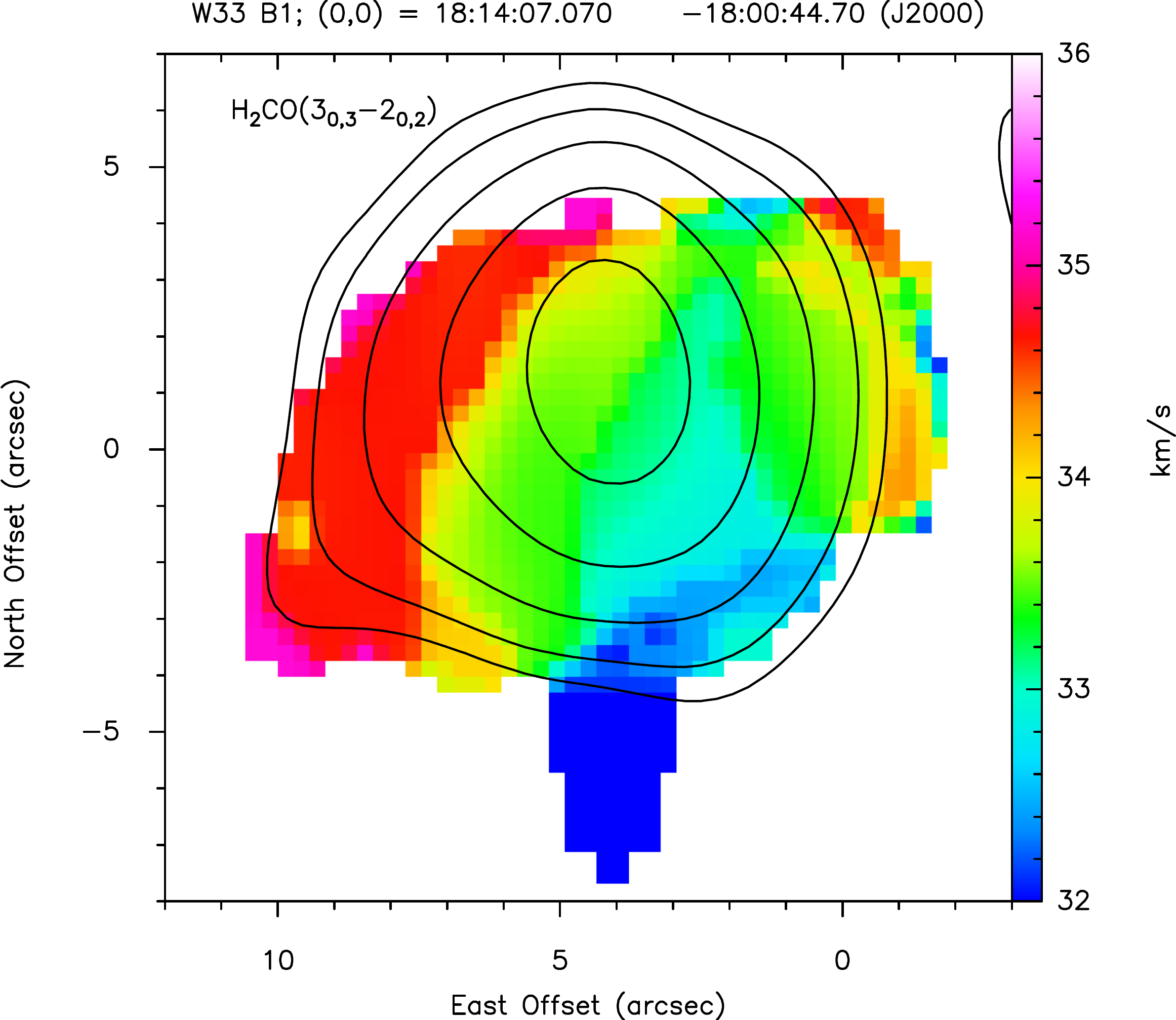}}
	\label{W33_SMA_VelGrad_All}
\end{figure*}

\addtocounter{figure}{-1}
\begin{figure*}
\caption{Continued.}
	\centering	
	\subfloat{\includegraphics[width=9cm]{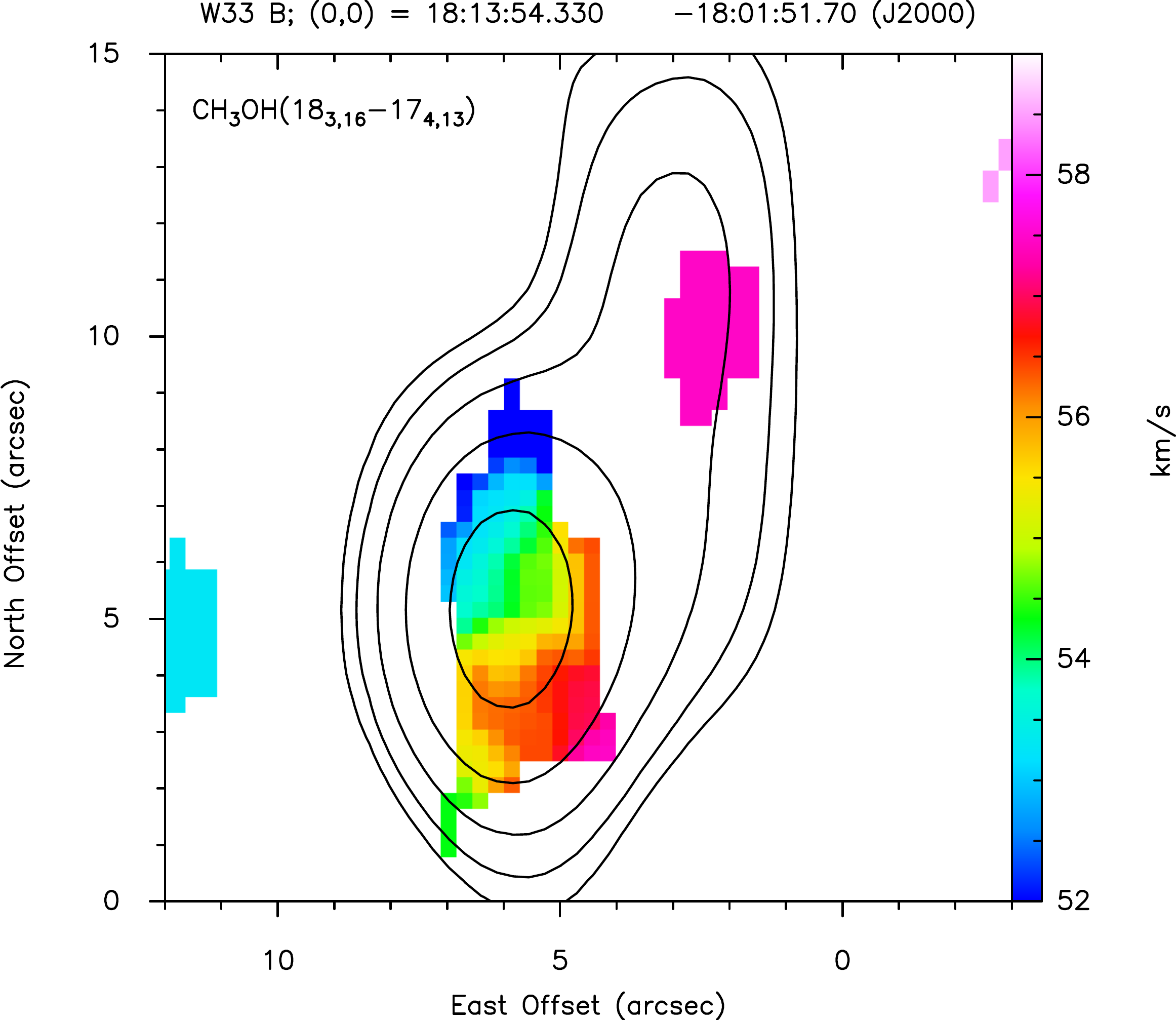}} \hspace{0.2cm}
	\subfloat{\includegraphics[width=9cm]{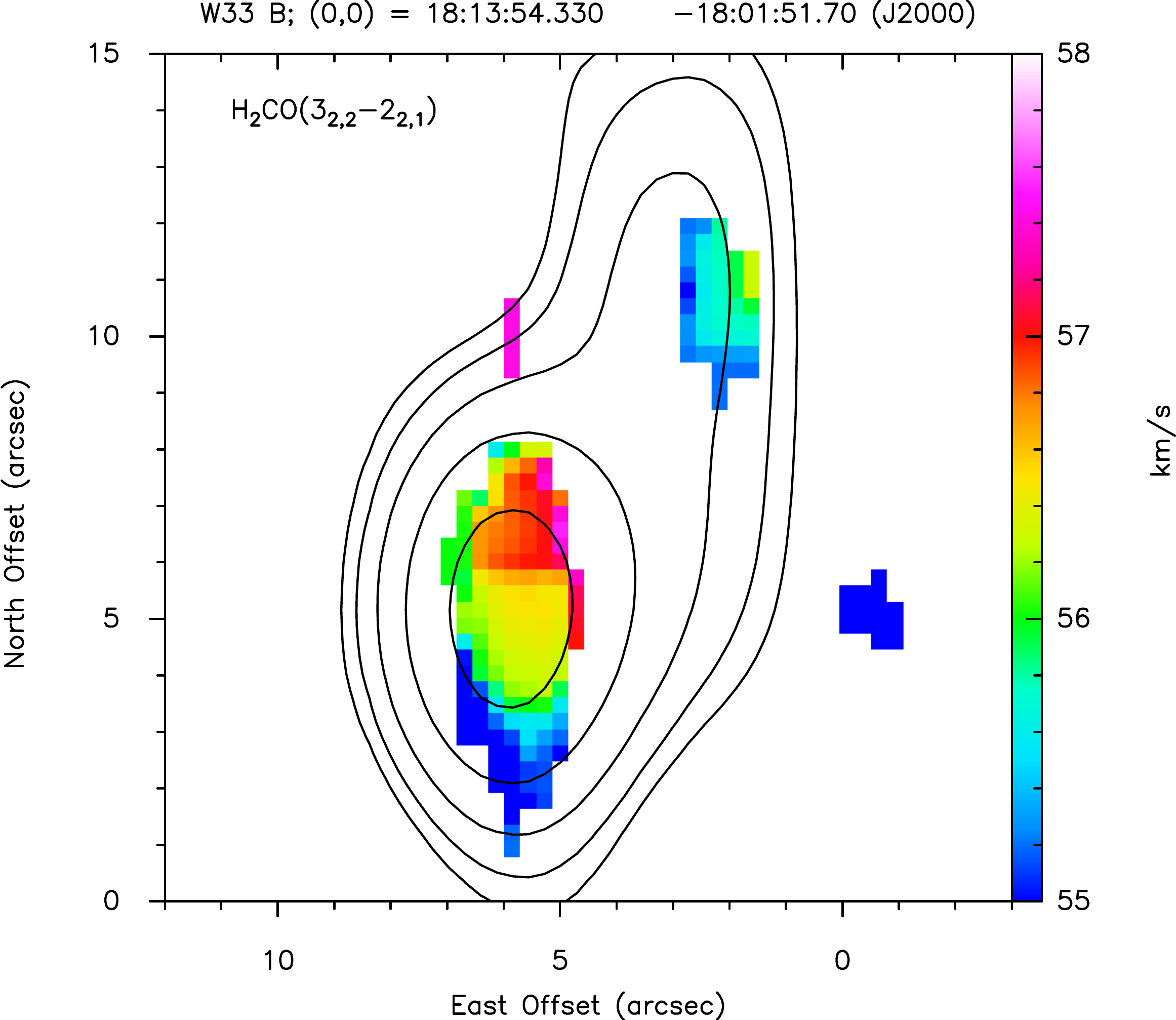}} \\
	\subfloat{\includegraphics[width=9cm]{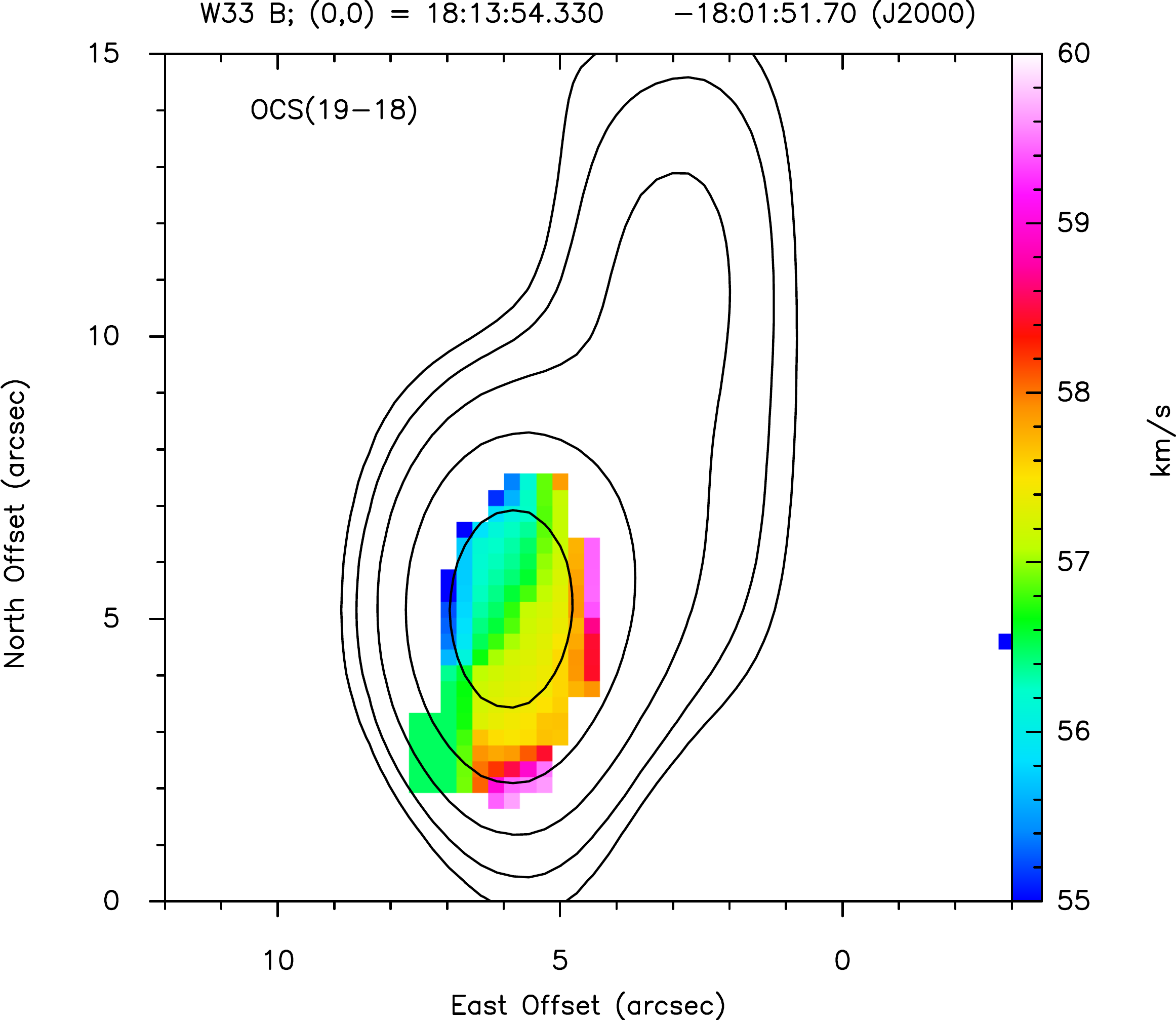}} \hspace{0.2cm}
	\subfloat{\includegraphics[width=9cm]{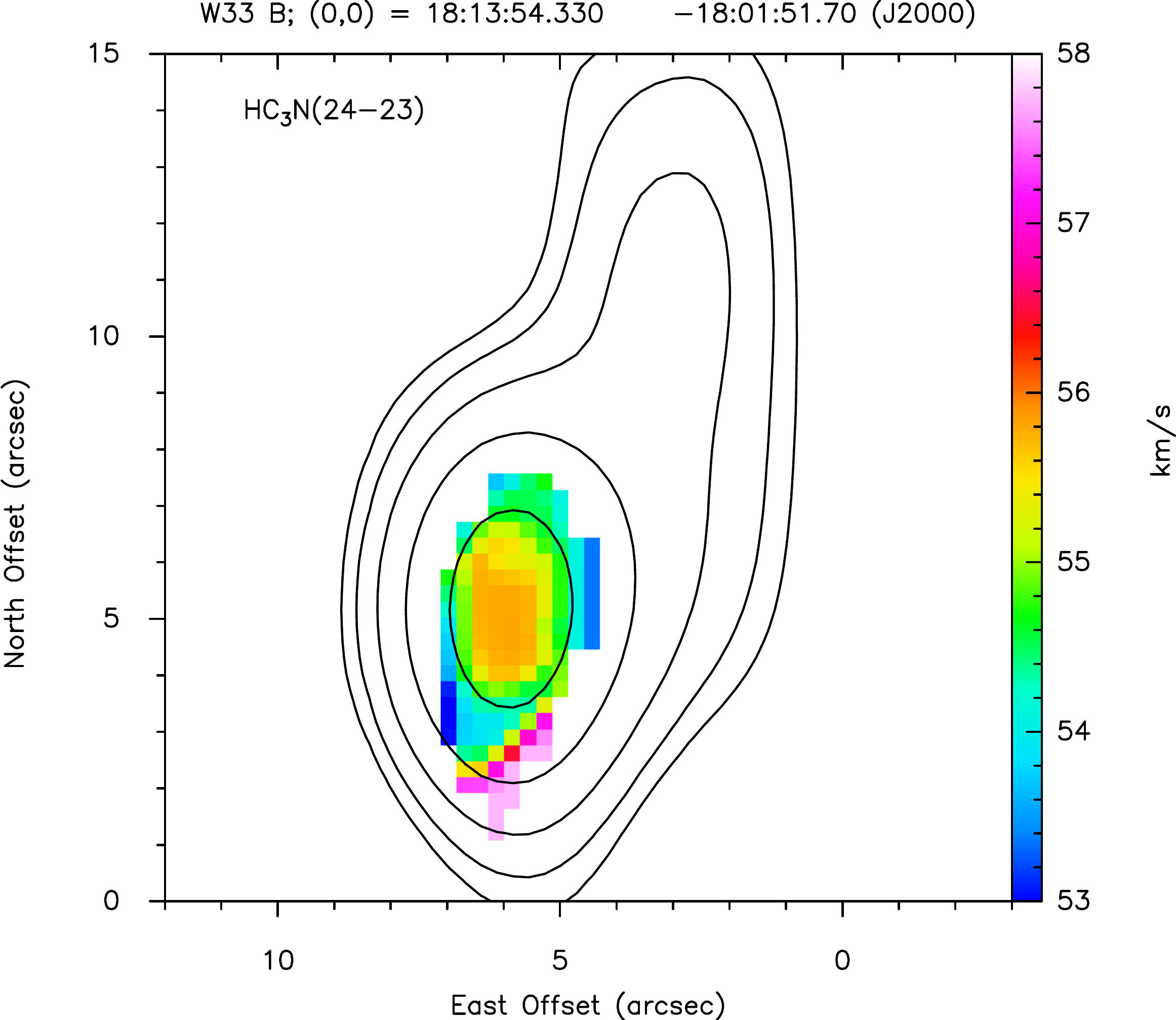}} \\
	\subfloat{\includegraphics[width=9cm]{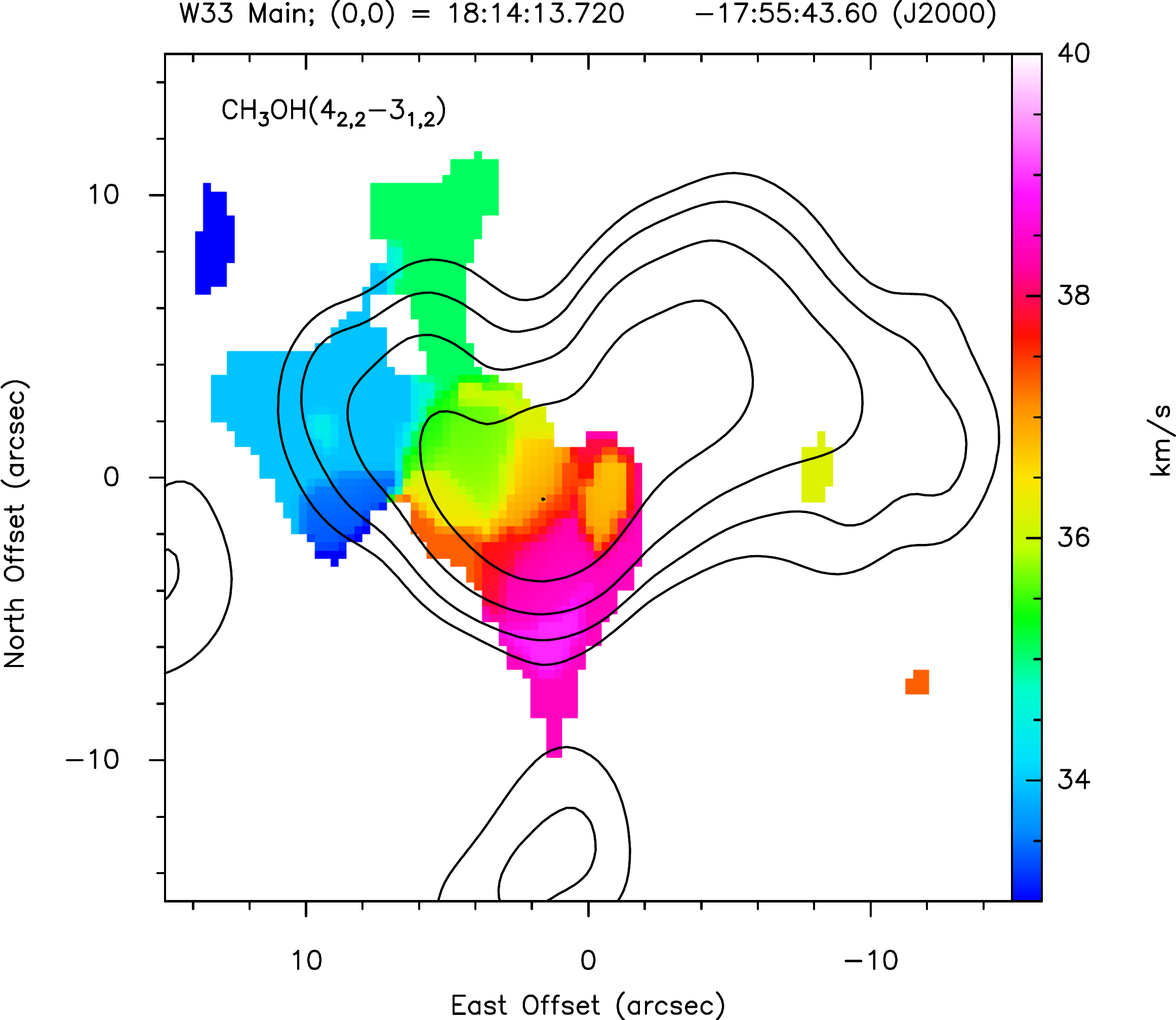}}\hspace{0.2cm} 
	\subfloat{\includegraphics[width=9cm]{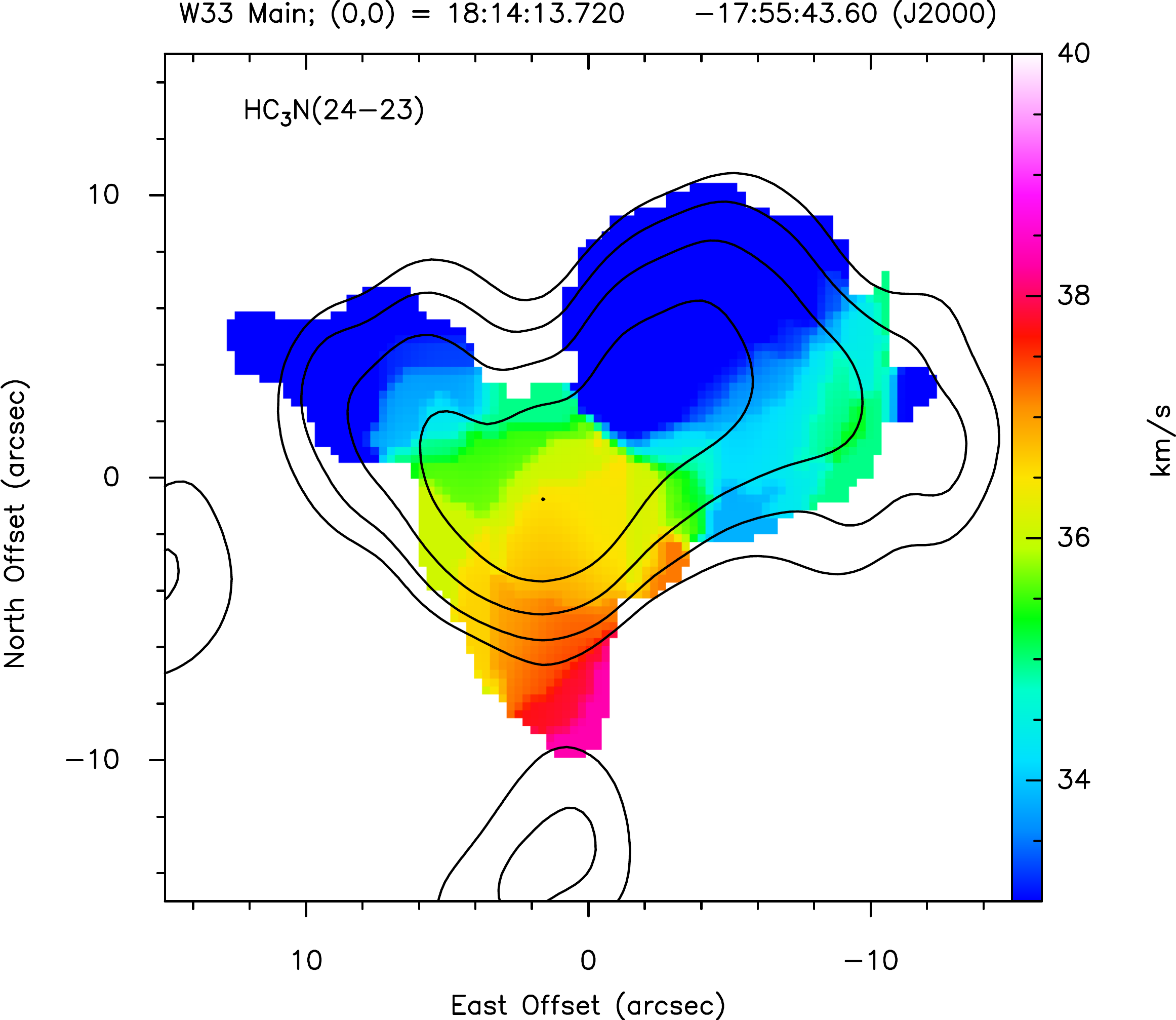}} 
\end{figure*}

\begin{figure*}
\caption{Velocity integrated CO emission in W33\,Main1, W33\,A1, W33\,B1, W33\,B, and W33\,Main. The background shows the 230 GHz continuum emission. Blue and red contours correspond to the most blueshifted and redshifted $^{12}$CO or C$^{18}$O emission in the five sources. In W33\,Main1, the redshifted and blueshifted emission is integrated over velocity ranges of 36$-$42 km s$^{-1}$ and 27$-$31 km s$^{-1}$, respectively (contour levels: 4--23 Jy beam$^{-1}$ km s$^{-1}$ in steps of 3 Jy beam$^{-1}$ km s$^{-1}$). In W33\,A1, the redshifted and blueshifted emission is integrated over velocity ranges of 40--47 km s$^{-1}$ and 25--32 km s$^{-1}$, respectively (contour levels: 6--28 Jy beam$^{-1}$ km s$^{-1}$ in steps of 3 Jy beam$^{-1}$ km s$^{-1}$). In W33\,B1, the redshifted and blueshifted emission is integrated over velocity ranges of 40--45 km s$^{-1}$ and 24--30 km s$^{-1}$, respectively (contour levels: 5--35 Jy beam$^{-1}$ km s$^{-1}$ in steps of 5 Jy beam$^{-1}$ km s$^{-1}$). In W33\,B, the redshifted and blueshifted emission is integrated over velocity ranges of 64--69 km s$^{-1}$ and 44--48 km s$^{-1}$, respectively (contour levels: 10--70 Jy beam$^{-1}$ km s$^{-1}$ in steps of 5 Jy beam$^{-1}$ km s$^{-1}$). In W33\,Main, the redshifted and blueshifted emission (from the IRAM30m+SMA data) is integrated over 39--42 km s$^{-1}$ and 28--32 km s$^{-1}$, respectively (contour levels: 10--25 Jy beam$^{-1}$ km s$^{-1}$ in steps of 2.5 Jy beam$^{-1}$ km s$^{-1}$). The synthesised beams are shown in the lower left corners of the images.}
	\centering
	\subfloat[$^{12}$CO emission in W33\,Main1]{\includegraphics[width=6.5cm]{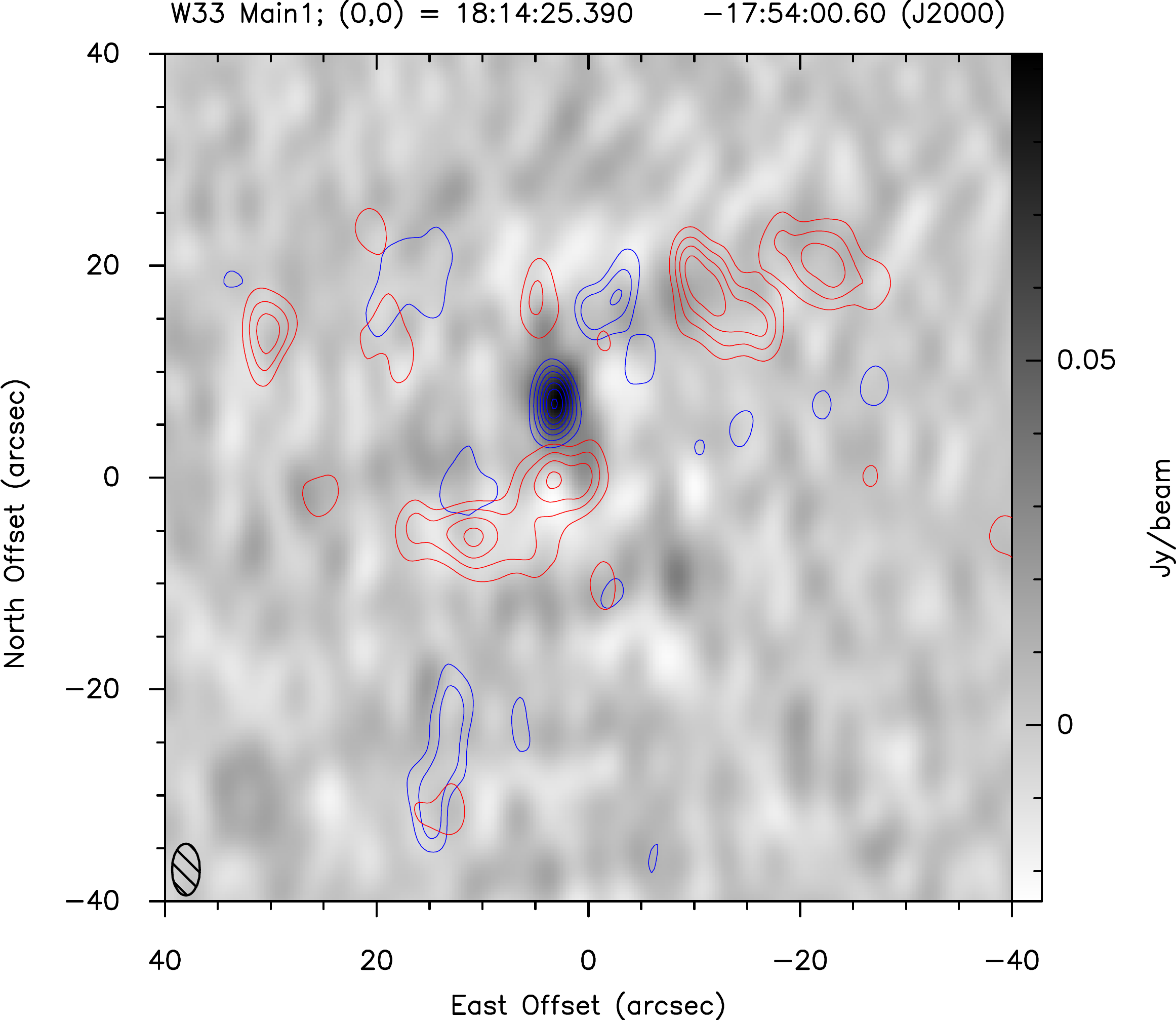}\label{W33M1_12CO}} \hspace{0.2cm}
	\subfloat[$^{12}$CO emission in W33\,A1]{\includegraphics[width=6.5cm]{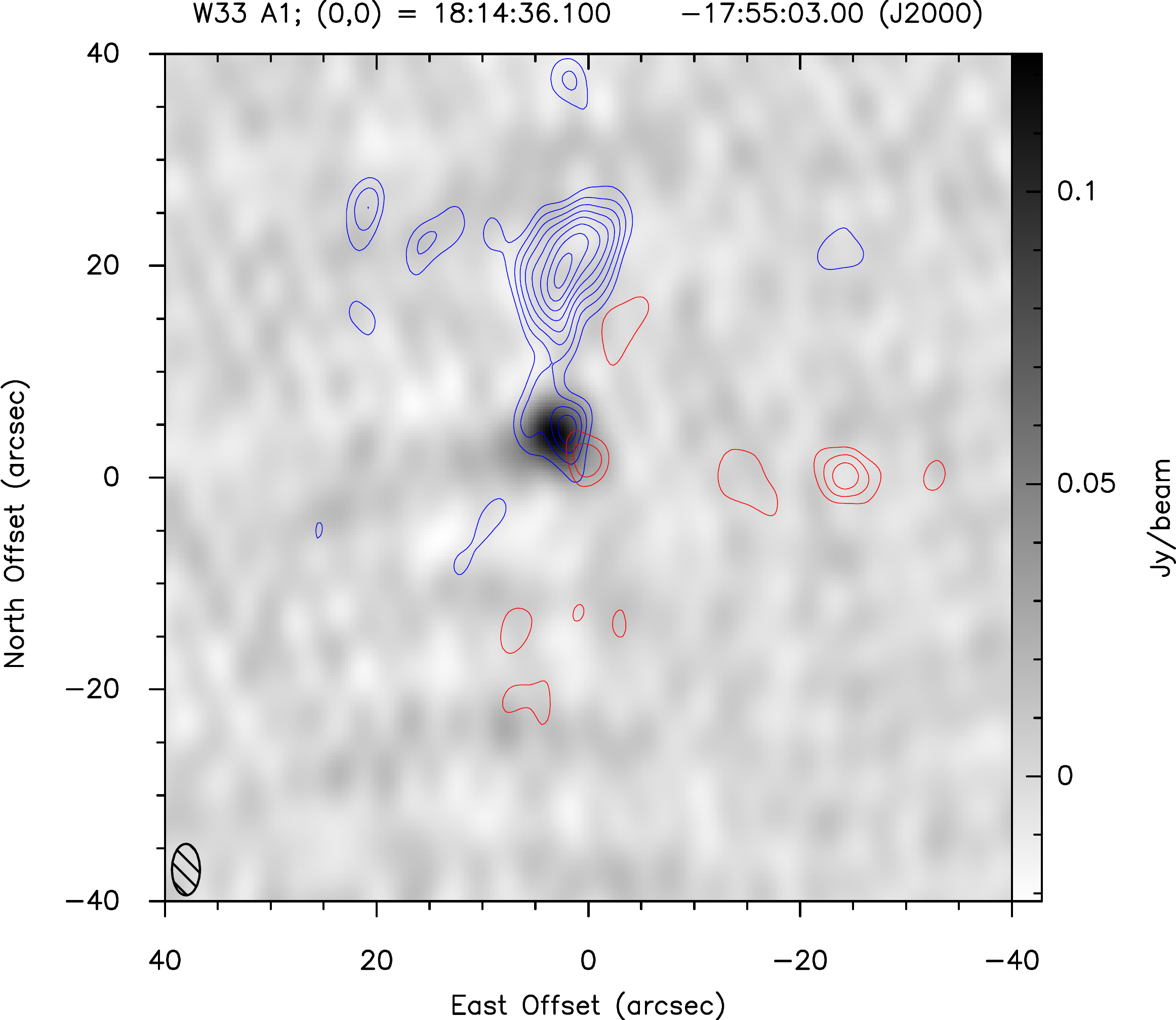}\label{W33A1_12CO}} 	\\
	\subfloat[$^{12}$CO emission in W33\,B1]{\includegraphics[width=6.5cm]{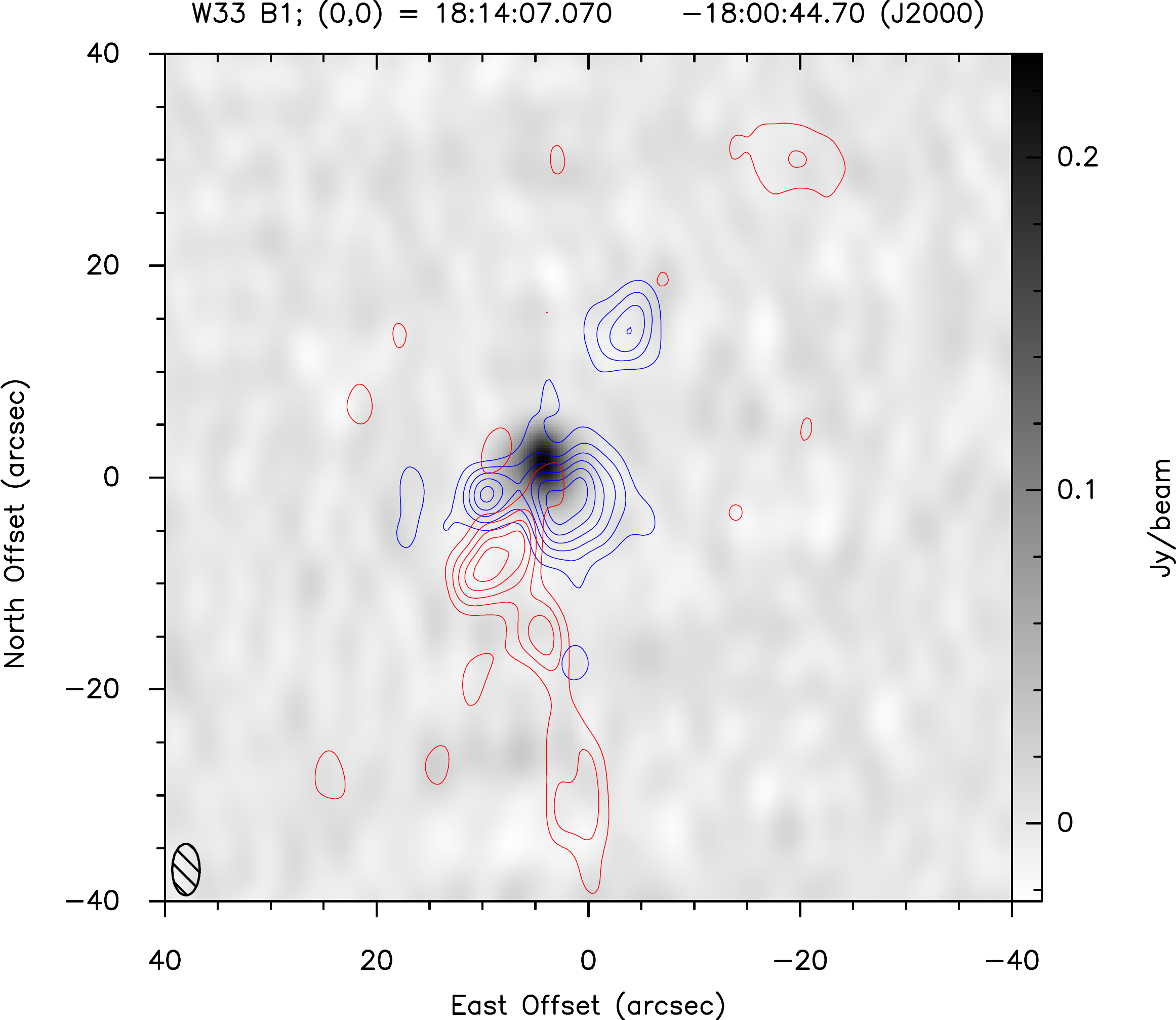}\label{W33B1_12CO}}\hspace{0.2cm}
	\subfloat[$^{12}$CO emission in W33\,B]{\includegraphics[width=6.5cm]{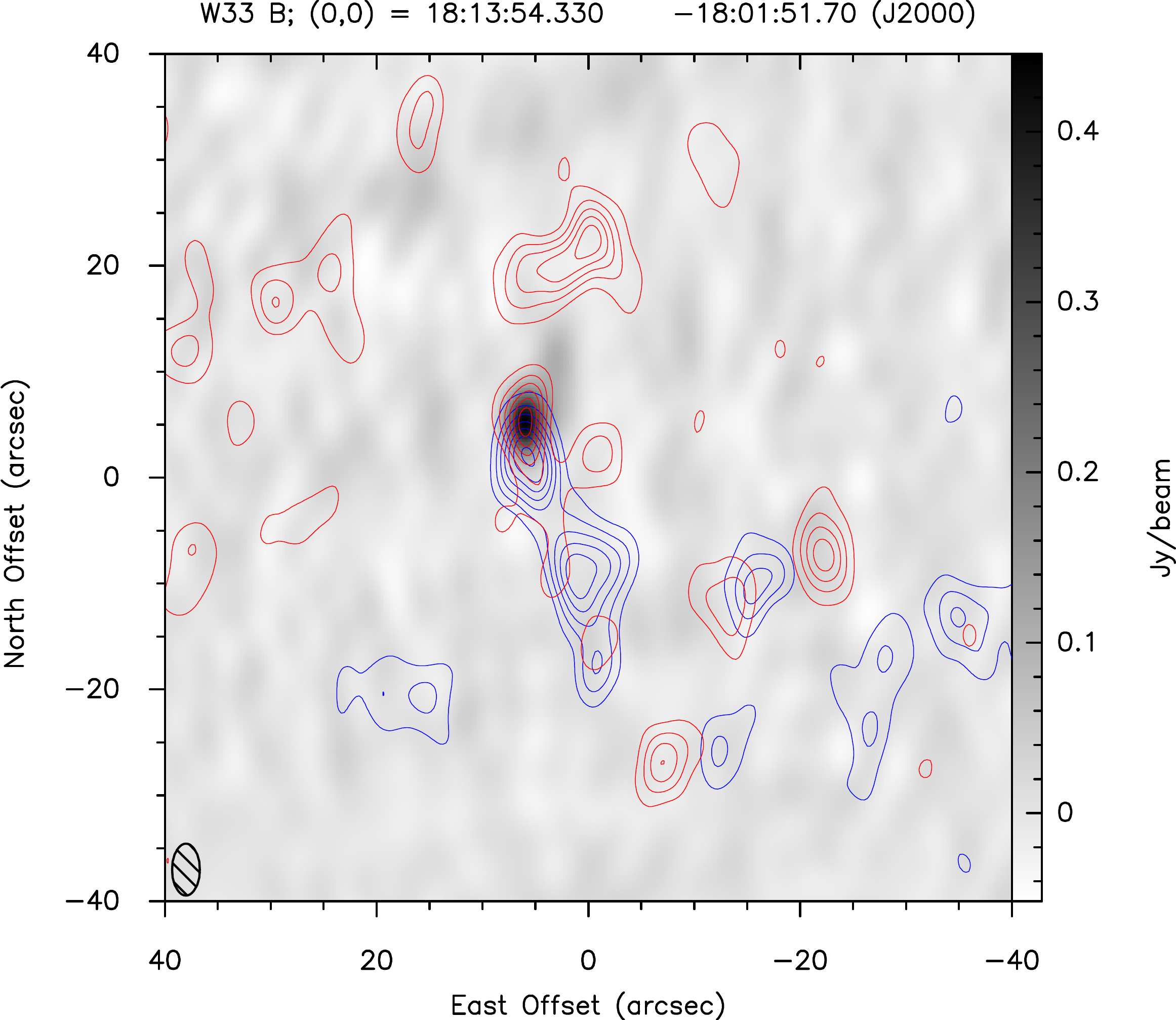}\label{W33B_12CO}} \\
	\subfloat[C$^{18}$O emission in W33\,Main]{\includegraphics[width=6.5cm]{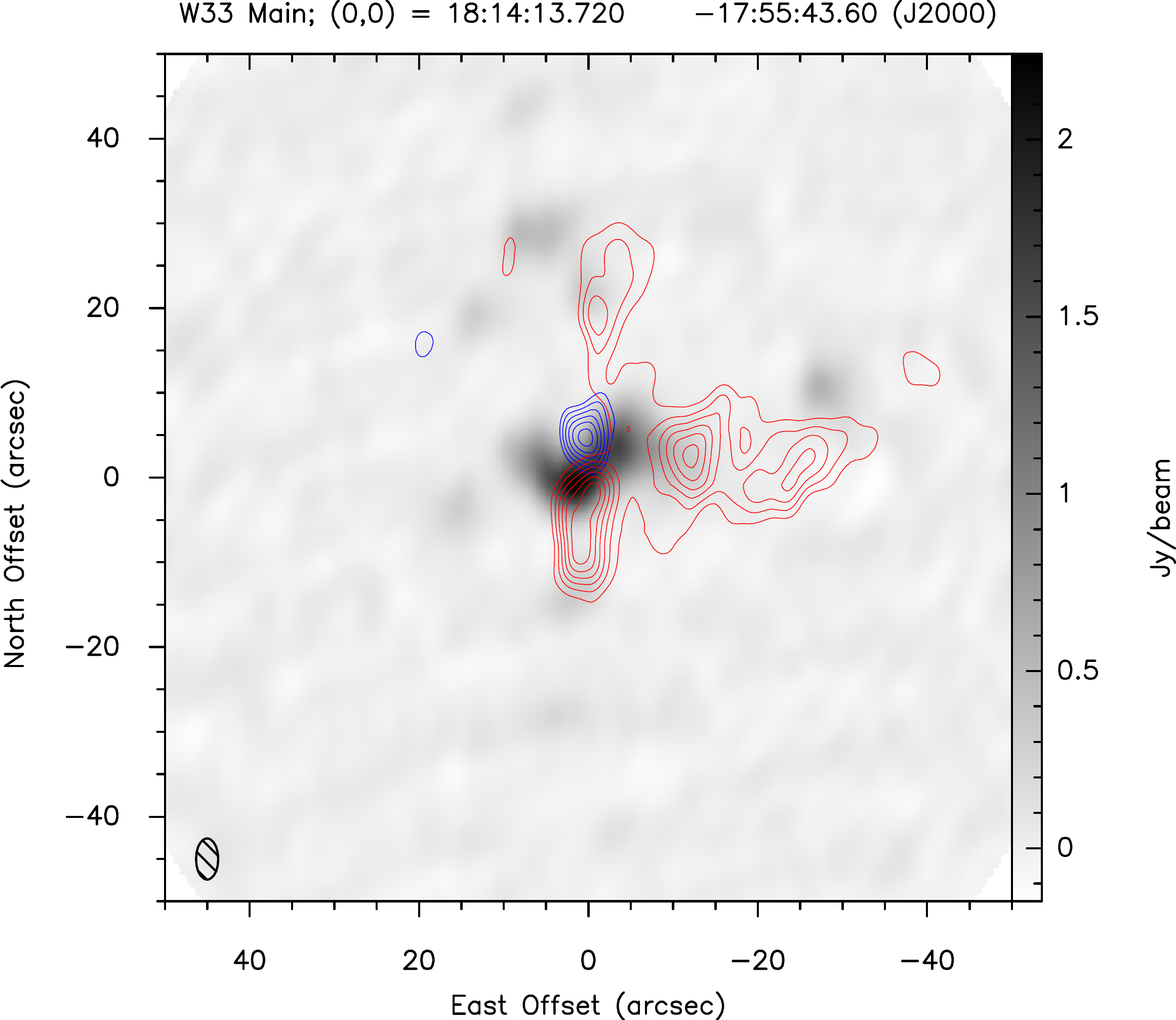}\label{W33M_C18O}} 
	\label{W33_SMA_12CO}
\end{figure*}

\begin{figure*}
\caption{Velocity integrated emission of $^{13}$CO and C$^{18}$O in W33\,Main. The images in the left panels show the data from the SMA observations while the images in the right panels are combinations of the IRAM30m and SMA data of these two transitions to correct for missing short spacings. The contours show the 230 GHz continuum emission in W33\,Main (contour levels same as in Fig. \ref{W33_SMA_CH0}). The $^{13}$CO and C$^{18}$O emission are integrated over velocity ranges of 27--46 km~s$^{-1}$ and 30--42 km~s$^{-1}$, respectively.}
	\centering
	\subfloat[SMA $^{13}$CO emission]{\includegraphics[width=6.4cm]{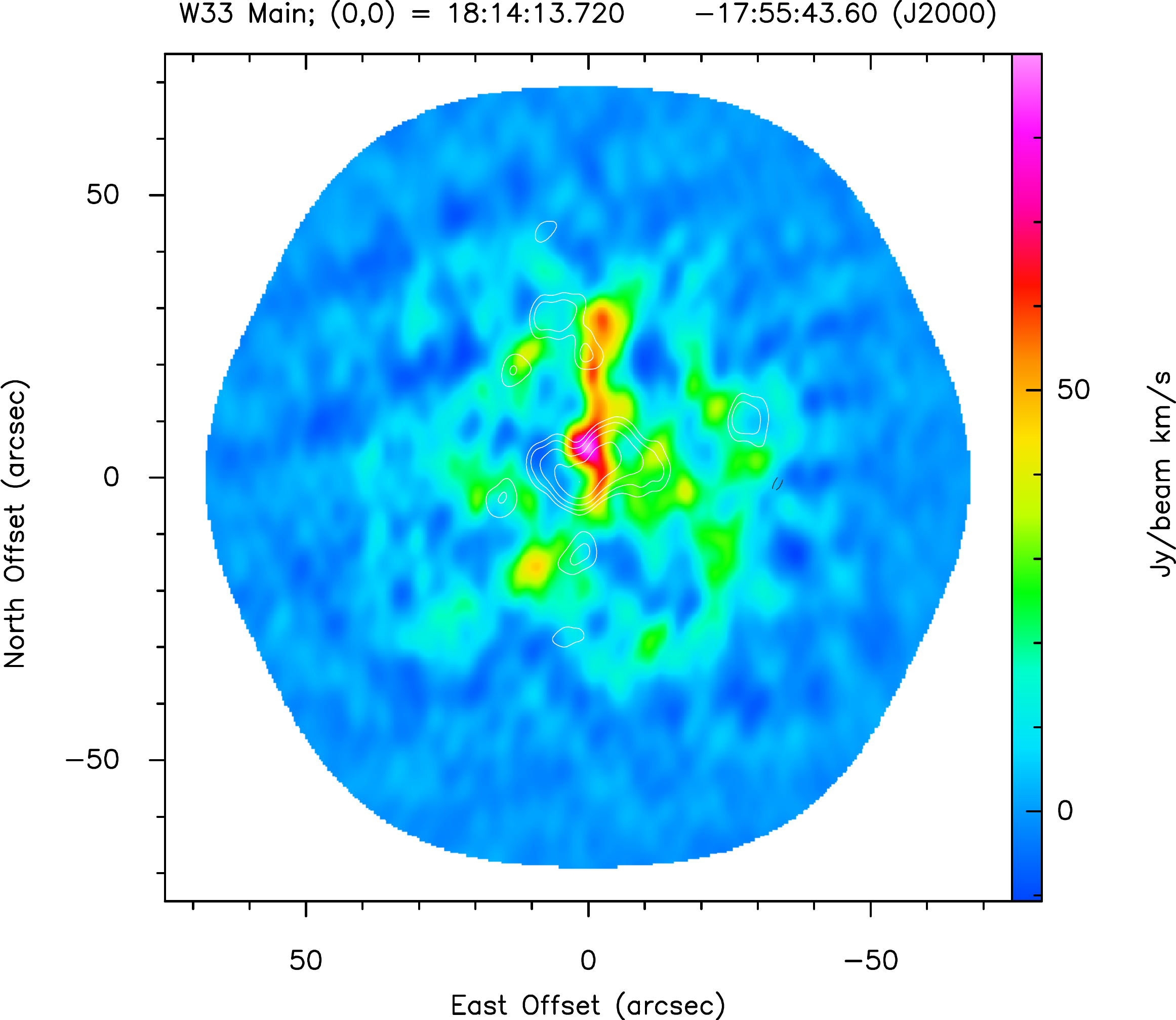}} \hspace{0.2cm}	
	\subfloat[Combined $^{13}$CO emission from SMA and IRAM30m]{\includegraphics[width=6.4cm]{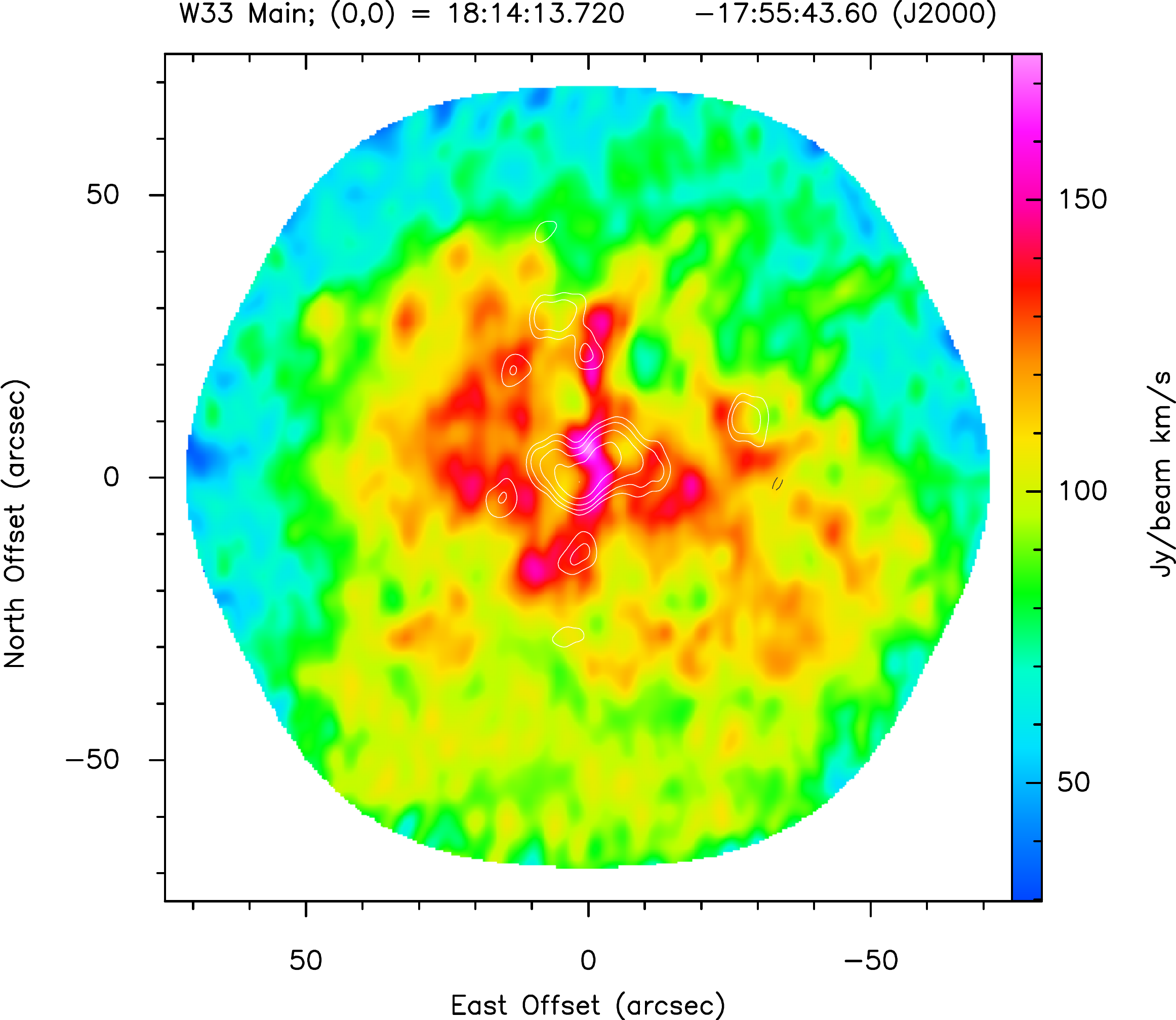}} \hspace{0.2cm}	\\
	\subfloat[SMA C$^{18}$O emission]{\includegraphics[width=6.4cm]{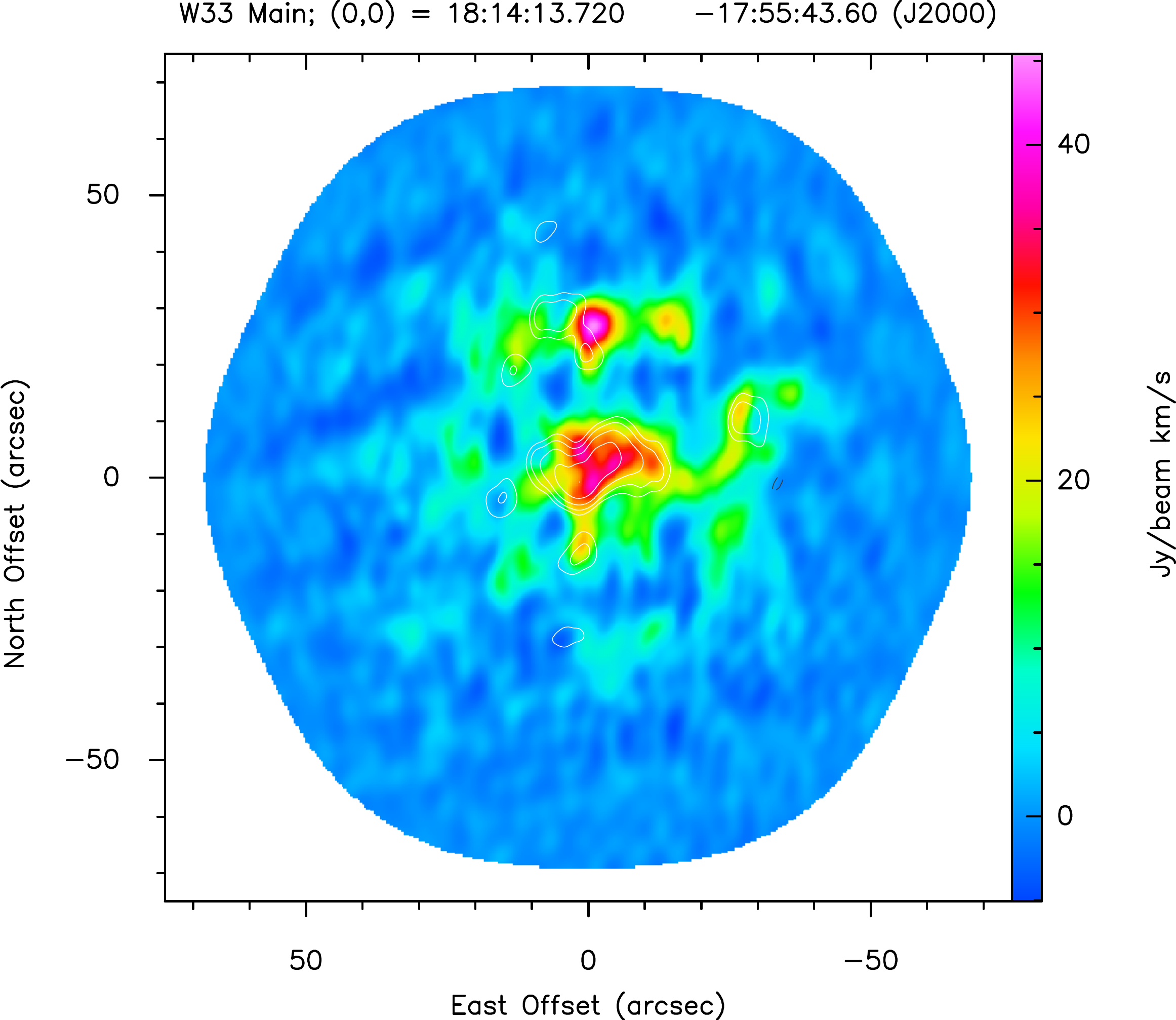}} \hspace{0.2cm}	
	\subfloat[Combined C$^{18}$O emission from SMA and IRAM30m]{\includegraphics[width=6.4cm]{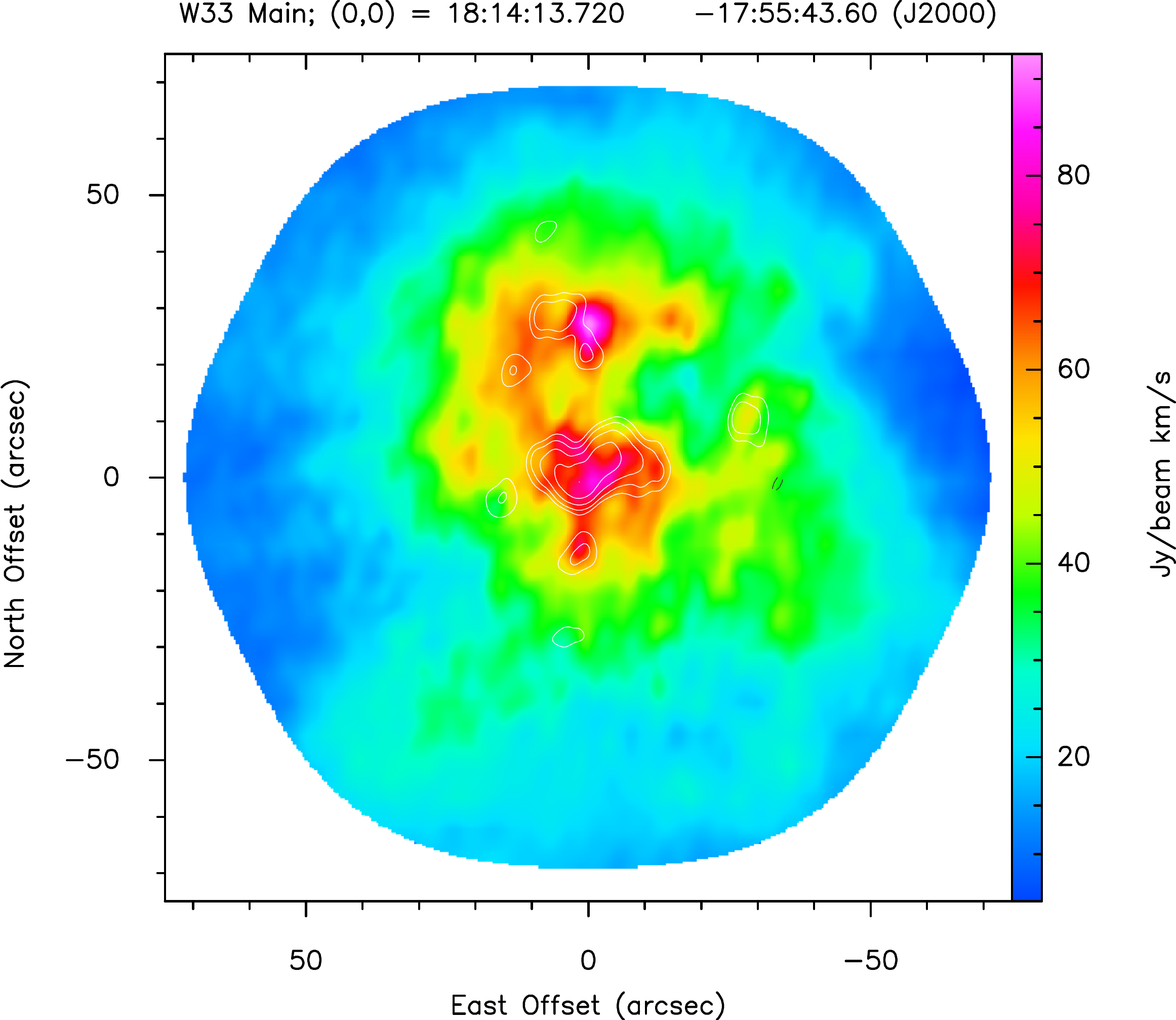}} \hspace{0.2cm}	
	\label{W33M_COMB_CO}
\end{figure*}

\begin{figure*}
\caption{$^{13}$CO spectra, integrated over one synthesised beam at the continuum peak of W33\,Main from the SMA data only (left) and from the combination of the SMA and the IRAM30m data (right).}
	\centering
	\subfloat[SMA $^{13}$CO spectrum]{\includegraphics[width=8.5cm]{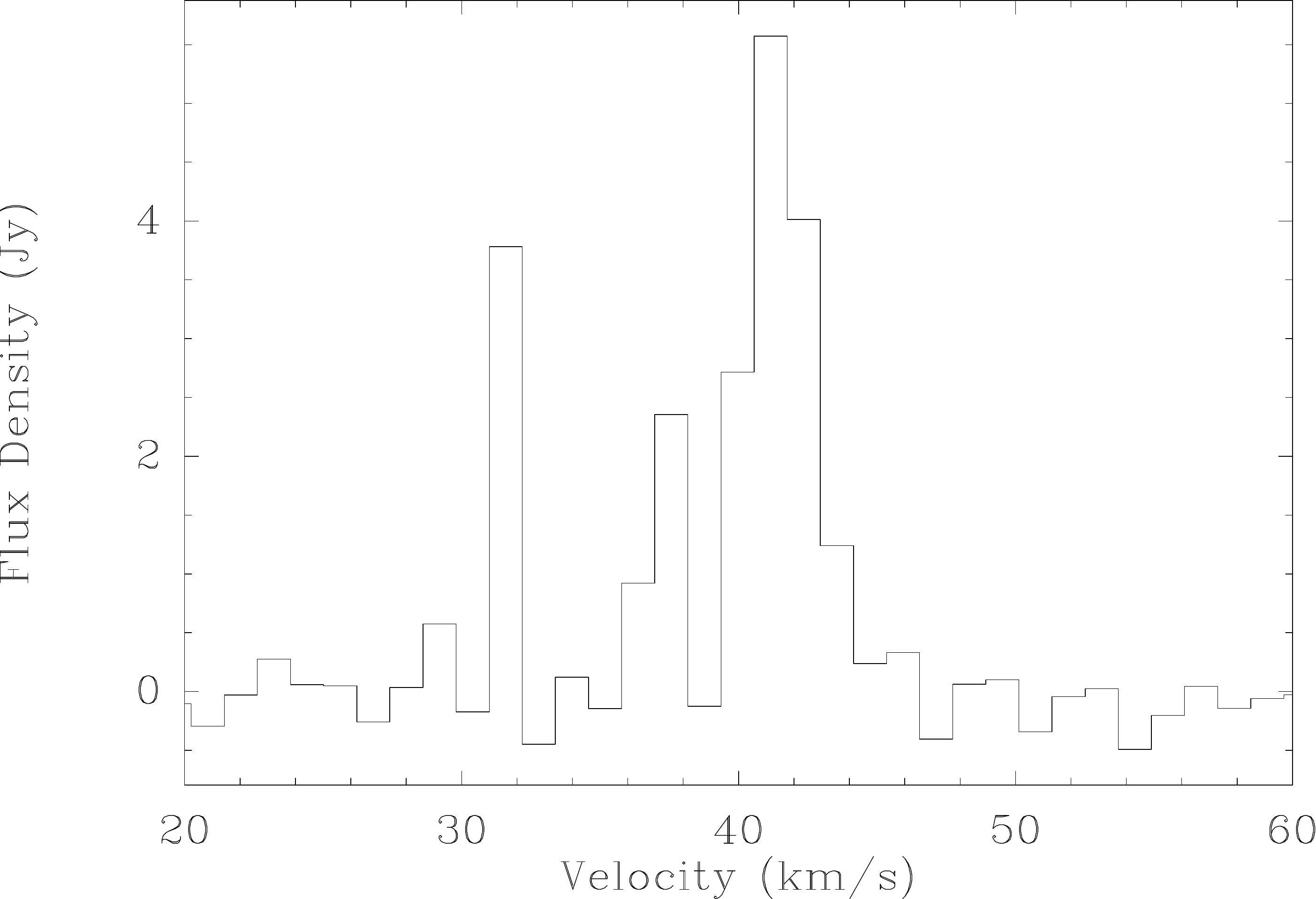}} \hspace{0.2cm}	
	\subfloat[SMA+IRAM30m $^{13}$CO spectrum]{\includegraphics[width=8.5cm]{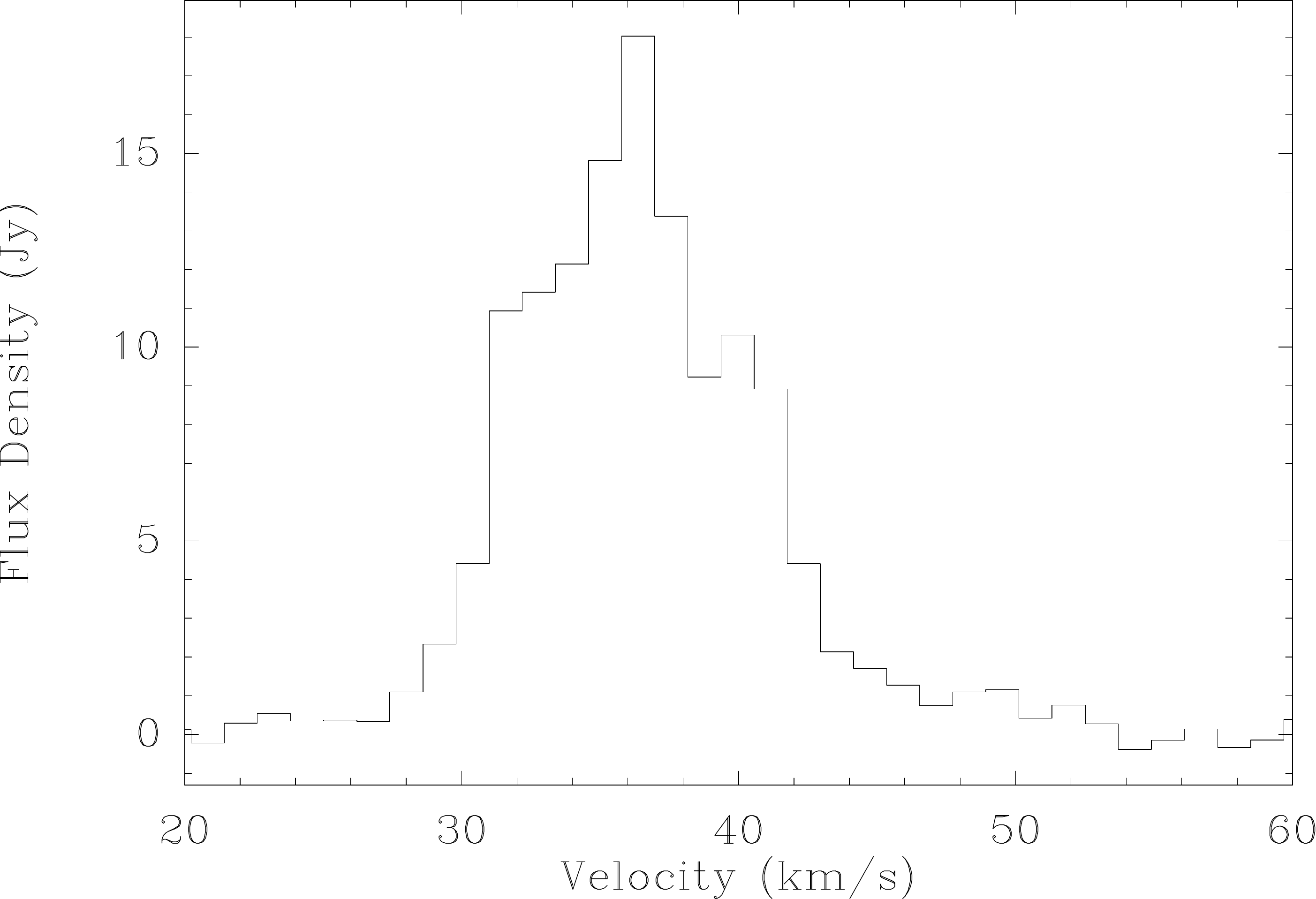}} \hspace{0.2cm}	
	\label{W33M_COMB_CO_Spectra}
\end{figure*}

\clearpage

\section{Weeds modelling}
\label{WeedsApp}

\paragraph{APEX observations}

A good fit of the 46~K H$_{2}$CO line in W33\,Main1 is achieved with a temperature of 40~K and a column density of 3.4~$\cdot$~10$^{13}$~cm$^{-2}$. The synthetic spectrum slightly underestimates the 35 and 141~K transitions and slightly overestimates the two 82~K transitions.

In W33\,A1, the synthetic spectrum, based on a kinetic temperature of 40~K and a column density of 4.1~$\cdot$~10$^{13}$~cm$^{-2}$, gives a good fit to the 46~K transition. As in W33\,Main1, the emission of the 35~K and the 141~K lines is a bit underestimated and the emission of the 82~K lines a bit overestimated. A slightly better fit for all transitions is achieved if we fit the two components with 25~K and 2.0~$\cdot$~10$^{13}$~cm$^{-2}$ and 55~K and 2.5~$\cdot$~10$^{13}$~cm$^{-2}$. 

The synthetic spectrum, computed from the RTD results of W33\,B1, is a good fit for the transitions with E$_{u}$ above 80~K but underestimate the emission of the 35~K and 46~K lines. We conclude that a low-temperature component is missing in the construction of the synthetic spectrum. Thus, a new synthetic spectrum was produced, based on two components with 30~K and 2.0~$\cdot$~10$^{13}$~cm$^{-2}$ and 60~K and 2.5~$\cdot$~10$^{13}$~cm$^{-2}$. The synthetic spectrum yields a good fit for the 46~K and 141~K lines but underestimates the emission of the 35~K transition and slightly overestimates 
the emission of the 82~K line.

In W33\,B, we produce a synthetic spectrum with two components of 30~K and 2.3~$\cdot$~10$^{13}$~cm$^{-2}$ and 100~K and 5.9~$\cdot$~10$^{13}$~cm$^{-2}$. This spectrum reproduces the transitions of the upper sideband well but underestimates the emission of the 46~K line.
The synthetic spectrum for W33\,A, constructed from two components of 40~K and 9.0~$\cdot$~10$^{13}$~cm$^{-2}$ and 100~K and 1.6~$\cdot$~10$^{14}$~cm$^{-2}$, yields a good fit for the high-energy and the 35~K transitions but overestimates the emission of the 46~K line by about 25\%.

\begin{figure}
\caption{Observed (black) and synthetic (red) spectra of CH$_{3}$CCH in W33\,Main.}
	\centering
	\includegraphics[width=8.5cm]{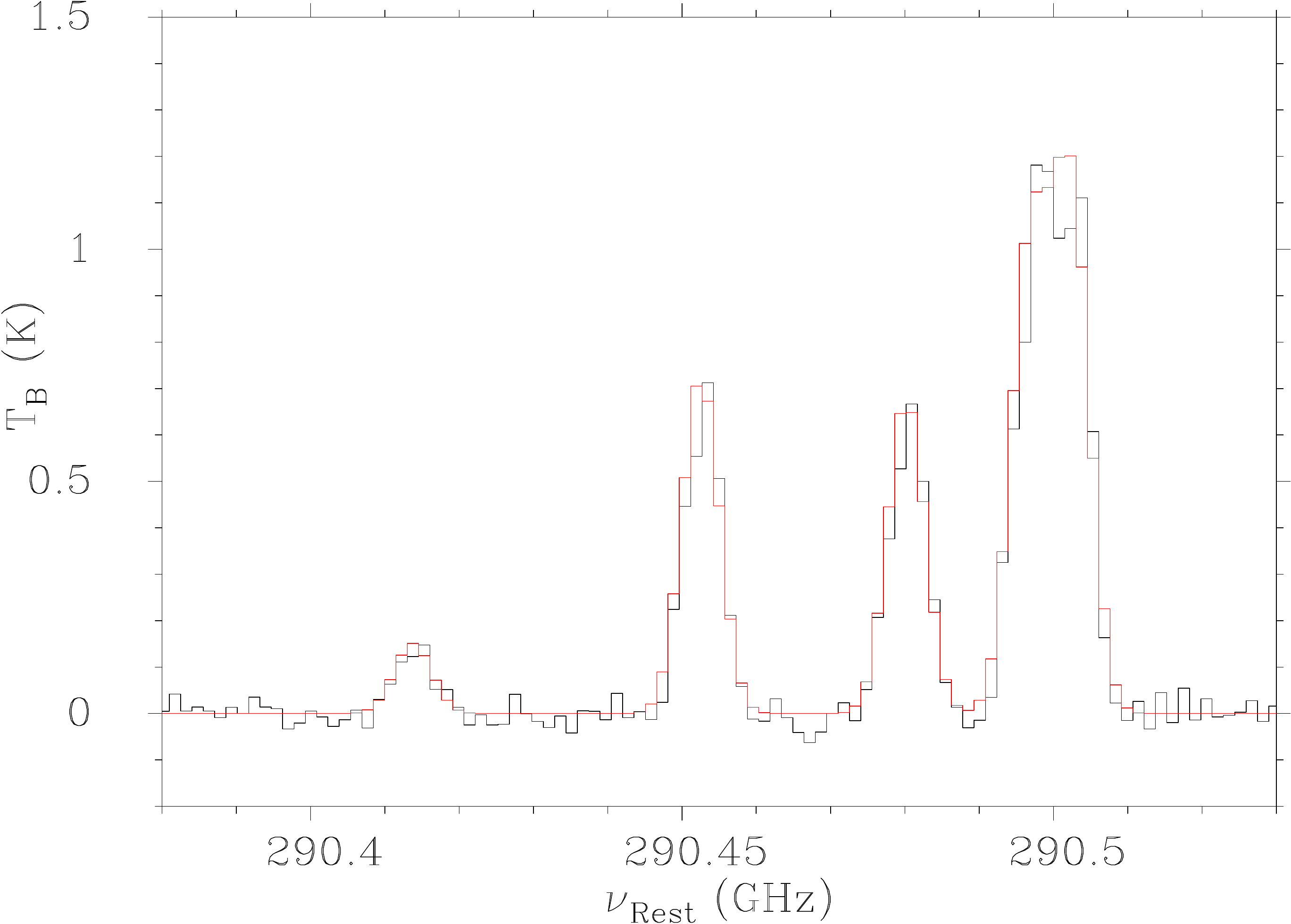}	
	\label{W33M_CH3CCH}
\end{figure}

Two components with 50~K and 2.0~$\cdot$~10$^{14}$~cm$^{-2}$ and 100~K and 4.2~$\cdot$~10$^{14}$~cm$^{-2}$ give a synthetic spectrum for W33\,Main, which fits all transitions except the 46 K very well. However, the emission of the 46~K line is overestimated by about 33\%.
In W33\,Main, we also produced synthetic spectra for CH$_{3}$OH and CH$_{3}$CCH. The RTD results of CH$_{3}$CCH give a synthetic spectrum that underestimates the emission of all detected transitions. A better fit is achieved using a kinetic temperature of 59~K and a column density of 3.4~$\cdot$~10$^{15}$~cm$^{-2}$ for the construction of the synthetic spectrum. The observed and the synthetic spectrum of CH$_{3}$CCH are shown in Fig. \ref{W33M_CH3CCH}.
For CH$_{3}$OH, we find a fairly good fit of the transitions for one component with 40~K and 1.3~$\cdot$~10$^{15}$~cm$^{-2}$. However, the synthetic spectrum underestimates the emission of CH$_{3}$OH(9$_{\textnormal{$-$1,9}}$$-$8$_{\textnormal{0,8}}$). 

\paragraph{SMA observations}

We tried to compute synthetic spectra from one component for the H$_{2}$CO transitions in W33\,A1, W33\,B1, and W33\,B but it was not possible to fit all three transitions well. The synthetic spectra show that the peak line ratio between H$_{2}$CO(3$_{\textnormal{2,1}}$$-$2$_{\textnormal{2,0}}$) and H$_{2}$CO(3$_{\textnormal{2,2}}$$-$2$_{\textnormal{2,1}}$) should be close to 1, and the peak of the H$_{2}$CO(3$_{\textnormal{0,3}}$$-$2$_{\textnormal{0,2}}$) line is larger than the peaks of the other two transitions. The peak line ratios H$_{2}$CO(3$_{\textnormal{2,1}}$$-$2$_{\textnormal{2,0}}$)/H$_{2}$CO(3$_{\textnormal{2,2}}$$-$2$_{\textnormal{2,1}}$) in W33\,A1 and W33\,B1 are 0.6 and 2.0. In W33\,B, the line ratio is close to 1, but the peak of H$_{2}$CO(3$_{\textnormal{0,3}}$$-$2$_{\textnormal{0,2}}$) is smaller than the peaks of the other two transitions. This indicates that the assumption of a single temperature relating the level populations is not justified for H$_{2}$CO in all three sources.

We constructed a synthetic spectrum for HNCO in W33\,B for one component with the RTD results. The synthetic spectrum underestimates the emission of all transitions. A good fit of the observed spectrum is achieved with 280~K and 2.5~$\cdot$~10$^{15}$~cm$^{-2}$. However, the emission of the transition at 220.58 GHz is slightly overestimated, while the emission of the transition at 218.98 GHz is slightly underestimated. Through comparison with the synthetic spectrum, we found two additional transitions of HNCO at 219.66 GHz (HNCO(10$_{\textnormal{3,8}}$$-$9$_{\textnormal{3,7}}$), HNCO(10$_{\textnormal{3,7}}$$-$9$_{\textnormal{3,6}}$)), which are blended and a 2$\sigma$ detection in our data.

The synthetic spectrum of the CH$_{3}$CN emission in W33\,B, which was compiled from the RTD results, is only a good fit for the CH$_{3}$CN(12$_{\textnormal{3}}$$-$11$_{\textnormal{3}}$) transition. A better fitting synthetic spectrum is constructed from a kinetic temperature of 280~K and a column density of 1.3~$\cdot$~10$^{15}$~cm$^{-2}$. However, the emission of the CH$_{3}$CN(12$_{\textnormal{3}}$$-$11$_{\textnormal{3}}$) transition is now overestimated. The comparison of synthetic and observed spectrum yield the identification of another transition of the CH$_{3}$CN ladder (CH$_{3}$CN(12$_{\textnormal{7}}$$-$11$_{\textnormal{7}}$)), which is a 2$\sigma$ detection in our data. 
The RTD results of CH$_{3}$OH in W33\,B yield a synthetic spectrum, which underestimates the emission of all CH$_{3}$OH transitions. The synthetic spectrum shows that at least two components are necessary to fit the observed spectrum. However, we do not find two combinations of temperatures and column densities that fit all transitions well.

\end{appendix}

\end{document}